\documentclass[12pt]{report}
\usepackage{setspace}
\usepackage[a4paper, left=35mm, top=25mm, right=25mm, bottom=25mm]{geometry}
\usepackage[utf8]{inputenc}
\usepackage{graphicx}
\usepackage{adjustbox}
\graphicspath{ {images/} }
\usepackage{pslatex}
\usepackage[T1]{fontenc}
\usepackage{epsfig}
\usepackage{longtable}
\usepackage{float}
\usepackage{calc}
\usepackage{ifthen}
\usepackage{amsmath}
\usepackage{cancel}
\usepackage{amssymb}
\usepackage{hyperref}
\usepackage{titling}
\usepackage[english]{babel}
\usepackage{lmodern}

\usepackage{color}

\usepackage{xcolor}

\def\eV{{\rm \ eV}}
\def\GeV{{\rm \ GeV}}
\def\MeV{{\rm \ MeV}}
\def\keV{{\rm \ keV}}

\def\Mpc{{\rm \ Mpc}}
\def\K{{\rm \ K}}
\def\kg{{\rm \ kg}}
\def\km{{\rm \ km}}
\def\s{{\rm \ s}}
\def\meV{{\rm \ meV}}

\makeatletter
    \def\thebibliography#1{\chapter*{References\@mkboth
      {REFERENCES}{REFERENCES}}\list
      {[\arabic{enumi}]}{\settowidth\labelwidth{[#1]}\leftmargin\labelwidth
	\advance\leftmargin\labelsep
	\usecounter{enumi}}
	\def\newblock{\hskip .11em plus .33em minus .07em}
	\sloppy\clubpenalty4000\widowpenalty4000
	\sfcode`\.=1000\relax}
    \makeatother

\begin{document}

\onehalfspacing

\title{ 
{Aspects of Inflation and Cosmology in Non-Minimally Coupled and $R^{2}$ Palatini Gravity}\\
\vspace{1cm}
{\large Lancaster University \\ Physics Department}\\
\vspace{1cm}
{\includegraphics[clip = true, width=0.5\textwidth]{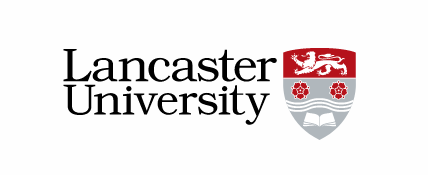} }
}

\author{Amy Lloyd-Stubbs}
\date{August 2022 \endgraf \bigskip
This thesis is submitted for the degree of Doctor of Philosophy. \endgraf}

\maketitle

\thispagestyle{plain}
\begin{center}

\Large Aspects of Inflation and Cosmology in Non-Minimally Coupled and $R^{2}$ Palatini Gravity

\vspace{0.4cm}
\large Amy Lloyd-Stubbs
       
\vspace{0.9cm}
\large Abstract

\end{center}

\vspace{0.5cm}

This thesis presents research exploring aspects of inflation and cosmology in the context of inflation models in which an inflaton is non-minimally coupled to the Ricci scalar, or is considered in conjunction with a term quadratic in the Ricci scalar. 
We consider a $\phi^{2}$ Palatini inflation model in $R^{2}$ gravity and investigate whether this model can overcome some of the problems of the original $\phi^{2}$ chaotic inflation model. We investigate the compatibility of this model with the observed CMB when treated as an effective theory of inflation in quantum gravity by examining the constraints on the model parameters arising due to Planck-suppressed potential corrections and reheating. Additionally, we consider two possible reheating channels and assess their viability in relation to the constraints on the size of the coupling to the $R^{2}$ term.
We present an application of the Affleck-Dine mechanism, in which quadratic $B$-violating potential terms generate the asymmetry, with a complex inflaton as the Affleck-Dine field. We derive the $B$ asymmetry generated in the inflaton condensate analytically and numerically. We use the present-day asymmetry to constrain the size of the $B$-violating mass term and derive an upper bound on the inflaton mass in order for the Affleck-Dine dynamics to be compatible with non-minimally coupled inflation in the metric and Palatini formalisms. The baryon isocurvature fraction generated in this model is also examined against observational constraints.
We demonstrate the existence of a new class of inflatonic Q-balls in a non-minimally coupled Palatini inflation model, through an analytical derivation of the Q-ball equation and numerical confirmation of the existence of solutions, and derive a range of the inflaton mass squared within which the model can inflate and produce Q-balls. We derive analytical estimates of the properties of these Q-balls, explore the effects of curvature, and discuss observational signatures of the model.

\pagenumbering{roman}

\tableofcontents

\addcontentsline{toc}{chapter}{List of Tables}
\listoftables

\addcontentsline{toc}{chapter}{List of Figures}
\listoffigures

\addcontentsline{toc}{chapter}{Acknowledgements}
\chapter*{Acknowledgements}
Firstly I would like to thank my supervisor Dr John McDonald for always answering my questions, engaging in all the discussions of cosmological subtleties I started, for continuously passing on references which would prove to be very useful someday, and generally for being a great supervisor throughout my PhD.

I would like to thank the staff and students of the Theoretical Particle Cosmology group at Lancaster who have been a part of the group during my studies, thank you for all of the post-seminar and office discussions and for being a supportive environment. My thanks also go to the Physics Department at Lancaster University for allowing me to complete my PhD in such a great department, and for all of the opportunities for training and teaching experience I was able to have while studying.

While undertaking my PhD I was supported by a Science and Technology Facilities Council (STFC) postgraduate studentship, without which I would not have been able to do the research I have done.

To anyone who ever asked me a question about my work at a conference, or answered my questions about theirs, thank you. It was an honour talking to you.

Additional thanks must be given to Dr David Sloan and Professor Ed Copeland for all of your excellent questions and interesting discussion during the exam for this PhD, and for making the viva experience a great dissemination exercise - I look forward to talking to you both about research again, hopefully many more times. Thank you also to Dr David Burton for overseeing the exam.

\addcontentsline{toc}{chapter}{Declaration}
\chapter*{Declaration}
This thesis is my own work and no portion of the work referred to in this thesis has been submitted in support of an application for another degree or qualification at this or any other institute of learning.

\chapter{Introduction}

\pagenumbering{arabic}

This thesis focuses on the cosmology of the early Universe, namely inflation and the embedding of inflation into the Hot Big Bang model of cosmology. Additionally we explore how a chosen model of inflation can be used as a basis for solving other problems in cosmology, or can exist in conjunction with other phenomena important to cosmology. It is important to consider this, as the consequences of an inflation model when incorporated into a more complete cosmological model can be far reaching, for both observational compatibility of the model, observability of the model, and the impacts of the physics of the model on the evolution of the Universe following inflation and other physical processes. 

In Chapter 2 of this thesis, we present an overview of Big Bang Cosmology, including inflation and the generation of the density perturbations following inflation. Chapter 3 presents some important general results used throughout Chapters 4-6 of the thesis from field theory, particle physics and non-minimally coupled inflation.

Chapters 4-6 present three pieces of original research. Chapter 4 presents a study of an inflation model in the Palatini formalism with a $\phi^{2}$ potential and an $R^{2}$ term in the gravitational part of the action. We present the slow-roll parameters and the inflationary observables in the Einstein frame and examine the observational compatibility of the model using three different physical constraints on the size of the $R^{2}$ term. Reheating in the model is also explored, for two specific reheating channels, and the constraints each reheating mechanism places on the model parameters are discussed.

In Chapter 5 we introduce a model of Affleck-Dine baryogenesis in the context of non-minimally coupled inflation with quadratic symmetry-breaking potential terms. We analytically derive expressions for the asymmetry generated in the inflaton condensate, and the asymmetry transferred to the Standard Model. We then test the analytical result numerically, and calculate the baron-to-entropy ratio generated in this model. The conditions for compatibility with non-minimally coupled inflation, isocurvature fluctuations, and the treatment of the asymmetry generated in this model in quantum terms are also discussed.

Chapter 6 presents a model of Q-balls within a non-minimally coupled inflation model in the Palatini formalism. The Q-ball equation is derived and constraints on the inflaton potential from inflation and the existence of Q-balls are used to derive a mass range for the inflaton compatible with the existence of Q-ball solutions. We solve the Q-ball equation both analytically and numerically and compare the predicted properties of the Q-balls. We also estimate what the effects of curvature on these Q-balls could be and explore possibilities for the formation of these Q-balls, as well as possible observational signatures of the model and the possible implications for these Q-balls on cosmology in the broader sense.

In Chapter 7 we present our conclusions.

\section{Connection to Published Work by the Author}

In the course of completing the research in this thesis, the results of the work were published.

The research presented in Chapter 4 resulted in the publication: 

\begin{itemize}

\item \textit{Sub-Planckian $\phi^{2}$ Inflation in the Palatini Formulation of Gravity with an $R^{2}$ Term}; Amy Lloyd-Stubbs and John McDonald; Physical Review D, Volume 101 (2020) 12, 123515; e-print: 2002.08324 [hep-ph].
\end{itemize}

The research presented in Chapter 5 resulted in the publication:

\begin{itemize}
\item \textit{A Minimal Approach to Baryogenesis via Affleck-Dine and Inflaton Mass Terms}; Amy Lloyd-Stubbs and John McDonald; Physical Review D, Volume 103 (2021), 123514; e-print: 2008.04339 [hep-ph],
\end{itemize}

with the exception of Figures \ref{figure:32} - \ref{figure:311} and the content of Sections \ref{section:37} and \ref{section:39} - \ref{section:310} which will form a part of a paper currently in progress.

The research presented in Chapter 6 resulted in the publication:
\begin{itemize}
\item \textit{Q-balls in Non-Minimally Coupled Palatini Inflation and their Implications for Cosmology}; A. K. Lloyd-Stubbs and J. McDonald; Physical Review D, Volume 105 (2022) 10, 103532; e-print: 2112.09121 [hep-th].
\end{itemize}

\section{Notation and Conventions}

Throughout this thesis, units where $\hbar = c = k_{B} = 1$ are used. Planck masses are left in explicitly, and should be taken as the reduced Planck mass $M_{pl}^{2} = 1/8\pi G$.

In Chapters 2 and 3, the mostly minus convention is used for the metric signature, $(+, -, -, -)$ unless specified otherwise. In Chapters 4-6 the metric signature used in each chapter is stated at the beginning of each individual chapter.

While care has been taken in this thesis to use different symbols for unrelated quantities, usage of the same symbol is sometimes unavoidable in places owing to convention within the fields of cosmology and particle physics, or for consistency with published work. Throughout this thesis symbols are defined in the text when their corresponding quantity is first introduced in the chapter, and any relation to other quantities with the same symbol should not be assumed unless directed.

\chapter{Cosmological Background}
In this chapter, we introduce the Big Bang Model of Cosmology as the basis for the research presented in this thesis, introduce inflation and the formation of the density perturbations following inflation, and discuss the conditions needed for baryogenesis to occur. 

\section{Big Bang Cosmology}\label{section:01}

In standard cosmology, the timeline of the evolution of the Universe begins with a singularity, after which the General Relativity description of gravity becomes applicable and we mark the beginning of cosmological time. The era prior to this is known as the Planck epoch, wherein temperatures exceeded the Planck energy $\sim 10^{19}\GeV$. The Universe evolved to the universe we observe today as it expanded and cooled over a period of 13.8 billion years, and the purpose of cosmology is to study how we reached the point we are at now, using the observations we now have access to.

By means of an introduction, we first present a timeline of the evolution of the Universe from the initial singularity to the present.

\subsection{Timeline of Cosmological Evolution}\label{section:011}

\begin{itemize}
\item Initial Singularity \, $\cdot$ \, "Big Bang" \, $t = 0-10^{-43}$s, $T \gtrsim 10^{19}$GeV, $T \gtrsim 10^{32}$K.

Taken to be the beginning of cosmological time, signifies the end of the Planck epoch and the transition to a regime in which General Relativity is a valid description of gravity.

\item  Inflation \, $\cdot$ \,  $t = 10^{-34}\s$.

An era of supercooled exponential expansion of space, in which the Universe expands by a factor of about $10^{28}$.

\item Reheating \, $\cdot$ \, $t > 10^{-34}\s$.

Phase of the end of inflation where the inflaton field decays and transfers its energy to the particles of the Standard Model. Transition from the inflaton field dominating the energy density to the Universe being filled with a thermal plasma of particles. Onset of epoch of radiation domination. Reheating temperature depends on the inflation model.

\item Electroweak Phase Transition \, $\cdot$ \,  $t = 2.0 \times 10^{-11}\s$,  $z = 10^{15}$,  $T = 159\GeV$, \\ $T = 1.9 \times 10^{15}\K$.

Weak interaction becomes significant. Higgs field develops an expectation value and the Higgs mechanism gives mass to particles.

\item Quark - Hadron Phase Transition \, $\cdot$ \,  $t = 2.0 \times 10^{-5}\s$, $z = 10^{12}$, $T = 150\MeV$, $T=1.7 \times 10^{12}\K$. 

Strong interaction becomes significant. Quarks and gluons combine into hadrons. Onset of confinement era, no free quarks at temperatures lower than this.  

\item Neutrino Decoupling \, $\cdot$ \, $ t = 1.0$s, $ z = 6.0 \times 10^{9}$, $T = 0.8\MeV$, $T=9.3 \times 10^{9}\K$.

Neutrinos decouple from the thermal plasma.

\item Electron - Positron Annihilation \, $\cdot$ \, $t = 6.0$s, $z = 2.0 \times 10^{9}$, $T = 500\keV$, \\ $T=5.8 \times 10^{9}\K$.

Electrons and positrons in the thermal plasma annihilate, and the energy from the annihilations is transferred to the photons still coupled to the thermal plasma. This "heats" the photons but not the neutrinos, which have already decoupled from the thermal plasma. Seed of the temperature of the photon background today, $T_{\gamma}$, being greater than the neutrino background temperature, $T_{\nu}$.

\item Big Bang Nucleosynthesis \, $\cdot$ \,  $t = 3.0$ minutes, $z = 4.0 \times 10^{8}$, $T = 100\keV$, \\ $T=1.16 \times 10^{9}\K$.

Formation of the nuclei of the light elements; namely H, $^{2}$H, $^{3}$He and $^{4}$He with trace amounts of higher proton number elements, following the freeze out of neutrons from equilibrium.

\item Matter - Radiation Equality \, $\cdot$ \, $t = 6.0 \times 10^{4}$ years,   $z = 3400$,  $T = 0.75\eV$, \\ $T=8706\K$. 

Transition of the Universe from an era of relativistic matter and radiation being the dominant component of the energy density to non-relativistic matter being the dominant component. 

\item Recombination \, $\cdot$ \, $t = (2.6 - 3.0) \times 10^{5}$ years, $z = 1200 - 1400$,  \\ $T = 0.30 - 0.33 \eV$, $T= 3500 - 3800\K$.

Formation of neutral hydrogen atoms from hydrogen nuclei and free electrons. Free electron density falls sharply and interactions between electrons and photons fall off. 

\item Decoupling \, $\cdot$ \, $t \simeq 3.5 \times 10^{5}$ years; $z \simeq 1100$, $T = 0.26\eV$, $T \simeq 3000\K$

Photons decouple from thermal equilibrium and begin freely streaming through the Universe. These photons form the Cosmic Microwave Background today. 

\item Reionisation \, $\cdot$ \,  $t = (1.0 - 4.0) \times 10^{8}$ years,  $z = 11 - 30$,  $T = 2.6 - 7.0 \meV$, $T=30 - 80\K$.

Neutral hydrogen left over from recombination is reionised by ultraviolet light from the first stars. The photon scattering from reionised hydrogen can contribute to temperature anisotropies in the sky. 

\item Dark Energy - Matter Equality \, $\cdot$ \,  $t = 9.0 \times 10^{9}$ years,  $z = 0.4$, $T = 0.33\meV$, $T=3.83\K$.

Transition from an era of non-relativistic matter being the dominant component of the Universe to an era of dark energy being the dominant contribution to the total energy density of the Universe. Epoch of accelerated expansion.

\item Present \, $\cdot$ \, $t = 1.38 \times 10^{10}$ years,  $z = 0$, $T = 0.24 \meV$, $T=2.79\K$.

The Universe as it is now. Epoch of vacuum energy dominated expansion.

\end{itemize}

\subsection{Homogeneity and Isotropy}\label{section:012}
In standard cosmology, we work to a set of rules which determine how we view the Universe and our place in it, referred to as the Cosmological Principle. These rules are namely that we treat the Universe as being isotropic and homogeneous on large scales. Homogeneity refers to the fact that matter is evenly distributed throughout the Universe when viewed on large scales, and the Universe therefore looks the same at each point. Isotropy refers to the fact that the Universe has no dynamical centre and therefore looks the same in all directions; there are no privileged observers. This combined with the assumption that the Universe is approximately smooth (average density of luminous matter is almost the same in every observed direction) on large scales comprises the Cosmological Principle. (Approximate large scale smoothness was confirmed by galaxy surveys 2dF Redshift Survey \cite{2df} and the Sloan Digital Sky Survey \cite{sdss}). The Universe is not entirely homogeneous and isotropic throughout its volume, but it is an implicit assumption when doing cosmological calculations that the Hubble volume we occupy can be considered to be isotropic, homogeneous and smooth.

\subsection{The Cosmic Microwave Background}\label{section:013}
The Cosmic Microwave Background (CMB) is microwave radiation which pervades all of the observable Universe today, discovered by radio astronomers in 1964 \cite{cmb} after the prediction of a 'relic radiation' from the Big Bang by \cite{gamow, alpher}, and interpreted as such by \cite{dprw}. This microwave radiation originates from the end of the epoch of recombination, when nuclei of the light elements and free electrons from the particle plasma formed neutral atoms for the first time. This caused interactions between photons and electrons - namely Thomson scattering - to fall off and for the photons to decouple from thermal equilibrium and start freely streaming through the Universe. It is at this point that the Universe became transparent to electromagnetic radiation, referred to as the surface of last scattering, beyond which the Universe is not observable from the perspective of Earth observers. The CMB has a temperature of $2.73 \K (2.14 \times 10^{-14}\GeV)$ today \cite{cmbtemp}, and has been shown by experiments COBE \cite{cobe}, WMAP \cite{wmapcmb} and later the Planck satellite \cite{planckcmb} to be isotropic to around one part in $10^{5}$.
\section{Hubble Expansion}\label{section:02}

In the 1920s, it was observed by Edwin Hubble that the Universe is expanding \cite{hubble}, and that galaxies further away were accelerating away at a faster rate. This relationship was parameterised using Hubble's Law:

\begin{equation}\label{eqn:1}
\textbf{v} = H \textbf{d},
\end{equation}

\noindent where $\textbf{v}$ is the recession velocity of the galaxy/distant object in question, $\textbf{d}$ is the proper distance from the observer to the object and the Hubble parameter $H$ evolves with time. Hubble's law is valid for distances of up to about $100\Mpc$ - beyond which the relationship becomes less well defined - and still applies today. The velocity of recession is directed along the distance vector $\textbf{r}$ in the direction of the proper distance to the observer

\begin{equation}\label{eqn:2}
\textbf{v} = \frac{d\textbf{r}}{dt}.
\end{equation}

\noindent In an expanding universe the proper distance from an observer to an object is given by $r = a\left(t \right) x$, where $a\left(t \right)$ is a quantity known as the scale factor which will be introduced more formally in Section \ref{section:03}, and $x$ corresponds to the comoving coordinates $x^{i} = \left( x^{1}, x^{2}, x^{2}\right)$. Comoving coordinates remain fixed in an expanding background (distances between objects or events remain constant over time), while proper coordinates are defined on the expanding space (distances increase with expansion). Equation (\ref{eqn:2}) can be written

\begin{equation}\label{eqn:3}
\textbf{v} = \frac{\left|\dot{\textbf{r}}\right|}{\left|\textbf{r}\right|}\textbf{r} = \frac{\dot{a}\left|\textbf{x}\right|}{a\left|\textbf{x}\right|}\textbf{r} = \frac{\dot{a}}{a}\textbf{r}.
\end{equation}

\noindent From Hubble's law (\ref{eqn:1}), this leads to the definition 

\begin{equation}\label{eqn:4}
H = \frac{1}{a}\frac{da}{dt},
\end{equation}

\noindent of the Hubble parameter in terms of the scale factor. From this definition, we can establish

\begin{equation}\label{eqn:5}
H = \frac{d\left(\ln a \right)}{dt} \Rightarrow a\left(t\right) \propto e^{\int H dt},
\end{equation}

\noindent as the time-dependence of the scale factor provided $\dot{a} > 0$.  

The Hubble parameter today is $H_{0} = 67.4 \km \s^{-1}\Mpc^{-1}$, as measured most recently by the Planck Satellite \cite{planck184}. It was originally discovered in the 1990s by two different research teams \cite{riess, perlmutter} studying Type 1a supernovae that the Universe is expanding now, and that the rate of expansion is accelerating. The data found in these studies was found to be consistent with the presence of a cosmological constant, $\Lambda$, which is referred to in the present era as Dark Energy.

Although we will use the value of the Hubble parameter today from the CMB predictions as the reference value throughout this thesis, there are conflicting observational values of $H_{0}$, between the measurements of the parameter from CMB experiments (see e.g. \cite{planck184} for the most recent result) and the measurement of the parameter from Type 1a supernovae data. More recent late-Universe measurements of $H_{0}$ from Cepheid variable data give a value of $H_{0} = 73.2 \pm 1.3 \km \s^{-1}\Mpc^{-1}$, which is a $4.2\sigma$ difference with the prediction on $H_{0}$ from the Planck experiment \cite{riess20}. This open problem in cosmology is known as the Hubble tension\footnote{Whether this tension in the measurements of $H_{0}$ is physically motivated is a current topic of research in cosmology, and the reduction of systematic effects in different methods of local measurement is an important facet of this. For detailed discussion of this see \cite{freedman21} and references therein. }.

\subsection{Horizons in Cosmology}\label{section:021}

There are a number of important definitions which are central to discussing cosmological history in relation to cosmological observations. Namely, these are the different types of horizon used in determining the connection between cosmological events. 

\noindent In order to discuss horizons in cosmology, it is convenient to use the definition of conformal time

\begin{equation}\label{eqn:6}
d\tau = \frac{dt}{a\left(\tau\right)},
\end{equation}

\noindent and comoving distance

\begin{equation}\label{eqn:7}
\chi = \int d\tau = \int \frac{dt}{a\left(\tau\right)}.
\end{equation}

\noindent In an isotropic expanding universe in comoving coordinates, the line element (introduced more formally in Section \ref{section:03}) is given by 

\begin{equation}\label{eqn:8}
ds^{2} = a^{2}\left( \tau \right) \left[ d\tau^{2} - d\chi^{2} \right].
\end{equation}

\noindent For lightlike (null) geodesics, $ds^{2} = 0$, and the path of the light photons is defined by

\begin{equation}\label{eqn:9}
\Delta \chi\left(\tau \right) = \pm \Delta \tau,
\end{equation}

\noindent where the sign convention $+ \left(- \right)$ corresponds to outgoing (incoming) photons from the perspective of the observer. Comoving coordinates therefore allow events to be placed at a precise location in space and time from the perspective of an observer in static space. 

\begin{itemize}

\item Particle Horizon

\noindent The particle horizon is defined as being the region within which past events can be observed by a given observer, or alternatively as the region of the past lightcone of an event or observer which causal influences come from. This is defined as

\begin{equation}\label{eqn:10}
\chi_{p}\left(\tau \right) = \tau - \tau_{i} = \int_{t_{i}}^{t} \frac{dt}{a\left(t\right)},
\end{equation}

\noindent where $\tau > \tau_{i}$ is some location in conformal time further in time from the spacelike surface $\tau = \tau_{i}$, corresponding to the Big Bang singularity, before which no signals can be observed by an observer in the present.

\item Event Horizon

\noindent The event horizon is defined as representing the furthest comoving distance an observer or event in the present can influence or observe events in the future. More formally

\begin{equation}\label{eqn:11}
\chi_{e}\left( \tau \right) = \tau_{f} - \tau = \int^{t_{f}}_{t} \frac{dt}{a\left(t\right)},
\end{equation}

\noindent where $\chi \left(\tau \right) > \chi_{e}\left(\tau \right)$. $\chi_{e}\left(\tau \right)$ defines a spatial surface beyond which an observer will never be able to observe events in, or receive signals from, the present. It denotes the region of the future light cone beyond the causal influence of the present.

\item Hubble Horizon

\noindent The Hubble horizon is defined as the distance of an object from a observer when it is expanding away at light speed at a given time. It can roughly be described as the region surrounding a particle within which it can receive signals or communicate with other particles in a given instant. The comoving Hubble horizon is defined as

\begin{equation}\label{eqn:12}
\chi_{H}\left(t \right) = \left(aH \right)^{-1},
\end{equation}

\noindent and can be approximated as the particle horizon at a given moment in time $t_{0}$. If the distance between two particles at the present time $t = t_{0}$, $d_{0}$, is greater than the Hubble horizon $\left(a\left(t_{0}\right)H\left(t_{0}\right)\right)^{-1}$, then the particles cannot communicate in the present but that may have been different in the past. If $d_{0}$ is greater than the particle horizon then the two particles have never been in causal contact and will never have been able to communicate.

\end{itemize}
\section{Friedmann-Lema\^{\i}tre-Robertson-Walker Spacetime}\label{section:03}
Expanding spacetime is described by the Friedmann-Lema\^{\i}tre-Robertson-Walker (FLRW) metric, 

\begin{equation}
ds^{2} = dt^{2} - a^{2}\left(t\right)g_{ij}dx^{i}dx^{j},
\end{equation}

\noindent given in polar coordinates by

\begin{equation}\label{eqn:13}
ds^{2} = dt^{2} - a^{2}\left(t\right)\left[\frac{dr^{2}}{1 - kr^{2}} + r^{2}d\theta^{2} + r^{2}\sin^{2}\theta d\phi^{2} \right],
\end{equation}

\noindent where $(r, \theta, \phi)$ are comoving coordinates. This spacetime can be visualised as spatial slices of constant $\left( r, \theta, \phi \right)$ at each point on the time axis, which are "counted" by the time coordinate. The factor $k$ can be chosen to be $+1$, $0$ or $-1$ depending on whether the universe in question has positive, zero, or negative curvature respectively (the choices of curvature are discussed in more depth in Section \ref{section:061}). The quantity $a\left(t\right)$ is known as the scale factor and parameterises the expansion of space. The scale factor can be treated as a dimensionless quantity normalised to $a(t_{0}) = a_{0} = 1$ in the present day, in which case $r$ has units of length and $k$ has units of inverse length squared, and we use this convention in this thesis.

\noindent In order to discuss the dynamics of the FLRW Universe, there are some important geometric quantities which need to be defined. The first of these is the Riemann curvature tensor

\begin{equation}\label{eqn:14}
R^{\sigma}_{\mu \rho \nu} = \partial_{\rho}\Gamma^{\sigma}_{\nu \mu} - \partial_{\nu}\Gamma^{\sigma}_{\rho \mu} + \Gamma^{\sigma}_{\rho \lambda}\Gamma^{\lambda}_{\nu \mu} - \Gamma^{\sigma}_{\nu \lambda}\Gamma^{\lambda}_{\rho \mu},
\end{equation}

\noindent which describes the curvature of each point on the spacetime manifold. It is built from the affine connection $\Gamma$, which in Riemannian geometry is given by the Levi-Civita connection

\begin{equation}\label{eqn:15}
\Gamma^{\sigma}_{\mu \rho} = \frac{1}{2}g^{\sigma \nu}\left[\partial_{\mu}g_{\nu \rho} + \partial_{\rho}g_{\mu \nu} - \partial_{\nu}g_{\mu \rho} \right],
\end{equation}

\noindent where the $g_{\mu \nu}$ are the metric tensor of the spacetime geometry. 

The non-zero components of the Levi-Civita connection (Christoffel symbols) in FLRW spacetime are \cite{kolbturner}

\begin{equation}\label{eqn:16}
\Gamma^{0}_{ij} = - \frac{\dot{a}}{a}g_{ij},
\end{equation}

\begin{equation}\label{eqn:17}
\Gamma^{i}_{0j} = \frac{\dot{a}}{a}\delta^{i}_{j},
\end{equation}

\noindent and 

\begin{equation}\label{eqn:18}
\Gamma^{i}_{j k} = \frac{1}{2}g^{il}\left[\partial_{j}g_{kl} + \partial_{k}g_{lj} - \partial_{l}g_{jk} \right].
\end{equation}

The Riemann tensor can be contracted on its first and third indices to give the Ricci tensor

\begin{equation}\label{eqn:19}
R^{\sigma}_{\mu \rho \nu} \rightarrow R^{\sigma}_{\mu \sigma \nu} \equiv R_{\mu \nu},
\end{equation}

\noindent which can be contracted using the metric tensor to give the Ricci scalar

\begin{equation}\label{eqn:20}
R_{\mu \nu} \rightarrow g^{\mu \nu}R_{\mu \nu} = R^{\nu}_{\nu} \equiv R.
\end{equation}

\noindent In FLRW spacetime, the non-zero components of the Ricci tensor are

\begin{equation}\label{eqn:21}
R_{00} = -\frac{3\ddot{a}}{a},
\end{equation}

\begin{equation}\label{eqn:22}
R_{ij} = -\left[\frac{\ddot{a}}{a} + 2\left(\frac{\dot{a}}{a}\right)^{2} + \frac{2k}{a^{2}} \right]g_{ij},
\end{equation}

\noindent and the Ricci scalar is

\begin{equation}\label{eqn:23}
R = -6\left[\frac{\ddot{a}}{a} + \left(\frac{\dot{a}}{a}\right)^{2} + \frac{k}{a^{2}} \right].
\end{equation}

\noindent These will be used in Section \ref{section:05}.

\section{Redshift} \label{section:04}
An important phenomenon used in cosmological observations is that of redshift, which is the process by which the wavelength of electromagnetic radiation may be shifted slightly in the spectrum to either a longer wavelength (red-shifted) or to a shorter wavelength (blue-shifted). 

In the classical description of redshift, the emitted light from distant objects is treated as plane waves. Let an object be at a distance $r = r_{1}$ from an observer at $r = 0$ emit a wave of light at a time $t = t_{1}$ which is received by an observer at $t = t_{0}$. The coordinate distance and time for lightlike geodesics $\left( ds^{2} = 0 \right)$ are related by

\begin{equation}\label{eqn:24}
\int^{t_{0}}_{t_{1}} \frac{dt}{a\left(t\right)} = \int^{r_{1}}_{0} \frac{dr}{\sqrt{1 - kr^{2}}}. 
\end{equation}

\noindent If a subsequent plane wave is emitted at $ t = t_{1} + \delta t_{1}$ and received at $ t = t_{0} + \delta t_{0}$, then in a comoving coordinate system we have that

\begin{equation}\label{eqn:25}
\int^{t_{0}}_{t_{1}} \frac{dt}{a\left(t\right)} = \int^{t_{0} + \delta t_{0}}_{t_{1} + \delta t_{1}} \frac{dt}{a\left(t\right)}
\end{equation}

\noindent due to the fact that the position of the source relative to the observer is fixed and the right hand side of (\ref{eqn:24}) is constant. If the time elapsed between emission and detection is the same for both waves, then we can say that the time delay between the first and second wave is also the same for both emission and detection, and we can write this as

\begin{equation}\label{eqn:26}
\int^{t_{1} +\delta t_{1} }_{t_{1}} \frac{dt}{a\left(t\right)} = \int^{t_{0} + \delta t_{0}}_{t_{0}} \frac{dt}{a\left(t\right)}.
\end{equation}

\noindent If $\delta t$ is a small timespan which measures the time between successive wavecrests we can say that it corresponds to the wavelength of the light, and that $a\left(t \right)$ is constant over the integration, giving 

\begin{equation}\label{eqn:27}
\frac{\delta t_{1}}{a\left(t_{1} \right)} = \frac{\delta t_{0}}{a\left(t_{0} \right)} \Rightarrow \frac{\lambda_{1}}{a\left(t_{1} \right)} = \frac{\lambda_{0}}{a\left(t_{0} \right)},
\end{equation}

\begin{equation}\label{eqn:28}
\Rightarrow \frac{\lambda_{0}}{\lambda_{1}} = \frac{a\left(t_{0} \right)}{a\left(t_{1} \right)} \equiv 1 + z,
\end{equation}

\noindent where the quantity $z$ is defined as the redshift of the photons. Since the redshift of an object is defined in terms of the ratio of the detected to the emitted wavelength, it is clear that an increase in the scale factor leads to an increase in the wavelength of the light from distant sources, and the greater the $a(t)$, the further away the source and the greater the redshift. 

In the case of treating light quantum mechanically, the light received by an observer as emitted by a distant object is treated as a stream of freely propagating photons. For radiation in quantised wavepackets, the wavelength of the light is inversely proportional to its momentum. From the geodesic equation of a lightlike trajectory- or more simplistically from (\ref{eqn:3}) - in FLRW spacetime, it can be shown that the three-momentum is inversely proportional to the scale factor, $ p \propto 1/a$. This means that the wavelengths of photons emitted at a source at time $t_{1}$, $\lambda\left(t_{1}\right) = \lambda_{1}$, and then received by the observer at a time $t_{0}$, $\lambda\left(t_{0}\right) = \lambda_{0}$ are given by

\begin{equation}\label{eqn:29}
\lambda_{1} = \frac{h}{p\left(t_{1}\right)}, \; \; \lambda_{0} = \frac{h}{p\left(t_{0}\right)}.
\end{equation}

\noindent Letting $p\left(t_{0,1}\right) = p_{0,1}$, the ratio of the emitted and observed wavelengths is

\begin{equation}\label{eqn:30}
\frac{\lambda_{0}}{\lambda_{1}} = \frac{h}{p_{0}}\frac{p_{1}}{h}.
\end{equation}

\noindent Since $p_{0} \propto 1/a\left(t_{0}\right), p_{1} \propto 1/a\left(t_{1}\right)$, this means that

\begin{equation}\label{eqn:31}
\frac{\lambda_{0}}{\lambda_{1}} = \frac{a\left(t_{0}\right)}{a\left(t_{1}\right)} = 1 + z,
\end{equation}

\noindent where the quantity $z$ is defined as the redshift of the photons. This shows that as the Universe expands, the wavelength of any freely propagating photon increases- as all proper length scales in an expanding universe do as the expansion progresses. The redshifting of light therefore occurs because the Universe expands between the light being emitted by its source and received by an observer.

\section{The Friedmann Equations}\label{section:05}

In order to consider the dynamics of the Universe in FLRW spacetime fully, we need the equations of motion of the cosmological model. The action of a physical theory is composed of the gravitational action and the "matter" action. In General Relativity, the gravitational action is given by the Einstein-Hilbert action \cite{einsteinhilbert}

\begin{equation}\label{eqn:32}
S_{EH} = -\int d^{4}x \sqrt{-g} \; \; \frac{1}{2}M_{pl}^{2}\left( R + 2\Lambda \right),
\end{equation}

\noindent where $g = det\left(g_{\mu \nu}\right)$ is the determinant of the metric tensor, $M_{pl}^{2}$ is the reduced Planck mass squared, $R$ is the Ricci scalar as defined in (\ref{eqn:20}) and $\Lambda$ is a cosmological constant term which we include for completeness. The matter part of the action is given by

\begin{equation}\label{eqn:33}
S_{matter} = \int d^{4}x \sqrt{-g} \; \; \mathcal{L}\left(\phi_{a}, \partial_{\mu}\phi_{a} \right),
\end{equation}

\noindent where $\phi_{a}$ is a sum over the matter fields of the theory.

The gravitational equations of motion of the theory can be derived by varying the action with respect to the spacetime metric, with solutions existing when the action is stationary under these variations $\delta S/\delta g_{\mu \nu} = 0$. Extremising the Einstein-Hilbert action in this way gives the Einstein equations 

\begin{equation}\label{eqn:34}
G_{\mu \nu} = 8\pi G T_{\mu \nu} + \Lambda g_{\mu \nu},
\end{equation}

\noindent where 

\begin{equation}\label{eqn:35}
G_{\mu \nu} = R_{\mu \nu} -  \frac{1}{2}R g_{\mu \nu},
\end{equation}

\noindent is the Einstein tensor. $T_{\mu \nu}$ is the energy-momentum tensor and this is defined by \cite{lythliddle}

\begin{equation}\label{eqn:36}
T^{\mu \nu} = 2\frac{\partial \mathcal{L}_{matter}}{\partial g_{\mu \nu}} - g^{\mu \nu}\mathcal{L}_{matter}.
\end{equation}

\noindent The Einstein tensor (left hand side of the Einstein equations) is generally recognised as describing the curvature of the spacetime, while the energy-momentum tensor (right hand side of the Einstein equations) is considered to describe the matter content of the Universe. 

\noindent The rules of FLRW spacetime impose certain constraints on the form that the energy-momentum tensor can take. Firstly, since the metric is symmetric and diagonal, the energy-momentum tensor itself must also be diagonal. This, and the assumption of isotropy, means that the off-diagonal components must be zero, so we have that 

\begin{equation}\label{eqn:37}
T^{0i} = T^{j0} = 0.
\end{equation}

\noindent Secondly, under the assumption of isotropy, the spatial components must be equal. The simplest realisation of an energy-momentum tensor with this form is that of a perfect fluid with energy density, $\rho(t)$, and pressure, $p(t)$, as seen by a comoving observer 

\begin{equation}\label{eqn:38}
T^{\mu}_{ \nu} = g^{\mu \lambda}T_{\lambda \nu} = diag\left( \rho, -p, -p, -p\right),
\end{equation}

\noindent where the non-zero elements more formally are given by

\begin{equation}\label{eqn:39}
T_{00} = \rho\left(t \right), \; \; \; T_{ij} = -p\left(t\right) g_{ij}\left(t, \textbf{x}\right),
\end{equation}

\noindent and $g_{ij}\left(t, \textbf{x}\right)$ is the spatial metric tensor.

In General Relativity, the conservation law for the energy-momentum tensor is given by

\begin{equation}\label{eqn:40}
\nabla_{\mu}T^{\mu}_{\nu} = \partial_{\mu}T^{\mu}_{\nu} + \Gamma^{\mu}_{\mu \lambda}T^{\lambda}_{\nu} - \Gamma^{\lambda}_{\mu \nu}T^{\mu}_{\lambda} =0,
\end{equation}

\noindent and the evolution of the energy density in the Universe is governed by the $\nu = 0$ equation from this conservation law. Using the fact that $T^{i}_{0}$ vanishes, the $\nu = 0$ equation is 

\begin{equation}\label{eqn:41}
\frac{d\rho}{dt} = \Gamma^{\mu}_{\mu 0}\rho - \Gamma^{\lambda}_{\mu 0}T^{\mu}_{\lambda} = 0.
\end{equation}

\noindent Using the symmetry of the Levi-Civita tensor on its second and third indices under zero torsion, we can use (\ref{eqn:17}) and find that the only surviving Christoffel symbols are the $\Gamma^{i}_{j 0}$ for $i = j$. Substituting $\Gamma^{i}_{i 0} = \dot{a}/a$ and $T^{i}_{i} = -p $ the $\nu = 0$ equation is

\begin{equation}\label{eqn:42}
\dot{\rho} + \frac{3\dot{a}}{a}\left( \rho + p \right) = 0.
\end{equation}

This equation can be solved using an equation of state solution, $p = w\rho$, where $w$ is a time independent parameter and its value changes depending on the nature of the fluid. This is a solution to the fluid equation (\ref{eqn:42}) if the energy density evolves as

\begin{equation}\label{eqn:43}
\rho \propto a^{-3\left(1 + w\right)}.
\end{equation}

\noindent In standard cosmology, there are three different types of fluid which may come to dominate the energy density of the Universe at a given point in its lifetime up until the present. These are 

\begin{enumerate}
\item Non-relativistic, pressureless matter - referred to as "matter".
\item Relativistic matter and radiation - referred to collectively as "radiation".
\item Vacuum energy/cosmological constant - referred to as "dark energy" or "$\Lambda$" in the context of the present state of the Universe.
\end{enumerate}

\noindent We will now explore the properties of each of these fluid types as solutions to the fluid equation.

\subsection{Matter}\label{section:051}

For the case of non-relativistic matter, in the equation of state we have that

\begin{equation}\label{eqn:44}
w = 0 \Rightarrow p = 0,
\end{equation}

\noindent which gives a fluid equation of 

\begin{equation}\label{eqn:45}
\dot{\rho} + \frac{3\dot{a}}{a}\rho = 0.
\end{equation}

\noindent This can be rewritten as

\begin{equation}\label{eqn:46}
\dot{\rho} + \frac{3\dot{a}}{a}\rho = \frac{1}{a^{3}}\frac{d}{dt}\left(\rho a^{3}\right) = 0 \Rightarrow \frac{d}{dt}\left(\rho a^{3}\right) = 0.
\end{equation}

\noindent Upon integration of this equation, we can see that $\rho a^{3} =$ constant and therefore

\begin{equation}\label{eqn:47}
\rho \propto \frac{1}{a^{3}}.
\end{equation}

\noindent This means that in a matter-dominated universe, the energy density decays away inversely proportional to the volume of the universe as it expands. This makes sense as, since $p = 0$, there is no pressure force from the fluid contributing to the expansion, and the fluid simply dilutes throughout the universe as its volume increases with time.

In both the fluid equation (\ref{eqn:42}) and the Friedmann equations (\ref{eqn:55}) - (\ref{eqn:57}), only the factor $\dot{a}/a$ appears in the equations provided $k=0$, so they are left unchanged if the scale factor is multiplied by a constant. As such, the energy density of a matter dominated universe is often normalised in terms of the energy density and scale factor today

\begin{equation}\label{eqn:48}
\rho = \rho_{0}\left(\frac{a_{0}}{a}\right)^{3},
\end{equation}

\noindent where the scale factor today is typically defined as $a_{0} = 1$, and this convention is followed throughout this thesis.

Note that "matter" in this context may refer to conventional matter as described by the Standard Model of particle physics, or it may refer to non-relativistic dark matter - Cold Dark Matter (CDM) - when discussing the matter components present in the Universe today. Dark matter, unlike conventional "luminous" matter does not reflect, emit or absorb electromagnetic radiation and is therefore difficult to detect but accounts for a significant fraction of the matter energy density of the Universe today (see Section \ref{section:061}).

\subsection{Radiation}\label{section:052}
In the case of a radiation dominated universe, we have that

\begin{equation}\label{eqn:49}
w = \frac{1}{3} \Rightarrow p = \frac{\rho}{3}.
\end{equation}

\noindent Substituting this into the fluid equation (\ref{eqn:42}), we have

\begin{equation}\label{eqn:50}
\dot{\rho} + \frac{4\dot{a}}{a}\rho = 0.
\end{equation}

\noindent This can be rewritten as 

\begin{equation}\label{eqn:51}
\dot{\rho} + \frac{4\dot{a}}{a}\rho = \frac{1}{a^{4}}\frac{d}{dt}\left(\rho a^{4}\right) = 0 \Rightarrow \frac{d}{dt}\left(\rho a^{4}\right) = 0.
\end{equation}

\noindent Integrating shows that $\rho a^{4} =$ constant, and that therefore

\begin{equation}\label{eqn:52}
\rho \propto \frac{1}{a^{4}} \Rightarrow \rho = \rho_{0}\left(\frac{a_{0}}{a}\right)^{4}.
\end{equation}

Radiation dilutes away faster with expansion than non-relativistic matter. This is because the positive pressure of radiation does work on the Universe as it expands, causing slower expansion and the radiation to lose energy with expansion faster than matter would.

\subsection{Vacuum Energy}\label{eqn:053}
In the case of a universe dominated by a cosmological constant, we have that

\begin{equation}\label{eqn:53}
w = -1 \Rightarrow p = -\rho,
\end{equation}

\noindent which gives a fluid equation of 

\begin{equation}\label{eqn:54}
\dot{\rho} = 0.
\end{equation}

\noindent The energy density is therefore constant in a universe dominated by a cosmological constant, and its negative pressure drives the expansion itself. In the most recent results from the Planck satellite (2018) \cite{planck184} it was found that the dark energy equation of state of the Universe was $w_{0} = -1.03 \pm 0.03$, consistent with the Universe currently being in an epoch of vacuum energy dominated expansion.

\subsection{The Friedmann Equation}\label{eqn:054}

The equations which govern the dynamics of the Universe can be derived by considering the Einstein equations in FLRW spacetime. Using the non-zero components of the Ricci tensor and the Ricci scalar given in (\ref{eqn:21}) - (\ref{eqn:23}), substituting into the Einstein equations (\ref{eqn:34}), and taking $g^{\mu \nu}T_{\mu \nu} = diag\left(\rho, -p, -p, -p \right)$, the $0-0$ component of the Einstein equations is 

\begin{equation}\label{eqn:55}
\left(\frac{\dot{a}}{a}\right)^{2} + \frac{k}{a^{2}} = \frac{8\pi G}{3}\rho + \frac{\Lambda}{3}.
\end{equation}

\noindent This is the first Friedmann equation, typically just referred to as the Friedmann equation. Taking the $i-i$ component of the Einstein equations in FLRW gives the second Friedmann equation

\begin{equation}\label{eqn:56}
\frac{2\ddot{a}}{a} + \left(\frac{\dot{a}}{a}\right)^{2} + \frac{k}{a^{2}} = -8\pi Gp + \Lambda,
\end{equation}

\noindent and subtracting (\ref{eqn:55}) from (\ref{eqn:56}) gives the acceleration equation

\begin{equation}\label{eqn:57}
\frac{\ddot{a}}{a} = -\frac{4\pi G}{3}\left(\rho + 3p\right) + \frac{\Lambda}{3}.
\end{equation}

Since $H = \dot{a}/a$, the Friedmann equation can be written in terms of the Hubble parameter

\begin{equation}\label{eqn:58}
H^{2} + \frac{k}{a^{2}} = \frac{8\pi G}{3}\rho + \frac{\Lambda}{3},
\end{equation}

\noindent as can the acceleration equation

\begin{equation}\label{eqn:59}
H^{2} + \dot{H} = -\frac{4\pi G}{3}\left(\rho + 3p\right) + \frac{\Lambda}{3}.
\end{equation}

\subsection{Time Dependence of the scale factor in different solutions of the Friedmann equation}\label{section:0541}

From the Friedmann equation, we can derive the time dependence of the scale factor for energy densities corresponding to the different types of matter. 

\begin{itemize}

\item Matter

\noindent For a flat universe $k = 0$ and an insignificant cosmological constant, the Friedmann equation for conventional matter energy density (\ref{eqn:48})

\begin{equation}\label{eqn:60}
\dot{a}^{2} = \frac{8\pi G}{3}\frac{a_{0}^{3}}{a}\rho_{0}.
\end{equation}

\noindent This equation can be solved by using a power law ansatz, $a \propto t^{q}$. Substituting this into (\ref{eqn:60}) and solving for $q$ we find that $q = 2/3$ and therefore for an energy density dominated by ordinary matter

\begin{equation}\label{eqn:61}
a \propto t^{\frac{2}{3}}.
\end{equation}

\noindent In terms of the Hubble parameter (\ref{eqn:4}) this means that

\begin{equation}\label{eqn:62}
H = \frac{2}{3t} \Rightarrow t = \frac{2}{3}H^{-1},
\end{equation}

\noindent in a matter dominated universe.

\item Radiation

\noindent Similarly for a radiation dominated universe, the Friedmann equation using (\ref{eqn:52}) is

\begin{equation}\label{eqn:63}
\dot{a}^{2} = \frac{8\pi G}{3}\frac{a_{0}^{4}}{a^{2}}\rho_{0},
\end{equation}

\noindent and using the same power law ansatz as we did for non-relativistic matter, we find that $q = 1/2$ and therefore

\begin{equation}\label{eqn:64}
a \propto t^{\frac{1}{2}},
\end{equation}

\noindent for a radiation dominated universe. The time dependence of the Hubble parameter is then

\begin{equation}\label{eqn:65}
H = \frac{1}{2t} \Rightarrow t = \frac{1}{2}H^{-1}.
\end{equation}

\item Vacuum Energy

\noindent For vacuum energy domination, $w = -1$, and using (\ref{eqn:43}) we find $\rho \propto a^{0}$ and is therefore constant. We have that in this case, the time dependence of the scale factor is 

\begin{equation}\label{eqn:66}
a \propto e^{\int H dt} \propto e^{Ht} \Rightarrow t = H^{-1}\ln \left(\frac{a}{a_{0}}\right).
\end{equation}

\end{itemize}

\section{Critical Energy Density of the Universe and Curvature}\label{section:06}

The Friedmann equation in terms of $H$ (\ref{eqn:58}) can be rewritten

\begin{equation}\label{eqn:67}
\frac{k}{H^{2} a^{2}} = \frac{8\pi G}{3H^{2}}\rho - 1.
\end{equation}

\noindent If we define the critical density of the Universe, $\rho_{c}$ to be

\begin{equation}\label{eqn:68}
\rho_{c} = \frac{3H^{2}}{8\pi G},
\end{equation}

\noindent and the ratio of the energy density to the critical density is given by

\begin{equation}\label{eqn:69}
\Omega = \frac{\rho}{\rho_{c}},
\end{equation}

\noindent then we can write (\ref{eqn:67}) as

\begin{equation}\label{eqn:70}
\frac{k}{H^{2} a^{2}} = \Omega - 1.
\end{equation}

\noindent This expression is valid for all times, although $\rho_{c}$ and $\Omega - 1$ are not constant and evolve as the Universe expands. The total energy density today $\Omega_{0}$ was measured to be $\Omega_{0} = 1.003 \pm 0.010$ by the Sloan Digital Sky Survey in 2006 \cite{sdss06}, so it has been confirmed observationally that the Universe is very close to critical density today. We discuss the implications of this in Sections \ref{section:061} and \ref{section:072}.

Since $\dot{a} \geq 0$ today, then $\left(aH\right)^{2} \geq 0$, and this means that the sign of $k$ affects the sign of $\Omega - 1$.

\subsection{Three Values of $k$}\label{section:061}
In FLRW spacetime, the parameter $k$ can take values of $+1$, $0$ or $-1$, depending on the curvature of the Universe.

\begin{itemize}
\item \textbf{Flat Geometry, $k=0$, $\Omega = 1$.} \\
$k=0$ corresponds to a flat universe, where the spatial geometry is three-dimensional Euclidean. Angles of a triangle add up to $180^{\circ}$ and the circumference of a circle of radius $r$ is given by $2\pi r$. Universe is infinite and expands in all directions.
\item \textbf{Spherical Geometry, $k = +1$, $\Omega > 1$.} \\
$k = +1$ corresponds to a closed universe of positive curvature, where the spatial geometry is spherical. Angles of a triangle add up to $>180^{\circ}$ and the circumference of a circle of radius $r$ is given by $< 2\pi r$. Observers in this kind of universe exist on the surface of the spatial three-sphere. Universe is finite but without a boundary, expands as a physical three-sphere of increasing radius.
\item \textbf{Hyperbolic Geometry, $k = -1$, $\Omega < 1$.} \\
$k = -1$ corresponds to an open universe of negative curvature, where the spatial geometry is hyperbolic. Angles of a triangle add up to $<180^{\circ}$ and the circumference of a circle of radius $r$ is given by $> 2\pi r$, parallel lines never meet. Universe can be visualised as a "saddleback".
\end{itemize}

\noindent To illustrate the interconnectedness of the $k$ parameter, the curvature of space and $\Omega$ we examine the spatial Ricci scalar in three dimensions, $^{(3)}R$ \cite{kolbturner}

\begin{equation}\label{eqn:71}
^{(3)}R = \frac{6k}{a^{2}} = 6H^{2}\left(\Omega - 1\right).
\end{equation}

\noindent The "radius of curvature" of the Universe is defined as \cite{kolbturner}

\begin{equation}\label{eqn:72}
r_{curv} = \frac{a\left(t\right)}{\sqrt{\left| k \right|}} = \sqrt{\frac{6}{\left| ^{(3)}R \right|}},
\end{equation}

\noindent and this can be written in terms of the Hubble parameter as

\begin{equation}\label{eqn:73}
r_{curv} = \frac{1}{H\sqrt{\left| \Omega - 1 \right|}}.
\end{equation}

\noindent From this expression for the radius of curvature of the Universe, we can see that for $\left| \Omega - 1 \right| \sim 1$, $r_{curv} \sim H^{-1}$, and for $\left| \Omega - 1 \right| << 1$, $r_{curv} >> H^{-1}$. This means that a universe close to critical density is very flat, and since we have $\left| \Omega - 1 \right| \sim 1$ today, it means that $\left| \Omega - 1 \right|$ must have been very small at early times and the Universe therefore must have been very near to critical density at early epochs. This means that it is safe to ignore spatial curvature when studying cosmology in the very early Universe.

In a generic universe of curvature $k$, the total energy density at a given time is 

\begin{equation}\label{eqn:74}
\rho =\frac{3H_{0}^{2}}{8\pi G}\left[\Omega_{\Lambda} + \Omega_{m}\left(\frac{a_{0}}{a}\right)^{3} + \Omega_{r}\left(\frac{a_{0}}{a}\right)^{4}\right],
\end{equation}

\begin{equation}\label{eqn:75}
\Rightarrow \frac{\rho}{\rho_{0,c}} = \Omega_{0} = \Omega_{\Lambda} + \Omega_{m}\left(\frac{a_{0}}{a}\right)^{3} + \Omega_{r}\left(\frac{a_{0}}{a}\right)^{4}.
\end{equation}

\noindent where $\Omega_{\Lambda}$ is defined as

\begin{equation}\label{eqn:76}
\Omega_{\Lambda} = \frac{\Lambda}{3H^{2}},
\end{equation}

\noindent corresponding to a cosmological constant energy density, $\rho_{\Lambda} = \Lambda/8\pi G$. The Friedmann equation at today's time is

\begin{equation}\label{eqn:77}
a_{0}^{2}H_{0}^{2} + k = \frac{8\pi G}{3}\rho a_{0}^{2},
\end{equation}

\noindent dividing through by $\left(a_{0}H_{0}\right)^{2}$ gives

\begin{equation}\label{eqn:78}
\frac{8\pi G}{3H_{0}^{2}}\rho = \frac{k}{a_{0}^{2}H_{0}^{2}} + 1.
\end{equation}

\noindent Given this and (\ref{eqn:74}), and taking $\rho = \rho + \rho_{\Lambda}$ to be the energy density of today, we can write 

\begin{equation}\label{eqn:79}
\Omega_{\Lambda, 0} + \Omega_{m, 0} + \Omega_{r, 0} + \Omega_{k, 0} = 1,
\end{equation}

\noindent where 

\begin{equation}\label{eqn:80}
\Omega_{k, 0} = - \frac{k}{a_{0}^{2}H_{0}^{2}}.
\end{equation}

From the 2018 results from the Planck satellite \cite{planck184} it was found that the matter density of the Universe is $\Omega_{m, 0} = 0.3166 \pm 0.0084$ ($\Omega_{m, 0}h^{2} = 0.315\pm 0.007$ ), of which the baryonic matter content is $\Omega_{b, 0}h^{2} = 0.0224 \pm 0.0001$ and the Cold Dark Matter (CDM) content is $\Omega_{c, 0}h^{2} = 0.120 \pm 0.001$, where the quantity $h$ is defined as the normalisation from the Hubble parameter today, defined as $H_{0}= 100 h \km \s^{-1} \Mpc^{-1}$. The dark energy density is $\Omega_{\Lambda, 0} = 0.6847 \pm 0.0073$. This means that the Universe today is about $30\%$ matter (luminous and dark) and $70\%$ dark energy, consistent with a vacuum energy dominated Universe.

\section{The Problems of the Hot Big Bang}\label{section:07}
There are a number of problems which arose during the conceptualisation of the Hot Big Bang model in the process of outlining a standard model of cosmological evolution, which were not explained by the model as it was at the time. These problems are as follows.
\subsection{The Horizon Problem}
The horizon problem refers to the issue of the uniformity of the Cosmic Microwave Background. As discussed in Section \ref{section:013} the Cosmic Microwave Background is a uniform temperature of $T_{CMB} = 2.4 \times 10^{-14}\GeV (2.725 \K)$ and is very nearly isotropic. Temperature uniformity across different regions of space is indicative of these regions of space at one point being in thermal equilibrium. In order to do so, the radiation in these currently causally disconnected regions of space would have to have interacted and thermalised.
The CMB was created at photon decoupling, after recombination, which means that the radiation from it has been travelling towards us since then. The fact that this has only just reached us means that, in the lifetime of the Universe, this radiation could not have travelled as far as the other side of the observable Universe during this time. In order to have interacted and thermalised enough to be near-isotropic in temperature, all regions of the observable Universe would need to have been in contact well before decoupling in order to establish thermal equilibrium and thus explain the temperature isotropy.
There is also the matter of the anisotropies in the CMB. At about one part in 100,000 the CMB is anisotropic, and contains small fluctuations first detected by the COBE satellite. These small fluctuations also pose another facet of the horizon problem. These irregularities could not be created within the CMB after these regions of space had thermalised, so it stands to reason that they must have been present to begin with. The model therefore required a mechanism which not only brought all regions of the observable Universe into contact well before decoupling, but also provided a means for the fluctuations in the CMB to be present at its formation.

\subsection{The Flatness Problem}\label{section:072}
In studies of Type 1a supernovae data in \cite{riess, perlmutter} and later studies of the CMB fluctuations from the BOOMERANG \cite{boomerang} and WMAP \cite{wmap5} experiments, it was shown that the curvature of the Universe is very close to being flat. Most recent observations from the Planck satellite combined with measurements from baryon acoustic oscillations (BAO) show the spatial curvature of the Universe to be $\Omega_{k, 0} = 0.001\pm 0.002$ \cite{planck184}. If we examine the Friedmann equation in terms of $\Omega$ (\ref{eqn:70}), and substitute the scale factor-time relations for matter and radiation (\ref{eqn:48}) and (\ref{eqn:52}) we find that 

\begin{equation}\label{eqn:81}
\left| \Omega_{tot} - 1\right| \propto t,
\end{equation}
\noindent for radiation, and 

\begin{equation}\label{eqn:82}
\left| \Omega_{tot} - 1\right| \propto t^{\frac{2}{3}},
\end{equation}
\noindent for ordinary matter. From this we can deduce that in a universe dominated by either matter or radiation the total density is an increasing function of time, and that the universe should move further away from critical density - and therefore further away from flat geometry - over its lifetime. 

\noindent In order for $\Omega$ to lie in the extremely narrow range it does today the Universe must have been extremely close to critical density at the beginning, but there is no obvious explanation as to why or how exactly the Universe has remained as close to flat geometry as it has while containing matter content which should drive it to increasing curvature over time. 

\subsection{The Relic Abundances Problem}\label{section:073}
In the majority of Grand Unified Theories of particle physics, heavy exotic objects such as magnetic monopoles are predicted, with energies $E \sim 10^{16}\GeV$. These have not been observed and standard cosmology as it was did not provide a mechanism for why this is the case, given that they should have been produced in abundance in the high temperatures of the early Universe. If they had been produced in the early Universe, even in a small amount, they should have come to dominate the Universe quickly after the radiation decayed away with expansion. It has been surmised that there must be some mechanism which diluted all of these massive, highly non-relativistic particles away very quickly before they could come to dominate the Universe, while also providing a possible explanation for why they have not been observed today.

\section{Inflation}\label{section:08}
A mechanism which solves all three of the aforementioned problems in the Hot Big Bang was first proposed by Alan Guth in 1981 \cite{guth} (extended by Linde in 1982 \cite{lindeinf} and originally proposed without the broader cosmological context by Starobinsky in 1979 \cite{starop1, starop2}), whereby the Universe initially expanded very rapidly by a large amount before continuing to expand through Hubble expansion as we understand it today. This initial epoch of exponential expansion was dubbed "inflation", and is defined as an epoch in which the scale factor was accelerating

\begin{equation}\label{eqn:83}
\ddot{a}\left(t\right) > 0 \Rightarrow \frac{d}{dt}\left(\dot{a}\right) >0 \Rightarrow \frac{d}{dt}\left(aH\right) > 0.
\end{equation}

\noindent From the acceleration equation (\ref{eqn:57}), in order for $\ddot{a} > 0$ to be possible, the following must be true

\begin{equation}\label{eqn:84}
\rho + 3p < 0 \Rightarrow p < -\frac{\rho}{3}, w < -\frac{1}{3},
\end{equation}

\noindent which shows that the energy density of the Universe must be composed of a fluid of negative pressure in order to drive expansion this rapid.

\noindent Inflation means that during the Planck epoch, all causally disconnected regions of space today were originally in causal contact, and were very rapidly separated during the era of exponential expansion. This allows that all regions of space we observe today had the same initial conditions, which allows for all regions of space which are outside of causal contact with each other today to have evolved to the same temperature, as well as providing an explanation for the large scale smoothness observed in the Universe today. This will be discussed in more detail in Section \ref{section:11}.

When referencing the fluid equation (\ref{eqn:42}), the Friedmann equations (\ref{eqn:55}) - (\ref{eqn:56}), and the acceleration equation (\ref{eqn:57}) from this point on in the thesis, it should be taken as implicit that the cosmological constant term is not included, and that $k=0$ unless it is included explicitly on page.

\noindent In addition to accelerated expansion and negative pressure, there are a number of other features which arise as a result of (\ref{eqn:83}) and (\ref{eqn:84}), and can also be used to constrain whether or not inflation persists in a model.

\begin{itemize}

\item Slowly-varying Hubble parameter

\noindent The condition for accelerated expansion (\ref{eqn:83}) can also be interpreted as a shrinking comoving Hubble horizon

\begin{equation}\label{eqn:85}
\frac{d}{dt}\left(aH\right)^{-1} < 0,
\end{equation}

\noindent where this condition can be written as

\begin{equation}\label{eqn:86}
\frac{d}{dt}\left(aH\right)^{-1} = -\frac{1}{a}\frac{\left( a \dot{H} + \dot{a}H\right)}{aH^{2}} = -\frac{1}{a}\left(1 - \epsilon \right) < 0,
\end{equation}

\noindent which is true if and only if $\epsilon < 1$, where we define the Hubble slow-roll parameter $\epsilon_{H}$ as

\begin{equation}\label{eqn:87}
\epsilon_{H} = -\frac{\dot{H}}{H^{2}}.
\end{equation}

\noindent This is an essential condition for inflation and will be referenced extensively throughout this chapter.

\item Quasi de-Sitter Expansion

\noindent If $\epsilon_{H} = 0$, the spacetime becomes de Sitter

\begin{equation}\label{eqn:88}
ds^{2} = dt^{2} - e^{2Ht}d\textbf{x}^{2},
\end{equation}

\noindent and the Universe continues accelerated expansion forever, so in order for inflation to end we need $\epsilon_{H} \neq 0$ and $\epsilon_{H}$ to be a small, finite number less than unity.

\item Constant Energy Density

\noindent Combining the fluid equation (\ref{eqn:42}) and the acceleration equation (\ref{eqn:57}) produces the following equation

\begin{equation}\label{eqn:89}
\frac{\dot{\rho}}{\rho H} = -3 \left(1 + \frac{p}{\rho} \right),
\end{equation}

\noindent and $\epsilon_{H}$ can be rewritten using the Friedmann equations (\ref{eqn:55}), (\ref{eqn:56}) to derive the condition

\begin{equation}\label{eqn:90}
\epsilon_{H} = -\frac{\dot{H}}{H^{2}} = \frac{3}{2}\left(1 + \frac{p}{\rho} \right) < 1.
\end{equation}

\noindent Combining (\ref{eqn:90}) with (\ref{eqn:89}) we have that

\begin{equation}\label{eqn:91}
\frac{\dot{\rho}}{\rho H} \equiv 2\epsilon_{H}.
\end{equation}

\noindent Rewriting the left hand side, we can write (\ref{eqn:91}) as

\begin{equation}\label{eqn:92}
\left|\frac{d\ln \rho}{d\ln a}\right| = 2\epsilon_{H} < 1.
\end{equation}

\noindent This shows that provided $\epsilon_{H}$ is small, the energy density can be regarded as essentially constant during accelerated expansion. It also shows that the negative pressure fluid dominating the Universe throughout inflation cannot be conventional matter, since conventional matter dilutes away with expansion as $\rho \propto a^{-3}$. The nature of the dominant energy density during inflation will be discussed in the next section.
\end{itemize}

\subsubsection{Interpretation of $\epsilon_{H}$ and $\eta_{H}$.}\label{section:080}

From (\ref{eqn:87}), we can write the condition for inflation as 

\begin{equation}\label{eqn:93}
\epsilon_{H} = -\frac{d\ln H}{dN} < 1,
\end{equation}

\noindent where we have defined

\begin{equation}\label{eqn:94}
dN = d\ln a = H dt,
\end{equation}

\noindent to be the change in the number of e-foldings (commonly abbreviated to "e-folds" in cosmology, this thesis will follow this convention from here) of accelerated expansion, such that $a \propto e^{N}$. The condition (\ref{eqn:93}) implies that the fractional change of the Hubble parameter per e-fold of inflation must be small in order to produce inflation ($\epsilon < 1$).

We will demonstrate in Section \ref{section:11} that inflation needs to persist for around 60 e-folds to solve the horizon problem. This means that $\epsilon_{H}$ must be kept smaller than one for at least as much time as necessary to achieve this.

\noindent In order to measure the change in $\epsilon_{H}$ we introduce a second Hubble slow-roll parameter

\begin{equation}\label{eqn:95}
\eta_{H} = \frac{\dot{\epsilon_{H}}}{H\epsilon_{H}} = \frac{d\ln \epsilon_{H}}{dN}.
\end{equation}

\noindent From this condition we can say that for $\left| \eta_{H} \right| < 1$, the fractional change of $\epsilon_{H}$ per e-fold is kept small and inflation persists. To summarise, $\epsilon_{H} < 1$ is the necessary condition for the Hubble parameter to vary sufficiently slowly, and for the energy density to comprise a fluid of negative pressure ($w < -1/3$), which allow accelerated expansion of space (inflation) to occur. $\left| \eta_{H} \right| < 1$ is the necessary condition to ensure that $\epsilon_{H}$ remains less than one long enough for there to be sufficient inflation to solve the horizon problem.

\subsection{Inflation Driven by a Scalar Field}\label{section:081}

In this subsection we explore the nature of the fluid constituting the energy density of the Universe during inflation. In order to produce accelerated expansion, we showed in Section \ref{section:08} that the energy density of the Universe must be dominated by a fluid of negative pressure. During inflation then, we must have that the matter content of the Universe is composed of such a fluid. 

\noindent In order to evaluate this in relation to the FLRW dynamics of the Universe, we introduce a scalar field varying in space and time $\phi \left(t, \textbf{x}\right)$ with a potential 
$V(\phi )$. This field is theorised to drive inflation and is known as the inflaton. In order for the inflaton field to be consistent with FLRW spacetime we require that the energy-momentum tensor of the inflaton matches with that of the perfect isotropic fluid required for FLRW cosmology.

The inflaton field Lagrangian may be written as a general scalar field Lagrangian with potential $V(\phi)$

\begin{equation}\label{eqn:96}
\mathcal{L} = \frac{1}{2}g^{\mu \nu}\partial_{\mu}\phi \partial_{\nu}\phi - V\left(\phi \right).
\end{equation}

\noindent in order to produce a massless scalar field theory comprised of scalar quanta of $\hbar\omega$ when the theory is quantised.

\noindent The energy-momentum tensor of a scalar field is given by 

\begin{equation}\label{eqn:97}
T_{\mu \nu} = \partial_{\mu}\phi \partial_{\nu}\phi - g_{\mu \nu}\left(\frac{1}{2}g^{\alpha \beta}\partial_{\alpha}\phi \partial_{\beta}\phi - V\left(\phi \right)\right),
\end{equation}

\noindent where $T^{\mu \nu}$ is derived for a scalar field in flat space ($g^{\mu \nu} = \eta^{\mu \nu}$) in Chapter 3, Section \ref{section:t34}. To match with the energy-momentum tensor of the perfect isotropic fluid of FLRW cosmology, we require that the $00$ component of (\ref{eqn:97}) must be $T^{0}_{0} = \rho(t)$, and that the spatial components must be $T^{i}_{j} = -p(t)\delta^{i}_{j}$ in accordance with (\ref{eqn:39}). Calculating these components of (\ref{eqn:97}), we find that

\begin{equation}\label{eqn:98}
T^{0}_{0} = \frac{1}{2}\dot{\phi}^{2} + V\left(\phi \right),
\end{equation}

\begin{equation}\label{eqn:99}
T^{i}_{j} = \frac{1}{2}\dot{\phi}^{2} - V\left(\phi \right),
\end{equation}

\noindent where in both calculations, the terms $\sim \left(\overrightarrow{\nabla}\phi \right)^{2}/a^{2}$ have been left out of both final results. This is because these gradient terms damp as $a^{-2} \propto e^{-2Ht}$ and will therefore become insignificant very quickly with expansion. In order for the inflaton field to drive the FLRW dynamics throughout inflation we therefore require the inflaton energy density and pressure to be

\begin{equation}\label{eqn:100}
\rho\left(t \right) = \frac{1}{2}\dot{\phi}^{2} + V\left(\phi \right),
\end{equation}

\begin{equation}\label{eqn:101}
p\left(t \right) = \frac{1}{2}\dot{\phi}^{2} - V\left(\phi \right),
\end{equation}

\noindent respectively. Comparing these to the negative energy condition (\ref{eqn:84}), we require that

\begin{equation}\label{eqn:102}
p = \frac{1}{2}\dot{\phi}^{2} - V\left(\phi \right) < -\frac{1}{6}\dot{\phi}^{2} - \frac{1}{3}V\left(\phi \right) = -\frac{\rho}{3},
\end{equation}

\noindent and we establish that 

\begin{equation}\label{eqn:103}
\dot{\phi}^{2} < V\left(\phi \right),
\end{equation}

\noindent leads to inflation. In other words the inflaton potential must dominate over its kinetic energy in order for the scalar field to act as the negative pressure fluid driving inflation.

Using (\ref{eqn:100}) we can rewrite the Friedmann equation (\ref{eqn:58}) as

\begin{equation}\label{eqn:104}
H^{2} = \frac{1}{3M_{pl}^{2}}\left[ \frac{1}{2}\dot{\phi}^{2} + V\left(\phi \right) \right],
\end{equation}

\noindent and using (\ref{eqn:100}) and (\ref{eqn:101}), the acceleration equation (\ref{eqn:59}) can be written as

\begin{equation}\label{eqn:105}
\dot{H} = -\frac{\dot{\phi}^{2}}{2M_{pl}^{2}}.
\end{equation}

\noindent Taking a time derivative of (\ref{eqn:104}), we obtain

\begin{equation}\label{eqn:106}
2H\dot{H} = \frac{1}{3M_{pl}^{2}}\left[ \ddot{\phi}\dot{\phi} + \frac{d V}{d \phi}\dot{\phi} \right].
\end{equation}

\noindent Substituting (\ref{eqn:105}) and dividing through by $\dot{\phi}$ we obtain

\begin{equation}\label{eqn:107}
\ddot{\phi} + 3H\dot{\phi} + \frac{d V}{d \phi} = 0.
\end{equation}

\noindent This is the classical evolution equation of the scalar field in a FLRW universe, also known as the "Klein-Gordon equation" of the scalar field. This is also the field equation governing the dynamics of the inflaton field which can be obtained by variation of the Einstein equations in FLRW spacetime with the matter content of the Universe dominated by the inflaton, or by conservation of the inflaton energy-momentum tensor. 

Examining the equation (\ref{eqn:107}), we can see that the expansion of the Universe provides friction through the Hubble term $3H\dot{\phi}$ and that the potential term $d V/d \phi$ acts like a force term. 

\subsection{Slow-Roll Inflation}\label{section:082}

In this section we will examine the conditions on inflation in light of the inflaton field driving the expansion. Using the definition of $\epsilon_{H}$ (\ref{eqn:87}) and (\ref{eqn:105}), we can write

\begin{equation}\label{eqn:108}
\epsilon_{H} = \frac{\dot{\phi}^{2}}{2M_{pl}^{2}H^{2}}.
\end{equation}

\noindent In order for inflation to occur we therefore require that

\begin{equation}\label{eqn:109}
\epsilon_{H} < 1 \Rightarrow \frac{1}{2}\dot{\phi}^{2} < M_{pl}^{2}H^{2} \Rightarrow \frac{1}{2}\dot{\phi}^{2} < \rho = 3M_{pl}^{2}H^{2},
\end{equation}

\noindent from the Friedmann equation (\ref{eqn:58}). This reiterates the point that in order for the inflaton to drive accelerated expansion, its kinetic energy must make a small contribution to the overall energy density. In order for inflation to proceed for sufficiently long, this must remain true for the duration, which requires that the acceleration of the inflaton field $\ddot{\phi}$ must be small, i.e $\dot{\phi}$ cannot increase significantly with time during the inflation period.

As a measure of this, we define the dimensionless acceleration per Hubble time

\begin{equation}\label{eqn:110}
\delta = -\frac{\ddot{\phi}}{H\dot{\phi}}.
\end{equation}

\noindent Taking a time derivative of (\ref{eqn:108}), we obtain

\begin{equation}\label{eqn:111}
\frac{d\epsilon_{H}}{dt} = \frac{\dot{\phi}\ddot{\phi}}{M_{pl}^{2}H^{2}} - \frac{\dot{\phi}^{2}\dot{H}}{M_{pl}^{2}H^{3}}.
\end{equation}

\noindent Combining this with (\ref{eqn:87}), (\ref{eqn:95}) and (\ref{eqn:110}), we can write the $\eta_{H}$ parameter as

\begin{equation}\label{eqn:112}
\eta_{H} = 2\left(\delta - \epsilon_{H} \right).
\end{equation}

\noindent This relation shows that if the acceleration of the inflaton field is small, and subsequently the kinetic energy of the inflaton field makes up a small proportion of the inflaton energy density, we have that

\begin{equation}\label{eqn:113}
\lbrace \epsilon_{H}, \left| \delta \right| \rbrace << 1 \Rightarrow \lbrace \epsilon_{H}, \left| \eta \right| \rbrace << 1.
\end{equation}

\noindent These conditions dictate that the inflaton can drive accelerated expansion for a sufficiently long time provided that the field slowly rolls down its potential. These conditions can be used to simplify the equations governing the dynamics of the inflaton field, and the expansion of the Universe in what we will refer to as the slow-roll approximation.

Firstly, if $\epsilon_{H} <<1$, then the implication is that $\dot{\phi}^{2}/2 << V\left(\phi \right)$. This can be used to simplify the Friedmann equation (\ref{eqn:58}) to

\begin{equation}\label{eqn:114}
H^{2} = \frac{V\left(\phi \right)}{3M_{pl}^{2}},
\end{equation}

\noindent during slow-roll, and the expansion of the Universe is therefore completely driven by the potential energy of the inflaton field. 

\noindent Secondly, the condition 

\begin{equation}\label{eqn:115}
\left| \delta \right| = \frac{\ddot{\phi}}{H\dot{\phi}} << 1 \Rightarrow \ddot{\phi} << H\dot{\phi},
\end{equation} 

\noindent can be used to simplify the Klein-Gordon equation of the scalar field (\ref{eqn:107}) to give

\begin{equation}\label{eqn:116}
3H\dot{\phi} \simeq -\frac{dV}{d\phi}.
\end{equation}

\noindent So during slow-roll, the gradient of the inflaton potential is proportional to the "speed" at which the inflaton field rolls down it. Combining this expression with (\ref{eqn:114}) and (\ref{eqn:108}) gives an expression for $\epsilon_{H}$ in terms of the inflaton potential

\begin{equation}\label{eqn:117}
\epsilon = \frac{M_{pl}^{2}}{2}\left(\frac{V_{,\phi}}{V}\right)^{2}.
\end{equation}

A similar expression can also be obtained for the $\eta_{H}$ parameter. Taking the time derivative of (\ref{eqn:116}) and dividing through by $3H^{2}$ gives

\begin{equation}\label{eqn:118}
\frac{V''}{3H^{2}} = -\frac{\ddot{\phi}}{\dot{\phi}H} - \frac{\dot{H}}{H^{2}} = \delta + \epsilon_{H}.
\end{equation}

\noindent Using the slow-roll Friedmann equation (\ref{eqn:114}), we can define the left-hand side of (\ref{eqn:118}) to be a parameter $\eta$, where

\begin{equation}\label{eqn:119}
\eta = M_{pl}^{2} \frac{V''}{V}.
\end{equation}

\noindent The expressions (\ref{eqn:117}) and (\ref{eqn:119}) are defined as being the potential slow-roll parameters, while the originally derived expressions for $\epsilon_{H}$ and $\eta_{H}$ ((\ref{eqn:87}) and (\ref{eqn:95}) respectively) are generally referred to as the Hubble slow roll parameters, and successful slow-roll proceeds for $\lbrace \epsilon, \left| \eta \right| \rbrace << 1$. The Hubble slow-roll parameters can be related to the potential slow-roll parameters through

\begin{equation}\label{eqn:120}
\epsilon \approx \epsilon_{H}, \; \; \eta \approx 2\epsilon_{H} - \frac{1}{2}\eta_{H}.
\end{equation}

\subsection{How Much Inflation?}\label{section:083}

The total number of e-folds of inflation is given by

\begin{equation}\label{eqn:121}
N_{tot} = \int^{a_{E}}_{a_{I}} d\ln a = \int^{t_{E}}_{t_{I}} H\left(t \right) dt,
\end{equation}

\noindent where $t_{E}$ and $t_{I}$ are defined as the end and the beginning of inflation respectively, or more formally as $\epsilon\left(t_{E}\right) = \epsilon\left(t_{I}\right) = 1$. In the slow-roll regime we can write

\begin{equation}\label{eqn:122}
H dt = \frac{H}{\dot{\phi}} d\phi = \frac{-1}{\sqrt{2\epsilon_{H}}}\frac{d\phi}{M_{pl}} \approx \frac{-1}{\sqrt{2\epsilon}}\frac{d\phi}{M_{pl}},
\end{equation}

\noindent using (\ref{eqn:108}) and (\ref{eqn:116}). Substituting in (\ref{eqn:117}) we can write the integrand (\ref{eqn:122}) into (\ref{eqn:121}) as

\begin{equation}\label{eqn:123}
N_{tot} = -\frac{1}{M_{pl}^{2}}\int^{\phi_{E}}_{\phi_{I}} \frac{V}{V'} d\phi.
\end{equation}

The largest scales observed in the CMB are produced about 60 e-folds before the end of inflation. A successful solution to the horizon problem therefore requires at least 60 e-folds of inflation (demonstrated in Section \ref{section:11}).

\subsection{Reheating}\label{section:084}

We now briefly discuss the end of inflation. Slow-roll inflation ends when $\epsilon, \left|\eta\right| =1$, and at this point the inflaton field begins rapidly rolling down its potential. The field gains kinetic energy as it does so, and inflation ends with the inflaton field transferring this energy to the particles of the Standard Model. This is known as reheating, and leads to the thermalisation of the Universe, signifying the beginning of the Hot Big Bang within Big Bang cosmology.

The inflaton potential therefore needs to be fairly flat for long enough in field space that the inflaton can slowly roll along it for a sufficient number of e-folds. At the end of inflation, the potential steepens and the inflaton field rolls down to the minimum of its potential, losing potential energy and gaining kinetic energy before it begins to oscillate about the minimum of its potential. The inflaton then transfers its energy to the Standard Model particles through damped oscillations about the minimum of its potential. 

If the inflaton potential can be approximated as $V\left(\phi \right) \approx \frac{1}{2}m^{2}\phi^{2}$ close to the minimum of its potential - as the inflaton potentials discussed in Chapters 3-5 of this thesis can - then the equation of motion for the scalar field from (\ref{eqn:107}) is

\begin{equation}\label{eqn:124}
\ddot{\phi} + 3H\dot{\phi} = -m^{2}\phi, 
\end{equation}

\noindent at the end of slow roll, where $m$ is the inflaton mass. This reduces to undamped oscillations of frequency $m$ once the expansion scale of the Universe becomes larger than the oscillation period $H << m$, and we can neglect the friction term. 

Still approximating the potential near its minimum to be quadratic, the continuity equation for the inflaton at the end of inflation can be written as

\begin{equation}\label{eqn:125}
\dot{\rho} + 3H\rho = -3Hp = -\frac{3}{2}H\left(m^{2}\phi^{2} - \dot{\phi}^{2}\right),
\end{equation}

\noindent where at the end of inflaton we have $\epsilon \sim 1, \frac{1}{2}\dot{\phi}^{2} \sim V\left(\phi \right)$, so the right hand side of the equation averages to zero very quickly and we are left with the fluid equation for conventional matter (\ref{eqn:45}). The oscillating inflaton field therefore decays away like conventional matter $\rho \propto a^{-3}$, and the decay of the inflaton field therefore leads to a decrease in the amplitude of the oscillations of the field.

The inflaton is coupled to the Standard Model particles, and produces them as it decays. These particles then interact and create new particles until the Universe is filled with a plasma of particles, which comes to reach thermal equilibrium. The point at which the energy density of the Universe becomes dominated by relativistic particles after inflation is defined by the temperature $T_{R}$, which is referred to as the reheating temperature. The temperature at which the particle plasma thermalises varies widely depending on the inflation model, and depends on the energy density at reheating, $\rho_{R}$. We require at least $T_{R} \simeq 1 \MeV$, and for the Universe to be thermalised by this point, in order for Big Bang Nucleosynthesis to proceed following inflation. If the inflaton decays very quickly, then we can approximate that $\rho_{R} \sim V\left(\phi \right)$ at the end of inflation as all of the energy density of the inflaton can be assumed to all be transferred to the Standard Model particles. This approximation is known as instantaneous reheating and will be referred to in Chapters 4-6. 

\section{Energy Density and Pressure}\label{section:09}

The energy density and pressure of the matter content of the Universe following inflation can be calculated from the phase space distributions of all of the species present. Upon reheating, the Universe is filled with a thermal plasma of all of the particle content of the Standard Model. Different particle species will fall out of equilibrium and become non-relativistic at different temperatures as the Universe cools and expands, namely once the rate of the particle interaction becomes smaller than the rate of expansion, $\Gamma < H$.

\noindent The total energy density and pressure as functions of temperature are therefore generally given as the energy density and pressure of the relativistic species in the plasma, since the energy density and pressure of the non-relativistic species ($m_{i} >> T$) is exponentially smaller than the energy density and pressure of the relativistic species ($m_{i} << T$) as calculated from the phase space distributions, so it is a good approximation of the total energy density and pressure of the thermal plasma \cite{kolbturner}. 

\noindent In this limit we have that \cite{kolbturner}

\begin{equation}\label{eqn:126}
\rho \left(T \right) = \frac{\pi^{2}}{30}g_{\ast} T^{4},
\end{equation}

\begin{equation}\label{eqn:127}
p \left(T \right) = \frac{\rho}{3} = \frac{\pi^{2}}{90}g_{\ast} T^{4},
\end{equation}

\noindent where

\begin{equation}\label{eqn:128}
g_{\ast} = \sum_{i = bosons} g_{i} \left(\frac{T_{i}}{T}\right)^{4} + \frac{7}{8} \sum_{i = fermions} g_{i} \left(\frac{T_{i}}{T}\right)^{4},
\end{equation}

\noindent is the total number of relativistic degrees of freedom present in the particle plasma, the factor of $7/8$ accounts for the difference in bosonic and fermionic statistics, and $T$ is the photon temperature at the time of calculation. At the end of inflation, all species of the Standard Model are relativistic and we have that $g_{\ast} = 106.75$ \cite{kolbturner}.

\section{Conservation of Entropy During Expansion}\label{section:10}
The Universe is in local thermal equilibrium for much of its history, from the end of inflation until decoupling, which means that the entropy in a comoving volume remains constant. The second law of thermodynamics can be written as

\begin{equation}\label{eqn:129}
T dS = d\left(\rho V \right) + pdV = d\left[\left(\rho + p\right)V \right] - Vdp,
\end{equation}

\noindent where $V \propto a^{3}$, and $\rho$ and $p$ are the energy density and pressure at thermal equilibrium. Using the fact that

\begin{equation}\label{eqn:130}
T\frac{dp}{dT} = \rho + p,
\end{equation}

\noindent (\ref{eqn:129}) can be written as

\begin{equation}\label{eqn:131}
dS = \frac{1}{T}d\left[\left( \rho + p \right)V \right] - \frac{\left( \rho + p \right)V }{T^{2}}dT = d\left[\frac{\left( \rho + p \right)V }{T} + constant \right].
\end{equation}

\noindent Up to a constant we therefore have that the entropy is

\begin{equation}\label{eqn:132}
S = \frac{a^{3}\left(\rho + p \right)}{T}.
\end{equation}

\noindent The first law of thermodynamics is 

\begin{equation}\label{eqn:133}
d\left[\left(\rho + p\right)V \right] = Vdp,
\end{equation}

\noindent and applying this to (\ref{eqn:129}) we find that $dS = 0$, and equivalently

\begin{equation}\label{eqn:134}
d\left[\frac{\left( \rho + p \right)V }{T} \right] = 0,
\end{equation}

\noindent meaning that the entropy of the Universe in thermal equilibrium in a given expanding volume is conserved, and that this is adiabatic expansion. 

The entropy density can be defined as 

\begin{equation}\label{eqn:135}
s = \frac{S}{V} = \frac{\rho + p}{T}.
\end{equation}

\noindent Since the energy density and pressure at equilibrium are dominated by relativistic species, this can be written as \cite{kolbturner}

\begin{equation}\label{eqn:136}
s = \frac{2\pi^{2}}{45}g_{\ast s}T^{3},
\end{equation}

\noindent where

\begin{equation}\label{eqn:137}
g_{\ast s} = \sum_{i = bosons} g_{i} \left(\frac{T_{i}}{T}\right)^{4} + \frac{7}{8} \sum_{i = fermions} g_{i} \left(\frac{T_{i}}{T}\right)^{4}.
\end{equation}

\noindent If the particle species have a common temperature at a given time in the history of the Universe then $g_{\ast s}$ is equivalent to $g_{\ast}$.

From (\ref{eqn:134}), the conservation of entropy implies that $s \propto a^{-3}$, which implies that $g_{\ast s}a^{3}T^{3} = constant$ as the Universe expands. This implies that $T \propto g_{\ast s}^{-\frac{1}{3}}a^{-1}$ or $T \propto 1/a$ if $g_{\ast s}^{-\frac{1}{3}}$ is constant as the Universe expands, meaning that the Universe cools when it expands.

\section{How Inflation Solves the Problems of the Hot Big Bang}\label{section:11}

In this section we will discuss how inflation solves the problems of the Hot Big Bang detailed in Section \ref{section:07}. 

\begin{itemize}
\item Horizon Problem

\noindent Inflation solves the horizon problem by providing a mechanism for the Universe to start off much smaller than it is now, in a smooth initial state, and expand to its present size. A smooth initial state requires causal contact between all regions of space in the Universe, and this therefore enables the causally disconnected regions of space today to initially be in causal contact, and evolve from the same set of initial conditions. This allows all space we observe today to thermalise well before decoupling, and explain the temperature isotropy of the CMB today.

\noindent In order to gauge the amount of inflation needed to solve the horizon problem we use the Hubble horizon, as it is a more conservative estimate than the particle horizon since $\chi_{p}\left(t \right) > \left(aH \right)^{-1}\left(t \right)$. In order to solve the horizon problem then, the observable Universe today must have at least been able to fit inside the comoving Hubble horizon at the beginning of inflation

\begin{equation}\label{eqn:138}
\left(a_{0}H_{0}\right)^{-1} < \left(a_{I}H_{I}\right)^{-1}.
\end{equation}

\noindent By means of a simplification, for the purposes of an estimate, we make the approximation that the Universe has been radiation dominated for most of its history since the end of inflation. We therefore have that

\begin{equation}\label{eqn:139}
H = \frac{1}{2t} \Rightarrow a\left(t \right) = a_{0}\left(\frac{t}{t_{0}}\right)^{\frac{1}{2}} \Rightarrow t \propto a^{2}, H \propto \frac{1}{a^{2}}.
\end{equation}

\noindent We can take the ratio of the comoving Hubble radius of the observable Universe today and the comoving Hubble radius at the end of inflation ($a = a_{E}$)

\begin{equation}\label{eqn:140}
\frac{a_{0}H_{0}}{a_{E}H_{E}} = \frac{a_{0}}{a_{E}}\left(\frac{a_{E}}{a_{0}} \right)^{2} =  \frac{a_{E}}{a_{0}} \sim \frac{T_{0}}{T_{E}},
\end{equation}

\noindent since $T \propto 1/a$ during inflation. Using $T_{0} \sim 10^{-3}\eV (2.7 \K)$ as the temperature of the CMB today, and using $T_{E} \sim 10^{15}\GeV$ as an estimate of the temperature at the end of inflation, we have that $T_{0}/T_{E} \sim 10^{-28}$. This means that 

\begin{equation}\label{eqn:141}
\left(a_{I}H_{I}\right)^{-1} > \left(a_{0}H_{0}\right)^{-1} \sim 10^{28}\left(a_{E}H_{E}\right)^{-1},
\end{equation}

\noindent and so in order to solve the horizon problem, the comoving Hubble horizon must shrink by a factor of around $10^{28}$ during inflation. In other words the Universe must have been smaller by a factor of $10^{-28}$ before inflation in order for the causally disconnected regions of space today to have been in causal contact at early times. In order for this to happen, we approximate that the Hubble parameter remains constant during inflation and the scale factor increases exponentially. For $H_{I} \approx H_{E}$, we then have

\begin{equation}\label{eqn:142}
\frac{a_{E}}{a_{I}} > 10^{28} \Rightarrow \ln \left(\frac{a_{E}}{a_{I}}\right) = N_{E} - N_{I} > 64.
\end{equation}

\noindent This implies that we need at least 60 e-folds of accelerated expansion in order for inflation to solve the horizon problem.

\item The Flatness Problem

\noindent In Section \ref{section:06}, we rewrote the Friedmann equation in terms of the total energy density $\Omega_{tot}$ (\ref{eqn:70}), and established in Section \ref{section:072} that for matter and radiation dominated universes, $\left| \Omega_{tot} - 1 \right|$ is an increasing function of time ((\ref{eqn:81}) and (\ref{eqn:82}) respectively). For accelerated expansion we have

\begin{equation}\label{eqn:143}
\ddot{a} > 0 \Rightarrow \frac{d}{dt}\left(\dot{a}\right) = \frac{d}{dt}\left(aH\right) > 0.
\end{equation}

\noindent This means that the right-hand side of (\ref{eqn:70}) is driven increasingly towards zero. If $H$ is approximately constant during inflation then

\begin{equation}\label{eqn:144}
\frac{\left| k \right|}{\left(aH\right)^{2}} \sim e^{-2Ht} \Rightarrow \left| \Omega_{tot} - 1 \right| \rightarrow 0,
\end{equation}

\noindent as inflation progresses. Inflation therefore pushes $\Omega_{tot}$ very close to one, such that all subsequent expansion after the end of inflation is not sufficient to drive it away from one, and the Universe consequently to a more curved evolution. Inflation therefore naturally predicts a universe very close to flat geometry and provides a mechanism for which the Universe today can have very close to a flat geometry.

\item The Relic Abundances Problem

\noindent Inflation was originally surmised to solve the relic abundances problem by diluting all of the heavy exotic particles produced in the very early Universe $(T \sim 10^{16}\GeV)$ such that they have not been observed in the Universe so far today. However, reheating temperatures can easily reach those needed to produce any Grand Unified scale particles, such as magnetic monopoles. Even if these heavy relics were diluted away by inflation, it is therefore possible that more would be created by the decay of the inflaton, and therefore inflation does not necessarily explain the non-observation of these particles. A solution to the relic abundances problem given the possibility of production during reheating, such as the dilution of the abundance of magnetic monopoles by the galactic magnetic field \cite{lazarides}, is a current area of research in cosmology.
\end{itemize}

\section{Density Perturbations from Inflation}\label{section:12}

In addition to solving the problems of the Hot Big Bang, inflation provides a natural mechanism for producing the primordial seeds for the larger structures of the Universe today (Large Scale Structure (LSS)) and also the temperature anisotropies of the CMB. In this section we briefly discuss the production of the density perturbations during inflation.

If we treat the inflaton as a quantum field, we can write it as an approximately spatially constant classical background which experiences small quantum fluctuations in space and time due to the Uncertainty Principle

\begin{equation}\label{eqn:145}
\phi \left(t, \textbf{x}\right) = \bar{\phi}\left( t \right) + \delta \phi \left(t, \textbf{x}\right).
\end{equation}

\noindent This means that different patches of space inflate by slightly different amounts, which results in local differences in energy density $ \delta \rho\left(t, \textbf{x} \right)$ once inflation ends, and eventually fluctuations in the temperature of the CMB, $\delta T$. This introduces irregularities into the spacetime which we have thus far treated as smooth and homogeneous. The dynamics of such a Universe can be examined by perturbing the Einstein equations (\ref{eqn:34}) on a given spatial slice of spacetime, which introduces perturbations in the spacetime metric, $g_{\mu \nu} = \bar{g}_{\mu \nu} + \delta g_{\mu \nu}$, and in the stress-energy content of the Universe, $\rho \left(t, \textbf{x} \right) = \bar{\rho}\left(t \right) + \delta \rho\left(t, \textbf{x} \right)$. 

Perturbing the Einstein equations results in the emergence of a quantity known as the comoving curvature perturbation $\mathcal{R}$. We will briefly outline how this quantity is obtained, and how it relates to the fluctuations of the inflaton field. This discussion closely follows that given in \cite{baumann}.

The first step in perturbing the Einstein equations is to perturb the FLRW metric. We start in flat FLRW space in conformal time

\begin{equation}\label{eqn:146}
ds^{2} = a^{2}\left(\tau \right)\left[d\tau^{2} - \delta_{ij}dx^{i}dx^{j}\right],
\end{equation}

\noindent and then perturb

\begin{equation}\label{eqn:147}
ds^{2} = a^{2}\left(\tau \right)\left[\left( 1 + 2A \right)d\tau^{2} - 2B_{i}dx^{i}d\tau - \left(\delta_{ij} + h_{ij}\right)dx^{i}dx^{j}\right],
\end{equation}

\noindent where $A, B$ and $h_{ij}$ are functions of space and time. $A$ is a scalar, $B_{i}$ is a three vector which can be decomposed into a scalar part and a divergenceless vector

\begin{equation}\label{eqn:148}
B_{i} = \partial_{i}B + \hat{B}_{i}, \; \partial^{i}\hat{B}_{i} = 0,
\end{equation}

\noindent and $h_{ij}$ is a rank-2 symmetric tensor which can be decomposed into scalar, vector and tensor parts

\begin{equation}\label{eqn:149}
h_{ij} = 2C\delta_{ij} + 2\partial_{\left(i \right.}\partial_{\left. j \right)}E + 2\partial_{\left(i \right.}\hat{E}_{\left. j \right)} + 2\hat{E}_{ij},
\end{equation}

\noindent where, as defined in \cite{baumann},

\begin{equation}\label{eqn:150}
\partial_{\left(i \right.}\partial_{\left. j \right)}E = \left( \partial_{i}\partial_{j} - \frac{1}{3}\delta_{ij}\nabla^{2} \right)E,
\end{equation}

\begin{equation}\label{eqn:151}
\partial_{\left(i \right.}\hat{E}_{\left. j \right)} = \frac{1}{2}\left( \partial_{i}\hat{E}_{j} + \partial_{j}\hat{E}_{i} \right),
\end{equation}

\noindent and $\partial^{i}\hat{E}_{i} = 0$, $\partial^{i}\hat{E}_{ij} = 0$ and $\hat{E}^{i}_{i} = 0$. Physically, the scalar perturbations lead to the density perturbations we observe in the Universe today, the vector perturbations aren't produced by inflation, and the tensor perturbations lead to gravitational waves. The scalar perturbations are those which we are the most concerned with when discussing inflation, and this discussion will therefore be limited to the generation of the scalar perturbations in the energy density of the Universe.

In order to simplify the treatment of the perturbations, we consider the induced metric on spatial slices of constant time

\begin{equation}\label{eqn:152}
\gamma_{ij} = a^{2}\left[ \left( 1 + 2C\right)\delta_{ij} + 2E_{ij} \right],
\end{equation}

\noindent which is the spatial part of (\ref{eqn:147}) and $E_{ij} =\partial_{\left(i \right.}\partial_{\left. j \right)}E$. The three-dimensional Ricci scalar on these spatial constant-time slices is  \cite{baumann}

\begin{equation}\label{eqn:153}
a^{2}R^{\left(3\right)} = -4\nabla^{2}\left( C - \frac{1}{3}\nabla^{2}E \right),
\end{equation}

\noindent where the terms in the bracket constitute what is known as the curvature perturbation. The comoving curvature perturbation \cite{baumann} is given by

\begin{equation}\label{eqn:154}
\mathcal{R} = C - \frac{1}{3}\nabla^{2}E  + \mathcal{H}\left( B + v \right),
\end{equation}

\noindent where $v$ is the comoving three-velocity and $\mathcal{H}$ is the comoving Hubble parameter. In the spatially flat gauge $C = E = 0$ and we have

\begin{equation}\label{eqn:155}
\mathcal{R} = \mathcal{H}\left( B + v \right).
\end{equation}

\noindent In order to derive an expression for the comoving curvature perturbation in terms of the fluctuations of the inflaton field, we examine the first-order off-diagonal contributions to the perturbed energy momentum tensor $\delta T^{0}_{j}$, the momentum fluxes. For a perfect fluid these are \cite{baumann}

\begin{equation}\label{eqn:156}
\delta T^{0}_{j} = - \left(\bar{\rho} + \bar{P}\right)\left(B_{j} + v_{j}\right),
\end{equation}

\noindent where $v_{j}$ is the coordinate velocity of the fluid. For a scalar field these are \cite{baumann}

\begin{equation}\label{eqn:157}
\delta T^{0}_{j} = g^{0\mu}\partial_{\mu}\phi \partial_{j}\delta \phi = \bar{g}^{00}\partial_{0}\bar{\phi}\partial_{j}\delta \phi = \frac{\bar{\phi}'}{a^{2}}\partial_{j}\delta \phi.
\end{equation}

\noindent Equating (\ref{eqn:156}) and (\ref{eqn:157}), we find that

\begin{equation}\label{eqn:158}
\left( B + v \right) = -\frac{\delta \phi}{\bar{\phi}'},
\end{equation}

\noindent and substituting this into (\ref{eqn:155}), we find that the comoving curvature perturbation relates to the fluctuations of the inflaton field by

\begin{equation}\label{eqn:159}
\mathcal{R} =  -\frac{\mathcal{H}}{\bar{\phi}'}\delta \phi,
\end{equation}

\noindent where this quantity is conserved on superhorizon scales.

We now briefly outline how the comoving curvature power spectrum is obtained. This outline is based on the derivation presented in \cite{baumann}. We start by defining a comoving field $f$, which relates to the fluctuations of the inflaton field through

\begin{equation}\label{eqn:160}
f = a\delta \phi. 
\end{equation}

\noindent This field obeys the Mukhanov-Sasaki equation

\begin{equation}\label{eqn:161}
f'' - \overrightarrow{\nabla}^{2}f - \frac{a''}{a}f = 0,
\end{equation}

\noindent which is derived from extremising the quadratic action in $f$ of the inflaton action in conformal time in de Sitter space. In terms of the Fourier modes of $f$, this is

\begin{equation}\label{eqn:162}
f_{k}'' + \left(k^{2} - \frac{a''}{a}\right)f_{k} = 0.
\end{equation}

\noindent On small scales (sub-horizon), the inflaton fluctuations can be modelled by a collection of harmonic oscillators. The $f$ field can therefore be canonically quantised, and written as a mode expansion in creation and annihilation operators

\begin{equation}\label{eqn:163}
\hat{f}\left(\tau, \textbf{x}\right) = \int \frac{d^{3}k}{\left(2\pi\right)^{\frac{3}{2}}}\left[ f_{k}(\tau)\hat{a}_{\textbf{k}} + f^{\ast}_{k}(\tau)\hat{a}^{\dagger}_{\textbf{k}}\right]e^{i\textbf{k}\cdot\textbf{x}}.
\end{equation}

\noindent In the limit that the modes of interest are deep inside the horizon, the Mukhanov-Sasaki equation (\ref{eqn:162}) reduces to

\begin{equation}\label{eqn:164}
f_{k}'' + k^{2}f_{k} \approx 0,
\end{equation}

\noindent which has the solutions $f_{k} \propto e^{\pm ik\tau}$, corresponding to a free field in Minkowski space. In order for the solution to match with the conventional field theory vacuum - with the vacuum state $\left| 0 \rangle \right.$ corresponding to the ground state of the Hamiltonian - we can only choose the positive frequency solution $f_{k} \propto e^{-ik\tau}$. This then defines the vacuum state for the modes during inflation on length scales much smaller than the horizon

\begin{equation}\label{eqn:165}
lim_{\tau \rightarrow -\infty}f_{k}\left(\tau \right) = \frac{1}{\sqrt{2k}}e^{-ik\tau},
\end{equation}

\noindent known as the Bunch-Davies vacuum. During slow-roll inflation, the exact solution to the Mukhanov-Sasaki equation is

\begin{equation}\label{eqn:166}
f_{k}\left(\tau \right) = \frac{e^{-ik\tau}}{\sqrt{2k}}\left( 1 - \frac{i}{k\tau}\right).
\end{equation}

\noindent This mode function is completely fixed by the initial condition (\ref{eqn:165}), and therefore the evolution of the modes corresponding to the inflaton fluctuations is also fixed, including the superhorizon evolution. 

In order to find the power spectrum of the inflaton fluctuations, we calculate the variance of the field $f$

\begin{equation}\label{eqn:167}
\left\langle 0 \left| \hat{f}^{\dagger}(\tau, \textbf{0}) \hat{f}(\tau, \textbf{0}) \right| 0 \right\rangle  = \int \frac{dk}{k}\mathcal{P}_{f}(k, \tau).
\end{equation}

\noindent Calculating the integrand on the right hand side, we find

\begin{equation}\label{eqn:168}
\mathcal{P}_{f}(k, \tau) = \frac{k^{3}}{2\pi^{2}}\left|f_{k}\left(\tau \right)\right|^{2},
\end{equation}

\noindent and then substituting (\ref{eqn:166}), this is

\begin{equation}\label{eqn:169}
\mathcal{P}_{f}(k, \tau) = \frac{k^{2}}{4\pi^{2}}\left( 1 + \frac{1}{k^{2}\tau^{2}}\right).
\end{equation}

\noindent Since, $\delta \phi = f/a$, we can use (\ref{eqn:169}) to define the power spectrum of the inflaton fluctuations

\begin{equation}\label{eqn:170}
\mathcal{P}_{\delta \phi}(k) = \frac{1}{a^{2}}\mathcal{P}_{f}(k, \tau) =  \frac{k^{2}}{4\pi^{2}a^{2}}\left( 1 + \frac{1}{k^{2}\tau^{2}}\right).
\end{equation}

Using the fact that conformal time is $\tau = -1/aH$, this can be written as

\begin{equation}\label{eqn:171}
\mathcal{P}_{\delta \phi}(k) = \frac{k^{2}}{4\pi^{2}a^{2}}\left( 1 + \frac{a^{2}H^{2}}{k^{2}}\right),
\end{equation}

\noindent where $k/a$ is the physical wavenumber equal to $2\pi/\lambda_{phys}$. For superhorizon perturbations, the physical wavelength of the fluctuation is larger than the Hubble horizon, $\lambda_{phys} \ge H^{-1}$, and we have that

\begin{equation}\label{eqn:172}
\frac{2\pi}{\lambda_{phys}} \le 2\pi H \Rightarrow \frac{k}{a} \le 2\pi H,
\end{equation}

\noindent Using this, we can rewrite (\ref{eqn:171}) as

\begin{equation}\label{eqn:173}
\mathcal{P}_{\delta \phi}(k) = \left(\frac{H}{2\pi}\right)^{2}\left( 1 + \left(\frac{k}{aH}\right)^{2}\right).
\end{equation}

\noindent Once perturbations exit the horizon, we have that $k << aH$ and their power spectrum  $\mathcal{P}_{\delta \phi}(k) \rightarrow \left(\frac{H}{2\pi}\right)^{2}$. Evaluated at $k = aH$, the power spectrum of the inflaton fluctuations is therefore

\begin{equation}\label{eqn:174}
\mathcal{P}_{\delta \phi}(k) = \left.\left(\frac{H}{2\pi}\right)^{2}\right|_{k = aH},
\end{equation}

\noindent where modes of a given $k$ exit the Hubble horizon at $k = aH$ when inflation stretches the wavelengths of these modes to superhorizon scales. From (\ref{eqn:159}) the variance of these fluctuations is related to the variance of the comoving curvature perturbation $\mathcal{R}$ by 

\begin{equation}\label{eqn:175}
\langle \left| \mathcal{R}_{k} \right| \rangle^{2} = \left(\frac{\mathcal{H}}{\bar{\phi}'}\right)^{2}\langle \left| \delta \phi_{k} \right|^{2} \rangle,
\end{equation}

\noindent where after horizon exit, the quantum expectation value of the inflaton fluctuations can be identified with the ensemble average of a classical field $\delta \phi$. It is a property of the comoving curvature perturbation $\mathcal{R}$ that it is conserved - does not evolve - on superhorizon scales. This means that the value of $\mathcal{R}$ at horizon crossing survives unaltered until much later times, and this gives an insight into the state of the curvature perturbations at the end of inflation.

From (\ref{eqn:174}) and (\ref{eqn:159}) the power spectrum of the comoving curvature perturbation is given by

\begin{equation}\label{eqn:176}
\mathcal{P}_{\mathcal{R}} \left(k \right) =  \left(\frac{\mathcal{H}}{\bar{\phi}'}\right)^{2} \mathcal{P}_{\delta \phi} \left.\left(k \right)\right|_{k = aH} = \left.\frac{H^{2}}{8\pi^{2}\epsilon M_{pl}^{2}}\right|_{k = aH}.
\end{equation}

\noindent at horizon crossing. At $k = aH$, if $H$ is constant, $\mathcal{P}_{\mathcal{R}}$ is independent of $k$ is therefore scale invariant. In actuality, the power spectrum of the curvature perturbation will be near-scale invariant since $H$ and $\epsilon$ are slowly-varying functions. The deviation from scale invariance can be measured by expressing $\mathcal{P}_{\mathcal{R}}$ as a power law

\begin{equation}\label{eqn:177}
\mathcal{P}_{\mathcal{R}}\left(k \right) = A_{s}\left(\frac{k}{k_{\ast}}\right)^{n_{s} - 1},
\end{equation}

\noindent where $A_{s} = 2.1 \times 10^{-9}$ is the measured amplitude of the power spectrum, as measured using a reference scale $k_{\ast}$, known as the pivot scale, of $k_{\ast} = 0.05\Mpc^{-1}$ \cite{planck184}. The quantity $n_{s}$ is referred to as the scalar spectral index, and the power spectrum

\begin{equation}\label{eqn:178}
n_{s} - 1 = \left.\frac{d\ln \mathcal{P}_{\mathcal{R}} }{d\ln k}\right|_{k = k_{\ast}},
\end{equation}

\noindent measures the deviation from scale invariance, where $n_{s} = 1$ corresponds to a scale invariant spectrum. Rewriting the right hand side of (\ref{eqn:178}) in terms of $\epsilon$ and $H$ ((\ref{eqn:93}) and (\ref{eqn:95}) respectively), we can express the deviation from scale invariance in terms of the slow-roll parameters

\begin{equation}\label{eqn:179}
n_{s} - 1 = -2\epsilon_{H} - \eta_{H} \Rightarrow n_{s} - 1 = 2\eta - 6\epsilon.
\end{equation}

The power spectrum of the tensor modes (gravitational waves) which arise from the perturbation of the geometry of the spacetime on a given slice of spacetime can also be measured \cite{baumann}

\begin{equation}\label{eqn:180}
\mathcal{P}_{t} \left(k \right) = \left.\frac{2H^{2}}{\pi^{2} M_{pl}^{2}}\right|_{k = aH} \Rightarrow \mathcal{P}_{t}\left(k \right) = A_{t}\left(\frac{k}{k_{\ast}}\right)^{n_{t}}.
\end{equation}

\noindent Analogously to $n_{s}$ for $\mathcal{P}_{\mathcal{R}}$, $n_{t}$ measures the deviation from scale invariance. An important quantity in observational cosmology is the amplitude of the tensor modes normalised with respect to the amplitude of the scalar modes

\begin{equation}\label{eqn:181}
r = \frac{A_{t}}{A_{s}} = 16\epsilon,
\end{equation}

\noindent known as the tensor-to-scalar ratio. 

\subsection{Adiabatic and Isocurvature Perturbations}\label{section:121}
In this section we examine the nature of the perturbations in the energy density arising during inflation.
Of the energy density fluctuations generated there are two different types, adiabatic and isocurvature perturbations, and the role they play in the evolution of the Universe is quite different. 

\begin{itemize}
\item Adiabatic perturbations

These are primordial perturbations in which the fractional perturbation of the number density of all conserved matter species, $i$, is equal to the fractional perturbation of the number density of photons

\begin{equation}\label{eqn:182}
\frac{\delta n_{i}}{n_{i}} = \frac{\delta n_{\gamma}}{n_{\gamma}}.
\end{equation}

\noindent The perturbations in the energy density during inflation are predicted to be adiabatic and can be visualised as parts of the expanding Universe where the expansion is slightly ahead or behind the average expansion. The expansion is adiabatic itself, which means that the total entropy within the expanding volume is conserved. An adiabatic perturbation therefore looks like a volume of space that has been adiabatically squeezed (squeezed while conserving the total entropy in the volume).

The total number of particles of a species $i$, $N_{i}$, in a volume $V$ is conserved. For small changes in number density, $\delta n_{i}$, and volume, $\delta V$, due to the adiabatic squeezing

\begin{equation}\label{eqn:183}
\delta N_{i} = \delta\left(n_{i}V\right) = 0 \Rightarrow \frac{dN_{i}}{dn_{i}}\delta n_{i} + \frac{dN_{i}}{dV}\delta V = 0,
\end{equation}

\noindent this means that for small changes in the number of a species in an expanding volume

\begin{equation}\label{eqn:184}
V\delta n_{i} + n_{i}\delta V = 0 \Rightarrow \frac{\delta n_{i}}{n_{i}} = -\frac{\delta V}{V},
\end{equation}

\noindent is true for all conserved matter species $i$.

The total entropy in a volume $V$ is proportional to the total number of photons in the volume. This is because the entropy density $s \propto T^{3}$ and so is the photon number density $n_{\gamma}$. Since entropy is conserved in an adiabatically expanding volume, the number of photons is therefore conserved, and the photon number density can be treated the same as the conserved matter species.

For adiabatic perturbations, for all species $i$ we therefore have

\begin{equation}\label{eqn:186}
\frac{\delta n_{1}}{n_{1}} = \frac{\delta n_{2}}{n_{2}} ... = ... \frac{\delta n_{\gamma}}{n_{\gamma}}.
\end{equation}

 Adiabatic fluctuations therefore refer to small changes in the total energy density of the Universe across all of its components. These adiabatic perturbations occur because all the particle densities, $\delta \rho_{i}$, originate from a single initial matter density, the inflaton, meaning that all fluctuations in the density of the species $i$ are equal to the inflaton fluctuations. This is a serendipitous side effect of inflation in that it predicts the generation of adiabatic density perturbations, since all particle species arise from the decay of the inflaton field. Inflation also predicts that different points in the Universe will look like they have expanded a bit more or a bit less than the average expansion, and we can interpret that a larger or smaller density of inflaton particles is therefore equivalent to less or more expansion relative to the average.

\item Isocurvature Fluctuations

The other kind of primordial perturbation which could be produced are isocurvature perturbations. In the case of isocurvature, the perturbations of individual conserved particle species, $\delta \rho_{i}$, can be non-zero but the total perturbation of the energy density of the Universe, $\delta \rho_{tot} = 0$.

This can be realised if there are two or more types of energy density contributing to the total energy density of the Universe, non-relativistic matter, $\delta \rho_{m}$, and radiation, $\delta \rho_{r}$, for instance

\begin{equation}\label{eqn:187}
\delta \rho_{m} = -\delta \rho_{r} \Rightarrow \delta \rho_{tot} = \delta \rho_{m} + \delta \rho_{r} = 0.
\end{equation}

A primordial density perturbation can be decomposed into the sum of an adiabatic part and an isocurvature part. The primordial perturbations observed in the Universe today are purely adiabatic, and have no observed isocurvature component, although it is possible that an unobserved isocurvature component could exist and this is a current topic of research in cosmology.

\end{itemize}
\section{Baryogenesis}\label{section:13}

Baryogenesis refers to the generation of the baryon-antibaryon asymmetry present in the Universe today, generally quantified in terms of the baryon-to-photon ratio. The baryon-to-photon ratio today is $\frac{n_{B}}{n_{\gamma}} = 6.06 \times 10^{-10}$ \cite{sdss06}. The presence and size of the baryon asymmetry is an unsolved problem in physics, and the precise mechanism which produces it is unknown. In 1967, Sakharov posited that the underlying processes leading to the generation of a baryon asymmetry must fulfill the following three criteria \cite{sakharov}

\begin{enumerate}

\item Baryon number ($B$) violation. \\
This refers to the fact that most particle interactions in the Standard Model conserve baryon number. In order for an asymmetry in baryon number to be produced there must be interactions which produce more baryons than antibaryons in the early Universe, possibly requiring physics not yet discovered by modern particle physicists. 
\item Charge Conjugation ($C$), and Charge-Parity ($CP$) Violation. \\
This refers to the fact that if all charges are reversed (Charge Conjugation) and the parity of any directional charges on particles are reversed (Charge-Parity: sign flip of all charges and spatial vectors combined) then particles should be indistinguishable from their antiparticles. Violation of both of these symmetries simultaneously enables the possibility of any excess in baryon number generated by particle interactions to not be wiped out by the corresponding antiparticle interactions.

As an example, consider a process $X \rightarrow Y + B$ in a theory which respects charge conjugation symmetry. The rate of this process is the same as the C-conjugate process $\bar{X} \rightarrow \bar{Y} + \bar{B}$

\begin{equation}\label{eqn:188}
\Gamma \left( X \rightarrow Y + B\right) = \Gamma \left(\bar{X} \rightarrow \bar{Y} + \bar{B}\right).
\end{equation}

\noindent The net rate of baryon production is proportional to the difference between these rates, because any excess baryon number could only be produced if the rates become different

\begin{equation}\label{eqn:189}
\frac{dB}{dt} \propto \Gamma \left(\bar{X} \rightarrow \bar{Y} + \bar{B}\right) - \Gamma \left( X \rightarrow Y + B\right).
\end{equation}

\noindent This vanishes when C-symmetry is respected. Charge conjugation symmetry violation alone is not enough to produce a baryon asymmetry however, and the additional violation of Charge-Parity (CP) symmetry is needed. If we suppose that $X$ is a heavy bosonic species and decays to either two right-handed or left-handed quarks \cite{clineb}; $X \rightarrow q_{L}q_{L}$, $X \rightarrow q_{R}q_{R}$, then under Charge Conjugation the decay products transform as

\begin{equation}\label{eqn:190}
q_{L} \rightarrow \bar{q}_{L},
\end{equation}

\noindent where the charges reverse but other quantum numbers remain the same. Under Charge-Parity

\begin{equation}\label{eqn:191}
q_{L} \rightarrow \bar{q}_{R},
\end{equation}

\noindent where the charge and the parity of the particles reverse. C-violation alone means that

\begin{equation}\label{eqn:192}
\Gamma \left( X \rightarrow q_{L}q_{L}\right) \neq \Gamma \left(\bar{X} \rightarrow \bar{q}_{L}\bar{q}_{L}\right),
\end{equation}

\noindent whereas CP-conservation implies

\begin{equation}\label{eqn:193}
\Gamma \left( X \rightarrow q_{L}q_{L}\right) = \Gamma \left(\bar{X} \rightarrow \bar{q}_{R}\bar{q}_{R}\right),
\end{equation}

\begin{equation}\label{eqn:194}
\Gamma \left( X \rightarrow q_{R}q_{R}\right) = \Gamma \left(\bar{X} \rightarrow \bar{q}_{L}\bar{q}_{L}\right),
\end{equation}

\noindent which gives the condition

\begin{equation}\label{eqn:195}
\Gamma \left( X \rightarrow q_{L}q_{L}\right) + \Gamma \left( X \rightarrow q_{R}q_{R}\right) = \Gamma \left(\bar{X} \rightarrow \bar{q}_{R}\bar{q}_{R}\right) + \Gamma \left(\bar{X} \rightarrow \bar{q}_{L}\bar{q}_{L}\right).
\end{equation}

\noindent An initial state of equal $X$ and $\bar{X}$ particles therefore produces no net asymmetry in quarks, and therefore no baryon asymmetry. CP-violation alone at most could generate an asymmetry in the handedness of the quarks, and C-violation is therefore needed additionally to generate a preference for the decay of $X$ or $\bar{X}$ particles to either left- or right-handed quarks in order to produce an excess of baryon number.

\item Departure from Thermal Equilibrium. \\
Thermal equilibrium means that all particle processes and their reverse processes have the same reaction rate. This means that any non-zero baryon number generated in particle interactions would be wiped out by the reverse reaction if the Universe were in thermal equilibrium. A departure from thermal equilibrium in this context means that forward or reverse reactions can become more favourable than the other direction, and a preference for generating baryon number can be established without it being washed out completely.

To illustrate this, consider again the example reaction above, $X \rightarrow Y + B$, where we now consider $X$ and $Y$ to have zero net baryon number and $B$ represents the excess baryon number generated. Thermal equilibrium means that this reaction and its reverse reaction occur at the same rate

\begin{equation}\label{eqn:196}
\Gamma\left( X \rightarrow Y + B\right) = \Gamma\left(Y + B \rightarrow X\right),
\end{equation}

\noindent and no net baryon asymmetry can be produced as any baryon number generated in the forward-reaction is depleted by the reverse-reaction. If the decay occurs out of equilibrium then $m_{X} > T$ at the time of decay, $\tau = 1/\Gamma$. The energy of the reverse process in this case is $\sim \mathcal{O}(T)$, which is insufficient to produce an $X$ particle, and we say that the rate of the reverse reaction is Boltzmann-suppressed $\Gamma\left(Y + B \rightarrow X\right) \sim e^{-m_{X}/T}$. Out-of-equilibrium decays therefore allow for a preferential rate of the forward-process, and an excess of baryon number can therefore be generated. 
\end{enumerate}

The baryon asymmetry must be in place before nucleosynthesis at around $1\MeV$. This means that baryogenesis must occur earlier, such as during the Electroweak Phase Transition, or as early as during inflation. The second possibility is the basis of the research presented in Chapter 5.

\chapter{Theoretical Background}

In this chapter we present some important background and relevant results in particle physics and quantum field theory which will be used in the original research presented in Chapters 4-6.

\section{Palatini and Metric Formalism}\label{section:t1}
In this thesis, the chapters which focus on inflation (namely Chapters 4 and 6) present work in which inflation is discussed in the Palatini formalism, as opposed to the conventionally used metric formalism used when dealing with non-minimally coupled inflation. In this section we will briefly outline what these formalisms are. 

\noindent General Relativity is described by the Einstein-Hilbert action \cite{einsteinhilbert}

\begin{equation}\label{eqn:b1}
S_{EH} = -\frac{1}{2}\int  \sqrt{-g} \; M_{pl}^{2}R.
\end{equation}

\noindent In this thesis, much of the work presented is based on models of non-minimally coupled inflation, where the gravitational action takes the form

\begin{equation}\label{eqn:b2}
S_{NMC} = -\frac{1}{2}\int  \sqrt{-g} \; M_{pl}^{2}\left( 1 + \frac{2\xi \left| \Phi \right|^{2}}{M_{pl}^{2}}\right)R,
\end{equation}

\noindent where in addition to the usual Einstein-Hilbert action there is an interaction between the scalar field $\phi$ and the scalar curvature $R$ \cite{buchbinder}. Additional terms quadratic in the Ricci scalar or the Ricci tensor, $R^{2}$, $R_{\mu \nu}R^{\mu \nu}$, or higher order may also be included, and can arise in the gravitational action as a result of quantum corrections or due to a UV completion depending on the model of gravity being used \cite{birrelldavies}. The inflation model discussed in Chapter 4 is based on the inclusion of an $R^{2}$ term.

In order to derive the equations of motion of a theory, the action must be extremised with respect to the parameters of the theory, $A_{i}$

\begin{equation}\label{eqn:b3}
\delta S = \frac{\delta}{\delta A_{i}}\left(\int d^{4}x \sqrt{-g} \mathcal{L} \right) = 0,
\end{equation}

\noindent where in the Palatini (affine) formalism no assumptions are initially made about the form of the affine connection, and as such the metric and the affine connection are treated as separate quantities and the action is varied with respect to these objects independently. The variation parameters, $A_{i}$, are therefore

\begin{equation}\label{eqn:b4}
A_{i} = \left( g, \partial g, \Gamma, \partial \Gamma, \phi_{a}, \partial \phi_{a} \right),
\end{equation}

\noindent i.e. the spacetime metric and its derivatives, the affine connection and its derivatives, and the matter fields of the theory and their derivatives, respectively. Conversely, in the (pure) metric formalism, the connection is assumed to be the Levi-Civita connection from the beginning, and the metric can therefore be varied using

\begin{equation}\label{eqn:b5}
A_{i} = \left( g, \partial g, \left(\partial g \right)^{2}, \phi_{a}, \partial \phi_{a} \right),
\end{equation}

\noindent since the connection is built from derivatives of the metric. 

At the level of the equations of motion, the two formalisms are equivalent in the absence of any fermion fields in the matter Lagrangian, or any matter fields coupled to the metric or the affine connection \cite{gravitation}. In the Palatini case, the condition that the covariant derivative of the metric must vanish arises when extremising the action \cite{gravitation}

\begin{equation}\label{eqn:b6}
\nabla_{\lambda}g_{\beta \gamma} = \frac{\partial g_{\beta \gamma}}{\partial x^{\lambda}} - g_{\gamma \sigma}\Gamma_{\beta \lambda}^{\sigma} - g_{\beta \sigma}\Gamma^{\sigma}_{\gamma \lambda} = 0,
\end{equation}

\noindent and solving for the Christoffel symbols then gives \cite{gravitation}

\begin{equation}\label{eqn:b7}
\Gamma_{\mu \nu}^{\rho} = \frac{1}{2}g^{\rho \sigma}\left[\partial_{\nu}g_{\mu \sigma} + \partial_{\mu}g_{\sigma \nu} - \partial_{\sigma}g_{\mu \nu} \right],
\end{equation}

\noindent which is the standard equation for the spacetime connection in Riemannian geometry, the Levi-Civita connection. In this case, both formalisms in Einstein-Hilbert gravity will produce the Einstein equations in a vacuum \cite{gravitation}. 

Differences arise in the presence of a non-minimal coupling of a matter field to gravity, or higher-order curvature terms in the gravitational action. If the non-minimal coupling is reparameterised in such a way that the gravitational action looks Einstein-Hilbert, then the two formalisms will produce equivalent equations of motion, but the reparameterised action of the field theory will be different. In the case of inflation, this also means that the predictions of the observables will be different depending on the formalism used.

\subsection{Non-Minimally Coupled Inflation in the Metric Formalism}\label{section:t11}

In this section we outline the treatment of non-minimally coupled inflation in the metric formalism, including the derivation of the inflationary observables, to allow comparison to the Palatini formalism used in Chapters 4 and 6. The metric convention used in this derivation is $\left(-, +, +, +\right)$.

\noindent The Jordan frame action of a general complex non-minimally coupled inflaton, $\Phi$, is given by

\begin{equation}\label{eqn:b8}
S_{J} = \int d^{4}x \sqrt{-g} \left[\frac{1}{2}M_{pl}^{2}R\Omega^{2} - \partial_{\mu}\Phi^{\dagger}\partial_{\mu}\Phi - V\left(\left|\Phi \right| \right) \right],
\end{equation}

\noindent where the conformal factor $\Omega^{2}$ is given by

\begin{equation}\label{eqn:b9}
\Omega^{2} = 1 + \frac{2\xi \left|\phi\right|^{2}}{M_{pl}^{2}}.
\end{equation}

\noindent $\xi$ parameterises the coupling between the inflaton and the Ricci scalar. In order to reparameterise the model in such a way that gravity appears as it does in conventional General Relativity, we perform a conformal transformation on the metric

\begin{equation}\label{eqn:b10}
g_{\mu \nu} \rightarrow \tilde{g}_{\mu \nu} = \Omega^{2}g_{\mu \nu},
\end{equation}

\begin{equation}\label{eqn:b11}
g^{\mu \nu} \rightarrow \tilde{g}^{\mu \nu} = \frac{1}{\Omega^{2}}g^{\mu \nu},
\end{equation}

\noindent to the Einstein frame. We follow the convention of performing calculations relating to inflation in the Einstein frame throughout this thesis, and we discuss the interpretation of the Jordan and Einstein frames in Section \ref{section:231}. From \cite{kaiser} we have that the transformation on the Ricci scalar in the metric formalism is

\begin{equation}\label{eqn:b12}
\tilde{R} = \frac{1}{\Omega^{2}}\left[R - \frac{2\left(D - 1\right)}{\Omega}\Box \Omega - \left( D - 1 \right)\left( D - 4 \right) \frac{1}{\Omega^{2}}g^{\mu \nu}\nabla_{\mu}\Omega \nabla_{\nu}\Omega \right],
\end{equation}

\noindent where

\begin{equation}\label{eqn:b13}
\Box\Omega = g^{\mu \nu}\nabla_{\mu}\nabla_{\nu}\Omega = \frac{1}{\sqrt{-g}}\partial_{\mu}\left[\sqrt{-g} g^{\mu \nu} \partial_{\nu}\Omega\right],
\end{equation}

\noindent and $D$ is the total number of dimensions. In four dimensions we have that the transformation on the Ricci scalar is

\begin{equation}\label{eqn:b14}
\tilde{R} = \frac{1}{\Omega^{2}}\left[R - \frac{6}{\Omega}\partial_{\mu}\partial^{\mu}\Omega \right].
\end{equation}

\noindent The integration measure transforms as

\begin{equation}\label{eqn:b15}
\sqrt{-g} \rightarrow \frac{\sqrt{-\tilde{g}}}{\Omega^{4}},
\end{equation}

\noindent and the derivatives on the kinetic term transform as

\begin{equation}\label{eqn:b16}
\partial_{\mu}\Phi^{\dagger}\partial^{\mu}\Phi = g_{\mu \nu}\partial^{\nu}\Phi^{\dagger}\partial^{\mu}\Phi \rightarrow \tilde{g}_{\mu \nu}\partial^{\nu}\Phi^{\dagger}\partial^{\mu}\Phi = g_{\mu \nu}\Omega^{2} \partial^{\nu}\Phi^{\dagger}\partial^{\mu}\Phi = \Omega^{2}\partial_{\mu}\Phi^{\dagger}\partial^{\mu}\Phi.
\end{equation}

\noindent It is also instructive at this stage to rewrite the complex field as

\begin{equation}\label{eqn:b17}
\Phi = \frac{\phi}{\sqrt{2}}e^{i\theta}.
\end{equation}

\noindent Assuming $\theta$ is constant throughout inflation such that the dynamical evolution of the inflaton is purely radial, (\ref{eqn:b14}) - (\ref{eqn:b17}) can be used to transform (\ref{eqn:b8}) and the Einstein frame action is

\begin{equation}\label{eqn:b18}
S_{E} = \int d^{4}x \; \frac{\sqrt{-\tilde{g}}}{\Omega^{4}} \left[\frac{1}{2}M_{pl}^{2}\Omega^{2}\left[\Omega^{2}\tilde{R} + \frac{6}{\Omega}\partial_{\mu}\partial^{\mu}\Omega \right] - \frac{\Omega^{2}}{2}\partial_{\mu}\phi \partial^{\mu}\phi - V\left(\phi \right) \right].
\end{equation}

\noindent Integrating the $\Omega$ kinetic term by parts, this becomes

\begin{equation}
S_{E} = \int d^{4}x \; \sqrt{-\tilde{g}} \left[\frac{1}{2}M_{pl}^{2}\left[\tilde{R} - \frac{6}{\Omega^{2}}\tilde{g}^{\mu \nu}\partial_{\mu}\Omega\partial^{\nu}\Omega \right] - \frac{1}{2\Omega^{2}}\partial_{\mu}\phi \partial^{\mu}\phi - \frac{V\left(\phi \right)}{\Omega^{4}} \right],
\end{equation}

\noindent and the $\Omega$ kinetic term can be written in terms of $\Omega^{2}$ such that

\begin{equation}\label{eqn:b19}
S_{E} = \int d^{4}x  \; \sqrt{-\tilde{g}} \left[\frac{1}{2}M_{pl}^{2}\tilde{R} - \frac{3M_{pl}^{2}}{4\Omega^{4}}\partial_{\mu}\Omega^{2}\partial^{\mu}\Omega^{2}  - \frac{1}{2\Omega^{2}}\partial_{\mu}\phi \partial^{\mu}\phi - \frac{V\left(\phi \right)}{\Omega^{4}} \right].
\end{equation}

\noindent Rewriting the $\left(\partial_{\mu}\Omega^{2}\right)^{2}$ term in terms of $\phi$, the Einstein frame action is

\begin{equation}\label{eqn:b20}
S_{E} = \int d^{4}x \; \sqrt{-\tilde{g}} \left[\frac{1}{2}M_{pl}^{2}\tilde{R} - \frac{3M_{pl}^{2}}{4\Omega^{4}}\left(\frac{2\xi \phi}{M_{pl}^{2}} \right)^{2}\partial_{\mu}\phi \partial^{\mu}\phi  - \frac{1}{2\Omega^{2}}\partial_{\mu}\phi \partial^{\mu}\phi - V_{E}\left(\phi \right) \right],
\end{equation}

\noindent where 

\begin{equation}\label{eqn:b21}
V_{E}\left(\phi \right) = \frac{V\left(\phi \right)}{\Omega^{4}},
\end{equation}

\noindent is the Einstein frame potential. The kinetic terms can be rewritten

\begin{equation}\label{eqn:b22}
\left[\frac{1}{2\Omega^{2}} +  \frac{6\xi^{2}\phi^{2}}{2\Omega^{4}M_{pl}^{2}}\right]\partial_{\mu}\phi \partial^{\mu}\phi = \frac{1}{2}\left[\frac{\Omega^{2} + 6\xi^{2}\phi^{2}/M_{pl}^{2}}{\Omega^{4}}\right]\partial_{\mu}\phi \partial^{\mu}\phi.
\end{equation}

\noindent In order to rewrite the inflaton kinetic terms into canonical form, we perform a field rescaling \cite{bellido}

\begin{equation}\label{eqn:b23}
\frac{d\chi}{d\phi} = \pm \sqrt{\frac{\Omega^{2} + 6\xi^{2}\phi^{2}/M_{pl}^{2}}{\Omega^{4}}} = \pm \sqrt{\frac{1 + \xi\left( 1 + 6\xi\right)\phi^{2}/M_{pl}^{2}}{\left(1 + \xi \phi^{2}/M_{pl}^{2} \right)^{2}}}.
\end{equation}
 
\noindent Choosing the positive solution, and integrating we find that \cite{bellido}

\begin{equation}\label{eqn:b24}
\frac{\sqrt{\xi}}{M_{pl}}\chi \left(\phi \right) = \sqrt{1 + 6\xi}\sinh^{-1}\left( \sqrt{1 + 6\xi} u \right) - \sqrt{6\xi}\sinh^{-1}\left(\sqrt{6\xi} \frac{u}{\sqrt{1 + u^{2}}} \right),
\end{equation}

\noindent where $u = \sqrt{\xi}\phi/M_{pl}$. 

To have successful inflation, we require that $\xi  >>1$, and therefore $1 + 6\xi \thickapprox 6\xi$. Using this, and the fact that 

\begin{equation}\label{eqn:b25}
\sinh^{-1}x = \ln \left( x + \sqrt{x^{2} + 1}\right),
\end{equation}

\noindent we can approximate (\ref{eqn:b24}) as

\begin{equation}\label{eqn:b26}
\frac{\sqrt{\xi}}{M_{pl}}\chi \left(\phi \right) \thickapprox \sqrt{6\xi}\ln \left( \sqrt{1 + u^{2}}\right).
\end{equation}

\noindent The canonically normalised scalar $\chi$ as a function of the inflaton $\phi$ is then

\begin{equation}\label{eqn:b27}
\chi \left(\phi \right) = \sqrt{6}M_{pl}\ln \left(\sqrt{1 + \frac{\xi \phi^{2}}{M_{pl}^{2}}} \right),
\end{equation}

\noindent and the inverse relation is

\begin{equation}\label{eqn:b28}
\phi \left( \chi \right) = \frac{M_{pl}}{\sqrt{\xi}}\sqrt{\exp\left(\sqrt{\frac{2}{3}}\frac{\chi}{M_{pl}} \right) - 1}.
\end{equation}

\noindent Since the inflaton field is large during inflation we can approximate

\begin{equation}\label{eqn:b29}
\phi^{2}\left( \chi \right) = \frac{M_{pl}^{2}}{\xi}\exp\left(\sqrt{\frac{2}{3}}\frac{\chi}{M_{pl}} \right). 
\end{equation}

\noindent The Einstein frame action in terms of the canonical field $\chi$ is then

\begin{equation}\label{eqn:b30}
S = \int d^{4}x \sqrt{-\tilde{g}} \left[ \frac{1}{2}M_{pl}^{2}\tilde{R} - \frac{1}{2}\partial_{\mu}\chi \partial^{\mu}\chi - V_{E}\left(\chi \right) \right].
\end{equation}

\noindent For the purposes of this discussion we choose the Jordan frame potential to be a purely quartic potential, as is the conventional choice for non-minimally coupled metric inflation

\begin{equation}\label{eqn:b31}
V\left(\phi \right) = \frac{\lambda}{4}\phi^{4},
\end{equation}

\noindent and the Einstein frame potential from (\ref{eqn:b21}) is therefore

\begin{equation}\label{eqn:b32}
V_{E}\left(\phi \right) = \frac{\lambda \phi^{4}}{4\left( 1 + \frac{\xi \phi^{2}}{M_{pl}^{2}}\right)^{2}} = \frac{\lambda \phi^{4}}{4\left(\frac{\xi \phi^{2}}{M_{pl}^{2}}\right)^{2}\left( 1 + \frac{M_{pl}^{2}}{\xi \phi^{2}}\right)^{2}}.
\end{equation}

\noindent During inflation, while the inflaton is well on the plateau of its potential, we have $\xi \phi^{2}/M_{pl}^{2} >>1$, therefore we can expand the potential (\ref{eqn:b32}) as

\begin{equation}\label{eqn:b33}
V_{E}\left(\phi \right) = \frac{\lambda \phi^{4}}{4}\left(\frac{M_{pl}^{2}}{\xi \phi^{2}}\right)^{2}\left[ 1 - \frac{2M_{pl}^{2}}{\xi \phi^{2}} \right],
\end{equation}

\noindent and the Einstein frame potential is therefore

\begin{equation}\label{eqn:b34}
V_{E}\left(\phi \right) = \frac{\lambda M_{pl}^{4}}{4\xi^{2}}\left[ 1 - \frac{2M_{pl}^{2}}{\xi \phi^{2}} \right].
\end{equation}

\noindent Using (\ref{eqn:b29}) we can write the Einstein frame potential in terms of the canonical scalar field $\chi$

\begin{equation}\label{eqn:b35}
V_{E}\left(\phi \right) = \frac{\lambda M_{pl}^{4}}{4\xi^{2}}\left[ 1 - 2\exp\left(-\sqrt{\frac{2}{3}}\frac{\chi}{M_{pl}}\right) \right].
\end{equation}

In order to calculate the slow-roll parameters and the inflationary observables, we require the first and second derivatives of the Einstein frame potential (\ref{eqn:b35})

\begin{equation}\label{eqn:b36}
V_{E}'\left(\phi \right) = \frac{\lambda M_{pl}^{3}}{2\xi^{2}}\sqrt{\frac{2}{3}}\exp\left(-\sqrt{\frac{2}{3}}\frac{\chi}{M_{pl}}\right),
\end{equation}

\begin{equation}\label{eqn:b37}
V_{E}''\left(\phi \right) = -\frac{\lambda M_{pl}^{2}}{3\xi^{2}}\exp\left(-\sqrt{\frac{2}{3}}\frac{\chi}{M_{pl}}\right),
\end{equation}

\noindent which gives the slow-roll parameters to be

\begin{equation}\label{eqn:b38}
\epsilon = \frac{M_{pl}^{2}}{2}\left(\frac{V_{E}'}{V_{E}}\right)^{2} = \frac{4}{3}\exp\left(-2\sqrt{\frac{2}{3}}\frac{\chi}{M_{pl}}\right),
\end{equation}

\begin{equation}\label{eqn:b39}
\eta = M_{pl}^{2} \frac{V_{E}''}{V_{E}} = -\frac{4}{3}\exp\left(-\sqrt{\frac{2}{3}}\frac{\chi}{M_{pl}}\right).
\end{equation}

\noindent The number of e-folds of inflation is given by (from \ref{eqn:123})

\begin{equation}\label{eqn:b40}
N = -\frac{1}{M_{pl}^{2}}\int^{\chi_{end}}_{\chi} \frac{V_{E}}{V_{E}'} d\chi.
\end{equation}

\noindent Approximating $V_{E} \approx \lambda M_{pl}^{4}/4\xi^{2}$ on the plateau, this integrates to

\begin{equation}\label{eqn:b41}
N = -\frac{3}{4}\left.\exp\left(\sqrt{\frac{2}{3}}\frac{\chi}{M_{pl}}\right)\right|^{\chi_{end}}_{\chi}.
\end{equation}

\noindent Assuming that the field is much smaller at the end of inflation, $\chi_{end} << \chi $, the number of e-folds of inflation is given by

\begin{equation}\label{eqn:b42}
N\left(\chi \right) = \frac{3}{4}\exp\left(\sqrt{\frac{2}{3}}\frac{\chi}{M_{pl}}\right),
\end{equation}

\noindent and the canonically normalised scalar as a function of the number of e-folds $N$ is therefore

\begin{equation}\label{eqn:b43}
\chi \left(N \right) = M_{pl}\sqrt{\frac{3}{2}}\ln \left(\frac{4N}{3}\right).
\end{equation}

\noindent Substituting this into (\ref{eqn:b38}) and (\ref{eqn:b39}) we find that the slow-roll parameters are

\begin{equation}\label{eqn:b44}
\epsilon = \frac{3}{4N^{2}},
\end{equation}

\begin{equation}\label{eqn:b45}
\eta = -\frac{1}{N}.
\end{equation}

\noindent From (\ref{eqn:179}) the scalar spectral index is 

\begin{equation}\label{eqn:b46}
n_{s} \simeq 1 + 2\eta  = 1 - \frac{2}{N},
\end{equation}

\noindent and the tensor-to-scalar ratio (\ref{eqn:181}) is 

\begin{equation}\label{eqn:b47}
r \approx 16\epsilon = \frac{12}{N^{2}}.
\end{equation}

\noindent The primordial curvature power spectrum from (\ref{eqn:176}) is given by 

\begin{equation}\label{eqn:b48}
P_{\mathcal{R}} = \frac{V_{E}}{24\pi^{2}\epsilon M_{pl}^{4}},
\end{equation}

\noindent substituting in (\ref{eqn:b44}) and working to the plateau approximation we can write this in terms of the number of e-folds

\begin{equation}\label{eqn:b49}
P_{\mathcal{R}} = \frac{\lambda N^{2}}{72\pi^{2}\xi^{2}}.
\end{equation}

\noindent The amplitude of the power spectrum is $A_{s} = 2.1 \times 10^{-9}$ \cite{Planck18}. For $N = 60$ this gives the value of the non-minimal coupling to be $\xi = 1.6 \times 10^{4}$ (as opposed to $\xi \sim 10^{9}$ in the Palatini formalism as we will demonstrate in later chapters). The scalar spectral index at $N = 60$ is $n_{s} = 0.9667$, which is in good agreement with the 2018 Planck result of $n_{s} = 0.9649 \pm 0.0042$ ($1-\sigma$) \cite{Planck18}, and the tensor-to-scalar ratio is $r = 3.3 \times 10^{-3}$ which is in good agreement with the result from the BICEP-2/KECK array of $r < 0.036$ at $95\%$ confidence \cite{bicep}, and will be within the detectable range of future CMB experiments \cite{cmbobserve}, such as LiteBIRD \cite{litebird}, as opposed to the typical Palatini result $r \sim 10^{-13}$, as we will show later.

\section{Unitarity}\label{section:t2}

In this section we discuss the concept of unitarity, how specifically this relates to non-minimally coupled inflation, and why it is important when building or studying non-minimally coupled inflation models.

\noindent Unitarity is defined as the requirement that the probabilities of an event happening must be equal to one. The condition for the conservation of probability can be derived using the $S$-matrix. The $S$-matrix is an operator which maps an initial state $i$ to a final state $f$

\begin{equation}\label{eqn:b50}
\mid i \rangle \rightarrow \hat{S} \mid i \rangle = \mid f \rangle,
\end{equation}

\noindent where the matrix elements of the $\hat{S}$ operator $\langle f \mid \hat{S} \mid i \rangle$ are the scattering amplitudes of all possible processes resulting in the final state $f$ from the initial state $i$. If the initial state $ \mid i \rangle$ is normalised such that  $\langle i  \mid i \rangle = 1$, then for a complete set of states $\mid n \rangle$ the probability that $\mid i \rangle$ evolves into $\mid n \rangle$ summed over all $\mid n \rangle$ must be equal to unity

\begin{equation}\label{eqn:b51}
\sum_{n} \left|\langle n \left| \hat{S} \right| i \rangle \right|^{2} = 1.
\end{equation}

\noindent Since $\mid n \rangle$ forms a complete set of states we have that

\begin{equation}\label{eqn:b52}
\sum_{n} \left| n \rangle  \langle n \right| =1,
\end{equation}

\noindent and we can write

\begin{equation}\label{eqn:b53}
\sum_{n} \langle i \vert \hat{S}^{\dagger} \vert n \rangle  \langle n \vert \hat{S} \vert i \rangle = \langle i \vert \hat{S}\hat{S}^{\dagger} \vert i \rangle = 1.
\end{equation}

\noindent We therefore have that

\begin{equation}\label{eqn:b54}
\hat{S}\hat{S}^{\dagger} = \hat{S}^{\dagger}\hat{S} = 1,
\end{equation}

\noindent which means that the $S-$matrix operator is manifestly unitary, and this is an expression of the conservation of probabilities in scattering experiments. 

\noindent The $S$-matrix can be written in terms of the transition matrix, or $T$-matrix

\begin{equation}\label{eqn:b55}
\hat{S} = 1 + i\hat{T},
\end{equation}

\noindent where the first term is unity because the only non-zero amplitude in the first term of the expansion corresponds to a process for which the initial and final states are identical. In terms of the $T$-matrix the unitarity condition of the $S$-matrix becomes \cite{maggiore}

\begin{equation}\label{eqn:b56}
-i\left(\hat{T} - \hat{T}^{\dagger} \right) = \hat{T}\hat{T}^{\dagger}.
\end{equation}

\noindent Writing the matrix element $\langle f \left| \hat{T} \right| i \rangle$ by $T_{fi}$ and inserting a complete set of states we can write (\ref{eqn:b56}) as

\begin{equation}\label{eqn:b57}
-i\left(T_{fi} - T^{\ast}_{if} \right) = \sum_{n} T_{fn}T^{\ast}_{in},
\end{equation}

\noindent where for $i = f$ this can be written as \cite{maggiore}

\begin{equation}\label{eqn:b58}
2Im T_{ii} = \sum_{n}\left|T_{in}\right|^{2}.
\end{equation}

\noindent This is the requirement for unitarity conservation in scattering processes. 

The $T$-matrix can be written in terms of an invariant amplitude $\mathcal{M}$. For a $2 \rightarrow 2$ scattering process this is given by \cite{lancasterblundell}

\begin{equation}\label{eqn:b59}
\langle p_{1f}p_{2f} \left| i\hat{T} \right| p_{2i}p_{1i} \rangle = \left(2\pi \right)^{4} \delta^{\left(4\right)}\left(p_{1f} + p_{2f} - p_{2i} - p_{1i} \right) i \mathcal{M},
\end{equation}

\noindent where $(p_{1f}, p_{2f})$, $(p_{1i}, p_{2i})$ are the final and initial momenta of the particles involved in the scattering. The invariant amplitude can be defined in terms of the differential scattering cross-section, which for a $2 \rightarrow 2$ scattering process is \cite{lancasterblundell}

\begin{equation}\label{eqn:b60}
\frac{d\sigma}{d\Omega} = \frac{\left|\mathcal{M}\right|^{2}}{64\pi^{2} E_{CM}^{2}},
\end{equation}

\noindent where $E_{CM}$ is the energy of the interaction in the centre-of-mass frame, and this cross section can be measured in scattering experiments.

We have that (\ref{eqn:b58}) can be rewritten (see e.g \cite{partialwave})

\begin{equation}\label{eqn:b61}
\left|Re\left(a_{l}\right)\right| \leq \frac{1}{2},
\end{equation}

\noindent as an alternative formulation of the condition for unitarity conservation in elastic $2 \rightarrow 2$ scattering. The partial wave amplitudes, $a_{l}$, are defined by expanding the scattering amplitude $\mathcal{M}$ in terms of the scattering angle $\theta$

\begin{equation}\label{eqn:b62}
-i\mathcal{M} = 16\pi \sum^{\infty}_{l = 0} \left(2l + 1\right)P_{l}\left(\cos\theta\right)a_{l},
\end{equation}

\noindent where $P_{l}\left(\cos\theta\right)$ are the Legendre polynomials. Generally from this expansion, $\mathcal{M} \sim a_{0}$, and so the condition for unitarity to be conserved in elastic $2 \rightarrow 2$ scattering from this is, $\mathcal{M} \simeq 1$. We will make use of this expression of the unitarity condition in Chapter 4.

Perturbative unitarity can be violated in scattering processes at a given energy scale, $E$, if the requirement (\ref{eqn:b58}) is not met. This would indicate that there is either new physics entering around the unitarity cutoff scale, $\Lambda$, which would suppress any large contributions to the imaginary scattering amplitudes of the $T$-matrix, or it means that the process becomes non-perturbative at the energy scale, $\Lambda$, and that the perturbative treatment of the process is no longer valid. The latter may not signal a breakdown of the theory if the unitarity violating contributions at tree level are cancelled at higher loop orders \cite{selfheal}, or the theory becomes strongly coupled close to the cutoff scale \cite{2loop, natjr}.

Unitarity violation is particularly relevant to non-minimally coupled inflation because the interaction of scalars with gravitons (mediated by the non-minimal coupling to the Ricci scalar in (\ref{eqn:b2}) violates unitarity at tree level, so in any non-minimally coupled inflation theory there will be unitarity violation from the scattering of inflaton scalars via graviton exchange \cite{scaleofqg}. A full investigation of unitarity in non-minimally coupled inflation is not discussed in this thesis, however it is an important issue when embedding non-minimally coupled inflation models into a theory of particle physics.

\section{The Effective Potential and Coleman-Weinberg Corrections}\label{section:t3}

Internal loops in Feynman diagrams result in quantum corrections to the Lagrangian of a theory, arising both for interactions explicit in the Lagrangian and new interactions created as a result of the corrections themselves \cite{cw}. These quantum corrections can affect the nature of the vacuum of the theory, and so must be considered in order to ensure that we are working in a theory where the minima of the potential describe the ground state of the theory. 

In order to arrive at this potential - referred to as the effective potential \cite{cw} - we must first define the effective action. We first consider the $\phi^{4}-$theory of a single scalar field $\phi$, described by a Lagrangian density $\mathcal{L}\left(\phi, \partial_{\mu}\phi \right)$, and add a coupling to an external source, $J\left(x \right)$

\begin{equation}\label{eqn:b63}
\mathcal{L} = \frac{1}{2}\partial_{\mu}\phi \partial^{\mu}\phi - \frac{\lambda}{4!}\phi^{4} + J\left(x \right)\phi \left(x \right).
\end{equation}

\noindent The generating functional, $W\left[J\right]$, is defined using

\begin{equation}\label{eqn:b64}
e^{iW\left[J\right]} = \langle 0^{+} \mid 0^{-} \rangle_{J},
\end{equation}

\noindent where the right hand side is the transition amplitude between the vacuum in the far past and the vacuum in the far future in the presence of the source, $J$. The generating functional can be expanded as

\begin{equation}\label{eqn:b65}
W = \sum_{n} \frac{1}{n!}\int d^{4}x_{1}...d^{4}x_{n} \; G^{\left(n \right)}\left(x_{1}...x_{n}\right) J\left(x_{1}\right)...J\left(x_{n}\right),
\end{equation}

\noindent where $G^{n}$ are the connected Green's functions, a sum of the connected Feynman diagrams of the theory with $n$ external lines.

The effective action is defined by

\begin{equation}\label{eqn:b66}
\Gamma \left(\phi_{c}\right) = W\left[J\right] - \int d^{4}x J\left(x \right) \phi_{c}\left(x \right),
\end{equation}

\noindent where $\phi_{c}$ is defined by

\begin{equation}\label{eqn:b67}
\phi_{c}\left(x \right) = \frac{\delta W}{\delta J\left(x \right)} = \frac{\langle 0^{+}\left| \phi (x) \right| 0^{-} \rangle_{J}}{\langle 0^{+}\mid 0^{-} \rangle_{J}},
\end{equation}

\noindent and represents the classical state of the field in the theory. The effective action can be expanded in a functional Taylor series in powers of $\phi_{c}$

\begin{equation}\label{eqn:b68}
\Gamma = \sum_{n} \frac{1}{n!}\int d^{4}x_{1}...d^{4}x_{n} \; \Gamma^{\left(n \right)}\left(x_{1}...x_{n}\right) \phi_{c}\left(x_{1}\right)...\phi_{c}\left(x_{n}\right),
\end{equation}

\noindent where $\Gamma^{\left(n \right)}$ is a sum of the one-particle irreducible (1PI - connected diagrams that cannot be disconnected by cutting a single internal line) Feynman diagrams with $n$ external lines. Alternatively, expanding the effective action in powers of momentum gives \cite{cw}

\begin{equation}\label{eqn:b69}
\Gamma = \int d^{4}x \left[ -V\left(\phi_{c}\right) + \frac{1}{2}\left(\partial_{\mu}\phi_{c}\right)^{2}Z\left(\phi_{c}\right) + ... \right],
\end{equation}

\noindent where $Z$ is a function used in the wavefunction renormalisation and $V$ is the effective potential, whose $n$th derivative is the sum of all 1PI diagrams with $n$ zero-momentum external lines.

The following renormalisation conditions on the mass squared $m^{2}$ and the scalar self-coupling $\lambda$ are imposed \cite{cw}

\begin{equation}\label{eqn:b70}
m^{2} = \left.\frac{d^{2}V}{d\phi_{c}^{2}}\right|_{\phi_{c} = 0} = 0,
\end{equation}

\begin{equation}\label{eqn:b71}
\lambda = \left.\frac{d^{4}V}{d\phi_{c}^{4}}\right|_{\phi_{c} = \mu},
\end{equation}

\noindent where $\mu$ is some arbitrary mass scale, later used as a renormalisation scale. 

\noindent Following the renormalisation procedure \cite{cw}, we find that the effective potential in the scalar $\phi^{4}$ theory (\ref{eqn:b63}) is \cite{cw}

\begin{equation}\label{eqn:b72}
V = \frac{\lambda}{4!}\phi_{c}^{4} + \frac{\lambda^{2}\phi_{c}^{4}}{256\pi^{2}}\left[\ln\left(\frac{\phi_{c}^{2}}{\mu^{2}} - \frac{25}{6}\right) \right].
\end{equation}

\noindent The complete one-loop correction to the effective potential can be written as \cite{msher}

\begin{equation}\label{eqn:b73}
\Delta V_{CW}^{\left(1\right)} \left(\phi_{c}\right) = \sum_{i} \left(-1\right)^{F}n_{i} \frac{m_{i}^{4}\left(\phi_{c}\right)}{64\pi^{2}}\ln \left(\frac{m_{i}^{2}\left(\phi_{c}\right)}{\mu^{2}} \right),
\end{equation}

\noindent where the sum of $i$ is over all the fields of the theory, $n_{i}$ is the number of degrees of freedom of the field $i$, $F$ is even (odd) for bosons (fermions), and $m_{i}$ are the masses of the fields $i$ in the vacuum where $\phi = \phi_{c}$. $\mu$ is the renormalisation scale of the theory, which is chosen to minimise the logarithmic terms in the potential at a given scale.

This expression of the corrections to the effective potential (\ref{eqn:b73}) will be used in Chapter 4 when the effects of reheating on the inflation model presented therein are discussed.

\section{Noether's Theorem for a Scalar Field Theory}\label{section:t34}

In this section we introduce Noether's theorem for a scalar field theory and derive the energy-momentum tensor. Noether's theorem can be taken as the statement that every continuous symmetry of the Lagrangian of a field theory gives rise to an associated conserved quantity (current and charge). We will outline the theorem here for a scalar field theory in flat space as an illustrative example.

We begin this discussion with the action of a scalar field theory in Minkowski space

\begin{equation}\label{eqn:b74}
S = \int d^{4}x \; \; \mathcal{L}\left[\phi_{a}\left(x \right), \partial_{\mu}\phi_{a}\left(x\right) \right],
\end{equation}

\noindent where $\phi_{a}\left(x \right)$ are the scalar matter fields of the theory and the square root of the negative determinant is not shown explicitly in this calculation since in Minkowski space $\sqrt{-g} = \sqrt{-\eta} = 1$. If $\phi_{a}\left(x \right)$ are solutions to the equations of motion of the theory, then the action (\ref{eqn:b74}) will be extremised under variation with respect to $\phi_{a}\left(x \right)$ and their derivatives. The action (\ref{eqn:b74}) can also be written in terms of the spacetime coordinates, $x \equiv x^{\mu}$, by substituting the solutions $\phi_{a}\left(x \right)$

\begin{equation}\label{eqn:b75}
S = \int d^{4}x \; \; \mathcal{L}\left[x\right],
\end{equation}

\noindent where the two expressions of the action are equivalent. To illustrate the proof of Noether's theorem, we consider an infinitesimal symmetry transformation in the spacetime coordinates

\begin{equation}\label{eqn:b76}
x^{\mu} \rightarrow x'^{\mu} = x^{\mu} + \delta x^{\mu} \equiv x + \delta x,
\end{equation}

\noindent where for the purposes of this discussion we restrict the spacetime symmetry transformation to constant coordinate shifts, i.e. $\delta x^{\mu}$ represents a constant shift in the coordinates and is not a function of $x^{\mu}$. We also consider an infinitesimal internal symmetry transformation on the fields, $\phi_{a}\left(x \right)$, and we define

\begin{equation}\label{eqn:b77}
\phi_{a}\left(x \right) \rightarrow \phi_{a}'\left(x' \right) = \phi_{a}\left(x \right) + \delta \phi_{a}\left(x \right),
\end{equation}

\noindent as being the total symmetry transformation (both internal and spacetime symmetries) of the Lagrangian. For a purely spacetime transformation, $\phi_{a}'\left(x'\right) = \phi_{a}\left(x'\right)$.

A symmetry transformation means that the transformed fields $\phi_{a}'\left(x'\right)$ satisfy equations of motion of the same form as those satisfied by $\phi_{a}\left(x\right)$. This will be true if the Lagrangian $\mathcal{L}'\left[\phi_{a}'\left(x'\right)\right]$ which gives the equations of motion for $\phi_{a}'\left(x'\right)$ is of the form

\begin{equation}\label{eqn:b78}
\mathcal{L}'\left[\phi_{a}'\left(x'\right)\right] = \mathcal{L}\left[\phi_{a}'\left(x'\right)\right],
\end{equation}

\noindent where $\mathcal{L}\left[\phi_{a}\left(x\right)\right]$ is the original Lagrangian, and if 

\begin{equation}\label{eqn:b79}
S' = \int d^{4}x' \; \; \mathcal{L}\left[\phi_{a}'\left(x'\right)\right] = \int d^{4}x \; \left[\mathcal{L}\left[\phi_{a}\left(x\right)\right] + \partial_{\mu}F^{\mu}\right] \equiv S + \delta S,
\end{equation}

\noindent where $F^{\mu} = F^{\mu}\left[\phi_{a}\left(x\right)\right]$. The total derivative is a surface term so $\delta S$ gives zero under variations, therefore when $\phi_{a}\left(x\right)$ extremises $S$, $\phi_{a}'\left(x'\right)$ extremises $S'$.

\noindent For an internal symmetry transformation we have 

\begin{equation}\label{eqn:b80}
\mathcal{L}'\left[x\right] = \mathcal{L}\left[\phi_{a}'\left(x \right)\right] = \mathcal{L}\left[\phi_{a}\left(x \right)\right] = \mathcal{L}\left[x\right].
\end{equation}

\noindent We therefore have

\begin{equation}\label{eqn:b81}
\delta \mathcal{L} = \mathcal{L}'\left[x\right] - \mathcal{L}\left[x\right] = 0,
\end{equation}

\noindent  and the change of the action due to the internal symmetry transformation is thus

\begin{equation}\label{eqn:b82}
\delta S = \int d^{4}x \; \; \delta \mathcal{L} = \int d^{4}x \; \; \mathcal{L}'\left[x\right] - \mathcal{L}\left[x\right] = 0,
\end{equation}

\noindent when $\phi_{a}'\left(x\right)$ is a solution to the equations of motion which extremises the transformed action $S'$, meaning that $\phi_{a}\left(x \right)$ must be a solution to the equations of motion of the untransformed action which extremises the action $S$.

The change of the action due to the spacetime transformation is

\begin{equation}\label{eqn:b83}
\delta S = \int d^{4}x' \; \; \mathcal{L}'\left[x'\right] - \int d^{4}x \; \; \mathcal{L}\left[x \right].
\end{equation}

\noindent Since we are only considering constant coordinate shifts in this discussion, we have that $d^{4}x' = d^{4}x$, and given (\ref{eqn:b80}) we have that, since $\phi_{a}'\left(x'\right) = \phi_{a}\left(x'\right)$, in this case,

\begin{equation}\label{eqn:b84}
\mathcal{L}'\left[x'\right] = \mathcal{L}\left[\phi_{a}'\left(x'\right)\right] = \mathcal{L}\left[\phi_{a}\left( x + \delta x \right)\right] = \mathcal{L}\left[x + \delta x \right],
\end{equation}

\noindent if $\phi_{a}\left(x \right)$ is a solution to the equations of motion of the untransformed action. The change of the action under a spacetime symmetry transformation (\ref{eqn:b83}) is therefore

\begin{equation}\label{eqn:b85}
\delta S = \int d^{4}x \; \; \mathcal{L}\left[x + \delta x\right] - \mathcal{L}\left[x \right] = \int d^{4}x \; \; \frac{\partial \mathcal{L}}{\partial x^{\mu}}\delta x^{\mu}.
\end{equation}

\noindent Since the coordinate shifts are constant, we can write (\ref{eqn:b85}) as a total derivative

\begin{equation}\label{eqn:b86}
\delta S = \int d^{4}x \; \; \partial_{\mu}\left(\mathcal{L}\delta x^{\mu}\right),
\end{equation}

\noindent where due to the fact that the transformed Lagrangian density has the same form as the original Lagrangian density, we can say that (\ref{eqn:b86}) represents the total change of the action (\ref{eqn:b74}) under internal and spacetime symmetry transformations.

Equivalently, writing the total change of the action in terms of the corresponding changes of the fields, we have

\begin{equation}\label{eqn:b87}
\delta S = \int d^{4}x' \; \; \mathcal{L}'\left[x'\right] - \int d^{4}x \; \; \mathcal{L}\left[x\right],
\end{equation}

\begin{equation}\label{eqn:b88}
\Rightarrow \delta S = \int d^{4}x \; \; \mathcal{L}\left[\phi_{a}'\left(x' \right), \partial_{\mu}\phi_{a}'\left(x' \right)\right] - \mathcal{L}\left[\phi_{a}\left(x \right), \partial_{\mu}\phi_{a}\left(x \right)\right],
\end{equation}

\noindent where we have used $\mathcal{L}'\left[\phi_{a}'\left(x'\right), \partial_{\mu}\phi_{a}'\left(x'\right)\right] = \mathcal{L}\left[\phi_{a}'\left(x'\right), \partial_{\mu}\phi_{a}'\left(x'\right)\right]$ since we are assuming a symmetry. Therefore

\begin{equation}\label{eqn:b89}
\begin{split}
\delta S = \int d^{4}x  & \; \; \mathcal{L}\left[\phi_{a}\left(x \right) + \delta \phi_{a}\left(x \right), \partial_{\mu}\phi_{a}\left(x \right) + \delta \left(\partial_{\mu}\phi_{a}\left(x \right)\right)\right]  \\
& - \mathcal{L}\left[\phi_{a}\left(x \right), \partial_{\mu}\phi_{a}\left(x \right)\right],
\end{split}
\end{equation}

\begin{equation}\label{eqn:b90}
\Rightarrow \delta S = \int d^{4}x \; \; \frac{\partial \mathcal{L}}{\partial \phi_{a}}\delta \phi_{a} + \frac{\partial \mathcal{L}}{\partial \left(\partial_{\mu}\phi_{a}\right)}\delta \left(\partial_{\mu}\phi_{a}\right).
\end{equation}

We now have that (\ref{eqn:b86}) and (\ref{eqn:b90}) are both expressions of the total change of the action under internal and spacetime symmetry transformations, and they must therefore be equal

\begin{equation}\label{eqn:b91}
\int d^{4}x \; \; \partial_{\mu}\left(\mathcal{L}\delta x^{\mu}\right) = \int d^{4}x \; \; \frac{\partial \mathcal{L}}{\partial \phi_{a}}\delta \phi_{a} + \frac{\partial \mathcal{L}}{\partial \left(\partial_{\mu}\phi_{a}\right)}\delta \left(\partial_{\mu}\phi_{a}\right),
\end{equation}

\noindent such that 

\begin{equation}\label{eqn:b92}
\int d^{4}x \; \; \frac{\partial \mathcal{L}}{\partial \phi_{a}}\delta \phi_{a} + \frac{\partial \mathcal{L}}{\partial \left(\partial_{\mu}\phi_{a}\right)}\delta \left(\partial_{\mu}\phi_{a}\right) - \partial_{\mu}\left(\mathcal{L}\delta x^{\mu}\right) = 0.
\end{equation}

\noindent Rewriting the second term in the integrand of (\ref{eqn:b92}) as half of a product rule, and rewriting the infinitesimal change in the derivative of the fields, $\delta \left(\partial_{\mu}\phi_{a}\right)$, we can write 

\begin{equation}\label{eqn:b93}
\frac{\partial \mathcal{L}}{\partial \left(\partial_{\mu}\phi_{a}\right)}\delta \left(\partial_{\mu}\phi_{a}\right) = \frac{\partial \mathcal{L}}{\partial \left(\partial_{\mu}\phi_{a}\right)}\partial_{\mu}\left(\delta \phi_{a}\right),
\end{equation}

\begin{equation}\label{eqn:b94}
\frac{\partial \mathcal{L}}{\partial \left(\partial_{\mu}\phi_{a}\right)}\partial_{\mu}\left(\delta \phi_{a}\right) = \partial_{\mu}\left[\frac{\partial \mathcal{L}}{\partial \left(\partial_{\mu}\phi_{a}\right)}\delta \phi_{a}\right] - \partial_{\mu}\left(\frac{\partial \mathcal{L}}{\partial \left(\partial_{\mu}\phi_{a}\right)}\right)\delta \phi_{a}.
\end{equation}

\noindent Thus (\ref{eqn:b92}) becomes

\begin{equation}\label{eqn:b95}
\int d^{4}x \; \; \left[\frac{\partial \mathcal{L}}{\partial \phi_{a}} - \partial_{\mu}\left(\frac{\partial \mathcal{L}}{\partial \left(\partial_{\mu}\phi_{a}\right)}\right)\right]\delta \phi_{a} + \partial_{\mu}\left[\frac{\partial \mathcal{L}}{\partial \left(\partial_{\mu}\phi_{a}\right)}\delta \phi_{a} - \mathcal{L}\delta x^{\mu}\right] = 0.
\end{equation}

\noindent We have

\begin{equation}\label{eqn:b96}
\frac{\partial \mathcal{L}}{\partial \phi_{a}} - \partial_{\mu}\left(\frac{\partial \mathcal{L}}{\partial \left(\partial_{\mu}\phi_{a}\right)}\right) = 0,
\end{equation}

\noindent for $\phi_{a}\left(x\right)$ being a solution to the equations of motion. Therefore, if $\phi_{a}\left(x\right)$ is a solution to the field equations, then due to the symmetry transformation it follows that

\begin{equation}\label{eqn:b97}
\partial_{\mu}\left[\frac{\partial \mathcal{L}}{\partial \left(\partial_{\mu}\phi_{a}\right)}\delta \phi_{a} - \mathcal{L}\delta x^{\mu}\right] = 0,
\end{equation}

\noindent Thus, $\partial_{\mu}j^{\mu} = 0$, where we define

\begin{equation}\label{eqn:b98}
j^{\mu} = \frac{\partial \mathcal{L}}{\partial \left(\partial_{\mu}\phi_{a}\right)}\delta \phi_{a} - \mathcal{L}\delta x^{\mu},
\end{equation}

\noindent as the conserved Noether currents arising from the symmetry transformations (\ref{eqn:b76}) and (\ref{eqn:b77}).

The conserved four-current $j^{\mu}$ is given by

\begin{equation}\label{eqn:b99}
j^{\mu} = \left(\rho_{Q}, -\textbf{j}\right),
\end{equation}

\noindent where the $0$-component of the Noether current corresponds to a charge density. This means that every conserved Noether current has an associated charge

\begin{equation}\label{eqn:b100}
Q = \int d^{3}x \; \; j^{0} = \int d^{3}x \; \; \rho_{Q},
\end{equation}

\noindent where the charge can be shown to be globally conserved by taking a time derivative of (\ref{eqn:b100})

\begin{equation}\label{eqn:b101}
\frac{dQ}{dt} = \int d^{3}x \; \; \frac{d j^{0}}{dt} =  -\int d^{3}x \; \; \overrightarrow{\nabla} \cdot \textbf{j} = 0.
\end{equation}

\subsubsection{The Energy-Momentum Tensor}

As a specific and relevant example of the application of Noether's theorem, we can consider the case of energy and momentum conservation in a scalar field theory in flat space. Conservation of energy and momentum requires that the Lagrangian of the theory be invariant under spacetime translations, $x^{\mu} \rightarrow x^{\mu} + \delta x^{\mu} = x + \delta x$, and the variation of the fields is therefore

\begin{equation}\label{eqn:b102}
\delta \phi_{a} = \phi_{a}\left(x + \delta x \right) - \phi_{a}\left(x\right) = \frac{\partial \phi_{a}}{\partial x^{\nu}}\delta x^{\nu},
\end{equation}

\noindent so the conserved current $j^{\mu}$ in this case is

\begin{equation}\label{eqn:b103}
j^{\mu} = \frac{\partial \mathcal{L}}{\partial \left(\partial_{\mu}\phi_{a}\right)}\frac{\partial \phi_{a}}{\partial x^{\nu}}\delta x^{\nu} - \mathcal{L}\delta x^{\nu}\delta^{\mu}_{\nu},
\end{equation}

\begin{equation}\label{eqn:b104}
j^{\mu} = \left(\frac{\partial \mathcal{L}}{\partial \left(\partial_{\mu}\phi_{a}\right)}\frac{\partial \phi_{a}}{\partial x^{\nu}} - \mathcal{L}\delta^{\mu}_{\nu}\right)\delta x^{\nu}.
\end{equation}

\noindent The energy-momentum tensor is then defined as 

\begin{equation}\label{eqn:b105}
T^{\mu}_{\nu} = \frac{\partial \mathcal{L}}{\partial \left(\partial_{\mu}\phi_{a}\right)}\frac{\partial \phi_{a}}{\partial x^{\nu}} - \mathcal{L}\delta^{\mu}_{\nu},
\end{equation}

\noindent where 

\begin{equation}\label{eqn:b106}
\partial_{\mu}j^{\mu} = 0 \Rightarrow \partial_{\mu} T^{\mu}_{\nu} = 0,
\end{equation}

\noindent and this is the statement of energy and momentum conservation.

\subsection{Noether's Theorem for a Complex Scalar Field Theory Charged Under a $U(1)$ Symmetry.}\label{section:t35}

Chapters 5 and 6 both deal with complex scalar field theories charged under a $U(1)$ symmetry, so it is pertinent here to address the general treatment of a $U(1)$ complex scalar field theory. For consistency, we will continue with this example in flat space.

A $U(1)$ field transformation for a complex field is given by

\begin{equation}\label{eqn:b107}
\Phi \rightarrow e^{i\alpha}\Phi, \; \; \; \Phi^{\dagger} \rightarrow e^{-i\alpha}\Phi^{\dagger},
\end{equation}

\noindent where a Lagrangian is $U(1)$-symmetric if it is invariant under these transformations. The general $U(1)$-symmetric Lagrangian density we will address in this work is

\begin{equation}\label{eqn:b108}
\mathcal{L} = \partial_{\mu}\Phi^{\dagger} \partial^{\mu}\Phi - V\left( \left| \Phi \right| \right).
\end{equation}

\noindent For small $\alpha$, the $U(1)$ transformation (\ref{eqn:b107}) is

\begin{equation}\label{eqn:b109}
\Phi \rightarrow \left(1 + i\alpha \right)\Phi = \Phi + \delta \Phi, \; \; \; \Phi^{\dagger} \rightarrow \left(1 - i\alpha \right)\Phi^{\dagger} = \Phi^{\dagger} + \delta \Phi^{\dagger},
\end{equation}

\noindent and in the case of a $U(1)$ symmetry, $\delta x^{\mu} = 0$, so there are no spacetime symmetries and we only need to deal with the variation in the fields when extremising the action of the theory. 

In order to extremise the action, we must vary the action with respect to the fields undergoing the transformations, in this case $\Phi$ and $\Phi^{\dagger}$, so the variation of the action with Lagrangian density (\ref{eqn:b108}) should then be

\begin{equation}\label{eqn:b110}
\delta S = \int d^{4}x \; \; \frac{\delta \mathcal{L}}{\delta \Phi}\delta \Phi + \frac{\delta \mathcal{L}}{\delta \Phi^{\dagger}}\delta \Phi^{\dagger} + \frac{\delta \mathcal{L}}{\delta \left(\partial_{\mu}\Phi\right)}\delta \left(\partial_{\mu} \Phi \right) + \frac{\delta \mathcal{L}}{\delta \left(\partial_{\mu}\Phi^{\dagger} \right)}\delta \left(\partial_{\mu} \Phi^{\dagger} \right).
\end{equation}

\noindent However, $\Phi$ and $\Phi^{\dagger}$ cannot be varied independently because they are conjugates of the same field. The issue can be expressed more clearly if we write the fields as

\begin{equation}\label{eqn:b111}
\Phi = \frac{1}{\sqrt{2}}\left( \phi_{1} + i\phi_{2}\right), \; \; \; \; \Phi^{\dagger} = \frac{1}{\sqrt{2}}\left(\phi_{1} - i\phi_{2}\right).
\end{equation}

\noindent Intuitively, it would seem that these fields cannot be varied independently of each other. In order to verify the method of variation for a scalar field theory we will first demonstrate Coleman's proof for the variation of complex fields \cite{colemanqft} to confirm this. 

The third and fourth terms of the integrand in (\ref{eqn:b110}) can be rewritten

\begin{equation}\label{eqn:b112}
\frac{\partial \mathcal{L}}{\partial \left(\partial_{\mu}\Phi\right)}\partial_{\mu} \left(\delta \Phi \right) = \partial_{\mu}\left(\frac{\partial \mathcal{L}}{\partial \left(\partial_{\mu}\Phi\right)}\delta \Phi \right) - \partial_{\mu}\left(\frac{\partial \mathcal{L}}{\partial \left(\partial_{\mu}\Phi\right)} \right)\delta \Phi,
\end{equation}

\begin{equation}\label{eqn:b113}
\frac{\partial \mathcal{L}}{\partial \left(\partial_{\mu}\Phi^{\dagger} \right)}\partial_{\mu} \left(\delta \Phi^{\dagger} \right) = \partial_{\mu}\left(\frac{\partial \mathcal{L}}{\partial \left(\partial_{\mu}\Phi^{\dagger} \right)}\delta \Phi^{\dagger} \right) - \partial_{\mu}\left(\frac{\partial \mathcal{L}}{\partial \left(\partial_{\mu}\Phi^{\dagger}\right)} \right)\delta \Phi^{\dagger}.
\end{equation}

\noindent The Lagrangian density can therefore be written as a sum of a $\Phi$ variation and a $\Phi^{\dagger}$ variation

\begin{equation}\label{eqn:b114}
\delta S = \int d^{4}x \left[ B\delta \Phi + B^{\dagger}\delta \Phi^{\dagger} \right],
\end{equation}

\noindent where $B$ and $B^{\dagger}$ are expressions to be determined. As aforementioned, $\Phi$ and $\Phi^{\dagger}$ are Hermitian conjugates, so the variations of each field are not independent of each other, and the fields cannot be varied as though they are independent.

Instead, we can use the forms of the field (\ref{eqn:b111}) and vary the action in terms of the real and imaginary parts of the complex fields

\begin{equation}\label{eqn:b115}
\delta \Phi = \frac{1}{\sqrt{2}}\left( \delta \phi_{1} + i\delta \phi_{2}\right), \; \; \; \delta \Phi^{\dagger} = \frac{1}{\sqrt{2}}\left( \delta \phi_{1} - i\delta \phi_{2}\right),
\end{equation}

\begin{equation}\label{eqn:b116}
Re\left[\delta \Phi \right] = \frac{\delta \phi_{1}}{\sqrt{2}} = Re\left[\delta \Phi^{\dagger} \right],
\end{equation}

\begin{equation}\label{eqn:b117}
Im\left[\delta \Phi \right] = \frac{\delta \phi_{2}}{\sqrt{2}} = -\left( -\frac{\delta \phi_{2}}{\sqrt{2}}\right) = -Im\left[\delta \Phi^{\dagger} \right].
\end{equation}

\noindent Writing the variation of the action in terms of $\phi_{1}$ and $\phi_{2}$, we have

\begin{equation}\label{eqn:b118}
\begin{split}
\delta S = \int d^{4}x & \left[\frac{B}{\sqrt{2}}\left(\delta \phi_{1} + i\delta \phi_{2} \right) + \frac{B^{\dagger}}{\sqrt{2}}\left(\delta \phi_{1} - i\delta \phi_{2} \right) \right] \\
& = \int d^{4}x \left[\frac{\delta \phi_{1}}{\sqrt{2}}\left(B + B^{\dagger} \right) + \frac{i\delta \phi_{2}}{\sqrt{2}}\left(B - B^{\dagger} \right) \right].
\end{split}
\end{equation}

\noindent The variations of the field are non-zero, so we need that

\begin{equation}\label{eqn:b119}
\frac{\delta \phi_{1}}{\sqrt{2}} \neq 0 \Rightarrow B + B^{\dagger} = 0 \Rightarrow B = -B^{\dagger},
\end{equation}

\begin{equation}\label{eqn:b120}
\frac{i\delta \phi_{2}}{\sqrt{2}} \neq 0 \Rightarrow B - B^{\dagger} = 0 \Rightarrow B = B^{\dagger},
\end{equation}

\noindent in order for $\delta S = 0$, as well as the total derivative terms in (\ref{eqn:b112}) and (\ref{eqn:b113}) being equal to zero. In order for (\ref{eqn:b119}) and (\ref{eqn:b120}) to be true, we must have

\begin{equation}\label{eqn:b121}
B = B^{\dagger} = 0.
\end{equation}

\noindent This gives independent Euler-Lagrange equations for $\Phi$ and $\Phi^{\dagger}$

\begin{equation}\label{eqn:b122}
B = \frac{\partial \mathcal{L}}{\partial \Phi} - \partial_{\mu}\left( \frac{\partial \mathcal{L}}{\partial \left(\partial_{\mu}\Phi\right)}\right) = 0,
\end{equation}

\begin{equation}\label{eqn:b123}
B^{\dagger} = \frac{\partial \mathcal{L}}{\partial \Phi^{\dagger}} - \partial_{\mu}\left( \frac{\partial \mathcal{L}}{\partial \left(\partial_{\mu}\Phi^{\dagger}\right)}\right) = 0,
\end{equation}

\noindent which gives independent field equations for $\Phi$ and $\Phi^{\dagger}$. This shows that, although technically the fields $\Phi$ and $\Phi^{\dagger}$ cannot be varied independently of each other, it is acceptable to treat them as independent fields when deriving the field equations for the $U(1)$ complex scalar field theory because the end result gives independent field equations for $\Phi$ and $\Phi^{\dagger}$.

Substituting (\ref{eqn:b122}) and (\ref{eqn:b123}) into (\ref{eqn:b114}) we find that the remaining terms in the extremised action are the total derivatives

\begin{equation}\label{eqn:b124}
\partial_{\mu}\left(\frac{\partial \mathcal{L}}{\partial \left(\partial_{\mu}\Phi\right)}\delta \Phi + \frac{\partial \mathcal{L}}{\partial \left(\partial_{\mu}\Phi^{\dagger}\right)}\delta \Phi^{\dagger} \right) = 0.
\end{equation}

\noindent The terms inside the bracket correspond to a conserved current 

\begin{equation}\label{eqn:b125}
j^{\mu} = \frac{\partial \mathcal{L}}{\partial \left(\partial_{\mu}\Phi\right)}\delta \Phi + \frac{\partial \mathcal{L}}{\partial \left(\partial_{\mu}\Phi^{\dagger}\right)}\delta \Phi^{\dagger},
\end{equation}

\noindent which for $\delta S = 0$ gives conservation of the $U(1)$ current $\partial_{\mu}j^{\mu} = 0$. This is the conserved Noether current associated with the $U(1)$ symmetry of the action.

The $0$-component of the conserved current is equal to the charge density $\rho_{Q}$ of the conserved $U(1)$ current, 

\begin{equation}\label{eqn:b126}
j^{0} = \frac{\partial \mathcal{L}}{\partial \left(\partial_{t}\Phi \right)} i\alpha\Phi - \frac{\partial \mathcal{L}}{\partial \left(\partial_{t} \Phi^{\dagger}\right)} i\alpha\Phi^{\dagger},
\end{equation}

\noindent where the transformation (\ref{eqn:b109}) has been substituted. For a general $U(1)$ Lagrangian of a complex scalar field theory

\begin{equation}\label{eqn:b127}
\mathcal{L} = \partial_{\mu}\Phi^{\dagger}\partial^{\mu}\Phi - V\left(\left|\Phi \right| \right),
\end{equation}

\noindent (\ref{eqn:b126}) is then

\begin{equation}\label{eqn:b128}
j^{0} = \rho_{Q} = i\alpha \left( \Phi \partial_{t}\Phi^{\dagger} - \Phi^{\dagger}\partial_{t}\Phi \right),
\end{equation}

\noindent and the global $U(1)$ charge of the scalar field is given by

\begin{equation}\label{eqn:b129}
Q = i\alpha \int d^{3}x  \left( \Phi \partial_{t}\Phi^{\dagger} - \Phi^{\dagger}\partial_{t}\Phi \right).
\end{equation}

\section{Global U(1) Charge as the charge asymmetry}\label{section:t36}

In this section we derive the charge of a complex scalar field charged under a $U(1)$ symmetry. In order to do this we must correctly normalise the $U(1)$ charge, $Q$, so that each complex scalar has charge $\pm1$. Non-boldfaced coordinate arguments (with the exception of time) denote four-vectors. 

For a canonically quantised scalar field theory, we upgrade the fields, $\phi$, to operators, $\hat{\phi}$, and  define the canonical conjugate momentum, $\Pi \left(t, \textbf{x} \right)$

\begin{equation}\label{eqn:b130}
\hat{\Pi} \left(t, \textbf{x} \right) = \frac{\partial \mathcal{L}}{\partial \left(\partial_{\mu}\hat{\Phi} \right)} = \hat{\dot{\Phi}} \left(t, \textbf{x} \right).
\end{equation}

\noindent We impose the equal time canonical commutation rule

\begin{equation}\label{eqn:b131}
\left[ \hat{\Phi} \left(t, \textbf{x} \right), \hat{\Pi} \left(t, \textbf{y} \right)\right] = i\delta^{\left(3\right)}\left(\textbf{x} - \textbf{y}\right),
\end{equation}

\noindent where at equal time we also have

\begin{equation}\label{eqn:b132}
\left[ \hat{\Phi} \left(t, \textbf{x} \right), \hat{\Phi} \left(t, \textbf{y} \right)\right] = \left[ \hat{\Pi} \left(t, \textbf{x} \right), \hat{\Pi} \left(t, \textbf{y} \right)\right] = 0.
\end{equation}

\noindent We can write the scalar field operator $\hat{\Phi} \left(t, \textbf{x} \right)$ as a mode expansion of time-independent creation and annihilation operators, $\hat{a}^{\dagger}_{\textbf{p}}$ and $\hat{a}_{\textbf{p}}$ \cite{maggiore}

\begin{equation}\label{eqn:b133}
\hat{\Phi}\left( x \right) = \int \frac{d^{3}p}{\left(2\pi\right)^{3} \sqrt{2E_{\textbf{p}}}} \left[\hat{a}_{\textbf{p}} e^{-ip\cdot x} + \hat{b}^{\dagger}_{\textbf{p}} e^{ip\cdot x}  \right],
\end{equation}

\begin{equation}\label{eqn:b134}
\hat{\Phi}^{\dagger}\left( x \right) = \int \frac{d^{3}q}{\left(2\pi\right)^{3} \sqrt{2E_{\textbf{q}}}} \left[\hat{a}^{\dagger}_{\textbf{q}} e^{iq\cdot x} + \hat{b}_{\textbf{q}} e^{-iq\cdot x}  \right],
\end{equation}

\noindent where the $\hat{a}$ operators correspond to particles and the $\hat{b}$ operators correspond to antiparticles, and $p$ and $q$ are the four momenta of the $\Phi$ and $\Phi^{\dagger}$ particles respectively. 

\noindent The annihilation operator defines the vacuum by 

\begin{equation}\label{eqn:b135}
\hat{a} \mid 0 \rangle = 0,
\end{equation}

\noindent and excited particle states are produced by repeat application of the creation operator

\begin{equation}\label{eqn:b136}
\mid n \rangle = \frac{1}{\sqrt{n\!}}\left(\hat{a}^{\dagger}\right)^{n}\mid 0 \rangle.
\end{equation}

\noindent Substituting (\ref{eqn:b133}) and (\ref{eqn:b134}) into (\ref{eqn:b130}) and (\ref{eqn:b131}) we find that the commutation relation in terms of  the creation and annihilation operators is

\begin{equation}\label{eqn:b137}
\left[\hat{a}_{\textbf{p}}, \hat{a}^{\dagger}_{\textbf{q}} \right] = \left[\hat{b}_{\textbf{p}}, \hat{b}^{\dagger}_{\textbf{q}} \right] = \left(2\pi\right)^{3}\delta^{\left(3\right)}\left(\textbf{p} - \textbf{q}\right),
\end{equation}

\noindent and commutators of all other combinations of creation and annihilation operators are zero.

The $U(1)$ charge $Q$ is given by (\ref{eqn:b129}). Substituting in (\ref{eqn:b133}) and (\ref{eqn:b134}) the charge operator can be defined as

\begin{multline}\label{eqn:b138}
\hat{Q} = i\alpha \int d^{3}x  \frac{d^{3}p}{\left(2\pi\right)^{3} \sqrt{2E_{\textbf{p}}}}\frac{d^{3}q}{\left(2\pi\right)^{3} \sqrt{2E_{\textbf{q}}}} \left[\left(\hat{a}^{\dagger}_{\textbf{q}} e^{iq\cdot x} - \hat{b}_{\textbf{q}} e^{-iq\cdot x}\right)\partial_{t}\left( \hat{a}_{\textbf{p}} e^{-ip\cdot x} + \hat{b}^{\dagger}_{\textbf{p}} e^{ip\cdot x}\right)\right. \\ 
\left. - \partial_{t}\left(\hat{a}^{\dagger}_{\textbf{q}} e^{iq\cdot x} + \hat{b}_{\textbf{q}} e^{-iq\cdot x} \right) \left(\hat{a}_{\textbf{p}} e^{-ip\cdot x} - \hat{b}^{\dagger}_{\textbf{p}} e^{ip\cdot x}\right)\right].
\end{multline}

\noindent The four-momenta $p,q$ are given by $p = \left(E_{\textbf{p}}, -\textbf{p}\right), q = \left(E_{\textbf{q}}, -\textbf{q}\right)$ so the time derivatives of the exponentials give

\begin{equation}\label{eqn:b139}
\partial_{t} e^{ip\cdot x} = iE_{\textbf{p}}e^{ip\cdot x},
\end{equation}

\noindent and the charge operator is then

\begin{multline}\label{eqn:b140}
\hat{Q} = \alpha \int d^{3}x  \frac{d^{3}p}{\left(2\pi\right)^{3} \sqrt{2E_{\textbf{p}}}}\frac{d^{3}q}{\left(2\pi\right)^{3} \sqrt{2E_{\textbf{q}}}} \left[E_{\textbf{q}}\left(\hat{a}^{\dagger}_{\textbf{q}} e^{iq\cdot x} + \hat{b}_{\textbf{q}} e^{-iq\cdot x}\right)\left( \hat{a}_{\textbf{p}} e^{-ip\cdot x} + \hat{b}^{\dagger}_{\textbf{p}} e^{ip\cdot x}\right) \right. \\
\left.- E_{\textbf{p}}\left(\hat{a}^{\dagger}_{\textbf{q}} e^{iq\cdot x} + \hat{b}_{\textbf{q}} e^{-iq\cdot x} \right) \left(\hat{a}_{\textbf{p}} e^{-ip\cdot x} + \hat{b}^{\dagger}_{\textbf{p}} e^{ip\cdot x}\right)\right].
\end{multline}

\noindent Expanding this and integrating over $x$, the $\hat{a}\hat{b}$ terms acquire delta functions of the type $\left(2\pi\right)^{3}\delta^{3}\left(\textbf{p} + \textbf{q}\right)$. For $\left| \textbf{p}\right| = \left|\textbf{q}\right|$, we have $E_{\textbf{p}} = E_{\textbf{q}}$ and these terms cancel upon performing the integration over $q$. The remaining terms ($\hat{a}^{\dagger}\hat{a}, \hat{b}\hat{b}^{\dagger}$) acquire delta functions of the form $\left(2\pi\right)^{3}\delta^{3}\left(\textbf{p} - \textbf{q}\right)$ from the integration of the exponential factors. The remaining terms are then

\begin{equation}\label{eqn:b141}
\hat{Q} = \alpha \int \frac{d^{3}p}{\left(2\pi\right)^{3} \sqrt{2E_{\textbf{p}}}}\frac{1}{\left(2\pi\right)^{3} \sqrt{2E_{\textbf{p}}}} E_{\textbf{p}} \left[ 2 \hat{a}_{\textbf{p}}^{\dagger}\hat{a}_{\textbf{p}}\left(2\pi\right)^{3} - 2\hat{b}_{\textbf{p}}\hat{b}^{\dagger}_{\textbf{p}}\left(2\pi\right)^{3} \right],
\end{equation}

\begin{equation}\label{eqn:b142}
\Rightarrow \hat{Q} = \alpha \int \frac{d^{3}p}{\left(2\pi\right)^{3}}\left[ \hat{a}_{\textbf{p}}^{\dagger}\hat{a}_{\textbf{p}} - \hat{b}_{\textbf{p}}\hat{b}^{\dagger}_{\textbf{p}} \right].
\end{equation}

\noindent Normally ordering this expression we find

\begin{equation}\label{eqn:b143}
\hat{Q} = \alpha \int \frac{d^{3}p}{\left(2\pi\right)^{3}}\left[ \hat{a}_{\textbf{p}}^{\dagger}\hat{a}_{\textbf{p}} - \hat{b}^{\dagger}_{\textbf{p}}\hat{b}_{\textbf{p}} \right].
\end{equation}

\noindent Since $\hat{a}^{\dagger}\hat{a}$ is the number operator for a harmonic oscillator 

\begin{equation}\label{eqn:b144}
\hat{N}_{a} = \int \frac{d^{3}p}{\left(2\pi\right)^{3}} \hat{a}_{\textbf{p}}^{\dagger}\hat{a}_{\textbf{p}},
\end{equation}

\noindent where

\begin{equation}\label{eqn:b145}
\hat{N} \mid \textbf{p}_{1}, ..., \textbf{p}_{n} \rangle = n \mid \textbf{p}_{1}, ..., \textbf{p}_{n} \rangle \Rightarrow \hat{N} \mid n \rangle = n \mid n \rangle,
\end{equation}

\noindent we find that

\begin{equation}\label{eqn:b146}
\hat{Q} = \alpha\left(\hat{N}_{a} - \hat{N}_{b}\right).
\end{equation}

\noindent We have that the $U\left(1 \right)$ charge is given by the difference between the number of particles produced by the creation operator $\hat{a}^{\dagger}_{\textbf{p}}$ and the number of antiparticles created by $\hat{b}^{\dagger}_{\textbf{p}}$ integrated over all momenta $p$, multiplied by $\alpha$.

\noindent Therefore, for $\alpha = 1$, (\ref{eqn:b129}) gives the correctly normalised charge for particles with $Q = \pm 1$. This will be used in Chapter 5 in order to calculate the charge asymmetry present in a condensate of $U(1)$ charged scalars.

\section{Q-balls}\label{section:t37}

A Q-ball is a non-topological soliton which minimises the energy of the field for a fixed global charge. The derivation for conventional Q-balls in flat space as originally derived in \cite{coleman85} will be outlined here to allow a comparison to the derivation presented in Chapter 6 of a new class of Q-ball obtained as part of the original work of this thesis.

In order to derive the Q-ball equation we begin by using the method of Lagrange multipliers to minimise the energy of the scalar field $\Phi$ for a fixed conserved $U(1)$ charge $Q$ carried by the scalar field. This is expressed as 

\begin{equation}\label{eqn:b147}
E_{Q} = E + \omega \left( Q - \int d^{3}x \; \rho_{Q} \right),
\end{equation}

\noindent where $\omega$ is the Lagrange parameter, whose physical interpretation will be revisited later. From (\ref{eqn:b100}) the global charge $Q$ is

\begin{equation}\label{eqn:b148}
Q = \int d^{3}x \;  j^{0} = \int d^{3}x \; \rho_{Q},
\end{equation}

\noindent and the global energy is given by the temporal component of the energy-momentum tensor integrated over all space

\begin{equation}\label{eqn:b149}
E = \int d^{3}x \; T^{00} = \int d^{3}x \; \rho_{E},
\end{equation}

\noindent The energy-momentum tensor is given by

\begin{equation}\label{eqn:b150}
T^{\mu \nu} = \frac{\partial \mathcal{L}}{\partial \left(\partial_{\mu}\phi_{a}\right)}\eta^{\nu \rho}\partial_{\rho}\phi_{a} - \delta^{\mu}_{\rho} \eta^{\nu \rho}\mathcal{L}
\end{equation}

\noindent where $\eta^{\nu \rho}$ is the Minkowski metric. The energy density is therefore

\begin{equation}\label{eqn:b151}
\rho_{E} = T^{00} = \partial_{t}\Phi^{\dagger} \partial_{t}\Phi + \partial_{i}\Phi^{\dagger} \partial^{i}\Phi + V(\mid \Phi \mid).
\end{equation}

\noindent The conserved $U(1)$ current is given by

\begin{equation}\label{eqn:b152}
j^{\mu} = \frac{\partial \mathcal{L}}{\partial \left(\partial_{\mu}\phi_{a}\right)}\delta \phi_{a},
\end{equation}

\noindent where $\partial_{\mu} j^{\mu} = 0$ is the condition for the current conservation. The $0$ component of the conserved current is

\begin{equation}\label{eqn:b153}
j^{0} = \frac{\partial \mathcal{L}}{\partial \left(\partial_{t}\Phi \right)} i\Phi - \frac{\partial \mathcal{L}}{\partial \left(\partial_{t} \Phi^{\dagger}\right)} i\Phi^{\dagger},
\end{equation}

\noindent and the charge density is therefore

\begin{equation}\label{eqn:b154}
\rho_{Q} = i\left( \Phi \partial_{t}\Phi^{\dagger} - \Phi^{\dagger}\partial_{t}\Phi \right).
\end{equation}

\noindent Substituting (\ref{eqn:b151}) and (\ref{eqn:b154}) into (\ref{eqn:b147}) we obtain

\begin{equation}\label{eqn:b155}
E_{Q} = \int d^{3}x \left[ \partial_{t}\Phi^{\dagger} \partial_{t}\Phi + \partial_{i}\Phi^{\dagger} \partial^{i}\Phi + V(\mid \Phi \mid) - i\omega\left( \Phi \partial_{t}\Phi^{\dagger} - \Phi^{\dagger}\partial_{t}\Phi \right) \right] + \omega Q.
\end{equation}

\noindent Expanding the time derivatives, we can write

\begin{equation}\label{eqn:b156}
\left| \partial_{t}\Phi - i\omega \Phi \right|^{2} = \partial_{t}\Phi \partial_{t}\Phi^{\dagger} + i\omega \Phi^{\dagger} \partial_{t}\Phi - i\omega \Phi \partial_{t} \Phi^{\dagger} - \omega^{2}\Phi^{\dagger}\Phi.
\end{equation}

\noindent Let

\begin{equation}\label{eqn:b157}
I = \partial_{t}\Phi^{\dagger}\partial_{t}\Phi + i\omega \Phi^{\dagger}\partial_{t}\Phi - i\omega \Phi \partial_{t}\Phi^{\dagger} =  \mid \partial_{t}\Phi - i\omega \Phi \mid^{2} - \omega^{2} \mid \Phi \mid^{2},
\end{equation}

\noindent and substitute into (\ref{eqn:b155}) to give the Q-ball energy functional 

\begin{equation}\label{eqn:b158}
E_{Q} = \int d^{3}x \left[ \mid \partial_{t}\Phi - i \omega \Phi \mid^{2} -  \omega^{2} \mid \Phi \mid^{2} + \partial_{i}\Phi ^{\dagger} \partial^{i} \Phi + V\left(\mid \Phi \mid \right) \right] + \omega Q.
\end{equation}

\noindent In order to extremise $E_{Q}$ we write

\begin{equation}\label{eqn:b159}
\Phi \left(x, t \right) = \Phi \left( x \right) e^{i \omega t} \longrightarrow \partial_{t} \Phi = i\omega \Phi.
\end{equation}

\noindent Imposing this condition on (\ref{eqn:b158}) we have

\begin{equation}\label{eqn:b160}
E_{Q} = \int d^{3}x \left[\mid \overrightarrow{\nabla} \Phi \mid^{2} + V\left(\left| \Phi \right| \right) - \omega^{2} \left| \Phi \right|^{2} \right],
\end{equation}

\noindent and henceforth define

\begin{equation}\label{eqn:b161}
V_{\omega} \left( \mid \Phi \mid \right) = V \left( \mid \Phi \mid \right) - \omega^{2}\mid \Phi \mid^{2},
\end{equation}

\noindent to be the Q-ball potential. Assuming that the Q-balls are spherically symmetric, we make the ansatz

\begin{equation}\label{eqn:b162}
\Phi \left(\textbf{x}\right) = \frac{\phi \left(\textbf{r}\right)}{\sqrt{2}} e^{i\omega t} = \frac{\phi\left(r\right) \hat{r}}{\sqrt{2}}e^{i\omega t},
\end{equation}

\noindent and the gradient operator is reduced to

\begin{equation}\label{eqn:b163}
\overrightarrow{\nabla}\Phi = \frac{\partial \Phi}{\partial r}\hat{r}\longrightarrow \mid \overrightarrow{\nabla}\Phi \mid = \left| \frac{1}{\sqrt{2}}\frac{\partial \phi}{\partial r}\hat{r} \right|^{2} = \frac{1}{2}\left(\frac{\partial \phi}{\partial r}\right)^{2}.
\end{equation}

\noindent since the coordinate dependence of the field is purely radial. Recasting the integral (\ref{eqn:b160}) into spherical polar coordinates, we have

\begin{equation}\label{eqn:b164}
E_{Q} = \int dr \; 4\pi r^{2} \left[ \frac{1}{2}\left(\frac{\partial \phi}{\partial r}\right)^{2} + V_{\omega} \left(\phi \right) \right] + \omega Q.
\end{equation}

\noindent The function $\mathcal{L}_{Q}$ to be extremised is then

\begin{equation}\label{eqn:b165}
\mathcal{L}_{Q}= 4\pi r^{2} \left[ \frac{1}{2}\left(\frac{\partial \phi}{\partial r}\right)^{2} + V_{\omega} \left(\phi \right) \right].
\end{equation}

\noindent This can be realised through the application of the Euler-Lagrange equations

\begin{equation}\label{eqn:b166}
\frac{\partial \mathcal{L}_{Q}}{\partial \phi} - \frac{d}{dr}\left(\frac{\partial \mathcal{L}_{Q}}{\partial \left(\partial_{r}\phi\right)}\right) = 0.
\end{equation}

\noindent Where we find that the second term is 
\begin{equation}\label{eqn:b167}
\frac{d}{dr}\left(\frac{\partial \mathcal{L}_{Q}}{\partial \left(\partial_{r}\phi\right)}\right) = 4\pi r^{2} \frac{\partial \phi}{\partial r}
\vspace{3mm}
\frac{d}{dr}\left(\frac{\partial \mathcal{L}_{Q}}{\partial \left(\partial_{r}\phi\right)}\right) = 8\pi r\frac{\partial \phi}{\partial r} + 4 \pi r^{2}\frac{\partial^{2}\phi}{\partial r^{2}},
\end{equation}

\noindent and the first term is

\begin{equation}\label{eqn:b168}
\frac{\partial \mathcal{L}_{Q}}{\partial \phi} = 4\pi r^{2} \frac{\partial V_{\omega}}{\partial \phi}.
\end{equation}

\noindent Substituting these in to (\ref{eqn:b166}) and dividing through by a factor of $4\pi r^{2}$ gives the field equation

\begin{equation}\label{eqn:b169}
\frac{\partial^{2}\phi}{\partial r^{2}} + \frac{2}{r}\frac{\partial \phi}{\partial r} = \frac{\partial V_{\omega}}{\partial \phi}.
\end{equation}

\noindent This is the field equation which produces Q-balls in a canonical complex scalar $U(1)$ theory, as derived in \cite{coleman85}. We will generalise this to the case of Q-balls with non-canonical kinetic terms in Chapter 6.

\chapter{Sub-Planckian $\Phi^{2}$ Inflation with an $R^{2}$ Term in Palatini gravity}

In this chapter we present a study of a $\phi^{2}$ inflation model in the Palatini formalism with an $R^{2}$ term. This model allows inflation based on a minimal $\phi^{2}$ potential to be compatible with observations, in contrast to the case of conventional $\phi^{2}$ inflation, which predicts a too large tensor-to-scalar ratio. Our aims are to test if Palatini $\phi^{2}$ chaotic inflation can also be compatible with a sub-Planckian inflaton and with the Planck scale suppressed potential corrections expected from a complete quantum gravity theory, as well as to show that a viable transition to conventional Big Bang cosmology following inflation is achievable. We first calculate the slow roll parameters and the inflationary observables for this model in the Einstein frame, and verify that it is possible to successfully inflate this model with a sub-Planckian inflaton. We test the robustness of the predictions of this model against Planck-suppressed potential corrections, in the general case and in the case of an approximate shift symmetry, and present the constraints this places on the model parameter $\alpha$ by considering the $\eta$-shift due to these corrections, and its contributions to the scalar spectral index. We consider two possibilities of the dominant reheating channel in this model: reheating via annihilation of inflaton scalars through the Higgs portal interaction, and reheating via the decay to right-handed neutrinos. We investigate the constraints that the shift in the scalar spectral index places on the couplings of these interactions, and also explore the predictions of the $R^{2}$ Palatini model with a $\phi^{2}$ potential in each reheating scenario in relation to the different constraints on the $\alpha$ parameter. We also show that fragmentation of the inflaton condensate is likely in this model and comment on the possible observable signatures from this model, as well as present and future observability.

\section{Chaotic Inflation and Starobinsky Inflation}\label{section:22}
First let us mention some important precursors to the Palatini quadratic inflation model we investigate here. Quadratic inflation was first proposed in Andrei Linde's theory of Chaotic Inflation in 1983 \cite{linde83}. This theory does not just apply to quadratic or even polynomial potentials, $V\left(\phi \right) \sim \phi^{n}$, however the quadratic potential is the simplest and only has one model parameter, the mass parameter $m$, so it is straightforward to calculate the inflationary observables and make predictions in the model. The general idea of chaotic inflation is that the Universe begins during the Planck epoch of size, $l \sim M_{pl}^{-1}$, with energy density, $\rho \sim M_{pl}/M_{pl}^{-3} \sim M_{pl}^{4}$ \cite{linde07}. From the stage when the energy density begins to fall below this point, $\rho \le M_{pl}^{4}$, the Universe can be treated as classical. We can infer from this idea that a basic initial condition for the Universe is \cite{linde07}

\begin{equation}\label{eqn:r1}
\frac{1}{2}\dot{\phi}^{2} + \frac{1}{2}\left(\partial_{i}\phi\right)^{2} + V\left(\phi \right) \sim M_{pl}^{4},
\end{equation}

\noindent from which we can say that inflation may proceed when $\frac{1}{2}\dot{\phi}^{2} + \frac{1}{2}\left(\partial_{i}\phi\right)^{2} \le V\left(\phi \right)$, and continue if the derivative terms of the inflaton field continue to decrease below the potential, assuming as a starting point $V\left(\phi \right) \sim M_{pl}^{4}$ \cite{linde07}. The chaotic inflation setup is therefore compatible with any inflation model with a potential which can increase to at least $V\left(\phi \right) \sim M_{pl}^{4}$.

In general, we expect that the inflation model should be treated as an effective theory, and this introduces non-renormalisable Planck scale-suppressed potential corrections - which can modify the predictions of the inflationary observables and move the model away from observational compatibility.

Although chaotic inflation is a widely compatible setup which provides a straightforward approach to the problem of initial conditions in inflation, it does raise some questions regarding the predictive power of an inflation model which follows this approach. We can assume $V\left(\phi \right) \sim M_{pl}^{4}$ as a reasonable starting point in order for the inflation to proceed, however this does also assume that the inflaton field itself must start from $\phi \gtrsim M_{pl}$ in a chaotic inflation scenario. In this case we need a quantum theory of gravity to allow us to make any meaningful physical predictions about what may happen above the Planck scale. 

Indeed, simple chaotic inflation models require a super-Planckian inflaton in order to explain inflation. The $\phi^{2}$ chaotic inflation model also predicts a tensor-to-scalar ratio of $r \sim 0.13$ which is above the limit set by the 2018 Planck satellite results and of the BICEP-2/KECK array of $r < 0.06$ at a pivot scale of $k=0.002\Mpc^{-1}$ \cite{Planck18}. However, by considering inflation in the Palatini formalism and the addition of the $R^{2}$ term, it is possible to suppress the tensor-to-scalar ratio of a $\phi^{2}$ model and bring it into the observational parameter space.

It is also befitting to mention Starobinsky inflation here, on account of the similarity of the inflaton action in the Starobinsky model compared to the Jordan frame action in the $R^{2}$ Palatini model with a $\phi^{2}$ potential. Starobinsky inflation was proposed in 1980 by Alexei Starobinsky \cite{staro80}, which considers the action

\begin{equation}\label{eqn:r2}
S = \frac{1}{2}\int d^{4}x \sqrt{-g} \left[ M_{pl}^{2}R + \frac{R^{2}}{6M^{2}}\right].
\end{equation}

\noindent This result emerges from considering quantum corrections to the Einstein-Hilbert action. The presence of $R^{2}$ corrections in the Einstein equations, given large curvature, produces an effective cosmological constant, interpreted as an era of de-Sitter inflationary expansion. The predictions from Starobinsky-type inflation models are in good agreement with observations \cite{kehagias14}, and it is interesting to consider a similar model in the Palatini formalism \cite{enckell18} \cite{antoniadis18} to compare the differences in predicted observables as well as the underlying physical mechanism producing the expansion.

\subsection{$R^{2}$ Palatini Inflation}\label{section: 221}
A number of studies have been done for $R^{2}$ inflation in the Palatini formalism \cite{enckell18}-\cite{gialamas20}. In particular, Enckell et al (2018) \cite{enckell18} consider the case for an unspecified potential and inflaton, and demonstrate that the presence of an $R^{2}$ term can suppress the height of the inflaton potential irrespective of its nature, such that any scalar potential will be transformed into a plateau - or hilltop-type - potential and the tensor-to-scalar ratio will be suppressed accordingly (see \cite{enckell18}-\cite{gialamas20}). In our research we are interested to study this effect on a simple quadratic potential, in particular whether the $R^{2}$ Palatini model with a $\phi^{2}$ potential could inflate successfully with a sub-Planckian inflaton field while maintaining its good agreement with observations; whether the model is robust against Planck-suppressed potential corrections from a quantum gravity completion; whether the model can reheat successfully, and the implications of this for the calculated spectral index; and what constraints these requirements would put on the model parameters.

\section{The Model}\label{section:23}
We begin by introducing the model. For the purposes of these calculations, the explicit factors of the reduced Planck mass have been left in and we use the $( -, +, +, +)$ signature for the metric. The inflaton action in the Jordan frame is

\begin{equation}\label{eqn:r3}
S_{J} = \int d^{4}x \sqrt{-g} \left[ \frac{M_{pl}^{2}}{2}R + \frac{\alpha}{4}R^{2} - \frac{1}{2}\partial_{\mu} \phi \partial^{\mu} \phi - V\left(\phi \right) \right].
\end{equation}

\noindent It is instructive to rewrite the action in terms of an auxiliary field $\chi$, which gives the action as \cite{antoniadis18}

\begin{equation}\label{eqn:r4}
S_{J} = \int d^{4}x \sqrt{-g} \left[ \frac{1}{2}\left( M_{pl}^{2} + \alpha \chi^{2} \right)R - \frac{\alpha}{4}\chi^{4} - \frac{1}{2}\partial_{\mu}\phi \partial^{\mu}\phi - V\left( \phi \right)\right].
\end{equation}

\subsection{Conformal Transformation}\label{section:231}
We now perform a conformal transformation on our spacetime metric and proceed with all of our calculations in the Einstein frame unless explicitly stated otherwise. The conformal transformation serves the purpose of reparameterising the $R^{2}$ term and its effects, such that we can undertake the inflationary calculations and interpret the results in a framework that we have a good understanding of. In this work the Jordan and Einstein frames are defined such that the Jordan frame is the "particle physics" frame, in the sense that it is the frame in which the model and its symmetries are defined, and the Einstein frame is then the "physics" frame in which the treatment of gravity can be considered equivalent to conventional General Relativity.

Henceforth, we define 

\begin{equation}\label{eqn:r5}
\Omega^{2} = 1 + \frac{\alpha \chi^{2}}{M_{pl}^{2}},
\end{equation}

\noindent as the conformal factor, and the conformal transformation of the spacetime metric is

\begin{equation}\label{eqn:r6}
g_{\mu \nu} \longrightarrow \tilde{g}_{\mu \nu} = \Omega^{2} g_{\mu \nu},
\end{equation}

\noindent and

\begin{equation}\label{eqn:r7}
g^{\mu \nu} \longrightarrow \tilde{g}^{\mu \nu} = \frac{1}{\Omega^{2}} g^{\mu \nu},
\end{equation}

\noindent where Einstein frame quantities will be denoted with a tilde. The negative determinant of the metric must then be transformed

\begin{equation}\label{eqn:r8}
\sqrt{-g} = \sqrt{-det g_{\mu\nu}} \longrightarrow \sqrt{-det \left(\frac{\tilde{g}_{\mu\nu}}{\Omega^{2}}\right)}.
\end{equation}

\noindent To do this, we use the fact that

\begin{equation}\label{eqn:r9}
det \left(c M \right) = c^{D} det(M),
\end{equation}

\noindent where $c$ is a scalar, $M$ is a matrix and $D$ is the number of spacetime dimensions. We consider $D=4$ in this work, and so the transformation is 

\begin{equation}\label{eqn:r10}
\sqrt{-det \left(\frac{\tilde{g}_{\mu\nu}}{\Omega^{2}}\right)} = \sqrt{-\frac{1}{\Omega^{8}}det\left(\tilde{g}_{\mu\nu}\right)} = \frac{1}{\Omega^{4}}\sqrt{-det \tilde{g}_{\mu\nu}} = \frac{\sqrt{-\tilde{g}}}{\Omega^{4}}.
\end{equation}

\noindent As aforementioned, the conformal transformation acts upon the metric, including any implicit factors of the metric used in index contractions, so while the derivatives of the $\partial_{\mu} \phi$ term themselves are unaffected, the term itself still needs to be considered. The inflaton derivative term is

\begin{equation}\label{eqn:r11}
\frac{1}{2}\partial_{\mu}\phi \partial^{\mu}\phi = \frac{1}{2}g_{\mu \nu}\partial^{\nu}\phi \partial^{\mu}\phi,
\end{equation}

\noindent and in the Einstein frame this becomes

\begin{equation}\label{eqn:r12}
\frac{1}{2}\tilde{g}_{\mu \nu}\partial^{\nu}\phi \partial^{\mu}\phi = \frac{1}{2}g_{\mu \nu}\Omega^{2}\partial^{\nu}\phi \partial^{\mu}\phi = \frac{1}{2}\Omega^{2}\partial_{\mu}\phi \partial^{\mu}\phi.
\end{equation}

\subsection{Transforming the Ricci Scalar}\label{section:232}

In the Palatini formalism, the spacetime connection $\Gamma$ is defined independently of the spacetime metric $g$. This means that $\Gamma$ and anything built from it, namely the Riemann tensor and its contractions, are treated as having no intrinsic dependence on the metric. In the Einstein frame this means

\begin{equation}\label{eqn:r13}
\tilde{R}^{\gamma}_{\mu \delta \nu} = R^{\gamma}_{\mu \delta \nu},
\end{equation}

\noindent This can then be contracted on its first and third indices to the Ricci tensor, $R_{\mu \nu}$,

\begin{equation}\label{eqn:r14}
R^{\gamma}_{\mu \delta \nu} \rightarrow  R^{\gamma}_{\mu \gamma \nu} = R_{\mu \nu},
\end{equation}

\noindent which can then be contracted to the Ricci scalar, $R$,

\begin{equation}\label{eqn:r15}
R_{\mu \nu} \longrightarrow g^{\mu \nu}R_{\mu \nu} = R^{\mu}_{\mu} \equiv R.
\end{equation}

\noindent We must transform the factor of the metric used to make the contraction, and thus

\begin{equation}\label{eqn:r16}
g^{\mu \nu}R_{\mu \nu} \longrightarrow  \tilde{g}^{\mu \nu}R_{\mu \nu} = \frac{1}{\Omega^{2}} g^{\mu \nu} R_{\mu \nu} = \frac{1}{\Omega^{2}}R.
\end{equation}

\noindent The transformation on the Ricci scalar in the Palatini formulation is therefore

\begin{equation}\label{eqn:r17}
R \longrightarrow \tilde{R} = \frac{1}{\Omega^{2}}R.
\end{equation}

\noindent This is in contrast to the transformation in the metric formalism, where the spacetime connection is assumed to be the Levi-Civita connection, which is built from derivatives of the spacetime metric, resulting in the transformation (\ref{eqn:b14}).

\subsection{Eliminating the Auxiliary field}\label{section:233}
Substituting (\ref{eqn:r10}), (\ref{eqn:r12}) and (\ref{eqn:r17}) into (\ref{eqn:r4}), we find that the Einstein frame action in terms of the auxiliary field $\chi$ in the Palatini formalism is

\begin{multline}\label{eqn:r18}
S_{E} =  \int d^{4}x \; \frac{\sqrt{-\tilde{g}}}{\Omega^{4}} \left[ \frac{1}{2}M_{pl}^{2}\tilde{R}\Omega^{2} \Omega^{2} -\frac{\alpha \chi^{4}}{4} - \frac{1}{2}\Omega^{2} \partial_{\mu}\phi \partial^{\mu} \phi - V\left( \phi \right) \right]  \\
 = \int d^{4}x \sqrt{-\tilde{g}} \left[ \frac{1}{2}M_{pl}^{2}\tilde{R} - \frac{\alpha \chi^{4}}{4\Omega^{4}} - \frac{1}{2\Omega^{2}} \partial_{\mu}\phi \partial^{\mu} \phi - \frac{V\left( \phi \right)}{\Omega^{4}} \right].
\end{multline}

\noindent The auxiliary field must be eliminated from the action in order to have the model represented in terms of the inflaton $\phi$ again. The first step is to extremise the action in terms of $\chi^{2}$ to find an equation of motion for $\chi^{2}$ in terms of the inflaton. This procedure was demonstrated in \cite{antoniadis18}.

Varying (\ref{eqn:r18}) with respect to $\chi^{2}$, we have

\begin{multline}\label{eqn:r19}
\delta S_{E} =  \int d^{4}x \; \sqrt{-\tilde{g}} \left[ -\frac{1}{2}\frac{\partial}{\partial \chi^{2}} \left(\frac{1}{\Omega^{2}}\right)\partial_{\mu}\phi \partial^{\mu}\phi \delta \chi^{2} - \frac{\alpha \chi^{4}}{4} \frac{\partial}{\partial \chi^{2}}\left(\frac{1}{\Omega^{4}}\right) \delta \chi^{2} \right. \\ 
 \left. - \frac{\alpha}{4 \Omega^{4}} \frac{\partial \chi^{4}}{\partial \chi^{2}} \delta \chi^{2}  - V\left( \phi \right) \frac{\partial}{\partial \chi^{2}}\left(\frac{1}{\Omega^{4}}\right) \delta \chi^{2} \right].
\end{multline}

\noindent Since $\frac{\delta S_{E}}{\delta \chi^{2}} = 0$, and the Einstein frame Lagrangian only depends on $\chi^{2}$, then it must be true that

\begin{equation}\label{eqn:r20}
\frac{\partial \mathcal{L}_{E}}{\partial \chi^{2}} = 0.
\end{equation}

\noindent Taking the derivative in (\ref{eqn:r20}) we have

\begin{equation}\label{eqn:r21}
\frac{\partial \mathcal{L}_{E}}{\partial \chi^{2}} = -\frac{2\alpha \chi^{2}}{4\Omega^{4}} - \frac{\alpha \chi^{4}}{4}\frac{\partial}{\partial \chi^{2}}\left(\frac{1}{\Omega^{4}}\right) - \frac{1}{2} \frac{\partial}{\partial \chi^{2}} \left(\frac{1}{\Omega^{2}}\right) \partial_{\mu} \phi \partial^{\mu}\phi - \frac{\partial}{\partial \chi^{2}}\left(\frac{1}{\Omega^{4}}\right)V\left(\phi \right),
\end{equation}

\noindent where

\begin{equation}\label{eqn:r22}
\frac{\partial}{\partial \chi^{2}} \left( \frac{1}{\Omega^{2}}\right) = -\frac{\alpha}{M_{pl}^{2} \Omega^{4}},
\end{equation}

\noindent and

\begin{equation}\label{eqn:r23}
\frac{\partial}{\partial \chi^{2}} \left( \frac{1}{\Omega^{4}}\right) = -\frac{2\alpha}{M_{pl}^{2} \Omega^{6}}.
\end{equation}

\noindent Therefore (\ref{eqn:r21}) is

\begin{equation}\label{eqn:r24}
\frac{\partial \mathcal{L}_{E}}{\partial \chi^{2}} = \frac{1}{\Omega^{6}} \left[ -\frac{\alpha \chi^{2}}{2}\Omega^{2} + \frac{\alpha \chi^{4}}{2} \frac{\alpha}{M_{pl}^{2}} + \frac{\alpha}{M_{pl}^{2}}\frac{\Omega^{2}}{2} \partial_{\mu} \phi \partial^{\mu}\phi + 2V\left(\phi \right) \frac{\alpha}{M_{pl}^{2}} \right].
\end{equation}

\noindent and from (\ref{eqn:r20}) we can write

\begin{equation}\label{eqn:r25}
 -\frac{\alpha \chi^{2}}{2}\Omega^{2} + \frac{\alpha \chi^{4}}{2} \frac{\alpha}{M_{pl}^{2}} + \frac{\alpha}{M_{pl}^{2}}\frac{\Omega^{2}}{2} \partial_{\mu} \phi \partial^{\mu}\phi + 2V\left(\phi \right) \frac{\alpha}{M_{pl}^{2}} = 0.
\end{equation}

\noindent Expanding the conformal factors explicitly and rearranging gives

\begin{equation}\label{eqn:r26}
\chi^{2} = \frac{\partial_{\mu} \phi \partial^{\mu}\phi + 4V}{\left[M_{pl}^{2} - \frac{\alpha}{M_{pl}^{2}}\partial_{\mu}\phi \partial^{\mu}\phi \right]}.
\end{equation}

\noindent For the remainder of this calculation we will use the shorthand $D = \partial_{\mu}\phi \partial^{\mu}\phi $ for brevity. $\chi^{2}$ is then

\begin{equation}\label{eqn:r27}
\chi^{2} = \frac{D + 4V}{\left[M_{pl}^{2} - \frac{\alpha D}{M_{pl}^{2}}\right]}.
\end{equation}

\noindent Substituting this back into the Einstein frame Lagrangian (subscript 'E' dropped for the remainder of the calculation) we have

\begin{equation}\label{eqn:r28}
\mathcal{L} = -\frac{1}{\Omega^{4}}\left[ \frac{\alpha}{4}\frac{\left(D + 4V\right)^{2}}{\left[M_{pl}^{2} - \frac{\alpha D}{M_{pl}^{2}}\right]^{2}} + \frac{\Omega^{2}}{2}D + V\left(\phi \right)\right].
\end{equation}

\noindent Let $I = \left[...\right]$, such that (\ref{eqn:r28}) is $\mathcal{L} = I/\Omega^{4}$. Substituting (\ref{eqn:r5}) and (\ref{eqn:r27}) into the second term in $I$ and expanding gives

\begin{equation}\label{eqn:r29}
\frac{\Omega^{2}}{2}D = \frac{D}{2}\left[ 1 + \frac{\alpha}{M_{pl}^{2}}\frac{\left( 4V + D \right)}{\left[ M_{pl}^{2} - \frac{\alpha D}{M_{pl}^{2}}\right]}\right] = \frac{D}{2}\left[\frac{\left(M_{pl}^{2} - \frac{\alpha D}{M_{pl}^{2}}\right) + \frac{\alpha}{M_{pl}^{2}}\left( 4V + D \right)}{\left( M_{pl}^{2} - \frac{\alpha D}{M_{pl}^{2}}\right)}\right].
\end{equation}

\noindent Expanding the numerator of (\ref{eqn:r29}) gives

\begin{equation}\label{eqn:r30}
\begin{split}
\frac{\Omega^{2}}{2}D = \frac{D}{2}\frac{1}{\left( M_{pl}^{2} - \frac{\alpha D}{M_{pl}^{2}} \right)} & \left[ M_{pl}^{2} - \frac{\alpha D}{M_{pl}^{2}} + \frac{4\alpha V}{M_{pl}^{2}} + \frac{\alpha D}{M_{pl}^{2}} \right] \\
& =  \frac{D}{2}\frac{1}{\left( M_{pl}^{2} - \frac{\alpha D}{M_{pl}^{2}} \right)}\left[ M_{pl}^{2} + \frac{4\alpha V}{M_{pl}^{2}} \right].
\end{split}
\end{equation}

\noindent $I$ can therefore be written

\begin{multline}\label{eqn:r31}
I = \frac{1}{\left( M_{pl}^{2} - \frac{\alpha D}{M_{pl}^{2}} \right)^{2}}\left[ V\left(\phi \right) \left(M_{pl}^{2} - \frac{\alpha D}{M_{pl}^{2}}\right)^{2} + \frac{\alpha}{4}\left( 4V + D \right)^{2}\right. \\
\left.+ \frac{D}{2}\left(M_{pl}^{2} - \frac{\alpha D}{M_{pl}^{2}}\right) \left( M_{pl}^{2} + \frac{4\alpha V}{M_{pl}^{2}} \right) \right].
\end{multline}

\noindent Expanding all terms in the square brackets gives

\begin{equation}\label{eqn:r32}
I = \frac{1}{\left(M_{pl}^{2} - \frac{\alpha D}{M_{pl}^{2}}\right)^{2}}\left[ V\left( M_{pl}^{4} + 2\alpha D - \frac{\alpha^{2} D^{2}}{M_{pl}^{4}}\right) + 4\alpha V^{2} + \frac{D}{2}M_{pl}^{4} - \frac{\alpha D^{2}}{4}\right],
\end{equation}

\noindent where the terms inside the square brackets can be written

\begin{equation}\label{eqn:r33}
I_{2} = \left(\frac{M_{pl}^{4}}{2}D - \frac{\alpha D^{2}}{4} \right)\left(1 + \frac{4\alpha V}{M_{pl}^{4}} \right) + M_{pl}^{2}V\left(M_{pl}^{2} + \frac{4\alpha V}{M_{pl}^{2}} \right).
\end{equation}

\noindent Substituting (\ref{eqn:r27}) into (\ref{eqn:r5}), the conformal factor is

\begin{equation}\label{eqn:r34}
\Omega^{2} = 1 + \frac{\alpha}{M_{pl}^{2}}\frac{\left(D + 4V\right)}{\left(M_{pl}^{2} - \frac{\alpha D}{M_{pl}^{2}}\right)},
\end{equation}

\noindent and can be written as

\begin{equation}\label{eqn:r35}
\Omega^{2} = \frac{1}{\left(M_{pl}^{2} - \frac{\alpha D}{M_{pl}^{2}}\right)}\left[M_{pl}^{2} - \frac{\alpha D}{M_{pl}^{2}} + \frac{\alpha}{M_{pl}^{2}}\left(D + 4V\right) \right] = \frac{1}{\left(M_{pl}^{2} - \frac{\alpha D}{M_{pl}^{2}}\right)}\left[M_{pl}^{2} + \frac{4\alpha V}{M_{pl}^{2}}\right].
\end{equation}

Substituting (\ref{eqn:r32}) back into (\ref{eqn:r28}) and using (\ref{eqn:r35}), the Einstein frame Lagrangian is

\begin{equation}\label{eqn:r36}
\mathcal{L} = -\frac{1}{\left[ M_{pl}^{2} + \frac{4\alpha V}{M_{pl}^{2}}\right]^{2}}\left[ V\left( M_{pl}^{4} + 2\alpha D - \frac{\alpha^{2} D^{2}}{M_{pl}^{4}}\right) + 4\alpha V^{2} + \frac{D}{2}M_{pl}^{4} - \frac{\alpha D^{2}}{4} \right].
\end{equation}

\noindent Using (\ref{eqn:r33}) this can be written as 

\begin{equation}\label{eqn:r37}
\mathcal{L} = -\frac{1}{\left[ M_{pl}^{2} + \frac{4\alpha V}{M_{pl}^{2}}\right]^{2}}\left[ \left(\frac{M_{pl}^{4}}{2}D - \frac{\alpha D^{2}}{4}\right) \left(1 + \frac{4\alpha V}{M_{pl}^{4}} \right) + M_{pl}^{2}V\left( M_{pl}^{2} + \frac{4\alpha V}{M_{pl}^{2}}\right)\right],
\end{equation}

\noindent and pulling out a factor of $\left( M_{pl}^{2} + 4\alpha V/M_{pl}^{2}\right)$ and simplifying we obtain

\begin{equation}\label{eqn:r38}
\mathcal{L} = -\frac{D}{2\left(1 + \frac{4\alpha V}{M_{pl}^{4}} \right)} + \frac{\alpha}{4 M_{pl}^{4}}\frac{D^{2}}{\left( 1 + \frac{4\alpha V}{M_{pl}^{4}}\right)} - \frac{V\left( \phi \right)}{\left( 1 + \frac{4\alpha V}{M_{pl}^{4}}\right)}.
\end{equation}

\noindent Substituting back for $D = \partial_{\mu}\phi \partial^{\mu}\phi$, the Einstein frame Lagrangian is

\begin{equation}\label{eqn:r39}
\mathcal{L} = -\frac{1}{2}\frac{\partial_{\mu} \phi \partial^{\mu}\phi}{\left( 1 + \frac{4\alpha V}{M_{pl}^{4}}\right)} + \frac{\alpha}{4 M_{pl}^{4}} \frac{\left(\partial_{\mu} \phi \partial^{\mu} \phi \right)^{2}}{\left( 1 + \frac{4\alpha V}{M_{pl}^{4}}\right)} - \frac{V\left( \phi \right)}{\left( 1 + \frac{4\alpha V}{M_{pl}^{4}}\right)},
\end{equation}

\noindent and the Einstein frame inflaton action of the $R^{2}$ Palatini model is then

\begin{equation}\label{eqn:r40}
S = \int d^{4}x \sqrt{-\tilde{g}} \left[ \frac{1}{2}M_{pl}^{2}\tilde{R} -\frac{1}{2}\frac{\partial_{\mu} \phi \partial^{\mu}\phi}{\left( 1 + \frac{4\alpha V}{M_{pl}^{4}}\right)} + \frac{\alpha}{4 M_{pl}^{4}} \frac{\left(\partial_{\mu} \phi \partial^{\mu} \phi \right)^{2}}{\left( 1 + \frac{4\alpha V}{M_{pl}^{4}}\right)} - \frac{V\left( \phi \right)}{\left( 1 + \frac{4\alpha V}{M_{pl}^{4}}\right)} \right].
\end{equation}

The effect of the non-canonical kinetic terms is that - when recast using a canonical variable - the inflaton potential is flattened and stretched into a plateau. The effect is similar in principle to that of the non-canonical kinetic terms in $\alpha $-attractor inflation (see e.g. \cite{alinde} for a review of $\alpha$-attractor inflation and its predictions), although the bases of the two models are fundamentally different. Roughly speaking, the effect in $\alpha$-attractor models comes from directly modifying the kinetic term with the purpose of turning the inflaton potential into a plateau, which is possible for the majority of potentials $V(\phi)$, provided that they are non-singular. In the Palatini model, the effect of flattening and stretching the inflaton potential is a side-effect of the inclusion of the $R^{2}$ term in the gravitational action - it can be interpreted as an effect due to a gravitational interaction in the model recast as a non-canonical inflaton interaction.

\section{Canonical Rescaling of the Inflaton Field}\label{section:24}

Before we can calculate any of the inflationary observables we must canonically normalise the inflaton kinetic term. The conformal factor is now, from (\ref{eqn:r5}) and (\ref{eqn:r35})

\begin{equation}\label{eqn:r41}
\Omega^{2} = \frac{1 + \frac{4\alpha V}{M_{pl}^{4}}}{1 - \frac{\alpha}{M_{pl}^{4}}\partial_{\mu}\phi \partial^{\mu}\phi}.
\end{equation}

\noindent In the limit where the slow-roll approximation is valid, when the field is high on the inflationary plateau, we will show later that the derivatives are negligible and we can say for now

\begin{equation}\label{eqn:r42}
\Omega^{2} = 1 + \frac{4\alpha V}{M_{pl}^{4}} = 1 + \frac{2\alpha m^{2} \phi^{2}}{M_{pl}^{4}}.
\end{equation}

\noindent We define a canonically normalised scalar field $\sigma$ as follows

\begin{equation}\label{eqn:r43}
\left(\frac{d \sigma}{d \phi}\right)^{2} = \frac{1}{1 + \frac{4\alpha V}{M_{pl}^{4}}} = \frac{1}{1 + \frac{2\alpha m^{2} \phi^{2}}{M_{pl}^{4}}} \Rightarrow \frac{d \sigma}{d \phi} = \pm \frac{1}{\sqrt{1 + \frac{2\alpha m^{2} \phi^{2}}{M_{pl}^{4}}}}.
\end{equation}

\noindent We choose the positive solution and integrate

\begin{equation}\label{eqn:r44}
\sigma \left(\phi \right) = \int \frac{d\phi}{\sqrt{1 + K\phi^{2}}} = \frac{1}{\sqrt{K}}\sinh^{-1}\left(\sqrt{K}\phi \right) + C = \frac{1}{\sqrt{K}}\log\left( \sqrt{1 + K\phi^{2}} + \sqrt{K}\phi \right) + C,
\end{equation}

\noindent where $C$ is a constant of integration and $K = 2\alpha m_{\phi}^{2}/M_{pl}^{4}$.

The $\partial_{\mu}\phi$ terms in (\ref{eqn:r40}) - referred to henceforth as kinetic terms - can be rewritten 

\begin{equation}\label{eqn:r45}
-\frac{1}{2} \frac{\partial_{\mu}\phi \partial^{\mu}\phi}{1 + \frac{4\alpha V}{M_{pl}^{4}}} + \frac{\alpha}{4 M_{pl}^{4}}\frac{\left( \partial_{\mu}\phi \partial^{\mu}\phi \right)^{2}}{1 + \frac{4\alpha V}{M_{pl}^{4}}} = -\frac{1}{2}\partial_{\mu}\sigma \partial^{\mu}\sigma + \frac{\alpha}{4 M_{pl}^{4}}\left( 1 + \frac{4\alpha V}{M_{pl}^{4}} \right)\left(\partial_{\mu}\sigma \partial^{\mu}\sigma \right)^{2},
\end{equation}

\noindent in terms of the canonically normalised scalar $\sigma$. If we are well up on the inflationary plateau, then we can safely assume 

\begin{equation}\label{eqn:r46}
\frac{4\alpha V}{M_{pl}^{4}} >>1,
\end{equation}

\noindent and we can write the canonical kinetic terms from (\ref{eqn:r45}) as

\begin{equation}\label{eqn:r47}
-\frac{1}{2}\partial_{\mu}\sigma \partial^{\mu}\sigma + \frac{\alpha}{4 M_{pl}^{4}}\left( 1 + \frac{4\alpha V}{M_{pl}^{4}} \right)\left(\partial_{\mu}\sigma \partial^{\mu}\sigma \right)^{2}.
\end{equation}

\noindent We will show later that the $\left(\partial \sigma\right)^{4}$ term will make a negligible contribution during slow roll, so it is safe to ignore for this discussion. It may be that this four-point inflaton interaction will become significant after inflation, and we will discuss this in the context of unitarity violation in a later section.

We therefore proceed with the Einstein frame action

\begin{equation}\label{eqn:r48}
S_{E} = \int d^{4}x \sqrt{-\tilde{g}} \left[ \frac{1}{2}M_{pl}^{2}\tilde{R} -\frac{1}{2}\partial_{\mu} \sigma \partial^{\mu}\sigma  - \frac{V\left(\phi \right)}{\left( 1 + \frac{4\alpha V}{M_{pl}^{4}}\right)} \right].
\end{equation}

\noindent where $\phi = \phi \left(\sigma \right)$. From here the "inflaton" in the Einstein frame can be taken to mean the $\sigma$ field unless stated otherwise. Our Einstein frame potential is henceforth defined to be

\begin{equation}\label{eqn:r49}
V_{E}\left(\phi \right) = \frac{V\left( \phi \right)}{1 + \frac{4\alpha V}{M_{pl}^{4}}},
\end{equation}

\noindent where the Jordan frame potential is the simple quadratic potential

\begin{equation}\label{eqn:r50}
V\left(\phi \right) = \frac{1}{2}m^{2}\phi^{2}.
\end{equation}

\noindent We define $\Omega \approx 1$ as the informal end of the plateau region of the potential and the beginning of the small field region. If

\begin{equation}\label{eqn:r51}
\Omega \approx \sqrt{\frac{2\alpha m_{\phi}^{2}\phi^{2}}{M_{pl}^{4}}} \approx 1,
\end{equation}

\noindent using (\ref{eqn:r46}) then

\begin{equation}\label{eqn:r52}
 \frac{2\alpha m_{\phi}^{2}\phi^{2}}{M_{pl}^{4}} \approx 1.
\end{equation}

\noindent Using this, we define a threshold in the Einstein frame potential which approximately separates the large and small field regions of the potential

\begin{equation}\label{eqn:r53}
\phi_{0} = \frac{M_{pl}^{2}}{\sqrt{2\alpha}m_{\phi}}.
\end{equation}

\noindent From (\ref{eqn:r44}) the canonical field is 

\begin{equation}\label{eqn:r54}
\sigma \left(\phi \right) = \frac{M_{pl}^{2}}{\sqrt{2\alpha}m_{\phi}} \ln \left( \sqrt{1 + \frac{2\alpha m_{\phi}^{2}}{M_{pl}^{4}}\phi^{2}} + \frac{\sqrt{2\alpha}m_{\phi}}{M_{pl}^{2}}\phi \right) = \phi_{0} \ln \left( \sqrt{1 + \frac{\phi^{2}}{\phi_{0}^{2}}} + \frac{\phi}{\phi_{0}}\right),
\end{equation}

\noindent where for $\phi << \phi_{0}$, we have that

\begin{equation}\label{eqn:r55}
\sigma \approx \phi_{0} \ln \left( \sqrt{1} + \frac{\phi}{\phi_{0}} \right) \approx \phi_{0} \cdot \frac{\phi}{\phi_{0}} = \phi,
\end{equation}

\noindent using $\ln \left(1 + x \right) \approx x$ for $\mid x \mid << 1$. In the large field limit, $\phi >> \phi_{0}$, we have 

\begin{equation}\label{eqn:r56}
\sigma \approx \phi_{0}\ln \left( \sqrt{\frac{\phi^{2}}{\phi_{0}^{2}}} + \frac{\phi}{\phi_{0}} \right) \approx \phi_{0}\ln \left(\frac{2\phi}{\phi_{0}} \right) = \frac{M_{pl}^{2}}{\sqrt{2\alpha}m_{\phi}}\ln \left(\frac{2\sqrt{2\alpha}m_{\phi}}{M_{pl}^{2}}\phi \right).
\end{equation}

\noindent This shows that there are two distinct regimes of the potential in terms of the canonical field

\begin{equation}\label{eqn:r57}
\sigma \approx \phi ; \; \; \phi < \phi_{0},
\end{equation}

\begin{equation}\label{eqn:r58}
\sigma \approx \frac{M_{pl}^{2}}{\sqrt{2\alpha}m_{\phi}}\ln \left(\frac{2\sqrt{2\alpha}m_{\phi}}{M_{pl}^{2}}\phi \right) ; \; \; \phi > \phi_{0},
\end{equation}

\noindent and this approximation will be used throughout this exposition in order to study the dynamics of the model in the large and small field limits. 

The Einstein frame potential is

\begin{equation}\label{eqn:r59}
V_{E}\left(\phi \right) = \frac{V\left( \phi \right)}{1 + \frac{4\alpha V}{M_{pl}^{4}}},
\end{equation}

\noindent which can be rewritten

\begin{equation}\label{eqn:r60}
V_{E} = \frac{V\left( \phi \right)}{\frac{4\alpha V}{M_{pl}^{4}}\left( 1 + \frac{M_{pl}^{4}}{4\alpha V}\right)} = \frac{M_{pl}^{4}}{4\alpha}\left(1 + \frac{M_{pl}^{4}}{4\alpha V}\right)^{-1}.
\end{equation}

\noindent In the small field limit, $\phi << \phi_{0}$, we have $\frac{4\alpha V}{M_{pl}^{4}} << 1$ and 

\begin{equation}\label{eqn:r61}
V_{E} \approx V\left(\phi \right) = \frac{1}{2}m_{\phi}^{2}\phi^{2}.
\end{equation}

\noindent So at small $\sigma = \phi$, the Einstein frame potential is approximately the Jordan frame potential, and in the quadratic regime of the potential we therefore have $m_{\sigma}^{2} \approx m_{\phi}^{2}$.

In the large field limit, $\phi >> \phi_{0}$ we have $\frac{M_{pl}^{4}}{4\alpha V} << 1$, and so

\begin{equation}\label{eqn:r62}
V_{E} \approx \frac{M_{pl}^{4}}{4\alpha}\left(1 - \frac{M_{pl}^{4}}{4\alpha V}\right) = \frac{M_{pl}^{4}}{4\alpha}\left(1 - \frac{\phi_{0}^{2}}{\phi^{2}}\right).
\end{equation}

\noindent From (\ref{eqn:r56}) we have that

\begin{equation}\label{eqn:r63}
\sigma \left( \phi \right) = \frac{M_{pl}^{2}}{\sqrt{2\alpha}m_{\phi}}\ln \left(\frac{2\sqrt{2\alpha}m_{\phi}}{M_{pl}^{2}}\phi \right) = \phi_{0}\ln \left(\frac{2\phi}{\phi_{0}} \right),
\end{equation}

\begin{equation}\label{eqn:r64}
\Rightarrow \phi \left( \sigma \right) = \frac{\phi_{0}}{2}\exp \left(\frac{\sigma}{\phi_{0}} \right) = \frac{M_{pl}^{2}}{2\sqrt{2\alpha}m_{\phi}}\exp \left(\frac{\sqrt{2\alpha}m_{\phi}}{M_{pl}^{2}}\sigma \right),
\end{equation}

\noindent and therefore

\begin{equation}\label{eqn:r65}
\frac{1}{\phi^{2}} =  \frac{4\left(2\alpha\right)m_{\phi}^{2}}{M_{pl}^{4}}\exp \left(-\frac{2\sqrt{2\alpha}m_{\phi}}{M_{pl}^{2}}\sigma \right).
\end{equation}

\begin{figure}[H]
\begin{center}
\includegraphics[clip = true, width=\textwidth, angle = 360]{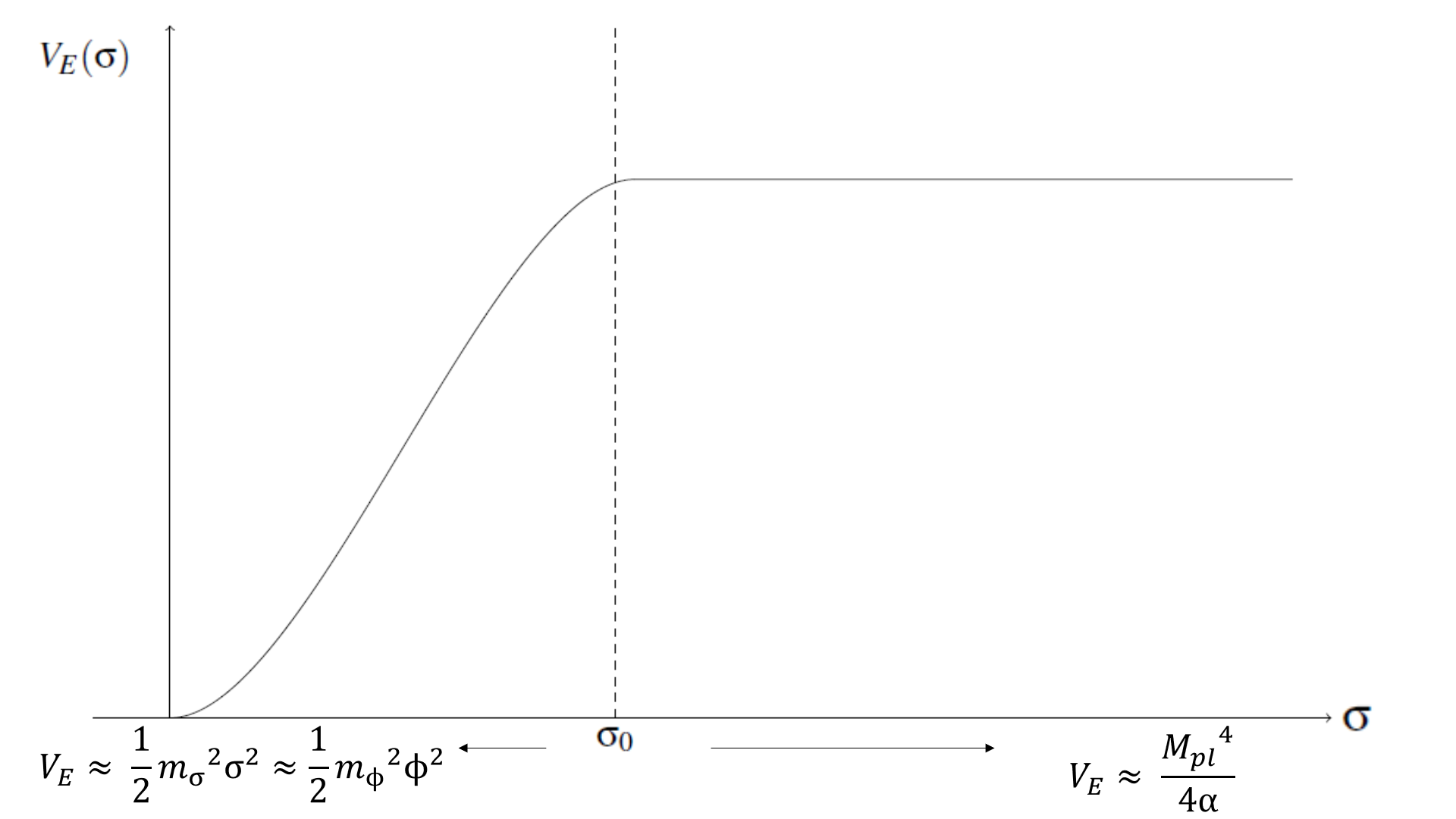}
\caption{Schematic of the inflaton potential in the Einstein frame illustrating the threshold approximation used to distinguish the large and small field limits in this model. } 
\label{figure:n1}
\end{center}
\end{figure}

\noindent Substituting (\ref{eqn:r65}) into the Einstein frame potential (\ref{eqn:r62}) we obtain

\begin{equation}\label{eqn:r66}
V_{E}\left(\sigma \right) \approx \frac{M_{pl}^{4}}{4\alpha}\left[ 1 - \frac{M_{pl}^{4}}{2\alpha m_{\phi}^{2}\phi^{2}} \right] \approx \frac{M_{pl}^{4}}{4\alpha}\left[ 1 - 4\exp \left(-\frac{2\sqrt{2\alpha}m_{\phi}}{M_{pl}^{2}}\sigma \right) \right],
\end{equation}

\noindent where the inflaton potential in relation to the threshold approximation is shown in Figure \ref{figure:n1}.

The number of e-folds of inflation is given by

\begin{equation}\label{eqn:r67}
N\left( \sigma \right) = -\frac{1}{M_{pl}^{2}} \int^{\sigma_{end}}_{\sigma} \frac{V_{E}}{V'_{E}} d\sigma,
\end{equation}

\noindent where $\sigma_{end}$ is the value of the field at the end of slow-roll inflation. On the inflationary plateau the Einstein frame potential is approximately flat, so we can say that at these high field values 

\begin{equation}\label{eqn:r68}
V_{E} \approx \frac{M_{pl}^{4}}{4\alpha},
\end{equation}

\noindent from (\ref{eqn:r62}) and 

\begin{equation}\label{eqn:r69}
V'_{E} = \frac{\partial V_{E}}{\partial \sigma} = 2\sqrt{\frac{2}{\alpha}}M_{pl}^{2} m_{\phi} \exp \left(-\frac{2\sqrt{2\alpha}m_{\phi}}{M_{pl}^{2}}\sigma \right).
\end{equation}

\noindent Substituting (\ref{eqn:r68}) and (\ref{eqn:r69}) into (\ref{eqn:r67}) we have

\begin{equation}\label{eqn:r70}
N\left( \sigma \right) = -\frac{1}{M_{pl}^{2}} \int^{\sigma_{end}}_{\sigma} \frac{M_{pl}^{4}}{4\alpha} \frac{1}{2} \sqrt{\frac{\alpha}{2}}\frac{1}{M_{pl}^{2}m_{\phi}}\exp \left(\frac{2\sqrt{2\alpha}m_{\phi}}{M_{pl}^{2}}\sigma \right)  d\sigma.
\end{equation}

\noindent Performing the integration of (\ref{eqn:r70}) gives

\begin{equation}\label{eqn:r71}
N\left( \sigma \right) =  \frac{M_{pl}^{2}}{32\alpha m_{\phi}^{2}} \exp \left(\frac{2\sqrt{2\alpha}m_{\phi}}{M_{pl}^{2}}\sigma \right) -  \frac{M_{pl}^{2}}{32\alpha m_{\phi}^{2}} \exp \left(\frac{2\sqrt{2\alpha}m_{\phi}}{M_{pl}^{2}}\sigma_{end} \right).
\end{equation}

\noindent If we make the assumption that the inflationary plateau is long, such that $\sigma_{end} << \sigma$, then the number of e-folds in terms of the canonical inflaton is

\begin{equation}\label{eqn:r72}
N\left( \sigma \right) \approx \frac{M_{pl}^{2}}{32\alpha m_{\phi}^{2}} \exp \left(\frac{2\sqrt{2\alpha}m_{\phi}}{M_{pl}^{2}}\sigma \right),
\end{equation}

\noindent and so

\begin{equation}\label{eqn:r73}
\sigma \left(N \right) = \frac{M_{pl}^{2}}{2\sqrt{2\alpha} m_{\phi}} \ln \left( \frac{32\alpha m_{\phi}^{2}}{M_{pl}^{2}} N \right).
\end{equation}

\noindent Substituting (\ref{eqn:r73}) into (\ref{eqn:r64}), we find that the corresponding $\phi \left(N\right)$ is

\begin{equation}\label{eqn:r74}
\phi \left( N \right) = 2\sqrt{N} M_{pl}.
\end{equation}

\noindent We are now ready to calculate the quantities which will allow us to assess how viable this inflation model is.

\section{Slow-Roll Parameters and Observables}\label{section:25}

There are a number of quantities which are useful in assessing the viability of an inflation model (see Section \ref{section:12}), namely the scalar spectral index $n_{s}$, the tensor-to-scalar ratio, $r$, and the running of the spectral index $\alpha_{s}$. These observable quantities are calculated in terms of the slow-roll parameters $\epsilon$ and $\eta$, which are given in the Einstein frame by

\begin{equation}\label{eqn:r75}
\epsilon = \frac{M_{pl}^{2}}{2} \left( \frac{V'_{E}}{V_{E}} \right)^{2},
\end{equation}

\begin{equation}\label{eqn:r76}
\eta = M_{pl}^{2} \frac{V''_{E}}{V_{E}},
\end{equation}

\noindent where the primes denote a derivative with respect to the inflaton field, $\sigma$, in the Einstein frame. We have from (\ref{eqn:r69}) that

\begin{equation}\label{eqn:r77}
\frac{\partial V_{E}}{\partial \sigma} = \left( -4 \right) \frac{M_{pl}^{4}}{4\alpha}\left(-\frac{2\sqrt{2\alpha}m_{\phi}}{M_{pl}^{2}} \exp \left( -\frac{2\sqrt{2\alpha}m_{\phi}}{M_{pl}^{2}}\sigma \right)\right),
\end{equation}

\noindent and differentiating again gives

\begin{equation}\label{eqn:r78}
\frac{\partial^{2} V_{E}}{\partial \sigma^{2}} =  \left( -4 \right) \frac{M_{pl}^{4}}{4\alpha}\left(-\frac{2\sqrt{2\alpha}m_{\phi}}{M_{pl}^{2}}\right)^{2} \exp \left( -\frac{2\sqrt{2\alpha}m_{\phi}}{M_{pl}^{2}}\sigma \right).
\end{equation}

\noindent Using $V_{E} \approx M_{pl}^{4}/4\alpha$, $\epsilon$ then is given by

\begin{equation}\label{eqn:r79}
\epsilon = \frac{64\alpha m_{\phi}^{2}}{M_{pl}^{2}} \exp \left(-\frac{4\sqrt{2\alpha}m_{\phi}}{M_{pl}^{2}} \sigma \right),
\end{equation}

\noindent and $\eta$ is

\begin{equation}\label{eqn:r80}
\eta = -\frac{32\alpha m_{\phi}^{2}}{M_{pl}^{2}} \exp \left(-\frac{2\sqrt{2\alpha}m_{\phi}}{M_{pl}^{2}}\sigma \right).
\end{equation}

\noindent We can substitute in the expression for $\sigma \left( N \right)$ (\ref{eqn:r73}) to (\ref{eqn:r79}) and (\ref{eqn:r80}) to give the slow-roll parameters in terms of the number of e-folds of inflation

\begin{equation}\label{eqn:r81}
\epsilon = \frac{M_{pl}^{2}}{16\alpha m_{\phi}^{2}} \frac{1}{N^{2}},
\end{equation}

\noindent and 

\begin{equation}\label{eqn:r82}
\eta = -\frac{1}{N}.
\end{equation}

\noindent The scalar spectral index is given by

\begin{equation}\label{eqn:r83}
n_{s} = 1 + 2\eta - 6\epsilon \approx 1 - \frac{2}{N},
\end{equation}

\noindent where we assume that the contribution from the $\epsilon$ term is negligible compared to the $\eta$ term. This is sensible if we note that

\begin{equation}\label{eqn:r84}
\epsilon = \frac{M_{pl}^{2}}{16\alpha m_{\phi}^{2}} \frac{1}{N^{2}} = \frac{1}{2N}\frac{M_{pl}^{2}}{8\alpha m_{\phi}^{2}}\frac{1}{N} = \frac{1}{2N}\frac{M_{pl}^{4}}{2\alpha m_{\phi}^{2}} \frac{1}{4N M_{pl}^{2}} = \frac{1}{2N} \frac{\phi_{0}^{2}}{\phi \left(N \right)^{2}},
\end{equation}

\noindent and $\phi_{0}^{2}<<\phi^{2}\left(N\right)$ during slow-roll inflation. The tensor to scalar ratio is then 

\begin{equation}\label{eqn:r85}
r = 16\epsilon \approx \frac{M_{pl}^{2}}{\alpha m_{\phi}^{2}} \frac{1}{N^{2}},
\end{equation}

\noindent and running of the spectral index is given by 

\begin{equation}\label{eqn:r86}
\alpha_{s} = -\frac{d n_{s}}{d N} \approx -\frac{2}{N^{2}}.
\end{equation}

\noindent The end of slow-roll inflation is determined by the condition

\begin{equation}\label{eqn:r87}
\left| \eta \left(\sigma \right) \right|  \approx 1,
\end{equation}

\noindent where

\begin{equation}\label{eqn:r88}
\left| \eta \left(\sigma \right) \right| \; = \; \left| -\frac{32\alpha m_{\phi}^{2}}{M_{pl}^{2}} \exp \left(-\frac{2\sqrt{2\alpha}m_{\phi}}{M_{pl}^{2}}\sigma \right) \right|.
\end{equation}

\noindent The condition for slow-roll inflation to end is thus

\begin{equation}\label{eqn:r89}
\frac{32\alpha m_{\phi}^{2}}{M_{pl}^{2}} \exp \left(-\frac{2\sqrt{2\alpha}m_{\phi}}{M_{pl}^{2}}\sigma \right) \approx 1,
\end{equation}

\noindent and the inflaton field at the end of slow-roll inflation is therefore

\begin{equation}\label{eqn:r90}
\sigma_{end} \approx \frac{M_{pl}^{2}}{2\sqrt{2\alpha}m_{\phi}} \ln \left(\frac{32\alpha m_{\phi}^{2}}{M_{pl}^{2}} \right).
\end{equation}

\noindent Using (\ref{eqn:r64}), the Jordan frame inflaton at the end of slow-roll inflation is then

\begin{multline}\label{eqn:r91}
\phi_{end} = \frac{M_{pl}^{2}}{2\sqrt{2\alpha}m_{\phi}}\exp \left(\frac{\sqrt{2\alpha}m_{\phi}}{M_{pl}^{2}}\sigma_{end} \right) \\
 = \frac{M_{pl}^{2}}{2\sqrt{2\alpha}m_{\phi}}\exp \left(\frac{\sqrt{2\alpha}m_{\phi}}{M_{pl}^{2}}\frac{M_{pl}^{2}}{2\sqrt{2\alpha}m_{\phi}} \ln \left(\frac{32\alpha m_{\phi}^{2}}{M_{pl}^{2}} \right) \right),
\end{multline}

\noindent which gives a value of the inflaton at the end of slow-roll inflation in the Jordan frame to be

\begin{equation}\label{eqn:r92}
\phi_{end} = 2M_{pl}.
\end{equation}

\noindent The primordial curvature power spectrum is given by

\begin{equation}\label{eqn:r93}
\mathcal{P}_{\mathcal{R}} = \frac{V_{E}}{24\pi^{2} \epsilon M_{pl}^{4}} = \frac{m_{\phi}^{2}N^{2}}{6\pi^{2} M_{pl}^{2}},
\end{equation}

\noindent using (\ref{eqn:r68}) and (\ref{eqn:r81}). This can be used to find an estimate for the inflaton mass $m_{\phi}$. Using $N=60$ as an estimate for the pivot scale (this will be discussed in more detail later), and the amplitude of the primordial curvature power spectrum from the Planck satellite results, $A_{s} = 2.1\times 10^{-9}$ \cite{Planck18}, we find that

\begin{equation}\label{eqn:r94}
m_{\phi} = \frac{\sqrt{6}\pi M_{pl}\sqrt{2.1 \times 10^{-9}}}{N} = 1.4 \times 10^{13} \GeV,
\end{equation}

\noindent which provides an estimate of the inflaton mass which we will utilise throughout this work.

\subsection{Equivalence of $n_{s}$ and $\mathcal{P}_{\mathcal{R}}$ in Palatini $R^{2}$ Quadratic Inflation and Conventional $\phi^{2}$ Chaotic Inflation}\label{section:251}
In this section we demonstrate the equivalence of the scalar spectral index, $n_{s}$, and the curvature power spectrum, $\mathcal{P}_{\mathcal{R}}$, between those calculated in the Einstein frame in the Palatini $\phi^{2}$ inflation model with an $R^{2}$ term and conventional chaotic $\phi^{2}$ inflation. This equivalence was first derived by Enckell et al in \cite{enckell18}.

We have that the Einstein frame potential $V_{E}$ in the Palatini $\phi^{2}$ inflation model with an $R^{2}$ term in terms of the Jordan frame potential $V$ is given by

\begin{equation}\label{eqn:r95}
V_{E}\left(\phi \right) = \frac{V\left(\phi \right)}{1 + \frac{4\alpha V }{M_{pl}^{4}}},
\end{equation}

\noindent differentiated with respect to $\phi$ this is 

\begin{equation}\label{eqn:r96}
\frac{\partial V_{E}}{\partial \phi} = \frac{V'}{\left( 1 + \frac{4\alpha V}{M_{pl}^{4}}\right)^{2}}.
\end{equation}

\noindent This can be converted to a derivative in terms of the canonical scalar $\sigma$ using the fact that

\begin{equation}\label{eqn:r97}
\frac{\partial V_{E}}{\partial \sigma} = \frac{\partial V_{E}}{\partial \phi}\frac{\partial \phi}{\partial \sigma} = \sqrt{1 + \frac{4\alpha V}{M_{pl}^{4}}}\frac{\partial V_{E}}{\partial \phi},
\end{equation}

\noindent from (\ref{eqn:r43}). This gives

\begin{equation}\label{eqn:r98}
\frac{\partial V_{E}}{\partial \sigma} = \frac{V'}{\left( 1 + \frac{4\alpha V}{M_{pl}^{4}}\right)^{\frac{3}{2}}},
\end{equation}

\noindent and differentiating again gives

\begin{equation}\label{eqn:r99}
\frac{\partial^{2}V_{E}}{\partial \sigma^{2}} = \frac{\partial}{\partial \sigma}\left(\sqrt{1 + \frac{4\alpha V}{M_{pl}^{4}}}\frac{\partial V_{E}}{\partial \phi} \right) = \sqrt{1 + \frac{4\alpha V}{M_{pl}^{4}}}\frac{\partial}{\partial \phi}\left(\sqrt{1 + \frac{4\alpha V}{M_{pl}^{4}}}\frac{\partial V_{E}}{\partial \phi} \right),
\end{equation}

\noindent using (\ref{eqn:r97}). Evaluating (\ref{eqn:r99}) gives

\begin{equation}\label{eqn:r100}
\frac{\partial^{2}V_{E}}{\partial \sigma^{2}} = \frac{V''}{\left( 1 + \frac{4\alpha V}{M_{pl}^{4}}\right)} - \frac{6\alpha V'^{2}}{M_{pl}^{4}\left( 1 + \frac{4\alpha V}{M_{pl}^{4}}\right)^{2}}.
\end{equation}

From (\ref{eqn:r76}), the $\eta$ parameter using (\ref{eqn:r100}) and (\ref{eqn:r95}) is

\begin{equation}\label{eqn:r101}
\eta = \frac{M_{pl}^{2}}{V}\left(1 + \frac{4\alpha V}{M_{pl}^{4}} \right)\left[\frac{V''}{\left( 1 + \frac{4\alpha V}{M_{pl}^{4}}\right)} - \frac{6\alpha V'^{2}}{M_{pl}^{4}\left( 1 + \frac{4\alpha V}{M_{pl}^{4}}\right)^{2}} \right],
\end{equation}

\begin{equation}\label{eqn:r102}
\Rightarrow \eta = M_{pl}^{2}\frac{V''}{V} - \frac{12\alpha V}{M_{pl}^{4}\left(1 + \frac{4\alpha V}{M_{pl}^{4}} \right)}\frac{M_{pl}^{2}}{2}\left(\frac{V'}{V}\right)^{2}.
\end{equation}

\noindent Defining $\bar{\eta}$ and $\bar{\epsilon}$ as the slow-roll parameters for the Jordan frame potential $V(\phi )$, we have that the $\eta$ parameter in the Einstein frame is

\begin{equation}\label{eqn:r103}
\eta = \bar{\eta} - \frac{12\alpha V}{M_{pl}^{4}\left(1 + \frac{4\alpha V}{M_{pl}^{4}} \right)}\bar{\epsilon}.
\end{equation}

\noindent The $\epsilon$ parameter in the Einstein frame from (\ref{eqn:r75}), using (\ref{eqn:r95}) and (\ref{eqn:r98}), is

\begin{equation}\label{eqn:r104}
\epsilon = \frac{M_{pl}^{2}}{2}\frac{\left(1 + \frac{4\alpha V}{M_{pl}^{4}} \right)^{2}}{V^{2}}\frac{V'^{2}}{\left(1 + \frac{4\alpha V}{M_{pl}^{4}} \right)^{3}},
\end{equation}

\begin{equation}\label{eqn:r105}
\Rightarrow \epsilon = \frac{M_{pl}^{2}}{2}\frac{V'^{2}}{V^{2}}\frac{1}{\left(1 + \frac{4\alpha V}{M_{pl}^{4}} \right)} = \frac{\bar{\epsilon}}{\left(1 + \frac{4\alpha V}{M_{pl}^{4}} \right)}.
\end{equation}

The scalar spectral index in the Einstein frame is

\begin{equation}\label{eqn:r106}
n_{s} = 1 + 2\eta - 6\epsilon,
\end{equation}

\noindent and substituting (\ref{eqn:r103}) and (\ref{eqn:r105}) into (\ref{eqn:r106}) $n_{s}$ is

\begin{equation}\label{eqn:r107}
n_{s} = 1 + 2\left[\bar{\eta} - \frac{12\alpha V}{M_{pl}^{4}\left(1 + \frac{4\alpha V}{M_{pl}^{4}} \right)}\bar{\epsilon}\right] - \frac{6\bar{\epsilon}}{\left(1 + \frac{4\alpha V}{M_{pl}^{4}} \right)},
\end{equation}

\begin{equation}\label{eqn:r108}
\Rightarrow n_{s} = 1 + 2\bar{\eta} - \frac{4\alpha V}{M_{pl}^{4}}\frac{6\bar{\epsilon}}{\left(1 + \frac{4\alpha V}{M_{pl}^{4}} \right)} - \frac{6\bar{\epsilon}}{\left(1 + \frac{4\alpha V}{M_{pl}^{4}} \right)},
\end{equation}

\begin{equation}\label{eqn:r109}
 \Rightarrow n_{s} = n_{s} \equiv 1 + 2\bar{\eta} - 6\bar{\epsilon}.
\end{equation}

\noindent We therefore have that the scalar spectral index in the Einstein frame for Palatini $R^{2}$ quadratic inflation, provided that the quartic derivative terms are negligible during slow-roll inflation, is equal to the scalar spectral index in the Jordan frame for conventional chaotic $\phi^{2}$ inflation, despite the fact that the slow-roll parameters (\ref{eqn:r103}), (\ref{eqn:r105}) and the resulting tensor-to-scalar ratio, $r$, are different.

The primordial curvature power spectrum in the Einstein frame (\ref{eqn:r93}) is 

\begin{equation}\label{eqn:r110}
\mathcal{P}_{\mathcal{R}} = \frac{V_{E}}{24\pi^{2}\epsilon M_{pl}^{4}},
\end{equation}

\noindent substituting (\ref{eqn:r95}) and (\ref{eqn:r105}) into the expression gives

\begin{equation}\label{eqn:r111}
\mathcal{P}_{\mathcal{R}} =  \frac{V}{\left(1 + \frac{4\alpha V}{M_{pl}^{4}} \right)}\frac{\left(1 + \frac{4\alpha V}{M_{pl}^{4}} \right)}{24\pi^{2}\bar{\epsilon} M_{pl}^{4}} \equiv \frac{V}{24\pi^{2}\bar{\epsilon} M_{pl}^{4}},
\end{equation}

\noindent and we have that the primordial curvature power spectrum for $\phi^{2}$ Palatini inflation with an $R^{2}$ term is equal to the primordial curvature power spectrum for conventional chaotic $\phi^{2}$ inflation in this case.

The Jordan frame potential using (\ref{eqn:r74}) is 

\begin{equation}\label{eqn:r112}
V\left(\phi \right) = 2m_{\phi}^{2}N M_{pl}^{2},
\end{equation}

\noindent and using (\ref{eqn:r75}) and (\ref{eqn:r76}), the slow-roll parameters in the Jordan frame are

\begin{equation}\label{eqn:r113}
\bar{\eta} = \bar{\epsilon} = \frac{1}{2N}.
\end{equation}

\noindent The scalar spectral index in the Jordan frame as a function of $N$ is then

\begin{equation}\label{eqn:r114}
n_{s} = 1 - \frac{2}{N},
\end{equation}

\noindent as it is in the Einstein frame (\ref{eqn:r83}) for $\left| \epsilon \right| << \left| \eta \right|$ during slow-roll. The primordial curvature power spectrum (\ref{eqn:r111}) as a function of N using (\ref{eqn:r112}) and (\ref{eqn:r113}) is then

\begin{equation}\label{eqn:r115}
\mathcal{P}_{\mathcal{R}} = \frac{m_{\phi}^{2}N^{2}}{6\pi^{2}M_{pl}^{2}},
\end{equation}

\noindent the same as (\ref{eqn:r93}) in the Einstein frame.

It is now instructive to check the equivalence of the number of e-folds in the two frames. In the Einstein frame the expression for the number of e-folds $N\left(\sigma \right)$ is

\begin{equation}\label{eqn:r116}
N = -\frac{1}{M_{pl}^{2}}\int^{\sigma_{end}}_{\sigma} \frac{V_{E}}{V_{E}'} d\sigma.
\end{equation}

\noindent Using (\ref{eqn:r95}), (\ref{eqn:r98}) and the fact that

\begin{equation}\label{eqn:r117}
d\sigma = \frac{d\sigma}{d\phi} d\phi = \frac{d\phi}{\sqrt{1 + \frac{4\alpha V}{M_{pl}^{4}}}},
\end{equation}

\noindent we have

\begin{equation}\label{eqn:r118}
N = -\frac{1}{M_{pl}^{2}}\int^{\phi\left(\sigma_{end}\right)}_{\phi} \frac{V}{\left(1 + \frac{4\alpha V}{M_{pl}^{4}} \right)}\frac{\left(1 + \frac{4\alpha V}{M_{pl}^{4}} \right)^{\frac{3}{2}}}{V'} \frac{d\phi}{\sqrt{1 + \frac{4\alpha V}{M_{pl}^{4}}}},
\end{equation}

\begin{equation}\label{eqn:r119}
\Rightarrow N = -\frac{1}{M_{pl}^{2}}\int^{\phi\left(\sigma_{end}\right)}_{\phi} \frac{V}{V'} d\phi,
\end{equation}

\noindent which is equivalent to the expression for the number of e-folds in conventional chaotic $\phi^{2}$ inflation provided that the field at the end of slow-roll inflation is very small $\phi\left(\sigma_{end}\right) << \phi\left(N\right)$, since $\phi \left(N_{end}\right)$ will be different depending on which frame the number of e-folds is calculated in, $\left| \eta \left(N_{end}\right) \right| = 1$ will generally occur at a different $\phi \left(N_{end}\right)$ to $\left| \bar{\eta} \left(N_{end}\right) \right| = 1$.

\section{Sub-Planckian Inflaton}\label{section:26}
As discussed in Section \ref{section:22}; one of the issues with traditional chaotic $\phi^{2}$ inflation is the size of the inflaton field needed in order to account for inflation, $\phi \sim 15M_{pl}$ at $N \approx 50$. In the Einstein frame, $M_{pl}$ is the scale of quantum gravity, since in the Einstein frame the description of gravity corresponds to the General Relativity description (in the Jordan frame the scale of quantum gravity is rescaled according to the conformal factor $\Omega$). A super-Planckian inflaton can be problematic when the model is considered as part of a quantum gravity theory, which we assume that the Standard Model of particle physics and the Standard Cosmological Model are effective theory descriptions of. This is because extending inflation up to a complete theory of gravity is expected to introduce Planck scale-suppressed corrections to the inflaton potential, which can alter the predictions of the model. We will explore the effects of this in greater depth later.

Firstly, we examine whether the $\phi^{2}$ Palatini inflation model with an $R^{2}$ term can produce successful inflation if the inflaton remains sub-Planckian for the duration of inflation. This constraint looks like

\begin{equation}\label{eqn:r120}
\sigma \left(N \right) < M_{pl}.
\end{equation}

\noindent The inflaton field in terms of the number of e-folds is given by

\begin{equation}\label{eqn:r121}
\sigma \left(N \right) = \frac{M_{pl}^{2}}{2\sqrt{2\alpha} m_{\phi}} \ln \left( \frac{32\alpha m_{\phi}^{2}}{M_{pl}^{2}} N \right).
\end{equation}

\noindent Substituting (\ref{eqn:r121}) into (\ref{eqn:r120}) we find the condition for a sub-Planckian inflaton to be

\begin{equation}\label{eqn:r122}
\ln \left(\frac{32 \alpha m_{\phi}^{2}N}{M_{pl}^{2}} \right) < \frac{2\sqrt{2\alpha}m_{\phi}}{M_{pl}},
\end{equation}

\noindent which can be expressed as

\begin{equation}\label{eqn:r123}
\ln\left( \alpha \right) + \ln\left(N\right) + \ln\left(\frac{32m_{\phi}^{2}}{M_{pl}^{2}}\right) \lesssim \frac{2\sqrt{2\alpha}m_{\phi}}{M_{pl}}.
\end{equation}

Using $M_{pl} = 2.4 \times 10^{18} \GeV$ and $m_{\phi} = 1.4 \times 10^{13} \GeV$ the condition (\ref{eqn:r123}) becomes

\begin{equation}\label{eqn:r124}
\ln \left(\alpha \right) - 16.5 < 1.65 \times 10^{-5}\sqrt{\alpha}.
\end{equation}

\noindent We find that this constraint is satisfied for $\alpha \gtrsim 1.0 \times 10^{12}$. Thus, we arrive at the condition in order for the inflaton field to remain sub-Planckian throughout inflation in this model is

\begin{equation}\label{eqn:r125}
\alpha \gtrsim 10^{12}.
\end{equation}

\noindent Although it is well-defined only in the Einstein frame, it is interesting to note that a similar (but not identical) condition can be derived in the Jordan frame by considering the size of the effective Planck mass during inflation, $ M_{pl,eff}^{2} = \Omega^{2}M_{pl}^{2}$, where the condition is then

\begin{equation}\label{eqn:r126}
\phi < M_{pl,eff}.
\end{equation}

\noindent We can interpret $M_{pl, eff}^{2} = \Omega^{2}M_{pl}^{2}$ as being due to the nature of the complete theory of quantum gravity. It tells us that the scale of the UV complete theory of gravity is $\phi$ dependent in the Jordan frame.

The plateau limit applies throughout the duration of inflation, so we can say

\begin{equation}\label{eqn:r127}
\frac{2\alpha m_{\phi}^{2}\phi^{2}}{M_{pl}^{4}} >>1 \Rightarrow \Omega^{2} \thickapprox \frac{2\alpha m_{\phi}^{2}\phi^{2}}{M_{pl}^{4}},
\end{equation}

\noindent and the effective Planck mass is therefore

\begin{equation}\label{eqn:r128}
M_{pl, eff}^{2} = \Omega^{2}M_{pl}^{2} \thickapprox \frac{2\alpha m_{\phi}^{2}\phi^{2}}{M_{pl}^{2}}.
\end{equation}

\noindent If we require that $\phi < M_{pl, eff}$ then we require the constraint on $\alpha$ to be 

\begin{equation}\label{eqn:r129}
\alpha > \frac{M_{pl}^{2}}{2 m_{\phi}^{2}},
\end{equation}

\noindent then for $m_{\phi} = 1.4 \times 10^{13}\GeV$ this gives

\begin{equation}\label{eqn:r130}
\alpha > 1.5 \times 10^{10},
\end{equation}

\noindent which is automatically satisfied if the Einstein frame constraint (\ref{eqn:r125}) is satisfied. This is a weaker constraint, and it is more appropriate in this context to consider the sub-Planckian constraint in the Einstein frame where the scalar $\sigma$ is canonically normalised and where the relationship between the Planck mass and scale of quantum gravity is well-defined since gravity corresponds to conventional General Relativity in the Einstein frame.

\section{Effective Theory Corrections from Quantum Gravity}\label{section:27}
\subsection{Planck-Suppressed Potential Corrections and the $\eta$-shift}\label{section:271}
It is at this point that we consider the robustness of the predictions of this inflation model against Planck scale-suppressed corrections to the potential from quantum gravity, and examine what kind of constraints these place on the model - namely the size of the $\alpha$ parameter. Since the majority of the observable quantities calculated in inflation models are directly derived from the inflaton potential, these quantities are highly sensitive to any modifications which may be made to the potential, and therefore any corrections added to the potential may significantly alter the predictions of the model. It is particularly important therefore to ensure that an inflation model can maintain good predictions for the inflationary observables when embedded into a theory of quantum gravity. 

In this case we will consider the lowest order corrections arising due to the existence of a UV completion of gravity itself, both generally and in the presence of an approximate shift symmetry in the model. The test of the 'tolerance' of an inflation model with respect to quantum corrections will be established by calculating what we refer to as the $\eta$-shift. That is, the amount by which the $\eta$ parameter is modified by the potential corrections. This shift can be constrained by the bounds on the scalar spectral index derived from experiments, and in order for the presence of such corrections to not significantly affect the value of the scalar spectral index in this model, we will require that

\begin{equation}\label{eqn:r131}
\left| \Delta \eta \right| < 0.001,
\end{equation}

\noindent in order to respect the bounds on $n_{s}$ given in \cite{planck184}.

The $\eta$-shift will be calculated by considering Planck-suppressed corrections to the Einstein frame potential. The effective Lagrangian of an inflation theory embedded as an effective theory of a quantum gravity completion is expected to be of the form \cite{infleff}

\begin{equation}\label{eqn:r132}
\mathcal{L}_{eff} = \mathcal{L} + \sum_{n} k_{n}\frac{\phi^{n}}{\Lambda^{n-4}},
\end{equation}

\noindent where $\Lambda$ is the scale of quantum gravity, $k_{n}$ are dimensionless constants, and the sum is over all operators consistent with the symmetries of the complete theory. In the $R^{2}$ Palatini inflation model with a $\phi^{2}$ potential in the Einstein frame, $\Lambda = M_{pl}$ is the scale at which perturbative quantum gravity breaks down, and the lowest order correction we can consider, assuming the corrections in the Einstein frame obey the existing $\phi \rightarrow -\phi$ symmetry of the Jordan frame potential, corresponding to a $\sigma \rightarrow -\sigma$ symmetry in the Einstein frame, would be

\begin{equation}\label{eqn:r133}
\Delta V_{E} = \frac{k_{6} \sigma^{6}}{M_{pl}^{2}},
\end{equation}

\noindent where we assume that $k_{6}$ is an order one number and we will drop the "6" subscript from here. The Einstein frame potential is then

\begin{equation}\label{eqn:r134}
V_{TOT} = V_{E} + \Delta V_{E} = V_{E} + \frac{k \sigma^{6}}{M_{pl}^{2}} + higher \;  order \;  corrections.
\end{equation}

\noindent Since $\epsilon << \left| \eta \right|$ during slow roll inflation, we can approximate the scalar spectral index as $n_{s} \approx 1 + 2\eta$. With the shift in the potential we then have that

\begin{equation}\label{eqn:r135}
\eta = M_{pl}^{2} \frac{V_{TOT}''}{V_{TOT}} = M_{pl}^{2} \frac{V_{E}'' + \Delta V_{E}''}{V_{E} + \Delta V_{E}}.
\end{equation}

\noindent Extracting a factor of $V_{E}$ from the denominator gives

\begin{equation}\label{eqn:r136}
\eta = \frac{M_{pl}^{2}}{V_{E}}\frac{V_{E}'' + \Delta V_{E}''}{\left(1 + \frac{\Delta V_{E}}{V_{E}} \right)}.
\end{equation}

\noindent Taking $\Delta V_{E} << V_{E}$ we can binomially expand the denominator as

\begin{equation}\label{eqn:r137}
\left(1 + \frac{\Delta V_{E}}{V_{E}} \right)^{-1} \thickapprox 1 - \frac{\Delta V_{E}}{V_{E}}.
\end{equation}

\noindent $\eta$ is then

\begin{equation}\label{eqn:r138}
\eta = \frac{M_{pl}^{2}}{V_{E}} \left(V_{E}'' + \Delta V_{E}'' \right) \left(1 - \frac{\Delta V_{E}}{V_{E}} \right) = M_{pl}^{2} \left(\frac{V_{E}'' }{V_{E}}+ \frac{\Delta V_{E}''}{V_{E}} \right) \left(1 - \frac{\Delta V_{E}}{V_{E}} \right),
\end{equation}

\noindent expanding this gives

\begin{equation}\label{eqn:r139}
\eta = M_{pl}^{2} \left( \frac{V_{E}''}{V_{E}} - \frac{\Delta V_{E}}{V_{E}^{2}}V_{E}'' + \frac{\Delta V_{E}''}{V_{E}} + \mathcal{O}\left( \Delta^{2} \right) \right),
\end{equation}

\noindent and to leading order the modified $\eta$ is therefore

\begin{equation}\label{eqn:r140}
\eta = M_{pl}^{2} \left( \frac{V_{E}''}{V_{E}} - \frac{\Delta V_{E}}{V_{E}^{2}}V_{E}'' + \frac{\Delta V_{E}''}{V_{E}} \right).
\end{equation}

\noindent To simplify this expression, we first want to show that

\begin{equation}\label{eqn:r141}
\left| \frac{\Delta V_{E}''}{V_{E}} \right| >> \left| \frac{V_{E}''}{V_{E}} \right| \left| \frac{\Delta V_{E}}{V_{E}} \right|, 
\end{equation}

\noindent where 

\begin{equation}\label{eqn:r142}
\Delta V_{E}'' = \frac{30k \sigma^{4}}{M_{pl}^{2}},
\end{equation}

\noindent from (\ref{eqn:r133}). Substituting (\ref{eqn:r142}) into (\ref{eqn:r141}), we find

\begin{equation}\label{eqn:r143}
\frac{30k \sigma^{4}}{M_{pl}^{2}} \frac{1}{V_{E}} >>  \left| \frac{V_{E}''}{V_{E}} \right| \frac{1}{V_{E}} \frac{k \sigma^{6}}{M_{pl}^{2}}.
\end{equation}

\noindent Following some cancellations the condition (\ref{eqn:r141}) becomes

\begin{equation}\label{eqn:r144}
\frac{\sigma^{2}}{M_{pl}^{2}}< < \frac{30}{M_{pl}^{2}\left| \frac{V_{E}''}{V_{E}} \right|} = \frac{30}{\left| \eta \right|}.
\end{equation}

\noindent For $n_{s} \thickapprox 0.966$, we have $2\eta = 0.034 \Rightarrow \eta = 0.017$. Substituting this into (\ref{eqn:r144}) gives 

\begin{equation}\label{eqn:r145}
\frac{\sigma^{2}}{M_{pl}^{2}} << 1765.
\end{equation}

\noindent This is generally satisfied for $\sigma < M_{pl}$, and we can therefore safely ignore the second term in (\ref{eqn:r140}). The shift in $\eta$ due to the $\sigma^{6}$ corrections is therefore

\begin{equation}\label{eqn:r146}
\Delta \eta \approx M_{pl}^{2} \frac{\Delta V_{E}''}{V_{E}}.
\end{equation}

\noindent The Einstein potential during slow roll can be approximated as 

\begin{equation}\label{eqn:r147}
V_{E} = \frac{M_{pl}^{4}}{4\alpha},
\end{equation}

\noindent and using (\ref{eqn:r142}) the $\eta$-shift is then given by

\begin{equation}\label{eqn:r148}
\Delta \eta = \frac{120k\alpha \sigma^{4}}{M_{pl}^{4}}.
\end{equation}

\noindent Substituting the expression for $\sigma (N)$, (\ref{eqn:r73}), into (\ref{eqn:r148}) to get the $\eta$-shift in terms of the number of e-folds we have

\begin{equation}\label{eqn:r149}
\Delta \eta = \frac{30kM_{pl}^{4}}{16\alpha m_{\phi}^{4}}\left[\ln\left(\frac{32\alpha m_{\phi}^{2}}{M_{pl}^{2}}N \right)\right]^{4},
\end{equation}

\noindent and using the properties of logarithms we can write this as

\begin{equation}\label{eqn:r150}
\Delta \eta = \frac{30kM_{pl}^{4}}{\alpha m_{\phi}^{4}}\left[\ln\left(\frac{4\sqrt{2\alpha} m_{\phi}}{M_{pl}}\sqrt{N} \right)\right]^{4}.
\end{equation}

\noindent The condition for the $\sigma^{6}$ potential corrections to not significantly affect the value of the spectral index then becomes

\begin{equation}\label{eqn:r151}
\frac{30kM_{pl}^{4}}{\alpha m_{\phi}^{4}}\left[\ln\left(\frac{4\sqrt{2\alpha} m_{\phi}}{M_{pl}}\sqrt{N} \right)\right]^{4} < 0.001.
\end{equation}

\noindent This allows us to reframe the requirement that the Planck-suppressed potential corrections do not significantly affect the model prediction of the spectral index as a constraint on the $\alpha$ parameter. We find that at $N=60$, using the inflaton mass of $m_{\phi} = 1.4 \times 10^{13}\GeV$, this constraint is satisfied for 

\begin{equation}\label{eqn:r152}
\alpha \gtrsim 1.5 \times 10^{31}, 
\end{equation}

\noindent assuming $k \simeq 1$. \\

This is however assuming that $\sigma \left( N \right)$ itself doesn't experience a correction from the $N$ integral, since the number of e-folds itself is calculated from the ratio $V_{TOT}/V_{TOT}'$ which should include the potential correction. We next consider this effect.

The number of e-folds including the potential corrections is given by

\begin{equation}\label{eqn:r153}
N = - \frac{1}{M_{pl}^{2}} \int_{\sigma}^{\sigma_{end}} \frac{V_{E} + \Delta V_{E}}{V_{E}' + \Delta V_{E}'} d\sigma = - \frac{1}{M_{pl}^{2}} \int_{\sigma}^{\sigma_{end}} \frac{V_{E}\left(1 + \frac{\Delta V_{E}}{V_{E}}\right)}{V_{E}' \left( 1 + \frac{\Delta V_{E}'}{V_{E}'}\right)} d\sigma.
\end{equation}

\noindent Assuming $\Delta V_{E}' << V_{E}'$ we can binomially expand the denominator of the integrand to give

\begin{equation}\label{eqn:r154}
N = - \frac{1}{M_{pl}^{2}} \int_{\sigma}^{\sigma_{end}} \frac{V_{E}}{V_{E}'}\left(1 + \frac{\Delta V_{E}}{V_{E}}\right) \left( 1 - \frac{\Delta V_{E}'}{V_{E}'}\right) d\sigma.
\end{equation}

\noindent Expanding the integrand $I$ gives

\begin{equation}\label{eqn:r155}
I =  \frac{V_{E}}{V_{E}'} \left[1 - \frac{\Delta V_{E}'}{V_{E}'} + \frac{\Delta V_{E}}{V_{E}} + \mathcal{O}\left(\Delta^{2}\right) \right].
\end{equation}

\noindent We work to leading order in potential corrections, and in order to simplify the calculation we first show that we can neglect the $\Delta V_{E}/V_{E}$ term. This can be neglected if 

\begin{equation}\label{eqn:r156}
\left| \frac{\Delta V_{E}}{V_{E}'} \right| << \left| \frac{\Delta V_{E}'}{V_{E}'} \right| \left| \frac{V_{E}}{V_{E}'} \right|.
\end{equation}

\noindent Here $\Delta V_{E}' = 6k\sigma^{5}/M_{pl}^{2}$, and the condition (\ref{eqn:r156}) becomes

\begin{equation}\label{eqn:r157}
\frac{k\sigma^{6}}{M_{pl}^{2}} << \frac{6k\sigma^{5}}{M_{pl}^{2}}\left| \frac{V_{E}}{V_{E}'} \right| \Rightarrow \sigma << 6 \left| \frac{V_{E}}{V_{E}'} \right|. 
\end{equation}

\noindent Using the expression for $\epsilon$, (\ref{eqn:r75}), we can write

\begin{equation}\label{eqn:r158}
\sqrt{\epsilon} = \frac{M_{pl}}{\sqrt{2}} \left| \frac{V_{E}'}{V_{E}} \right| \Rightarrow \left| \frac{V_{E}}{V_{E}'} \right| = \frac{M_{pl}}{\sqrt{2\epsilon}},
\end{equation}

\noindent and the condition on $\sigma$ for the $\frac{\Delta V_{E}}{V_{E}}$ term being negligible is then

\begin{equation}\label{eqn:r159}
\sigma << \frac{6M_{pl}}{\sqrt{2\epsilon}}.
\end{equation}

\noindent Since $\epsilon << 1$ during slow-roll this is generally satisfied, as we have $\sigma < M_{pl}$ for $\alpha \geq 10^{12}$.

We can therefore ignore the $\frac{\Delta V_{E}}{V_{E}}$ term. The number of e-folds accounting for the shift due to the potential correction is then

\begin{equation}\label{eqn:r160}
N = - \frac{1}{M_{pl}^{2}} \int_{\sigma}^{\sigma_{end}} \frac{V_{E}}{V_{E}'}\left( 1 - \frac{\Delta V_{E}'}{V_{E}'}\right) d\sigma.
\end{equation}

\noindent This can be written as a relative shift in the number of e-folds of inflation due to the potential correction, $N_{TOT} = N + \Delta N$, where

\begin{equation}\label{eqn:r161}
\Delta N = \frac{1}{M_{pl}^{2}} \int_{\sigma}^{\sigma_{end}} \frac{V_{E}}{V_{E}'}\frac{\Delta V_{E}'}{V_{E}'} d\sigma.
\end{equation}

\noindent If there is a shift in the first derivative of

\begin{equation}\label{eqn:r162}
\left| \frac{\Delta V_{E}'}{V_{E}'} \right| \thickapprox 0.1,
\end{equation}

\noindent during inflation then $N$ shifts by a factor of approximately $0.1$ and $\left| \Delta N/N \right| \approx 0.1$. In this model, $\left| \eta \right| \approx \left| 1/N \right| \Rightarrow \left|\Delta \eta \right| \approx \left| \Delta \left(1/N \right) \right|$. If $\Delta N << N$ we can write

\begin{equation}\label{eqn:r163}
\frac{1}{N} \rightarrow \frac{1}{N}\left( 1 + \frac{\Delta N}{N} \right)^{-1} =  \frac{1}{N}\left( 1 - \frac{\Delta N}{N} \right) \equiv \frac{1}{N} + \Delta \left(\frac{1}{N}\right).
\end{equation}

\noindent The $\eta$-shift is then

\begin{equation}\label{eqn:r164}
\left| \Delta \eta \right| = \left| \Delta \left(\frac{1}{N} \right) \right| = \frac{\left| \Delta N \right|}{N^{2}}.
\end{equation}

\noindent If $\left| \Delta N/N \right| \approx 0.1$ and $N=60$ then the $\eta$-shift is $\approx 0.002$. In order for the $N$-shift to make a small enough modification to the prediction of the scalar spectral index such that it remains within the parameter space put forward in \cite{Planck18}, the $\eta$-shift needs to be $0.001$ or less. The $N$-shift which fulfills this requirement on the $\eta$-shift is $0.06$ or less. This means that the condition we will use to constrain $\alpha$ is that during inflation

\begin{equation}\label{eqn:r165}
\left| \frac{\Delta V_{E}'}{V_{E}'} \right| \lesssim 0.06.
\end{equation}

\noindent We have that $\Delta V_{E}' = 6k\sigma^{5}/M_{pl}^{2}$ and using

\begin{equation}\label{eqn:r166}
V_{E}' = \frac{2\sqrt{2}}{\sqrt{\alpha}} m_{\phi}M_{pl}^{2} \exp \left(-\frac{2\sqrt{2\alpha}m_{\phi}}{M_{pl}^{2}}\sigma \right),
\end{equation}

\noindent we have

\begin{equation}\label{eqn:r167}
\frac{\Delta V_{E}'}{V_{E}'} = \frac{6k\sigma^{5}}{M_{pl}^{2}} \frac{\sqrt{\alpha}}{2\sqrt{2}m_{\phi}M_{pl}^{2}} \exp \left(\frac{2\sqrt{2\alpha}m_{\phi}}{M_{pl}^{2}}\sigma \right).
\end{equation}

\noindent Substituting in $\sigma \left(N \right)$ (\ref{eqn:r73}), we find

\begin{equation}\label{eqn:r168}
\frac{\Delta V_{E}'}{V_{E}'} = \frac{6k N}{16 \alpha} \frac{M_{pl}^{4}}{m_{\phi}^{4}} \left[ \ln \left( \frac{32\alpha m_{\phi}^{2}}{M_{pl}^{2}} N\right) \right]^{5}  \\ \Rightarrow \frac{3k N}{8 \alpha} \frac{M_{pl}^{4}}{m_{\phi}^{4}} \left[ \ln \left( \frac{32\alpha m_{\phi}^{2}}{M_{pl}^{2}} N\right) \right]^{5} \lesssim 0.06,
\end{equation}

\noindent and the constraint on $\alpha$ from the shift in the number of e-folds due to Planck-suppressed potential corrections is therefore

\begin{equation}\label{eqn:r169}
\alpha \gtrsim \frac{50k N}{8} \frac{M_{pl}^{4}}{m_{\phi}^{4}} \left[ \ln \left( \frac{32\alpha m_{\phi}^{2}}{M_{pl}^{2}} N\right) \right]^{5}.
\end{equation}

Taking $N=60$ and $k \simeq 1$, as expected for the dimensionless constants, we find that the lower bound on $\alpha$ needed for a sufficiently small $\eta$-shift due to Planck-suppressed potential corrections is

\begin{equation}\label{eqn:r170}
\alpha \gtrsim 2.2 \times 10^{32}.
\end{equation}

In general we therefore require that $\alpha \gtrsim 10^{32}$ for the observed scalar spectral index to be consistent with the Planck results \cite{planck184} if we want to treat this inflation model as an effective model in a quantum gravity theory.

\subsection{Planck-Suppressed Potential Corrections with a Broken Shift Symmetry}\label{section:272}

A shift symmetry is a global internal symmetry characterised by

\begin{equation}\label{eqn:r171}
\phi \rightarrow \phi + constant,
\end{equation}

\noindent invariance within the theory. An exact shift symmetry forbids any potential terms dependent on the scalar field, and only remains unbroken in the case of a constant or zero potential. In this model the renormalisable part of the potential is an explicit mass term, $\frac{1}{2}m_{\sigma}^{2}\sigma^{2} \approx \frac{1}{2}m_{\phi}^{2}\phi^{2}$, which means that a shift symmetry in the $R^{2}$ Palatini inflation model with a $\phi^{2}$ potential would be an approximate symmetry which would become significantly broken when the potential enters into the quadratic regime, $\frac{1}{2}m_{\sigma}^{2}\sigma^{2} \approx \frac{1}{2}m_{\phi}^{2}\phi^{2}$. The symmetry is restored in the limit that $m_{\phi}^{2} \rightarrow 0$. If we assume that the shift symmetry-breaking parameter of the complete theory is the inflaton mass squared, $m_{\phi}^{2}$, then this means that any corrections added to the potential must be proportional to $m_{\phi}^{2}$ in order to respect the symmetry.

This is significant, because in the absence of symmetries, scalar masses receive corrections of the form \cite{infleff}

\begin{equation}\label{eqn:r172}
\Delta m^{2} \propto \Lambda^{2},
\end{equation}

\noindent where in the $R^{2}$ Palatini model with a $\phi^{2}$ potential in the Einstein frame $\Lambda = M_{pl}$. Corrections of this size would produce a large renormalisation of the $\eta$ parameter, and the model would not be able to inflate successfully. In the presence of an approximate shift symmetry, the potential is protected from receiving corrections of this size because all corrections must be proportional to the symmetry breaking parameter. This means that the scalar mass corrections instead take the form

\begin{equation}\label{eqn:r173}
\Delta m^{2} \propto m^{2},
\end{equation}

\noindent which can protect the model from a large renormalisation of the inflaton mass.

In the case of a shift-symmetry broken by $m_{\phi}^{2}$, the Planck-suppressed correction also acquires a factor of $m_{\phi}^{2}$, and is therefore of the form

\begin{equation}\label{eqn:r174}
\Delta V_{E} \approx \frac{m_{\phi}^{2} \sigma^{6}}{M_{pl}^{4}},
\end{equation}

\noindent which corresponds to $k \approx m_{\phi}^{2}/M_{pl}^{2}$, in the previous analysis.

\noindent The $\eta$-shift is then 

\begin{equation}\label{eqn:r175}
\Delta \eta = M_{pl}^{2} \frac{\Delta V_{E}''}{V_{E}},
\end{equation}

\noindent where

\begin{equation}\label{eqn:r176}
\Delta V_{E}' \approx \frac{6m_{\phi}^{2}\sigma^{5}}{M_{pl}^{4}} \hspace{10mm} \Delta V_{E}'' \approx \frac{30 m_{\phi}^{2}\sigma^{4}}{M_{pl}^{4}}.
\end{equation}

Using the plateau approximation for $V_{E} \approx M_{pl}^{4}/4\alpha$ and substituting in the expression for $\sigma \left( N \right)$ (\ref{eqn:r73}), we find that the constraint on the $\eta$-shift (\ref{eqn:r131}) becomes

\begin{equation}\label{eqn:r177}
\Delta \eta = \frac{30M_{pl}^{2}}{\alpha m_{\phi}^{2}} \ln \left[\frac{4\sqrt{2\alpha}m_{\phi}}{M_{pl}} \sqrt{N} \right]^{4} \leq 0.001.
\end{equation}

\noindent This gives a bound on $\alpha$ 

\begin{equation}\label{eqn:r178}
\alpha \geq \frac{\left(3.0 \times 10^{4}\right) M_{pl}^{2}}{m_{\phi}^{2}}\left[\ln \left(\frac{4\sqrt{2\alpha}m_{\phi}}{M_{pl}}\sqrt{N} \right) \right]^{4},
\end{equation}

\noindent which is satisfied for $\alpha \gtrsim 3.6 \times 10^{19}$. Thus an approximate shift symmetry alleviates the effect of Planck-suppressed corrections and allows for a much smaller value of $\alpha$.

For completeness, we will also check the contribution from a shift in the number of e-foldings for the case of Planck-suppressed potential corrections with a broken shift symmetry. The constraint that the $\eta$-shift be kept to $0.001$ or less, assuming that the potential corrections cause a shift in the number of e-folds of inflation, is given by (\ref{eqn:r165}). In the case of a broken shift symmetry this constraint is

\begin{equation}\label{eqn:r179}
\frac{\Delta V_{E}'}{V_{E}'} = \frac{3M_{pl}^{2}N}{8\alpha m_{\phi}^{2}}\left[\ln\left(\frac{32\alpha m_{\phi}^{2}}{M_{pl}^{2}} N\right) \right]^{5} \le 0.06,
\end{equation}

\begin{equation}\label{eqn:r180}
\Rightarrow \alpha \ge \frac{50M_{pl}^{2}N}{8 m_{\phi}^{2}}\left[\ln\left(\frac{32\alpha m_{\phi}^{2}}{M_{pl}^{2}} N\right) \right]^{5},
\end{equation}

\noindent and we find that this constraint is satisfied for

\begin{equation}\label{eqn:r181}
\alpha \gtrsim 3.0 \times 10^{20}.
\end{equation}

We therefore have three different constraints on the $\alpha$ parameter ($\alpha \gtrsim 10^{12}, 10^{20}, 10^{32}$) which could apply to this inflationary paradigm, depending on the form of the corrections due to the full quantum gravity completion, and what the minimum requirements of the model are chosen to be.

\subsection{Contribution of the $\epsilon$-shift to the $n_{s}$-shift}\label{section:273}

Before moving on it is worth clarifying that the shift of the $\eta$ parameter due to the inclusion of Planck-suppressed potential corrections is the dominant effect on the scalar spectral index resulting from the inclusion of these corrections. In this section we therefore demonstrate that the shift of the $\epsilon$ parameter is small enough to be neglected in the shift of the scalar spectral index compared to the $\eta$-shift.

Using the expression of the scalar spectral index (\ref{eqn:r83}), we can write that the shift on $n_{s}$ arising due to the $\epsilon$ parameter is 

\begin{equation}\label{eqn:r182}
\Delta n_{s} = -6 \Delta \epsilon,
\end{equation}

\noindent where in terms of the total potential $\epsilon$ is 

\begin{equation}\label{eqn:r183}
\epsilon = \frac{M_{pl}^{2}}{2}\left(\frac{V_{E}'}{V_{E}}\right)^{2} = \frac{M_{pl}^{2}}{2}\left(\frac{V_{E}' + \Delta V_{E}'}{V_{E} + \Delta V_{E}}\right)^{2},
\end{equation}

\begin{equation}\label{eqn:r184}
\Rightarrow \epsilon = \epsilon_{0}\frac{\left(1 + \frac{\Delta V_{E}'}{V_{E}'} \right)^{2}}{\left(1 + \frac{\Delta V_{E}}{V_{E}} \right)^{2}},
\end{equation}

\noindent where $\epsilon_{0} = M_{pl}^{2}/2\left(V_{E}'/V_{E}\right)^{2}$. Assuming $\Delta V_{E}' << V_{E}'$ and $\Delta V_{E} << V_{E}$, this can be expanded as

\begin{equation}\label{eqn:r185}
\epsilon = \epsilon_{0} \left( 1 + \frac{2\Delta V_{E}'}{V_{E}'} \right)\left(1 - \frac{2\Delta V_{E}}{V_{E}} \right).
\end{equation}

\noindent To $\mathcal{O}\left(\Delta \right)$ this is 

\begin{equation}\label{eqn:r186}
\epsilon = \epsilon_{0} \left( 1 + \frac{2\Delta V_{E}'}{V_{E}'} - \frac{2\Delta V_{E}}{V_{E}} \right),
\end{equation}

\noindent and the $\epsilon$-shift is therefore

\begin{equation}\label{eqn:r187}
\Delta \epsilon = 2\epsilon_{0}\left(\frac{\Delta V_{E}'}{V_{E}'} - \frac{\Delta V_{E}}{V_{E}} \right). 
\end{equation}

\noindent From (\ref{eqn:r165}) we can write the first term in (\ref{eqn:r187}) as

\begin{equation}\label{eqn:r188}
\Delta \epsilon \lesssim 0.12\epsilon_{0}, 
\end{equation}

\noindent and (\ref{eqn:r156}) can be rearranged to give

\begin{equation}\label{eqn:r189}
\left|\frac{\Delta V_{E}}{V_{E}} \right| << \left|\frac{\Delta V_{E}'}{V_{E}'} \right|.
\end{equation}

\noindent It is therefore safe to neglect the second term in (\ref{eqn:r187}) relative to the first and we are left with

\begin{equation}\label{eqn:r190}
\Delta \epsilon \lesssim 0.12\epsilon_{0},
\end{equation}

\noindent as the upper bound on the $\epsilon$-shift due to the inclusion of Planck-suppressed potential corrections. Using (\ref{eqn:r81}) as our expression for $\epsilon_{0}$ we can calculate the $\epsilon-$shift for each $\alpha$ derived in Sections \ref{section:26} - \ref{section:27}. Using $m_{\phi} = 1.4 \times 10^{13}\GeV$ as an estimate of the inflaton mass and measuring the shift at $N = 60$ we find that

\begin{equation}\label{eqn:r191}
\alpha \ge 10^{12} \Rightarrow \Delta \epsilon \le 6.1 \times 10^{-8},
\end{equation}

\noindent as a minimum requirement if we impose that the inflaton is sub-Planckian (\ref{eqn:r120}),

\begin{equation}\label{eqn:r192}
\alpha \ge 10^{20} \Rightarrow \Delta \epsilon \le 6.1 \times 10^{-16},
\end{equation}

\noindent in the case of Planck-suppressed corrections added to the potential with a broken shift symmetry, and 

\begin{equation}\label{eqn:r193}
\alpha \ge 10^{32} \Rightarrow \Delta \epsilon \le 6.1 \times 10^{-28},
\end{equation}

\noindent for the case of general Planck-suppressed potential corrections. In every case the $\epsilon$-shift is extremely small compared to the $\eta$-shift $\le 10^{-3}$, and it is therefore reasonable to use the $\eta$-shift as a measure of the $n_{s}-$shift due to any additional corrections to the inflaton potential.

\subsection{The End of Inflation and Reheating}\label{section:274}
We have demonstrated that in order for the $\phi^{2}$ Palatini inflation model with an $R^{2}$ term to be compatible with Planck-scale suppressed potential corrections, very large values of $\alpha$ are required. We must therefore investigate whether such models can produce a post-inflation cosmology compatible with the Hot Big Bang model, and thus we next consider the dynamics at the end of inflation and reheating. 
The first question in order to examine this is to establish in which regime of the potential slow-roll inflation ends: while the field is still on the plateau or as the field is entering the $\sigma^{2}$ regime. In order to do this we need to check the size of the field at the end of slow-roll inflation, $\sigma_{end}$, against the size of the field at the point where the potential transitions to a $\sigma^{2}$ regime, $\sigma_{0}$.

At the point at which the potential moves from the plateau regime to the quadratic regime we can approximate that the $\phi$ field and the canonical inflaton $\sigma$ are equivalent, so at the transition the inflaton takes the value

\begin{equation}\label{eqn:r194}
\phi_{0} = \frac{M_{pl}^{2}}{\sqrt{2\alpha}m_{\phi}} = \sigma_{0}.
\end{equation}

\noindent As established earlier, the canonical inflaton at the end of inflation is given by (\ref{eqn:r90})

\begin{equation}\label{eqn:r195}
\sigma_{end} = \frac{M_{pl}^{2}}{2\sqrt{2\alpha}m_{\phi}}\ln \left(\frac{32\alpha m_{\phi}^{2}}{M_{pl}^{2}} \right),
\end{equation}

\noindent and we can take the ratio of these two values of $\sigma$ to find

\begin{equation}\label{eqn:r196}
\frac{\sigma_{end}}{\sigma_{0}} = \frac{1}{2}\ln \left(\frac{32\alpha m_{\phi}^{2}}{M_{pl}^{2}}\right).
\end{equation}

\noindent For $\alpha \sim 10^{12} - 10^{32}$ using the mass estimate (\ref{eqn:r94}) we obtain

\begin{equation}\label{eqn:r197}
\frac{\sigma_{end}}{\sigma_{0}} = 3.5 - 26.5.
\end{equation}

\noindent It is clear then that $\sigma_{0} < \sigma_{end}$. This confirms that slow-roll inflation ends while the field is still on the plateau. The Hubble parameter at $\sigma_{end}$ can thus be calculated using the slow-roll approximation on the plateau

\begin{equation}\label{eqn:r198}
\tilde{H}^{2} = \frac{V_{E}}{3 M_{pl}^{2}},
\end{equation}

\noindent where the Einstein frame potential is $V_{E} = M_{pl}^{4}/4\alpha $ on the plateau and so

\begin{equation}\label{eqn:r199}
\tilde{H} = \frac{M_{pl}}{\sqrt{12\alpha}}.
\end{equation}

During slow-roll inflation the energy density of the Universe is dominated by the potential energy of the inflaton, $\tilde{\rho} = V_{E}$. As we have established, slow-roll inflation ends while the field is still on the plateau, and here the field subsequently enters into oscillations about the potential minimum with a large oscillation amplitude. We will initially assume that the inflaton field decays rapidly after it enters into oscillations, and that the energy density which was previously dominated by the inflaton potential decays completely to radiation with essentially no loss of energy due to expansion before the inflaton condensate decays completely. This scenario can be described as the approximation of instantaneous reheating, whereupon all of the inflaton field completely decays within a few oscillations about the minimum of its potential to radiation, and reheats the Universe within a few e-folds of expansion after the end of inflation - which we approximate here as being equivalent to the end of slow-roll. Specific reheating mechanisms will be discussed in Sections \ref{section:29} - \ref{section:211}.

We can use this approximation to provide an upper bound on the reheating temperature of the model, and of the number of e-folds of expansion at the CMB pivot scale. The energy density at the end of inflation is given by

\begin{equation}\label{eqn:r200}
\tilde{\rho} = V_{E} = 3M_{pl}^{2}\tilde{H}^{2},
\end{equation}

\noindent and the energy density of radiation - which $\tilde{\rho}$ is converted to - is given by

\begin{equation}\label{eqn:r201}
\tilde{\rho_{r}} = \frac{\pi^{2}}{30}g\left(T\right) T^{4}.
\end{equation}

\noindent At the end of inflation we have that the energy density of the inflaton potential is all converted to radiation, so we can equate $\tilde{\rho} = \tilde{\rho_{r}}$ at the end of inflation, giving

\begin{equation}\label{eqn:r202}
3M_{pl}^{2}\tilde{H}^{2} = \frac{\pi^{2}}{30}g\left(T_{R_{max}}\right) T^{4}_{R_{max}},
\end{equation}

\noindent where we are defining $T_{R_{max}}$ as the reheating temperature in the case of instantaneous reheating, corresponding to the maximum possible reheating temperature. $g\left(T_{R_{max}}\right)$ is the effective number of relativistic degrees of freedom contributing to the total energy density; in the Standard Model this is given by $g\left(T_{R_{max}}\right) = 106.75$.

\noindent Substituting $\tilde{H}$ from (\ref{eqn:r199}), the upper bound on the reheating temperature is given by

\begin{equation}\label{eqn:r203}
T_{R_{max}} = \left(\frac{15}{2\pi^{2} g\left(T_{R_{max}}\right)}\right)^{\frac{1}{4}} \frac{M_{pl}}{\alpha^{\frac{1}{4}}},
\end{equation}

\noindent which is a function of the $\alpha$ parameter. 

From here we can calculate the number of e-folds of inflation corresponding to horizon exit of the CMB pivot scale. Using the fact that the entropy remains constant during slow roll inflation,

\begin{equation}\label{eqn:r204}
a^{3}g_{s}\left(T \right) T^{3} = constant,
\end{equation}

\noindent and since slow-roll inflation ends while the field is still on the plateau, we can write

\begin{equation}\label{eqn:r205}
\frac{a_{1}}{a_{2}} =  \left(\frac{g_{s}\left(T_{2}\right)}{g_{s}\left(T_{1}\right)}\right)^{\frac{1}{3}} \frac{T_{2}}{T_{1}}.
\end{equation}

\noindent We have that, in an expanding spacetime, the wavelength of the perturbations varies with scale factor as

\begin{equation}\label{eqn:r206}
\lambda = \frac{a}{a_{0}}\lambda_{0}.
\end{equation}

\noindent Introducing the scale factor at the end of slow-roll inflation $a_{end}$, this becomes

\begin{equation}\label{eqn:r207}
\lambda = \frac{a}{a_{end}}\frac{a_{end}}{a_{0}}\lambda_{0} \Rightarrow \lambda = e^{-N}\frac{a_{end}}{a_{0}}\lambda_{0},
\end{equation}

\noindent where $a_{0}$ is the scale factor today. This means

\begin{equation}\label{eqn:r208}
\frac{a_{end}}{a_{0}} = \frac{\lambda}{\lambda_{0}}e^{N} = \left[\frac{g_{s}\left(T_{0}\right)}{g_{s}\left(T_{end}\right)} \right]^{\frac{1}{3}} \frac{T_{0}}{T_{end}},
\end{equation}

\noindent using (\ref{eqn:r205}). Using $T_{end} = T_{R_{max}}$, with some rearrangement this becomes

\begin{equation}\label{eqn:r209}
e^{N} = \left[\frac{g_{s}\left(T_{0}\right)}{g_{s}\left(T_{R_{max}}\right)} \right]^{\frac{1}{3}} \frac{T_{0}}{T_{R_{max}}}\frac{\lambda_{0}}{\lambda}.
\end{equation}

\noindent Since slow-roll inflation ends while the field is solidly on the plateau, the mode corresponding to the wavelength $\lambda$ exits the horizon when $\lambda \approx \tilde{H}^{-1}$. This gives a final expression for the number of e-folds of inflation at the pivot scale $\lambda_{0}$ to be 

\begin{equation}\label{eqn:r210}
N = \ln\left[ \left(\frac{g_{s}\left(T_{0}\right)}{g_{s}\left(T_{R_{max}}\right)}\right)^{\frac{1}{3}} \frac{T_{0}\lambda_{0} \tilde{H}}{T_{R_{max}}}\right].
\end{equation}

\noindent Using the earlier derived expression for the maximum reheating temperature (\ref{eqn:r203}), we can write this expression in terms of the $\alpha$ parameter

\begin{equation}\label{eqn:r211}
N = \ln\left[ \left(\frac{g_{s}\left(T_{0}\right)}{g_{s}\left(T_{R_{max}}\right)}\right)^{\frac{1}{3}} \left[ \frac{2\pi^{2} g\left( T_{R_{max}}\right)}{15}\right]^{\frac{1}{4}} \frac{\alpha^{\frac{1}{4}}}{M_{pl}} T_{0}\lambda_{0} \tilde{H}\right].
\end{equation}

\noindent Then, using the slow roll expression for $\tilde{H}$ (\ref{eqn:r199}), this can be written as

\begin{equation}\label{eqn:r212}
N = \ln\left[ \left(\frac{g_{s}\left(T_{0}\right)}{g_{s}\left(T_{R_{max}}\right)}\right)^{\frac{1}{3}} \left[ \frac{\pi^{2} g\left( T_{R_{max}}\right)}{1080\alpha}\right]^{\frac{1}{4}} T_{0}\lambda_{0}\right],
\end{equation}

\noindent where $T_{0}$ as the present day temperature of the CMB, given by $T_{0} = 2.4 \times 10^{-13}\GeV$, and the Standard Model effective number of relativistic degrees of freedom in equilibrium at today's temperature is $g_{s}\left(T_{0}\right) = 3.91$. At the end of inflation, $T = T_{R_{max}}$ and the energy density of the Universe is contained within a plasma at equilibrium with no relativistic species decoupled, so $g_{s}\left(T_{R_{max}}\right) = g\left(T_{R_{max}}\right)$, and we use the Standard Model value of $106.75$ as an estimate. $\lambda_{0}$ is the wavelength of the mode which exits the horizon at the pivot scale. For observations by the Planck experiment the pivot scale is $k_{0} = 0.05\Mpc^{-1} = 3.2 \times 10^{-40}\GeV$, which gives $\lambda_{0} = 1.96 \times 10^{40}\GeV^{-1}$. Substituting these numbers into the expression for $N$ (\ref{eqn:r212}) we get

\begin{equation}\label{eqn:r213}
N = 62.61 - \frac{1}{4} \ln\left(\alpha \right).
\end{equation}

\noindent We will use this to calculate the predicted number of e-folds at the pivot scale given each constraint on $\alpha$, presenting the results in Section \ref{section:28}.

\subsubsection{End of Slow-Roll Inflation}

The end of slow-roll inflation corresponds to $\left|\eta \right| = 1$. From (\ref{eqn:r82}) we have that $\left|\eta \right| = 1$ corresponds to $N = 1$. In order to to check the validity of this inflation model it is instructive to check the size of the $\epsilon$ parameter at the end of slow-roll inflation. Using (\ref{eqn:r81}) at $N = 1$ we find that the value of $\epsilon$ at the end of slow-roll inflation for each bound on $\alpha$ derived is 

\begin{equation}
\alpha = 10^{12} \Rightarrow \epsilon = 1.84 \times 10^{-3},
\end{equation}

\begin{equation}
\alpha = 10^{20} \Rightarrow \epsilon = 1.84 \times 10^{-11},
\end{equation}

\begin{equation}
\alpha = 10^{32} \Rightarrow \epsilon = 1.84 \times 10^{-23},
\end{equation}

\noindent and we have that $\epsilon << \eta$ consistently for each regime of $\alpha$ we consider. We therefore have that the slow-roll treatment of the inflation model is consistent for all regimes of $\alpha$ we consider.

\subsection{Unitarity Violation and the Significance of the $\left(\partial \sigma \right)^{4}$ Term}\label{section:275}

As discussed in Section \ref{section:t2}, unitarity violation is an ongoing issue in the field of non-minimally coupled scalar field inflation, and is therefore worthy of discussion in the context of this model. Here we examine the $\left(\partial_{\mu} \phi \right)^{4}$ term in the inflaton action (denoted \textit{I} here)

\begin{equation}\label{eqn:r214}
\textit{I} = \frac{\alpha}{4M_{pl}^{4}} \frac{\left(\partial_{\mu}\phi \partial^{\mu}\phi \right)^{2}}{1 + \frac{4\alpha V\left(\phi \right)}{M_{pl}^{4}}}.
\end{equation}

\noindent This corresponds to a $2 \rightarrow 2$ scattering of particles. We express this in terms of the canonically normalised scalar $\sigma$. Using the substitution (\ref{eqn:r43}), we find that

\begin{equation}\label{eqn:r215}
\left(\partial_{\mu}\phi \partial^{\mu}\phi \right)^{2} = \left( 1 + \frac{4\alpha V\left(\phi \right)}{M_{pl}^{4}}\right)^{2} \left(\partial_{\mu}\sigma \partial^{\mu}\sigma \right)^{2},
\end{equation}

\noindent which means that the $\left(\partial \sigma \right)^{4}$ term becomes

\begin{equation}\label{eqn:r216}
\textit{I} = \frac{\alpha}{4M_{pl}^{4}}\left( 1 + \frac{4\alpha V\left(\phi \right)}{M_{pl}^{4}}\right)\left(\partial_{\mu}\sigma \partial^{\mu}\sigma \right)^{2}.
\end{equation}

\noindent Since we are studying the effects of this term in the inflationary regime, we expand the canonical inflaton about its classical background

\begin{equation}\label{eqn:r217}
\sigma \left( \textbf{x}, t \right) = \bar{\sigma}\left(t \right) + \delta \sigma \left( \textbf{x}, t \right),
\end{equation}

\noindent where $\bar{\sigma}$ corresponds to the constant classical background of the field and $\delta \sigma$ are the quantum fluctuations of the field about this background, corresponding to particles. Since the background can be regarded as varying slowly enough in time to be constant during slow roll inflation, we can approximate $\partial_{\mu} \sigma = \partial_{\mu} \delta \sigma$, and write the interaction term as 

\begin{equation}\label{eqn:r218}
\textit{I} = \frac{\alpha}{4M_{pl}^{4}}\left(\partial_{\mu}\delta \sigma \partial^{\mu}\delta \sigma \right)^{2} \left( 1 + \frac{4\alpha V\left(\bar{\phi}\right)}{M_{pl}^{4}}\right),
\end{equation}

\noindent where $\bar{\phi} = \phi\left(\bar{\sigma}\right)$. This term describes the interaction of $\delta \sigma \delta \sigma \rightarrow \delta \sigma \delta \sigma$, which we can interpret as the scattering of inflaton particles during inflation. 

In order to establish whether this interaction may cause any issues for unitarity violation in the model, we must examine the amplitude of the process. Using the Feynman rules for $\delta \sigma \delta \sigma \rightarrow \delta \sigma \delta \sigma$ scattering, we find that each derivative term in momentum space contributes a four-momentum, $\partial_{\mu} \rightarrow k_{\mu}$, and since these particles are effectively massless during slow-roll we can write $\left| \textbf{k} \right|^{2} = \tilde{E}^{2}$, where $\tilde{E}$ is the energy of the scattering particles calculated in the Einstein frame. Dimensionally, we therefore have that the tree-level amplitude for $\delta \sigma \delta \sigma \rightarrow \delta \sigma \delta \sigma$ scattering is

\begin{equation}\label{eqn:r219}
\left| \mathcal{M} \right| \approx \frac{\alpha}{4} \frac{\tilde{E}^{4}}{M_{pl}^{4}} \left( 1 + \frac{4\alpha V\left(\bar{\phi}\right)}{M_{pl}^{4}}\right),
\end{equation}

\noindent where $\tilde{E}$ corresponds to the energy scale of the perturbations of the inflaton during slow roll. 

This interaction violates unitarity when $\left| \mathcal{M} \right| \gtrsim 1$. This happens when the energy of the interaction exceeds the unitarity cutoff of the model, $\tilde{E} \gtrsim \tilde{\Lambda}$, defined in the Einstein frame. The unitarity cutoff is therefore given by

\begin{equation}\label{eqn:r220}
\tilde{\Lambda} \approx \frac{\sqrt{2}M_{pl}}{\alpha^{\frac{1}{4}}}\left(1 + \frac{4\alpha V\left(\bar{\phi}\right)}{M_{pl}^{4}} \right)^{-\frac{1}{4}}.
\end{equation}

\noindent We can write this in terms of the number of e-folds of inflation $N$. Substituting $\phi\left(N \right)$ (\ref{eqn:r74}) into the Jordan frame potential $V\left(\phi \right)$ (\ref{eqn:r50}) gives 

\begin{equation}\label{eqn:r221}
V\left( \bar{\phi}\right) = 2m_{\phi}^{2}M_{pl}^{2}N.
\end{equation}

\noindent Substituting this into $\tilde{\Lambda}$ and assuming $4\alpha V\left(\phi \right)/M_{pl}^{4} >> 1$ during inflation, the unitarity cutoff in the Einstein frame for $\delta \sigma \delta \sigma \rightarrow \delta \sigma \delta \sigma$ scattering is 

\begin{equation}\label{eqn:r222}
\tilde{\Lambda} = \frac{M_{pl}^{2}}{\sqrt{\alpha} \left(2M_{pl}^{2} m_{\phi}^{2}N \right)^{\frac{1}{4}}}.
\end{equation}

The minimum requirement for unitarity conservation in the context of this interaction is that the energy scale of the quantum fluctuations does not exceed the unitarity cutoff. During inflation we can say that the energy scale of the quantum fluctuations are $\sim \tilde{H}$ in the Einstein frame, so we can write this constraint as

\begin{equation}\label{eqn:r223}
\tilde{H} \lesssim \tilde{\Lambda}.
\end{equation}

\noindent Using the fact that $\tilde{H} = M_{pl}/\sqrt{12\alpha }$ we can calculate the constraint

\begin{equation}\label{eqn:r224}
\frac{M_{pl}}{\sqrt{12\alpha}} \lesssim \frac{M_{pl}^{2}}{\sqrt{\alpha} \left(2M_{pl}^{2} m_{\phi}^{2}N \right)^{\frac{1}{4}}}\Rightarrow \left(\frac{m_{\phi}^{2}}{M_{pl}^{2}}\right)N \lesssim 72.
\end{equation}

\noindent Since $m_{\phi}^{2}/M_{pl}^{2} = 3.4 \times 10^{-11}$ using (\ref{eqn:r94}), this constraint is easily satisfied for any $N$ in the range $N \sim \mathcal{O}\left( 1 - 10^{12} \right)$, and so for all $\alpha$ in this model this constraint is easily satisfied. This means that as a tree-level estimate, unitarity is not violated in non-minimally coupled $R^{2}$ Palatini inflation with a $\phi^{2}$ potential for any regime of $\alpha$.

For interest we can also consider the unitarity cutoff for this interaction in the present vacuum. We calculate this by setting $4\alpha V\left(\phi \right)/M_{pl}^{4} = 0, \left(\Omega = 1 \right)$, which gives

\begin{equation}\label{eqn:r225}
\Lambda = \frac{\sqrt{2} M_{pl}}{\alpha^{\frac{1}{4}}}.
\end{equation}

\noindent This can be interpreted as the cutoff of the $R^{2}$ Palatini inflation model with a $\phi^{2}$ potential as an effective theory. In order for this model to be consistent with unitarity, there must either be some new physics that enters at a $\phi$ particle scattering energy below $\Lambda$, or the scattering must become non-perturbative but unitary at this energy.

\subsection{The Effect of the $\left(\partial \sigma \right)^{4}$ Term on Slow-Roll Inflation}\label{section:276}

Before we move on to examining the predictions of the $R^{2}$ Palatini model with a $\phi^{2}$ potential, we will first confirm that the $\left(\partial \sigma \right)^{4}$ term is not significant during slow-roll inflation, which we have assumed in our analysis. From (\ref{eqn:r40}), the Einstein frame kinetic terms are:

\begin{multline}\label{eqn:r226}
\mathcal{D} = -\frac{1}{2\Omega^{2}}\partial_{\mu}\phi \partial^{\mu}\phi + \frac{\alpha}{4\Omega^{2}M_{pl}^{4}}\left(\partial_{\mu}\phi \partial^{\mu}\phi \right)^{2} \\
= -\frac{1}{2}\left[1 - \frac{2\alpha}{4M_{pl}^{4}}\left(1 + \frac{4\alpha V}{M_{pl}^{4}} \right)\partial_{\mu}\sigma \partial^{\mu}\sigma \right]\partial_{\mu}\sigma \partial^{\mu}\sigma.
\end{multline}

\noindent In the plateau limit this is

\begin{equation}\label{eqn:r227}
\mathcal{D} = -\frac{1}{2}\left[1 - \frac{2\alpha^{2}V}{M_{pl}^{8}}\partial_{\mu}\sigma \partial^{\mu}\sigma \right]\partial_{\mu}\sigma \partial^{\mu}\sigma.
\end{equation}

\noindent The classical field depends only on time, so the term in the brackets $\mathcal{B} = \left[...\right]$ is

\begin{equation}\label{eqn:r228}
\mathcal{B} = 1 + \frac{2\alpha^{2}V}{M_{pl}^{8}}\dot{\sigma}^{2},
\end{equation}

\noindent where $\sigma = \sigma \left(N\right)$ is given by (\ref{eqn:r73}). We can write

\begin{equation}\label{eqn:r229}
\dot{\sigma} = \frac{d\sigma}{dN}\frac{dN}{dt},
\end{equation}

\noindent where

\begin{equation}\label{eqn:r230}
\frac{d\sigma}{dN} = \frac{M_{pl}^{2}}{2\sqrt{2\alpha} m_{\phi}}\frac{1}{N},
\end{equation}

\noindent and 

\begin{equation}\label{eqn:r231}
\frac{dN}{dt} = \tilde{H} = \frac{M_{pl}}{\sqrt{12\alpha}}.
\end{equation}

\noindent We then have

\begin{equation}\label{eqn:r232}
\frac{2\alpha^{2}V}{M_{pl}^{8}}\dot{\sigma}^{2} = \frac{\phi^{2}}{8 M_{pl}^{2}}\frac{1}{12 N^{2}},
\end{equation}

\noindent and using $\phi^{2} = 4NM_{pl}^{2}$ (\ref{eqn:r74}), the kinetic terms are

\begin{equation}\label{eqn:r233}
\mathcal{D} = -\frac{1}{2}\left[1 + \frac{1}{24N}\right]\partial_{\mu}\sigma \partial^{\mu}\sigma.
\end{equation}

\noindent The $\left(\partial \sigma \right)^{4}$ term can therefore be neglected relative to the $\left(\partial \sigma\right)^{2}$ term in (\ref{eqn:r40}) during slow-roll, as it will make a negligible contribution to the dynamics and does not need to be factored into the calculation of the observables. It is however still important to consider this term in relation to unitarity violation, and it may also be that the $\left(\partial \sigma \right)^{4}$ interaction could become significant after inflation and alter the treatment of reheating in the $R^{2}$ quadratic Palatini inflation model.

The condition (\ref{eqn:r233}) is also the condition for the derivative term in (\ref{eqn:r41}) to be negligible during inflation as this is the condition for the quartic terms in $\partial_{\mu}\phi$ to be negligible in the action (\ref{eqn:r40}).

\section{Observational Compatibility: A first pass}\label{section:28}
At this point we will examine the predictions of the $R^{2}$ quadratic inflation model in the different regimes of $\alpha$ so far, before we consider possible reheating channels. Table \ref{table:21} shows the values of the scalar spectral index, the tensor-to-scalar ratio, the number of e-folds corresponding to pivot scale horizon exit, the estimate of instantaneous reheating temperature, the value of the field, the Hubble parameter in the Einstein frame, the unitarity cutoff in the Einstein frame, calculated using the pivot scale $k = 0.05 \Mpc^{-1}$ (in accordance with the 2018 Planck results \cite{planck184}), as well as the effective theory cutoff in the present vacuum, for each regime of $\alpha$ we have examined.

\begin{table}[H]\label{table:21}
\begin{center}
\begin{adjustbox}{max width=\textwidth}
\begin{tabular}{| c | c | c | c | c | c | c | c | c |}
\hline
$\alpha$ & $n_{s}$ & $r$ & $N$ & $T_{R_{MAX}}/ \GeV $ & $\sigma \left(N \right)/ \GeV $ & $ \tilde{H}/ \GeV $ & $\tilde{\Lambda}/ \GeV $ & $\Lambda / \GeV $ \\
\hline
$ 10^{12}$ & $0.9641$ & $9.5 \times 10^{-6}$ & $55.7$ & $7.0 \times 10^{14}$ & $1.5 \times 10^{18}$ & $6.9 \times 10^{11}$ & $3.1 \times 10^{14}$ & $3.4 \times 10^{15}$ \\
\hline
$10^{20}$ & $0.9609$ & $1.1 \times 10^{-13}$ & $51.1$ & $7.0 \times 10^{12}$ & $4.3 \times 10^{14}$ & $6.9 \times 10^{7}$ & $3.1 \times 10^{10}$ & $3.4 \times 10^{13}$ \\
\hline
$10^{32}$ & $0.9548$ & $1.5 \times 10^{-25}$ & $44.2$ & $7.0 \times 10^{9}$ & $8.0 \times 10^{8}$ & $70.0$ & $3.3 \times 10^{4}$ & $3.4 \times 10^{10}$ \\
\hline
\end{tabular}
\end{adjustbox}
\caption{Table showing the scalar spectral index, $n_{s}$, the tensor-to-scalar ratio, $r$, the number of e-folds corresponding to pivot scale horizon exit, $N$, the temperature for instantaneous reheating, $T_{R_{MAX}}$, the value of the inflaton at the pivot scale,  $\sigma \left(N \right)$, the Hubble parameter in the Einstein frame, $\tilde{H}$, the unitarity cutoff scale in the Einstein frame, $\tilde{\Lambda}$, and the unitarity cutoff in the present vacuum, $\Lambda$, for the different regimes of $\alpha$ derived in Sections \ref{section:26} - \ref{section:27}.}
\end{center}
\end{table}

\begin{figure}[H]
\begin{center}
\includegraphics[clip = true, width=0.75\textwidth, angle = 360]{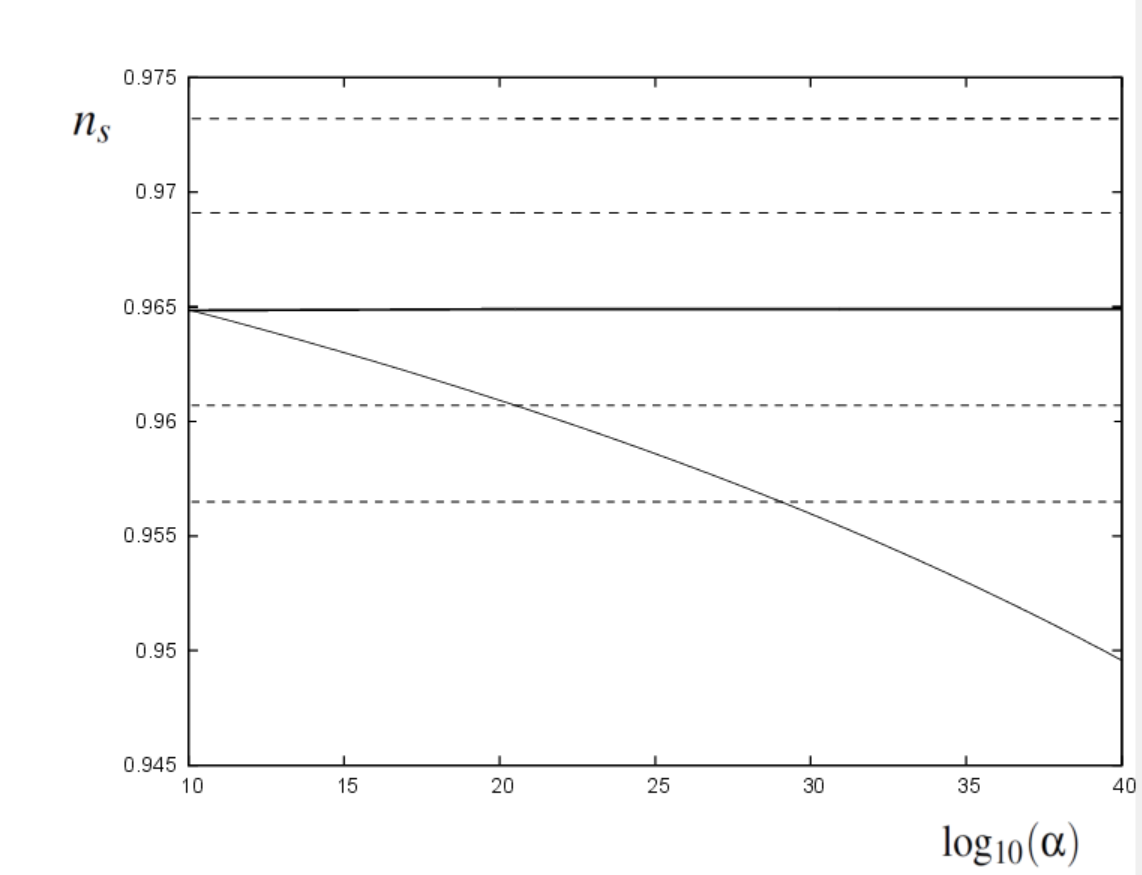}
\caption{Plot showing the scalar spectral index $n_{s}$ as a function of $\alpha$. The thick central line shows the 2018 Planck result for $n_{s}$ and the dotted lines indicate the $1-\sigma$ and $2-\sigma$ bounds on the Planck result. } 
\label{figure:21}
\end{center}
\end{figure} 

From (\ref{eqn:r213}), we have that the number of e-folds at the pivot scale is a function of $\alpha$, and that for each $\alpha$, the pivot scale $N$ is different, as shown in Table \ref{table:21}. This means that for each pivot scale $N\left(\alpha \right)$, by (\ref{eqn:r94}) the pivot scale inflaton mass will technically also be modified for each $\alpha$. For the values of $\alpha$ considered in this work we find that the modification is very small, such that the inflaton mass is $m_{\phi} \simeq 10^{13}\GeV$ for all values of $N\left(\alpha \right)$ considered here. For this reason, we use the $N = 60$ pivot scale mass estimate $m_{\phi} = 1.4 \times 10^{13}\GeV$ when calculating all mass-dependent quantities listed in Table \ref{table:21}, with the understanding that if the pivot scale masses were used for each $\alpha$, these quantities would be modified within their present order of magnitude.

Using the results from the Planck satellite \cite{Planck18}, the scalar spectral index is given by $n_{s} = 0.9649 \pm 0.0042$ ($1-\sigma$), assuming $\Lambda CDM$ and no running of the spectral index. We can see from Table \ref{table:21}, and from Figure \ref{figure:21}, that the scalar spectral index in the case of a sub-Planckian inflaton, $\alpha \gtrsim 10^{12}$, and for Planck-suppressed potential corrections with a broken shift-symmetry, $\alpha \gtrsim 10^{20}$, are in good agreement with the Planck result. The case of Planck-suppressed potential corrections without a shift symmetry, $\alpha \gtrsim 10^{32}$ sits slightly outside of the $2-\sigma$ lower bound on $n_{s}$ ($n_{s}> 0.9565$). An additional correction to the potential could bring the $\alpha \gtrsim 10^{32}$ case into agreement with $n_{s}$ observations. For example, it is possible that quantum corrections resulting from couplings to Standard Model particles arising from reheating could serve this purpose, and we will explore the possibility of this in the next section. Alternatively, the assumption $k \approx 1$ could be an overestimate since $k \approx 1/6! \approx 10^{-3}$ is expected if the strength of the $\sigma^{6}$ interactions corresponding to the Feynman rule is $\approx 1/M_{pl}^{2}$. We find that the bound on $\alpha$ is then modified to $\alpha \gtrsim 10^{29}$, and the scalar spectral index is within the $2-\sigma$ Planck bound on $n_{s}$, and it is also possible that Planck-suppressed corrections at $\alpha \approx 10^{32}$ could themselves modify $n_{s}$.

The $R^{2}$ quadratic Palatini inflation model predicts a highly suppressed tensor-to-scalar ratio, with $r \lesssim 10^{-5}$ for $\alpha \gtrsim 10^{12}$, all of which are below the observational limit of $r \sim 10^{-3}$ in the next generation of CMB experiments \cite{cmbobserve,litebird}. This can distinguish Palatini $\phi^{2}$ inflation from metric Higgs Inflation, which predicts $r \approx 0.004$ \cite{convenhiggs}.

The maximum possible reheating temperatures range from $10^{9}-10^{14}\GeV$, which are quite conventional. In practice, the reheating temperatures could be lower than this if we consider reheating mechanisms beyond the instantaneous approximation, and we explore this in the next Sections \ref{section:29} - \ref{section:211}.

The Hubble parameter predictions are quite low, although the minimum condition for unitarity conservation (\ref{eqn:r224}) is satisfied for each $\alpha$ regime we consider. A stronger condition would be that the field at the pivot scale be less than the unitarity cutoff, $\sigma (N) < \tilde{\Lambda}$, in order that new physics corrections due to a unitarity conserving completion can be neglected, however this is not satisfied in this model. Therefore, we need the potential corrections due to new physics at $\tilde{\Lambda}$ to be suppressed, either because the new physics does not introduce corrections to the inflaton potential, or because unitarity is conserved non-perturbatively at energies greater than $\tilde{\Lambda}$, thereby not requiring any new physics.

\section{Specific Reheating Mechanisms}\label{section:29}

So far we have considered $R^{2}$ quadratic Palatini inflation in the context of instantaneous reheating, which means that the inflaton condensate immediately decays to radiation at the end of (slow-roll) inflation. This is a useful case to consider as it gives an upper estimate on the reheating temperature. However, if we are interested in considering $R^{2}$ quadratic Palatini inflation in the broader context of cosmology and particle physics, it is useful to consider some possible reheating mechanisms and how the inclusion of these interactions of other particles with the inflaton could affect the predictions of the model. 

With the inclusion of a defined reheating channel come quantum corrections to the inflaton potential, and in turn these can alter the predicted value of the scalar spectral index and the reheating temperature. These quantum corrections come in the form of Coleman-Weinberg (CW) corrections (Section \ref{section:t3})

\begin{equation}\label{eqn:r234}
\Delta V_{CW}\left(\phi \right) = \sum_{i} \pm \frac{m_{i}^{4}\left(\phi \right)}{64 \pi^{2}}\ln \left( \frac{m_{i}^{2}\left(\phi \right)}{\mu^{2}} \right),
\end{equation}

\noindent which gives the 1-loop CW effective potential. The sum $i$ is over the number of particle degrees of freedom, and the plus (minus) sign corresponds to bosonic (fermionic) degrees of freedom. $m_{i}$ are the masses of these degrees of freedom and $\mu$ is the chosen renormalisation scale for each mass scale.

The possible reheating channels we consider in this model are reheating via decay to right handed (RH) neutrinos, 

\begin{equation}\label{eqn:r235}
V_{N}\left(\phi \right) = \frac{\lambda_{\phi N}}{2}\phi \bar{N}_{R}^{c} N_{R} + h.c.,
\end{equation}

\noindent and reheating via the Higgs portal coupling

\begin{equation}\label{eqn:r236}
V_{H}\left(\phi \right) = \frac{\lambda_{\phi H}}{2} \phi^{2} \left| H \right|^{2}.
\end{equation}

The inflaton effective potential in the Jordan frame can be written as 

\begin{equation}\label{eqn:r237}
V_{TOT}\left(\phi \right) = V\left(\phi \right) + \Delta V_{CW}\left(\phi \right),
\end{equation}

\noindent where $ \Delta V_{CW}\left(\phi \right)$ is the correction corresponding to the chosen reheating mechanism: either decay to RH neutrinos or annihilation through the Higgs portal coupling. Up until this point we have done all of the inflationary calculations in the Einstein frame, but here we calculate the CW corrections and the effective potential in the Jordan frame. We can use the CW corrections due to the reheating channels to calculate the shift in the scalar spectral index, $n_{s}$, from these corrections in a similar manner to the shift in the spectral index due to the Planck-suppressed corrections in Section \ref{section:27}. To do this, we will use the equivalence between $n_{s}$ computed with the Jordan frame potential in terms of $\bar{\eta}$ and $\bar{\epsilon}$, and $n_{s}$ computed in the Einstein frame with $\eta$ and $\epsilon$, outlined in Section \ref{section:251}. The spectral index in the Jordan frame is given by

\begin{equation}\label{eqn:r238}
n_{s} = 1 + 2\bar{\eta} - 6\bar{\epsilon},
\end{equation}

\noindent so we must first consider the corrections to the slow-roll parameters from the Jordan frame effective potential. $\bar{\eta}$ is given by

\begin{equation}\label{eqn:r239}
\bar{\eta} = M_{pl}^{2}\frac{V_{TOT}''}{V_{TOT}} = M_{pl}^{2} \frac{V'' + \Delta V_{CW}''}{V + \Delta V_{CW}},
\end{equation}

\noindent where $V_{TOT}\left(\phi \right)$ is the Jordan frame potential. Taking out a factor of $V''/V$ gives

\begin{equation}\label{eqn:r240}
\bar{\eta} = M_{pl}^{2}\frac{V''}{V}\frac{\left( 1 + \frac{\Delta V_{CW}''}{V''}\right)}{\left(1 + \frac{\Delta V_{CW}}{V} \right)}.
\end{equation}

\noindent Assume that $\left| \Delta V_{CW} \right|/V <<1$, and $\left|\Delta V_{CW}''/V''\right| <<1$. In this case, we then expand

\begin{equation}\label{eqn:r241}
\left( 1 + \frac{\Delta V_{CW}}{V}\right)^{-1} \approx 1 - \frac{\Delta V_{CW}}{V},
\end{equation}

\begin{equation}\label{eqn:r242}
\Rightarrow \bar{\eta} = M_{pl}^{2}\frac{V''}{V}\left( 1 + \frac{\Delta V_{CW}''}{V''}\right)\left( 1 - \frac{\Delta V_{CW}}{V}\right).
\end{equation}

\noindent To $\mathcal{O}\left(\Delta \right)$, $\bar{\eta}$ is then

\begin{equation}\label{eqn:r243}
\bar{\eta} = M_{pl}^{2}\frac{V''}{V}\left[1 - \frac{\Delta V_{CW}}{V} + \frac{\Delta V_{CW}''}{V''} \right]. 
\end{equation}

\noindent We can write this as

\begin{equation}\label{eqn:r244}
\bar{\eta}_{TOT} = \bar{\eta} + \Delta \bar{\eta},
\end{equation}

\noindent and thus define the shift on the $\bar{\eta}$ parameter as

\begin{equation}\label{eqn:r245}
\Delta \bar{\eta} = M_{pl}^{2} \frac{V''}{V}\left[ \frac{\Delta V_{CW}''}{V''} -  \frac{\Delta V_{CW}}{V}\right] = \eta \left[ \frac{\Delta V_{CW}''}{V''} -  \frac{\Delta V_{CW}}{V}\right].
\end{equation}

\noindent The $\bar{\epsilon}$ parameter can be treated similarly

\begin{equation}\label{eqn:r246}
\bar{\epsilon} = \frac{M_{pl}^{2}}{2}\frac{\left(V' + \Delta V_{CW}' \right)^{2}}{\left(V + \Delta V_{CW} \right)^{2}},
\end{equation}

\begin{equation}\label{eqn:r247}
\Rightarrow \bar{\epsilon} =  \frac{M_{pl}^{2}}{2}\left(\frac{V'}{V}\right)^{2}\frac{\left(1 + \frac{\Delta V_{CW}'}{V'} \right)^{2}}{\left(1 + \frac{\Delta V_{CW}}{V} \right)^{2}}.
\end{equation}

Both the numerator and the denominator can be expanded as follows, using $\left| \Delta V_{CW} \right|/V << 1$ and $\left|\Delta V_{CW}'/V'\right| <<1$

\begin{equation}\label{eqn:r248}
\left( 1 + \frac{\Delta V_{CW}}{V} \right)^{-2} \approx 1 - \frac{2\Delta V_{CW}}{V},
\end{equation}

\begin{equation}\label{eqn:r249}
\left( 1 + \frac{\Delta V_{CW}'}{V'}\right)^{2} \approx 1 + \frac{2\Delta V_{CW}'}{V'},
\end{equation}

\noindent so $\bar{\epsilon}$ can be written as

\begin{equation}\label{eqn:r250}
\bar{\epsilon} \approx \frac{M_{pl}^{2}}{2}\left(\frac{V'}{V} \right)^{2}\left(1 + \frac{2\Delta V_{CW}'}{V'} \right) \left(1 - \frac{2\Delta V_{CW}}{V} \right).
\end{equation}

\noindent Expanded to $\mathcal{O}\left( \Delta \right)$ this is

\begin{equation}\label{eqn:r251}
\bar{\epsilon} = \frac{M_{pl}^{2}}{2}\left(\frac{V'}{V} \right)^{2} \left[1 - \frac{2\Delta V_{CW}}{V} + \frac{2\Delta V_{CW}'}{V'} \right] = \bar{\epsilon} + \Delta \bar{\epsilon},
\end{equation}

\noindent with the $\bar{\epsilon}$-shift given by

\begin{equation}\label{eqn:r252}
\Delta \epsilon = \frac{M_{pl}^{2}}{2}\left(\frac{V'}{V} \right)^{2} \left[\frac{2\Delta V_{CW}'}{V'}  - \frac{2\Delta V_{CW}}{V} \right] = \epsilon \left(\frac{2\Delta V_{CW}'}{V'}  - \frac{2\Delta V_{CW}}{V} \right).
\end{equation}

\noindent Using these, we can write that the shift on the scalar spectral index is then

\begin{equation}\label{eqn:r253}
\Delta n_{s} = 2\Delta \bar{\eta} - 6\Delta \bar{\epsilon},
\end{equation}

\begin{equation}\label{eqn:r254}
\Rightarrow \Delta n_{s} = 2 \bar{\eta} \left( \frac{\Delta V_{CW}''}{V''} -  \frac{\Delta V_{CW}}{V}\right) - 6\bar{\epsilon} \left(\frac{2\Delta V_{CW}'}{V'}  - \frac{2\Delta V_{CW}}{V} \right).
\end{equation}

\noindent In the Jordan frame we have

\begin{equation}\label{eqn:r255}
V = \frac{1}{2}m_{\phi}^{2}\phi^{2}, \; \; V' = m_{\phi}^{2}\phi, \; \; V'' = m_{\phi}^{2}.
\end{equation}

\noindent The bare slow-roll parameters are then

\begin{equation}\label{eqn:r256}
\bar{\eta} = \frac{2M_{pl}^{2}}{\phi^{2}}, \; \; \; \bar{\epsilon} = \frac{2M_{pl}^{2}}{\phi^{2}}.
\end{equation}

\noindent Using the fact that $\phi \left(N \right) = 2\sqrt{N}M_{pl}$, we have that

\begin{equation}\label{eqn:r257}
\bar{\epsilon} = \bar{\eta} = \frac{1}{2N}.
\end{equation}

\noindent The $n_{s}$-shift (\ref{eqn:r254}) is thus

\begin{equation}\label{eqn:r258}
\Delta n_{s} = \frac{2}{2N} \left( \frac{\Delta V_{CW}''}{V''} -  \frac{\Delta V_{CW}}{V}\right) - \frac{6}{2N} \left(\frac{2\Delta V_{CW}'}{V'}  - \frac{2\Delta V_{CW}}{V} \right).
\end{equation}

\noindent Fully expanded, the shift on the scalar spectral index to leading order is therefore

\begin{equation}\label{eqn:r259}
\Delta n_{s} = \frac{1}{N} \left[\frac{\Delta V_{CW}''}{V''} - \frac{6\Delta V_{CW}'}{V'} + \frac{5\Delta V_{CW}}{V} \right],
\end{equation}

\noindent and we are now ready to calculate the shift in the spectral index in the case of a specific reheating channel in $R^{2}$ quadratic Palatini inflation.

\subsection{Reheating via Decay to Right-Handed Neutrinos}\label{section:291}
Firstly we want to examine the shift in the scalar spectral index which would occur if the $R^{2}$ quadratic Palatini inflation model were to reheat via the decay of the inflaton to RH neutrinos. The general 1-loop CW corrections are given by 

\begin{equation}\label{eqn:r260}
\Delta V_{CW} \left(\phi \right) = \sum_{i} \pm \frac{m_{i}^{4}\left(\phi \right)}{64\pi^{2}} \ln \left( \frac{m_{i}^{2}\left(\phi \right)}{\mu^{2}} \right).
\end{equation}

\noindent The inflaton coupling to RH neutrinos contributes two fermionic degrees of freedom to the 1-loop CW corrections, which appear in the effective potential as

\begin{equation}\label{eqn:r261}
\Delta V_{CW_{N}} = -\frac{m_{N}^{4}}{32\pi^{2}}\ln \left(\frac{m_{N}^{2}}{\mu^{2}}\right).
\end{equation}

\noindent Examining the effective potential RH neutrino coupling term (\ref{eqn:r235}), we can define

\begin{equation}\label{eqn:r262}
m_{N} = \lambda_{\phi N} \phi,
\end{equation}

\noindent and this can be substituted into $\Delta V_{CW_{N}}$ (\ref{eqn:r261}) to give

\begin{equation}\label{eqn:r263}
\Delta V_{CW_{N}} = -\frac{\lambda_{\phi N}^{4}\phi^{4}}{32\pi^{2}}\ln \left(\frac{\lambda_{\phi N}^{2}\phi^{2}}{\mu^{2}}\right).
\end{equation}

\noindent The first and second derivatives with respect to $\phi$ are

\begin{equation}\label{eqn:r264}
\Delta V_{CW_{N}}' = -\frac{4\lambda_{\phi N}^{4}\phi^{3}}{32\pi^{2}}\ln \left(\frac{\lambda_{\phi N}^{2}\phi^{2}}{\mu^{2}} \right) - \frac{2\lambda_{\phi N}^{4}\phi^{3}}{32\pi^{2}},
\end{equation}

\begin{equation}\label{eqn:r265}
\Delta V_{CW_{N}}'' = -\frac{12\lambda_{\phi N}^{4}\phi^{2}}{32\pi^{2}}\ln \left( \frac{\lambda_{\phi N}^{2} \phi^{2}}{\mu^{2}}\right) - \frac{14\lambda_{\phi N}^{4}\phi^{2}}{32\pi^{2}}.
\end{equation}

\noindent Substituting (\ref{eqn:r261}), (\ref{eqn:r264}) and (\ref{eqn:r265}) into the expression for the $n_{s}$-shift (\ref{eqn:r259}) we have

\begin{equation}\label{eqn:r266}
\Delta n_{s_{N}} = -\frac{\lambda_{\phi N}^{4}\phi^{2}}{16\pi^{2}m_{\phi}^{2}N}\left[1 - \ln \left(\frac{\lambda_{\phi N}^{2} \phi^{2}}{\mu^{2}} \right) \right].
\end{equation}

\noindent Using (\ref{eqn:r74}), $\phi^{2} = 4NM_{pl}^{2}$, $\Delta n_{s_{N}}$ can be rewritten as

\begin{equation}\label{eqn:r267}
\Delta n_{s_{N}} = -\frac{\lambda_{\phi N}^{4}M_{pl}^{2}}{4\pi^{2}m_{\phi}^{2}}\left[1 - \ln \left(\frac{4\lambda_{\phi N}^{2} N M_{pl}^{2}}{\mu^{2}} \right) \right].
\end{equation}

Given that we know the shift on the scalar spectral index due to the quantum corrections to the potential from the decay of the inflaton to RH neutrinos, we can use this to constrain the size of the coupling of the inflaton to RH neutrinos, $\lambda_{\phi N}$.

\subsection{Reheating via the Higgs Portal}\label{section:292}

We now consider the treatment of reheating in the $R^{2}$ quadratic Palatini inflation via a Higgs portal coupling. The Higgs portal coupling to the inflaton contributes four bosonic degrees of freedom to the 1-loop CW effective potential, and these corrections are given by

\begin{equation}\label{eqn:r268}
\Delta V_{CW_{H}} = \frac{m_{H}^{4}}{16\pi^{2}}\ln \left(\frac{m_{H}^{2}}{\mu^{2}}\right).
\end{equation}

\noindent Examining the portal coupling term in the inflaton potential (\ref{eqn:r236}), we can define

\begin{equation}\label{eqn:r269}
m_{H}^{2} = \lambda_{\phi H}\phi^{2},
\end{equation}

\noindent and rewrite $\Delta V_{CW_{H}}$ (\ref{eqn:r268}) as

\begin{equation}\label{eqn:r270}
\Delta V_{CW_{H}} = \frac{\lambda_{\phi H}^{2}\phi^{4}}{16\pi^{2}}\ln \left(\frac{\lambda_{\phi H}\phi^{2}}{\mu^{2}}\right).
\end{equation}

\noindent The first and second derivatives of $\Delta V_{CW_{H}}$ are given by

\begin{equation}\label{eqn:r271}
\Delta V_{CW_{H}}' = \frac{4\lambda_{\phi H}^{2}\phi^{3}}{16\pi^{2}}\ln \left(\frac{\lambda_{\phi H}\phi^{2}}{\mu^{2}}\right) + \frac{2\lambda_{\phi H}^{2}\phi^{3}}{16\pi^{2}},
\end{equation}

\begin{equation}\label{eqn:r272}
\Delta V_{CW_{H}}'' = \frac{12\lambda_{\phi H}\phi^{2}}{16\pi^{2}}\ln \left(\frac{\lambda_{\phi H}\phi^{2}}{\mu^{2}}\right) + \frac{14\lambda_{\phi H}^{2} \phi^{2}}{16\pi^{2}}.
\end{equation}

Substituting (\ref{eqn:r270}), (\ref{eqn:r271}) and (\ref{eqn:r272}) into the expression for the $n_{s}$-shift (\ref{eqn:r259}) we have

\begin{equation}\label{eqn:r273}
\Delta n_{s_{H}} = \frac{\lambda_{\phi H}^{2}\phi^{2}}{8\pi^{2} m_{\phi}^{2}N} \left[1 - \ln\left( \frac{\lambda_{\phi H}\phi^{2}}{\mu^{2}}\right) \right].
\end{equation}

\noindent Replacing $\phi$ by $\phi \left(N\right)$ (\ref{eqn:r74}) we obtain

\begin{equation}\label{eqn:r274}
\Delta n_{s_{H}} = \frac{\lambda_{\phi H}^{2}M_{pl}^{2}}{2\pi^{2} m_{\phi}^{2}} \left[1 - \ln\left( \frac{4\lambda_{\phi H} N M_{pl}^{2}}{\mu^{2}}\right) \right].
\end{equation}

\noindent We are now able to examine the constraints on the size of the coupling of the Higgs to the inflaton in a similar way to its coupling to RH neutrinos.

\subsection{Choice of the Renormalisation Scale $\mu$}\label{section:293}
Before we proceed with calculating the effects of these reheating interactions on the predictions of the model, we must eliminate the logarithms in each of the CW corrections $\Delta V_{CW}$. In order to do so we will choose an appropriate renormalisation scale $\mu$, such that the logarithmic terms in each case must be equal to zero.

In each reheating scenario, we therefore choose the renormalisation scale to be equal to the effective mass parameter of the particle interacting with the inflaton, i.e.

\begin{equation}\label{eqn:r275}
\mu_{H}^{2} \equiv m_{H}^{2} = \lambda_{\phi H}\phi^{2} \Rightarrow \ln \left(\frac{\lambda_{\phi H}\phi^{2}}{\mu_{H}^{2}}\right) = 0,
\end{equation}

\noindent for the case of the Higgs portal coupling and

\begin{equation}\label{eqn:r276}
\mu_{N}^{2} \equiv m_{N}^{2} = \lambda_{\phi N}^{2}\phi^{2} \Rightarrow \ln \left(\frac{\lambda_{\phi N}^{2}\phi^{2}}{\mu_{N}^{2}}\right) = 0,
\end{equation}

\noindent for the case of a coupling to RH neutrinos, where in both cases $\mu$ is defined at $\phi = \phi \left(N \right)$ with $N$ being the pivot scale. $m_{\phi}$ is also defined as its pivot scale value (\ref{eqn:r94}) in the next stage of these calculations. Setting $\mu_{i}^{2} = m_{i}^{2}$ in each scenario gives the respective $\Delta n_{s_{i}}$ to be

\begin{equation}\label{eqn:r277}
\Delta n_{s_{H}} = \frac{\lambda_{\phi H}^{2}M_{pl}^{2}}{2\pi^{2} m_{\phi}^{2}},
\end{equation}

\begin{equation}\label{eqn:r278}
\Delta n_{s_{N}} = -\frac{\lambda_{\phi N}^{4}M_{pl}^{2}}{4\pi^{2}m_{\phi}^{2}}.
\end{equation}

\subsection{Constraints on $\lambda_{\phi N}$ and $\lambda_{\phi H}$ from the $n_{s}$-shift}\label{section:094}
We will now calculate the constraints on the couplings between the inflaton and the Higgs and RH neutrinos arising from the requirement that the shift of the scalar spectral index resulting from the quantum corrections due to these couplings not be too large, as to not spoil the predictions of the model. Using $\left| \Delta n_{s} \right| < 0.001$ as the upper bound on the shift of the spectral index, we can write

\begin{equation}\label{eqn:r279}
\left| \frac{\lambda_{\phi H}^{2}M_{pl}^{2}}{2\pi^{2} m_{\phi}^{2}} \right| < 0.001,
\end{equation}

\begin{equation}\label{eqn:r280}
\left| -\frac{\lambda_{\phi N}^{4}M_{pl}^{2}}{4\pi^{2}m_{\phi}^{2}} \right| < 0.001.
\end{equation}

\noindent These can be rearranged to give the following constraints on the couplings

\begin{equation}\label{eqn:r281}
\lambda_{\phi H}^{2} < \frac{0.001 \times 2\pi^{2}m_{\phi}^{2}}{M_{pl}^{2}},
\end{equation}

\begin{equation}\label{eqn:r282}
\lambda_{\phi N}^{4} < \frac{0.001 \times 4\pi^{2}m_{\phi}^{2}}{M_{pl}^{2}}.
\end{equation}

\noindent Using $m_{\phi} = 1.4 \times 10^{13} \GeV$ as an estimate, the upper bounds on the couplings are 

\begin{equation}\label{eqn:r283}
\lambda_{\phi H} < 8.2 \times 10^{-7},
\end{equation}

\begin{equation}\label{eqn:r284}
\lambda_{\phi N} < 1.1 \times 10^{-3}.
\end{equation}

\section{Reheating via Higgs Portal Annihilation and Fragmentation of the Inflationary Condensate}\label{section:210}
Reheating via the Higgs portal involves the process of the annihilation of two $\phi$ scalars to two Higgs particles. There are two main possibilities for how reheating can proceed through this channel: rapid preheating and fragmentation of the inflationary condensate.

Rapid preheating corresponds to the case where the condensate scalars annihilate to relativistic Higgs particles with momentum $p = m_{\phi}$ \cite{traschen} \cite{big3}. This causes further annihilation events to be Bose-enhanced, and preheating can then proceed very quickly. This mechanism can only occur if the inflaton condensate does not fragment, and is only viable if the Bose-enhancement is strong and the condensate annihilates very rapidly. 

Condensate fragmentation occurs when the inflaton condensate breaks apart into a number of discrete clumps of scalars. This phenomenon can lead to the formation of oscillons or other breather states in the case of a real inflaton \cite{john01}, \cite{amin11},\cite{lozanov17}, and can lead to the formation of non-topological solitons (NTS) such as Q-balls if the scalar field is complex and carries a conserved Noether charge (a discussion of Q-balls is presented in Chapter 6). We will not speculate too much on the precise nature of what exactly these condensate fragments are here, beyond the fact that the inflaton field in this case is real. It is however common practice in inflationary cosmology that the objects formed from the fragmentation of a real inflaton condensate are referred to as oscillons, whether or not the corresponding field solution is an oscillon solution. We adopt this convention here, with the statement that the nature of these objects is an ongoing area of study in cosmology and a possibility for future investigation in the case of quadratic $R^{2}$ Palatini inflation specifically.

An inflaton condensate will fragment when there is sufficient growth of the inflaton perturbations within the condensate, and if a field solution corresponding to the objects the condensate will fragment into is present in the theory\footnote{This is by no means a complete checklist, condensate fragmentation is a non-trivial and highly non-linear process, which is an active area of study in inflationary cosmology. This is a simplified picture for the purpose of discussion here.}. In this case, supposing the theory admits an oscillon solution, the condensate will fragment into oscillons of typical diameter $\sim m_{\phi}^{-1}$ \cite{osclife}. This means that the inflaton scalars will be bound into objects in their zero-momentum state, and upon annihilation within the oscillon the Higgs particles created will quickly escape from the volume of the oscillon. The annihilation cannot, therefore, be Bose-enhanced if the condensate fragments, and preheating cannot proceed as it would if the inflaton condensate were to remain intact. 

Another consequence of fragmentation is that, assuming the oscillons are in a stable (or at least metastable) state while the annihilation is occurring and not decaying themselves, the number density of the scalars, $n_{\phi}$, within the volume of the oscillons is constant \cite{john01}. This is in contrast to the case of a coherent condensate, where the number density $n_{\phi} = \rho_{\phi}/m_{\phi}$, decays away as matter with expansion, $n_{\phi} \propto 1/a^{3}$. In order for the annihilation process to reheat the Universe, it must occur faster than the rate of expansion, a condition which can be expressed by 

\begin{equation}\label{eqn:r285}
\Gamma \gtrsim \tilde{H},
\end{equation}

\noindent which will eventually be satisfied if the scalars are bound up in oscillons (fragmentation occurs), as the Hubble rate decreases with expansion while the annihilation rate does not. On the other hand, in the case of a coherent condensate (no fragmentation) we have that 

\begin{equation}\label{eqn:r286}
\Gamma_{\phi H} \propto n_{\phi} \propto \frac{1}{a^{3}},
\end{equation}

\noindent while the Hubble rate changes with expansion as

\begin{equation}\label{eqn:r287}
\tilde{H} \propto \rho_{\phi}^{\frac{1}{2}} \propto \frac{1}{a^{\frac{3}{2}}}.
\end{equation}

\noindent This means that it is not possible for the annihilation rate to exceed the Hubble rate, unless it does so immediately after inflation ends, as it will decrease faster with expansion throughout. As a result, we find that in our analysis of $R^{2}$ quadratic Palatini inflation, annihilation via the Higgs portal is not a viable reheating mechanism unless the condensate undergoes fragmentation, or proceeds fast enough to reheat immediately.

In order to provide a revised estimate of the reheating temperature given annihilation to Higgs particles as the reheating mechanism, we must check whether or not the condensate undergoes fragmentation. There is a simple way to check this, and the results used here were first presented in \cite{jinsu17} and built upon in \cite{jinsu22}. 

We remark that this check provides a general estimate of fragmentation and does not rely on a particular underlying mechanism. It is possible in $R^{2}$ quadratic Palatini inflation that the condensate could fragment as a result of tachyonic preheating \cite{tomberg211}, since inflation ends while the inflaton is still on the inflationary plateau and begins rapidly rolling towards the minimum of its potential. This would result in faster fragmentation than this method estimates for, since we assume here that fragmentation will occur after the formation of the coherently oscillating condensate. This will mean that the bound obtained on $\alpha$ using this method will be a sufficient condition for fragmentation in this model.

The renormalisable potential at small $\phi$ in a general inflation model with a $\phi \rightarrow -\phi$ symmetry will have the form

\begin{equation}\label{eqn:r288}
V\left(\phi \right) = \frac{1}{2}m_{\phi}^{2} \phi^{2} - A\phi^{4},
\end{equation}

\noindent where the potential is dominated by the $\phi^{2}$ term. At $\phi < \phi_{0}$, we are in the quadratic dominated regime of the potential where $\sigma \approx \phi$ and the Einstein frame is given by 

\begin{equation}\label{eqn:r289}
V_{E} = \frac{V\left(\phi \right)}{\left(1 + \frac{4\alpha V\left(\phi \right)}{M_{pl}^{4}}\right)} \approx V\left(\phi \right)\left( 1 - \frac{4\alpha V\left(\phi \right)}{M_{pl}^{4}}\right),
\end{equation}

\begin{equation}\label{eqn:r290}
V_{E} \approx  V\left(\phi \right) - \frac{4\alpha V\left(\phi \right)^{2}}{M_{pl}^{4}} = \frac{1}{2}m_{\phi}^{2}\phi^{2} - \frac{\alpha m_{\phi}^{4}\phi^{4}}{M_{pl}^{4}}.
\end{equation}

\noindent We therefore have that

\begin{equation}\label{eqn:r291}
A = \frac{\alpha m_{\phi}^{4}}{M_{pl}^{4}},
\end{equation}

\noindent comparing (\ref{eqn:r290}) to (\ref{eqn:r288}). From \cite{jinsu17}, \cite{jinsu22} we have that the sufficient condition for the fragmentation of the condensate is

\begin{equation}\label{eqn:r292}
A > \frac{100}{r_{v}}\frac{m_{\phi}^{2}}{M_{pl}^{2}},
\end{equation}

\noindent where $r_{v}$ is a number less than one which corresponds to the ratio of the quartic part of the potential to the quadratic part at the point where the oscillations begin. Since we are assuming that fragmentation occurs when the quadratic part of the potential dominates we can set $r_{v}=0.1$, and this gives

\begin{equation}\label{eqn:r293}
\frac{\alpha m_{\phi}^{2}}{M_{pl}^{2}} > 1000.
\end{equation}

\noindent Using $m_{\phi} = 1.4 \times 10^{13} \GeV$ (\ref{eqn:r94}), this gives 

\begin{equation}\label{eqn:r294}
\alpha > 2.9 \times 10^{13},
\end{equation}

\noindent as the sufficient condition for fragmentation in $R^{2}$ quadratic Palatini inflation. This is essentially the same as the bound on $\alpha$ which was later derived in \cite{tomberg211} for fragmentation from tachyonic preheating in Palatini $R^{2}$ inflation. This is strongly satisfied for the larger values of $\alpha$ from Planck-suppressed potential corrections but not for $\alpha = 10^{12}$, which is our minimal constraint that the inflaton field $\sigma$ be sub-Planckian. The cases where there are Planck-suppressed potential corrections strongly favour fragmentation of the inflaton condensate. 

If the fragmentation is rapid, then we can approximate that all of the energy density from the inflaton potential is transferred to the oscillons such that

\begin{equation}\label{eqn:r295}
\rho_{\phi} \approx 3\tilde{H}^{2}M_{pl}^{2} \approx V_{E} \approx \frac{M_{pl}^{4}}{4\alpha}.
\end{equation}

\noindent The number density of $\phi$ scalars in the oscillons is then

\begin{equation}\label{eqn:r296}
n_{\phi} = \frac{\rho_{\phi}}{m_{\phi}} = \frac{M_{pl}^{4}}{4\alpha m_{\phi}},
\end{equation}

\noindent which we take to be constant. The zero momentum scalars annihilate to the real Higgs scalars, $\phi \phi \rightarrow h_{i} h_{i} \;\; \left(i = 1, ..., 4\right)$, with a cross section times relative velocity of \cite{higgsann}

\begin{equation}\label{eqn:r297}
<\sigma_{\phi H} v_{rel}> = \frac{\lambda_{\phi H}^{2}}{16\pi m_{\phi}^{2}},
\end{equation}

\noindent giving the perturbative annihilation rate to be

\begin{equation}\label{eqn:r298}
\Gamma_{\phi H} = n_{\phi}<\sigma_{\phi H} v_{rel}> = \frac{\lambda_{\phi H}^{2}M_{pl}^{4}}{64\pi \alpha m_{\phi}^{3}}.
\end{equation}

\noindent The reheating temperature is then obtained by

\begin{equation}\label{eqn:r299}
\Gamma_{\phi H} = \tilde{H} = \frac{k_{T}T_{R}^{2}}{M_{pl}}, \; \; \; \; k_{T}^{2} = \frac{\pi^{2}g\left(T \right)}{90},
\end{equation}

\noindent with the assumption that all of the energy density bound up in the oscillons goes into the relativistic Higgs particles upon annihilation. Taking $g\left(T \right) \approx 100$ gives $k_{T} = 3.3$, and we can write the reheating temperature as

\begin{equation}\label{eqn:r300}
T_{R} = 3.46 \times 10^{28} \lambda_{\phi H} \left(\frac{10^{32}}{\alpha m_{\phi}^{3}}\right)^{\frac{1}{2}} \GeV,
\end{equation}

\noindent or by setting $m_{\phi} = 1.4 \times 10^{13}\GeV$,

\begin{equation}\label{eqn:r301}
T_{R} \le 6.61 \times 10^{8} \lambda_{\phi H} \left(\frac{10^{32}}{\alpha}\right)^{\frac{1}{2}} \GeV.
\end{equation}

\noindent Using (\ref{eqn:r301}) we find, with $\lambda_{\phi H} < 8.2 \times 10^{-7}$ from (\ref{eqn:r283}), an upper bound on the reheating temperature from decay of the inflaton via the Higgs portal is 

\begin{equation}\label{eqn:r302}
T_{R} \le 541.6\left(\frac{10^{32}}{\alpha}\right)^{\frac{1}{2}} \GeV.
\end{equation}

\noindent The upper bounds on $T_{R}$ for the cases of interest are then

\begin{equation}\label{eqn:r303}
T_{R} \le 5.42 \times 10^{12} \GeV, \; \; \; \alpha \simeq 10^{12};
\end{equation}

\begin{equation}\label{eqn:r304}
T_{R} \le 5.42 \times 10^{8} \GeV, \; \; \; \alpha \simeq 10^{20};
\end{equation}

\begin{equation}\label{eqn:r305}
T_{R} \le 541.6 \GeV \; \; \; \alpha \simeq 10^{32}.
\end{equation}

\noindent Successful reheating can therefore proceed via the Higgs portal annihilation channel, assuming that fragmentation occurs. All the reheating temperatures predicted in this case are lower than those for the case of instantaneous reheating described in Section \ref{section:274} and presented in Section \ref{section:28}. In practice, a lower reheating temperature would likely result in a lower value of the scalar spectral index, $n_{s}$, in each case. Given that the value of the $n_{s}$ for the case of $\alpha \gtrsim 10^{32}$ is already in tension with the $2-\sigma$ lower bound from the Planck results \cite{Planck18}, this would suggest that reheating via annihilation of inflaton scalars to Higgs particles is observationally disfavoured in this case.

These estimates rely on the assumption that the condensate fragments are stable, or at least metastable throughout the duration of reheating, such that the number density of the inflaton scalars within the oscillons in constant. In practice oscillons may decay or disperse, leading to a decrease in $n_{\phi}$ over the lifetime of the oscillon, which would reduce the annihilation rate and so decrease $T_{R}$. The assumption that the oscillons are stable therefore favours the annihilation process, and means that we can treat the calculated reheating temperatures for this reheating channel in each regime of $\alpha$ as being the maximum estimate for each case given reheating via the Higgs portal. There is also the possibility that after slow roll inflation, during reheating, the $\left(\partial \sigma \right)^{4}$ term may become significant and scattering of the inflaton scalars may contribute to the post-inflation dynamics. Although we do not consider this case here, it is possible that these interactions could become significant and could modify the fragmentation analysis.

The inflaton mass in this model in the present electroweak vacuum is $\mathcal{O}\left(10^{13}\GeV\right)$, and as such there are no direct particle physics constraints on the Higgs portal coupling to the inflaton which are useful in this model. For scalars with a mass less than half of the Higgs boson mass, it is possible to pair produce such scalars through Higgs boson decay, which can provide a general constraint on the portal coupling of the Higgs to an inflaton in the case of light scalars. The strongest constraint on this interaction comes from the branching ratio of Higgs decay to invisible particles. The strongest present LHC bounds require that the portal coupling to such scalars is less than approximately $0.05$ (see Figure 2 of \cite{heisig}), which is larger than the upper bound calculated for the Higgs-inflaton coupling in the Palatini $\phi^{2} R^{2}$ model using the shift on the scalar spectral index, 
$\lambda_{\phi H} < \mathcal{O}\left(10^{-7}\right)$. This is a small value, which is not unusual when coupling the inflaton to Standard Model particles since weakly coupling the inflaton to the Standard Model naturally protects the inflaton potential from quantum corrections which could affect the ability of the model to inflate successfully. While it may be possible, therefore, to constrain the reheating via the Higgs portal of a model with a light inflaton using collider bounds on the Higgs-portal coupling, it is not possible in this model since the inflaton mass is too large for any present particle physics constraints to be meaningful.

\section{Reheating via Decay to Right-Handed Neutrinos}\label{section:211}
We will now consider results from the decay of the inflaton to right-handed neutrinos. Assuming the mass of the neutrinos is small compared to the inflaton mass, the decay rate of the scalars to right-handed neutrinos is given by \cite{nrdecay}

\begin{equation}\label{eqn:r306}
\Gamma_{\phi \rightarrow NN} = \frac{\lambda_{\phi N}^{2}m_{\phi}}{16\pi}.
\end{equation}

\noindent In order for reheating to proceed instantaneously via this channel we require that

\begin{equation}\label{eqn:r307}
\Gamma_{\phi \rightarrow NN} > \tilde{H} \Rightarrow \frac{\lambda_{\phi N}^{2}m_{\phi}}{16\pi} > \frac{M_{pl}}{\sqrt{12\alpha}}.
\end{equation}

\noindent This gives the constraint on $\lambda_{\phi N}$

\begin{equation}\label{eqn:r308}
\lambda_{\phi N} > \left(\frac{64\pi^{2} M_{pl}^{2}}{3 m_{\phi}^{2}\alpha}\right)^{\frac{1}{4}},
\end{equation}

\noindent which gives

\begin{equation}\label{eqn:r309}
\lambda_{\phi N} > \frac{1}{\alpha^{\frac{1}{4}}}\left(\frac{ 3.48 \times 10^{19}\GeV}{m_{\phi}}\right).
\end{equation}

\noindent From considering the shift on the scalar spectral index due to the presence of the right-handed neutrino interaction term in the inflaton potential we have that $\lambda_{\phi N} < 1.1 \times 10^{-3}$ in order to not significantly alter the observational predictions of the Palatini $R^{2}$ quadratic inflation model. This is useful here as it allows us to examine for which values of $\alpha$ reheating through decay to right-handed neutrinos can proceed. From (\ref{eqn:r309}) we find 

\begin{equation}\label{eqn:r310}
\alpha = 10^{12}, \; \; \; \lambda_{\phi N} >1.58;
\end{equation}

\begin{equation}\label{eqn:r311}
\alpha = 10^{20}, \; \; \;  \lambda_{\phi N} > 0.016;
\end{equation}

\begin{equation}\label{eqn:r312}
\alpha = 10^{32} \; \; \; \lambda_{\phi N} > 1.6 \times 10^{-5}.
\end{equation}

\noindent using (\ref{eqn:r94}) as the mass estimate of the inflaton. It is clear from this that only the case of $\alpha = 10^{32}$, corresponding to $R^{2}$ quadratic Palatini inflation with general Planck-suppressed potential corrections, can instantaneously reheat via the decay to right-handed neutrinos without introducing corrections to the potential too large to maintain the prediction of $n_{s}$.

If the decay were not instantaneous, then reheating will proceed at a lower reheating temperature, and a consequently lower value of the scalar spectral index. As long as $\alpha$ is small compared to $10^{32}$ this is fine, as for these smaller values of $\alpha$ the scalar spectral index is still well within the $2-\sigma$ bounds from the results of the Planck satellite (2018). Reheating via decay to RH neutrinos at a lower temperature is therefore possible for the case of sub-Planckian inflaton, requiring $\alpha \ge 10^{12}$, and Planck-suppressed corrections with a broken shift symmetry, requiring $\alpha \ge 10^{20}$. It would however rule out the case of $R^{2}$ quadratic Palatini inflation with general Planck-suppressed potential corrections observationally, as for $\alpha > 10^{32}$, the scalar spectral index is already on the $2-\sigma$ boundary of the Planck bounds in the case of instantaneous reheating.

\section{Summary}\label{section:212}

In this chapter we have presented an analysis of the $R^{2}$ Palatini inflation model with a simple $\phi^{2}$ Jordan frame potential. We derived three different constraints on the $\alpha$ parameter, depending upon the form of the quantum gravity corrections. We find that in order for the canonical inflaton field to be sub-Planckian, we require that $\alpha \gtrsim 10^{12}$; if there are Planck-suppressed potential corrections arising from a quantum gravity completion with an approximate shift symmetry then we require $\alpha \gtrsim 10^{20}$, and if there are general Planck-suppressed potential corrections arising from a quantum gravity completion then we require $\alpha \gtrsim 10^{32}$. The constraints arising from the cases where there are Planck-suppressed potential corrections were derived using the fact that in order for the model to maintain good agreement with the observations of the inflationary observables, these potential corrections must not cause a shift in the $\eta$ parameter larger than 0.001. These values of $\alpha$ are very large, however it is typical for $R^{2}$ models to require large values of the dimensionless coupling in order to inflate successfully. Conventional Starobinsky-type models typically require $\alpha \approx 10^{10}$ \cite{kehagias14}, and $\alpha \gtrsim 10^{8}$ for $R^{2}$ quadratic Palatini inflation to occur at all \cite{enckell182}.

The large coefficient on the $R^{2}$ term in the gravitational action raises the question of the UV-compatibility of the model in the case of higher order $R^{n}, n > 2$ terms, if the gravitational action contained some function $F(R)$ expanded as a Taylor series in the Ricci scalar for example, or if higher order in $R$ terms were to arise due to quantum corrections to the model. If the first case applies, and we are allowing the coefficient of the $R^{2}$ - presumably a low-order - term to be very large then we are opening the model to the possibility that there could be higher-order $R$ terms in the action with comparably large coefficients - rendering them non-negligible, and introducing the possibility of an infinite series of $R$ terms with large coefficients in the gravitational action. On a superficial level, any additional higher-order $R$ terms have the potential to spoil the predictions of the inflation\footnote{see e.g. \cite{bekov} for a recent study of the effects of these terms in the Jordan frame, the effects in the Einstein frame would depend on the transformation used to recast the $R^{n}$ terms and how these would manifest in the inflaton sector of the action.}, and an infinite series of non-negligible $R^{n}, n > 2$ terms would render the model UV-incompatible and therefore not useful as a gravity theory. While it is possible that there could be higher order $R$ terms present in a model like this, either from a series expansion in the underlying gravity theory or arising due to quantum corrections, the model we consider in this work is \textit{defined} as an Einstein-Hilbert plus $\alpha R^{2}$ gravitational action, and we find that the model can produce successful inflation in this framework for certain values of the $\alpha$ parameter. 

We initially considered the $R^{2}$ quadratic Palatini model in the case of instantaneous reheating, and we use this as an estimate on the upper bound of the reheating temperature for the model. The instantaneous reheating temperature is in the range $10^{9} -10^{14}\GeV$ for the three constraints of $\alpha$ we consider, and in all cases the model reheats to a temperature sufficient for Big Bang Nucleosynthesis and common models of baryogenesis. However, the reheating temperature decreases as $\alpha$ increases, which also means that there are fewer e-folds of expansion after inflation as $\alpha$ gets larger, which also means a decrease in scalar spectral index $n_{s}$.

The prediction of the scalar spectral index in the cases of a sub-Planckian inflaton, $\alpha \gtrsim 10^{12}$, and Planck-suppressed potential corrections in the presence of an approximate shift symmetry, $\alpha \gtrsim 10^{20}$, are in agreement with the observations of the scalar spectral index from the 2018 Planck results to within $1-\sigma$ \cite{Planck18}, while the $n_{s}$ prediction for the case of general Planck-suppressed potential corrections in a completion of quantum gravity, $\alpha \gtrsim 10^{32}$, sits below but very close to the $2-\sigma$ bound on $n_{s}$ from the 2018 Planck results. In order to bring the model into agreement with the observed value of $n_{s}$, an additional small correction would need to be added to the potential in order to raise the predicted value of $n_{s}$ into the observational bounds, or a smaller coefficient of the non-renormalisable Planck-suppressed correction would be needed. It is possible that the quantum corrections arising from the interactions with particles during reheating can serve this purpose. Nevertheless, if the model is consistent with Planck-suppressed corrections with no additional symmetry suppression, then we expect $n_{s}$ to be close to the $2-\sigma$ lower bound.

The $R^{2}$ quadratic Palatini inflation model predicts a highly suppressed tensor-to-scalar ratio $ r \sim 10^{-25}-10^{-6}$ which is a general prediction of $R^{2}$ Palatini inflation. All of these predictions for all $\alpha$ in this study are below the observational limit of $r \sim 10^{-3}$ in the next generation of CMB experiments \cite{cmbobserve,litebird}.

We also investigated the tree level unitarity in the model and found that in each constraint of $\alpha$, the minimum requirement for unitarity conservation, $\tilde{H} < \tilde{\Lambda}$ is satisfied in every case, although the stronger constraint of $\sigma < \tilde{\Lambda}$ is not. Either the introduction of new physics at a scale before or at the unitarity cutoff $\tilde{\Lambda}$ which does not introduce non-renormalisable corrections to the potential at $\sigma > \tilde{\Lambda}$, or interactions that are non-perturbative but unitary at scales above $\tilde{\Lambda}$ and do not modify the potential at $\sigma > \tilde{\Lambda}$, would be necessary to satisfy this constraint.

We considered two specific reheating mechanisms: reheating via annihilation of inflaton scalars to Higgs bosons, and reheating via decay of inflaton scalars to right-handed neutrinos. We use the constraint on the $\eta$-shift in order to constrain the size of the Coleman-Weinberg corrections to the potential which arise as a result of the reheating interactions, and in doing so we calculate the upper bounds on the couplings of these interactions. We find that $\lambda_{\phi H} < 8.2 \times 10^{-7}$ and $\lambda_{\phi N} < 1.1 \times 10^{-3}$.

We find that the inflaton condensate is likely to fragment in this model, either after the condensate has begun coherent oscillations, or earlier if tachyonic preheating occurs, provided that $\alpha > 2.9 \times 10^{13}$. Fragmentation can therefore occur in the model in the presence of Planck-suppressed potential corrections from a quantum gravity completion, but may not do so if we consider a purely sub-Planckian inflaton.  

In the case of annihilation via the Higgs portal, unless the annihilation proceeds very quickly there is unlikely to be preheating in this scenario, and the reheating will be slower. This results in a lower reheating temperature than in the case of instantaneous reheating, $T_{R} = 10^{2} - 10^{12}\GeV$, where $T_{R} \sim 10^{2}\GeV$ corresponds to $\alpha \approx 10^{32}$. This very low reheating temperature will likely result in a very low value of the scalar spectral index, and so it appears that the case of general Planck-suppressed potential corrections from a quantum gravity completion is observationally disfavoured for reheating via the Higgs portal.

For reheating through the decay to right-handed neutrinos, we use the fact that in order to reheat the Universe instantaneously we must have $\Gamma_{\phi NN} > \tilde{H}$ at the end of inflation to derive a lower bound on the coupling of the inflaton to right-handed neutrinos, $\lambda_{\phi N}$. We then compare this for each $\alpha$ regime in order to determine a range for $\lambda_{\phi N}$ which successfully reheats the model without causing an unacceptably large shift in the scalar spectral index. We find that instantaneous reheating via decay to right-handed neutrinos disfavours $\alpha \approx 10^{12}, 10^{20}$. The model cannot reheat instantaneously through this channel for these values of $\alpha$ without causing an unacceptably large $n_{s}$-shift. The case of general Planck-suppressed potential corrections with larger values of $\alpha$ can reheat instantaneously through this channel without creating too large of an $n_{s}$-shift. Reheating could proceed for smaller $\alpha$ if the decay is not instantaneous but this would lower the reheating temperature, and consequently the predicted value of $n_{s}$. This could cause some tension with the observed value if this smaller $\alpha$ is still close to $10^{32}$, but could likely maintain good predictions for $\alpha <<10^{32}$.

Although it is a small part of this chapter, it is worth coming back to the possibility of oscillon formation from fragmentation of the inflaton condensate, which is an active field of research ( see e.g. \cite{amin11}, \cite{lozanov17}, \cite{amin10} - \cite{cyncynates21}) and could have observable consequences, including gravitational waves \cite{antusch17}, \cite{lozanov19}. Indeed an interesting consideration for observability in the $R^{2}$ quadratic Palatini inflation model is the potential for gravitational wave sources. Many studies have been done which illustrate that phases of non-linear dynamics in the early Universe can generate their own stochastic gravitational wave background (see e.g. \cite{khlebnikov97} - \cite{amin14}), as can increasingly non-linear interactions between perturbations during a preheating phase (e.g. \cite{assadullahi09} - \cite{jedamzik102}). There is also the possibility that an early epoch of oscillon domination after inflation could leave a gravitational wave signature \cite{alabidi13}. These are very speculative ideas, however it may be that these signatures could be observable in the future \cite{aggarwal18}, and could help to differentiate $R^{2}$ quadratic Palatini inflation observationally.

In conclusion, we have shown that the $R^{2}$ quadratic Palatini model can inflate successfully with a sub-Planckian inflaton, and its predictions for the scalar spectral index can be robust against Planck-suppressed potential corrections from a quantum gravity completion for sufficiently large $\alpha$, while predicting a much smaller tensor-to-scalar ratio than conventional chaotic $\phi^{2}$ inflation. In addition, it can reheat to a sufficiently high temperature to produce a viable post-inflation cosmology and can serve as a basis for the formation of oscillons after inflation. Therefore, the Palatini $R^{2}$ framework allows a minimal quadratic potential to overcome the obstacles encountered by conventional quadratic chaotic inflation and to serve as a consistent basis for standard Hot Big Bang cosmological evolution following inflation.

\chapter{Quadratic Term Affleck-Dine Inflation and Non-Minimally Coupled Inflation Models}

In this work we present an application of the Affleck-Dine mechanism via $U(1)$-violating quadratic potential terms in a simple non-minimally coupled inflation model with a complex inflaton. We present a detailed description of the dynamics of the generation of the asymmetry, and perform an analytical calculation of its evolution from the inflaton condensate through its transfer to the Standard Model particles. We illustrate that this model can produce the observed baryon-to-entropy ratio, confirm the validity of the analytical approximation, and determine the limits of its validity using a numerical computation. We examine the compatibility of this model with the dynamics of a non-minimally coupled inflation model, and we also consider the baryon isocurvature fraction produced in this model as compared to the observational limit. We address the possibility of washout of the generated baryon asymmetry for the general case of the decay of the inflaton condensate to fermions, and we also discuss the validity of the classical treatment of the baryon asymmetry given the quantum nature of the scalar fields.

\section{Affleck-Dine Mechanism}\label{section:32}
The underlying mechanism for generating the baryon asymmetry in this model is based on Affleck-Dine baryogenesis \cite{affleckdine}. Affleck-Dine baryogenesis is often employed as part of a supersymmetric model of particle physics \cite{dine952, dine95}, where the Affleck-Dine field is one of the flat directions of the potential of a supersymmetric theory (SUSY). This can also occur in conjunction with the formation of Q-balls and the generation of dark matter \cite{enqvist98} - \cite{doddato113} . These 'flat directions' of the potential are a common feature of supersymmetric models. The breaking of underlying symmetries that the flat direction scalars - commonly known as Affleck-Dine (AD) fields - are charged under sets the scale of the potential and generates non-zero vacuum expectation values for the scalar fields which may be away from the potential minima. When the effective mass of the AD field in a given direction becomes comparable to the Hubble rate, $m_{AD} \sim H$, the field begins to oscillate about the minimum of its lifted potential. The AD fields therefore form a zero-momentum coherently oscillating condensate which decays away with expansion like non-relativistic matter, $\rho_{AD} \sim a^{-3}$.

Affleck-Dine baryogenesis has also been considered beyond SUSY models, in the context of chaotic inflation \cite{charng} - \cite{babichev2}\footnote{The authors in \cite{babichev1, babichev2} consider a chaotic inflation model where a second complex scalar field carrying $U(1)$ charge is coupled to a function of the inflaton field, which explicitly breaks the $U(1)$ symmetry in the later stages of inflation and generates a baryon number asymmetry. In \cite{babichev1}, the authors consider the possibility of a non-minimal coupling of the complex scalar in the context of the suppression of isocurvature perturbations, and in \cite{babichev2} the complex scalar potential uses quadratic $U(1)$-violating terms leading to subsequent AD baryogenesis alongside the inflation.}, and non-minimally coupled inflation with quartic symmetry-breaking potential terms \cite{cline1, cline2}, with the inflaton being the AD field. This is the basis of the new model we consider in this chapter, where the inflaton field is a complex scalar charged under a global $U(1)$ symmetry. Specifically, we consider a renormalisable $U(1)$-symmetric inflaton potential with quadratic $U(1)$ symmetry-breaking terms. This differs from the more conventional cubic and quartic $U(1)$-breaking terms considered previously in inflaton AD models. The baryon asymmetry is generated when the inflaton field rolls down its potential and begins to oscillate about its minimum, forming a coherent scalar condensate, which subsequently decays. Oscillations of the field, $\Phi \leftrightarrow \Phi^{\dagger}$, as it is decaying means that a symmetry-conserving decay of the Affleck-Dine field can lead to an asymmetry being produced in the decay products, and subsequently in the baryon asymmetry of the Universe. This mechanism initially produces a much larger asymmetry in the inflaton condensate than that needed to produce the observed baryon-to-entropy ratio. This overproduction can be suppressed by averaging over the oscillations of the asymmetry as it is transferred to the particle content of the Standard Model, in order for the resulting asymmetry in baryon number to produce the observed baryon-to-entropy ratio.  For Affleck-Dine baryogenesis in general, it may be that this asymmetry is then transferred to the baryon content of the Universe via some other process, e.g. sphalerons in the case where the asymmetry initially generated through the decay of the Affleck-Dine condensate is lepton number (leptogenesis) \cite{sphalerons} \cite{maybe}. As we are concerned with the quadratic term AD mechanism in general in this discussion, we do not consider a specific decay route or mechanism for the transfer of the $U(1)$ asymmetry in this work, however we assume that the inflaton condensate decays to unspecified Majorana fermions $\psi, \bar{\psi}$. An interesting application would be leptogenesis via the decay of the inflaton to right-handed neutrinos. The quadratic AD mechanism presented here as part of work published in \cite{usad} was subsequently applied by Mohapatra and Okada to a model producing neutron - antineutron oscillations \cite{mo1}, and the generation of neutrino mass and Dark Matter \cite{mo2} - \cite{mo4}. When discussing the underlying symmetry, and referring to the transferred asymmetry after the formation of the coherent condensate in this work, we will refer to the quantum number in question as "$B$", with the understanding that the eventual resulting asymmetry will be in baryon number, irrespective of the exact mechanism leading to the formation of the $B$-asymmetry from the $U(1)$ asymmetry.

\section{The Model}\label{section:33}
The starting point for this Affleck-Dine model is to consider the $U(1)$ symmetric inflaton potential

\begin{equation}\label{eqn:ad1}
V\left(\left|\Phi \right| \right) = m_{\Phi}^{2}\left|\Phi \right|^{2} + \lambda_{\Phi}\left|\Phi \right|^{4},
\end{equation}

\noindent where the inflaton field $\Phi$ is also the Affleck-Dine field in this model. The renormalisable $U(1)$-violating terms of the AD field potential in general are given by

\begin{equation}\label{eqn:ad2}
V_{\bcancel{U\left(1\right)}} = A\left( \Phi^{2} + \Phi^{\dagger^{2}} \right) + B\left( \Phi^{3} + \Phi^{\dagger^{3}}\right) + C\left(\Phi^{4} + \Phi^{\dagger^{4}}\right),
\end{equation}

\noindent where $A = B = C =0$ conserves the $U(1)$ symmetry. This potential is naturally compatible with non-minimally coupled inflation models based on $\phi^{4}$ potentials. The presence of the $U(1)$-breaking terms change the trajectory of the field in the complex plane by introducing a dependence on the phase $\theta$, moving the field into an elliptical orbit away from the purely symmetric motion of the field. We impose the $\Phi \leftrightarrow -\Phi$ symmetry respected by the  $U(1)-$symmetric potential, which eliminates the $B$ term. Previous studies of Affleck-Dine baryogenesis with a non-minimally coupled inflaton as the AD field \cite{babichev2,cline1} consider symmetry breaking terms of cubic order or higher so here we will also set $C=0$ in order to exclusively study the effects of the quadratic $U(1)-$violating $A$ term and the asymmetry subsequently generated. 

\subsection{The Potential}\label{section:331}
The full $U(1)$-violating  potential we consider is therefore

\begin{equation}\label{eqn:ad3}
V\left(\Phi \right) = m_{\Phi}^{2}\left| \Phi \right|^{2} + \lambda_{\Phi} \left| \Phi \right|^{4} - A\left(\Phi^{2} + \Phi^{\dagger^{2}}\right),
\end{equation}

\noindent where we can assume without loss of generality that $A$ is real. We can write the scalar field as the sum of two real fields

\begin{equation}\label{eqn:ad4}
\Phi = \frac{1}{\sqrt{2}}\left(\phi_{1} + i\phi_{2}\right); \; \; \; \; \Phi^{\dagger} = \frac{1}{\sqrt{2}}\left(\phi_{1} - i\phi_{2}\right),
\end{equation}

\noindent which gives the potential in terms of the components $\phi_{1}$ and $\phi_{2}$ to be

\begin{equation}\label{eqn:ad5}
V\left(\phi_{1}, \phi_{2}\right) = \frac{1}{2}m_{\Phi}^{2}\left(\phi_{1}^{2} + \phi_{2}^{2}\right) + \frac{1}{4}\lambda_{\Phi}\left(\phi_{1}^{2} + \phi_{2}^{2}\right)^{2} - A\left(\phi_{1}^{2} - \phi_{2}^{2}\right).
\end{equation}

\noindent This leads to the following field equations

\begin{equation}\label{eqn:ad6}
\ddot{\phi}_{1} + 3H\dot{\phi}_{1} = -m_{1}^{2}\phi_{1} - \lambda_{\Phi}\left(\phi_{1}^{2} + \phi_{2}^{2}\right)\phi_{1},
\end{equation}

\begin{equation}\label{eqn:ad7}
\ddot{\phi}_{2} + 3H\dot{\phi}_{2} = -m_{2}^{2}\phi_{2} - \lambda_{\Phi}\left(\phi_{1}^{2} + \phi_{2}^{2}\right)\phi_{2},
\end{equation}

\noindent where we have absorbed the symmetry-breaking mass term into the mass term for $\Phi$ by defining

\begin{equation}\label{eqn:ad8}
m_{1}^{2} = m_{\Phi}^{2} - 2A ; \; \; \; \; \; m_{2}^{2} = m_{\Phi}^{2} + 2A.
\end{equation}

In the limit that $\lambda_{\Phi} \rightarrow 0$, the equations for $\phi_{1}$ and $\phi_{2}$ are completely decoupled from each other, and their field equations will have solutions corresponding to two coherently oscillating fields with angular frequencies $m_{1}$ and $m_{2}$ respectively.

\subsection{Threshold Approximation of the Asymmetry} \label{section:332}
In order to solve the field equations (\ref{eqn:ad6}) and (\ref{eqn:ad7}) analytically we use a threshold approximation for the potential, such that 

\begin{equation}\label{eqn:ad9}
V\left(\Phi \right) = \lambda_{\Phi}\left| \Phi \right|^{4} \; \; ; \; \; \phi > \phi_{\ast},
\end{equation}

\begin{equation}\label{eqn:ad10}
V\left(\Phi \right) = m_{\Phi}^{2}\left| \Phi \right|^{2} - A\left( \Phi^{2} + \Phi^{\dagger^{2}}\right) \; \; ; \; \; \phi < \phi_{\ast},
\end{equation}

\noindent where we write the field as $\Phi = \phi e^{i\theta}/\sqrt{2}$. The threshold $\phi_{\ast} = m_{\Phi}/\sqrt{\lambda_{\Phi}}$ is the value of $\phi$ at which the quartic term dominates the gradient of the potential $\partial V_{\phi}/\partial \phi$, and we have assumed when calculating the threshold that the $A$ term can be neglected against the scalar mass term as an estimate of the dominant dynamics in the $\phi < \phi_{\ast}$ region of the potential.

\begin{figure}[H]
\begin{center}
\includegraphics[clip = true, width=\textwidth, angle = 360]{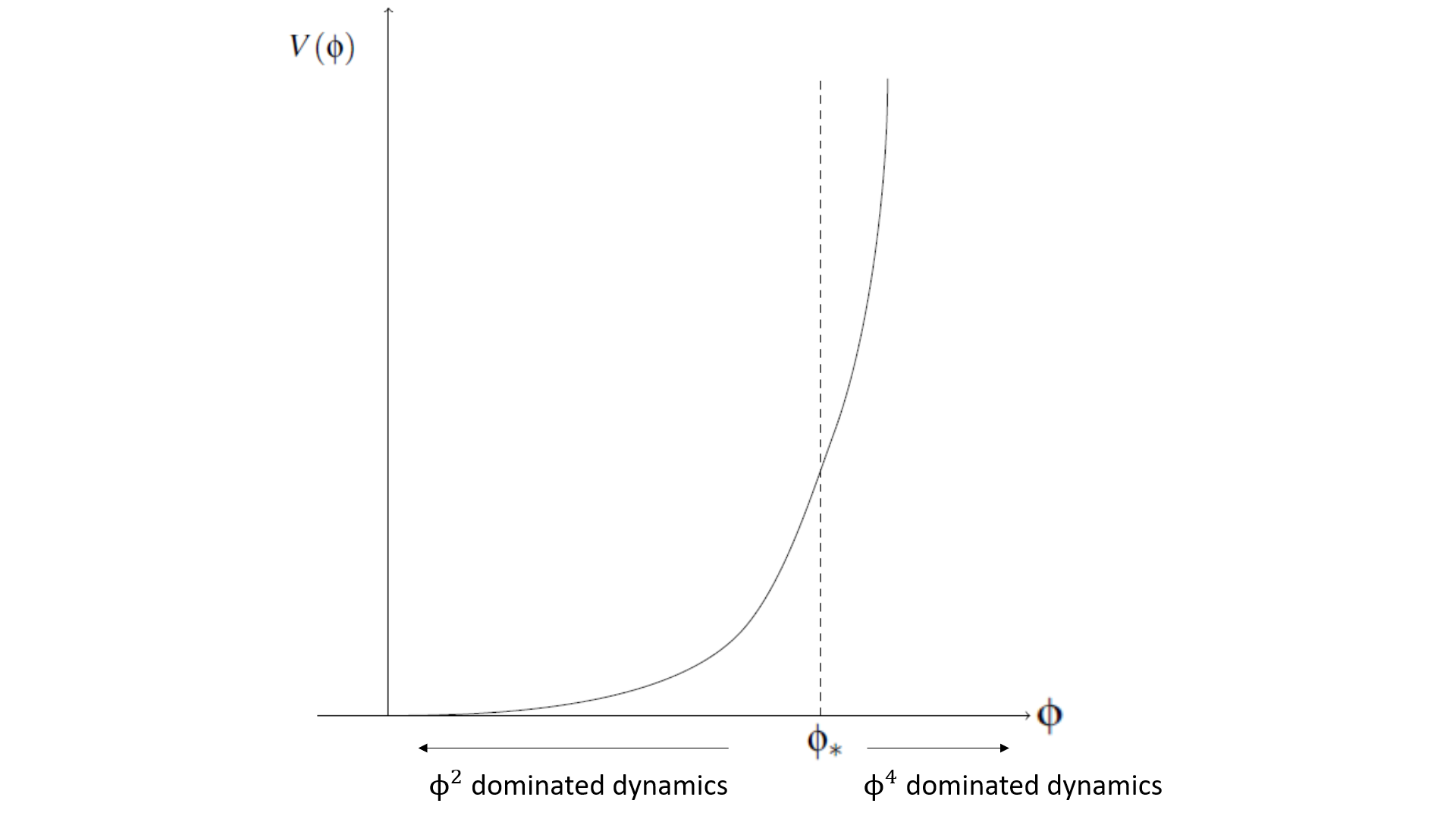}
\caption{Schematic of the inflaton potential illustrating the threshold approximation. } 
\label{figure:n15}
\end{center}
\end{figure} 

The potential is initially dominated strongly by the $\left|\Phi \right|^{4}$ term (see Figure \ref{figure:n15}), and the energy density of the coherent oscillations in the $\left|\Phi\right|^{4}$ potential during this phase is decaying away as $\rho \propto a^{-4}$ \cite{turner}. Since $\rho = V\left(\phi \right)$ during slow-roll when $\phi$ equals the amplitude of the oscillations, we have that the amplitude decays as $\phi \propto 1/a$ as it oscillates in the $\left|\Phi\right|^{4}$ part of the potential \cite{turner}. Let the initial amplitudes of $\phi_{1}$ and $\phi_{2}$ be $\phi_{1i}$ and $\phi_{2i}$ respectively, and let $\phi_{1\ast}$ and $\phi_{2\ast}$ be the respective values of the fields at the threshold. We also assume that initially the $\phi_{1}$ and $\phi_{2}$ fields are oscillating in phase such that there is no initial asymmetry. Using the inverse proportionality of the field to the scale factor, we can write the threshold values as

\begin{equation}\label{eqn:ad11}
\phi_{1\ast} = \left(\frac{a_{i}}{a_{\ast}}\right)\phi_{1i} = \left(\frac{\phi_{\ast}}{\phi_{i}}\right)\phi_{1i},
\end{equation}

\begin{equation}\label{eqn:ad12}
\phi_{2\ast} = \left(\frac{a_{i}}{a_{\ast}}\right)\phi_{2i} = \left(\frac{\phi_{\ast}}{\phi_{i}}\right)\phi_{2i},
\end{equation}

\noindent where $\phi_{i} = \left( \phi_{1i}^{2} + \phi_{2i}^{2}\right)^{\frac{1}{2}}$, and the field evolves purely due to the mass-squared terms once $a > a_{\ast}$. 

We assume that upon crossing the threshold the field enters into rapid $\Phi^{2}$ coherent oscillations, and that $m_{1}, m_{2} >>H$, such that we can neglect the effects of expansion during these oscillations. We also assume that the asymmetry will be generated during the $\Phi^{2}$ oscillations of the field, and therefore we need to consider solutions to the equations 

\begin{equation}\label{eqn:ad13}
\ddot{\phi}_{1} + m_{1}^{2}\phi_{1} = 0,
\end{equation}

\begin{equation}\label{eqn:ad14}
\ddot{\phi}_{2} + m_{2}^{2}\phi_{2} = 0,
\end{equation}

\noindent when $\phi < \phi_{\ast}$, in order to calculate the asymmetry. The solutions of these equations are of the form

\begin{equation}\label{eqn:ad15}
\phi \left( t \right) = D \cos \left( \omega t  + \delta \right),
\end{equation}

\noindent where the phase $\delta$ and the amplitude $D$ are determined by the initial conditions of the field. In our case $\omega = m_{1}, m_{2}$ for $\phi_{1}$ and $\phi_{2}$ respectively, and the initial phase can be set to zero, since there is no initial asymmetry.

\noindent Since we are interested in the asymmetry produced by the oscillations once the potential has crossed the threshold $\phi_{\ast}$ into the quadratic regime, we can set $t \rightarrow t - t_{\ast}$ as the time variable in the cosine function (\ref{eqn:ad15}). This means that at $t = t_{\ast}$ we need 

\begin{equation}\label{eqn:ad16}
\phi_{1,2}\left(t_{\ast}\right)  \rightarrow \phi_{1,2}\left(0\right) = \phi_{1\ast, 2\ast}.
\end{equation}

\noindent In the region of the potential dominated by the mass-squared terms, the energy density is decaying with the scale factor like non-relativistic matter \cite{turner}, $\rho \propto a^{-3}$, which means that the field is expected to decay as $\phi \propto a^{-3/2}$. This means that for any $a > a_{\ast}$, the amplitude of the field will be 

\begin{equation}\label{eqn:ad17}
\phi_{1,2} = \left(\frac{a_{\ast}}{a}\right)^{\frac{3}{2}}\phi_{1\ast, 2\ast},
\end{equation}

\noindent and the solutions to the field equations (\ref{eqn:ad13}) and (\ref{eqn:ad14}) at $t > t_{\ast}$ are therefore, to a good approximation (as we will confirm later),

\begin{equation}\label{eqn:ad18}
\phi_{1}\left(t \right) = \phi_{1\ast}\left(\frac{a_{\ast}}{a}\right)^{\frac{3}{2}}\cos \left( m_{1} \left( t - t_{\ast} \right) \right),
\end{equation}

\begin{equation}\label{eqn:ad19}
\phi_{2}\left(t \right) = \phi_{2\ast}\left(\frac{a_{\ast}}{a}\right)^{\frac{3}{2}}\cos \left( m_{2} \left( t - t_{\ast} \right) \right).
\end{equation}

\section{Analytical Derivation of the Asymmetry}\label{section:34}

We are now ready to calculate the $U(1)$ asymmetry. We will write $\phi_{1,2}\left( t \right) = \phi_{1,2}$ in this calculation for brevity.

\subsection{Noether's Theorem for a Scalar $U(1)$ Theory}\label{section:341}
We first derive the expression for the charge density. The $U(1)-$ invariant action of the theory in Minkowski space is

\begin{equation}\label{eqn:ad20}
S = \int d^{4}x \; \; \left[\partial_{\mu}\Phi^{\dagger}\partial^{\mu}\Phi - V\left(\left| \Phi \right| \right)\right],
\end{equation}

\noindent where we use the flat space approximation to calculate the form of the asymmetry here since we are assuming for the purposes of the analytical calculation that the effects of expansion will be negligible on the dynamics of the asymmetry. The effects of expansion on the formation and evolution of the asymmetry are examined in the numerical calculation of the asymmetry discussed in Section \ref{section:37}. We follow the procedure for the variation of the action in a $U(1)$-symmetric theory outlined in Chapter 3, and vary the action while treating $\Phi$ and $\Phi^{\dagger}$ as independent fields (see Section \ref{section:t35})

\begin{equation}\label{eqn:ad21}
\delta S = \int d^{4}x \; \left[\frac{\partial \mathcal{L}}{\partial \Phi}\delta \Phi + \frac{\partial \mathcal{L}}{\partial \left(\partial_{\mu}\Phi\right)}\delta\left(\partial_{\mu} \Phi\right)  + \frac{\partial \mathcal{L}}{\partial \Phi^{\dagger}}\delta \Phi^{\dagger} + \frac{\partial \mathcal{L}}{\partial \left(\partial_{\mu}\Phi^{\dagger}\right)}\delta\left(\partial_{\mu} \Phi^{\dagger}\right) \right],
\end{equation}

\noindent where we can rewrite the second and fourth terms of the Lagrangian density as half of the product rule for total derivatives to give

\begin{multline}\label{eqn:ad22}
\delta S = \int d^{4}x \; \left[\left(\frac{\partial \mathcal{L}}{\partial \Phi}  - \partial_{\mu}\left(\frac{\partial \mathcal{L}}{\partial \left(\partial_{\mu}\Phi\right)} \right)\right)\delta \Phi + \partial_{\mu}\left(\frac{\partial \mathcal{L}}{\partial \left(\partial_{\mu}\Phi \right)}\delta \Phi \right)\right.  \\ 
\left.+ \left(\frac{\partial \mathcal{L}}{\partial \Phi^{\dagger}} - \partial_{\mu}\left(\frac{\partial \mathcal{L}}{\partial \left(\partial_{\mu} \right)}\right)\right)\delta \Phi^{\dagger} + \partial_{\mu}\left(\frac{\partial \mathcal{L}}{\partial \left(\partial_{\mu}\Phi^{\dagger} \right)}\delta \Phi^{\dagger} \right) + \frac{\partial \mathcal{L}}{\partial \left(\partial_{\mu}\Phi^{\dagger}\right)}\delta \left(\partial_{\mu} \Phi^{\dagger}\right) \right].
\end{multline}

\noindent The first and third terms are zero from (\ref{eqn:b122}) and (\ref{eqn:b123}), and this leaves

\begin{equation}\label{eqn:ad23}
\delta S = \int d^{4}x \; \; \partial_{\mu}\left(\frac{\partial \mathcal{L}}{\partial \left(\partial_{\mu}\Phi\right)}\delta \Phi + \frac{\partial \mathcal{L}}{\partial \left(\partial_{\mu}\Phi^{\dagger}\right)}\delta \Phi^{\dagger} \right),
\end{equation}

\noindent which must be equal to zero in order for $\delta S = 0$. This means that $\partial_{\mu}j^{\mu} = 0$, where 

\begin{equation}\label{eqn:ad24}
j^{\mu} = \frac{\partial \mathcal{L}}{\partial \left(\partial_{\mu}\Phi\right)}\delta \Phi + \frac{\partial \mathcal{L}}{\partial \left(\partial_{\mu}\Phi^{\dagger}\right)}\delta \Phi^{\dagger}
\end{equation}

\noindent is the conserved $U(1)$ current of the theory. Evaluating this gives

\begin{equation}\label{eqn:ad25}
j^{\mu} = \partial_{\mu}\Phi^{\dagger}\delta^{\nu}_{\mu}\eta^{\mu \nu} \delta \Phi + \delta^{\mu}_{\mu}\partial^{\mu}\Phi \delta \Phi^{\dagger}.
\end{equation}

We have from (\ref{eqn:b128}), that the timelike component of the conserved current $j^{\mu} = \left( \rho, \overrightarrow{j}\right)$ is equal to the charge density of a theory with a conserved Noether charge

\begin{equation}\label{eqn:ad26}
Q = \int d^{3}x \; \; \rho_{Q} = \int d^{3}x \; \; j^{0},
\end{equation}

\noindent and so the charge density of the $U(1)$ theory is

\begin{equation}\label{eqn:ad27}
j^{0} = \dot{\Phi}^{\dagger}\delta \Phi + \dot{\Phi} \delta \Phi^{\dagger}.
\end{equation}

\noindent From the $U(1)$ transformation (\ref{eqn:b109}), we have 

\begin{equation}\label{eqn:ad28}
\delta \Phi = i\alpha \Phi, \; \; \; \delta \Phi^{\dagger} = -i\alpha \Phi^{\dagger},
\end{equation}

\noindent which gives the charge density to be 

\begin{equation}\label{eqn:ad29}
\rho_{Q} = i\alpha \left[\Phi \dot{\Phi}^{\dagger} - \Phi^{\dagger}\dot{\Phi} \right].
\end{equation}

In a $U(1)$ theory where the charge of each particle is normalised to $Q\left[\Phi \right] = +1$ and $Q\left[\Phi^{\dagger} \right] = -1$, we normalise $\alpha = 1$ to give

\begin{equation}\label{eqn:ad30}
\rho_{Q} = i\left[\Phi \dot{\Phi}^{\dagger} - \Phi^{\dagger}\dot{\Phi} \right].
\end{equation}

\noindent Evaluating the global charge upon quantising the fields (see Section \ref{section:t36}) reveals that the total $U(1)$ charge is determined by difference of the number density of particles to the number density of antiparticles in the system. The charge density of the system is therefore also determined by the difference in the number of particles and antiparticles in the system. We can therefore say that the number asymmetry of $\Phi$ and $\Phi^{\dagger}$ particles in the system is given by

\begin{equation}\label{eqn:ad31}
n\left(t \right) = i\left( \Phi \dot{\Phi}^{\dagger} - \Phi^{\dagger} \dot{\Phi} \right).
\end{equation}

\subsection{Analytical Expression of the Asymmetry}\label{section:342}
Using (\ref{eqn:ad4}), (\ref{eqn:ad31}) can be written in terms of $\phi_{1}$ and $\phi_{2}$ as

\begin{equation}\label{eqn:ad32}
n \left(t \right) = \dot{\phi}_{1}\phi_{2} - \dot{\phi}_{2}\phi_{1}.
\end{equation}

\noindent In order to evaluate the dynamics of the asymmetry, we must evaluate (\ref{eqn:ad32}) in terms of the solutions for $\phi_{1}$ and $\phi_{2}$, (\ref{eqn:ad18}) and (\ref{eqn:ad19}) respectively. Differentiating these solutions with respect to time we find

\begin{equation}\label{eqn:ad33}
\dot{\phi}_{1,2} = \phi_{1\ast, 2\ast} \left(\frac{a_{\ast}}{a}\right)^{\frac{3}{2}}\left[-\frac{3}{2}H \cos\left(m_{1,2}\left(t - t_{\ast}\right) \right) - m_{1,2}\sin \left(m_{1,2}\left(t - t_{\ast}\right) \right) \right].
\end{equation}

\noindent Since the frequency of oscillation of the fields is much faster than the rate of expansion, $m_{1,2} >>H$, we can approximate (\ref{eqn:ad33}) by the sine term and use

\begin{equation}\label{eqn:ad34}
\dot{\phi}_{1,2} = -\phi_{1,2 \ast} \left(\frac{a_{\ast}}{a}\right)^{\frac{3}{2}} m_{1,2}\sin \left(m_{1,2}\left(t - t_{\ast}\right) \right).
\end{equation}

\noindent Substituting (\ref{eqn:ad18}), (\ref{eqn:ad19}) and (\ref{eqn:ad34}) into the asymmetry (\ref{eqn:ad32}) we have

\begin{multline}\label{eqn:ad35}
n\left( t \right) = \phi_{1\ast}\phi_{2\ast} \left(\frac{a_{\ast}}{a}\right)^{3}\left[ m_{2}\sin \left(m_{2} \left( t - t_{\ast} \right) \right) \cos \left(m_{1} \left( t - t_{\ast} \right) \right) \right.\\
\left.- m_{1} \sin \left(m_{1} \left( t - t_{\ast} \right) \right) \cos \left(m_{2} \left( t - t_{\ast} \right) \right) \right].
\end{multline}

\noindent Using the definitions of the field masses squared (\ref{eqn:ad8}), and assuming that $2A << m_{\Phi}^{2}$, we can write

\begin{equation}\label{eqn:ad36}
m_{1} = \left( m_{\Phi}^{2} - 2A \right)^{\frac{1}{2}} = m_{\Phi}\left( 1 - \frac{2A}{m_{\Phi}^{2}} \right)^{\frac{1}{2}} \approx m_{\Phi}\left( 1 - \frac{A}{m_{\Phi}^{2}} \right),
\end{equation}

\begin{equation}\label{eqn:ad37}
m_{2} = \left( m_{\Phi}^{2} + 2A \right)^{\frac{1}{2}} = m_{\Phi}\left( 1 + \frac{2A}{m_{\Phi}^{2}} \right)^{\frac{1}{2}} \approx m_{\Phi}\left( 1 + \frac{A}{m_{\Phi}^{2}} \right),
\end{equation}

\noindent and the asymmetry can then be written as

\begin{multline}\label{eqn:ad38}
n\left( t \right) = \phi_{1\ast}\phi_{2\ast} \left(\frac{a_{\ast}}{a}\right)^{3}\left[ \left(m_{\Phi} + \frac{A}{m_{\Phi}} \right)\sin \left(m_{2} \left( t - t_{\ast} \right) \right) \cos \left(m_{1} \left( t - t_{\ast} \right) \right) \right.  \\ 
\left. - \left(m_{\Phi} - \frac{A}{m_{\Phi}} \right) \sin \left(m_{1} \left( t - t_{\ast} \right) \right) \cos \left(m_{2} \left( t - t_{\ast} \right) \right) \right].
\end{multline}

\noindent Separating the $m_{\Phi}$ and the $A/m_{\Phi}$ terms this is

\begin{equation}\label{eqn:ad39}
\begin{split}
n\left( t \right) = &\phi_{1\ast}\phi_{2\ast} \left(\frac{a_{\ast}}{a}\right)^{3}\left[m_{\Phi}\left[\sin \left(m_{2}\left( t - t_{\ast}\right) \right)\cos\left(m_{1} \left( t - t_{\ast} \right) \right)\right. \right.\\
& \left. \left. - \sin \left(m_{1}\left( t - t_{\ast}\right) \right)\cos\left(m_{2} \left( t - t_{\ast} \right) \right) \right] + \frac{A}{m_{\phi}}\left[\sin \left(m_{2}\left( t - t_{\ast}\right) \right)\cos\left(m_{1} \left( t - t_{\ast} \right) \right) \right. \right. \\
& \left.\left.+ \sin \left(m_{1}\left( t - t_{\ast}\right) \right)\cos\left(m_{2} \left( t - t_{\ast} \right) \right) \right] \right].
\end{split}
\end{equation}

\noindent These terms can be condensed using the trigonometric identities

\begin{multline}\label{eqn:ad40}
\sin \left(m_{2}\left( t - t_{\ast}\right) \right)\cos\left(m_{1} \left( t - t_{\ast} \right) \right) - \sin \left(m_{1}\left( t - t_{\ast}\right) \right)\cos\left(m_{2} \left( t - t_{\ast} \right) \right) \\
= \sin \left[m_{2} \left( t - t_{\ast} \right) - m_{1} \left( t - t_{\ast} \right) \right],
\end{multline}

\noindent and 

\begin{multline}\label{eqn:ad41}
\sin \left(m_{2}\left( t - t_{\ast}\right) \right)\cos\left(m_{1} \left( t - t_{\ast} \right) \right) + \sin \left(m_{1}\left( t - t_{\ast}\right) \right)\cos\left(m_{2} \left( t - t_{\ast} \right) \right) \\
= \sin \left[m_{2} \left( t - t_{\ast} \right) + m_{1} \left( t - t_{\ast} \right) \right],
\end{multline}

\noindent which means that the asymmetry is

\begin{multline}\label{eqn:ad42}
n\left( t \right) = \phi_{1\ast}\phi_{2\ast} \left(\frac{a_{\ast}}{a}\right)^{3}\left[m_{\Phi}\sin \left[m_{2} \left( t - t_{\ast} \right) - m_{1} \left( t - t_{\ast} \right) \right] + \right.\\
\left.\frac{A}{m_{\Phi}}\sin \left[m_{2} \left( t - t_{\ast} \right) + m_{1} \left( t - t_{\ast} \right) \right] \right].
\end{multline}

\noindent From (\ref{eqn:ad8}) we have that

\begin{equation}\label{eqn:ad43}
m_{2} - m_{1} = \frac{2A}{m_{\Phi}}, \; \; \; m_{2} + m_{1} = 2m_{\Phi}
\end{equation}

\noindent and approximating $m_{1} \approx m_{2} \approx m_{\Phi}$ to work to first order in $A$, the asymmetry in the inflaton condensate becomes

\begin{equation}\label{eqn:ad44}
n\left( t \right) = \phi_{1\ast}\phi_{2\ast} \left(\frac{a_{\ast}}{a}\right)^{3}\left[m_{\Phi}\sin \left(\frac{2A}{m_{\Phi}} \left( t - t_{\ast} \right) \right) + \frac{A}{m_{\Phi}}\sin \left(2m_{\Phi} \left( t - t_{\ast} \right) \right) \right].
\end{equation}

\noindent Since $m_{\Phi} >>H$, the oscillation period $m_{\Phi}^{-1}$ is much smaller than the timescale of expansion, so we can treat the scale factor $a$ as a constant when examining the asymmetry averaged over a large number of coherent oscillations. In the same limit, the second term in (\ref{eqn:ad44}) averages to zero and the final expression for the condensate asymmetry is then

\begin{equation}\label{eqn:ad45}
n\left( t \right) = \phi_{1\ast}\phi_{2\ast} \left(\frac{a_{\ast}}{a}\right)^{3}m_{\Phi}\sin \left(\frac{2A}{m_{\Phi}} \left( t - t_{\ast} \right) \right).
\end{equation}

\noindent The condensate asymmetry for $t > t_{\ast}$ can then be written in terms of the initial field values using (\ref{eqn:ad11}), (\ref{eqn:ad12}) as follows

\begin{equation}\label{eqn:ad46}
n\left( t \right) = \phi_{1i}\phi_{2i}\left(\frac{\phi_{i}}{\phi_{\ast}}\right) \left(\frac{a_{i}}{a}\right)^{3}m_{\Phi}\sin \left(\frac{2A}{m_{\Phi}} \left( t - t_{\ast} \right) \right).
\end{equation}

\noindent We will later confirm numerically that this expression gives the asymmetry of the inflaton condensate due to the potential (\ref{eqn:ad3}).

\subsection{Condensate Asymmetry \& Dynamics}\label{section:343}

We now define the comoving inflaton condensate asymmetry as

\begin{equation}\label{eqn:ad47}
n_{c}\left(t \right) = \left(\frac{a\left(t \right)}{a_{i}}\right)^{3} n\left( t \right),
\end{equation}

\noindent which is constant when there is no production or decay of the asymmetry. For $t > t_{\ast}$ we therefore have

\begin{equation}\label{eqn:ad48}
n_{c}\left( t \right) = \phi_{1i}\phi_{2i} \left(\frac{\phi_{i}}{\phi_{\ast}}\right) m_{\Phi} \sin \left( \frac{2A\left(t - t_{\ast} \right)}{m_{\Phi}}\right),
\end{equation}

\noindent where we define $n_{c}\left( t \right) = 0$ at $t=t_{\ast}$. Since the symmetry-breaking term in (\ref{eqn:ad3}) is small during the regime when the potential is $\left|\Phi\right|^{4}$ dominated, we assume that any asymmetry generated for $t < t_{\ast}$ is negligibly small, and only study the threshold asymmetry generated in the $\Phi^{2}$ regime. We will later determine numerically the condition under which this assumption is valid, and consider the effect of the symmetry breaking dynamics becoming significant during $\left|\Phi\right|^{4}$ dominated oscillations.

The asymmetry is initially zero and then evolves as an oscillating function. As the inflaton scalars decay away and reheat the Universe, they periodically undergo a phase rotation between the $\Phi$ $\left( Q_{U\left(1\right)} = +1\right)$ and $\Phi^{\dagger}$ $\left( Q_{U\left(1\right)} = -1\right)$ states. The reason that this can generate a net asymmetry when the inflaton condensate decays is that as the condensate is decaying, scalars are leaving the system continuously while the phase of the inflaton field oscillates between its $\Phi$ and $\Phi^{\dagger}$ states. This means that the number of scalars present in the condensate will decrease, and therefore the maximum amount of $+$ or $-$ charge that can be present will also decrease with each phase oscillation the field undergoes. This means that there is no longer an exact cancellation between the amount of $+$ or $-$ charge present in the condensate during each half-cycle in the phase oscillation. This leaves a residual amount of non-zero asymmetry transferred via decays to the Standard Model sector for each cycle.

The condensate decays away completely after a time $\tau_{\Phi} \approx \Gamma_{\Phi}^{-1}$, where this time period corresponds to the lifetime of the $\Phi$ scalars, and the point where the condensate has completely decayed is denoted $t_{R}$, corresponding to the end of reheating. During this decay process the asymmetry generated within the condensate is transferred to the Standard Model particle content of the Universe through the continuous $B$-conserving decays of the inflaton particles from $t_{\ast}$ to $t_{R}$, until the condensate has completely decayed.

\subsection{Transferred Asymmetry \& Dynamics}\label{section:344}
We now consider the asymmetry as it appears in the particle plasma from the onset of the condensate decay - referred to as the "transferred asymmetry". To illustrate the process, we ignore for now any decay of the condensate itself which may have occurred for $t < t_{\ast}$. We then have that the comoving asymmetry transferred to the particle plasma is

\begin{equation}\label{eqn:ad49}
\hat{n}_{c}\left(t \right) = \int^{t}_{t_{\ast}} \Gamma_{\Phi} n_{c}\left(t \right) dt.
\end{equation}

\noindent Using (\ref{eqn:ad48}) we find that the comoving asymmetry transferred to the Standard Model particle content is 

\begin{equation}\label{eqn:ad50}
\hat{n}_{c}\left(t \right) = \frac{\Gamma_{\Phi} \phi_{1i} \phi_{2i} m_{\Phi}^{2}}{2A}\left(\frac{\phi_{i}}{\phi_{\ast}}\right) \left[1 - \cos \left(\frac{2A\left( t - t_{\ast}\right)}{m_{\Phi}}\right) \right].
\end{equation}

\noindent At $t = t_{\ast}$, $\hat{n}_{c}\left(t \right) = 0$ since the condensate has not begun to decay yet. The transferred asymmetry increases quadratically with $t - t_{\ast}$ until $t - t_{\ast} \approx \pi m_{\Phi}/4A$, and then over time oscillates between a minimum and maximum value, $\hat{n}_{c}\left(t \right) = 0$ at $t - t_{\ast} = n\pi m_{\Phi}/2A$, and $\hat{n}_{c}\left(t \right) = \frac{\Gamma_{\Phi} \phi_{1i} \phi_{2i} m_{\Phi}^{2}}{A}\left(\frac{\phi_{i}}{\phi_{\ast}}\right)$  at $t - t_{\ast} = n\pi m_{\Phi}/4A$ for integer $n$, respectively. We denote the period of oscillation by $T_{asy} = \pi m_{\Phi}/A$. 

\begin{figure}[H]
\includegraphics[clip = true, width=\textwidth, angle = 360]{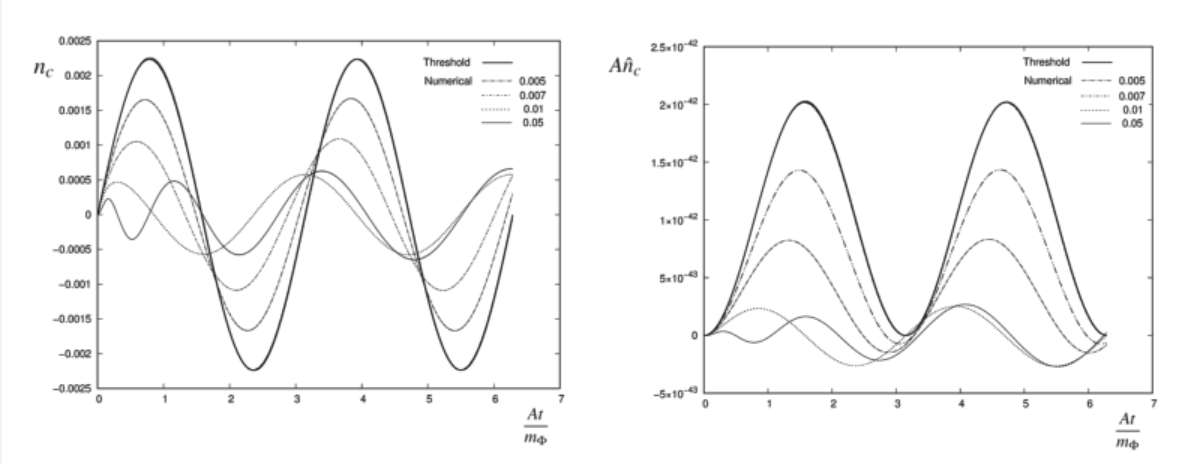}
\caption{\small Plot showing the condensate asymmetry (left) and the transferred asymmetry (right) for $m_{\Phi} = 10^{16}\GeV$, $\lambda_{\Phi} = 0.1$ and $T_{R} = 10^{8}\GeV$. In this case $A_{th}^{\frac{1}{2}}/m_{\Phi} = 0.003$, and the threshold asymmetry in each case is given by the solid line on each plot ((\ref{eqn:ad48}) and (\ref{eqn:ad50}) respectively) and the numerical results for $A^{\frac{1}{2}}/m_{\Phi} = 0.001, 0.005, 0.007, 0.01$ and $0.05$ are shown. We find that the numerical result for $A^{\frac{1}{2}}/m_{\Phi} = 0.001$ is exactly the result from the threshold asymmetry, as expected from (\ref{eqn:ad100}). } 
\label{figure:31}
\end{figure}

The transferred asymmetry is periodically zero but never crosses zero (see right panel of Figure \ref{figure:31}). Physically, this is because as the condensate oscillates between its $\Phi$ and $\Phi^{\dagger}$ states, any asymmetry generated by the decay of the condensate exactly cancels the asymmetry generated in the previous half-cycle. This means that the asymmetry is conserved as it is transferred to the Standard Model particles from the decay of the condensate, and averages to a non-zero value over long periods of time. 

A more accurate examination of the evolution of the transferred asymmetry requires including an exponential decay factor in (\ref{eqn:ad49}) to account for the $B$-conserving decay of the condensate asymmetry

\begin{equation}\label{eqn:ad51}
\hat{n}_{c}\left(t \right) = \int^{t}_{t_{\ast}} \Gamma_{\Phi} e^{-\Gamma_{\Phi}\left(t - t_{\ast}\right)}n_{c}\left(t \right) dt.
\end{equation}

\noindent Substituting (\ref{eqn:ad48}), we have

\begin{equation}\label{eqn:ad52}
\hat{n}_{c}\left(t \right) = \Gamma_{\Phi}\phi_{1i}\phi_{2i}\left(\frac{\phi_{i}}{\phi_{\ast}}\right)m_{\Phi}\int^{t}_{t_{\ast}} e^{-\Gamma_{\Phi}\left(t - t_{\ast}\right)}\sin\left(\frac{2A\left(t - t_{\ast}\right)}{m_{\Phi}}\right) dt.
\end{equation}

\noindent For a general integral of the form (\ref{eqn:ad52}) we have that

\begin{multline}\label{eqn:ad53}
\int^{x}_{0} e^{-ax}\sin\left(bx\right) dx = \left[-\frac{e^{-ax}\left[a\sin\left(bx\right) + b\cos\left(bx\right)\right]}{a^{2} + b^{2}} \right]^{x}_{0} \\
= -\frac{e^{-ax}\left[a\sin\left(bx\right) + b\cos\left(bx\right)\right]}{a^{2} + b^{2}} + \frac{b^{2}}{a^{2} + b^{2}}.
\end{multline}

\noindent Averaging over many oscillations gives the limit $x \rightarrow \infty$, and we have

\begin{equation}\label{eqn:ad54}
\int^{x}_{0} e^{-ax}\sin\left(bx\right) dx \rightarrow \frac{b^{2}}{a^{2} + b^{2}}.
\end{equation}

\noindent In (\ref{eqn:ad53}) in our case we have $x = t - t_{\ast}$, $a = \Gamma_{\Phi}$ and $b = 2A/m_{\Phi}$, and the integral is

\begin{equation}\label{eqn:ad55}
\int^{t}_{t_{\ast}} e^{-\Gamma_{\Phi}\left(t - t_{\ast}\right)}\sin\left(\frac{2A\left(t - t_{\ast}\right)}{m_{\Phi}}\right) dt = \frac{\frac{2A}{m_{\Phi}}}{\left(\Gamma_{\Phi}^{2} + \frac{4A^{2}}{m_{\Phi}^{2}} \right)} = \frac{m_{\Phi}}{2A\left[ 1 + \left(\frac{m_{\Phi}\Gamma_{\Phi}}{2A}\right)^{2}\right]}.
\end{equation}

\noindent We have therefore that the comoving transferred asymmetry averaged over many oscillations is 

\begin{equation}\label{eqn:ad56}
\hat{n}_{c}\left(t \right) = \frac{\Gamma_{\Phi}\phi_{1i}\phi_{2i}m_{\Phi}^{2}}{2A}\left(\frac{\phi_{i}}{\phi_{\ast}}\right)\left[ 1 + \left(\frac{m_{\Phi}\Gamma_{\Phi}}{2A}\right)^{2}\right]^{-1}.
\end{equation}

\noindent We will later confirm the validity of this result both analytically and numerically by solving the field equations. 

Thus, for $t > t_{R}$, the condensate has completely decayed, and the transferred comoving asymmetry is constant. This means that all of the asymmetry from the oscillating condensate has been transferred by the decay of the inflaton scalars, and that the particle plasma is left carrying a finite non-zero asymmetry as a result. This is a promising result, and it demonstrates clearly that this method of asymmetry generation in the Standard Model by the transfer of an asymmetry due to inflaton mass terms through the decay of the inflaton condensate is a viable method for generating the observed baryon asymmetry. We now consider whether this mechanism can produce the observed baryon-to-entropy ratio.

\section{Generating the Present Day Baryon Asymmetry from the Decay of the Affleck-Dine Condensate}\label{section:35}

We consider two scenarios here when examining the ability of the Affleck-Dine baryogenesis model with a quadratic symmetry breaking term to produce the observed baryon asymmetry. We firstly consider the case where the lifetime of the inflaton condensate, $\tau_{\Phi}$, is much larger than the oscillation period of the asymmetry, $T_{asy}$, such that $2A/m_{\Phi}\Gamma_{\Phi}>>1$. Physically this means that the asymmetry oscillates a large number of times before the condensate decays away, and that the final asymmetry transferred to the Standard Model particles will be averaged over a large number of oscillations. In this case, the comoving transferred asymmetry (\ref{eqn:ad56}) can be approximated

\begin{equation}\label{eqn:ad57}
\hat{n}_{c}\left(t \right) = \frac{\phi_{1i}\phi_{2i} \Gamma_{\Phi} m_{\Phi}^{2}}{2A}\left( \frac{\phi_{i}}{\phi_{\ast}} \right)\left[ 1 - \left(\frac{\Gamma_{\Phi} m_{\Phi}}{2A}\right)^{2}\right] \approx \frac{\phi_{1i}\phi_{2i} \Gamma_{\Phi} m_{\Phi}^{2}}{2A}\left( \frac{\phi_{i}}{\phi_{\ast}} \right).
\end{equation}

\noindent At $t = t_{i}$ we have that

\begin{equation}\label{eqn:ad58}
\Phi = \frac{\phi_{i} e^{i\theta}}{\sqrt{2}} = \frac{1}{\sqrt{2}}\left( \phi_{1i} + i\phi_{2i}\right),
\end{equation}

\begin{equation}\label{eqn:ad59}
\Phi^{\dagger} = \frac{\phi_{i} e^{-i\theta}}{\sqrt{2}} = \frac{1}{\sqrt{2}}\left( \phi_{1i} - i\phi_{2i}\right),
\end{equation}

\noindent which gives the relations

\begin{equation}\label{eqn:ad60}
\phi_{1i} = \frac{\phi_{i}}{2}\left(e^{i\theta} + e^{-i\theta} \right) = \phi_{i}\cos \theta,
\end{equation}

\begin{equation}\label{eqn:ad61}
\phi_{2i} = \frac{\phi_{i}}{2i}\left(e^{i\theta} - e^{-i\theta} \right) = \phi_{i}\sin \theta,
\end{equation}

\noindent and we can write

\begin{equation}\label{eqn:ad62}
\phi_{1i}\phi_{2i} = \phi_{i}^{2}\sin\theta \cos \theta = \frac{\phi_{i}^{2}}{2}\sin 2\theta,
\end{equation}

\noindent where $\theta$ is the initial angle of the oscillating $\Phi$ field. Substituting (\ref{eqn:ad62}) into (\ref{eqn:ad57}), the comoving transferred asymmetry for $\tau_{\Phi} > T_{asy}$ is therefore

\begin{equation}\label{eqn:ad63}
\hat{n}_{c, tot} = \frac{\phi_{i}^{2}\sin 2\theta \Gamma_{\Phi} m_{\Phi}^{2}}{4A}\left( \frac{\phi_{i}}{\phi_{\ast}} \right).
\end{equation}

We assume that by the end of reheating all of the condensate has decayed away, and all of the asymmetry has been transferred to the Standard Model particles. This means that the asymmetry at $a = a_{R}$, assuming the asymmetry is not washed out by $B$-violating interactions of the inflaton or its decay products, will be the value of the asymmetry that we measure today. It is therefore the value of the transferred asymmetry at reheating that we need to calculate. This value is

\begin{equation}\label{eqn:ad64}
\hat{n}_{tot} = \left(\frac{a_{i}}{a_{R}}\right)^{3}\hat{n}_{c,tot} =   \left(\frac{a_{i}}{a_{R}}\right)^{3} \frac{\phi_{i}^{2}\sin 2\theta \Gamma_{\Phi} m_{\Phi}^{2}}{4A}\left( \frac{\phi_{i}}{\phi_{\ast}} \right),
\end{equation}

\noindent where $\hat{n}_{tot}$ is the physical transferred asymmetry at $t_{R}$, obtained by rescaling the comoving transferred asymmetry according to (\ref{eqn:ad47}).

\noindent For $a_{i} = a < a_{\ast}, a \propto 1/\phi$, and for $a_{R} = a > a_{\ast}, a \propto \phi^{-\frac{2}{3}}$, therefore

\begin{equation}\label{eqn:ad65}
\left(\frac{a_{i}}{a_{R}}\right)^{3} = \left(\frac{a_{i}}{a_{\ast}}\cdot \frac{a_{\ast}}{a_{R}}\right)^{3} =  \left(\frac{\phi_{\ast}}{\phi_{i}}\right)^{3}\frac{\phi_{R}^{2}}{\phi_{\ast}^{2}},
\end{equation}

\noindent the physical transferred asymmetry is then

\begin{equation}\label{eqn:ad66}
\hat{n}_{tot} = \frac{\phi_{R}^{2}\sin 2\theta \Gamma_{\Phi} m_{\Phi}^{2}}{4A}.
\end{equation}

\noindent At reheating we have that $V = m_{\Phi}^{2}\phi_{R}^{2}/2 = 3M_{pl}^{2}H_{R}^{2}$, and $H_{R} = \Gamma_{\Phi}$, which means that the final asymmetry at reheating is then

\begin{equation}\label{eqn:ad67}
\hat{n}_{tot} = \frac{3M_{pl}^{2}\sin 2\theta \Gamma_{\Phi}^{3} m_{\Phi}^{2}}{2A}.
\end{equation}

\noindent Assuming that all of the $U(1)$ asymmetry is transferred to the Standard Model particles upon decay of the inflaton condensate, and that the asymmetry is not washed out by later-time $B$-violating processes, then the transferred asymmetry at reheating will become the baryon asymmetry today. We can then say that

\begin{equation}\label{eqn:ad68}
\frac{n_{B}}{s} = \frac{\hat{n}_{tot}}{s},
\end{equation}

\noindent where $s$ is the entropy density, and in calculation here we will take this to be the entropy density at reheating. The entropy density is (\ref{eqn:136})

\begin{equation}\label{eqn:ad69}
s = \frac{2\pi^{2}}{45}g\left(T \right)T^{3},
\end{equation}

\noindent and at $T_{R}$ we have that 

\begin{equation}\label{eqn:ad70}
s = 4k_{T_{R}}^{2}T_{R}^{3}; \; \; k_{T_{R}}^{2} = \frac{\pi^{2}g\left(T_{R} \right)}{90},
\end{equation}

\noindent where $g\left(T\right)$ is the total number of relativistic degrees of freedom. Since $\Gamma_{\Phi} = H_{R} = k_{T_{R}}T_{R}^{2}/M_{pl}$ \cite{kofmanre}, we can write the predicted baryon to entropy ratio using (\ref{eqn:ad67}) and (\ref{eqn:ad70}) as

\begin{equation}\label{eqn:ad71}
\frac{n_{B}}{s} = \frac{\hat{n}_{tot}}{s} = \frac{3 \sin 2\theta k_{T_{R}}T_{R}^{3}}{8A M_{pl}}.
\end{equation}

\noindent Normalising the inflaton mass and the reheating temperature using $T_{R} = 10^{8} \GeV$ (this value of the reheating temperature is within the range which may be calculable from the detectable spectrum of primordial gravitational waves \cite{nakayama}), $m_{\Phi} = 10^{13} \GeV$, and $g\left(T \right) = 106.75$ for the Standard Model degrees of freedom, we have $k_{T_{R}} = 3.42$, and substituting into (\ref{eqn:ad71}), $n_{B}/s$ can be expressed as

\begin{equation}\label{eqn:ad72}
\frac{n_{B}}{s} = 5.35 \times 10^{-21}\frac{m_{\Phi}^{2}}{A}\left(\frac{T_{R}}{10^{8}\GeV}\right)^{3}\left(\frac{10^{13} \GeV}{m_{\Phi}}\right)^{2}\sin 2\theta.
\end{equation}

The observed baryon-to-entropy ratio is \cite{planck184}

\begin{equation}\label{eqn:ad73}
\left(\frac{n_{B}}{s}\right)_{obs} = \left(0.861 \pm 0.005\right) \times 10^{-10}.
\end{equation}

\noindent Therefore we find from (\ref{eqn:ad72}) that

\begin{equation}\label{eqn:ad74}
\frac{A^{\frac{1}{2}}}{m_{\Phi}} = 7.88 \times 10^{-6}\sin^{\frac{1}{2}}2\theta \left(\frac{10^{13}\GeV}{m_{\Phi}}\right)\left(\frac{T_{R}}{10^{8}\GeV}\right)^{\frac{3}{2}},
\end{equation}

\noindent is needed in order to produce the observed asymmetry today using the threshold estimate (\ref{eqn:ad72}) of the asymmetry transferred to the Standard Model. 

It is useful to check the size of the maximum predicted baryon-to-entropy ratio which can be produced from the generation of an asymmetry through quadratic symmetry breaking terms in an Affleck-Dine condensate. We do this by considering (\ref{eqn:ad56}) in the case $A = A_{max} = \Gamma_{\Phi}m_{\Phi}/2$, which corresponds to a maximum comoving transferred asymmetry,

\begin{equation}\label{eqn:ad75}
\hat{n}_{c,max} = \frac{\phi_{1i}\phi_{2i} m_{\Phi}}{2}\left(\frac{\phi_{i}}{\phi_{\ast}}\right) = \frac{\phi_{i}^{2}\sin 2\theta m_{\Phi}}{4}\left(\frac{\phi_{i}}{\phi_{\ast}}\right),
\end{equation}

\noindent where (\ref{eqn:ad62}) is used in the second equality. The maximum transferred asymmetry is then

\begin{multline}\label{eqn:ad76}
\hat{n}_{max} =  \left(\frac{a_{i}}{a_{R}}\right)^{3}\frac{\phi_{i}^{2}\sin 2\theta m_{\Phi}}{4}\left(\frac{\phi_{i}}{\phi_{\ast}}\right) = \left(\frac{a_{i}}{a_{\ast}}\cdot\frac{a_{\ast}}{a_{R}}\right)^{3}\left(\frac{\phi_{i}}{\phi_{\ast}}\right)\frac{\phi_{i}^{2}\sin 2\theta m_{\Phi}}{4}  \\
= \left(\frac{\phi_{\ast}}{\phi_{i}}\right)^{3}\left(\frac{\phi_{R}}{\phi_{\ast}}\right)^{2}\left(\frac{\phi_{i}}{\phi_{\ast}}\right)\frac{\phi_{i}^{2}\sin 2\theta m_{\Phi}}{4}.
\end{multline}

\noindent Using again the fact that at reheating $V = m_{\Phi}^{2}\phi_{R}^{2}/2 = 3M_{pl}^{2}H_{R}^{2}$, and  $H_{R} = \Gamma_{\Phi}$, we have that the maximum asymmetry which could be transferred is

\begin{equation}\label{eqn:ad77}
\left(\frac{\hat{n}_{B}}{s}\right)_{max} = \frac{3M_{pl}^{2}\Gamma_{\Phi}^{2}\sin 2\theta}{2m_{\Phi}}.
\end{equation}

\noindent Taking $\Gamma_{\Phi}^{2} = k_{T_{R}}^{2}T_{R}^{4}/M_{pl}^{2}$, the maximum predicted baryon-to-entropy ratio is then

\begin{equation}\label{eqn:ad78}
\left(\frac{n_{B}}{s}\right)_{max} = \frac{3T_{R} \sin 2\theta}{8m_{\Phi}}.
\end{equation}

\noindent Using the earlier normalisation of scalar mass, $m_{\Phi} = 10^{13}\GeV$, and reheating temperature, $T_{R} = 10^{8}\GeV$, we can write this as

\begin{equation}\label{eqn:ad79}
\frac{\hat{n}_{Bmax}}{s} = \left(3.75 \times 10^{-6}\right)\sin 2 \theta \left(\frac{T_{R}}{10^{8}\GeV}\right)\left(\frac{10^{13}\GeV}{m_{\Phi}}\right).
\end{equation}

\noindent This can easily be larger than the observed baryon-to-entropy ratio (\ref{eqn:ad73}), which implies that the suppression of the asymmetry arising from averaging over asymmetry oscillations can play an important role in this model when $\tau_{\Phi} > T_{asy}$.

We also want to consider whether it would be possible to generate the required present-day baryon asymmetry in the case where the condensate decays away very quickly, before the asymmetry can undergo many oscillations, i.e. $\tau_{\Phi} < T_{asy}$. This corresponds to the limit $\Gamma_{\Phi}m_{\Phi}/2A >>1$, which when applied to the comoving transferred asymmetry over large times (\ref{eqn:ad56}) gives 

\begin{equation}\label{eqn:ad80}
\hat{n}_{c,tot} = \frac{\Gamma_{\Phi}\phi_{1i}\phi_{2i}m_{\Phi}^{2}}{2A}\left(\frac{\phi_{i}}{\phi_{\ast}}\right)\left(\frac{2A}{\Gamma_{\Phi}m_{\Phi}}\right)^{2} = \frac{A\phi_{i}^{2}\sin 2\theta }{\Gamma_{\Phi}}\left(\frac{\phi_{i}}{\phi_{\ast}}\right).
\end{equation}

\noindent The total physical transferred asymmetry at reheating is then

\begin{multline}\label{eqn:ad81}
\hat{n}_{tot} = \left(\frac{a_{i}}{a_{R}}\right)^{3}\frac{A\phi_{i}^{2}\sin 2\theta }{\Gamma_{\Phi}}\left(\frac{\phi_{i}}{\phi_{\ast}}\right) = \left(\frac{a_{i}}{a_{\ast}}\cdot\frac{a_{\ast}}{a_{R}}\right)^{3}\left(\frac{\phi_{i}}{\phi_{\ast}}\right)\frac{A\phi_{i}^{2}\sin 2\theta }{\Gamma_{\Phi}} \\
= \left(\frac{\phi_{\ast}}{\phi_{i}}\right)^{3}\left(\frac{\phi_{R}}{\phi_{\ast}}\right)^{2}\left(\frac{\phi_{i}}{\phi_{\ast}}\right)\frac{A\phi_{i}^{2}\sin 2\theta }{\Gamma_{\Phi}} = \left(\frac{\phi_{R}}{\phi_{i}}\right)^{2}\frac{A\phi_{i}^{2}\sin 2\theta }{\Gamma_{\Phi}}.
\end{multline}

\noindent Using $V = m_{\Phi}^{2}\phi_{R}^{2}/2 = 3M_{pl}^{2}H_{R}^{2}$, and  $H_{R} = \Gamma_{\Phi}$, we can rewrite the total transferred asymmetry at reheating as

\begin{equation}\label{eqn:ad82}
\hat{n}_{tot} = \frac{6M_{pl}^{2}A \sin 2\theta \Gamma_{\Phi}}{m_{\Phi}^{2}}.
\end{equation}

\noindent For $\tau_{\Phi} < T_{asy}$, dividing by (\ref{eqn:ad70}) we find the predicted baryon-to-entropy ratio to be

\begin{equation}\label{eqn:ad83}
\frac{n_{B}}{s} = \frac{3A M_{pl} \sin 2\theta}{2 k_{T_{R}}T_{R}m_{\Phi}^{2}}.
\end{equation}

\noindent Equating this to $\left(n_{B}/s \right)_{obs}$, (\ref{eqn:ad73}), we find 

\begin{equation}\label{eqn:ad84}
\frac{A^{\frac{1}{2}}}{m_{\Phi}} = 9.05 \times 10^{-11} \left(\frac{T_{R}}{10^{8}\GeV}\right)^{\frac{1}{2}}\left(\frac{1}{\sin 2\theta}\right)^{\frac{1}{2}},
\end{equation}

\noindent is needed in order to account for the observed value of the baryon-to-entropy ratio in the case $\tau_{\Phi} < T_{asy}$. 

This value is typically much smaller than (\ref{eqn:ad74}), where we considered a condensate lifetime much larger than the oscillation period of the asymmetry. Physically this makes sense because for  $\tau_{\Phi} < T_{asy}$, the condensate decays before any oscillations of the inflaton condensate asymmetry take place, therefore there is no additional suppression of the final baryon asymmetry due to averaging over inflaton condensate oscillations. Therefore a greater suppression of $A^{\frac{1}{2}}/m_{\Phi}$ is required in this case to account for the observed baryon asymmetry. 

We can conclude that it is possible to produce the baryon-to-entropy ratio observed today from the model of Affleck-Dine baryogenesis via quadratic inflaton symmetry-breaking terms.

Dynamically this version of Affleck-Dine baryogenesis is quite different from conventional Affleck-Dine baryogenesis based on cubic or higher order terms, where a constant asymmetry is generated in the condensate, with the higher order B-violating potential terms becoming negligible relative to the $\Phi^{2}$ terms at late times.

\section{Validity of the Threshold Approximation}\label{section:36}
Thus far we have used a threshold approximation in order to derive the $B$-asymmetry generated in this model, and to predict the baryon-to-entropy ratio which could be produced as a result. This assumes that there is no significant asymmetry generated for $\phi > \phi_{\ast}$, which in turn assumes that the $A$ terms in the inflaton potential are not dynamically significant in the generation of the condensate asymmetry during the $\left|\Phi\right|^{4}$ dominated evolution. 

To study this we need to consider the dynamics of the angular field, $\theta$. If the inflaton phase oscillates at a frequency greater than the Hubble rate when the potential is $\left|\Phi\right|^{4}$ dominated then the angular field becomes dynamical and could affect the validity of the threshold approximation. We therefore need to examine the phase dynamics of the inflaton field to establish whether they could significantly impact the evolution of the asymmetry. We will first consider this analytically and then check the condition numerically.

The potential of the theory is

\begin{equation}\label{eqn:ad85}
V\left(\Phi \right) = m_{\Phi}^{2}\left| \Phi \right|^{2} + \lambda_{\Phi}\left|\Phi \right|^{4} - A\left(\Phi^{2} + \Phi^{\dagger^{2}} \right).
\end{equation}

\noindent Rewriting the field as $\Phi = \phi e^{i\theta}/\sqrt{2}$, this becomes

\begin{equation}\label{eqn:ad86}
V\left(\phi \right) = \frac{1}{2}m_{\Phi}^{2}\phi^{2} + \frac{\lambda_{\Phi}}{4}\phi^{4} - \frac{A\phi^{2}}{2}\left[ e^{i2\theta} + e^{-i2\theta}\right] = \frac{1}{2}m_{\Phi}^{2}\phi^{2} + \frac{\lambda_{\Phi}}{4}\phi^{4} - A\phi^{2}\cos 2\theta,
\end{equation}

\noindent where $\theta = 0$ corresponds to no asymmetry, and the potential has a minimum at $\theta = 0$. This shows that the symmetry breaking term introduces an oscillating phase to the inflaton field, and this can cause the Affleck-Dine field to undergo damped oscillations along an elliptical path in field space, which can affect the evolution of the asymmetry. If the field has zero phase, then the field evolves exclusively in the radial direction towards the minimum of its potential. 

Since the field naturally evolves along its minimum, we can treat the phase $\theta$ as a perturbation about zero. If $\theta$ is small then we can write

\begin{equation}\label{eqn:ad87}
\cos 2\theta \approx 1 - \frac{1}{2}\left(2\theta \right)^{2} = 1 - 2\theta^{2},
\end{equation}

\noindent making the potential (\ref{eqn:ad86})

\begin{equation}\label{eqn:ad88}
V\left(\phi \right) = \frac{1}{2}m_{\Phi}^{2}\phi^{2} + \frac{\lambda_{\Phi}}{4}\phi^{4} - A\phi^{2}\left(1 - 2\theta^{2} \right),
\end{equation}

\noindent which enables us to define the $\theta$-potential as

\begin{equation}\label{eqn:ad89}
V_{\theta} = 2A\phi^{2} \theta^{2}.
\end{equation}

\noindent The derivative term in the inflaton Lagrangian density in terms of the radial and angular fields is

\begin{equation}\label{eqn:ad90}
\mathcal{L}_{K} =\partial_{\mu}\Phi^{\dagger}\partial^{\mu}\Phi = \frac{1}{2}\left[\partial_{\mu}\phi \partial^{\mu}\phi + \phi^{2}\partial_{\mu}\theta \partial^{\mu}\theta \right],
\end{equation}

\noindent and it is therefore possible to write a $\theta$ Lagrangian density which describes the dynamics of the angular component of the complex field

\begin{equation}\label{eqn:ad91}
\mathcal{L}_{\theta} = \frac{\phi^{2}}{2}\partial_{\mu}\theta \partial^{\mu}\theta - 2A\phi^{2} \theta^{2}.
\end{equation}

\noindent It is difficult to discern the effect of the $\theta$ field on its own, since the dynamics of the angular component are coupled to the radial field. A way to approximately remedy this is to average over the $\phi$ oscillations such that

\begin{equation}\label{eqn:ad92}
\phi^{2} \longrightarrow \langle \phi^{2} \rangle,
\end{equation}

\begin{equation}\label{eqn:ad93}
\Rightarrow \mathcal{L}_{\theta} = \frac{\langle \phi^{2}\rangle}{2}\partial_{\mu}\theta \partial^{\mu}\theta - 2A\langle \phi^{2}\rangle \theta^{2}.
\end{equation}

\noindent This is a reasonable approximation to make if $\theta$ doesn't change very much during one $\phi$ oscillation. Assuming this is satisfied, we can treat $\langle \phi^{2} \rangle$ as a constant, where we neglect the effects of the expansion for now. We then define the perturbations of the angular field about the minimum as

\begin{equation}\label{eqn:ad94}
\delta \theta = \langle \phi^{2} \rangle^{\frac{1}{2}} \theta.
\end{equation}

\noindent The Lagrangian density describing the perturbations of the angular field is then

\begin{equation}\label{eqn:ad95}
\mathcal{L}_{\delta \theta} = \frac{1}{2}\partial_{\mu}\delta \theta \partial^{\mu}\delta \theta - 2A \delta \theta^{2},
\end{equation}

\noindent where the second term is a mass term of the $\theta$-perturbations, and the mass of these perturbations is given by

\begin{equation}\label{eqn:ad96}
m_{\delta \theta} = 2\sqrt{A}.
\end{equation}

$m_{\delta \theta}$ gives the angular frequency of the oscillation of the phase field. For $m_{\delta \theta} \gtrsim H$ the phase field becomes dynamical when the radial $\Phi$ oscillations are $\left|\Phi\right|^{4}$ dominated. Below this threshold the angular component does not contribute significantly to the evolution of the field. Therefore $2\sqrt{A} < H$ gives the condition for the $\left|\Phi\right|^{4}$-dominated regime to not become significant in the generation of the asymmetry, such that the phase $\theta$ does not evolve until the potential is solidly in the $\Phi^{2}$ regime. 

Assuming $A/m_{\Phi}^{2} <<1$, the potential at the threshold $\left( \phi = \phi_{\ast} = m_{\Phi}/\sqrt{\lambda_{\Phi}}\right)$  can be written as

\begin{equation}\label{eqn:ad97}
V\left(\phi_{\ast} \right) = \frac{1}{2}m_{\Phi}^{2}\phi_{\ast}^{2} + \frac{\lambda_{\Phi}}{4}\phi_{\ast}^{4} = \frac{3 m_{\Phi}^{4}}{4\lambda_{\Phi}},
\end{equation}

\noindent and the Hubble parameter at the threshold is 

\begin{equation}\label{eqn:ad98}
H_{\ast}^{2} = \frac{V_{\ast}}{3 M_{pl}^{2}} = \frac{m_{\Phi}^{4}}{4 M_{pl}^{2} \lambda_{\Phi}}.
\end{equation}

\noindent In order for the $\theta$ field to not become dynamical until $\phi < \phi_{\ast}$ we therefore require

\begin{equation}\label{eqn:ad99}
m_{\delta \theta} = 2\sqrt{A} < H = \frac{m_{\Phi}^{2}}{2 M_{pl} \sqrt{\lambda_{\Phi}}},
\end{equation}

\noindent which can be rewritten as the constraint on the ratio $A^{\frac{1}{2}}/m_{\Phi}$

\begin{equation}\label{eqn:ad100}
\frac{A^{\frac{1}{2}}}{m_{\Phi}} <  \frac{A_{th}^{\frac{1}{2}}}{m_{\Phi}} = \frac{m_{\Phi}}{4 M_{pl} \sqrt{\lambda_{\Phi}}}.
\end{equation}

\noindent The value of this ratio at the threshold, $A_{th}^{\frac{1}{2}}/m_{\Phi}$, can be normalised with respect to the inflaton mass. Using $m_{\Phi} = 10^{13}\GeV$ for consistency we have

\begin{equation}\label{eqn:ad101}
\frac{A^{\frac{1}{2}}}{m_{\Phi}} <  \frac{A_{th}^{\frac{1}{2}}}{m_{\Phi}} = 1.04 \times 10^{-6} \lambda_{\Phi}^{-\frac{1}{2}}\left(\frac{m_{\Phi}}{10^{13}\GeV}\right).
\end{equation}

\noindent This gives the condition for the analytic threshold approximation for the baryon asymmetry to be valid. This can easily be compatible with the established constraints on the ratio $A^{\frac{1}{2}}/m_{\Phi}$, (\ref{eqn:ad74}) and (\ref{eqn:ad84}), in order for the quadratic symmetry-breaking term to produce an asymmetry sufficient enough to result in the observed baryon-to-entropy ratio. This Affleck-Dine model with quadratic symmetry-breaking terms can therefore produce a sufficient baryon asymmetry - both in the case of averaging the generated asymmetry over many oscillations and in the case of rapid condensate decay - which is well described by a threshold approximation within the parameter space of $A^{\frac{1}{2}}/m_{\Phi}$ required by the observed baryon-to-entropy ratio today.

In Figure \ref{figure:31} we show the results of a comparison between the threshold approximation without decay for the condensate asymmetry (\ref{eqn:ad48}) (left panel) and the transferred asymmetry (\ref{eqn:ad50}) (right panel) and numerical calculations at different values of $A^{\frac{1}{2}}/m_{\Phi}$ for the condensate and transferred asymmetries in the limit that $\Gamma_{\Phi}\left(t - t_{\ast}\right) < < 1$, such that the time elapsed studying the asymmetries is less than the lifetime of the scalars and there is no significant decay of the condensate itself. The numerical condensate asymmetry oscillates about zero, as the threshold approximation corresponding to (\ref{eqn:ad48}) does, which corresponds to the phase rotation of the $\Phi$ scalars within the condensate.

From the right panel of Figure \ref{figure:31} we can see that the threshold transferred asymmetry oscillates with minima at zero without crossing zero, as outlined in Section \ref{section:344} for (\ref{eqn:ad50}). We can see that for increasing $A^{\frac{1}{2}}/m_{\Phi}$, the amplitude of the transferred asymmetry, $A\hat{n}_{c}$, decreases in amplitude with an increasing $A > A_{th}$, down to about $A^{\frac{1}{2}}/m_{\Phi} \approx 0.01$ where the amplitude becomes constant at a value about a factor of ten smaller than the amplitude of the threshold approximation of the transferred asymmetry. For $m_{\Phi} = 10^{16}\GeV$ and $\lambda_{\Phi} = 0.1$, from (\ref{eqn:ad100}) we find that $A_{th}^{\frac{1}{2}}/m_{\Phi} = 0.003$, and that for $A^{\frac{1}{2}}/m_{\Phi} = 0.001$ the numerical solutions are, as expected, in perfect agreement with the threshold approximation for both the condensate and the transferred asymmetries, which confirms that the threshold approximation of the asymmetries is valid within the constraint of (\ref{eqn:ad100}). This shows that for larger $A^{\frac{1}{2}}/m_{\Phi}$, the asymmetries are modified by the dynamics of the $\left| \Phi \right|^{4}$ dominated era of the potential, and we can see from the right panel of Figure \ref{figure:31} that for larger $A^{\frac{1}{2}}/m_{\Phi}$ the transferred asymmetry does cross zero while it oscillates. However, since the transferred asymmetry corresponds to the total asymmetry transferred to the Standard Model averaged over the decay of the condensate asymmetry, this shouldn't have an effect on the size of the baryon asymmetry transferred to the Standard Model, similar to the case of the transferred asymmetry in the threshold approximation. This is because the asymmetry will average to a small but finite number over many oscillations as the excess particle and antiparticle decay products will annihilate, leaving a small excess of baryons.

\section{Numerical Test and Results}\label{section:37}
We numerically tested the analytical prediction of the generation of the asymmetry and the resulting baryon-to-entropy ratio, in order to confirm the analytical results, and test the validity of the approximations used in deriving the asymmetry analytically. The numerical computations were done using a Fortran code which solves the field equations including terms corresponding to the decay of the fields $\phi_{1,2}$

\begin{equation}\label{eqn:ad102}
\ddot{\phi}_{1} + 3H\dot{\phi}_{1} + \Gamma_{\Phi}\dot{\phi}_{1} = -m_{1}^{2}\phi_{1},
\end{equation}

\begin{equation}\label{eqn:ad103}
\ddot{\phi}_{2} + 3H\dot{\phi}_{2} + \Gamma_{\Phi}\dot{\phi}_{2} = -m_{2}^{2}\phi_{2}.
\end{equation}

\noindent The asymmetry is calculated using $n\left(t \right) = \dot{\phi}_{1}\phi_{2} - \dot{\phi}_{2}\phi_{1}$ directly from the field equations, rather than by substituting the analytic solutions for $\phi_{1,2}$ (\ref{eqn:ad18}), (\ref{eqn:ad19}) and using the threshold approximation.

The main physical assumption made in the analytical calculation of the asymmetry, which is treated differently in the numerical computation, is in the calculation of the reheating temperature, $T_{R}$, from the approximation of instantaneous reheating. When deriving the baryon-to-entropy ratio from the transferred asymmetry, $\hat{n}$, analytically, the working assumption is that the condensate decays instantly to radiation at $t = t_{R}$, at which point $\Gamma_{\Phi} = H\left(T_{R}\right)$, and the reheating temperature is calculated using this assumption. The final comoving transferred asymmetry is therefore assumed to originate at $\Gamma_{\Phi} = H\left(T_{R}\right)$, and the baryon-to-entropy ratio is calculated using the entropy density at reheating (\ref{eqn:ad70}). In actuality, the condensate will decay over a time $\Delta t \sim \Gamma_{\Phi}^{-1}$, and in the numerical computation the reheating temperature is calculated exactly from the radiation density from the inflaton decay. We find that this has a small impact on the final $n_{B}/s$.

\subsection{Numerical Treatment of Radiation Density}\label{section:371}

In the numerical treatment of the asymmetry we take into account the loss of the energy density of radiation due to the decay of the condensate asymmetry. In this section, we demonstrate how to do this.

For a general potential $V\left(\phi_{1}, \phi_{2}\right)$, we have that the field equations for the scalars $\phi_{1,2}$ are

\begin{equation}\label{eqn:ad104}
\ddot{\phi}_{1} + 3H\dot{\phi}_{1} + \Gamma_{\Phi}\dot{\phi}_{1} = -\frac{\partial V}{\partial \phi_{1}},
\end{equation}

\begin{equation}\label{eqn:ad105}
\ddot{\phi}_{2} + 3H\dot{\phi}_{2} + \Gamma_{\Phi}\dot{\phi}_{2} = -\frac{\partial V}{\partial \phi_{2}}.
\end{equation}

\noindent The energy density of a field (\ref{eqn:ad4}) is given by

\begin{equation}\label{eqn:ad106}
\rho_{\Phi} = \frac{1}{2}\dot{\phi}_{1}^{2} + \frac{1}{2}\dot{\phi}_{2}^{2} + V\left(\phi_{1}, \phi_{2}\right),
\end{equation}

\noindent and the derivative of the energy density of the scalar field with respect to time is thus

\begin{equation}\label{eqn:ad107}
\frac{d\rho_{\Phi}}{dt} = \dot{\phi_{1}}\ddot{\phi_{1}} + \dot{\phi_{2}}\ddot{\phi_{2}} + \frac{\partial V}{\partial \phi_{1}}\dot{\phi_{1}} + \frac{\partial V}{\partial \phi_{2}}\dot{\phi_{2}}.
\end{equation}

\noindent Rearranging (\ref{eqn:ad104}), (\ref{eqn:ad105}) in terms of $\ddot{\phi}_{1,2}$ and substituting into the change in energy density of the scalar field over time (\ref{eqn:ad107}) we obtain

\begin{equation}\label{eqn:ad108}
\frac{d\rho_{\Phi}}{dt} = -3H\left(\dot{\phi}_{1}^{2} + \dot{\phi}_{2}^{2}\right) - \Gamma_{\Phi}\left(\dot{\phi}_{1}^{2} + \dot{\phi}_{2}^{2}\right),
\end{equation}

\noindent where the first term on the right-hand side corresponds to the loss of energy density of the scalar field due to expansion of the Universe, and the second term gives the loss of energy density due to the decay of the scalar condensate.

The change in the energy density of the condensate due to the decay of the inflaton scalars in our model is therefore

\begin{equation}\label{eqn:ad109}
\frac{d\rho_{\Phi}}{dt} = -\Gamma_{\Phi}\left(\dot{\phi}_{1}^{2} + \dot{\phi}_{2}^{2}\right),
\end{equation}

\noindent which means that the increase in the energy density of radiation due to the decay of the condensate to radiation is 

\begin{equation}\label{eqn:ad110}
\frac{d\rho_{r}}{dt} = \Gamma_{\Phi}\left(\dot{\phi}_{1}^{2} + \dot{\phi}_{2}^{2}\right),
\end{equation}

\noindent where the energy density of radiation decays away due to the expansion of the Universe as

\begin{equation}\label{eqn:ad111}
\rho_{r} \propto \frac{1}{a^{4}} \Rightarrow \frac{d\rho_{r}}{da} = -\frac{4}{a}\rho_{r} \Rightarrow \frac{d\rho_{r}}{dt} = -4H\rho_{r}.
\end{equation}

\noindent The total change in the energy density of radiation, taking into account both the increase due to the decay of the condensate and the dilution due to the expansion of the Universe is therefore

\begin{equation}\label{eqn:ad112}
 \frac{d\rho_{r}}{dt} = \Gamma_{\Phi}\left(\dot{\phi}_{1}^{2} + \dot{\phi}_{2}^{2}\right) - 4H\rho_{r}.
\end{equation}

The energy density of radiation is thus computed exactly from the field equations, and the temperature as a function of time can be obtained from this via (\ref{eqn:126})

\begin{equation}\label{eqn:ad113}
\rho_{r} = \frac{\pi^{2}}{30}g\left(T \right) T^{4}.
\end{equation}

As we did in the analytical calculation, the baryon asymmetry can be equated to the total transferred asymmetry, and the baryon-to-entropy ratio can be calculated as a function of time. In the next section we will compare the numerical calculations of the baryon-to-entropy ratio with our analytical estimates in the limits $\tau_{\Phi} = T_{asy}$ and $\tau_{\Phi} > T_{asy}$.

\subsection{Numerical Computation of the Baryon-to-Entropy Ratio for Large and Small Condensate Lifetimes}\label{section:372}

In this section we present the results of the numerical computation of the baryon-to-entropy ratio from the transferred asymmetry obtained directly from the field equations, for both the case of the lifetime of the $\Phi$ scalars being equal to the oscillation period of the asymmetry, and for the case of the lifetime of the $\Phi$ scalars being larger than the oscillation period of the asymmetry.

\subsubsection{Numerical Test Case $\tau_{\Phi} = T_{asy}$}

We first test the case where 

\begin{equation}\label{eqn:ad114}
\tau_{\Phi} = T_{asy} = \frac{\pi m_{\Phi}}{A},
\end{equation}

\noindent corresponding to the inflaton condensate decaying away completely within a single oscillation of the field.

\noindent For values $m_{\Phi} = 10^{16}\GeV$, $A^{\frac{1}{2}} = 10^{13}\GeV$, and $\sin 2\theta =1$, the predicted decay rate is 

\begin{equation}\label{eqn:ad115}
\Gamma_{\Phi} = \frac{1}{\tau_{\Phi}} = \frac{1}{T_{asy}} = \frac{A}{\pi m_{\Phi}} = 3.18 \times 10^{9}\GeV,
\end{equation}

\noindent the predicted reheating temperature is

\begin{equation}\label{eqn:ad116}
T_{R} = k_{T_{R}}^{-\frac{1}{2}}\left(\Gamma_{\Phi}M_{pl}\right)^{\frac{1}{2}} = 4.72 \times 10^{13}\GeV,
\end{equation}

\noindent and using (\ref{eqn:ad71}) the predicted baryon-to-entropy ratio is

\begin{equation}\label{eqn:ad117}
\frac{n_{B}}{s} = 5.62 \times 10^{-4}.
\end{equation}

\noindent The numerical computation gives a baryon-to-entropy ratio for $\tau_{\Phi} = T_{asy}$ with $m_{\Phi} = 10^{16}\GeV$, $A^{\frac{1}{2}} = 10^{13}\GeV$, and $\sin 2\theta =1$ of

\begin{equation}\label{eqn:ad118}
\frac{n_{B}}{s} = 5.28 \times 10^{-4},
\end{equation}

\noindent which is a very good agreement with the analytical prediction.

\begin{figure}[H]
\begin{center}
\includegraphics[clip = true, width=\textwidth, angle = 360]{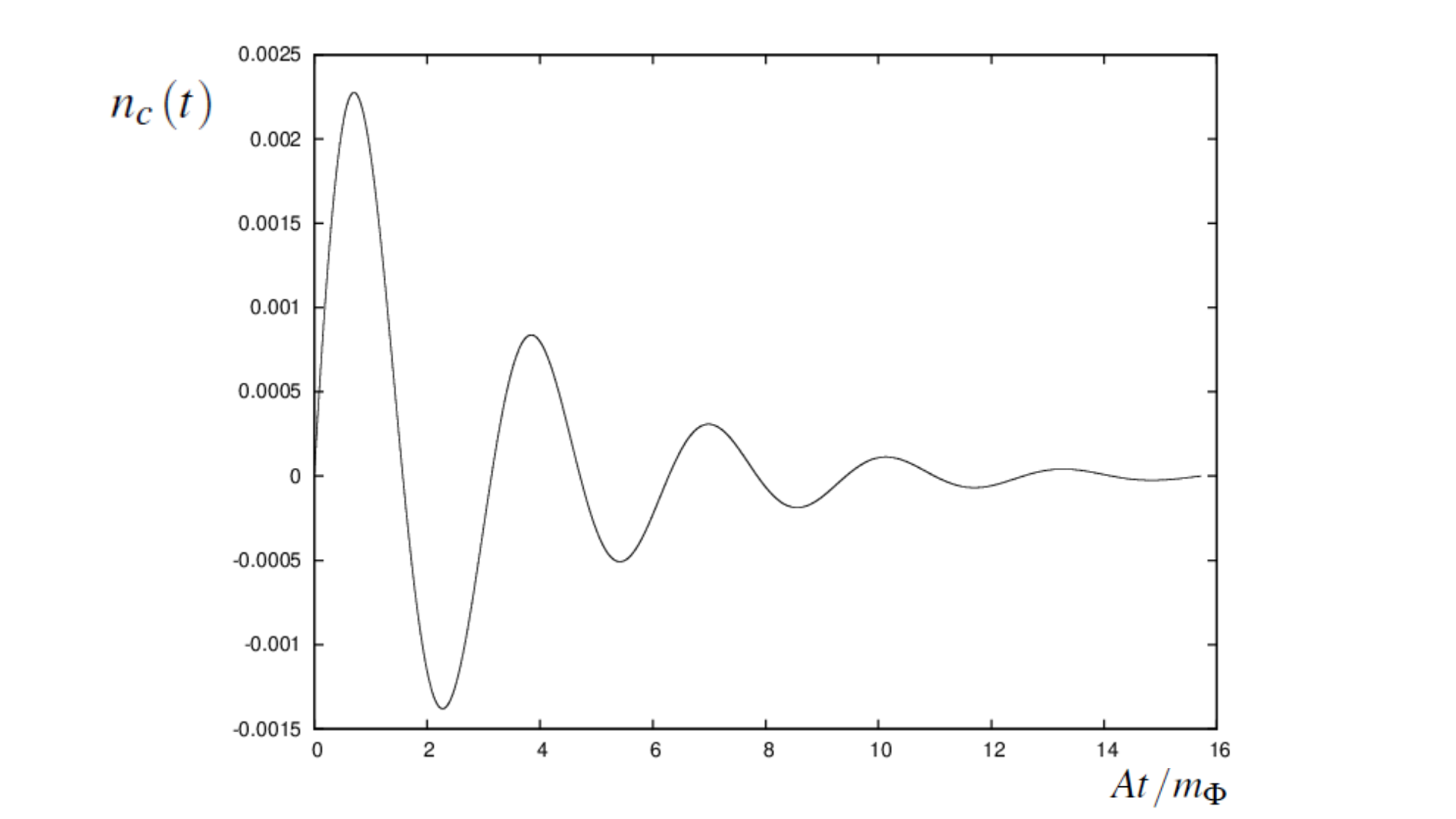}
\caption{Numerically calculated comoving condensate asymmetry, $n_{c}$, vs. $At/m_{\Phi}$ for $\tau_{\Phi} = T_{asy}$. } 
\label{figure:32}
\end{center}
\end{figure}

\begin{figure}[H]
\begin{center}
\includegraphics[clip = true, width=\textwidth, angle = 360]{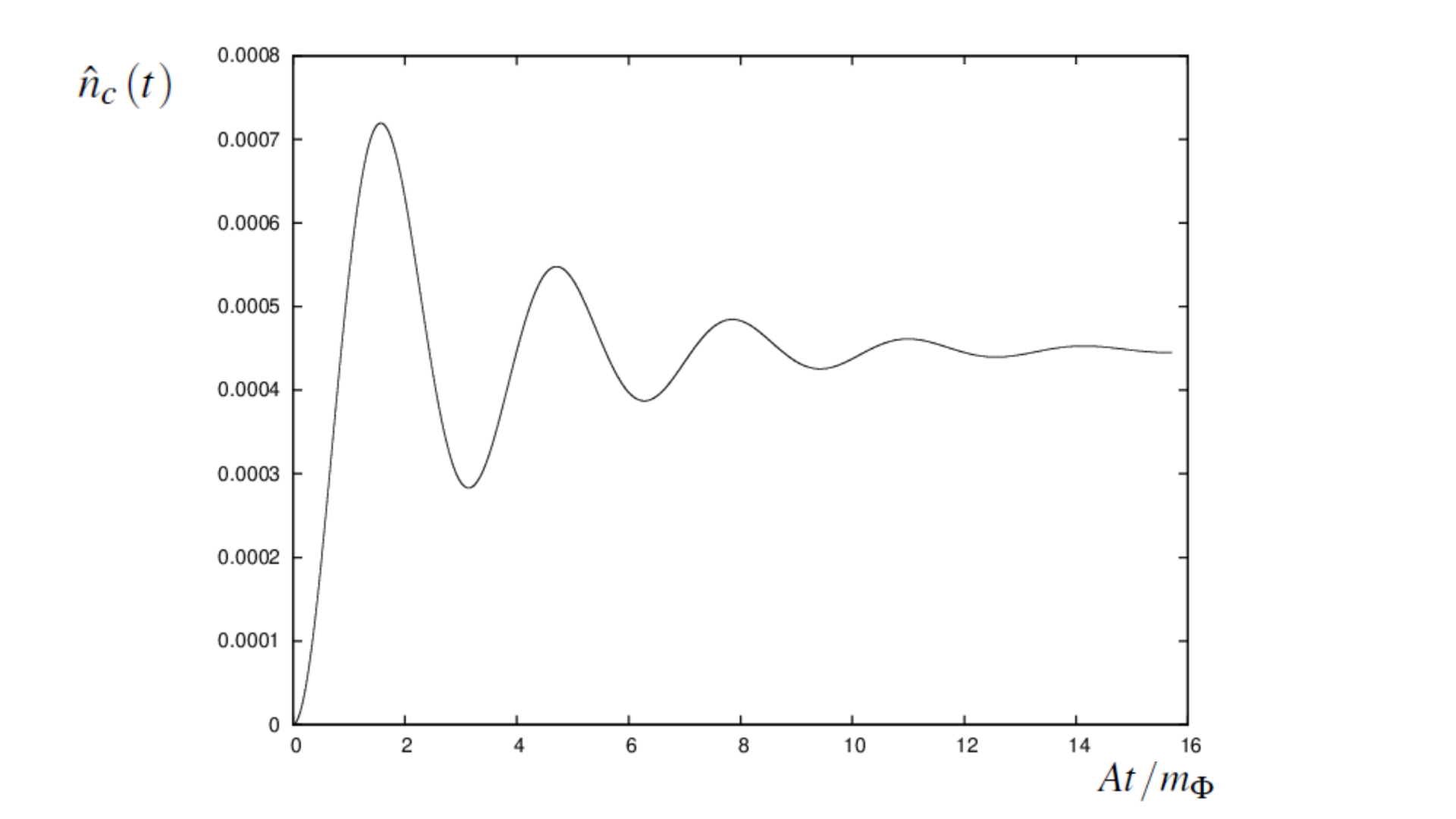}
\caption{Numerically calculated comoving transferred asymmetry, $\hat{n}_{c}$, vs. $At/m_{\Phi}$ for $\tau_{\Phi} = T_{asy}$. } 
\label{figure:33}
\end{center}
\end{figure}

\begin{figure}[H]
\begin{center}
\includegraphics[clip = true, width=\textwidth, angle = 360]{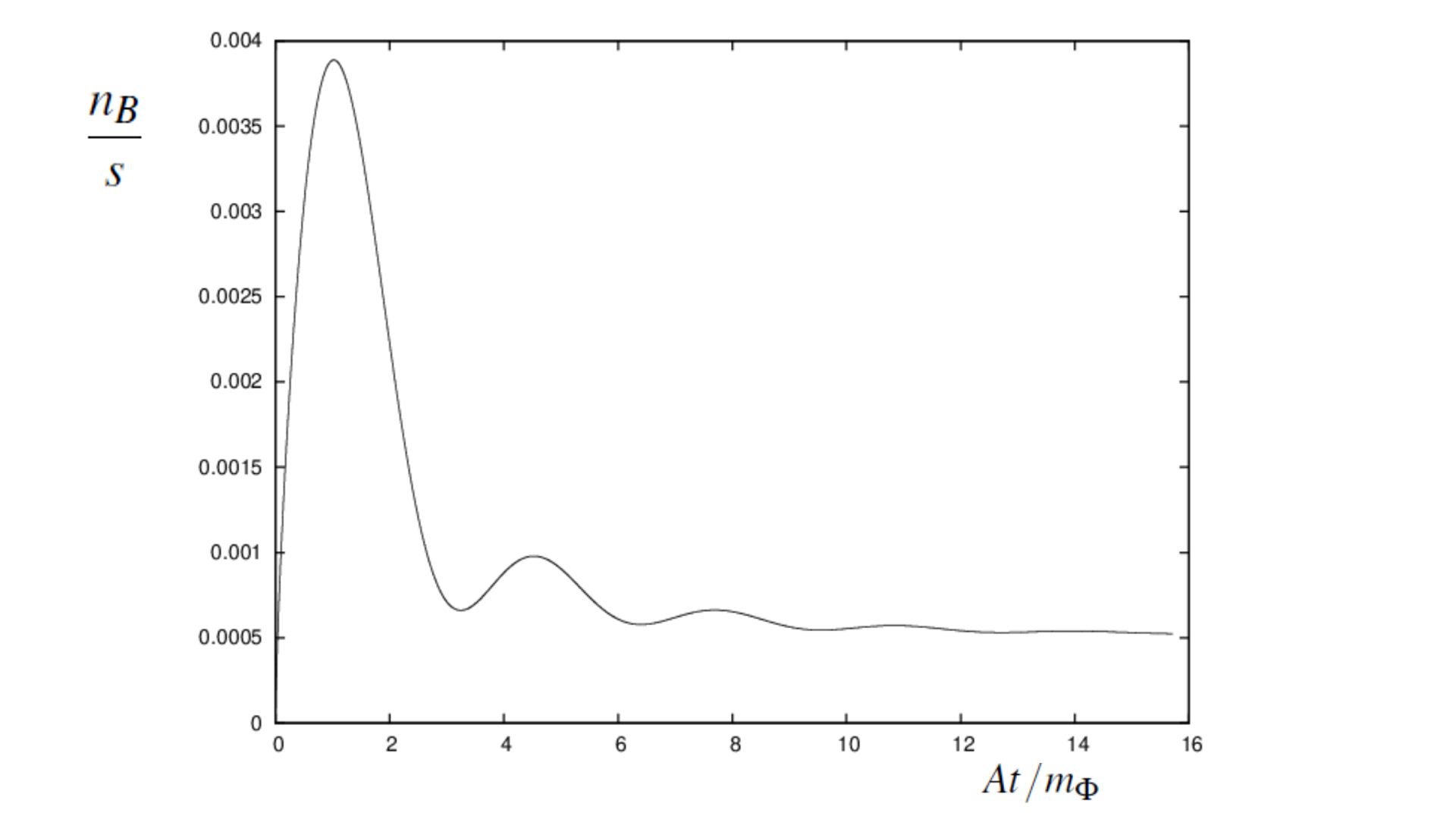}
\caption{Numerical baryon-to-entropy ratio vs. $At/m_{\Phi}$ for $\tau_{\Phi} = T_{asy}$. } 
\label{figure:34}
\end{center}
\end{figure}

Figures \ref{figure:32} - \ref{figure:34} show the evolution of the baryon-to-entropy ratio, the comoving condensate asymmetry and the transferred asymmetry over time in the  $\tau_{\Phi} = T_{asy}$ scenario.

As we can see from Figure \ref{figure:32}, the condensate asymmetry oscillates about zero with decreasing amplitude - corresponding to the inflaton field undergoing a phase oscillation between its $\Phi$ and $\Phi^{\dagger}$ states while the scalars themselves are decaying, until the condensate has completely decayed and the condensate asymmetry is therefore zero. This occurs over a small number of phase oscillations of the field, and therefore a small number of oscillations of the asymmetry.

From Figure \ref{figure:33} we can see that the transferred asymmetry increases as an asymmetry is initially generated and then oscillates with the oscillation of the condensate asymmetry while the inflaton scalars decay - transferring the asymmetry to the Standard Model - until all of the condensate has decayed, and the transferred asymmetry settles at a small, positive, finite value, giving the baryon-to-entropy ratio shown in Figure \ref{figure:34}.

\begin{figure}[H]
\begin{center}
\includegraphics[clip = true, width=\textwidth, angle = 360]{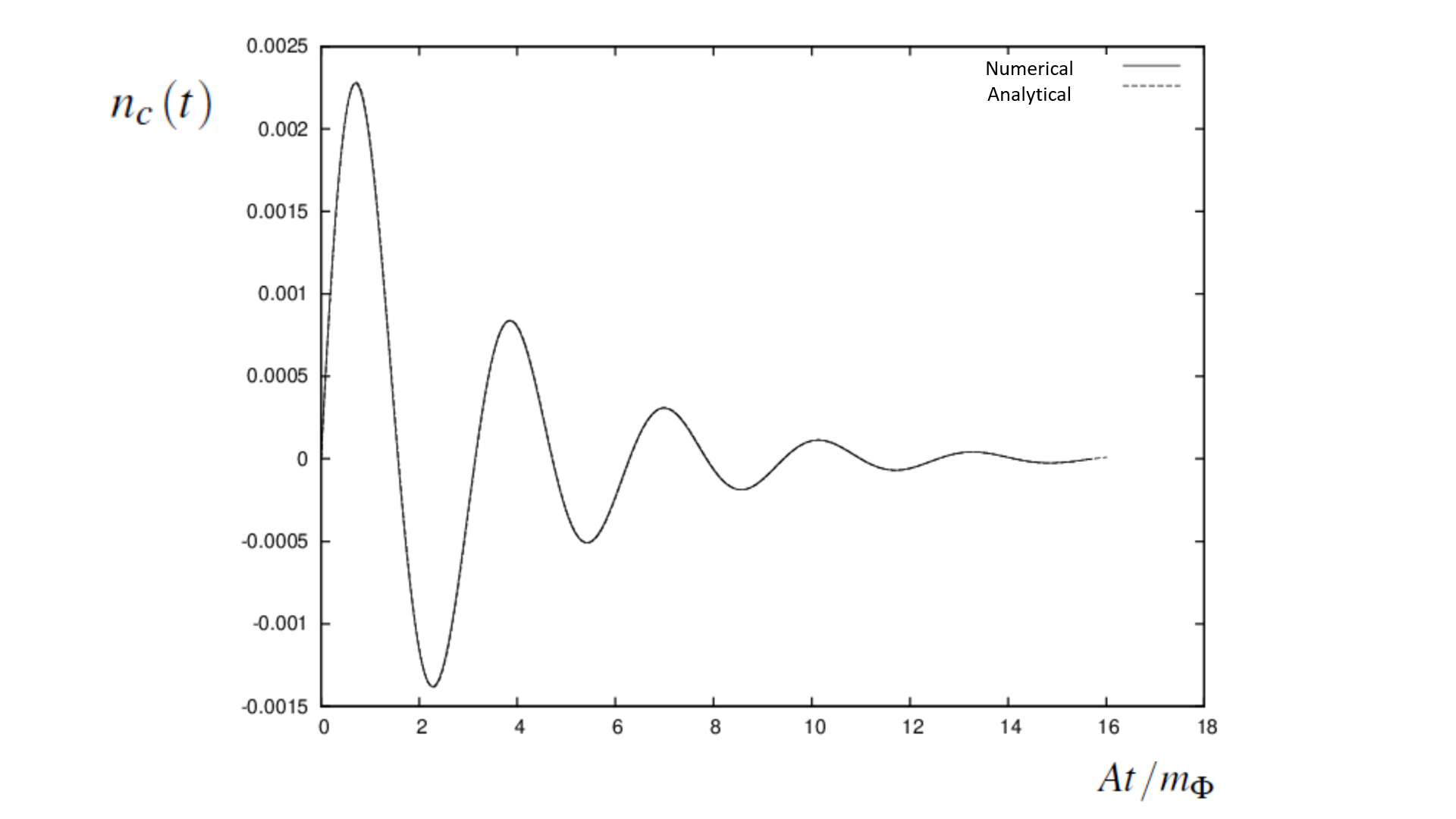}
\caption{Comparison of numerically calculated comoving condensate asymmetry with the analytical estimate from (\ref{eqn:ad119}), $n_{c}$, vs. $At/m_{\Phi}$ for $\tau_{\Phi} = T_{asy}$. } 
\label{figure:35}
\end{center}
\end{figure} 

\begin{figure}[H]
\begin{center}
\includegraphics[clip = true, width=\textwidth, angle = 360]{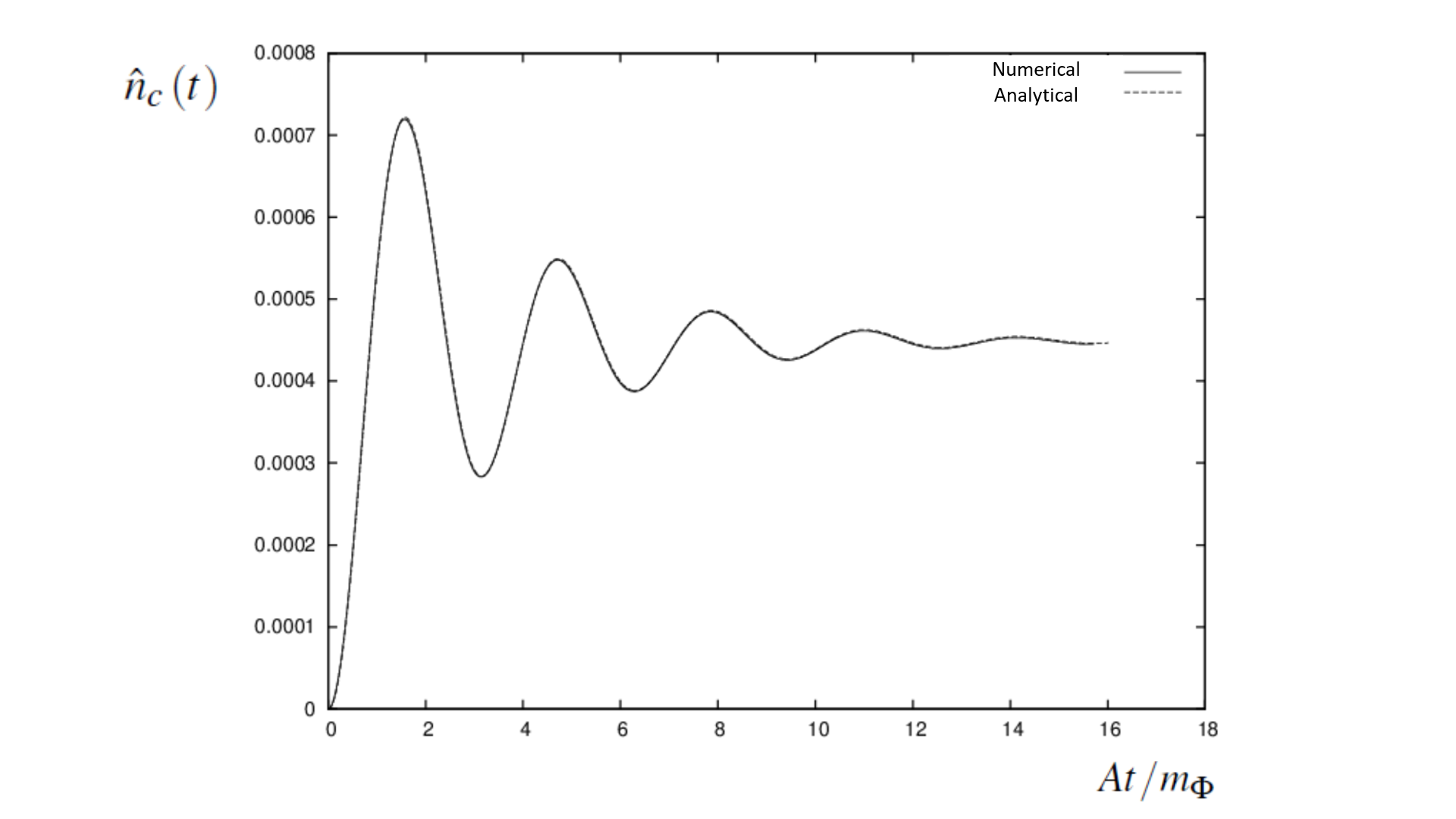}
\caption{Comparison of numerically calculated comoving transferred asymmetry with the analytical estimate from (\ref{eqn:ad51}), $\hat{n}_{c}$, vs. $At/m_{\Phi}$ for $\tau_{\Phi} = T_{asy}$. } 
\label{figure:36}
\end{center}
\end{figure} 

From the analysis of Figures \ref{figure:35} and \ref{figure:36}, we find that the numerical condensate and transferred asymmetries are in perfect agreement with the analytical integral (\ref{eqn:ad51}) with the comoving condensate asymmetry including decay given by (\ref{eqn:ad48}) multiplied by the decay factor $\exp(-\Gamma_{\Phi}(t - t_{\ast})$

\begin{equation}\label{eqn:ad119}
n_{c}\left(t \right) = \phi_{1i}\phi_{2i}\left(\frac{\phi_{i}}{\phi_{\ast}}\right)m_{\Phi}\exp\left(-\Gamma_{\Phi}(t - t_{\ast}) \right)\sin\left(\frac{2A(t-t_{\ast})}{m_{\Phi}}\right),
\end{equation}

\noindent when decay of the condensate is taken into account. We verify analytically that this treatment of the decay of the condensate when calculating the asymmetries is accurate in Section \ref{section:373}.

\subsubsection{Numerical Test Case $\tau_{\Phi} > T_{asy}$}
We also examine the case where 

\begin{equation}\label{eqn:ad120}
\tau_{\Phi} = 5T_{asy} = \frac{5\pi m_{\Phi}}{A}, 
\end{equation}

\noindent corresponding to the condensate decaying over many oscillations of the inflaton scalars.

\noindent For values $m_{\Phi} = 10^{16}\GeV$, $A^{\frac{1}{2}} = 10^{13}\GeV$, and $\sin 2\theta =1$, the predicted decay rate is 

\begin{equation}\label{eqn:ad121}
\Gamma_{\Phi} = \frac{1}{5T_{asy}} = \frac{A}{5\pi m_{\Phi}} = 6.37 \times 10^{8}\GeV,
\end{equation}

\noindent the predicted reheating temperature is

\begin{equation}\label{eqn:ad122}
T_{R} = 2.11 \times 10^{13}\GeV,
\end{equation}

\noindent and using (\ref{eqn:ad71}) the predicted baryon-to-entropy ratio is

\begin{equation}\label{eqn:ad123}
\frac{n_{B}}{s} = 5.05 \times 10^{-5}.
\end{equation}

\noindent The numerical computation gives a baryon-to-entropy ratio for $\tau_{\Phi} = 5T_{asy}$ with $m_{\Phi} = 10^{16}\GeV$, $A^{\frac{1}{2}} = 10^{13}\GeV$, and $\sin 2\theta =1$ of

\begin{equation}\label{eqn:ad124}
\frac{n_{B}}{s} = 4.84 \times 10^{-5},
\end{equation}

\noindent which is in very good agreement with the analytical prediction.

\begin{figure}[H]
\begin{center}
\includegraphics[clip = true, width=\textwidth, angle = 360]{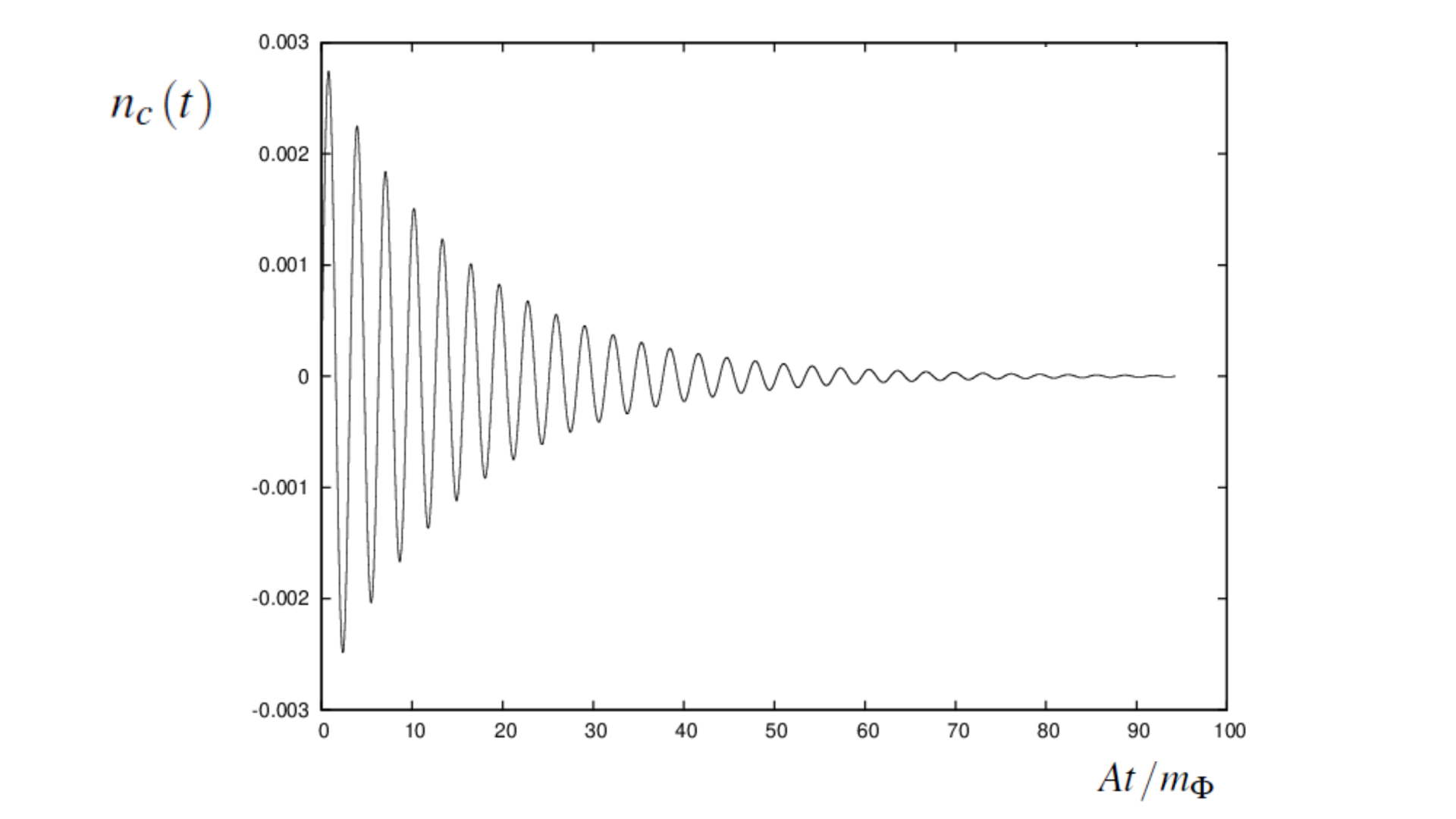}
\caption{Numerically calculated comoving condensate asymmetry, $n_{c}$, vs. $At/m_{\Phi}$ for $\tau_{\Phi} = 5T_{asy}$. } 
\label{figure:37}
\end{center}
\end{figure}

\begin{figure}[H]
\begin{center}
\includegraphics[clip = true, width=\textwidth, angle = 360]{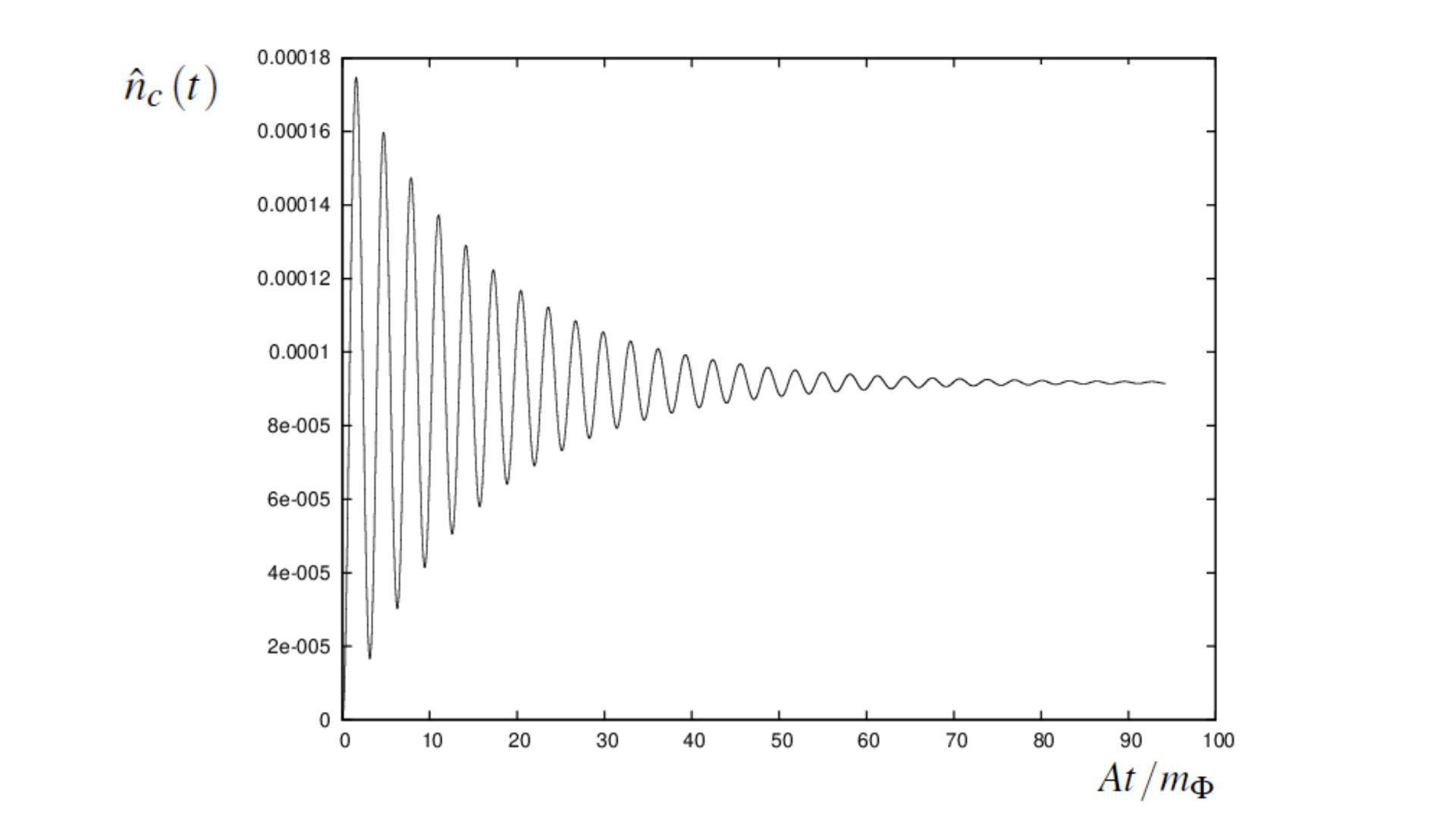}
\caption{Numerically calculated comoving transferred asymmetry, $\hat{n}_{c}$, vs. $At/m_{\Phi}$ for $\tau_{\Phi} = 5T_{asy}$. } 
\label{figure:38}
\end{center}
\end{figure}

\begin{figure}[H]
\begin{center}
\includegraphics[clip = true, width=\textwidth, angle = 360]{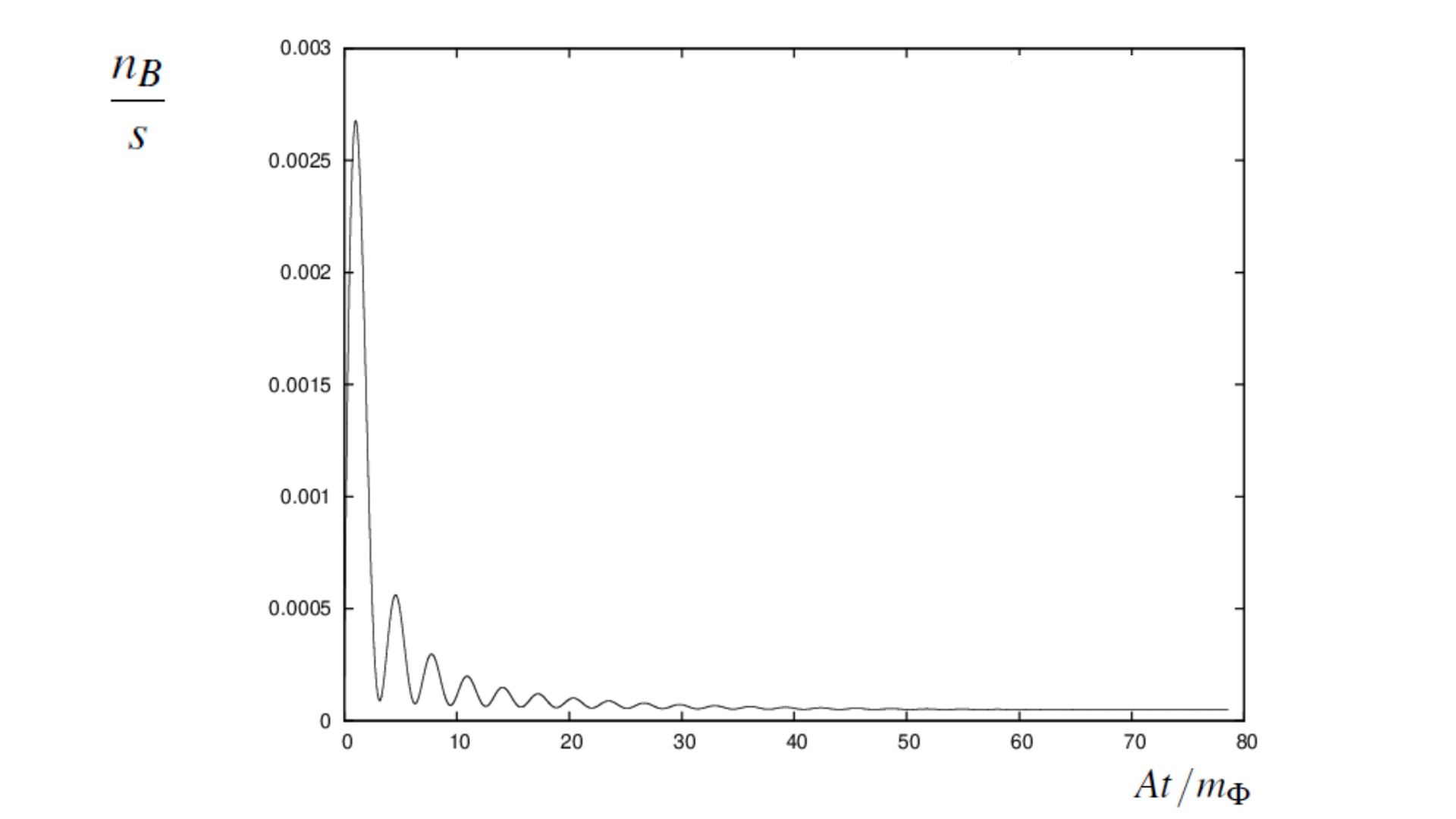}
\caption{Numerical baryon-to-entropy ratio vs. $At/m_{\Phi}$ for $\tau_{\Phi} = 5T_{asy}$. } 
\label{figure:39}
\end{center}
\end{figure}

Figures \ref{figure:37} - \ref{figure:39} show the evolution of the baryon-to-entropy ratio, the comoving condensate asymmetry, and the transferred asymmetry over time in the  $\tau_{\Phi} = 5T_{asy}$ scenario.

\begin{figure}[H]
\begin{center}
\includegraphics[clip = true, width=\textwidth, angle = 360]{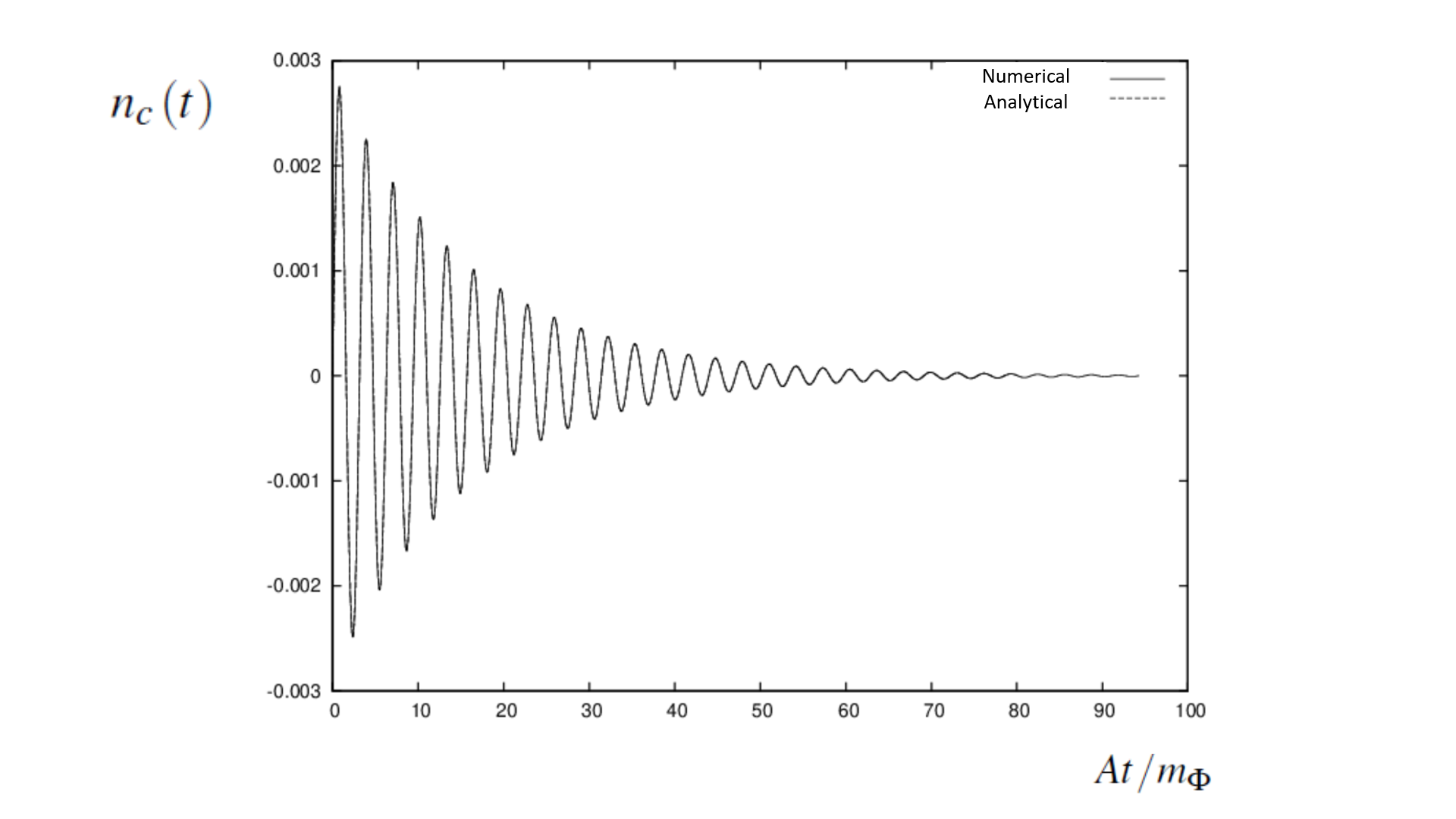}
\caption{Comparison of numerically calculated comoving condensate asymmetry with the analytical estimate from (\ref{eqn:ad119}), $n_{c}$, vs. $At/m_{\Phi}$ for $\tau_{\Phi} = 5T_{asy}$. } 
\label{figure:310}
\end{center}
\end{figure} 

\begin{figure}[H]
\begin{center}
\includegraphics[clip = true, width=\textwidth, angle = 360]{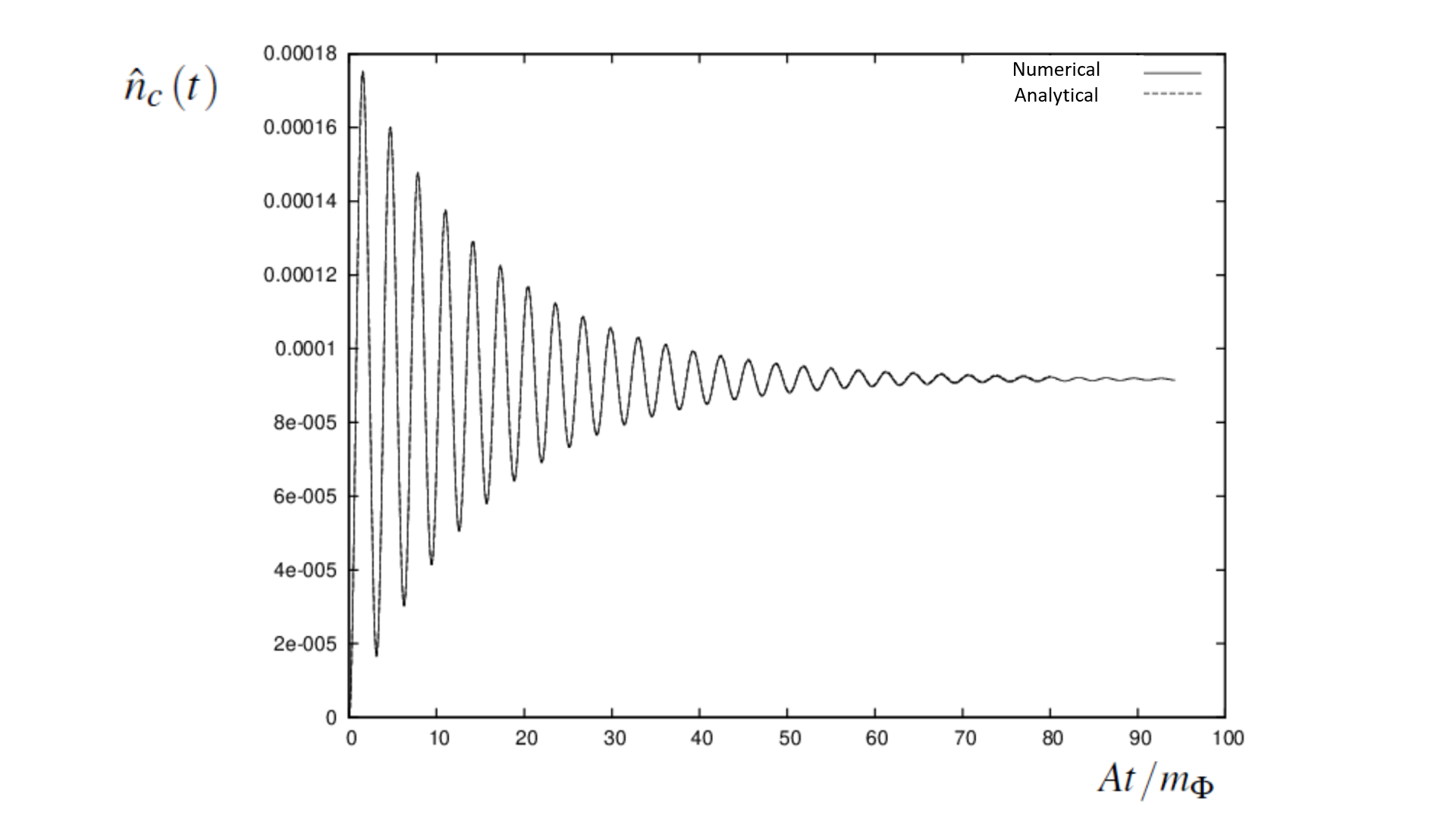}
\caption{Comparison of numerically calculated comoving transferred asymmetry with the analytical estimate using (\ref{eqn:ad51}), $\hat{n}_{c}$, vs. $At/m_{\Phi}$ for $\tau_{\Phi} = 5T_{asy}$. } 
\label{figure:311}
\end{center}
\end{figure}

Figure \ref{figure:37} shows the condensate asymmetry oscillating about zero with decreasing amplitude, until the condensate has completely decayed and the condensate asymmetry is zero. Since in this case we are considering $\tau_{\Phi} > T_{asy}$, this occurs over a large number of phase oscillations of the field, and therefore a large number of oscillations of the asymmetry. Similarly, Figure \ref{figure:38} shows the transferred asymmetry increasing as an asymmetry is initially generated in the condensate and transferred through the initial inflaton decays. The transferred asymmetry then continues oscillating while the inflaton scalars continue decaying, transferring the asymmetry to the Standard Model until the inflaton condensate has decayed completely. The transferred asymmetry then settles at a small, positive, finite value, corresponding to the total asymmetry transferred to the Standard Model, and thus the baryon-to-entropy ratio shown in Figure \ref{figure:39}. As in the case of $\tau_{\Phi} = T_{asy}$, we can see from Figures \ref{figure:310} and \ref{figure:311}, that the numerically calculated comoving condensate and transferred asymmetries are in perfect agreement with the comoving condensate and transferred asymmetries calculated analytically using the threshold approximation for $\tau_{\Phi} = 5T_{asy}$, demonstrating that the threshold approximation accurately models the behaviour of the condensate and transferred asymmetries, as well as accurately predicting the baryon-to-entropy ratio, both in the case of rapid decay of the inflaton condensate and decay over many oscillations of the asymmetry.

\subsection{Validity of the Analytical Approximation of the Decay of the Condensate Asymmetry}\label{section:373}
In this section we examine the validity of the analytical approximation of the decay of the condensate asymmetry, and thus the validity of the analytical treatment of the threshold approximation. In our treatment thus far, we have assumed that the condensate asymmetry decays as

\begin{equation}\label{eqn:ad125}
n_{c}\left(t \right) = n_{0}\left(t \right)e^{-\Gamma_{\Phi}\left(t - t_{\ast}\right)},
\end{equation}

\noindent where $n_{0}\left(t \right)$ is the condensate asymmetry without decay. We will show that this is true if

\begin{equation}\label{eqn:ad126}
\Gamma_{\Phi} << m_{\Phi}^{2},
\end{equation}

\noindent and 

\begin{equation}\label{eqn:ad127}
H\Gamma_{\Phi} << m_{\Phi}^{2}.
\end{equation}

\noindent We are considering rapid oscillations in the amplitude of the inflaton field, $\phi$, in this model, meaning that $m_{\Phi}^{2} >> H^{2}$, and up to the complete decay of the inflaton condensate we have that $\Gamma_{\Phi} \lesssim H$. From these two conditions, it follows that (\ref{eqn:ad126}), (\ref{eqn:ad127}) are satisfied. 

We can prove (\ref{eqn:ad125}) is true under these conditions. As a starting point, we assume that the field solutions in the case of condensate decay are

\begin{equation}\label{eqn:ad128}
\phi_{i}\left(t \right) = \phi_{i0}\left(t \right)e^{-\frac{\Gamma_{\Phi}t}{2}}, \; \; \; i = 1,2
\end{equation}

\noindent where $\phi_{i0}\left(t \right)$ is the solution to the field equations without the decay terms. The full field equations with the decay terms are

\begin{equation}\label{eqn:ad129}
\ddot{\phi}_{1} + 3H\dot{\phi}_{1} + \Gamma_{\Phi}\dot{\phi}_{1} = -m_{1}^{2}\phi_{1},
\end{equation}

\begin{equation}\label{eqn:ad130}
\ddot{\phi}_{2} + 3H\dot{\phi}_{2} + \Gamma_{\Phi}\dot{\phi}_{2} = -m_{2}^{2}\phi_{2},
\end{equation}

\noindent and in order to verify the analytical approximation of the decay of the condensate asymmetry we need to substitute (\ref{eqn:ad128}) into the field equations with the decay terms. Using (\ref{eqn:ad128}) we have that

\begin{equation}\label{eqn:ad131}
\dot{\phi}_{i}\left(t \right) = \left[ \dot{\phi}_{i0}\left(t \right) - \frac{\Gamma_{\Phi}}{2}\phi_{i0}\left(t \right) \right]e^{-\frac{\Gamma_{\Phi}t}{2}},
\end{equation}

\noindent and 

\begin{equation}\label{eqn:ad132}
\ddot{\phi}_{i}\left(t \right) = \left[\ddot{\phi}_{i0}\left(t\right) - \Gamma_{\Phi}\dot{\phi}_{i0}\left(t \right) + \frac{\Gamma_{\Phi}^{2}}{4} \right]e^{-\frac{\Gamma_{\Phi}t}{2}}.
\end{equation}

\noindent Substituting (\ref{eqn:ad131}), (\ref{eqn:ad132}) into the field equations (\ref{eqn:ad129}), (\ref{eqn:ad130}) we find that both sides are proportional to $e^{-\frac{\Gamma_{\Phi}t}{2}}$, and can therefore cancel the exponentials. The field equations with decay terms in terms of the field solution (\ref{eqn:ad128}) are then

\begin{equation}\label{eqn:ad133}
\ddot{\phi}_{i0}\left(t \right) + 3H\dot{\phi}_{i0}\left(t \right) = - \left[m_{i}^{2} - \frac{\Gamma_{\Phi}^{2}}{4} - \frac{3H\Gamma_{\Phi}}{2} \right]\phi_{i0}\left(t \right).
\end{equation}

\noindent These are the same as the field equations with no decay term if

\begin{equation}\label{eqn:ad134}
\frac{\Gamma_{\Phi}^{2}}{4} << m_{i}^{2}, \; \; \frac{3H\Gamma_{\Phi}}{2} << m_{i}^{2},
\end{equation}

\noindent and if these conditions are satisfied then we can infer that (\ref{eqn:ad126}) and (\ref{eqn:ad127}) are also satisfied, and are true for $\Gamma_{\Phi}^{2} << m_{i}^{2}$ and $H^{2} << m_{i}^{2}$. These conditions are generally satisfied for a rapidly oscillating field, which is a basic assumption of the dynamics of this model.

The condensate asymmetry is given by

\begin{equation}\label{eqn:ad135}
n\left(t \right) = \dot{\phi}_{1}\phi_{2} - \dot{\phi}_{2}\phi_{1}.
\end{equation}

\noindent Substituting (\ref{eqn:ad128}) and (\ref{eqn:ad131}) into the condensate asymmetry (\ref{eqn:ad135}) we find that

\begin{equation}\label{eqn:ad136}
n\left(t \right) = \left(\dot{\phi}_{10}\phi_{20} - \dot{\phi}_{20}\phi_{10}\right)e^{-\Gamma_{\Phi}t},
\end{equation}

\noindent which is simply

\begin{equation}\label{eqn:ad137}
n\left(t \right) = n_{0}\left(t \right)e^{-\Gamma_{\Phi}t}.
\end{equation}

\noindent The condensate asymmetry for a decaying condensate is therefore equal to the condensate asymmetry with no decay multiplied by $e^{-\Gamma_{\Phi}t}$, as put forward in (\ref{eqn:ad51}) in Section \ref{section:344}, and the analytical approximation of the decay of the condensate asymmetry is therefore valid in the limit of rapid oscillations, as we assume throughout this model.

\section{Baryon Asymmetry Washout by Inflaton Exchange}\label{section:38}
So far we have assumed that the condensate decays via $B$-conserving processes, and that the mean asymmetry formed in the inflaton condensate is transferred equally to the Standard Model particles. This means that we assume there are no additional processes happening which could remove the asymmetry from the particle plasma. In this section we will consider a $B$-violating process which follows the annihilation of two anti-fermions $\bar{\psi}$ to a virtual inflaton, which then decays to a pair of fermions $\psi$, which can subtract baryon number from the Universe and potentially wipe out the asymmetry. Such a process would arise from inflaton exchange combined with the interaction $\lambda_{\psi}\bar{\psi}^{c}\psi \Phi$, responsible for inflaton condensate decay. The rate of this process is given by

\begin{equation}\label{eqn:ad138}
\Gamma_{\Delta B} = n_{\Phi} \langle \sigma v \rangle,
\end{equation}

\noindent where we assume this process is relativistic, $v=1$. At reheating $t = t_{R}$, the number density of the inflaton particles is $n_{\Phi} \sim T_{R}^{3}$, and dimensionally the rate is given by

\begin{equation}\label{eqn:ad139}
\Gamma_{\Delta B} \sim \frac{\lambda_{\psi}^{4}A^{2}T_{R}^{5}}{m_{\Phi}^{8}},
\end{equation}

\noindent where inflaton exchange implies an amplitude proportional to $\lambda_{\psi}^{2}A$. This process will not be significant - and therefore washout due to inflaton exchange will not pose a problem for Affleck-Dine baryogenesis with quadratic symmetry-breaking terms - provided that $\Gamma_{\Delta B} < H\left(T_{R}\right)$. Using $H\left(T_{R}\right) \sim T_{R}^{2}/M_{pl}$, this requires that

\begin{equation}\label{eqn:ad140}
\frac{\lambda_{\psi}^{4}A^{2}T_{R}^{5}}{m_{\Phi}^{8}} < H\left(T_{R}\right) \sim \frac{T_{R}^{2}}{M_{pl}}.
\end{equation}

\noindent Rearranging (\ref{eqn:ad140}) gives the following constraint on the coupling $\lambda_{\psi}$

\begin{equation}\label{eqn:ad141}
\lambda_{\psi}^{4} \lesssim \frac{m_{\Phi}^{8}}{M_{pl}A^{2} T_{R}^{3}} \Rightarrow \lambda_{\psi}^{2} \lesssim \frac{m_{\Phi}^{4}}{M_{pl}^{\frac{1}{2}} A T_{R}^{\frac{3}{2}}}.
\end{equation}

\noindent Normalising the inflaton mass and the reheating temperature using $m_{\phi} = 10^{13}\GeV$ and $T_{R} = 10^{8}\GeV$, this can be expressed as

\begin{equation}\label{eqn:ad142}
\lambda_{\psi}^{2} \lesssim \left(6.5 \times 10^{4}\right)\left(\frac{m_{\Phi}^{2}}{A}\right)\left(\frac{m_{\Phi}}{10^{13}\GeV}\right)^{2}\left(\frac{10^{8} \GeV}{T_{R}}\right)^{\frac{3}{2}}.
\end{equation}

\noindent Provided that this condition is satisfied, the asymmetry will not be washed out by inflaton exchange. This can be rewritten as a constraint on the reheating temperature if we consider the fact that the inflaton decay rate is given by $\Gamma_{\Phi} = \lambda_{\psi}^{2}m_{\Phi}/4\pi $ \cite{infnr}, and $\Gamma_{\Phi} = H_{R}$. The reheating temperature is then

\begin{equation}\label{eqn:ad143}
T_{R} \approx \lambda_{\psi} \sqrt{m_{\Phi} M_{pl}},
\end{equation}

\noindent which allows the constraint (\ref{eqn:ad141}) to be expressed as 

\begin{equation}\label{eqn:ad144}
T_{R} \lesssim \left(\frac{m_{\Phi}^{2}}{A}\right)^{\frac{2}{7}}\left(\frac{M_{pl}}{m_{\Phi}}\right)^{\frac{1}{7}}m_{\Phi}.
\end{equation}

\noindent Provided this is satisfied, there is little risk of the asymmetry being washed out by inflaton exchange. Since $A << m_{\Phi}^{2}$ and $m_{\Phi} < M_{pl}$, then this is generally satisfied for Affleck-Dine baryogenesis with quadratic symmetry-breaking terms if $T_{R} < m_{\Phi}$. A more in-depth examination of washout would require a specific model of the inflaton decay processes and the transfer of the baryon asymmetry.

\section{Consistency with Non-Minimally Coupled Inflation}\label{section:39}

\begin{figure}[H]
\begin{center}
\includegraphics[clip = true, width=\textwidth, angle = 360]{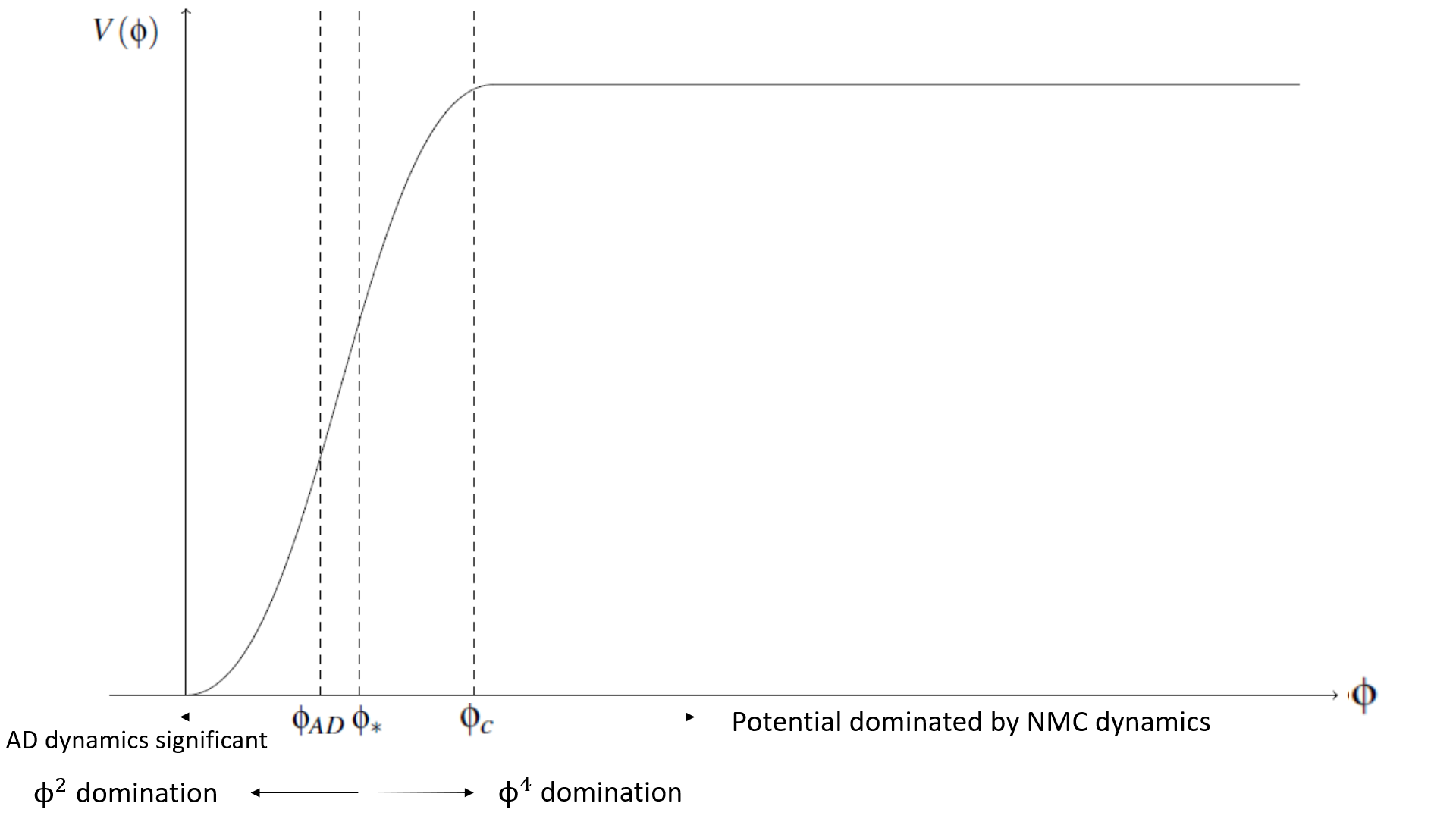}
\caption{Schematic of the inflaton potential illustrating the relationship between the different regimes of the potential as thresholds in the dominant dynamics of the inflaton field, assuming $\phi$ is canonically normalised, producing the characteristic plateau potential of non-minimally coupled models. } 
\label{figure:n25}
\end{center}
\end{figure}

In this section we consider the case of non-minimally coupled inflation with a quartic potential, which is naturally compatible with the potential of the model (\ref{eqn:ad3}). We explore the conditions needed for the analytic expressions of the baryon asymmetry we have derived in this work to be consistent with non-minimally coupled inflation. To achieve this we need that the inflaton dynamics stop being dominated by the non-minimal coupling before the Affleck-Dine dynamics become significant. This can be stated as requiring that the point of non-minimal coupling domination, $\phi_{c}$, must be larger than the threshold point $\phi_{\ast}$, $\phi_{\ast} < \phi_{c}$. From Section \ref{section:36}, we have that once $H^{2} = 4A$, at a point we denote as $\phi_{AD}$, the phase field $\theta$ becomes dynamical, and we need $\phi_{AD} < \phi_{\ast}$ in order for no asymmetry to be produced while the potential is in its $\left|\Phi\right|^{4}$ dominated regime and for the threshold approximation of the asymmetry to hold. We therefore require that $\phi_{AD} < \phi_{\ast} < \phi_{c}$ (see Figure \ref{figure:n25}) in order for this model of Affleck-Dine baryogenesis via quadratic $B$-violating terms as described by the threshold approximation to be compatible with the AD field being a non-minimally coupled inflaton.

In the Palatini formalism, the point at which the non-minimal coupling ceases to be the dominant contribution to the dynamics of the inflaton field is when $\Omega^{2} = 1$. This represents the point at which the canonical inflaton in the Einstein frame becomes equivalent to the Jordan frame inflaton, and the potential is approximately given by the Jordan frame potential. We have that

\begin{equation}\label{eqn:ad145}
\Omega^{2} = 1 + \frac{\xi \phi_{c}^{2}}{M_{pl}^{2}} \approx \mathcal{O}\left(1\right).
\end{equation}

\noindent Since this corresponds to the edge of the plateau, we have 

\begin{equation}\label{eqn:ad146}
\frac{\xi \phi_{c}^{2}}{M_{pl}^{2}} \sim 1 \Rightarrow \phi_{c} = \frac{M_{pl}}{\sqrt{\xi}},
\end{equation}

\noindent in the Palatini formalism. 

In the metric formalism, the condition for equivalence of the inflaton between frames is (from the definition of the canonical field transformation in Section \ref{section:t11})

\begin{equation}\label{eqn:ad147}
\frac{\Omega^{2} + 6\xi^{2}\phi_{c}^{2}/M_{pl}^{2}}{\Omega^{4}} =1,
\end{equation}

\noindent for $\Omega^{2} = 1$. Using the slow-roll inflation condition on the non-minimal coupling, $\xi >> 1$, we have 

\begin{equation}\label{eqn:ad148}
\frac{6\xi^{2}\phi_{c}^{2}}{M_{pl}^{2}} = 1 \Rightarrow \phi_{c} = \frac{M_{pl}}{\sqrt{6}\xi}.
\end{equation}

Using the condition $\phi_{\ast} = m_{\Phi}/\sqrt{\lambda_{\Phi}} < \phi_{c}$, in the Palatini formalism (\ref{eqn:ad146}) we therefore require

\begin{equation}\label{eqn:ad149}
\frac{m_{\Phi}}{\sqrt{\lambda_{\Phi}}} <  \frac{M_{pl}}{\sqrt{\xi}},
\end{equation}

\noindent which gives the condition on the inflaton mass for the model to be consistent with non-minimally coupled dynamics as 

\begin{equation}\label{eqn:ad150}
m_{\Phi} < \frac{\sqrt{\lambda_{\Phi}}M_{pl}}{\sqrt{\xi}}.
\end{equation}

The comoving curvature power spectrum for a quartic potential in the Palatini formalism is (from (\ref{eqn:q46})\footnote{In Chapter 6 we derive the primordial curvature power spectrum and the inflation observables in the Palatini formalism generalised to include an inflaton mass term. We find that the power spectrum and observables are the same as those for conventional Palatini inflation with a quartic inflaton potential. See e.g. \cite{tenkanengrav} for a recent review of the calculation of the observables in quartic Palatini inflation.})

\begin{equation}\label{eqn:ad151}
\mathcal{P}_{\mathcal{R}} = \frac{\lambda_{\Phi}N^{2}}{12\pi^{2}\xi },
\end{equation}

\noindent and using $N = 55$ as an estimate of the pivot scale and taking the amplitude of $\mathcal{P}_{\mathcal{R}}$ to be $A_{s} = 2.1 \times 10^{-9}$ \cite{planck184}, $\xi$ is

\begin{equation}\label{eqn:ad152}
\xi = 1.22\times 10^{10}\lambda_{\Phi}.
\end{equation}

\noindent Using $\xi$ normalised to the Palatini power spectrum $\mathcal{P}_{\mathcal{R}}$ given in (\ref{eqn:ad152}), the constraint on the inflaton mass (\ref{eqn:ad150}) becomes

\begin{equation}\label{eqn:ad153}
m_{\Phi} < 2.2 \times 10^{13}\GeV.
\end{equation}

For the case of the metric formalism (\ref{eqn:ad148}), for $\phi_{\ast} < \phi_{c}$, we have that

\begin{equation}\label{eqn:ad154}
\frac{m_{\Phi}}{\sqrt{\lambda_{\Phi}}} <  \frac{M_{pl}}{\sqrt{6}\xi},
\end{equation}

\noindent which gives

\begin{equation}\label{eqn:ad155}
m_{\Phi} < \frac{\sqrt{\lambda_{\Phi}}M_{pl}}{\sqrt{6}\xi}.
\end{equation}

\noindent The comoving curvature power spectrum in the metric formalism is given by (from (\ref{eqn:b49}))

\begin{equation}\label{eqn:ad156}
\mathcal{P}_{\mathcal{R}} = \frac{\lambda_{\Phi}}{72\pi^{2}\xi^{2} }N^{2},
\end{equation}

\noindent and using $N =55$ and $A_{s} = 2.1 \times 10^{-9}$, this gives the value of the non-minimal coupling to be

\begin{equation}\label{eqn:ad157}
\xi = 4.5 \times 10^{4}\sqrt{\lambda_{\Phi}}.
\end{equation}

\noindent Using $\xi$ normalised to the metric power spectrum $\mathcal{P}_{\mathcal{R}}$ given in (\ref{eqn:ad157}) this gives

\begin{equation}\label{eqn:ad158}
m_{\Phi} < 2.2 \times 10^{13}\GeV,
\end{equation}

\noindent in order for the model to be consistent with non-minimally coupled inflaton dynamics in the metric formalism, which is the coincidentally the same numerical constraint for the inflaton mass for the Palatini formalism. This means that this model is compatible with an embedding into a non-minimally coupled inflation model within a well defined upper bound on the inflaton mass.

\section{Baryon Isocurvature Perturbations in Non-Minimally Coupled Inflation}\label{section:310}

In this model, it is possible that the angular component of the inflaton field may undergo quantum fluctuations during inflation which result in density perturbations uncorrelated with the perturbations in the photon energy density. These are isocurvature fluctuations, as defined in Section \ref{section:121}. In Affleck-Dine baryogenesis models, the phase of the AD field can produce isocurvature fluctuations in the local baryon density, and therefore in baryon number density \cite{iso1} - \cite{iso5}. These isocurvature fluctuations in baryon number density can be observable, and this can lead to constraints on inflation models used in conjunction with the Affleck-Dine model in order for the model to be compatible with observations. In this section, we explore the baryon isocurvature perturbations generated in this model of Affleck-Dine inflation with quadratic $B$-violating terms, and examine whether this model is compatible with the bounds on the isocurvature fraction from the Planck experiment (2018). 

We define the isocurvature perturbations as

\begin{equation}\label{eqn:ad159}
I = \left(\frac{\delta \rho_{i}}{\rho_{i}}\right)_{iso},
\end{equation}

\noindent for a given species $i$. In order to compare the isocurvature perturbations in baryon energy density in this model to the isocurvature bounds defined using Planck data \cite{Planck18}, which are normalised to cold dark matter isocurvature perturbations, we begin with the isocurvature perturbations in cold dark matter density

\begin{equation}\label{eqn:ad160}
I = \left(\frac{\delta \rho_{CDM}}{\rho_{CDM}}\right)_{iso},
\end{equation}

\noindent and then use the fact that $\delta \rho_{CDM} \rightarrow \delta \rho_{B}$ to find the effective cold dark matter isocurvature perturbations due to baryon number perturbations

\begin{equation}\label{eqn:ad161}
I = \frac{\rho_{B}}{\rho_{CDM}}\left(\frac{\delta \rho_{B}}{\rho_{B}}\right)_{iso} = \frac{\Omega_{B}}{\Omega_{CDM}}\left(\frac{\delta \rho_{B}}{\rho_{B}}\right)_{iso},
\end{equation}

\noindent where the $\Omega$ here are the total energy densities as defined in Section \ref{section:061}. The isocurvature fraction $\beta_{iso}$ defined in \cite{Planck18} is

\begin{equation}\label{eqn:ad162}
\beta_{iso} = \frac{\mathcal{P}_{I}}{\mathcal{P}_{\mathcal{R}} + \mathcal{P}_{I}},
\end{equation}

\noindent assuming that the adiabatic and isocurvature fluctuations are uncorrelated. $\mathcal{P}_{\mathcal{R}} = A_{s}$ is the primordial curvature power spectrum, and to calculate $\mathcal{P}_{I}$ we need to calculate the power spectrum of the baryon isocurvature fluctuations $\delta \rho_{B}/\rho_{B}$ due to the fluctuations of the inflaton field, given by $\delta \theta$.

For non-minimally coupled inflation with a complex field, we have that

\begin{equation}\label{eqn:ad163}
\Phi = \frac{\phi}{\sqrt{2}}e^{i\theta} = \frac{1}{\sqrt{2}}\left(\phi_{1} + i\phi_{2}\right),
\end{equation}

\noindent where $\theta$ is an effectively massless field during inflation and the evolution of the inflaton field is purely along the radial direction, corresponding to $\theta = 0$ along the $\phi_{1}$ direction. Writing the inflaton as an effectively constant background plus a fluctuation in the field, we have that

\begin{equation}\label{eqn:ad164}
\Phi = \frac{\bar{\phi}}{\sqrt{2}}e^{i\delta\theta} \simeq  \frac{\bar{\phi}}{\sqrt{2}}\left( 1 + i\delta \theta \right) = \frac{1}{\sqrt{2}}\left(\bar{\phi}_{1} + i\delta \phi_{2}\right),
\end{equation}

\noindent for small $\delta \theta$, where

\begin{equation}\label{eqn:ad165}
\bar{\phi}_{1} = \bar{\phi}, \; \; \; \delta \phi_{2} = \bar{\phi}\delta \theta.
\end{equation}

We have that the inflaton kinetic term in both the metric and Palatini formalisms for a complex inflaton field is given by

\begin{equation}\label{eqn:ad166}
\frac{1}{\Omega^{2}}\partial_{\mu}\Phi^{\dagger}\partial^{\mu}\Phi = \frac{1}{2\left(1 + \frac{\xi \phi_{1}^{2}}{M_{pl}^{2}} + \frac{\xi \phi_{2}^{2}}{M_{pl}^{2}}\right)}\left( \partial_{\mu}\phi_{1}\partial^{\mu}\phi_{1} + \partial_{\mu}\phi_{2}\partial^{\mu}\phi_{2}\right),
\end{equation}

\noindent and during slow-roll inflation we have that 

\begin{equation}\label{eqn:ad167}
\frac{\xi \bar{\phi}_{1}^{2}}{M_{pl}^{2}} >> 1, \frac{\xi \phi_{2}^{2}}{M_{pl}^{2}},
\end{equation}

\noindent and the kinetic term for the angular field can therefore be written

\begin{equation}\label{eqn:ad168}
\frac{M_{pl}^{2}}{2\xi \bar{\phi}_{1}^{2}}\partial_{\mu}\phi_{2}\partial^{\mu}\phi_{2}.
\end{equation}

\noindent We define a canonical field $\chi_{2}$ such that

\begin{equation}\label{eqn:ad169}
\frac{d\chi_{2}}{d\phi_{2}} = \frac{M_{pl}}{\sqrt{\xi}\bar{\phi}_{1}} \Rightarrow \chi_{2} = \frac{M_{pl}}{\sqrt{\xi}\bar{\phi}_{1}} \phi_{2},
\end{equation}

\noindent and substituting this into (\ref{eqn:ad168}) we find the canonical kinetic term to be

\begin{equation}\label{eqn:ad170}
\frac{M_{pl}^{2}}{2\xi \bar{\phi}_{1}^{2}}\partial_{\mu}\phi_{2}\partial^{\mu}\phi_{2} \rightarrow \frac{1}{2}\partial_{\mu}\chi_{2}\partial^{\mu}\chi_{2}.
\end{equation}

\noindent In terms of fluctuations of the inflaton field we therefore have

\begin{equation}\label{eqn:ad171}
\delta \chi_{2} = \frac{M_{pl}}{\sqrt{\xi}\bar{\phi}_{1}} \delta \phi_{2} = \frac{M_{pl}}{\sqrt{\xi}\bar{\phi}}\bar{\phi}\delta \theta,
\end{equation}

\begin{equation}\label{eqn:ad172}
\Rightarrow \delta \theta = \frac{\sqrt{\xi}}{M_{pl}}\delta \chi_{2}.
\end{equation}

\noindent Since $\chi_{2}$ is a canonically normalised field, its power spectrum is the standard expression given in Section \ref{section:12}

\begin{equation}\label{eqn:ad173}
\mathcal{P}_{\delta \chi_{2}} = \left(\frac{H}{2\pi}\right)^{2},
\end{equation}

\noindent and from (\ref{eqn:ad172}) we have that the corresponding power spectrum in $\delta \theta$ is

\begin{equation}\label{eqn:ad174}
\mathcal{P}_{\delta \theta} = \frac{\xi}{M_{pl}^{2}}\mathcal{P}_{\delta \chi_{2}} \Rightarrow \mathcal{P}_{\delta \theta} = \frac{\xi H^{2}}{4\pi^{2}M_{pl}^{2}}.
\end{equation}

The fluctuations in the baryon energy density relate to fluctuations in baryon number density through $\rho_{B} = m_{B}n_{B}$ for a fixed mass of the baryonic species, and we have

\begin{equation}\label{eqn:ad175}
\frac{\delta \rho_{B}}{\rho_{B}} = \frac{\delta n_{B}}{n_{B}}.
\end{equation}

\noindent In this model we have that 

\begin{equation}\label{eqn:ad176}
n_{B} \propto \sin \left(2\theta \right),
\end{equation}

\noindent from (\ref{eqn:ad72}), (\ref{eqn:ad79}), (\ref{eqn:ad82}), for an initial phase angle $\theta$, and we can write the fluctuation in baryon number density as

\begin{equation}\label{eqn:ad177}
\frac{\delta n_{B}}{n_{B}} = \frac{d n_{B}}{d\theta}\frac{\delta \theta}{n_{B}}.
\end{equation}

\noindent Using (\ref{eqn:ad176}), we have that 

\begin{equation}\label{eqn:ad178}
\frac{d n_{B}}{d\theta} = 2\cos \left(2\theta \right),
\end{equation}

\noindent and using (\ref{eqn:ad176}) and (\ref{eqn:ad178}) substituted into (\ref{eqn:ad177}), the fluctuation in baryon number density in terms of the fluctuation in $\theta$ is

\begin{equation}\label{eqn:ad179}
\frac{\delta n_{B}}{n_{B}} \propto \frac{2\delta \theta}{\tan \left(2\theta \right)}.
\end{equation}

The isocurvature fluctuation in baryon number density from (\ref{eqn:ad161}) is then

\begin{equation}\label{eqn:ad180}
I = \frac{\Omega_{B}}{\Omega_{CDM}}\frac{\delta n_{B}}{n_{B}} = \frac{\Omega_{B}}{\Omega_{CDM}}\frac{2\delta \theta}{\tan \left(2\theta \right)},
\end{equation}

\noindent the power spectrum of the baryon isocurvature fluctuations is thus

\begin{equation}\label{eqn:ad181}
\mathcal{P}_{I} =  \left(\frac{\Omega_{B}}{\Omega_{CDM}}\right)^{2}\frac{4}{\tan^{2} \left(2\theta \right)}\mathcal{P}_{\delta \theta},
\end{equation}

\noindent and using (\ref{eqn:ad174}), we can write (\ref{eqn:ad181}) as

\begin{equation}\label{eqn:ad182}
\mathcal{P}_{I} =  \left(\frac{\Omega_{B}}{\Omega_{CDM}}\right)^{2}\frac{\xi H^{2}}{\tan^{2} \left(2\theta \right)\pi^{2}M_{pl}^{2}}.
\end{equation}

\noindent The ratio $\mathcal{P}_{I}/\mathcal{P}_{\mathcal{R}}$ from (\ref{eqn:ad162}) can be written in terms of the isocurvature fraction

\begin{equation}\label{eqn:ad183}
\frac{\mathcal{P}_{I}}{\mathcal{P}_{\mathcal{R}}} = \frac{\beta_{iso}}{1 - \beta_{iso}},
\end{equation}

\noindent and using the observational limit on the isocurvature fraction, $\beta_{iso, lim}$, \cite{Planck18} we can constrain this ratio

\begin{equation}\label{eqn:ad184}
\frac{\mathcal{P}_{I}}{\mathcal{P}_{\mathcal{R}}} = \left(\frac{\Omega_{B}}{\Omega_{CDM}}\right)^{2}\frac{\xi H^{2}}{\tan^{2} \left(2\theta \right)\pi^{2}M_{pl}^{2}\mathcal{P}_{\mathcal{R}}} < \frac{\beta_{iso, lim}}{1 - \beta_{iso, lim}}.
\end{equation}

\noindent With some rearrangement this can be recast as a constraint on the Hubble parameter during inflation

\begin{equation}\label{eqn:ad185}
H_{inf} < \frac{\Omega_{CDM}}{\Omega_{B}}\frac{M_{pl}\pi A_{s}^{\frac{1}{2}}}{\sqrt{\xi}}\tan \left(2\theta \right) \left(\frac{\beta_{iso, lim}}{1 - \beta_{iso, lim}} \right)^{\frac{1}{2}}.
\end{equation}

\noindent Using $A_{s} = 2.1 \times 10^{-9}$, $\Omega_{CDM}/\Omega_{B} = 5.3$ and $\beta_{iso, lim} = 0.038$ from \cite{Planck18} this is

\begin{equation}\label{eqn:ad186}
H_{inf} < 3.6\times 10^{14}\left(\frac{\tan\left(2\theta \right)}{\sqrt{\xi}}\right)\GeV,
\end{equation}

\noindent which is the requirement in order to produce an isocurvature fraction consistent with observations.

The Einstein frame potential on the plateau in both the metric and Palatini formalisms for a quartic inflaton potential (see e.g. \cite{tenkanengrav} for a review of the derivation of (\ref{eqn:ad187}) in both formalisms) is 

\begin{equation}\label{eqn:ad187}
V_{E} = \frac{\lambda_{\Phi}M_{pl}^{4}}{4\xi^{2}},
\end{equation}

\noindent and from (\ref{eqn:r198}) we have that

\begin{equation}\label{eqn:ad188}
H_{inf} = \left(\frac{\lambda_{\Phi}}{12}\right)^{\frac{1}{2}}\frac{M_{pl}}{\xi}.
\end{equation}

\noindent Using (\ref{eqn:ad188}) and (\ref{eqn:ad186}), we can write a constraint on the non-minimal coupling $\xi$ needed in order to produce an acceptable isocurvature fraction in this model,

\begin{equation}\label{eqn:ad189}
\left(\frac{\lambda_{\Phi}}{12}\right)^{\frac{1}{2}}\frac{M_{pl}}{\xi} < 3.6\times 10^{14}\left(\frac{\tan\left(2\theta \right)}{\sqrt{\xi}}\right)\GeV,
\end{equation}

\noindent therefore

\begin{equation}\label{eqn:ad190}
\xi > 3.7 \times 10^{6}\frac{\lambda_{\Phi}}{\tan^{2}\left(2\theta \right)}.
\end{equation}

We can now examine $\xi$ in the metric and Palatini formalisms and test whether the non-minimal coupling in each formalism satisfies the constraint (\ref{eqn:ad190}). The non-minimal coupling $\xi$ normalised from the comoving curvature power spectrum in the Palatini formalism is $\xi = 1.22\times 10^{10}\lambda_{\Phi}$ (\ref{eqn:ad152}) which easily satisfies the constraint (\ref{eqn:ad190}) provided that $\tan \left(2\theta \right)$ isn't exceptionally small, and also shows that the constraint on $\xi$ ultimately does not depend on the size of the inflaton self-coupling $\lambda_{\Phi}$. This shows that this model in the Palatini formalism produces an isocurvature perturbation in the baryon number density compatible with observations.

In the metric formalism we find $\xi = 4.5 \times 10^{4}\sqrt{\lambda_{\Phi}}$ (\ref{eqn:ad157}), and therefore

\begin{equation}\label{eqn:ad191}
\lambda_{\Phi} < 1.5\times 10^{-4}\tan^{4}\left(2\theta \right),
\end{equation}

\noindent is required in order for the constraint (\ref{eqn:ad190}) to be satisfied and isocurvature perturbations to be compatible with observations in the metric formalism.

Therefore in metric non-minimally coupled inflation, isocurvature perturbations place a significant constraint on the inflaton self-coupling. This also means that the baryon isocurvature fraction can be close to the present CMB bound if $\lambda_{\Phi} \sim 10^{-4}$, and so potentially observable as the bound improves, in contrast to the case of Palatini inflation where it will generally be much smaller than the CMB bound.

\section{Semi-Classical Treatment of the Coherently Oscillating Condensate}\label{section:311}

Throughout our derivation and analysis of the Affleck-Dine baryogenesis model with quadratic symmetry-breaking terms, we have assumed that $\phi_{1}$ and $\phi_{2}$ can be treated as classical fields. However, we must consider whether the classical treatment of the fields is valid in this context.

The inflaton field after inflation forms an oscillating coherent quantum condensate. In a second quantised theory, a coherent condensate can be treated as a classical field because the occupation number of the field states is much larger than one \cite{davidson} \cite{lozanovre}. In order for this to be satisfied, we require in the general sense that $X > m_{X}$ for the field $X$ which we are considering. However, in our model the condensate decays when $\phi < \phi_{\ast} = m_{\Phi}/\sqrt{\lambda_{\Phi}}$, and so $\phi < m_{\Phi}$ generally, meaning that the condensate can no longer be considered classically. Nevertheless, we will show that the classical calculation of the asymmetry is still correct.

The oscillating classical field corresponds to a coherent state of the field $| \phi_{i}\left(t \right) \rangle$. In general, the expectation value of the field operator in the coherent state $| \phi_{i}\left(t \right) \rangle$ is given by its classical value,

\begin{equation}\label{eqn:ad192}
\langle \phi_{i}\left(t \right) | \hat{\phi}_{i}| \phi_{i}\left(t \right) \rangle = \phi_{i,cl}\left(t \right),
\end{equation}

\noindent where $\phi_{i, cl}$ is the classical value of the field $\phi_{i}$ which satisfies the classical equations of motion. Since we are treating the components of the fields $\phi_{1}$ and $\phi_{2}$ as independent scalar fields, the coherent state of the complex field is the product of the coherent states of $\phi_{1}$ and $\phi_{2}$

\begin{equation}\label{eqn:ad193}
| \Phi \left( t \right) \rangle = | \phi_{1}\left(t \right) \rangle | \phi_{2}\left(t \right) \rangle.
\end{equation}

\noindent The baryon number density operator is given by

\begin{equation}\label{eqn:ad194}
\hat{n} = \hat{\dot{\phi}}_{1}\hat{\phi}_{2} - \hat{\dot{\phi}}_{2}\hat{\phi}_{1},
\end{equation}

\noindent and the expectation value of the baryon number density in the coherent state  $| \phi_{1}\left(t \right) \rangle | \phi_{2}\left(t \right) \rangle$ is given by

\begin{multline}\label{eqn:ad195}
\langle \Phi \left( t \right) | \hat{n} | \Phi \left( t \right) \rangle = \langle \Phi \left( t \right) | \hat{\dot{\phi}}_{1}\hat{\phi}_{2} - \hat{\dot{\phi}}_{2}\hat{\phi}_{1} | \Phi \left( t \right) \rangle = \langle \phi_{1}\left(t \right) | \langle \phi_{2}\left(t \right) | \hat{\dot{\phi}}_{1}\hat{\phi}_{2} - \hat{\dot{\phi}}_{2}\hat{\phi}_{1}| \phi_{1}\left(t \right) \rangle | \phi_{2}\left(t \right) \rangle \\
= \langle \phi_{1} \left| \hat{\dot{\phi}}_{1} \right| \phi_{1} \rangle \langle \phi_{2} \left| \hat{\phi}_{2} \right| \phi_{2} \rangle  - \langle \phi_{1} \left| \hat{\phi}_{1} \right| \phi_{1} \rangle \langle \phi_{2} \left| \hat{\dot{\phi}}_{2} \right| \phi_{2} \rangle.
\end{multline}

\noindent Thus the expectation value of the baryon asymmetry operator in the coherent state is given by

\begin{equation}\label{eqn:ad196}
\langle \Phi \left( t \right) | \hat{n} | \Phi \left( t \right) \rangle  = \dot{\phi}_{1}\phi_{2} - \dot{\phi}_{2}\phi_{1} \equiv n_{cl}.
\end{equation}

\noindent Therefore the expectation value of the baryon asymmetry operator in the coherent state $i$ is equal to the baryon number density, $n_{cl}$, calculated using the classical fields $\phi_{1}$ and $\phi_{2}$. This is true even if the classical oscillating field is no longer a good approximation to the coherent state.

Since $\phi_{i} < m_{\Phi}$ in this model, the variance of the expectation value of the field becomes large compared to the squared classical field, which means that the quantum fluctuations of the fields around their expectation values will be large, and the fields cannot be considered classical. However, the length scale of the field fluctuations at the onset of inflaton decay must be smaller than the horizon. The present observed Universe evolves from a spacetime volume much larger than the size of the horizon at inflaton decay, therefore the observed baryon asymmetry today will be given by its spatial average, which corresponds to the classical value of the asymmetry. This means that the observed baryon-to-entropy ratio is equivalent to the baryon asymmetry calculated using the classical fields, $n_{cl}$, in this case, despite the fact that the fields themselves cannot be considered classical. The classical calculation of the baryon-to-entropy ratio is therefore still correct in the Affleck-Dine baryogenesis model with quadratic symmetry-breaking terms.

\section{Summary}\label{section:312}
In this chapter we have considered an Affleck-Dine baryogenesis scenario resulting from a model of non-minimally coupled inflation with a complex inflaton charged under a global $U(1)$ symmetry, with the complex inflaton taking the role of the AD field. We considered inflation from a renormalisable $U(1)$-symmetric $\left| \Phi \right|^{2} + \left| \Phi \right|^{4}$-type potential, with a quadratic $U(1)$-violating term $A\left( \Phi^{2} + \Phi^{\dagger^{2}}\right)$. This term introduces a phase dependence, altering the field's trajectory from being purely radial (along the $\phi_{1}$ direction) into a varying elliptical orbit in the complex plane, corresponding to an oscillating asymmetry.

At the end of slow-roll inflation, the inflaton is coherently oscillating. Initially these are $\left|\Phi\right|^{4}$ dominated oscillations, and then later becomes $\Phi^{2}$ oscillations as the oscillations become damped and the inflaton field oscillates about its minimum deep in the $\Phi^{2}$ region of the potential. At this stage the coherent condensate decays in a $B$-conserving process and subsequently reheats the universe. While the field undergoes coherent oscillations, it undergoes a periodic phase oscillation between its eigenstates $\Phi \leftrightarrow \Phi^{\dagger}$. This means that while the condensate is decaying, it is also periodically rotating between its $\Phi$ and $\Phi^{\dagger}$ states. So for each half-cycle in the phase oscillation, either $\Phi$ or $\Phi^{\dagger}$ scalars will be dominantly decaying, and will therefore produce either a baryon or anti-baryon number contribution to the particle plasma. Due to the fact that the condensate is decaying away while this is happening, it means that the produced baryon number and anti-baryon number of each oscillation doesn't exactly cancel with the charge produced in the previous half-cycle. This results in an overall asymmetry in baryon number being transferred to the particle content of the Standard Model, and this is how the observed small but finite baryon asymmetry is generated.

We were interested in studying the baryon asymmetry generated in this Affleck-Dine baryogenesis model from a quadratic symmetry-breaking term. In order to do this we derived the $U(1)$ asymmetry using a threshold approximation, wherein the asymmetry is generated when the potential is strongly dominated by its quadratic terms. We modelled the asymmetry of the condensate, and its evolution into the asymmetry which is transferred to the Standard Model, both in the case where the lifetime of the condensate is much larger than the oscillation period of the inflaton phase, $\tau_{\Phi} > T_{asy}$, and in the case where the condensate decays away before a phase rotation can occur, $\tau_{\Phi} < T_{asy}$. We find that in both cases the asymmetry transferred to the Standard Model derived using the threshold approximation could account for the observed baryon-to-entropy ratio. For $\tau_{\Phi} < T_{asy}$ it is typically much greater, and for $\tau_{\Phi} > T_{asy}$ it can be much less, due to the effect of suppression of the asymmetry due to averaging over a large number of oscillations. Given that the typical baryon asymmetry generated is much larger than that observed, the suppression due to oscillations is advantageous, requiring a smaller suppression of the quadratic symmetry-breaking terms.

We examine the dynamics of the angular component of the inflaton field $\theta$, and determine that the perturbations of the $\theta$ field do not significantly evolve before the potential becomes $\Phi^{2}$ dominated, provided that the constraint (\ref{eqn:ad100}) is satisfied, which is easily compatible with the constraints on $A^{\frac{1}{2}}/m_{\Phi}^{2}$, (\ref{eqn:ad74}) and (\ref{eqn:ad84}), derived in order to generate the observed baryon-to-entropy ratio from the transferred asymmetry. We numerically verified the constraint (\ref{eqn:ad100}) and showed that when this is violated there is an additional suppression of the asymmetry due to the damping of the phase during $\Phi^{4}$ oscillations.

We verify the validity of the threshold approximation in the derivation of the asymmetry, and the calculation of the baryon-to-entropy ratio using a numerical computation. In the numerical computation, we include decay terms in the field equations to account for the decay of the condensate, and we also model the dilution of the radiation energy density due to the expansion of the Universe. In the analytical calculation we assume that at time $t = t_{R}$, the condensate instantly decays to radiation and use this to calculate the reheating temperature. In the numerical computation of Section \ref{section:37}, we model the condensate as decaying away over a time period $\sim \Gamma_{\Phi}^{-1}$, and then calculate the reheating temperature directly from the energy density of radiation. We verify that the threshold approximation is reasonable for the analytical calculation for cases of the asymmetry generated with $\tau_{\Phi} = T_{asy}$ and $\tau_{\Phi} > T_{asy}$, with the predictions of the baryon-to-entropy ratio obtained numerically being within $10\%$ of the analytical predictions and the behaviour of the comoving condensate and transferred asymmetries in each case when modelled numerically is in perfect agreement with the behaviour predicted analytically using the threshold approximation. We also demonstrated that washout of the asymmetry through $B$-violating inflaton exchange is unlikely to pose a problem in this model for the range of reheating temperatures we would consider. This shows that this model provides a robust general framework for studying Affleck-Dine baryogenesis due to quadratic terms in models with a complex inflaton as the Affleck-Dine field.

In Section \ref{section:39} we discuss the compatibility of the Affleck-Dine baryogenesis via quadratic $B$-violating potential terms model with the dynamics of non-minimally coupled inflation. We find that in order for this to be realised, we require the non-minimally coupled dynamics of the inflaton to become insignificant before the threshold of $\Phi^{2}$ domination, and before the Affleck-Dine dynamics become significant. In both the metric and Palatini formalisms of non-minimally coupled inflation, we find that the model presented in this chapter can fulfill this requirement if the mass of the inflaton is smaller than $2.2 \times 10^{13}\GeV$.

We discuss baryon isocurvature perturbations in Section \ref{section:310} and we derive the constraints needed to ensure that the isocurvature fraction generated by a non-minimally coupled inflaton in this framework is not too large, and therefore consistent with the isocurvature bounds given in the Planck results \cite{Planck18}. We find that the isocurvature fraction predicted in this model can be consistent with the observational limit in both the metric and Palatini formalisms subject to constraints on the non-minimal coupling, $\xi$, (\ref{eqn:ad152}), (\ref{eqn:ad157}) and the inflaton self-coupling in the metric case (\ref{eqn:ad191}). In particular, in the metric case there is a significant upper bound on the size of the inflaton self-coupling, $\lambda_{\Phi} \sim 10^{-4}$. This also means that in the metric case, the baryon isocurvature perturbations could potentially become observable with an improved CMB bound on the isocurvature fraction in the future.

We also examine the validity of the classical treatment of the transfer of the asymmetry to the Standard Model. Since the inflaton field initially forms a coherently oscillating condensate with a large occupation number, the inflaton field can be treated as a classical field. However, when the inflaton decays, its amplitude $\phi$ is less than $m_{\Phi}$ and the coherent state is no longer in the classical limit. This raises the question of whether the classical treatment of the asymmetry in order to generate the observed baryon-to-entropy ratio is necessarily correct. We find that due to the spatial averaging of the baryon asymmetry, it is equal to the expectation value of the transferred asymmetry operator, which is equal to its classical value. Therefore, while it may not be true that the asymmetry can be considered strictly classical, the baryon asymmetry calculated using the classical fields still gives the correct value of the asymmetry.

While in this work we have considered a general Affleck-Dine scenario without studying a specific decay path of the inflaton or resulting transfer route of the $U(1)$-asymmetry to baryon number, it is possible that this work could be applied to a leptogenesis scenario from the decay of the inflaton to right-handed neutrinos, and then a subsequent transfer of the asymmetry to baryon number via sphaleron processes. This would allow a more in-depth analysis of the possibility of baryon asymmetry washout, and could also provide a framework for a model of inflation, baryogenesis and dark matter. This study is in progress.

\chapter{Q-balls from Non-Minimally Coupled Inflation in Palatini gravity}

In this chapter we discuss the formulation of Q-balls in the framework of a non-minimally coupled Palatini inflation model. We present a study of the Palatini inflation model for a $\left|\Phi\right|^{2} + \left|\Phi\right|^{4}$-type potential in the Jordan frame (the first time an inflaton mass term has been included in Palatini inflation) and present the values for the inflationary observables. We derive the Q-ball field equation in a non-minimally coupled Palatini framework, discuss the existence conditions of these Q-balls and derive an inflaton mass range for which the model can both support inflation and produce Q-balls. In doing so, we will, for the first time, derive Q-ball solutions for the case of a complex scalar with a non-canonical kinetic term, corresponding to a new class of Q-ball. We derive the zeroes of the Q-ball equation and use these to solve the equation numerically. We verify the existence of Q-ball solutions over a range of field values and present some important properties of these Q-balls, including energy, charge and radius. We also present an analytical approximation of the Q-ball solution and obtain expressions for the energy and charge using this calculation. We show analytically that these Q-balls are stable and derive the energy-charge relation, both of which are then confirmed numerically. We consider the effects of curvature on these Q-balls and derive approximate relations between the inflaton self-coupling, $\lambda$, and the size and global energy of the Q-balls in order to predict, as a first approximation, the radius needed for one of these Q-balls to collapse to a black hole following formation in the presence of curvature. We discuss reheating in the case of Q-balls and Q-ball derived Black Holes, and speculate on the post-inflation cosmology which could result from an early matter-dominated era of Q-balls or their associated Black Holes. Finally the observability of the model is discussed, including the possibility of observable gravitational waves from Q-ball formation and/or decay and Q-ball dark matter. We also comment on the implications for the case of a real inflaton and oscillons.

\section{Q-balls in the Context of Non-Minimally Coupled Palatini Inflation}\label{section:42}

Q-balls are a subset of the class of field theory solutions known as non-topological solitons. Solitons are extended objects which can be visualised as droplets composed of many particles, held together by an attractive interaction between the particles. Their stability arises due to a conserved quantity of the theory, if the solitons are non-topological in nature. 

In the case of Q-balls, the objects are composed of complex scalar particles charged under a conserved Noether symmetry of the theory, and the objects themselves carry a large charge. Charged scalar field configurations stable against small perturbations were originally theorised to exist by Rosen \cite{rosen} in 1968. These ideas were then explored further and non-topological solitons were formally described by Friedberg, Lee et al  \cite{friedberg76,lee92} later in the 20th century. The formulation of Q-balls as we understand them today was developed by Coleman in 1985 \cite{coleman85}. 

Since then, Q-balls have been studied extensively in the context of supersymmetry (SUSY)  \cite{kasuya002} - \cite{doddato113},\cite{kusenko97,kusenko98}, generally forming from scalar condensates in the flat directions of supersymmetric potentials. They are often incorporated into supersymmetric theories in conjunction with Affleck-Dine baryogenesis \cite{affleckdine} - \cite{dine95}  as a dark matter candidate, either as complete Q-balls or as a source of particle dark matter when the Q-balls decay \cite{enqvist98} - \cite{kasuya001}. The globally charged nature of Q-balls makes them a good dark matter candidate \cite{fujii01} - \cite{doddato113}, \cite{kusenko97} - \cite{troitsky15}, \cite{kusenko01} - \cite{shoemaker092} both as compact objects or as the seeds of particle dark matter from decays. The possibility of dark matter being accounted for by Primordial Black Holes seeded from the collapse of supersymmetric Q-balls has also recently been considered \cite{kusenko21}.

Many numerical simulations have been performed to model the formation of Q-balls from scalar condensates \cite{kasuya004} - \cite{hiramatsu10}. In \cite{hiramatsu10}, Hiramatsu et al showed that the fragmentation of a neutral condensate in a model based on a complex scalar first forms oscillons which then fragment to $\pm$ Q-ball pairs. Using Hiramatsu et al's results \cite{hiramatsu10} as a benchmark, we will consider the possibility that the Q-balls in this model may also form from the fragmentation of the inflaton condensate into neutral lumps of scalars, which then decay into pairs of $\pm$ Q-balls.

The fragmentation of a scalar condensate is a complex process which depends heavily on the underlying particle physics and requires detailed numerical simulations to model. As such, a quantitative description of the process is beyond the scope of the research presented in this thesis. However, it is possible that the fragmentation of the inflaton condensate in this case could arise as a result of tachyonic preheating \cite{kofman01} - \cite{tomberg21}.

Tachyonic preheating is the process by which an inflaton condensate becomes unstable due to the amplified growth of specific wavelengths of perturbations arising due to a tachyonic instability of the inflaton potential. This can lead to the formation of overdensities within a previously homogeneous condensate, which - if soliton solutions are admissible in the underlying field theory - can lead to the condensate breaking apart to form solitons or other related objects \cite{amin102,kasuya03}, the precise nature of which depends on the nature of the inflaton field itself.

In the work this chapter is based on we were interested in exploring Q-balls in a non-SUSY context. More specifically, we were interested in whether Q-balls composed of inflaton scalars could be produced in a model of non-minimally coupled Palatini inflation, possibly as a result of tachyonic preheating, whether the model could inflate and reheat successfully in conjunction with the production of Q-balls, what the properties of these Q-balls would be, how they would compare to the Q-balls of Coleman's solution, and what the broader implications for cosmology could be from the presence of these Q-balls.

\section{The Model}\label{section:43}

In this section we introduce the non-minimally coupled Palatini inflation model within which we want to produce inflatonic Q-balls. This model describes inflation driven by a complex scalar inflaton $\Phi$ charged under a global $U(1)$ symmetry and non-minimally coupled to gravity. In particular, it differs from existing analyses of Palatini inflation by the inclusion of an inflaton mass term, which has not been previously considered. We use the $\left( +, -, -, - \right)$ convention for the metric and $M_{pl}$ should be taken to be the reduced Planck mass.

The Jordan frame inflaton action is given by

\begin{equation}\label{eqn:q1}
S_{J} = \int d^{4}x \sqrt{-g} \left[ -\frac{1}{2}M_{pl}^{2}\left( 1 + \frac{2\xi \mid \Phi \mid^{2}}{M_{pl}^{2}}\right)R + \partial_{\mu}\Phi^{ \dagger} \partial^{\mu}\Phi - V\left(\mid \Phi \mid \right) \right],
\end{equation}

\noindent where the Jordan frame potential is 

\begin{equation}\label{eqn:q2}
V\left( \mid \Phi \mid \right) = m^{2} \mid \Phi \mid^{2} + \lambda \mid \Phi \mid^{4},
\end{equation}

\noindent where $m$ is the inflaton mass and $\lambda$ is the inflaton self-coupling. In order to recast the model in terms of conventional General Relativity, we perform a conformal transformation on the spacetime metric, given by

\begin{equation}\label{eqn:q3}
g_{\mu \nu} \longrightarrow \tilde{g}_{\mu \nu} = \Omega^{2}g_{\mu \nu},
\end{equation}

\noindent where the conformal factor $\Omega^{2}\left(\Phi \right)$ is

\begin{equation}\label{eqn:q4}
\Omega^{2} = 1 + \frac{2\xi \left| \Phi \right|^{2}}{M_{pl}^{2}}.
\end{equation}

In the Palatini formulation, the Ricci scalar transforms as (from (\ref{eqn:r17}))

\begin{equation}\label{eqn:q5}
R \longrightarrow \tilde{R} = \frac{R}{\Omega^{2}},
\end{equation}

\noindent and the integration measure, $\sqrt{-g}$ in (\ref{eqn:q1}), transforms as 

\begin{equation}\label{eqn:q6}
\sqrt{-det \left(\frac{\tilde{g}_{\mu\nu}}{\Omega^{2}}\right)} = \sqrt{-\frac{1}{\Omega^{8}}det\left(\tilde{g}_{\mu\nu}\right)} = \frac{1}{\Omega^{4}}\sqrt{-det \tilde{g}_{\mu\nu}} = \frac{\sqrt{-\tilde{g}}}{\Omega^{4}}.
\end{equation}

The conformal transformation (\ref{eqn:q3}) acts purely on the metric of the theory. This means that the transformation does not act on coordinates and the transformation on the derivative term in (\ref{eqn:q1}) is due to the implicit factor of the metric used in the contraction of the indices

\begin{equation}\label{eqn:q7}
\partial_{\mu}\Phi^{\dagger} \partial^{\mu}\Phi = g_{\mu \nu}\partial^{\nu}\Phi^{\dagger} \partial^{\mu}\Phi,
\end{equation}

\begin{equation}\label{eqn:q8}
\Rightarrow \tilde{g}_{\mu \nu}\partial^{\nu}\Phi^{\dagger} \partial^{\mu}\Phi = g_{\mu \nu}\Omega^{2}\partial^{\nu}\Phi^{\dagger} \partial^{\mu}\Phi = \Omega^{2}\partial_{\mu}\Phi^{\dagger} \partial^{\mu}\Phi.
\end{equation}

\noindent The transformed action is then

\begin{equation}\label{eqn:q9}
S_{E} = \int d^{4}x \sqrt{-\tilde{g}} \left[ -\frac{1}{2}M_{pl}^{2}\tilde{R} + \frac{1}{\Omega^{2}}\partial_{\mu}\Phi^{ \dagger} \partial^{\mu}\Phi - V_{E}\left(\mid \Phi \mid \right) \right],
\end{equation}

\noindent where we are henceforth working in the Einstein frame (denoted by a subscript 'E' for the action or potential, or a tilde for geometric or other quantities). Our Einstein frame potential is then defined as

\begin{equation}\label{eqn:q10}
V_{E}\left(\mid \Phi \mid \right) = \frac{V\left(\mid \Phi \mid \right)}{\Omega^{4}}.
\end{equation}

As in Chapter 4, all calculations are performed in the Einstein frame in this chapter unless explicitly stated otherwise.\\

\section{Slow Roll Parameters and Inflationary Observables}\label{section:44}

In this section we calculate the observables for non-minimally coupled Palatini inflation for the potential (\ref{eqn:q10}) and check the compatibility of this model with the observations from the Planck satellite experiment. For ease of calculation we write the complex inflaton field in the form

\begin{equation}\label{eqn:q11}
\Phi = \frac{\phi}{\sqrt{2}}e^{i\theta},
\end{equation}

\noindent which means that the conformal factor (\ref{eqn:q4}) becomes

\begin{equation}\label{eqn:q12}
\Omega^{2} = 1 + \frac{\xi \phi^{2}}{M_{pl}^{2}},
\end{equation}

\noindent and the Jordan frame potential is 

\begin{equation} \label{eqn:q13}
V\left( \phi \right) = \frac{1}{2}m^{2}\phi^{2} + \frac{\lambda}{4}\phi^{4}.
\end{equation}

\noindent The Einstein frame potential in terms of (\ref{eqn:q11}) is thus

\begin{equation}\label{eqn:q14}
V_{E}\left( \phi \right) = \frac{1}{2\Omega^{4}}m^{2}\phi^{2} + \frac{\lambda}{4\Omega^{4}}\phi^{4} = \frac{m^{2}\phi^{2}}{2\left( 1 + \frac{\xi \phi^{2}}{M_{pl}^{2}}\right)^{2}} + \frac{\lambda \phi^{4}}{4\left( 1 + \frac{\xi \phi^{2}}{M_{pl}^{2}}\right)^{2}},
\end{equation}

\begin{equation}\label{eqn:q15}
\Rightarrow V_{E} = \frac{\lambda M_{pl}^{4}}{4 \xi^{2}\left( 1 + \frac{M_{pl}^{2}}{\xi \phi^{2}}\right)^{2}}\left[ 1 + \frac{2 m^{2}}{\lambda \phi^{2}}\right].
\end{equation}

\noindent Since we are assuming that inflation takes place upon the plateau of the potential, and we expect the field to be very large in this regime, we define what we will refer to as "the plateau limit", where

\begin{equation}\label{eqn:q16}
\frac{\xi \phi^{2}}{M_{pl}^{2}} >> 1.
\end{equation}

\noindent It is assumed throughout this chapter that this approximation is a reflection of the dominant dynamics during and at the instantaneous end of slow-roll inflation, and more generally while the inflaton is otherwise on its plateau. 

\noindent From (\ref{eqn:q16}), we have that $M_{pl}^{2}/\xi \phi^{2} << 1$, and can therefore approximate

\begin{equation}\label{eqn:q17}
\left( 1 + \frac{M_{pl}^{2}}{\xi \phi^{2}}\right)^{-2} \thickapprox 1 - 2\frac{M_{pl}^{2}}{\xi \phi^{2}},
\end{equation}

\begin{equation}\label{eqn:q18}
\Rightarrow V_{E} = \frac{\lambda M_{pl}^{4}}{4 \xi^{2}}\left[ 1 + \frac{2 m^{2}}{\lambda \phi^{2}}\right] \left( 1 - \frac{2M_{pl}^{2}}{\xi \phi^{2}} \right).
\end{equation}

\noindent To leading order in small quantities, the Einstein frame potential is,

\begin{equation}\label{eqn:q19}
V_{E} = \frac{\lambda M_{pl}^{4}}{4\xi^{2}}\left[ 1 + \frac{2m^{2}}{\lambda \phi^{2}} - \frac{2M_{pl}^{2}}{\xi \phi^{2}} \right],
\end{equation}

\noindent which can be rewritten as

\begin{equation}\label{eqn:q20}
V_{E} = \frac{\lambda M_{pl}^{4}}{4\xi^{2}}\left[ 1 - \frac{2M_{pl}^{2}}{\xi \phi^{2}}\beta \right],
\end{equation}

\noindent where $\beta$ is defined as

\begin{equation}\label{eqn:q21}
\beta = 1 - \frac{\xi m^{2}}{\lambda M_{pl}^{2}}.
\end{equation}

To canonically normalise the inflaton field, we perform the field rescaling

\begin{equation}\label{eqn:q22}
\frac{d\sigma}{d\phi} = \frac{1}{\sqrt{1 + \frac{\xi \phi^{2}}{M_{pl}^{2}}}},
\end{equation}

\begin{equation}\label{eqn:q23}
\Rightarrow \int d\sigma = \int  \frac{d\phi}{\sqrt{1 + \frac{\xi \phi^{2}}{M_{pl}^{2}}}}.
\end{equation}

\noindent Writing $a = \xi/M_{pl}^{2}$ we can use

\begin{equation}\label{eqn:q24}
\int  \frac{dx}{\sqrt{1 + ax^{2}}} = \frac{1}{\sqrt{a}}\sinh^{-1} \left(\sqrt{a}x\right) + C,
\end{equation}

\noindent and the rescaled scalar field is therefore

\begin{equation}\label{eqn:q25}
\sigma \left(\phi \right) = \frac{M_{pl}}{\sqrt{\xi}}\sinh^{-1} \left(\frac{\sqrt{\xi}}{M_{pl}}\phi\right),
\end{equation}

\noindent where $\sigma \rightarrow \phi$ as $\phi \rightarrow 0$. It is important to note that this is a field rescaling and not a truly canonically normalised scalar field, due to the complex nature of the field $\Phi$. When discussing inflation in this model specifically, it can be regarded as a "canonical inflaton" since the inflation dynamics are determined by the radial component, $\phi$, of the inflaton field. Elsewhere in this chapter, $\sigma$ should be taken as a field redefinition, and the rescaling (\ref{eqn:q22}) will be used multiple times in calculations throughout this chapter.

It follows from (\ref{eqn:q25}) that

\begin{equation}\label{eqn:q26}
\phi \left(\sigma \right) = \frac{M_{pl}}{\sqrt{\xi}}\sinh \left(\frac{\sqrt{\xi}}{M_{pl}}\sigma \right).
\end{equation}

\noindent The hyperbolic sine function is 

\begin{equation}\label{eqn:q27}
\sinh\left(x \right) = \frac{e^{x} - e^{-x}}{2},
\end{equation}

\noindent and we can therefore rewrite (\ref{eqn:q26}) as

\begin{equation}\label{eqn:q28}
\phi \left(\sigma \right) = \frac{M_{pl}}{2\sqrt{\xi}}\left(e^{\frac{\sqrt{\xi}}{M_{pl}}\sigma} - e^{-\frac{\sqrt{\xi}}{M_{pl}}\sigma} \right).
\end{equation}

\noindent The rescaled field will be large in the same regime as the physical inflaton field. From (\ref{eqn:q16}), we can say that the exponential arguments of (\ref{eqn:q28}) will be large during inflation. This means that the decreasing exponential will be significantly smaller than the increasing exponential while on the plateau. During inflation, we can therefore say that the inflaton field in terms of the canonically normalised scalar is

\begin{equation}\label{eqn:q29}
\phi = \frac{M_{pl}}{2\sqrt{\xi}} e^{\frac{\sqrt{\xi}}{M_{pl}}\sigma},
\end{equation}

\noindent to a good approximation. Substituting (\ref{eqn:q29}) into the Einstein frame potential (\ref{eqn:q20}) gives

\begin{equation}\label{eqn:q30}
V_{E}\left(\sigma \right) = \frac{\lambda M_{pl}^{4}}{4\xi^{2}}\left[ 1 - 8\beta e^{-2\frac{\sqrt{\xi}}{M_{pl}}\sigma}  \right].
\end{equation}

\begin{figure}[H]
\begin{center}
\includegraphics[clip = true, width=\textwidth, angle = 360]{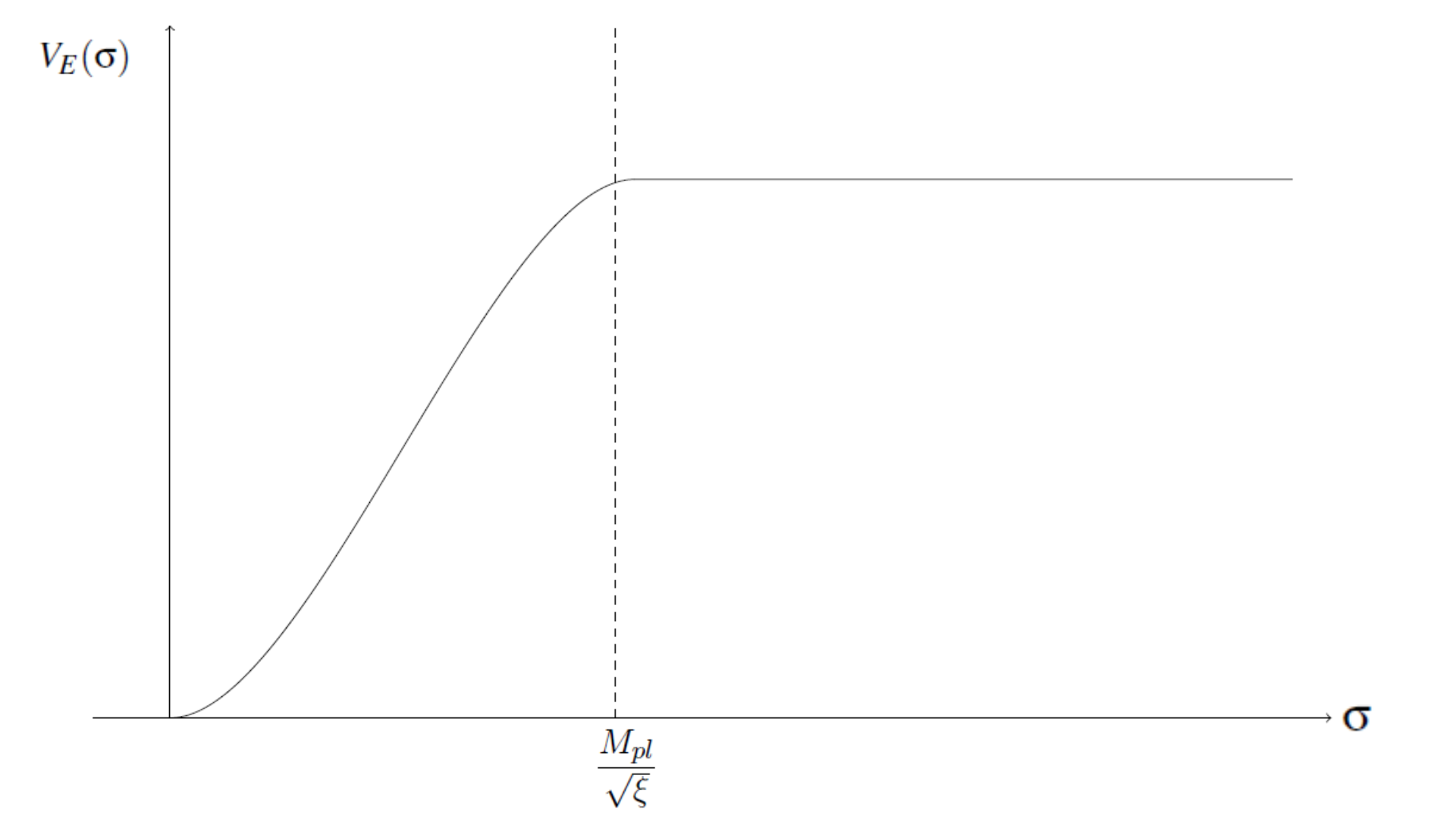}
\caption{Schematic of the inflaton potential in terms of the quasi-canonical field $\sigma$ indicating the edge of the plateau, as used in defining the plateau limit used in this research. } 
\label{figure:n16}
\end{center}
\end{figure}

\noindent This is the form of the inflaton potential in the Einstein frame which we will use in deriving the expressions for the inflationary observables (see Figure \ref{figure:n16}). In existing analyses of Palatini inflation, $m = 0$ and $\beta = 1$. Here we are generalising to $m \neq 0$ and $\beta \neq 1$.

The slow roll parameters are given by

\begin{equation}\label{eqn:q31}
\epsilon = \frac{M_{pl}^{2}}{2}\left(\frac{V_{E}'}{V_{E}} \right)^{2},
\end{equation}

\begin{equation}\label{eqn:q32}
\eta = M_{pl}\frac{V_{E}''}{V_{E}},
\end{equation}

\noindent where a prime $'$ denotes a derivative with respect to a field, $\phi$, in this case. Working with the assumption that $\phi$ is large we approximate 

\begin{equation}\label{eqn:q33}
V_{E} \thickapprox \frac{\lambda M_{pl}^{4}}{4\xi^{2}},
\end{equation}

\noindent while on the plateau. This is another approximation which will be used in this work when the plateau limit is invoked. The first and second derivatives of (\ref{eqn:q30}) are

\begin{equation}\label{eqn:q34}
\frac{\partial V_{E}}{\partial \sigma} = \frac{\lambda M_{pl}^{4}}{4\xi^{2}}\cdot \left(-8\beta \right) \cdot \left(-\frac{2\sqrt{\xi}}{M_{pl}}\right) e^{-2\frac{\sqrt{\xi}}{M_{pl}}\sigma} = \frac{\lambda M_{pl}^{4}}{4\xi^{2}}\left(\frac{16\beta \sqrt{\xi}}{M_{pl}}\right) e^{-2\frac{\sqrt{\xi}}{M_{pl}}\sigma},
\end{equation}

\begin{equation}\label{eqn:q35}
\frac{\partial^{2}V_{E}}{\partial \sigma^{2}} =  \frac{\lambda M_{pl}^{4}}{4\xi^{2}}\left(-\frac{32\beta \xi}{M_{pl}^{2}}\right) e^{-2\frac{\sqrt{\xi}}{M_{pl}}\sigma}.
\end{equation}

Using (\ref{eqn:q34}), (\ref{eqn:q35}) the slow roll parameters are thus

\begin{equation}\label{eqn:q36}
\epsilon = \frac{M_{pl}^{2}}{2}\left( \frac{4\xi^{2}}{\lambda M_{pl}^{4}} \cdot \frac{\lambda M_{pl}^{4}}{4\xi^{2}}\left(\frac{16\beta \sqrt{\xi}}{M_{pl}}\right) e^{-2\frac{\sqrt{\xi}}{M_{pl}}\sigma}\right)^{2} = 128 \beta^{2} \xi e^{-4\frac{\sqrt{\xi}}{M_{pl}}\sigma},
\end{equation}

\begin{equation}\label{eqn:q37}
\eta = M_{pl}^{2} \frac{4\xi^{2}}{\lambda M_{pl}^{4}}\cdot \frac{\lambda M_{pl}^{4}}{4\xi^{2}}\left(-\frac{32\beta \xi}{M_{pl}^{2}}\right) e^{-2\frac{\sqrt{\xi}}{M_{pl}}\sigma} = -32\beta \xi e^{-2\frac{\sqrt{\xi}}{M_{pl}}\sigma}.
\end{equation}

The number of e-folds of inflation is calculated using

\begin{equation}
\begin{split}\label{eqn:q38}
N = -\frac{1}{M_{pl}^{2}}\int^{\sigma_{end}}_{\sigma} & \frac{V_{E}}{V_{E}'} d\sigma  = -\frac{1}{M_{pl}^{2}}\int^{\sigma_{end}}_{\sigma} \frac{M_{pl}}{12\beta \sqrt{\xi}}\exp \left(\frac{2\sqrt{\xi}}{M_{pl}}\sigma \right) d\sigma \\
& = \frac{1}{32\xi \beta} \left[ \exp \left(\frac{2\sqrt{\xi}}{M_{pl}}\sigma \right) - \exp \left(\frac{2\sqrt{\xi}}{M_{pl}}\sigma_{end} \right)\right].
\end{split}
\end{equation}
\noindent Working with the assumption that $\sigma_{end} << \sigma $, the number of e-folds is given by

\begin{equation}\label{eqn:q39}
N = \frac{1}{32\xi \beta}\exp \left(\frac{2\sqrt{\xi}}{M_{pl}}\sigma \right).
\end{equation}

\noindent This can be rewritten to give an expression for $\sigma \left(N \right)$

\begin{equation}\label{eqn:q40}
\sigma \left(N \right) = \frac{M_{pl}}{2\sqrt{\xi}} \ln \left(32\xi \beta N \right).
\end{equation}

\noindent Substituting (\ref{eqn:q40}) into (\ref{eqn:q36}) and (\ref{eqn:q37}) to find $\eta$ and $\epsilon$ in terms of the number of e-folds of inflation, we find 

\begin{equation}\label{eqn:q41}
\epsilon = 128 \beta^{2} \xi \exp \left[-4\frac{\sqrt{\xi}}{M_{pl}} \left(\frac{M_{pl}}{2\sqrt{\xi}}\ln \left(32\xi \beta N \right)\right)\right] = \frac{1}{8\xi N^{2}},
\end{equation}

\begin{equation}\label{eqn:q42}
\eta = -32\beta \xi \exp \left[-2\frac{\sqrt{\xi}}{M_{pl}} \left(\frac{M_{pl}}{2\sqrt{\xi}} \ln \left(32\xi \beta N \right)\right)\right] = -\frac{1}{N}.
\end{equation}

\noindent These are the standard expressions for $\epsilon$ and $\eta$ in Palatini inflation, and are independent of the $\beta$ parameter. This means that $\eta$ and $\epsilon$ are not affected by the presence of the mass term during inflation, provided that $\beta > 0$.

It is instructive to examine the size of the $\epsilon$ parameter relative to the $\eta$ parameter in this model to check the consistency of the slow-roll approximation in this inflation model. From Section \ref{section:082} we have that slow-roll inflation ends when $\left| \eta \right| = 1$, and from (\ref{eqn:q42}) we can see that the end of slow-roll corresponds to $N = 1$. Using (\ref{eqn:q41}), we find that the value at the end of slow-roll inflation is $\epsilon = 1.03 \times 10^{-10}$, which is significantly smaller than the $\eta$ parameter. This shows that this model is consistent with slow-roll inflation, and that $\epsilon << \eta$ throughout.

The scalar spectral index, $n_{s}$ in terms of $\eta$ and $\epsilon$ is

\begin{equation}\label{eqn:q43}
n_{s} = 1 + 2\eta - 6\epsilon.
\end{equation}

\noindent Since $\epsilon << \eta$ in this case we approximate

\begin{equation}\label{eqn:q44}
n_{s} \approx 1 - \frac{2}{N}.
\end{equation}

\noindent The tensor-to-scalar ratio $r$ is 

\begin{equation}\label{eqn:q45}
r \approx 16\epsilon = 16 \cdot \frac{1}{8\xi N^{2}} = \frac{2}{\xi N^{2}}.
\end{equation}

\noindent  and the primordial curvature power spectrum is

\begin{equation}\label{eqn:q46}
\mathcal{P}_{\mathcal{R}} = \frac{V_{E}}{24\pi^{2} \epsilon M_{pl}^{4}} =  \frac{\lambda}{12\xi \pi^{2}} N^{2}.
\end{equation}

\noindent From the results of the Planck satellite experiment \cite{planck184} (2018), the observed amplitude of the power spectrum is $A_{s} = 2.1 \times 10^{-9}$. Using $\lambda = 0.1$ and $N = 55$ as an estimate of the self-coupling and the pivot scale, (\ref{eqn:q46}) gives a non-minimal coupling of  $\xi = 1.2163 \times 10^{9}$. Using the same estimate, the inflationary observables are $n_{s} = 0.9636$ and $r = 6.01 \times 10^{-13}$. The scalar spectral index is within the bounds of the 2018 results from the Planck satellite (assuming $\Lambda$CDM and no running of the spectral index), $n_{s} = 0.9649 \pm 0.0042$ (1-$\sigma$) \cite{planck184}, whilst the tensor-to-scalar ratio is heavily suppressed, as is typically the case in Palatini inflation models. These values of the observables are only an estimate of the predictions of the model since, as we will discuss further in Section \ref{section:481} - \ref{section:410}, the post-inflationary cosmology of this model may have an effect on the number of e-folds of inflation needed for the model to inflate successfully, and consequently the location of the pivot scale, leading to an adjustment in $n_{s}$ and $r$. 

The inflaton field can be expressed in terms of the number of e-folds as 

\begin{equation}\label{eqn:q47}
\phi \left(N \right) = \frac{M_{pl}}{2\sqrt{\xi}} \exp \left(\frac{\sqrt{\xi}}{M_{pl}} \left( \frac{M_{pl}}{2\sqrt{\xi}} \ln \left(32\xi \beta N \right)\right)\right) = 2\sqrt{2}M_{pl} \sqrt{\beta N},
\end{equation}

\noindent so the mass term directly influences the amplitude of the field throughout inflation. Defining the end of slow-roll inflation to be when $\left| \eta \right| \approx 1$, giving $N_{end} \approx 1$, the value of the field at the end of slow-roll inflation is then

\begin{equation}\label{eqn:q48}
\phi_{end} = 2\sqrt{2}M_{pl} \sqrt{\beta}.
\end{equation}

\noindent $\beta$ will be in the range $0.1-1$ for the values of inflaton mass relevant here. In the following section we consider limits on the inflaton mass in this model needed to produce successful inflation, and we consider how this governs the effect of the mass term on the value of the field at the end of slow-roll inflation.

\subsection{The Upper Bound on the Inflaton Mass }\label{section:441}

For the purposes of the research presented here, we are primarily interested in Q-ball solutions which are compatible with non-minimally coupled Palatini inflation. Since the Q-balls will be formed from inflaton scalars, the effects of the inflaton mass from the presence of the $\beta$ parameter on the Einstein frame potential, and subsequently on the existence of Q-ball solutions, must be established, including the range of inflaton masses for which inflation is possible. It is important to ascertain the limits placed on the inflaton mass by the necessity that the potential be compatible with inflation first, in order to later establish the range of inflaton masses for which Q-ball solutions are compatible with Palatini inflation.

We start from the Einstein frame potential 

\begin{equation}\label{eqn:q49}
V_{E}\left(\phi \right) = \frac{m^{2} \phi^{2}}{2\Omega^{4}} + \frac{\lambda \phi^{4}}{4\Omega^{4}},
\end{equation}

\noindent and differentiate to give

\begin{equation}\label{eqn:q50}
\frac{\partial V_{E}}{\partial \phi} = \frac{1}{\Omega^{6}}\left[m^{2}\phi + \left(\lambda - \frac{\xi m^{2}}{M_{pl}^{2}}\right) \phi^{3} \right].
\end{equation}

The inflaton potential must have a positive gradient with respect to the inflaton for inflation to occur, we therefore require

\begin{equation}\label{eqn:q51}
\frac{\partial V_{E}}{\partial \phi} > 0    \Rightarrow \lambda - \frac{\xi m^{2}}{M_{pl}^{2}} > 0.
\end{equation}

\noindent This means that, in order for inflation to occur in this non-minimally coupled Palatini model, the inflaton mass must obey the upper bound

\begin{equation} \label{eqn:q52}
m^{2} < \frac{\lambda M_{pl}^{2}}{\xi},
\end{equation}

\noindent where the upper limit on $m^{2}$ corresponds to $\beta =0$. For the purposes of studying these Q-balls numerically, we will consider masses which give a value for $\beta$ in the range $\sim 0.1 - 1$, which correspond to inflaton field values at the end of inflation (approximated as the end of slow-roll) in the range $\phi_{end} \sim \left(1 - 3 \right) M_{pl}$. The size of this mass term directly affects the shape of the potential, but will only have a significant effect on the value of the field at the end of inflation for inflation masses squared close to the upper limit (\ref{eqn:q52}).

\section{Derivation of the Q-ball Equation}\label{section:45}

In this section we derive the Q-ball equation in non-minimally coupled Palatini gravity. We work in the Einstein frame in flat space with the action

\begin{equation} \label{eqn:q53}
S = \int d^{4}x \; \; \frac{1}{\Omega^{2}}\partial_{\mu}\Phi^{\dagger} \partial^{\mu}\Phi - \frac{1}{\Omega^{4}}V\left(\left| \Phi \right| \right).
\end{equation}

We will show that the flat space calculation is a valid approximation for the Q-ball solutions that we will obtain. While it is true that gravitational effects can affect the stability of Q-balls within certain limits, or alter the size of the Q-balls (see e.g. \cite{multamaki022}, \cite{tamaki11}), the attractive interaction between scalars dominates over any gravitational effects for Q-balls of the size we are working with in this model. The effects of gravity, and indeed the presence of a non-minimal coupling of gravity to the scalar field in the Jordan frame are factors to consider in the post-inflationary cosmology of this model and may affect the survival of any relic Q-balls to the present day. We we will later consider a first-pass approximation of the effects of curvature on these Q-ball solutions in Section \ref{section:49} of this chapter, and proceed with the understanding that the purpose of this work is to ascertain the flat space properties of these Q-balls. Expansion could affect the formation of these Q-balls and their subsequent evolution, and this is something we touch upon in Sections \ref{section:48} - \ref{section:411}. For the purposes of establishing the existence of Q-balls made from scalars with a non-minimal coupling to gravity without the cosmological considerations of their existence we derive the Q-ball equation using the energy-momentum tensor for a scalar field in a non-expanding spacetime, and compare these Q-balls to the conventional flat space Q-balls as derived in \cite{coleman85}.

In order to derive the Q-ball equation we begin by using the method of Lagrange multipliers in order to minimise the energy of the inflaton field with respect to the conserved $U(1)$ charge of the field 

\begin{equation} \label{eqn:q54}
E_{Q} = E + \omega \left( Q - \int d^{3}x \rho_{Q} \right),
\end{equation}

\noindent which gives a "Q-ball action", $E_{Q}$, which we will refer to as the Q-ball energy functional. This can be extremised to obtain the field equations of the theory, the solutions of which - subject to a number of conditions which we discuss in Sections \ref{section:451} - \ref{section:453} - correspond to Q-balls. The global energy, $E$, and charge, $Q$, are

\begin{equation} \label{eqn:q55}
Q = \int d^{3}x \; j^{0} = \int d^{3}x \rho_{Q},
\end{equation}

\begin{equation} \label{eqn:q56}
E = \int d^{3}x \;  T^{00} = \int d^{3}x  \rho_{E},
\end{equation}

\noindent where $j^{0}$ is the temporal component of the conserved $U(1)$ Noether current, $j^{\mu}$, and $T^{00}$ is the temporal component of the energy-momentum tensor $T^{\mu \nu}$ for a complex scalar field, given by

\begin{equation} \label{eqn:q57}
T^{\mu \nu} = \frac{\partial \mathcal{L}}{\partial \left(\partial_{\mu}\phi_{a}\right)}\eta^{\nu \rho}\partial_{\rho}\phi_{a} - \delta^{\mu}_{\rho} \eta^{\nu \rho}\mathcal{L},
\end{equation}

\noindent where $\eta^{\nu \rho}$ is the Minkowski metric. The energy density of the $\Phi$ field is therefore

\begin{equation} \label{eqn:q58}
\rho_{E} = T^{00} = \frac{1}{\Omega^{2}}\partial_{t}\Phi^{\dagger} \partial_{t}\Phi + \frac{1}{\Omega^{2}}\partial_{i}\Phi^{\dagger} \partial^{i}\Phi + \frac{V(\mid \Phi \mid)}{\Omega^{4}}.
\end{equation}

\noindent This includes the effect of the non-canonical kinetic term of $\Phi$, which has not been previously studied in the context of Q-ball solutions.

The conserved Noether current $j^{\mu}$ of the $U(1)$ symmetry of the model is defined by

\begin{equation} \label{eqn:q59}
\partial_{\mu} j^{\mu} = 0, j^{\mu} = \frac{\partial \mathcal{L}}{\partial \left(\partial_{\mu}\phi_{a}\right)}\delta \phi_{a}.
\end{equation}

\noindent The temporal components of (\ref{eqn:q59}) are

\begin{equation} \label{eqn:q60}
j^{0} = \frac{\partial \mathcal{L}}{\partial \left(\partial_{t}\Phi \right)} i\Phi - \frac{\partial \mathcal{L}}{\partial \left(\partial_{t} \Phi^{\dagger}\right)} i\Phi^{\dagger},
\end{equation}

\noindent which from (\ref{eqn:q53}) gives the charge density of the inflaton field to be

\begin{equation} \label{eqn:q61}
\rho_{Q} = \frac{i}{\Omega^{2}}\left( \Phi \partial_{t}\Phi^{\dagger} - \Phi^{\dagger}\partial_{t}\Phi \right).
\end{equation}

\noindent Substituting (\ref{eqn:q58}) and (\ref{eqn:q61}) into the Q-ball energy functional (\ref{eqn:q54}) gives

\begin{equation} \label{eqn:q62}
E_{Q} = \int d^{3}x \left[ \frac{1}{\Omega^{2}}\partial_{t}\Phi^{\dagger} \partial_{t}\Phi + \frac{1}{\Omega^{2}}\partial_{i}\Phi^{\dagger} \partial^{i}\Phi + \frac{V(\left|\Phi \right|)}{\Omega^{4}} - \frac{\omega i}{\Omega^{2}}\left( \Phi \partial_{t}\Phi^{\dagger} - \Phi^{\dagger}\partial_{t}\Phi \right) \right] + \omega Q.
\end{equation}

\noindent The temporal derivative terms in (\ref{eqn:q62}) can be rewritten as

\begin{equation}
\begin{split} \label{eqn:q63}
\left(\partial_{t}\Phi - i\omega \Phi \right) \left(\partial_{t}\Phi^{\dagger} + i\omega \Phi^{\dagger} \right) 
& = \partial_{t}\Phi \partial_{t}\Phi^{\dagger} + i\omega \Phi^{\dagger} \partial_{t}\Phi - i\omega \Phi \partial_{t} \Phi^{\dagger} - \omega^{2}\Phi^{\dagger}\Phi \\
& = \mid \partial_{t}\Phi - i\omega \Phi \mid^{2},
\end{split}
\end{equation}

\noindent in order to rewrite the Q-ball energy functional in a more insightful way. If we define the time derivatives in the integrand of (\ref{eqn:q62}) as

\begin{equation} \label{eqn:q64}
I = \frac{1}{\Omega^{2}}\left[ \partial_{t}\Phi^{\dagger}\partial_{t}\Phi + i\omega \Phi^{\dagger}\partial_{t}\Phi - i\omega \Phi \partial_{t}\Phi^{\dagger} \right],
\end{equation}

\noindent these can be rewritten as in (\ref{eqn:q63})

\begin{equation} \label{eqn:q65}
I = \frac{1}{\Omega^{2}}\left[ \mid \partial_{t}\Phi - i\omega \Phi \mid^{2} - \omega^{2} \mid \Phi \mid^{2} \right].
\end{equation}

\noindent Substituting (\ref{eqn:q65}) back into the Q-ball energy functional (\ref{eqn:q62}) we have that

\begin{equation} \label{eqn:q66}
E_{Q} = \int d^{3}x \left[ \frac{1}{\Omega^{2}} \mid \partial_{t}\Phi - i \omega \Phi \mid^{2} - \frac{1}{\Omega^{2}} \omega^{2} \mid \Phi \mid^{2} + \frac{1}{\Omega^{2}}\partial_{i}\Phi ^{\dagger} \partial^{i} \Phi + \frac{1}{\Omega^{4}}V\left(\mid \Phi \mid \right) \right] + \omega Q,
\end{equation}

\noindent whereupon it is clear that a solution which extremises $E_{Q}$ is

\begin{equation} \label{eqn:q67}
\Phi \left(x,t \right) = \Phi \left( x \right) e^{i \omega t} \longrightarrow \partial_{t} \Phi = i\omega \Phi.
\end{equation}

\noindent Applying (\ref{eqn:q67}), the Q-ball energy functional (\ref{eqn:q62}) becomes

\begin{equation} \label{eqn:q68}
E_{Q} = \int d^{3}x \left[ \frac{1}{\Omega^{2}} \mid \overrightarrow{\nabla} \Phi \mid^{2} + \frac{1}{\Omega^{4}} V\left(\mid \Phi \mid \right) - \frac{1}{\Omega^{2}}\omega^{2} \mid \Phi \mid^{2} \right].
\end{equation}

\noindent Let

\begin{equation} \label{eqn:q69}
V_{\omega} \left( \mid \Phi \mid \right) = \frac{1}{\Omega^{4}}V \left( \mid \Phi \mid \right) - \frac{1}{\Omega^{2}}\omega^{2}\mid \Phi \mid^{2},
\end{equation}

\noindent be defined as the Q-ball potential. Assuming that the Q-balls are spherically symmetric, we assume that the coordinate dependence of the inflaton field, when extremised with respect to its charge, is purely radial, which gives 

\begin{equation} \label{eqn:q70}
\Phi \left(\textbf{x}\right) = \frac{\phi \left(\textbf{r}\right)}{\sqrt{2}} e^{i\omega t} = \frac{\phi\left(r\right) \hat{r}}{\sqrt{2}}e^{i\omega t},
\end{equation}

\noindent as the field solution corresponding to Q-balls.

Since the coordinate dependence of the field is purely radial, the gradient operator is reduced to

\begin{equation} \label{eqn:q71}
\overrightarrow{\nabla}\Phi = \frac{\partial \Phi}{\partial r}\hat{r} \\
\longrightarrow \frac{1}{\Omega^{2}} \left| \overrightarrow{\nabla}\Phi \right| = \frac{1}{\Omega^{2}}\left| \frac{1}{\sqrt{2}}\frac{\partial \phi}{\partial r}\hat{r} \right|^{2} = \frac{1}{2\Omega^{2}}\left(\frac{\partial \phi}{\partial r}\right)^{2}.
\end{equation}

\noindent Rewriting in spherical polar coordinates, we have that the Q-ball energy functional is

\begin{equation} \label{eqn:q72}
E_{Q} = \int dr \; 4\pi r^{2} \left[ \frac{1}{2\Omega^{2}}\left(\frac{\partial \phi}{\partial r}\right)^{2} + V_{\omega} \left(\phi \right) \right] + \omega Q.
\end{equation}

\noindent We define the function $\mathcal{L}_{Q}$ (which we refer to as the "Q-ball effective Lagrangian") to be extremised as

\begin{equation} \label{eqn:q73}
\mathcal{L}_{Q}= 4\pi r^{2} \left[ \frac{1}{2\Omega^{2}}\left(\frac{\partial \phi}{\partial r}\right)^{2} + V_{\omega} \left(\phi \right) \right],
\end{equation}

\noindent and apply the Euler-Lagrange equations (derived for a complex scalar field in flat space in (\ref{eqn:b96}))

\begin{equation} \label{eqn:q74}
\frac{\partial \mathcal{L}_{Q}}{\partial \phi} - \frac{d}{dr}\left(\frac{\partial \mathcal{L}_{Q}}{\partial \left(\partial_{r}\phi\right)}\right) = 0,
\end{equation}

\noindent in order to extremise (\ref{eqn:q72}).

\noindent Using (\ref{eqn:q73}), the second term of (\ref{eqn:q74}) is

\begin{multline} \label{eqn:q75}
\frac{d}{dr}\left(\frac{\partial \mathcal{L}_{Q}}{\partial \left(\partial_{r}\phi\right)}\right) = \frac{4\pi r^{2}}{\Omega^{2}} \frac{\partial \phi}{\partial r}\frac{d}{dr}\left(\frac{\partial \mathcal{L}_{Q}}{\partial \left(\partial_{r}\phi\right)}\right) \\
= \frac{8\pi r}{\Omega^{2}}\frac{\partial \phi}{\partial r} + \frac{4 \pi r^{2}}{\Omega^{2}}\frac{\partial^{2}\phi}{\partial r^{2}} - \frac{8 \pi r^{2}}{M_{pl}^{2}}\frac{\xi \phi}{\Omega^{4}}\left(\frac{\partial \phi}{\partial r}\right)^{2},
\end{multline}

\noindent and the first term of (\ref{eqn:q74}) is

\begin{equation} \label{eqn:q76}
\frac{\partial \mathcal{L}_{Q}}{\partial \phi} = 4\pi r^{2} \frac{\partial V_{\omega}}{\partial \phi} - \frac{4 \pi r^{2}}{M_{pl}^{2}}\frac{\xi \phi}{\Omega^{4}}\left(\frac{\partial \phi}{\partial r}\right)^{2}.
\end{equation}

\noindent Substituting (\ref{eqn:q75}) and (\ref{eqn:q76}) into (\ref{eqn:q74}), dividing through by a factor of $4\pi r^{2}$ and multiplying by $\Omega^{2}$ gives the Q-ball equation

\begin{equation} \label{eqn:q77}
\frac{\partial^{2}\phi}{\partial r^{2}} + \frac{2}{r}\frac{\partial \phi}{\partial r} - K\left(\phi \right) \left(\frac{\partial \phi}{\partial r}\right)^{2} = \Omega^{2}\frac{\partial V_{\omega}}{\partial \phi},
\end{equation}

\noindent where 

\begin{equation} \label{eqn:q78}
K(\phi) = \frac{\xi \phi}{M_{pl}^{2} \Omega^{2}}.
\end{equation}

Equation (\ref{eqn:q77}) corresponds to an entirely new and different form of the Q-ball equation than that for a scalar theory with canonical kinetic terms, and therefore to a new class of Q-ball. The result of the Q-ball equation for the minimally coupled case \cite{coleman85} can be recovered by setting $\Omega^{2} = 1 \left(\xi = 0 \right)$ 

\begin{equation}\label{eqn:q79}
\frac{\partial^{2} \phi}{\partial r^{2}} + \frac{2}{r}\frac{\partial \phi}{\partial r} = \frac{\partial V_{\omega}}{\partial \phi},
\end{equation}

\noindent where in this instance

\begin{equation}\label{eqn:q80}
V_{\omega} \left( \phi \right) = \frac{1}{2}m^{2}\phi^{2} + \frac{\lambda}{4}\phi^{4} - \frac{1}{2}\omega^{2}\phi^{2}.
\end{equation}

\noindent Where we see that the difference the presence of the non-minimal coupling for (\ref{eqn:q77}) makes as compared to (\ref{eqn:q79}) is in the additional gradient squared term, and the dependence of the conformal factor in the Q-ball potential and on the right-hand side. This shows that, although the non-minimal coupling of the scalar field to gravity is recast in the Einstein frame using the conformal transformation, its effects are still manifest in the construction of the Q-balls in the mathematical sense, in that the attractive interaction binding the scalars into Q-balls seems to contain an additional component due to the non-minimal coupling. Physically, in the Jordan frame, this would be gravitational in nature, whereas as the Q-balls are studied here in the Einstein frame, it is merely a component of the regular attractive interaction between the inflaton scalars.
\\

\subsection{Existence Conditions of Q-balls}\label{section:451}

The existence of Q-ball solutions in a given theory is heavily dependent on the form of the potential, and the realisation of physical Q-balls relies on the $\omega$ parameter being within an acceptable range. The physical reasons for this will be discussed in later sections. In this section we explore the form of the potential in this model of non-minimally coupled Palatini Q-balls, derive the constraints on the existence of Q-balls in this model from the potential and compare the results to those derived in the case of conventional Q-balls.\\

\subsubsection{Rescaling the Non-Canonical Q-ball Equation in Terms of a Quasi-Canonical Scalar}\label{section:4511}
The Palatini Q-ball equation in terms of the rescaled field $\sigma$ has a similar (but not the same) form as the general form of the Q-ball equation for the conventional case (\ref{eqn:q79}), and the dynamics can therefore be understood as being similar in this form.

To illustrate this, we begin with the non-canonical Q-ball equation derived at the beginning of the section

\begin{equation}\label{eqn:q81}
\frac{\partial^{2} \phi}{\partial r^{2}} + \frac{2}{r}\frac{\partial \phi}{\partial r} - \frac{\xi \phi}{M_{pl}^{2} \Omega^{2}} \left(\frac{\partial \phi}{\partial r}\right)^{2} = \Omega^{2}\frac{\partial V_{\omega}}{\partial \phi},
\end{equation}

\noindent and apply the rescaling

\begin{equation}\label{eqn:q82}
\frac{d\phi}{d r} \longrightarrow \Omega \frac{d\sigma}{d r},
\end{equation}

\noindent such that

\begin{equation}\label{eqn:q83}
\frac{d^{2}\phi}{d r^{2}} = \frac{d}{d r}\left(\frac{d\phi}{d r}\right) = \frac{d}{d r}\left(\Omega \frac{d\sigma}{d r}\right) = \Omega \frac{d^{2}\sigma}{d r^{2}} + \frac{d\Omega}{d r }\frac{d\sigma}{d r},
\end{equation}

\begin{equation}\label{eqn:q84}
\frac{d V_{\omega}}{d\phi} = \frac{d \sigma}{d \phi} \frac{d V_{\omega}}{d \phi} = \frac{1}{\Omega}\frac{d V_{\omega}}{d \sigma}.
\end{equation}

\noindent Substituting (\ref{eqn:q82}) - (\ref{eqn:q84}) into (\ref{eqn:q81}) gives the rescaled Q-ball equation

\begin{equation}\label{eqn:q85}
\Omega \frac{d^{2}\sigma}{d r^{2}} + \frac{2\Omega}{r} \frac{d\sigma}{d r} + \frac{d\Omega}{d r} \frac{d \sigma}{d r} - \frac{\xi \phi \Omega^{2}}{M_{pl}^{2} \Omega^{2}}\left(\frac{d \sigma}{d r}\right)^{2} = \frac{\Omega^{2}}{\Omega}\frac{d V_{\omega}}{d \sigma}.
\end{equation}

\noindent We can write

\begin{equation}\label{eqn:q86}
\frac{d\Omega}{d r } = \frac{d\phi}{d r} \frac{d\Omega}{d\phi} = \frac{d\sigma}{d r} \frac{d\phi}{d\sigma} \frac{d\Omega}{d\phi},
\end{equation}

\noindent and 

\begin{equation}\label{eqn:q87}
\frac{d\Omega}{d\phi} = \frac{d}{d\phi}\sqrt{1 + \frac{\xi \phi^{2}}{M_{pl}^{2}}} = \frac{1}{2} \frac{2\xi \phi}{M_{pl}^{2}}\left(1 + \frac{\xi \phi^{2}}{M_{pl}^{2}}\right)^{-\frac{1}{2}} = \frac{\xi \phi}{M_{pl}^{2} \Omega}.
\end{equation}

\noindent Therefore

\begin{equation}\label{eqn:q88}
\frac{d\Omega}{d r} = \frac{d\sigma}{d r} \frac{d\phi}{d\sigma} \frac{d\Omega}{d\phi} = \frac{d\sigma}{d r} \Omega \frac{\xi \phi}{M_{pl}^{2} \Omega} = \frac{d\sigma}{d r} \frac{\xi \phi}{M_{pl}^{2}}.
\end{equation}

\noindent Using (\ref{eqn:q88}) on the third term of (\ref{eqn:q85}), the rescaled Q-ball equation becomes

\begin{equation}\label{eqn:q89}
\Omega \frac{d^{2}\sigma}{d r^{2}} + \frac{2\Omega}{r} \frac{d\sigma}{d r} + \frac{\xi \phi}{M_{pl}^{2}}\left(\frac{d \sigma}{d r}\right)^{2} - \frac{\xi \phi}{M_{pl}^{2}}\left(\frac{d \sigma}{d r}\right)^{2} = \Omega\frac{d V_{\omega}}{d \sigma}.
\end{equation}

\noindent We can then cancel the gradient squared terms and divide through by $\Omega$ to obtain the Q-ball equation in terms of the $\sigma$ field

\begin{equation}\label{eqn:q90}
\frac{d^{2}\sigma}{d r^{2}} + \frac{2}{r}\frac{d\sigma}{d r} = \frac{d V_{\omega}}{d\sigma},
\end{equation}

\noindent which is exactly the same form as the Q-ball equation for $\phi$ in the $\Omega \rightarrow 1$ limit (\ref{eqn:q79}), but with $V_{\omega}$ given by (\ref{eqn:q69}). This shows that this is not the same as the equation for a canonically normalised field, as in that case we would have $\sigma^{2}$ rather than $\phi^{2}(\sigma)/\Omega^{2}(\sigma)$ in (\ref{eqn:q79}). Therefore (\ref{eqn:q90}) will produce significantly different Q-ball solutions compared to the conventional Q-ball solution, although it does suggest that there are some broad similarities between the two solutions.

It is at this point that we highlight the fact that it is not in general possible to transform a complex field such as $\Phi$ to a canonically normalised scalar in the strict sense, and that the transformation used in this case to compare the dynamics of the non-minimally coupled case to conventional Q-balls is quasi-canonical. The rescaled field $\sigma$ is therefore not interpreted as a physically meaningful field, but a function of the radial component $\phi$ of the physical complex inflaton used to illustrate some of the properties of the non-canonical Q-ball solution.

\subsubsection{Q-ball Existence and Coleman's Mechanical Analogy in the Case of Non-Minimally Coupled Palatini Q-balls}\label{section:4512}

In this section we discuss the application of Coleman's mechanical analogy \cite{coleman85}, originally used as a tool for understanding the precise nature of Q-ball solutions and the dependence of their existence on the form of the scalar potential, to Q-balls in the non-minimally coupled Palatini case. This analogy was originally derived for Q-balls in the canonical scalar field case, and the underlying dynamics in this model therefore look a little different, as we will discuss later. 

\begin{figure}[H]\label{figure:41}
\begin{center}
\includegraphics[clip = true, width=\textwidth, angle = 360]{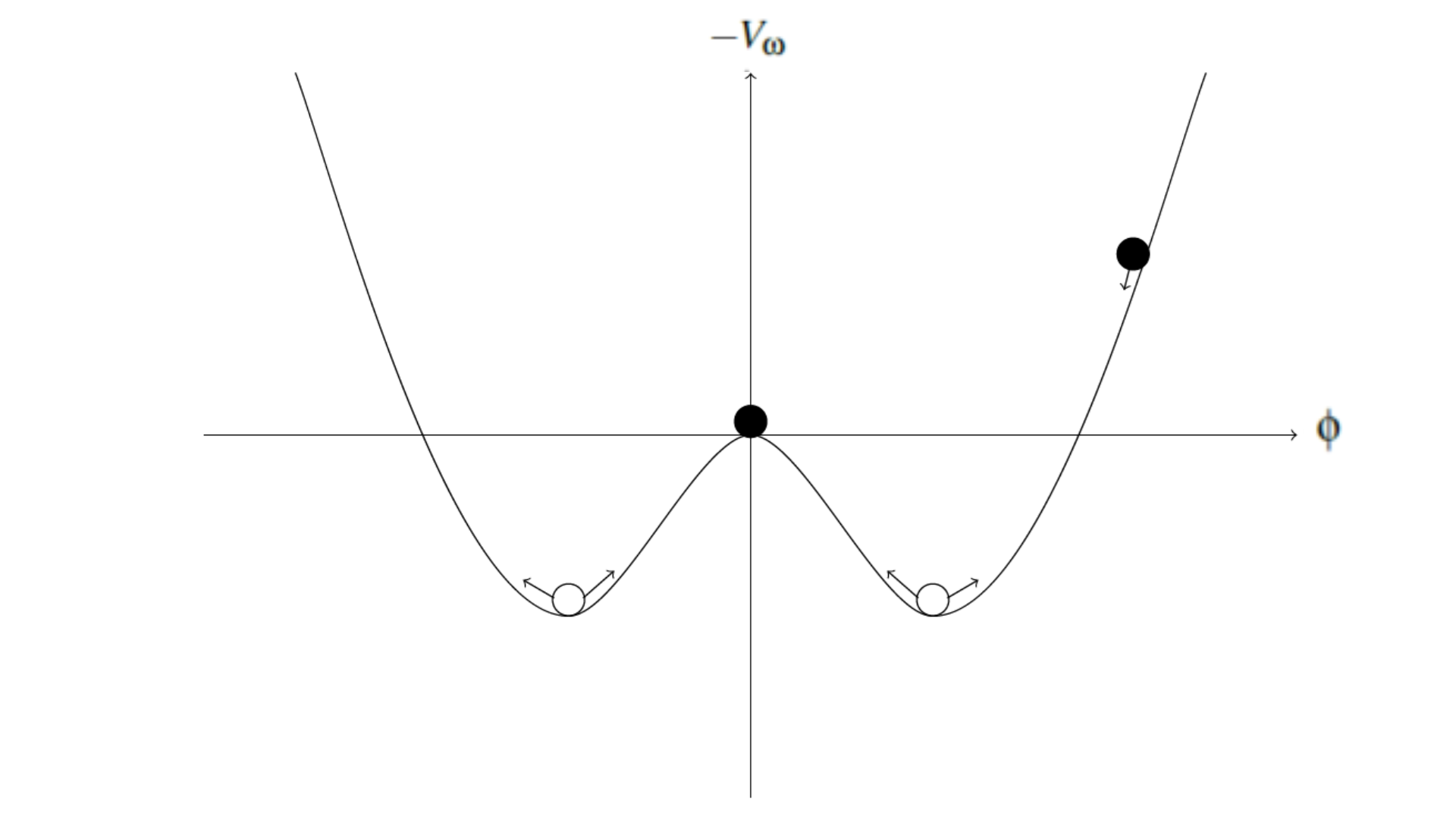}
\caption{Schematic potential and particle motion as described by (\ref{eqn:q79}) and (\ref{eqn:q90}).} 
\end{center}
\end{figure}

We first introduce the analogy in the context of the conventional case of a canonically normalised scalar, for which $\sigma = \phi$. If we interpret the field $\sigma$ $(\phi)$ as a position coordinate, and $r$ as a time coordinate, then (\ref{eqn:q90}) ((\ref{eqn:q79})) looks like it is describing the damped motion of a particle moving in the potential $-V_{\omega}$. Figure \ref{figure:41} schematically shows the potential $-V_{\omega}$ and the motion of the "particle" within it . \\

The "particle" begins its motion at the black circle on the right, at a starting point we will refer to as $\phi_{0}$. The local maximum at the origin of the potential $\phi = 0$ corresponds to the field in the vacuum at $r \rightarrow \infty$, and the idea of this analogy is that in order to obtain a Q-ball solution in a given potential $V_{\omega}$, the conditions must be precisely right such that the particle rolling down from $\phi_{0}$ on the right comes to rest exactly at the origin as illustrated in Figure \ref{figure:41}. One such condition is the initial placement, or starting location of the particle, $\phi_{0}$. If the particle is initially placed too close to $\phi = 0$ - $\phi_{0}$ is too small - then the particle will not gain enough momentum to reach the top of the local maximum at $\phi = 0$, and it will roll backwards to oscillate about the local minimum to the right in Figure \ref{figure:41} (clear circle on the right hand side). This is known as \textit{undershoot}, and can also occur if $\phi_{0}$ is sufficiently high but the gradient of the potential is too shallow. Conversely, if $\phi_{0}$ is too large (starting placement is too far from $\phi = 0$, or the potential itself is too steep on the approach to the positive $\phi$ local minimum, the field will gather too much momentum and will roll clear of the local maximum at $\phi = 0$ and into the negative $\phi$ regime, where it will oscillate about the negative $\phi$ local minimum on the left in Figure \ref{figure:41} (clear circle on the left hand side). This is known as \textit{overshoot}.\\

\begin{figure}[H]
\begin{center}
\includegraphics[clip = true, width=0.75\textwidth, angle = 360]{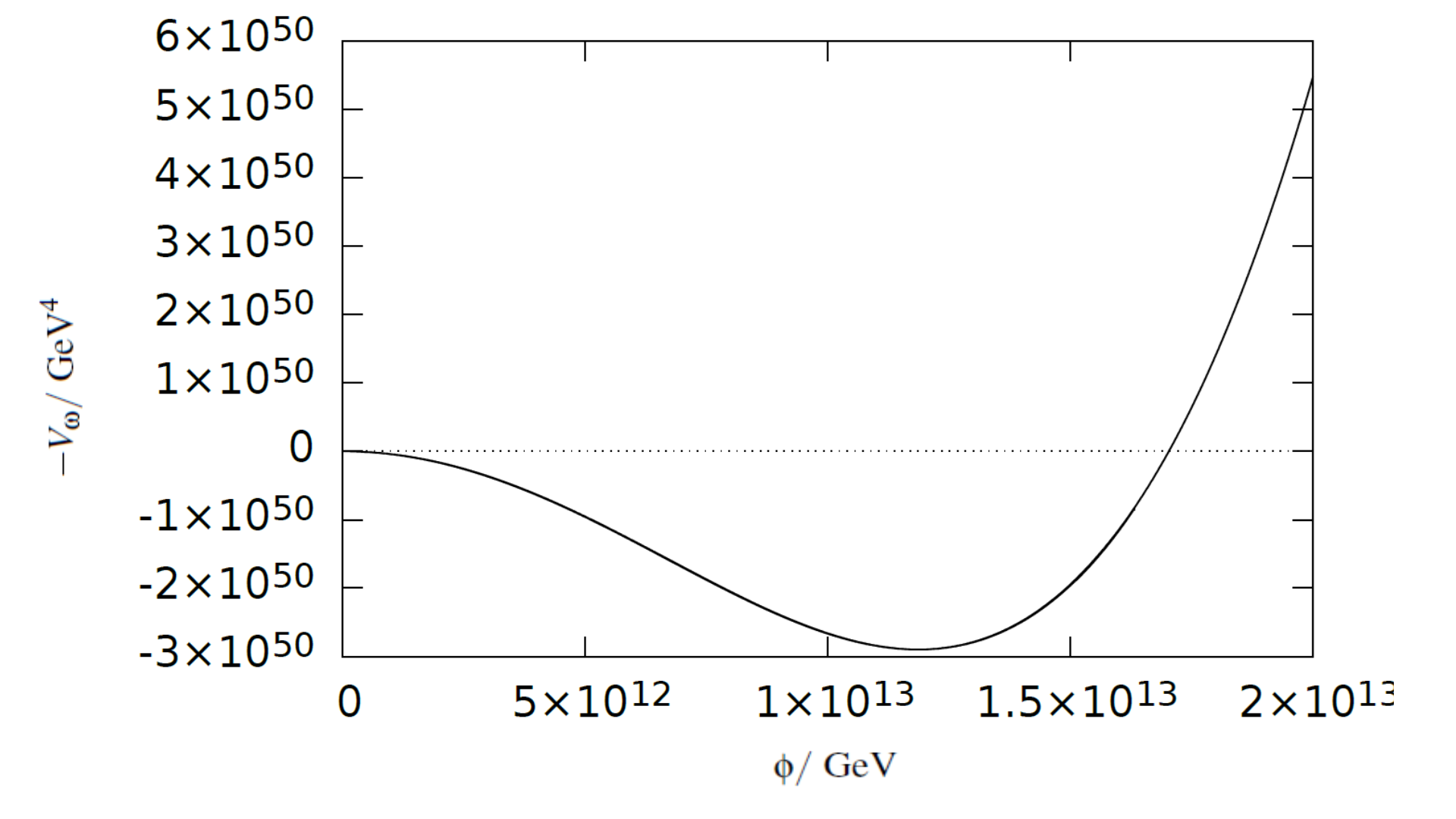}
\caption{$-V_{\omega}(\phi)$ for a non-minimally coupled Palatini Q-ball solution with large $\omega$ .} 
\label{figure:42}
\end{center}
\end{figure}

\begin{figure}[H]
\begin{center}
\includegraphics[clip = true, width=0.75\textwidth, angle = 360]{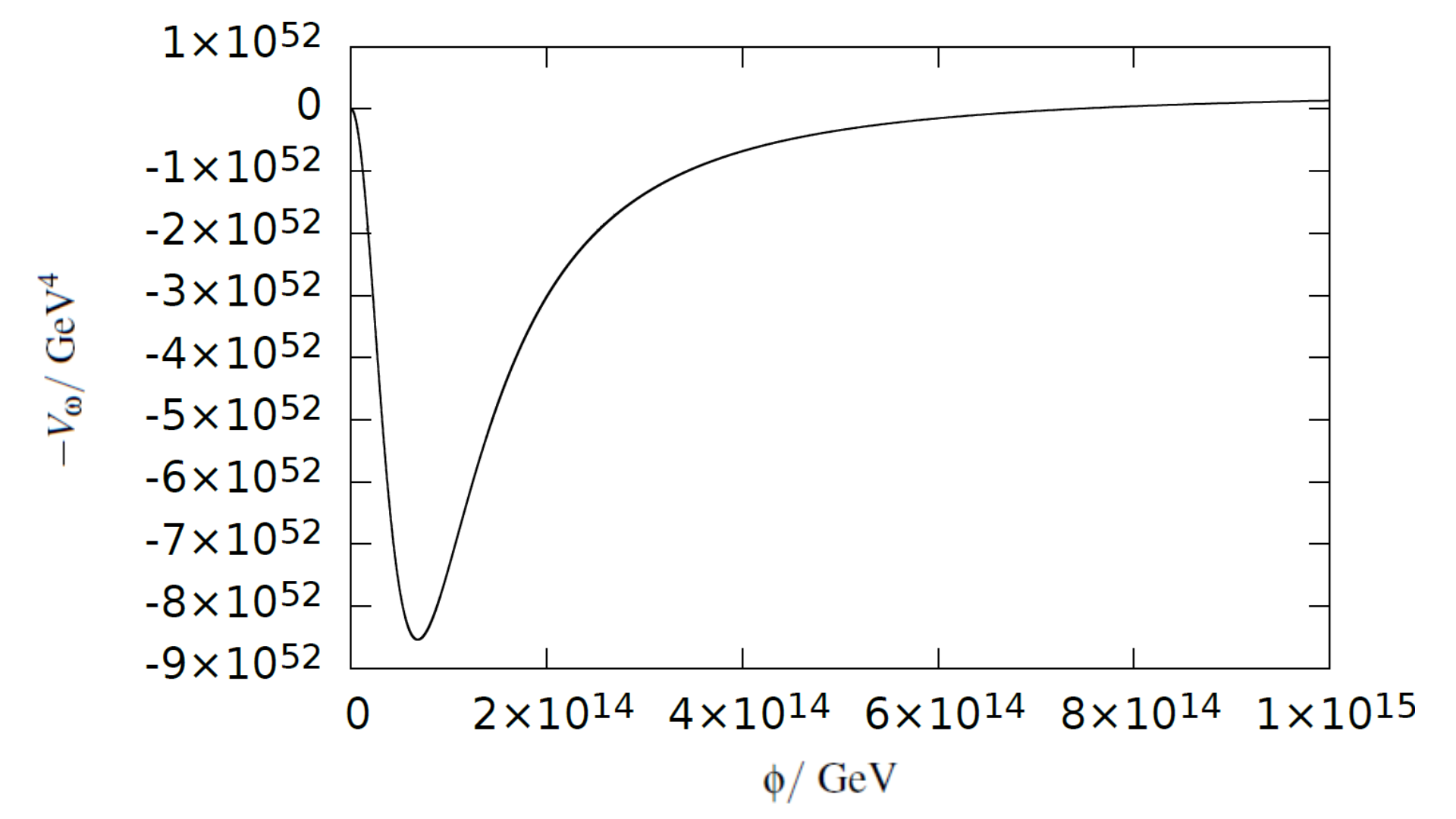}
\caption{$-V_{\omega}(\phi)$ for a non-minimally coupled Palatini Q-ball solution with small $\omega$.} 
\label{figure:43}
\end{center}
\end{figure}

Plots of $-V_{\omega}\left(\phi \right)$ for $\phi > 0$ are shown in Figure \ref{figure:42} and Figure \ref{figure:43}. Interestingly, the inverted potential before rescaling has the same form as the inverted potential in the conventional case (comparing to the $\phi > 0$ region of the schematic in Figure \ref{figure:41}) corresponding to (\ref{eqn:q79}). This demonstrates that the idea of Coleman's mechanical analogy can still be applied to explore the dependence of the existence of non-canonical Q-balls on the form of the scalar field potential in the underlying theory, although the underlying dynamics are different. As shown at the beginning of this section, in the non-minimally coupled Palatini case, the Q-ball equation in terms of $\phi$, (\ref{eqn:q77}), has an additional gradient squared term with a negative sign and a dependence on the conformal factor on the right-hand side, both explicitly and in the form of the Q-ball potential, $V_{\omega}$, in the non-minimally coupled case. The gradient squared term can be interpreted as some external energy input, which makes the motion of the particle dynamically different from the simple Newtonian motion described in (\ref{eqn:q79}), although the inverted potential looks very similar and the general idea as an analogy for the precise conditions on the potential and the field itself to produce Q-balls is still applicable. When considered in terms of the rescaled field $\sigma$, the Q-ball equation, has the same form as in the conventional case but with a modified potential $V_{\omega}$, (\ref{eqn:q69}). Equation (\ref{eqn:q90}) can therefore also be used as a means for comparison to the conventional case when considering the mechanical analogy with the understanding that, in our definition of the quasi-canonical field, $\phi$ is a function of $\sigma$ such that $V_{\omega}\left(\phi \right) = V_{\omega}\left(\phi \left(\sigma \right) \right)$.

We can now use the rules derived from this analogy \cite{coleman85} for conventional minimally-coupled Q-balls to derive an analogous set of rules for the case of non-minimally coupled Q-balls to ensure that the Q-ball potential is compatible with producing Q-ball solutions. To avoid undershoot, the initial placement of the particle ($\phi_{0}, -V_{\omega}\left(\phi_{0}\right)$), must sit at least at, or higher than zero in order for the particle to reach the local maximum of the origin, otherwise the particle will not gather sufficient momentum. This condition can be expressed as

\begin{equation}\label{eqn:q91}
max\left(-V_{\omega}\right) \geq 0,
\end{equation}

\noindent as it is in the conventional model of Q-balls \cite{coleman85}, and is significant because it implies that there is a lower bound on $\omega$ for which Q-balls can exist in a given potential. For the $V_{\omega}$ used in this model (\ref{eqn:q69}), this result does not appear to produce anything meaningful analytically, which is not an unexpected result as the inflaton potential in the Einstein frame corresponds to a plateau. It is possible that there is another way to attain a lower bound on the value of $\omega$ for non-minimally coupled Palatini Q-balls, and we will return to this possibility in Section \ref{section:46} when numerical solutions of the Q-balls are explored, although a definitive lower bound on $\omega$ in this model is at the point of submitting this thesis unconfirmed.

To avoid overshoot we require that there is an extremum at $\phi = 0$ which is a local maximum, so that the particle can come to rest in the vacuum and not continue rolling into the $\phi < 0$ region

\begin{equation}\label{eqn:q92}
\left.\frac{d^{2}\left(-V_{\omega} \right)}{d\phi^{2}}\right|_{\phi = 0} < 0,
\end{equation}

\noindent this condition may be satisfied if and only if 

\begin{equation}\label{eqn:q93}
\left.\omega^{2} < \frac{d^{2}V}{d\phi^{2}} \right|_{\phi = 0}.
\end{equation}

\noindent Calculating the left-hand side of (\ref{eqn:q92}) we have

\begin{equation}\label{eqn:q94}
\frac{d^{2}\left(-V_{\omega} \right)}{d\phi^{2}} = \frac{\omega^{2}}{\Omega^{2}} - \frac{m^{2}}{\Omega^{4}} - \frac{3\lambda \phi^{2}}{\Omega^{4}}.
\end{equation}

\noindent At $\phi = 0$, $\Omega =1$ and (\ref{eqn:q92}) therefore gives the result

\begin{equation}\label{eqn:q95}
\left.\frac{d^{2}\left(-V_{\omega} \right)}{d\phi^{2}}\right|_{\phi = 0} = \omega^{2} - m^{2} < 0 \Rightarrow \omega^{2} < m^{2},
\end{equation}

\noindent and from (\ref{eqn:q93}) we find

\begin{equation}\label{eqn:q96}
\left.\omega^{2} < \frac{d^{2}V}{d\phi^{2}} \right|_{\phi = 0} = m^{2}.
\end{equation}

\noindent This means that

\begin{equation}\label{eqn:q97}
\omega < m,
\end{equation}

\noindent is a hard constraint on the existence of non-minimally coupled Palatini Q-balls, and is the same result as that on the existence of conventional minimally-coupled Q-balls derived from the mechanical analogy in \cite{coleman85}, as the condition to avoid overshoot.\\

\subsection{Q-ball Stability}\label{section:452}

Stability of Q-balls is an important property, particularly when discussing Q-balls as cosmological objects. A Q-ball is said to be \textit{absolutely stable} if \cite{coleman85}

\begin{equation}\label{eqn:q98}
E < mQ,
\end{equation}

\noindent holds. $E$ is the energy of the Q-ball and $mQ$ can be interpreted as the energy of $Q$ free quanta of scalar particles of mass $m$ in the vacuum. This essentially means that in order to be \textit{absolutely stable} the energy of a Q-ball must be less than the sum of the energies of the individual component scalars it is composed of when treated as free quanta. This difference in energy corresponds to the binding energy of the Q-ball. \textit{Classical stability} implies a solution that is stable against small perturbations. As an approximation, the classical stability of a Q-ball can be gauged by the sign of the derivative of the charge with respect to $\omega$, i.e. if $\partial Q/\partial \omega <0$ then a Q-ball is classically stable \cite{friedberg76} \cite{lee92}. The absolute stability of the numerically generated non-minimally coupled Palatini Q-balls is discussed in Section \ref{section:46}.

\subsection{Mass Range of the Inflaton from Inflation and the Existence of Q-balls }\label{section:453}

In Section \ref{section:44}, we demonstrated that the form of the inflaton potential - and the state of the inflaton field at the end of inflation - depends explicitly on the size of the mass term (\ref{eqn:q20}), which means that the size of this mass term determines the initial conditions for Q-balls in non-minimally coupled Palatini inflation. It is therefore important that we establish the range of inflaton masses which are compatible with the existence condition (\ref{eqn:q97}) of these Q-balls. In Section \ref{section:441}, we derived the constraint (\ref{eqn:q52}) which gives an upper bound on the inflaton mass squared, from the condition that the inflaton potential must be of positive gradient in order for inflation to proceed. In this section, we use this, and the existence condition of non-minimally coupled Palatini Q-balls derived in the previous section, (\ref{eqn:q97}), to derive a range of inflaton masses for which both inflation can occur and inflatonic Q-balls can exist. We will refer to this range of compatible inflaton masses which produce both inflation and Q-balls as the "Q-ball window".

We begin with the non-canonical Q-ball potential 

\begin{equation}\label{eqn:q99}
V_{\omega} \left(\phi \right) =  \frac{1}{\Omega^{4}}\left(\frac{1}{2} m^{2} \phi^{2} + \frac{\lambda}{4} \phi^{4} \right)  - \frac{\omega^{2} \phi^{2}}{2\Omega^{2}},
\end{equation}

\noindent and expand to leading order in $M_{pl}^{2}/\xi \phi^{2}$ to give

\begin{equation}\label{eqn:q100}
V_{\omega} \left(\phi \right) = \frac{M_{pl}^{4}}{2\xi^{2}\phi^{2}}\left( m^{2} + \omega^{2} - \frac{2m^{2}M_{pl}^{2}}{\xi \phi^{2}} \right) - \frac{\lambda M_{pl}^{6}}{2 \xi^{3}\phi^{2}} + \frac{\lambda M_{pl}^{4}}{4\xi^{2}} - \frac{\omega^{2}M_{pl}^{2}}{2\xi}.
\end{equation}

\noindent In other words,

\begin{equation} \label{eqn:q101}
V_{\omega}\left(\phi \right) = \frac{M_{pl}^{4}}{2\xi^{2}\phi^{2}}\left[ m^{2} + \omega^{2} - \frac{\lambda M_{pl}^{2}}{\xi} \right] + \hspace{1mm}higher\hspace{1.0mm} order\hspace{1.0mm} and \hspace{1.0mm} constant ~\hspace{0.8mm} terms.
\end{equation}

\noindent  For (\ref{eqn:q77}) to yield a Q-ball solution, the solution $\phi \left(r \right)$ must decrease as $r$ increases from zero. This is true if the $\phi$ dependent term on the right hand side of (\ref{eqn:q101}) is positive. The existence condition for a Q-ball from the Q-ball potential (\ref{eqn:q101}) is therefore 

\begin{equation}\label{eqn:q102}
m^{2} + \omega^{2} > \frac{\lambda M_{pl}^{2}}{\xi}. 
\end{equation}

\noindent In Section \ref{section:461} we will analytically demonstrate that if $\omega < m$, as required in order for Q-balls to exist, then the condition (\ref{eqn:q102}) must also be satisfied in order for the right-hand side of the Q-ball equation to have zeroes for some $\phi \neq 0$. Both interpretations of (\ref{eqn:q102}) correspond to a necessary condition on the Q-ball potential for Q-ball solutions to exist in the model. 

\noindent We define

\begin{equation}\label{eqn:q103}
\omega_{c}^{2} = \frac{\lambda M_{pl}^{2}}{\xi},
\end{equation}

\noindent and use this to rewrite (\ref{eqn:q102})

\begin{equation} \label{eqn:q104}
m^{2} + \omega^{2} > \omega_{c}^{2}. 
\end{equation}

\noindent If we combine (\ref{eqn:q104})  with the condition for the inflaton potential to be compatible with inflation (\ref{eqn:q52}), which we can write as $m^{2} < \omega_{c}^{2}$, we find the condition for inflation and the existence of Q-balls in terms of the $\omega$ parameter and the inflaton mass to be

\begin{equation}\label{eqn:q105}
m^{2} < \omega_{c}^{2} < m^{2} + \omega^{2}.
\end{equation}

\noindent The existence of Q-balls also requires that $\omega < m$ (\ref{eqn:q97}), and the right-hand side of the inequality (\ref{eqn:q105}) can be constrained

\begin{equation}\label{eqn:q106}
\omega^{2} + m^{2} < 2m^{2}.
\end{equation}

\noindent (\ref{eqn:q105}) can therefore be stated as

\begin{equation}\label{eqn:q107}
m^{2} < \omega_{c}^{2} < 2m^{2},
\end{equation}

\noindent which may be equivalently written as an range of inflaton masses squared

\begin{equation} \label{eqn:q108}
\frac{\omega_{c}^{2}}{2} < m^{2} < \omega_{c}^{2}.
\end{equation}

\noindent This gives the range in inflaton mass squared (the Q-ball window) for which the inflaton potential is compatible with both inflation and with the existence of Q-balls. It also provides a mass window within which we can numerically search for Q-ball solutions which are compatible with inflation, the topic of the next section. From (\ref{eqn:q108}), the Q-ball window seems to favour inflaton masses close to the upper bound derived from inflation (\ref{eqn:q52}), and while Q-balls may form for inflaton masses larger than the upper bound (\ref{eqn:q52}) in this model, inflation would not be possible in conjunction with these Q-balls.

\section{Numerical Solutions of the Q-ball Equation}\label{section:46}

In this section we discuss and present the results of the numerical analysis of non-minimally coupled Palatini Q-balls. In this project we solved the non-canonical Q-ball equation (\ref{eqn:q77}) numerically for a fixed mass within the range specified in (\ref{eqn:q108}), over a range of $\omega$ for the following boundary conditions at $r=0$

\begin{equation} \label{eqn:q109}
\phi \left(r =0\right) = \phi_{0} \;\;\;\; ; \;\;\;
\frac{\partial \phi}{\partial r} \left(r=0\right) = 0.
\end{equation}

\noindent For each $\omega$ we ran the code for trial values of the inflaton field, corresponding to $\phi_{0}$, and scanned for Q-ball solutions. The details of this methodology are outlined in the following subsection. When discussing numerical results, the inflaton mass $m$ and the parameter $\omega$ are listed in units of $\omega_{c}$.

\subsection{Zeroes of the Q-ball Equation}\label{section:461}

\begin{figure}[H]
\begin{center}
\includegraphics[clip = true, width=0.75\textwidth, angle = 360]{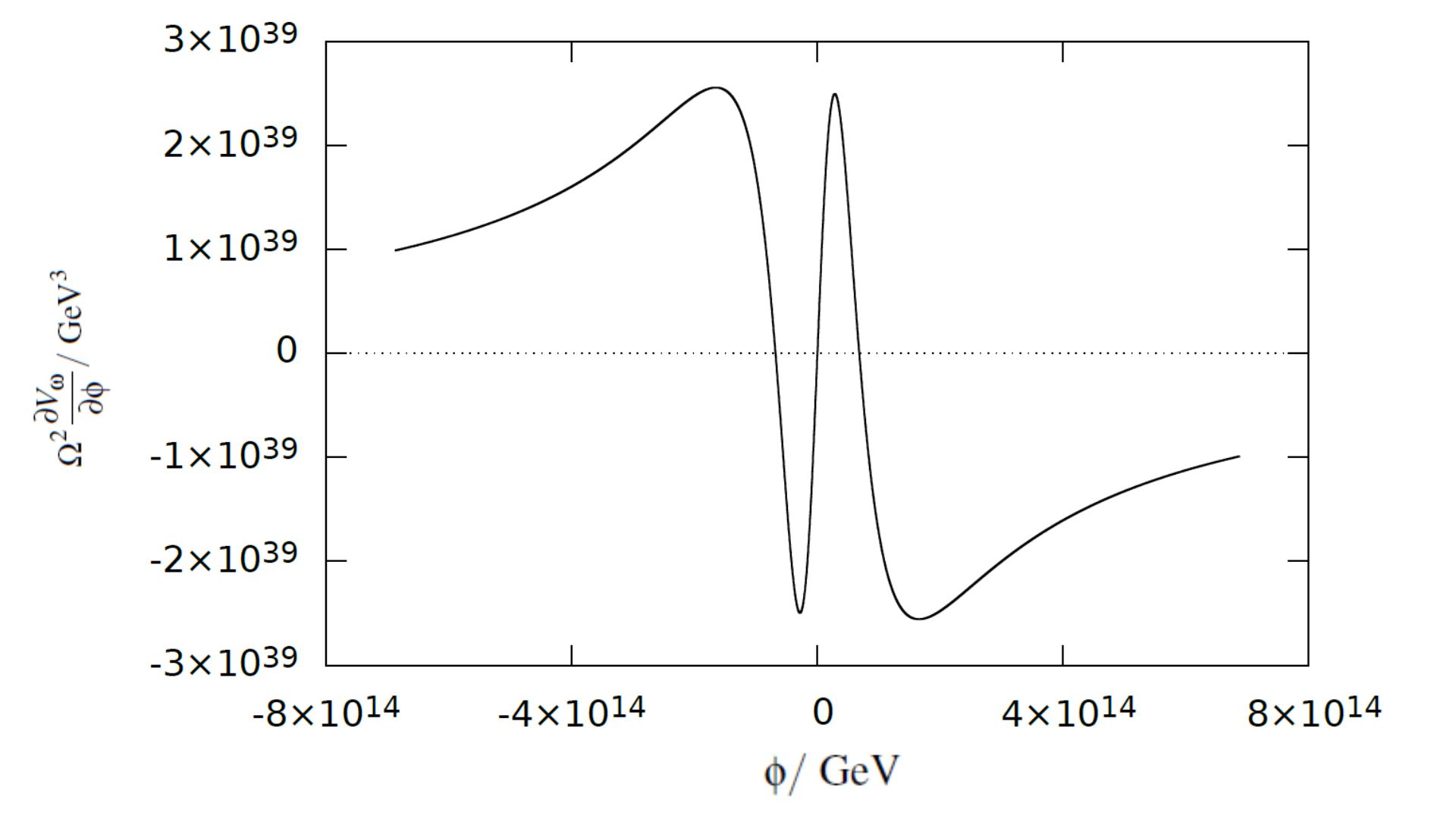}
\caption{Plot illustrating the zeroes of the Q-ball equation.} 
\label{figure:44}
\end{center}
\end{figure}

In this section we discuss the zeroes of the right-hand side Q-ball equation (\ref{eqn:q77}), and how they relate to the existence of Q-ball solutions. These zeroes are shown in Figure \ref{figure:44}, and correspond to fixed points of the Q-ball equation, which in turn correspond to the extrema of the Q-ball effective potential $V_{\omega}\left(\phi\right)$, shown schematically in Figure \ref{figure:41}. The fixed points of the Q-ball equation can be related to the mechanical analogy outlined in Section \ref{section:451} and more importantly, used to constrain $\phi_{0}$ when searching for Q-ball solutions numerically.

The non-minimally coupled Q-ball equation derived in this work (\ref{eqn:q77}) has three zeroes (shown in Figure \ref{figure:44}). These correspond to two stable fixed points, which are symmetric attractors about $\Omega^{2}\partial V_{\omega}/\partial \phi = 0$, and one unstable fixed point.

\vspace{-0.3cm}

\noindent Figure \ref{figure:44} shows an example of the right hand side of the Q-ball equation as a function of $\phi$. The points where the function crosses zero are the zeroes or fixed points of the Q-ball equation. The attractor fixed points are the furthest to the left (denoted $\phi_{-}$) in the negative region of $\phi$ and the furthest to the right (denoted $\phi_{+}$) in the positive regime of $\phi$. These provide a reference for determining the value of $\phi$ at $r = 0$, $\phi_{0}$, for which a Q-ball solution exists. If the field gets caught by - and starts to oscillate around -  the positive fixed point, $\phi_{+}$, as $r$ increases, then $\phi_{0}$ is too low to produce a Q-ball. The lowest $\phi_{0}$ which produces this result can then be used as a lower bound for the range of $\phi_{0}$ which could generate a Q-ball solution for a given $\omega$. In the opposite limit, if $\phi_{0}$ is too high then the field can drop below zero as $r$ increases and begin to oscillate around the negative fixed point, $\phi_{-}$. The lowest $\phi_{0}$ which produces this result can then act as an upper bound on the range of $\phi_{0}$ for which there could exist a Q-ball solution for the chosen $\omega$. The fixed points of the Q-ball equation therefore produce behaviour in the function of the field over distance, $\phi \left(r \right)$, which can be used to restrict the parameter space of $\phi_{0}$ used when searching for Q-ball solutions numerically. 

This can be understood in terms of the mechanical analogy of Q-balls outlined in Section \ref{section:451} by noting that the fixed points of the Q-ball equation correspond to the extrema of the inverted Q-ball effective potential $-V_{\omega}$, in the analogy. The local minimum in the $\phi >0$ region of Figure \ref{figure:41} corresponds to the positive attractor fixed point $\phi_{+}$ (first zero from the right in Figure \ref{figure:44}). The unstable fixed point corresponds to the local maximum at $\phi =0$ in Figure \ref{figure:41}, and is the central fixed point in Figure \ref{figure:44}. The local minimum in the $\phi < 0$ region of the potential in Figure \ref{figure:41} corresponds to the symmetric negative fixed point $\phi_{-}$ (furthest zero from the right in Figure \ref{figure:44}).

This illustrates how precise the values of the parameters need to be to obtain a Q-ball solution numerically, as well as how strict the conditions on the potential in order to produce Q-balls are. The field profile for a Q-ball solution is represented by $\phi\left(r\right)$ asymptoting to zero as $r \rightarrow \infty$, corresponding to the field coming to rest exactly at the top of the local maximum at $\phi = 0$ in Figure \ref{figure:41} from its starting point $\phi_{0}$ in the language of Coleman's analogy. It is also important to emphasise that the existence of the zeroes of the Q-ball equation determine the existence of the Q-ball solutions themselves, without the fixed points there cannot be Q-ball solutions.

We now derive the condition for the existence of zeroes of the Q-ball equation. The requirement for extrema to exist is that $\partial V_{\omega}/\partial \phi = 0$ for some $\phi \neq 0$. Using (\ref{eqn:q99}) and setting the first derivative to zero gives the location of the symmetric $\phi \neq 0$ zeroes to be 

\begin{equation} \label{eqn:q110}
\phi = \pm \frac{M_{Pl}}{\sqrt{\xi}} \frac{\sqrt{m^{2} - \omega^{2}}}{\sqrt{m^{2} + \omega^{2} - \omega_{c}^{2} }  }.
\end{equation}

\noindent In order for the Q-ball solution to be stable we require that $\omega^{2} < m^{2}$. This will be verified both numerically later in this section and analytically in Section \ref{section:472}. The range of values of $m^{2}$ for which relevant zeros of the Q-ball equation exist is therefore

\begin{equation}\label{eqn:q111}
m^{2} + \omega^{2} > \omega_{c}^{2},
\end{equation}

\noindent which is exactly the condition (\ref{eqn:q104}) derived from the Q-ball equation. We confirm numerically in Section \ref{section:462} that zeroes with $\phi \neq 0$ exist over the range of $\omega$ satisfying (\ref{eqn:q104}) and (\ref{eqn:q111}).

\subsection{Numerical Results}\label{section:462}

In this section we present the results of solving the non-minimally coupled Palatini Q-ball equation numerically, and discuss the properties of these Q-balls. We present ten Q-ball solutions for the inflaton field of mass $m = 0.9\omega_{c}$, which is within the Q-ball window (\ref{eqn:q108}), close to the upper bound of inflaton masses which inflate the model (\ref{eqn:q52}), for a range of $\omega$ from $0.89\omega_{c}$ to $0.707155\omega_{c}$. This produces Q-balls with initial field values in the range $\phi_{0} \sim 10^{13} - 10^{17} \GeV$, illustrating that Q-balls could be produced from the kind of field values predicted at the end of non-minimally coupled Palatini inflation in this model. We present the radii, charge and energy of the Q-balls, in addition to some other parameters in relation to stability of the Q-balls which we define later in this section. The bounds on $\omega$ are chosen because they lie close to the bounds for which a Q-ball solution can be obtained for $m = 0.9\omega_{c}$. The upper bound, $\omega = 0.89\omega_{c}$, lies close to the stability limit of $\omega = m$ for Q-balls of the chosen mass, whereas the lower limit was discovered numerically while searching for Q-ball solutions for $m = 0.9\omega_{c}$. It may be expected that there are Q-balls with $\phi_{0}$ larger than this - of particular interest being a $\phi_{0} \sim 10^{18}\GeV$ Q-ball, discussed in depth in Section \ref{section:49} - but the precision on $\omega$ would need to be even greater to find it. It is interesting to note that the lower limit is close to being $\omega \sim \omega_{c}/\sqrt{2}$, which may have some kind of theoretical relevance to the existence of Q-balls in this framework.

Since there is no fixed definition of the radius of the Q-ball, two different possibilities are considered in this work for the purposes of evaluating the numerical Q-ball solutions. One possibility considered is a definition we will call $r_{X}$ - referred to as "the $X$ point" when discussing the numerical procedure - and a second definition, $r_{Z}$ - similarly "the $Z$ point" - as two different possibilities for understanding the concept of the Q-ball's "edge" when evaluating the properties numerically. The $X$ point is defined as being at a distance $r = r_{X}$ from the centre of the Q-ball $r = 0$ at which the field has decreased to $1 \%$ of its initial value $\phi_{0}$. This definition allows the properties of the Q-ball solutions to be examined using a fixed definition of what the radius is for all $\phi_{0}$. The $Z$ point is defined as being the point at which the code used to find a Q-ball solution for a given $\phi_{0}$ and $\omega$ cuts off having found a Q-ball solution. This distance from the centre of the Q-ball appears in the Q-ball profile, $\phi \left(r \right)$, as the point at which the field stops asymptoting towards zero and begins to increase again, moving away from the expected behaviour of a Q-ball solution. This occurs at $r = r_{Z}$ and this point corresponds to the maximum value of $r$ for which the Q-ball solution is valid for a given $\phi_{0}$ and $\omega$. Although less well defined than the $X$ point, since it will be different for every Q-ball solution, the $Z$ point cutoff is the more useful definition for calculating the energy $E$ and charge $Q$ of the Q-balls since the $Z$ point corresponds to the upper value of $r$ to which the Q-ball equation is integrated.

Two tables of values for the Q-ball properties are therefore presented, corresponding to those calculated at the $X$ point and the $Z$ point, as an exploration into how the radius of the Q-ball should be interpreted numerically and whether it makes any significant difference to the predicted properties of the Q-balls. Quantities with a subscript $X$ are calculated at the $X$ point, and those with a $Z$ are calculated at the $Z$ point.

\begin{figure}[H]
\begin{center}
\includegraphics[clip = true, width=0.75\textwidth, angle = 360]{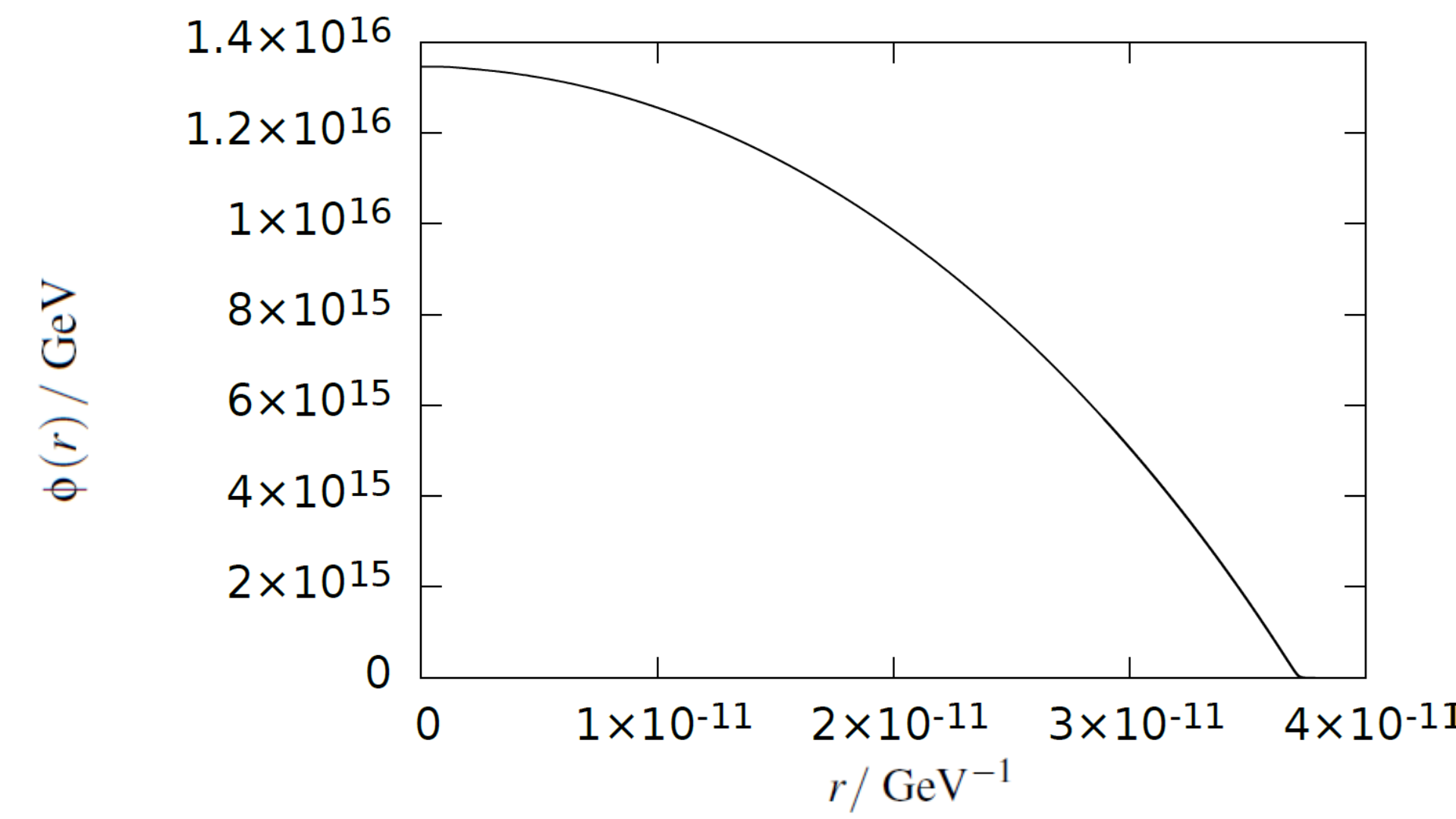}
\caption{Q-ball solution for $\omega = 0.709\omega_{c} $, $m=0.9\omega_{c}$ and $\phi_{0} = 1.3464098 \times 10^{16} \GeV$ obtained by solving (\ref{eqn:q77}) numerically.} 
\label{figure:45}
\end{center}
\end{figure}

\begin{figure}[H]
\begin{center}
\includegraphics[clip = true, width=0.75\textwidth, angle = 360]{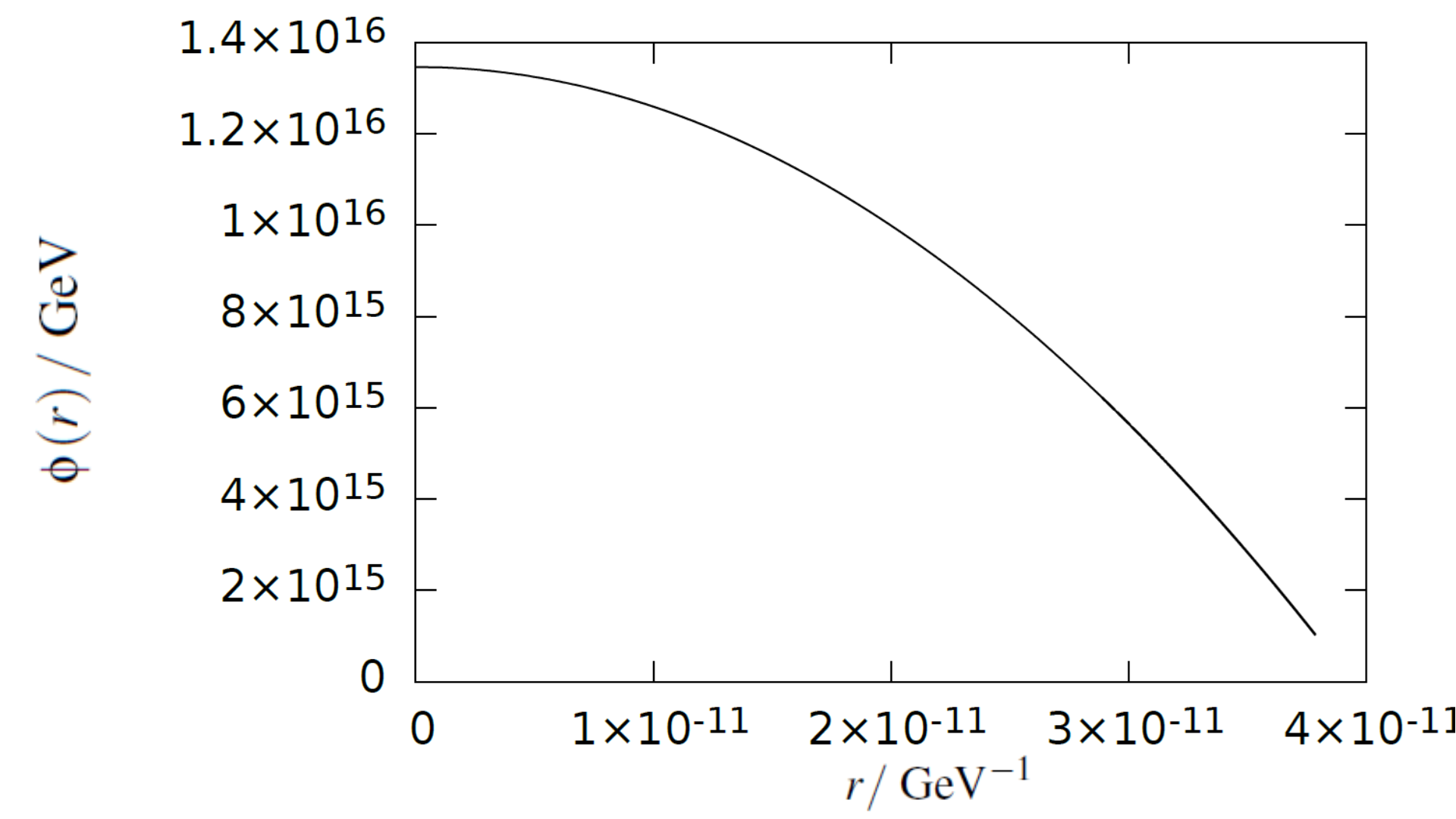}
\caption{Q-ball solution for  $\omega = 0.709\omega_{c} $, $m=0.9\omega_{c}$ and $\phi_{0} = 1.3464098 \times 10^{16} \GeV$ obtained using the analytical solution (\ref{eqn:q125}).} 
\label{figure:46}
\end{center}
\end{figure}

\begin{figure}[H]
\begin{center}
\includegraphics[clip = true, width=0.75\textwidth, angle = 360]{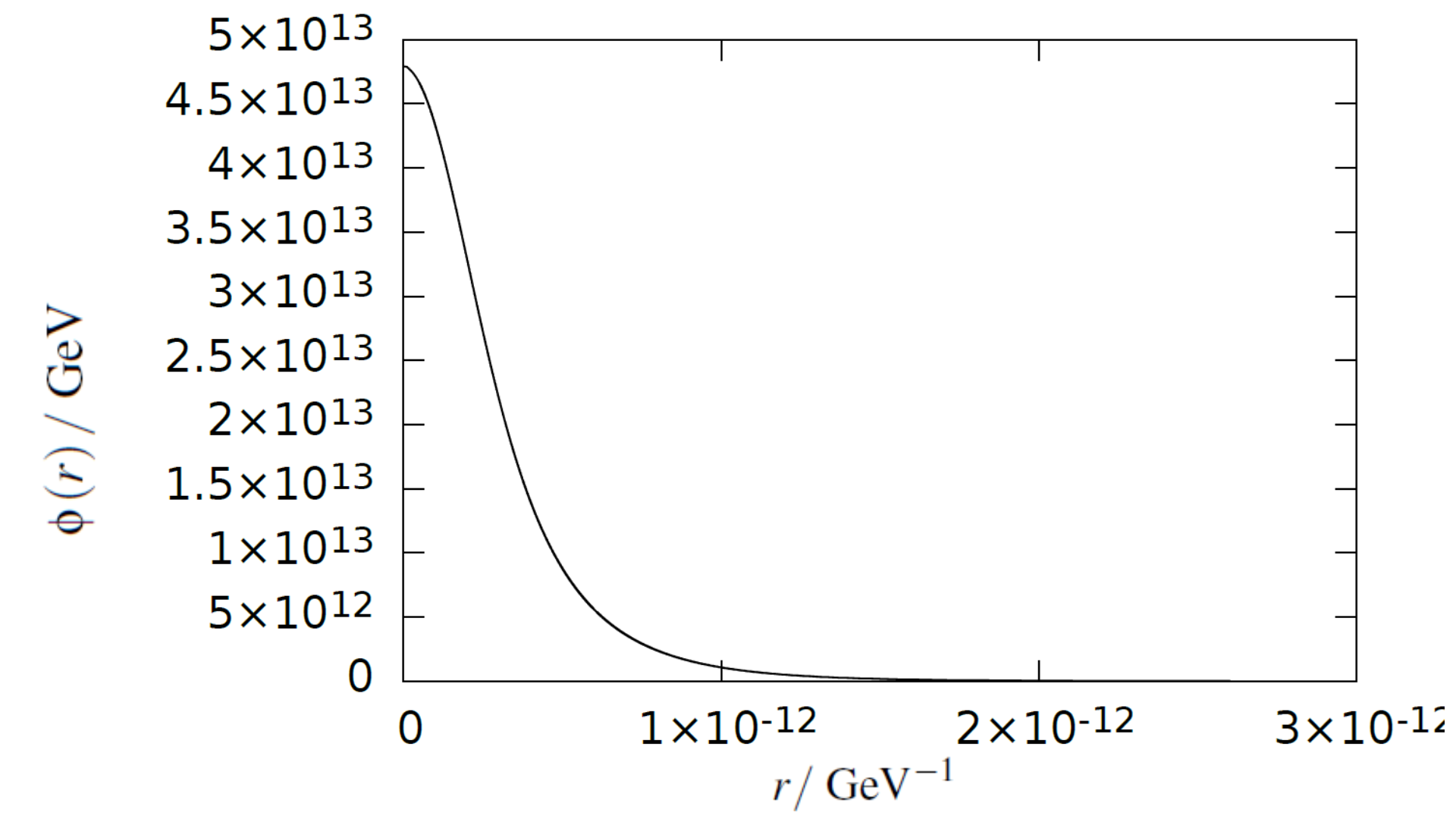}
\caption{Q-ball solution for $\omega = 0.89\omega_{c}$, $m=0.9\omega_{c}$ and $\phi_{0} = 4.7918\times 10^{13} \GeV$ obtained by solving (\ref{eqn:q77}) numerically. } 
\label{figure:47}
\end{center}
\end{figure}

\begin{figure}[H]
\begin{center}
\includegraphics[clip = true, width=0.75\textwidth, angle = 360]{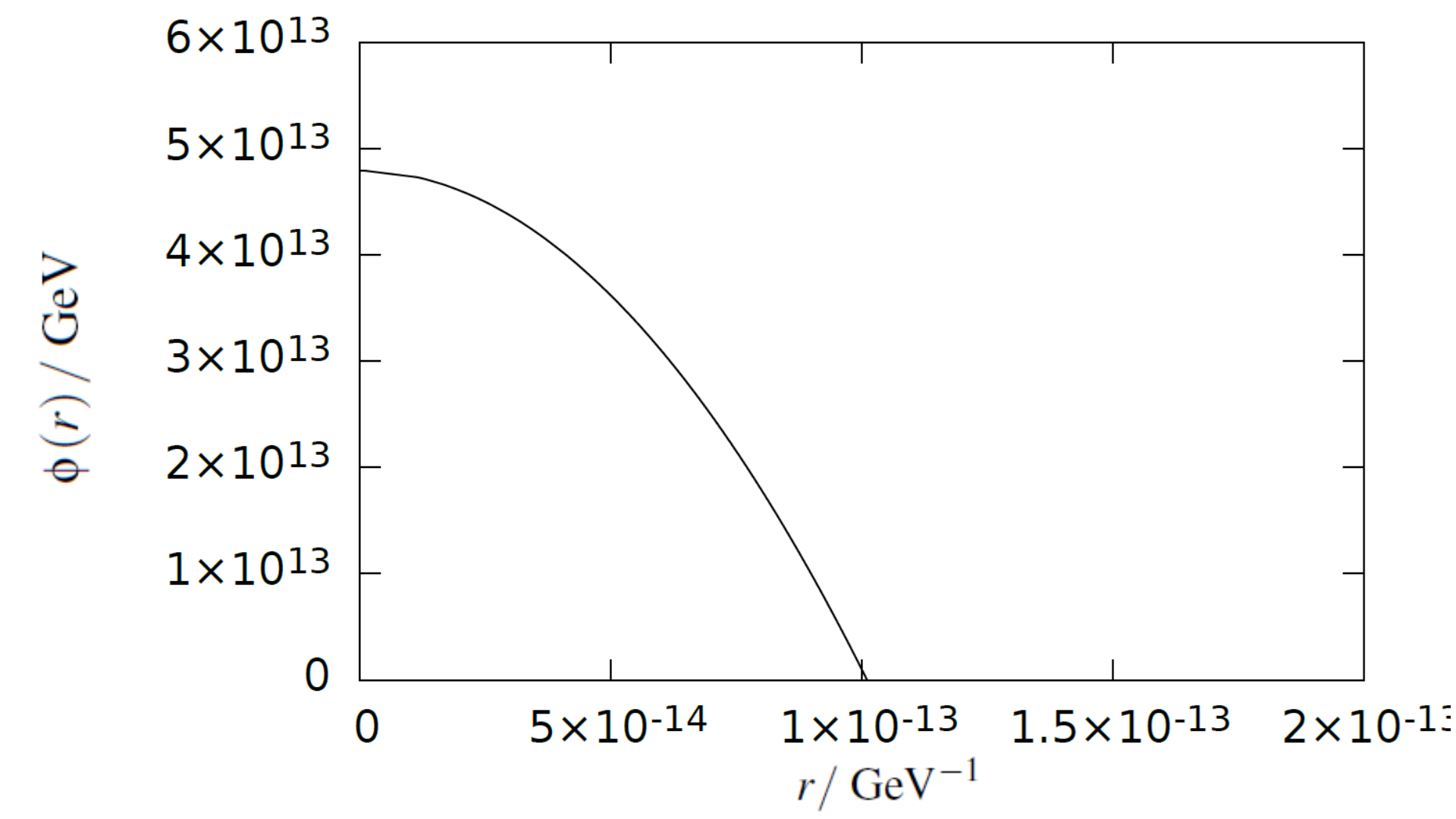}
\caption{Q-ball solution for $\omega = 0.89\omega_{c}$, $m=0.9\omega_{c}$ and $\phi_{0} = 4.7918\times 10^{13} \GeV$ obtained using the analytical approximation (\ref{eqn:q125}). } 
\label{figure:48}
\end{center}
\end{figure}

Figures \ref{figure:45} - \ref{figure:48} show two Q-ball solutions, calculated numerically and using an analytical approximation derived in Section \ref{section:47}, (\ref{eqn:q125}). Figures \ref{figure:45} and \ref{figure:46} show a Q-ball with $\phi_{0} \sim 10^{16}\GeV$, and the Figures \ref{figure:47} and \ref{figure:48} show a Q-ball with $\phi_{0} \sim 10^{13}\GeV$, with $\omega$ close to the upper bound of compatibility with the existence of Q-balls, $\omega = m$. The inflaton self-coupling is $\lambda = 0.1$ throughout and the corresponding non-minimal coupling is $\xi = 1.2163 \times 10^{9}$, using the estimate from (\ref{eqn:q46}) derived in Section \ref{section:44}.

In Tables \ref{table:41} and \ref{table:42} we present the numerical properties of the Q-ball solutions using the $r_{Z}$ and $r_{X}$ definitions of the Q-ball radius, respectively. Quantities recorded are the $\omega$ parameter; the value of the field at $r = 0$, $\phi_{0}$; the radius of the Q-ball, $r_{Z}$ or $r_{X}$; the value of the field at the defined edge of the Q-ball, and the total energy, $E$, charge, $Q$, and the energy-charge ratio, $E/Q$ of the Q-ball.

\noindent In order to assess whether the Q-balls found are stable, we introduce the quantities $\Delta_{\omega}$ and $\Delta_{m}$. $\Delta_{\omega}$ is defined by

\begin{equation}\label{eqn:q112}
\Delta_{\omega} = \frac{1}{\omega}\left(\frac{E}{Q} - \omega \right).
\end{equation}

\noindent This quantity measures the extent to which the $\omega$ parameter for each Q-ball solution is equal to the energy-charge ratio. The relationship between $\omega$ and the energy-charge ratio is established analytically for this model in Section \ref{section:47}.
 
$\Delta_{m}$ is defined by

\begin{equation}\label{eqn:q113}
\Delta_{m} = \frac{1}{m}\left(\frac{E}{Q} - m \right).
\end{equation}

\noindent This quantity provides a measure of absolute stability. As discussed in Section \ref{section:452}, absolute stability is defined by the energy of a Q-ball in relation to the energy of its component scalars if they were free particles, $E < mQ$. $\Delta_{m} < 0$ is therefore the condition for a Q-ball is to be absolutely stable.

\begin{table}[H]
\begin{center}
\begin{adjustbox}{max width=\textwidth}
\begin{tabular}{| c | c | c | c | c | c | c | c | c |}
\hline
$\omega / \omega_{c}$ & $\phi_{0} / \GeV$ & $\phi_{Z}/\GeV$ & $r_{Z}/\GeV^{-1}$ & $Q_{Z}$ & $E_{Z}/\GeV$ & $(E/Q)_{Z}/\GeV$ & $\Delta_{\omega, Z}$ & $ \Delta_{m, Z}$ \\
\hline
$0.707155$ & $3.2217991 \times 10^{17} $ & $1.61 \times 10^{10}$ & $8.94 \times 10^{-10}$ & $2.17 \times 10^{14}$ & $3.34 \times 10^{27}$ & $1.54 \times 10^{13}$ & $-2.79 \times 10^{-4}$ & $-0.21$ \\
\hline
$0.709$ & $1.3464098 \times 10^{16}$ & $1.63 \times 10^{10}$ & $3.79 \times 10^{-11}$ & $1.56 \times 10^{10}$ & $2.40 \times 10^{23}$ & $1.54 \times 10^{13}$ & $9.71 \times 10^{-4}$ & $-0.21$\\
\hline
$0.71$ & $8.855792 \times 10^{15}$ & $2.05 \times 10^{10}$ & $2.50 \times 10^{-11}$ & $4.37 \times 10^{9}$ & $6.76 \times 10^{22}$ & $1.55 \times 10^{13}$ & $1.64 \times 10^{-3}$ & $-0.21$ \\
\hline
$0.72$ & $1.960795 \times 10^{15}$ & $9.76 \times 10^{9}$ & $6.09 \times 10^{-12}$ &  $4.43 \times 10^{7}$ & $6.99 \times 10^{20}$ & $1.58 \times 10^{13}$ & $7.71 \times 10^{-3}$ & $-0.19$ \\
\hline
$0.73$ & $1.090258 \times 10^{15}$ & $6.60 \times 10^{9}$ & $3.73 \times 10^{-12}$ &  $7.09 \times 10^{6}$ & $1.14 \times 10^{20}$ & $1.61 \times 10^{13}$ & $1.30 \times 10^{-2}$ & $-0.18$ \\
\hline
$0.74$ & $7.45339 \times 10^{14}$ & $1.61 \times 10^{10}$ & $2.72 \times 10^{-12}$ &  $2.16 \times 10^{6}$ & $3.54 \times 10^{19}$ & $1.64 \times 10^{13}$ & $1.76 \times 10^{-2}$ & $-0.16$ \\
\hline
$0.75$ & $5.61953 \times 10^{14}$ & $1.14 \times 10^{10}$ & $2.28 \times 10^{-12}$ & $8.85 \times 10^{5}$ & $1.48 \times 10^{19}$ & $1.67 \times 10^{13}$ & $2.17 \times 10^{-2}$ & $-0.15$ \\
\hline
$0.80$ & $2.29632 \times 10^{14}$ & $1.96 \times 10^{10}$ & $1.47 \times 10^{-12}$ & $5.60 \times 10^{4}$ & $1.01 \times 10^{18}$ & $1.80 \times 10^{13}$ & $3.51 \times 10^{-2}$ & $-8.00 \times 10^{-2}$ \\
\hline
$0.85$ & $1.16877 \times 10^{14}$ & $1.57 \times 10^{10}$ & $1.49 \times 10^{-12}$ & $1.01 \times 10^{4}$ & $1.93 \times 10^{17}$ & $1.91 \times 10^{13}$ & $3.51 \times 10^{-2}$ & $-2.24 \times 10^{-2}$ \\
\hline
$0.89$ & $4.7918 \times 10^{13}$ & $8.82 \times 10^{9}$ & $2.70 \times 10^{-12}$ &  $3.98 \times 10^{3}$ & $7.82 \times 10^{16}$ & $1.97 \times 10^{13}$ & $1.50 \times 10^{-2}$ & $3.74 \times 10^{-3}$ \\
\hline
\end{tabular}
\end{adjustbox}
\caption{Table listing important properties of the $m=0.9\omega_{c}$ Q-balls, calculated using the $Z$ point definition of the Q-ball radius. $\phi_{0}$ denotes the value of the field at $r = 0$. $\phi_{Z}$, $E_{Z}$, $Q_{Z}$, $(E/Q)_{Z}$, $\Delta_{\omega, Z}$ and $\Delta_{m, Z}$ are calculated numerically at the point $r_{Z}$.}
\label{table:41}
\end{center}
\end{table}

\begin{table}[H]
\begin{center}
\begin{adjustbox}{max width=\textwidth}
\begin{tabular}{| c | c | c | c | c | c | c | c | c |}
\hline
$\omega / \omega_{c}$ & $\phi_{0}/\GeV$ & $\phi_{X}/\GeV$ & $r_{X}/\GeV^{-1}$ & $Q_{X}$ & $E_{X}/\GeV$ & $(E/Q)_{X}/\GeV$ & $\Delta_{\omega, X}$ & $\Delta_{m, X}$\\
\hline
$0.707155$ & $3.2217991 \times 10^{17} $ & $3.22 \times 10^{15}$ & $8.89 \times 10^{-10}$ & $2.15 \times 10^{14}$ & $3.30 \times 10^{27}$ & $1.54 \times 10^{13}$ & $-4.47 \times 10^{-4}$ & $-0.22$ \\
\hline
$0.709$ & $1.3464098 \times 10^{16}$ & $1.35 \times 10^{14}$ & $3.70 \times 10^{-11}$ & $1.54 \times 10^{10}$ & $2.38 \times 10^{23}$ & $1.54 \times 10^{13}$ & $ -1.27 \times 10^{-3}$ & $-0.21$ \\
\hline
$0.71$ & $8.855792 \times 10^{15}$ & $8.86 \times 10^{13}$ & $2.43 \times 10^{-11}$ & $4.34 \times 10^{9}$ & $6.70 \times 10^{22}$ & $1.54 \times 10^{13}$ & $-7.78 \times 10^{-4}$ & $-0.21$ \\
\hline
$0.72$ & $1.960795 \times 10^{15}$ & $1.96 \times 10^{13}$ & $5.38 \times 10^{-12}$ & $4.42 \times 10^{7}$ & $6.97 \times 10^{20}$ & $1.58 \times 10^{13}$ & $6.58 \times 10^{-3}$ & $-0.19$\\
\hline
$0.73$ & $1.090258 \times 10^{15}$ & $1.09 \times 10^{13}$ & $3.03 \times 10^{-12}$ & $7.08 \times 10^{6}$ & $1.14 \times 10^{20}$ & $1.61 \times 10^{13}$ & $1.23 \times 10^{-2}$ & $-0.18$\\
\hline
$0.74$ & $7.45339 \times 10^{14}$ & $7.45 \times 10^{12}$ & $2.13 \times 10^{-12}$ & $2.16 \times 10^{6}$ & $3.54 \times 10^{19}$ & $1.64 \times 10^{13}$ & $1.71 \times 10^{-2}$ & $-0.16$ \\
\hline
$0.75$ & $5.61953 \times 10^{14}$ & $5.62 \times 10^{12}$ & $1.66 \times 10^{-12}$ & $8.85 \times 10^{5}$ & $1.48 \times 10^{19}$ & $1.67 \times 10^{13}$ & $2.13 \times 10^{-2}$ & $-0.15$\\
\hline
$0.80$ & $2.29632 \times 10^{14}$ & $2.30 \times 10^{12}$ & $8.97 \times 10^{-13}$ & $5.60 \times 10^{4}$ & $1.01 \times 10^{18}$ & $1.80 \times 10^{13}$ & $3.48 \times 10^{-2}$ & $-8.02 \times 10^{-2}$ \\
\hline
$0.85$ & $1.16877 \times 10^{14}$ & $1.17 \times 10^{12}$ & $7.84 \times 10^{-13}$ & $1.01 \times 10^{4}$ & $1.93 \times 10^{17}$ & $1.91 \times 10^{13}$ & $3.49 \times 10^{-2}$ & $-2.26 \times 10^{-2}$ \\
\hline
$0.89$ & $4.7918 \times 10^{13}$ & $4.79 \times 10^{11}$ & $1.21 \times 10^{-12}$ & $3.96 \times 10^{3}$ & $7.79 \times 10^{16}$ & $1.97 \times 10^{13}$ & $1.50 \times 10^{-2}$ & $3.69 \times 10^{-3}$ \\
\hline
\end{tabular}
\end{adjustbox}
\caption{Table listing important properties for the $m=0.9\omega_{c}$ Q-balls calculated using the $X$ point definition of the Q-ball radius. $\phi_{0}$ is the value of the scalar field at $r = 0$. $\phi_{X}$, $E_{X}$, $Q_{X}$, $(E/Q)_{X}$, $\Delta_{\omega, X}$ and $\Delta_{m, X}$ are calculated numerically at the point $r_{X}$.}
\label{table:42}
\end{center}
\end{table}

The energy-charge ratio increases by less than an order of magnitude from the largest $\phi_{0}$ Q-ball to the smallest at $\omega = 0.89\omega_{c}$. Comparing the data in Tables \ref{table:41} and \ref{table:42} also shows that the calculated energy, $E$, charge $Q$, and energy-charge ratio, $E/Q$, using the $X$ point definition of radius are comparable to the values obtained for these quantities calculated using the $Z$ point definition for each Q-ball. The radii for each Q-ball are similarly close in magnitude for each definition, with the difference between $r\left(X\right)$ and $r\left(Z\right)$ becoming slightly more pronounced in the smaller $\phi_{0}$ Q-balls.

\begin{figure}[H]
\begin{center}
\includegraphics[clip = true, width=0.75\textwidth, angle = 360]{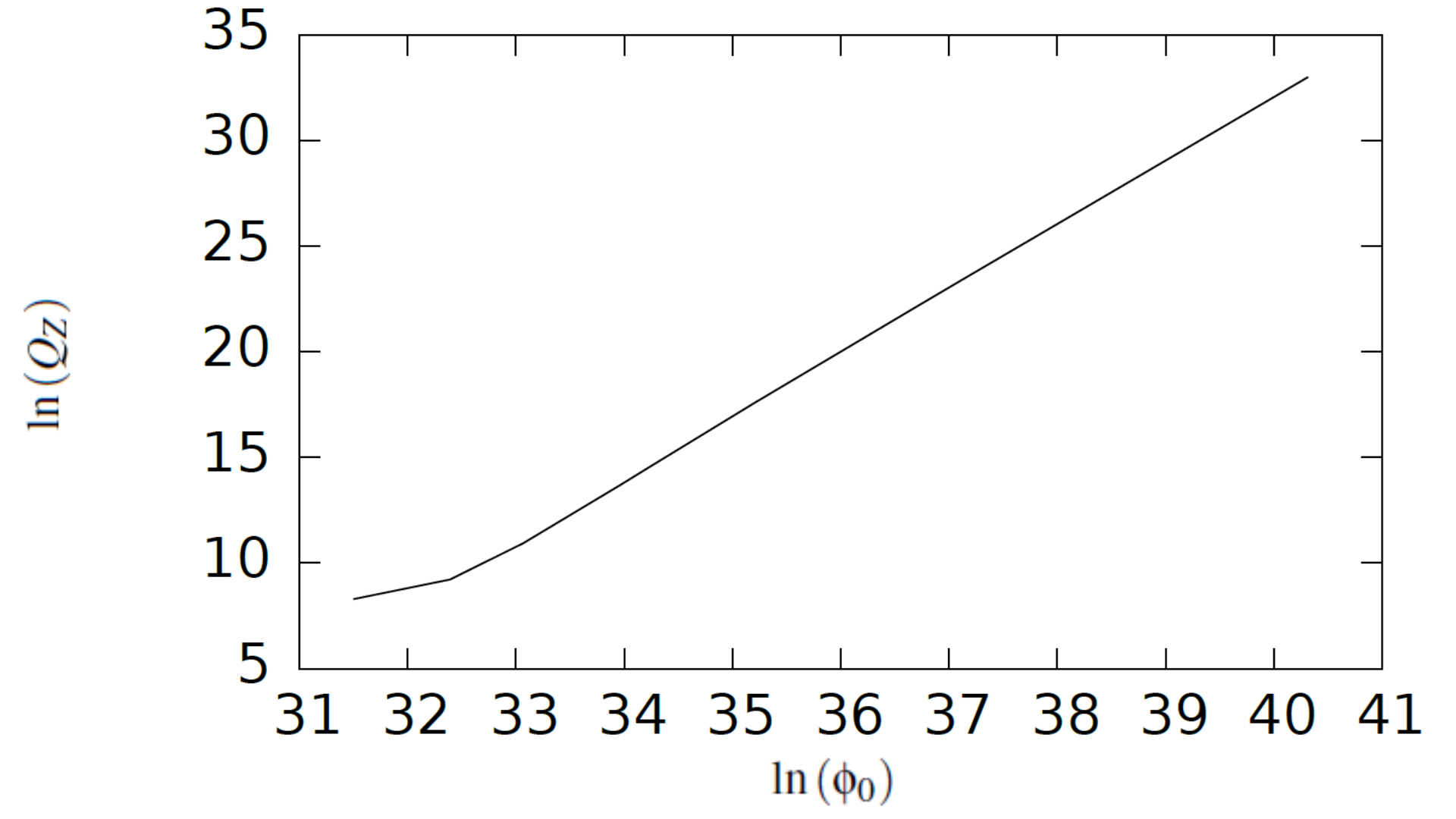}
\caption{Log-log plot of $\ln Q_{z}$, vs. $\ln \phi_{0}$ for the $m=0.9\omega_{c}$ Q-balls.}
\label{figure:49}
\end{center}
\end{figure}

\begin{figure}[H]
\begin{center}
\includegraphics[clip = true, width=0.75\textwidth, angle = 360]{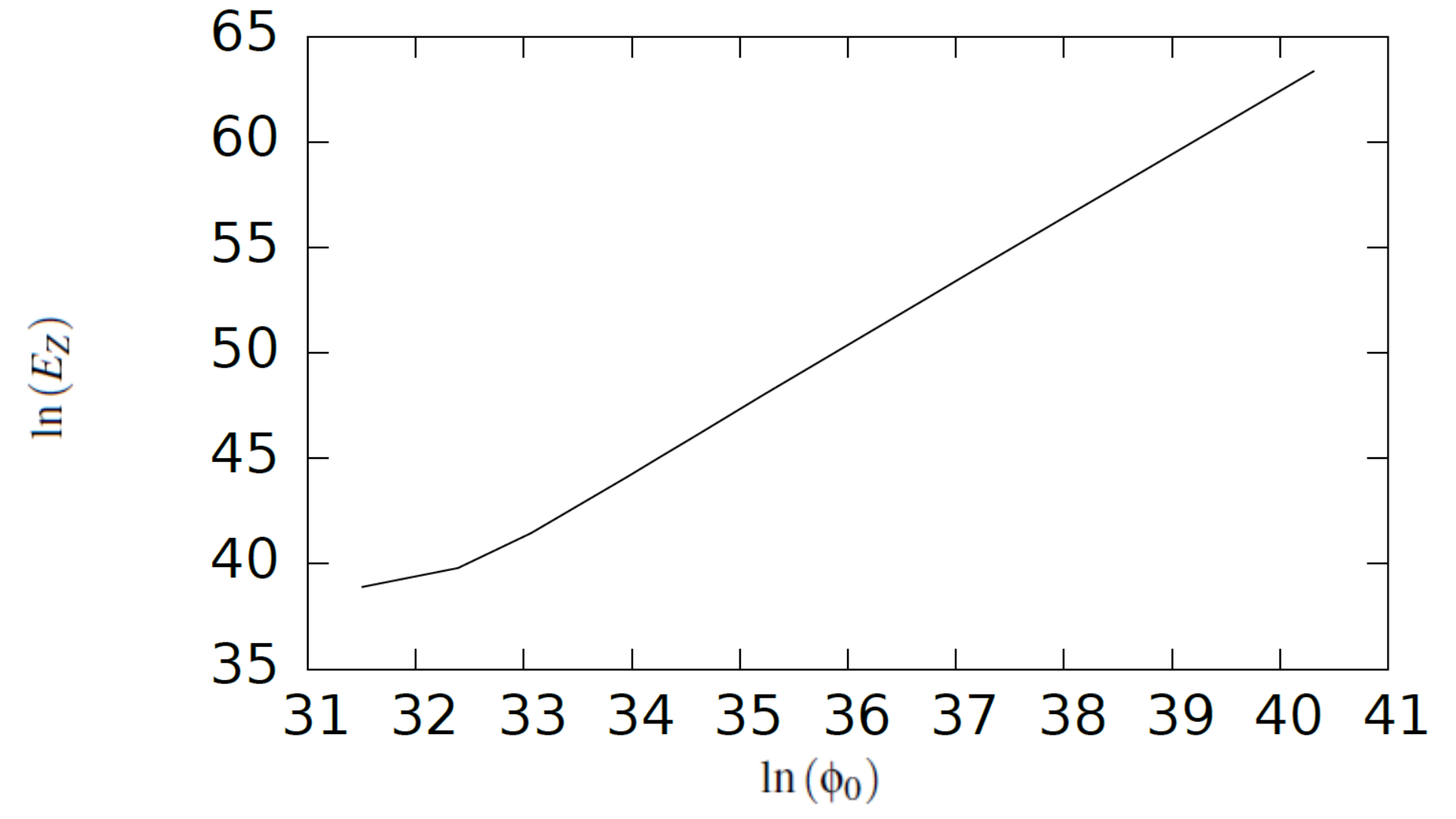}
\caption{Log-log plot of $\ln E_{z}$, vs. $\ln \phi_{0}$ for the $m=0.9\omega_{c}$ Q-balls.}
\label{figure:410}
\end{center}
\end{figure}

The larger $\phi_{0}$ Q-balls have generally larger radii and the increase in radius is linear with $\phi_{0}$ to a good approximation for $\phi_{0} > M_{Pl}/\sqrt{\xi}  \; ( \sim 6.9 \times 10^{13} \GeV)$. Increasing $\phi_{0}$ also increases the Q-ball energy and charge, as expected since a physically larger Q-ball (larger radius) will be composed of a larger number of scalars and will therefore carry a greater charge and energy.  Figures \ref{figure:49} and \ref{figure:410} show the relationships between the logarithm of the energy of the Q-balls and the logarithm of $\phi_{0}$, and the logarithm of the charge of the Q-balls and the logarithm of $\phi_{0}$ respectively. We can see from these plots that $E$ and $Q$ are proportional to $\phi_{0}^{3}$ to a good approximation for the larger $\phi_{0}$ Q-balls, and that the approximate proportionality becomes less representative for the smaller $\phi_{0}$ Q-balls. This proportionality is predicted by the analytical approximation derived in Section \ref{section:47}, and is therefore a useful test of the validity of the numerical Q-ball solutions.

\begin{figure}[H]
\begin{center}
\includegraphics[clip = true, width=0.75\textwidth, angle = 360]{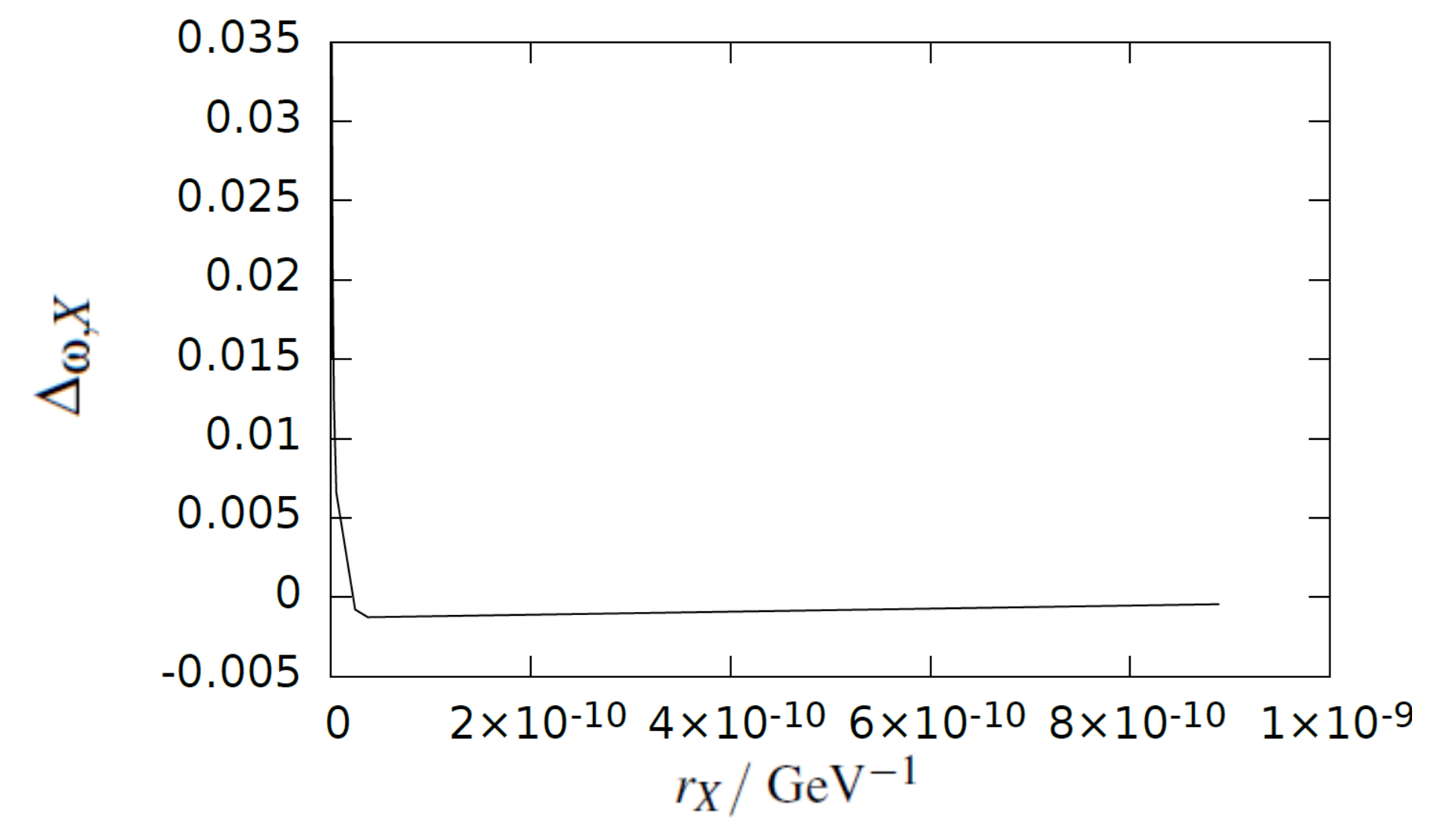}
\caption{$\Delta_{\omega,X}$ vs $r_{X}$ for the $m=0.9\omega_{c}$ Q-balls. } 
\label{figure:411}
\end{center}
\end{figure}

\noindent As we can see from Tables \ref{table:41} and \ref{table:42}, $\left| \Delta_{\omega} \right| \sim 10^{-4} - 10^{-2}$ in general. This shows that $\omega$ is close to the value of the energy-charge ratio, and therefore to a good approximation can be equated to the chemical potential for the Q-ball solutions considered in this work. $ \left| \Delta_{\omega} \right| $ increases as $\phi_{0}$ decreases, and from Figure \ref{figure:411} it is clear that that $ \Delta_{\omega} $ increases as Q-ball radius decreases in general. This is consistent with the patterns seen in the other quantities considered, in that the behaviour patterns of the properties for the smaller $\phi_{0}$ Q-balls are less well-defined than for the larger $\phi_{0}$ Q-balls. It is possible that this is a reflection of the Q-balls becoming unstable as the potential begins to deviate from the plateau ($\phi_{0} \rightarrow M_{Pl}/\sqrt{\xi}$).

\begin{figure}[H]
\begin{center}
\includegraphics[clip = true, width=0.75\textwidth, angle = 360]{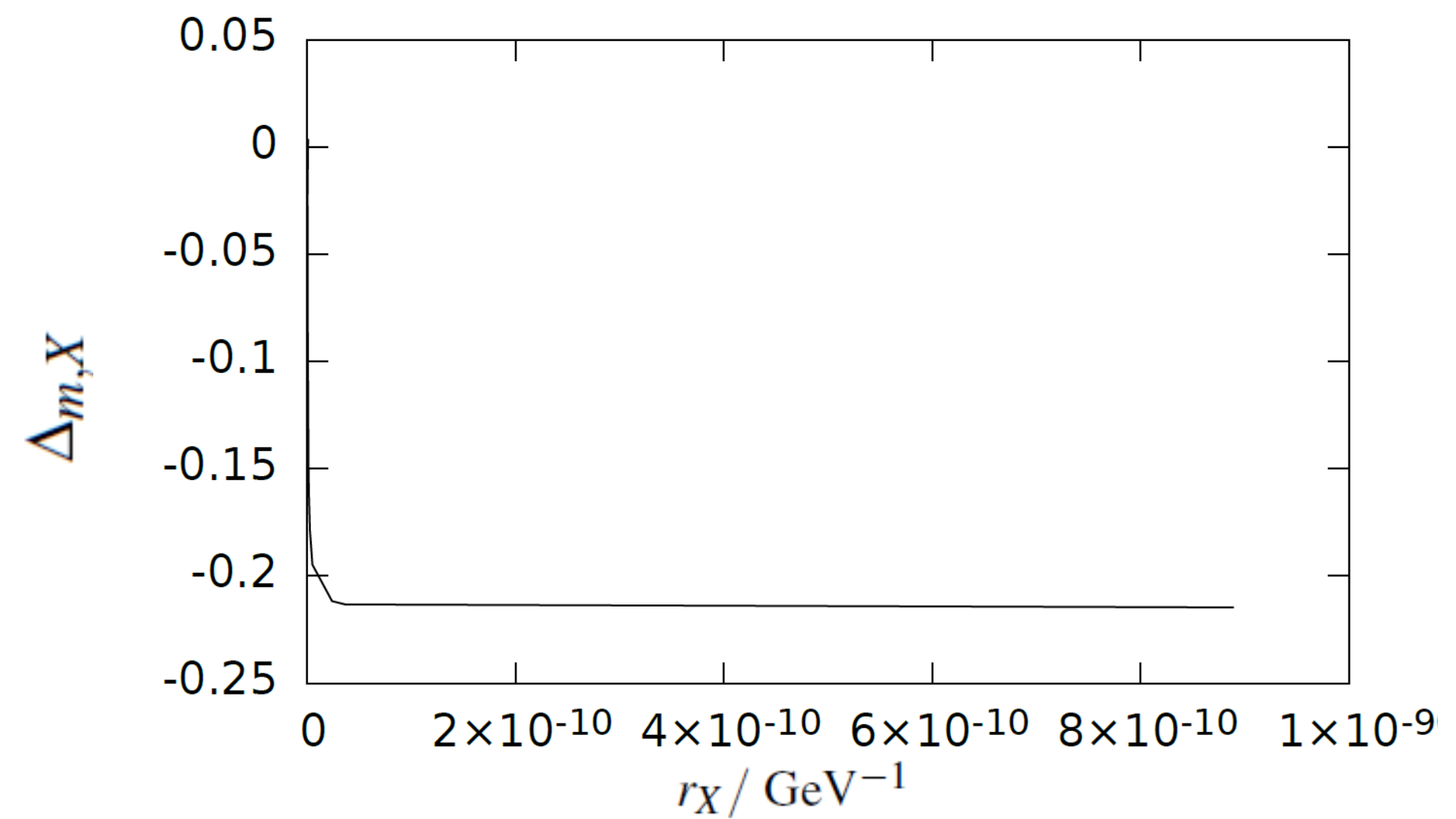}
\caption{$\Delta_{m,X}$ vs $r_{X}$ for the $m=0.9\omega_{c}$ Q-balls. } 
\label{figure:412}
\end{center}
\end{figure}

We can see from Tables \ref{table:41} and \ref{table:42}, and clearly in Figure \ref{figure:412}, that $\Delta_{m}$ is negative for all Q-ball solutions considered except the $\omega = 0.89\omega_{c}$ Q-ball, meaning that all of the Q-balls considered are absolutely stable except for the $\omega = 0.89\omega_{c}$ case. Since the magnitude of $\Delta_{m}$ for $\omega = 0.89\omega_{c}$ is still small although it is positive, it is possible that $\Delta_{m}$ is actually negative but smaller than the level of numerical errors in the Q-ball calculation.

The most significant result of the numerical analysis is that absolutely stable Q-balls of $\phi_{0} \geq 10^{17} \GeV$ can be generated in this model. This shows that the Q-ball mass window allows for the formation of Q-balls from field values we expect at the end of non-minimally coupled Palatini inflation (see Section \ref{section:44}). It is therefore possible that these Q-balls could form from the fragmentation of the inflaton condensate following tachyonic preheating at the end of non-minimally coupled Palatini inflation \cite{rubio19}. The presence of these Q-balls could significantly affect the post-inflation cosmology of the model, with consequences such as faster reheating and the production of gravitational waves from the decay of these Q-balls, leading to measurable effects. This is discussed further in Sections \ref{section:48} - \ref{section:410}.

\section{Analytical Approximation of the Q-ball Solution and Estimate of Q-ball Properties}\label{section:47}
In this section we derive an analytical approximation of the Q-ball solution, and calculate the energy and charge in this approximation as an estimate of the Q-ball properties. We compare these to the results obtained from the numerical solutions in Section \ref{section:46} and derive some important relations between energy and charge.

In this model we are assuming that inflation ends while the inflaton field is on the inflationary plateau. Using the plateau limit (\ref{eqn:q16}), we can approximate the right hand side of the Q-ball equation (\ref{eqn:q77}) to be 

\begin{equation}\label{eqn:q114}
\Omega^{2}\frac{\partial V_{\omega}}{\partial \phi} = - \frac{\gamma}{\phi},
\end{equation}

\noindent to leading order in $1/\phi^{2}$, where $\gamma$ is defined as

\begin{equation}\label{eqn:q115}
\gamma = \frac{M_{pl}^{2}}{\xi}\left( m^{2} + \omega^{2} - \frac{\lambda M_{pl}^{2}}{\xi}\right).
\end{equation}

\noindent In the plateau limit we have that $\phi >> M_{pl}/\sqrt{\xi}$, and we also assume that

\begin{equation}\label{eqn:q116}
\left| \frac{\partial^{2}\phi}{\partial r^{2}} + \frac{2}{r}\frac{\partial \phi}{\partial r} \right| >> K\left(\phi \right) \left(\frac{\partial \phi}{\partial r}\right)^{2},
\end{equation}

\noindent where $K\left(\phi \right) \sim 1/\phi$ on the plateau (we will confirm that (\ref{eqn:q116}) is consistent below). Using (\ref{eqn:q114}) and (\ref{eqn:q116}), the Q-ball equation can be approximated as

\begin{equation}\label{eqn:q117}
\frac{\partial^{2}\phi}{\partial r^{2}} + \frac{2}{r}\frac{\partial \phi}{\partial r} = - \frac{\gamma}{\phi},
\end{equation}

\noindent where $\gamma > 0$ for Q-ball solutions to exist. 

We assume that $\phi$ does not deviate from its initial value $\phi_{0}$ too much, and also that

\begin{equation}\label{eqn:q118}
\frac{\partial \phi}{\partial r} \rightarrow 0,
\end{equation}

\noindent as $r \rightarrow 0$.

\noindent Setting $\phi = \phi_{0}$ on the right hand side of (\ref{eqn:q117}), we make the following ansatz

\begin{equation}\label{eqn:q119}
\phi \left( r \right) = \phi_{0} - A r^{2},
\end{equation}

\noindent where $A$ is a coefficient to be determined. In order for the field profile to have the typical Q-ball shape, we require that the right hand side of the equation is negative. This means that

\begin{equation}\label{eqn:q120}
\gamma > 0 \longrightarrow m^{2} + \omega^{2} > \frac{\lambda M_{pl}^{2}}{\xi}.
\end{equation}

\noindent Substituting (\ref{eqn:q119}) into (\ref{eqn:q117}), we find that the left hand side becomes a constant, $-6A$, and that the approximation (\ref{eqn:q116}) holds because 

\begin{equation}\label{eqn:q121}
K\left(\phi \right) \left(\frac{\partial \phi}{\partial r}\right)^{2}  \approx \frac{4A^{2}r^{2}}{\phi_{0}},
\end{equation}

\noindent is a very small number for $\phi >> M_{pl}/\sqrt{\xi}$, and goes to zero as $r \rightarrow 0$. This leaves

\begin{equation}\label{eqn:q122}
-6A = -\frac{\gamma}{\phi_{0}} \Rightarrow A = \frac{\gamma}{6 \phi_{0}},
\end{equation}

\noindent and approximation of the field profile (\ref{eqn:q119}) is therefore

\begin{equation}\label{eqn:q123}
\phi \left( r \right) = \phi_{0} -  \frac{\gamma}{6 \phi_{0}}r^{2}.
\end{equation}

The assumption that $\phi \approx \phi_{0}$ holds to a good approximation if

\begin{equation}\label{eqn:q124}
r^{2} < \frac{r_{Q}^{2}}{4} = \frac{6 \phi_{0}^{2}}{4\gamma},
\end{equation}

\noindent since applying (\ref{eqn:q119}) at $r_{Q}/2$ gives $\phi \left(r_{Q}/2\right) = \frac{3}{4}\phi_{0}$.

\noindent To a good approximation, we can therefore say that the Q-ball solution follows

\begin{equation}\label{eqn:q125}
\phi \left( r \right) = \phi_{0}\left[ 1 - \left(\frac{r}{r_{Q}}\right)^{2} \right],
\end{equation}

\noindent for values of $r < r_{Q}/2$, where the assumption (\ref{eqn:q116}) is consistent with this solution. We can use this to estimate the Q-ball properties, such as energy and charge, which can be used as a check on the numerical solution and also to determine the parameter dependence of the numerical Q-ball solutions.

\subsection{Analytical Estimate of Q-ball Energy and Charge}\label{section:471}

In this section we derive analytical expressions for the energy, $E$, and the global charge, $Q$, of the non-minimally coupled Palatini Q-balls. Assuming $\phi \approx \phi_{0}$, and working in the plateau limit, we have that the extremised Q-ball energy $E$ is approximately (from (\ref{eqn:q72}))

\begin{equation}\label{eqn:q126}
E = \int^{r_{Q}/2}_{0} 4\pi r^{2} \; dr \left[\frac{1}{2\Omega^{2}}\left(\frac{\partial \phi}{\partial r}\right)^{2} + \frac{V\left(\phi \right)}{\Omega^{4}}\right].
\end{equation}

\noindent For $r < r_{Q}/2$, and assuming $\phi \approx \phi_{0}$ we approximate the Jordan frame potential to be

\begin{equation}\label{eqn:q127}
V\left(\phi \right) \approx \frac{\lambda}{4}\phi_{0}^{4},
\end{equation}

\noindent since $\lambda\phi_{0}^{4} >> m^{2}\phi_{0}^{2}$ in the Jordan frame in this case,  and the conformal factor is $\Omega^{2} \approx \xi \phi_{0}^{2}/M_{pl}^{2}$. Using (\ref{eqn:q125}), we have that

\begin{equation}\label{eqn:q128}
\frac{\partial \phi}{\partial r} = -\frac{2r \phi_{0}}{r_{Q}^{2}}.
\end{equation}

\noindent Using (\ref{eqn:q127}) and (\ref{eqn:q128}), we find that the Q-ball energy (\ref{eqn:q126}) becomes

\begin{equation}\label{eqn:q129}
E = \left( \frac{\omega^{2}\phi_{0}^{2}}{2\Omega^{2}} + \frac{\lambda \phi_{0}^{4}}{4\Omega^{4}}\right) \int^{r_{Q}/2}_{0} 4\pi r^{2} \; dr + \frac{2\phi_{0}^{2}}{r_{Q}^{4}}\int^{r_{Q}/2}_{0} 4\pi r^{4} \; dr.
\end{equation}

\noindent Performing the integration gives

\begin{equation}\label{eqn:q130}
E = \left( \frac{\omega^{2}\phi_{0}^{2}}{2\Omega^{2}} + \frac{\lambda \phi_{0}^{4}}{4\Omega^{4}}\right)\frac{4\pi}{3}\left(\frac{r_{Q}}{2}\right)^{3} + \frac{2\phi_{0}^{2}}{r_{Q}^{4}}\frac{4\pi}{5}\left(\frac{r_{Q}}{2}\right)^{5},
\end{equation}

\noindent and with some rearrangement this is

\begin{equation}\label{eqn:q131}
E = \frac{\pi M_{pl}^{2}}{2 \xi}\left[ \frac{1}{6}\left(\omega^{2} + \frac{\lambda M_{pl}^{2}}{2\xi}\right) + \frac{1}{10 r_{Q}^{2}}\right]r_{Q}^{3}.
\end{equation}

\noindent Using (\ref{eqn:q115}) and (\ref{eqn:q124}), we have that 

\begin{equation}\label{eqn:q132}
\frac{1}{r_{Q}^{2}} = \frac{\gamma}{6\phi_{0}^{2}} = \frac{M_{pl}^{2}}{6\xi \phi_{0}^{2}}\left(m^{2} + \omega^{2} - \omega_{c}^{2} \right),
\end{equation}

\noindent and since we are working in the plateau limit, $M_{pl}^{2}/\xi \phi_{0}^{2} << 1$, we can say that

\begin{equation}\label{eqn:q133}
\frac{M_{pl}^{2}}{6\xi \phi_{0}^{2}}\left(m^{2} + \omega^{2} - \omega_{c}^{2} \right) << \frac{1}{6}\left(m^{2} + \omega^{2} - \omega_{c}^{2} \right).
\end{equation}

\noindent This means that the $1/r_{Q}^{2}$ term is small compared to the $1/6$ term in (\ref{eqn:q131}), and the Q-ball energy can be approximated as

\begin{equation}\label{eqn:q134}
E = \frac{\pi M_{pl}^{2}}{12 \xi}\left[ \omega^{2} + \frac{\lambda M_{pl}^{2}}{2\xi}\right] r_{Q}^{3}.
\end{equation}

Note that this expression holds for $r < r_{Q}/2$ and is therefore a lower bound estimate on the Q-ball energies calculated numerically. Similarly, the Q-ball charge can be approximated analytically. The charge density of the Q-balls from (\ref{eqn:q61}) using the Q-ball ansatz (\ref{eqn:q70}) is 

\begin{equation}\label{eqn:q135}
\rho_{Q} = \frac{\omega \phi^{2}}{\Omega^{2}},
\end{equation}

\noindent where for $\phi \approx \phi_{0}$ and $\Omega^{2} \approx \xi \phi_{0}^{2}/M_{pl}^{2}$ this is 

\begin{equation}\label{eqn:q136}
\rho_{Q} = \frac{\omega M_{pl}^{2}}{\xi}.
\end{equation}

\noindent The global charge from (\ref{eqn:q55}) using (\ref{eqn:q136}) is thus

\begin{equation}\label{eqn:q137}
Q = \int^{r_{Q}/2}_{0} 4\pi r^{2} \; dr \frac{\omega M_{pl}^{2}}{\xi} = \frac{4\pi \omega M_{pl}^{2}}{\xi}\left( \frac{r_{Q}}{2}\right)^{3},
\end{equation}

\noindent and integrating for $r < r_{Q}/2$ gives the Q-ball charge to be

\begin{equation}\label{eqn:q138}
Q = \frac{\pi \omega M_{pl}^{2}}{6 \xi} r_{Q}^{3}.
\end{equation}

\noindent We can see from (\ref{eqn:q134}) and (\ref{eqn:q138}) that the energy and charge of the Q-balls are both proportional to $r_{Q}^{3}$, and therefore to the volume of the Q-ball. From (\ref{eqn:q124}), we find that $r_{Q} \propto \phi_{0}$, and $E, Q \propto \phi_{0}^{3}$ in the analytical approximation. From the energy and charge calculated for the numerical solutions in Section \ref{section:462}, we find that there is a proportionality between the Q-ball energy and charge and $\phi_{0}^{3}$, as illustrated more clearly in Figures \ref{figure:49} and \ref{figure:410}, and that the relationship is truest and essentially exact for the large $\phi_{0}$ Q-balls. Figures \ref{figure:46} and \ref{figure:48} in Section \ref{section:462} show the analytical approximate solution for $\phi$, (\ref{eqn:q125}), as a function of $r$ for large and small $\phi_{0}$. Comparing these with the numerical solutions for the same $\phi_{0}, \omega$ shown in Figures \ref{figure:45} and \ref{figure:47}, it is clear that the analytical approximation closely follows the numerical solution for small $r$. For $r < r_{Q}$ specifically, the analytical approximation (\ref{eqn:q125}) is a good fit to the exact solution calculated by solving (\ref{eqn:q77}) in the case of larger $\phi_{0}$ ($\phi_{0} >> M_{pl}/\sqrt{\xi}$), but the agreement is less apparent for smaller $\phi_{0} \rightarrow M_{Pl}/\sqrt{\xi}$. These results show that the analytical approximation of the Q-ball solution, and therefore the analytical approximations of the energy and charge, are a good approximation of the behaviour of the Q-ball solutions and the properties of the Q-balls themselves in the limit that $\phi_{0}$ is large and $r$ is small.

Taking the ratio of the analytical approximation of energy (\ref{eqn:q134}) and charge (\ref{eqn:q138}) gives

\begin{equation}\label{eqn:q139}
\frac{E}{Q} = \frac{1}{2}\left[ \omega + \frac{\omega_{c}^{2}}{2\omega}\right],
\end{equation}

\noindent as an estimate of the energy-charge ratio. In Section \ref{section:462} we found that the numerical results for the energy-charge ratio show that $E/Q \approx \omega$ when $\phi_{0}$ is large, with the deviation from $E/Q = \omega$ increasing for the smaller $\phi_{0}$ Q-balls. We see here that the analytical approximation of the energy-charge ratio does not produce this relation. It is likely that this is due to the fact that $E$ and $Q$ are only integrated to $r = r_{Q}/2$ in the analytical approximation. In Section \ref{section:472} we derive an exact expression for the energy-charge ratio in terms of $\phi \left(r \right)$, which will allow for a more accurate determination of the energy-charge ratio.

Table \ref{table:43} shows the estimates of the energy and charge calculated using the analytical approximation expressions (\ref{eqn:q134}) and (\ref{eqn:q138}), for each of the ten Q-ball solutions presented in Section \ref{section:462}.

\begin{table}[H]
\begin{center}
\begin{tabular}{| c | c | c | c | c |}
\hline
$\omega / \omega_{c}$ & $\phi_{0}/\GeV$ & $r_{Q}/\GeV^{-1}$ & $Q$  & $E/\GeV$ \\
\hline
$0.707155$ & $3.2217991 \times 10^{17} $ & $9.46 \times 10^{-10}$ & $3.23 \times 10^{13}$ & $4.97 \times 10^{26}$\\
\hline
$0.709$ & $1.3464098 \times 10^{16}$ & $3.94 \times 10^{-11}$ & $2.34 \times 10^{9}$ & $3.60 \times 10^{22}$\\
\hline
$0.71$ & $8.855792 \times 10^{15}$ & $2.58 \times 10^{-11}$ & $6.58 \times 10^{8}$ & $1.01 \times 10^{22}$\\
\hline
$0.72$ & $1.960795\times 10^{15}$ & $5.60 \times 10^{-12}$ & $6.82 \times 10^{6}$ & $1.05 \times 10^{20}$ \\
\hline
$0.73$ & $1.090258 \times 10^{15}$ & $3.05 \times 10^{-12}$ & $1.12 \times 10^{6}$ & $1.72 \times 10^{19}$ \\
\hline
$0.74$ & $7.45339 \times 10^{14}$ & $2.04 \times 10^{-12}$ & $3.39 \times 10^{5}$ & $5.22 \times 10^{18}$ \\
\hline
$0.75$ & $5.61953 \times 10^{14}$ & $1.51 \times 10^{-12}$ & $1.39 \times 10^{5}$ & $2.15 \times 10^{18}$ \\
\hline
$0.80$ & $2.29632 \times 10^{14}$ & $5.60 \times 10^{-13}$ & $7.58 \times 10^{3}$ & $1.18 \times 10^{17}$ \\
\hline
$0.85$ & $1.16877 \times 10^{14}$ & $2.62 \times 10^{-13}$ & $825$ & $1.29 \times 10^{16}$ \\
\hline
$0.89$ & $4.7918 \times 10^{13}$ & $1.01 \times 10^{-13}$ & $50$ & $7.82 \times 10^{14}$ \\
\hline
\end{tabular}
\caption{Table showing $r_{Q}$, $E$ and $Q$ as a function of $\phi_{0}$ from the analytical approximation from (\ref{eqn:q134}) and (\ref{eqn:q138}) for the $m=0.9\omega_{c}$ Q-ball solutions presented in Section \ref{section:46}.}
\label{table:43}
\end{center}
\end{table}

\noindent Comparing these results to the numerically calculated values of energy and charge in Tables \ref{table:41} and \ref{table:42} it is clear that the analytical estimates (\ref{eqn:q134}) and (\ref{eqn:q138}) consistently underestimate the energy and charge by about an order of magnitude for most of the Q-ball solutions. This makes sense when one considers the fact that the analytical expressions are only integrated to $r = r_{Q}/2$, as this is the limit at which the analytical approximation ceases to be a valid description of the behaviour of the Q-ball solution. The underestimate becomes greater as $\phi_{0}$ decreases, which is consistent with the analytical Q-ball profile $\phi(r)$ becoming a poorer fit to the numerical profile as $\phi_{0}$ approaches the edge of plateau, and therefore the limit of plateau description of the inflaton field dynamics. However, since the analytical approximation is a good approximation for the behaviour of the Q-ball solutions with larger $\phi_{0}$, it can be used as an estimate of the properties of these Q-balls and to examine the dependence of these quantities on the other parameters of the model.

\subsection{Analytical Derivation of the Q-ball Chemical Potential Relation}\label{section:472}
In this section we derive the energy per unit charge for Palatini non-minimally coupled Q-balls and demonstrate that this is can be equated to the $\omega$ parameter in this model. This discussion follows very closely the derivation in Heeck et. al. \cite{heeck21} for the case of conventional minimally coupled Q-balls.

\noindent As a starting point we consider the Q-ball energy functional (\ref{eqn:q72})

\begin{equation}\label{eqn:q140}
E_{Q} = \int 4\pi r^{2} dr \left[ \frac{\omega^{2}\phi^{2}}{2\Omega^{2}} + \frac{1}{2\Omega^{2}}\left(\frac{\partial \phi}{\partial r}\right)^{2} + \frac{V}{\Omega^{4}}\right] + \omega Q - \omega \int 4\pi r^{2} dr   \rho_{Q},
\end{equation}

\noindent where

\begin{equation}\label{eqn:q141}
\rho_{Q} = \frac{\omega \phi^{2}}{\Omega^{2}}.
\end{equation}

\noindent Absorbing the $\rho_{Q}$ integral in (\ref{eqn:q140}) into the first integral and rewriting the $\omega$ term, this becomes

\begin{equation}\label{eqn:q142}
E_{Q} = \int 4\pi r^{2} dr \left[\frac{1}{2\Omega^{2}}\left( \frac{\partial \phi}{\partial r}\right)^{2} + \frac{\omega^{2}\phi^{2}}{\Omega^{2}} - \frac{\omega^{2}\phi^{2}}{2\Omega^{2}} - \frac{\omega^{2}\phi^{2}}{\Omega^{2}} + \frac{V}{\Omega^{4}}\right] + \omega Q.
\end{equation}

\noindent We now differentiate (\ref{eqn:q142}) with respect to $\omega$, 

\begin{equation}\label{eqn:q143}
\frac{d E_{Q}}{d\omega} = \int 4\pi r^{2} dr \frac{\partial}{\partial \omega}\left[\frac{1}{2\Omega^{2}}\left( \frac{\partial \phi}{\partial r}\right)^{2} + \frac{\omega^{2}\phi^{2}}{\Omega^{2}} - \frac{\omega^{2}\phi^{2}}{2\Omega^{2}} - \frac{\omega^{2}\phi^{2}}{\Omega^{2}} + \frac{V}{\Omega^{4}}\right] + Q + \omega \frac{d Q}{d\omega},
\end{equation}

\noindent and rewrite the integral using the field rescaling outlined in Section \ref{section:451} to proceed with this part of the calculation in terms of $\sigma$

\begin{multline}\label{eqn:q144}
\frac{d E_{Q}}{d\omega} = \int 4\pi r^{2} dr \left[ \frac{1}{2}\frac{\partial}{\partial \omega}\left(\frac{\partial \sigma}{\partial r}\right)^{2} + \frac{\partial}{\partial \omega}\left(\frac{\omega^{2}\phi^{2}}{\Omega^{2}}\right) - \frac{\omega^{2}}{2}\frac{\partial}{\partial \omega}\left(\frac{\phi^{2}}{\Omega^{2}}\right) \right. \\
\left. - \frac{\partial}{\partial \omega}\left(\frac{\omega^{2} \phi^{2}}{\Omega^{2}}\right) + \frac{\partial}{\partial \omega}\left(\frac{V}{\Omega^{4}}\right)\right] + \omega \frac{d Q}{d\omega}.
\end{multline}

\noindent Rescaling the Q-ball equation in terms of $\sigma$ (\ref{eqn:q90}),we find that the left-hand side can be written

\begin{equation}\label{eqn:q145}
\frac{d^{2}\sigma}{d r^{2}} + \frac{2}{r}\frac{d \sigma}{d r} = \frac{1}{r^{2}}\frac{d}{dr}\left(r^{2}\frac{d\sigma}{dr}\right),
\end{equation}

\noindent which can then be substituted back into the Q-ball equation (\ref{eqn:q90}) and rearranged to give

\begin{equation}\label{eqn:q146}
\frac{\partial}{\partial \sigma}\left(\frac{V}{\Omega^{4}}\right) = \frac{1}{r^{2}}\frac{d}{dr}\left(r^{2}\frac{d\sigma}{dr}\right) + \frac{\partial}{\partial \sigma}\left(\frac{\omega^{2}\phi^{2}}{2\Omega^{2}}\right).
\end{equation}

\noindent We can write the left-hand side of (\ref{eqn:q146})

\begin{equation}\label{eqn:q147}
\frac{\partial}{\partial \omega}\left(\frac{V}{\Omega^{4}}\right) = \frac{\partial}{\partial \sigma}\left(\frac{V}{\Omega^{4}}\right)\frac{\partial \sigma}{\partial \omega},
\end{equation}

\begin{equation}\label{eqn:q148}
\Rightarrow \frac{\partial}{\partial \omega}\left(\frac{V}{\Omega^{4}}\right) = \frac{1}{r^{2}}\frac{d}{dr}\left(r^{2}\frac{d\sigma}{dr}\right)\frac{\partial \sigma}{\partial \omega} + \frac{\partial}{\partial \sigma}\left(\frac{\omega^{2}\phi^{2}}{2\Omega^{2}}\right)\frac{\partial \sigma}{\partial \omega}.
\end{equation}

\noindent Substituting (\ref{eqn:q148}) back into (\ref{eqn:q144}) gives

\begin{multline}\label{eqn:q149}
\frac{d E_{Q}}{d\omega} = \int 4\pi r^{2} dr \left[\frac{1}{2}\frac{\partial}{\partial \omega}\left(\frac{\partial \sigma}{\partial r}\right) \left(\frac{\partial \sigma}{\partial r}\right) -\frac{\omega^{2}}{2}\frac{\partial}{\partial \omega}\left(\frac{\phi^{2}}{\Omega^{2}}\right) + \frac{1}{r^{2}}\frac{d}{dr}\left(r^{2}\frac{d\sigma}{dr}\right)\frac{\partial \sigma}{\partial \omega} \right. \\
\left. + \frac{\partial}{\partial \sigma}\left(\frac{\omega^{2}\phi^{2}}{2\Omega^{2}}\right)\frac{\partial \sigma}{\partial \omega}\right] + \omega \frac{d Q}{d\omega},
\end{multline}

\noindent after canceling the $\pm \frac{\partial}{\partial \omega}\left(\frac{\omega^{2}\phi^{2}}{\Omega^{2}}\right)$ terms. Rewriting the second term in (\ref{eqn:q149})

\begin{equation}\label{eqn:q150}
\frac{\omega^{2}}{2}\frac{\partial}{\partial \omega}\left(\frac{\phi^{2}}{\Omega^{2}}\right) = \frac{\omega^{2}}{2}\frac{\partial}{\partial \sigma}\left(\frac{\phi^{2}}{\Omega^{2}}\right) \frac{\partial \sigma}{\partial \omega},
\end{equation}

\noindent and canceling with the fourth term in (\ref{eqn:q149}) leaves

\begin{equation}\label{eqn:q151}
\frac{d E_{Q}}{d\omega} = \int 4\pi r^{2} dr \left[\frac{1}{2}\frac{\partial}{\partial \omega}\left(\frac{\partial \sigma}{\partial r}\right) \left(\frac{\partial \sigma}{\partial r}\right) + \frac{1}{r^{2}}\frac{d}{dr}\left(r^{2}\frac{d\sigma}{dr}\right)\frac{\partial \sigma}{\partial \omega}\right] + \omega \frac{dQ}{d\omega}.
\end{equation}

\noindent We then integrate the second term by parts 

\begin{equation}\label{eqn:q152}
\int \frac{1}{r^{2}}\frac{d}{dr}\left(r^{2}\frac{d\sigma}{dr}\right)\frac{\partial \sigma}{\partial \omega} dr = \left. \frac{\partial \sigma}{\partial r}\frac{\partial \sigma}{\partial \omega}\right| - \int \frac{\partial \sigma}{\partial r} \frac{\partial^{2} \sigma}{\partial r \partial \omega},
\end{equation}

\noindent and obtain

\begin{equation}\label{eqn:q153}
\frac{d E_{Q}}{d\omega} = \int 4\pi r^{2} dr \left[\frac{1}{2}\frac{\partial}{\partial \omega}\left(\frac{\partial \sigma}{\partial r}\right) \left(\frac{\partial \sigma}{\partial r}\right) -  \frac{\partial \sigma}{\partial r} \frac{\partial^{2} \sigma}{\partial r \partial \omega} \right] + \omega \frac{dQ}{d\omega},
\end{equation}

\noindent where the boundary term from the integration is neglected due to the fact that the gradient of the scalar field goes to zero at infinity. The remaining terms in the integrand cancel and we are left with precisely

\begin{equation}\label{eqn:q154}
\frac{d E_{Q}}{d\omega} = \omega \frac{dQ}{d\omega}.
\end{equation}

\noindent This is the same result as derived in \cite{heeck21} for the case of a conventional canonical scalar, and shows that for Q-balls in non-minimally coupled Palatini gravity, the $\omega$ parameter can be equated to the chemical potential of the system 

\begin{equation}\label{eqn:q155}
\frac{d E_{Q}}{dQ} = \omega,
\end{equation}

\noindent and that $E_{Q}\left(\omega \right)$ and $Q\left( \omega \right)$ increase and decrease together, as in the case of conventional Q-balls \cite{coleman85}.

This makes sense in terms of $\omega$ being the Lagrange multiplier, since for a given $Q$ there will be a value of $\omega$ which extremises the energy of the scalar field subject to the constraint of constant charge. Hence $Q$ is a function of $\omega$, and extremising the scalar field for a fixed $Q(\omega)$ will produce an energy configuration $E_{Q}(\omega)$ corresponding to a Q-ball. 

It is also straightforward to show analytically that the $\omega$ parameter is approximately the ratio of the energy to the global charge. The following derivation closely follows that of the derivation of Eq.$12$ in Heeck et al. \cite{heeck21} for the case of conventional Q-balls.

From (\ref{eqn:q72}), the effective Q-ball action can be written

\begin{equation}\label{eqn:q156}
\int dr \; \mathcal{L_{Q}} = \int 4\pi r^{2} \; dr \left[\frac{1}{2}\left(\frac{d\sigma}{dr}\right)^{2} -\frac{\omega^{2}\phi^{2}}{2\Omega^{2}} + \frac{V}{\Omega^{4}}\right],
\end{equation}

\noindent where the inflaton is understood as being a function of the rescaled field $\sigma$ in this case. We then perform a coordinate rescaling $r \rightarrow \chi \tilde{r}$, and write (\ref{eqn:q156}) in terms of $\tilde{r}$

\begin{equation}\label{eqn:q157}
\int d\tilde{r} \; \mathcal{L_{Q}} = \int 4\pi \; d\tilde{r} \;  \tilde{r}^{2}  \;  \chi^{3} \, \left[\frac{1}{2\chi^{2}}\left(\frac{\partial \sigma}{\partial r}\right)^{2} - \frac{\omega^{2}\phi^{2}}{2\Omega^{2}} + \frac{V}{\Omega^{4}}\right].
\end{equation}

\noindent This (\ref{eqn:q157}) contains an explicit dependence on $\chi$ from the factors present in the integrand, and an implicit dependence on $\chi$, since the quasi-canonical field is a function of $r$ and transforms as $\sigma \left( r\right) \rightarrow \sigma \left(\chi \tilde{r} \right)$, meaning that a small variation in $\chi$ will cause a small variation in $\sigma$. This means that we can write the variation of the Q-ball effective Lagrangian with respect to $\chi$ as

\begin{equation}\label{eqn:q158}
\frac{d\mathcal{L_{Q}}}{d\chi} = \left(\frac{d\mathcal{L_{Q}}}{d\chi}\right)_{explicit} +  \left(\frac{d\mathcal{L_{Q}}}{d\chi}\right)_{implicit},
\end{equation}

\noindent where $\chi = 1$ corresponds to a Q-ball solution which extremises the Q-ball effective action. This means that $\int d\tilde{r} \left(\frac{d \mathcal{L}_{Q}}{d\chi}\right)_{implicit} = 0$ at $\chi =1$. Since the value of (\ref{eqn:q157}) is unchanged following the rescaling of the radial coordinate, we require that

\begin{equation}\label{eqn:q159}
\int d\tilde{r} \; \left(\frac{d \mathcal{L}_{Q}}{d\chi}\right)_{explicit} = 0,
\end{equation}

\noindent when $\chi = 1$. The left-hand side of this expression is 

\begin{multline}\label{eqn:q160}
\int d\tilde{r} \; \left(\frac{d \mathcal{L}_{Q}}{d\chi}\right)_{explicit} = 4\pi \int d\tilde{r} \; \frac{\partial}{\partial \chi}\left[ \frac{\chi}{2}\left(\frac{d\sigma}{dr}\right)^{2} - \frac{\chi^{3}\omega^{2}\phi^{2}}{2\Omega^{2}} + \frac{\chi^{3}V}{\Omega^{4}}\right] \\ 
= 4\pi \int \; d\tilde{r} \; \tilde{r}^{2} \left[\frac{1}{2}\left(\frac{d\sigma}{d\tilde{r}}\right)^{2} + 3\chi^{2}\left(\frac{V}{\Omega^{4}} - \frac{\omega^{2}\phi^{2}}{2\Omega^{2}}\right)\right],
\end{multline}

\noindent which means that for $\chi = 1, \tilde{r} \rightarrow r$, and in order for (\ref{eqn:q159}) to be satisfied we require that

\begin{equation}\label{eqn:q161}
4\pi \int dr \; r^{2} \left[\frac{1}{2}\left(\frac{d\sigma}{dr}\right)^{2} + 3\left(\frac{V}{\Omega^{4}} - \frac{\omega^{2} \phi^{2}}{2\Omega^{2}}\right)\right] = 0.
\end{equation}

\noindent This can be rewritten to obtain the condition

\begin{equation}\label{eqn:q162}
\begin{split}
4\pi \int dr \; r^{2} \left[ \frac{1}{2}\left( \frac{d\sigma}{dr} \right)^{2} + \frac{3V}{\Omega^{4} } \right] = 4\pi \int r^{2} & \; dr \frac{ 3\omega^{2} \phi^{2} }{ 2 \Omega^{2} } \\
& = 4\pi \cdot \frac{3}{2}\int r^{2} \; dr \frac{ \omega^{2} \phi^{2} }{ \Omega^{2} } = \frac{3}{2}\omega Q,
\end{split}
\end{equation}

\begin{equation}\label{eqn:q163}
\Rightarrow 4\pi \int dr \;  r^{2} \left[\frac{1}{2}\left(\frac{d\sigma}{dr}\right)^{2} + 3\left(\frac{V}{\Omega^{4} }\right) \right] = \frac{3}{2}\omega Q.
\end{equation}

\noindent The Q-ball energy functional (\ref{eqn:q72}) can be written 

\begin{equation}\label{eqn:q164}
E_{Q} = \omega Q + 4\pi \int dr \; r^{2} \left[\frac{1}{2}\left(\frac{d\sigma}{dr}\right)^{2} + \frac{V}{\Omega^{4}}\right] - 4\pi \int dr \; r^{2} \frac{\omega^{2}\phi^{2}}{2\Omega^{2}},
\end{equation}

\noindent where the last term is precisely 

\begin{equation}\label{eqn:q165}
4\pi \int dr \; r^{2} \frac{\omega^{2}\phi^{2}}{2\Omega^{2}} = \frac{1}{2}\omega Q.
\end{equation}

\noindent Using (\ref{eqn:q163}) we can write

\begin{equation}\label{eqn:q166}
\frac{1}{2}\omega Q = 4\pi \int dr \;  r^{2} \left[\frac{1}{6}\left(\frac{d\sigma}{dr}\right)^{2} + \frac{V}{\Omega^{4}}\right], 
\end{equation}

\noindent which, substituted into the Q-ball energy functional (\ref{eqn:q164}), gives

\begin{equation}\label{eqn:q167}
E_{Q} = \omega Q + 4\pi \int dr \; r^{2} \left[\frac{1}{2}\left(\frac{d\sigma}{dr}\right)^{2} + \frac{V}{\Omega^{4}}\right] - 4\pi \int dr \; r^{2} \left[\frac{1}{6}\left(\frac{d\sigma}{dr}\right)^{2} + \frac{V}{\Omega^{4}}\right], 
\end{equation}

\noindent which is the expression for the energy of the field extremised with respect to the global charge $Q$

\begin{equation}\label{eqn:q168}
E_{Q} = \omega Q + \frac{4\pi}{3}\int dr \; r^{2} \left(\frac{d\sigma}{dr}\right)^{2}.
\end{equation}

Thus, the energy of the Q-ball is, in terms of $\phi$,

\begin{equation}\label{eqn:q169}
E_{Q} = \omega Q + \frac{4\pi}{3}\int dr \; r^{2} \frac{1}{\Omega^{2}}\left(\frac{d\phi}{dr}\right)^{2},
\end{equation}

\noindent where the integral can be interpreted as a surface energy term. There are some interesting things which can be said about this relation. Dividing through by the charge gives

\begin{equation}\label{eqn:q170}
\frac{E_{Q}}{Q} = \omega  + \frac{4\pi}{3Q}\int dr \; r^{2} \frac{1}{\Omega^{2}}\left(\frac{d\phi}{dr}\right)^{2},
\end{equation}

\noindent which shows that the $\omega$ parameter is equal to the energy-charge ratio of the Q-ball up to a small correction due to a contribution from the surface energy of the Q-ball. In the limit $\Omega \rightarrow 1$, this expression is equal to that obtained for conventional Q-balls in \cite{heeck21}. It also provides some analytical verification of the condition for absolute stability. If the Q-ball is absolutely stable, $E_{Q} < mQ$, then

\begin{equation}\label{eqn:q171}
\frac{E_{Q}}{Q}  = \omega  + \frac{4\pi}{3Q}\int dr \; r^{2} \frac{1}{\Omega^{2}}\left(\frac{d\phi}{dr}\right)^{2} < m.
\end{equation}

\noindent Provided that the surface term is significantly smaller than $\omega$, we find that

\begin{equation}\label{eqn:q172}
\omega < m,
\end{equation}

\noindent is an alternative formulation of the absolute stability condition for these Q-balls. Since $\omega < m$ in order for Q-ball solutions to exist, it follows that if (\ref{eqn:q172}) holds then all Q-ball solutions will be stable. 

\subsection{Application of the Energy-Charge Relation to the Analytical Approximate Q-ball Solution}\label{section:473}

\noindent In this section we use the analytical Q-ball solution to show that the energy to charge ratio is close to $\omega$ for $\phi_{0} >> M_{pl}/\sqrt{\xi}$, confirming the numerical results in Figure \ref{figure:411}. The analytical approximation for the scalar field is given by (\ref{eqn:q125}) and the analytical expression for the Q-ball charge is given by (\ref{eqn:q138}). The analytical approximation of $\phi$ is valid provided that the field does not deviate too far from its initial value, $\phi \approx \phi_{0}$, which is true for $r < r_{Q}/2$. The analytical expression for the charge is calculated for $r < r_{Q}/2$, so the integral in this expression must be taken as far as $r = r_{Q}/2$.

\noindent Differentiating the inflaton field in the analytical approximation (\ref{eqn:q125}) gives

\begin{equation}\label{eqn:q173}
\frac{d\phi}{d r} = -\frac{2r \phi_{0}}{r_{Q}^{2}} \Rightarrow \left(\frac{d\phi}{d r}\right)^{2} = \frac{4r^{2}\phi_{0}^{2}}{r_{Q}^{4}}.
\end{equation}

\noindent Substituting this into (\ref{eqn:q171}) in the plateau limit $\Omega^{2} \approx \xi \phi_{0}^{2}/M_{pl}^{2}$, and the charge $Q$ from (\ref{eqn:q138}) gives

\begin{equation}\label{eqn:q174}
\frac{E}{Q} = \omega + \frac{4\pi \cdot 6\xi}{\pi \omega M_{pl}^{2}}\frac{1}{r_{Q}^{3}} \int^{r_{Q}/2}_{0} r^{2} \; dr \frac{M_{pl}^{2}}{3\xi \phi_{0}^{2}} \frac{4r^{2}\phi_{0}^{2}}{r_{Q}^{4}} = \omega + \frac{32}{\omega r_{Q}^{7}}\int^{r_{Q}/2}_{0} r^{4} dr.
\end{equation}

\noindent Performing the integration in (\ref{eqn:q174}) we obtain

\begin{equation}\label{eqn:q175}
\frac{E}{Q} = \omega + \frac{1}{5\omega}\frac{1}{r_{Q}^{2}}.
\end{equation}

\noindent Substituting in the expression for $r_{Q}^{2}$ from (\ref{eqn:q132}) gives

\begin{equation}\label{eqn:q176}
\frac{E}{Q} = \omega + \frac{1}{30\omega}\frac{M_{pl}^{2}}{\xi \phi_{0}^{2}}\left( m^{2} + \omega^{2} - \omega_{c}^{2}\right).
\end{equation}

\noindent For an inflaton mass within the Q-ball window (\ref{eqn:q108}), $m^{2} + \omega^{2} - \omega_{c}^{2} \le m^{2}$, and we have

\begin{equation}\label{eqn:q177}
\frac{E}{Q} \le \omega + \frac{\omega}{30}\frac{M_{pl}^{2}}{\xi \phi_{0}^{2}}\left[ 1 + \frac{m^{2}}{\omega^{2}}\right].
\end{equation}

\noindent Taking $m^{2} \approx \omega^{2}$, we have that, at most

\begin{equation}\label{eqn:q178}
\frac{E}{Q} = \omega \left[ 1 + \mathcal{O}\left( \frac{M_{pl}^{2}}{\xi \phi_{0}^{2}}\right) \right],
\end{equation}

\noindent and therefore

\begin{equation}\label{eqn:q179}
\frac{E}{Q} \thickapprox \omega,
\end{equation}

\noindent for $\phi_{0} >> M_{pl}/\sqrt{\xi}$. This means that we can safely say to a very good approximation that the $\omega$ parameter in this model is the energy-charge ratio for $\phi_{0} >> M_{pl}/\sqrt{\xi}$. 

\subsection{Analytical Dependence of Q-ball Energy and Radius on the Inflaton Self-Coupling}\label{section:474}

The $\gamma$ parameter as defined (\ref{eqn:q115}) is

\begin{equation}\label{eqn:q180}
\gamma = \frac{M_{pl}^{2}}{\xi}\left(m^{2} + \omega^{2} - \omega_{c}^{2} \right),
\end{equation}

\noindent and $\omega_{c}^{2} = \lambda M_{pl}^{2}/\xi$. If we approximate $m^{2} \approx \omega^{2} \approx \omega_{c}^{2}$, we have that

\begin{equation}\label{eqn:q181}
\gamma \approx \frac{M_{pl}^{2}}{\xi}\omega_{c}^{2} = \frac{\lambda M_{pl}^{4}}{\xi^{2}} \approx 4 V_{E},
\end{equation}

\noindent from (\ref{eqn:q33}) in the plateau approximation.

The primordial curvature power spectrum is given by

\begin{equation}\label{eqn:q182}
\mathcal{P}_{\mathcal{R}} = \frac{\lambda N^{2}}{12\pi^{2}\xi},
\end{equation}

\noindent and this can be rearranged to give an expression for the inflaton self-coupling, $\lambda$,

\begin{equation}\label{eqn:q183}
\lambda = \left(\frac{12\pi^{2} P_{R}}{N^{2}}\right)\xi \Rightarrow \lambda \propto \xi.
\end{equation}

\noindent Using this, from (\ref{eqn:q181}), we have that

\begin{equation}\label{eqn:q184}
\gamma \propto \frac{\lambda}{\lambda^{2}} \Rightarrow \gamma \sim \frac{1}{\lambda}.
\end{equation}

\noindent This means that, taking the expression for $r_{Q}$ from (\ref{eqn:q124}), we can establish the dependence of $r_{Q}$ on the inflaton self-coupling

\begin{equation}\label{eqn:q185}
r_{Q} = \frac{\sqrt{6}\phi_{0}}{\sqrt{\gamma}} \Rightarrow r_{Q} \propto \sqrt{\lambda}\phi_{0}.
\end{equation}

The analytical expression for the Q-ball energy is given by (\ref{eqn:q134}). Since in general $\omega^{2} \approx \omega_{c}^{2}$, which follows from $m^{2} \simeq \omega^{2}$ and (\ref{eqn:q108}), the energy of the Q-ball becomes

\begin{equation}\label{eqn:q186}
E \approx \frac{3}{2} \frac{\pi M_{pl}^{2}}{12\xi}\omega_{c}^{2}r_{Q}^{3} \propto \frac{\omega_{c}^{2}r_{Q}^{3}}{\xi} \sim \frac{\lambda}{\xi}\frac{r_{Q}^{3}}{\xi},
\end{equation}

\begin{equation}\label{eqn:q187}
\Rightarrow E \propto \sqrt{\lambda}\phi_{0}^{3}. 
\end{equation}

\noindent These relations will become important in Section \ref{section:49} when we consider the effects of curvature on non-minimally coupled Palatini Q-balls.\\

\section{Formation of Non-Minimally Coupled Palatini Q-balls Following Tachyonic Preheating}\label{section:48}
As discussed in Section \ref{section:42}, we will consider the possibility, indicated by the results in \cite{hiramatsu10}, that the Q-balls in this model may arise as a result of the fragmentation of the inflationary condensate, where the inflationary condensate breaks into smaller compact lumps of scalars which subsequently decay into $\pm$ Q-ball pairs. If the inflationary potential is “flatter than $\phi^{2}$”, then this in general results in an attractive interaction between the scalars which creates a negative pressure within the inflaton condensate. This is one of the conditions which serves as a prerequisite for condensate fragmentation.\\

A requirement is that there is a tachyonic instability in the inflationary potential (second derivative of the effective potential/effective mass squared term is less than zero). This is satisfied on the plateau of the inflaton potential. When the field passes the tachyonic instability, quantum fluctuations of the field will be amplified as the field rolls quickly down its effective potential from the top of the local maximum. This tachyonic growth of certain modes during inflation can cause the condensate to fragment, provided that the perturbations in the energy density of the condensate are not diluted away by expansion before the tachyonic growth becomes significant. However, if there are no non-topological soliton solutions present in the scalar field theory of the inflaton then the condensate will not fragment and any large perturbations induced from tachyonic preheating will be smoothed out by expansion.

It has been shown in \cite{rubio19} for Palatini inflation with a real scalar field that the inflaton fluctuations grow rapidly due to tachyonic preheating at the end of slow-roll inflation, where this tachyonic growth occurs within a single inflaton oscillation for non-minimal couplings of the size relevant to this study. The non-linear fluctuations generated then become the dominant contribution to the dynamics of the inflaton field after growing very rapidly, and it is these amplified non-linear fluctuations which may cause the inflaton condensate to become unstable and fragment. In \cite{rubio19}, the analysis is undertaken for a $\phi^{4}$ inflaton potential, which means that any amplified perturbations will be smoothed out somewhat more than in our case due to the repulsive interaction between the scalars dominating the dynamics of the inflaton within the condensate.

This analysis from \cite{rubio19} can therefore be applied to our model to provide a lower estimate on the range of perturbation wavelengths which undergo tachyonic growth and become non-linear. This provides a useful check on whether tachyonic preheating could be the underlying mechanism behind the formation of Q-balls in non-minimally coupled Palatini inflation, since if this is the case then the range of wavelengths of the amplified non-linear perturbations in this model should overlap with the radii of the relevant Q-balls generated in our numerical analysis (see Table \ref{table:42}). Namely, we consider the radius of the $\phi_{0} = 3.2217991 \times 10^{17} \GeV$ Q-ball, since this is within the range of estimated field values at the end of Palatini inflation, in order to estimate whether these Q-balls are likely to form from the fragmentation of the inflaton condensate following tachyonic preheating. 

The inflaton begins to undergo oscillations when slow-roll inflation ends. We find that the value of the inflaton field at the onset of these oscillations in the model outlined in Section \ref{section:44} is 

\begin{equation}\label{eqn:q188}
\phi_{end} = 2\sqrt{2} M_{pl} \sqrt{\beta},
\end{equation}

\noindent where $\beta = 0.1-1$, which gives $\phi_{end} \sim (1-3)M_{pl}$ for the inflaton masses relevant to this work. When the non-linear fragments form we can expect that the value of the field will actually be less than this due to expansion taking place while tachyonic preheating is occurring. Field values in the range $10^{17}-10^{18} \GeV$ are therefore a reasonable estimate. 
For the purpose of this study we utilise a quantity introduced in \cite{rubio19} defined as the perturbation wavenumber 

\begin{equation}\label{eqn:q189}
\kappa = \sqrt{\frac{\xi}{\lambda}}\frac{k}{M_{pl}},
\end{equation}

\noindent which is a rescaling of the physical wavenumber $k$. The range of perturbation wavelengths for which strong tachyonic growth of the perturbations occurs is given by $\kappa_{min} < \kappa < \kappa_{max}$, where the authors in \cite{rubio19} define $\kappa_{max} \thickapprox 0.4a$ \cite{rubio19}, where $a$ is the scale factor. $\kappa_{min}$ corresponds to the size of the horizon. This range in perturbation wavenumber corresponds to a range of physical wavelengths, $\lambda_{min} < \lambda < \lambda_{max}$, for which the perturbations are expected to become non-linear. $\lambda_{min}$ is determined by $\kappa_{max}$ and $\lambda_{max} = 2 \pi H^{-1}$.

\noindent The maximum physical wavenumber $k/a$ from (\ref{eqn:q189}) is 

\begin{equation}\label{eqn:q190}
\frac{k_{max}}{a}  = \sqrt{\frac{\lambda}{\xi} } \frac{\kappa_{max}}{a} M_{Pl}  = \frac{2 \pi}{\lambda_{min}},
\end{equation}

\noindent and using $\kappa_{max} = 0.4 a$, we find that the minimum physical wavelength of the perturbations which will become non-linear due to tachyonic preheating is

\begin{equation}\label{eqn:q191}
\lambda_{min}  = \frac{2\pi}{0.4}\sqrt{\frac{\xi}{\lambda}}M_{pl}^{-1} \equiv 15.7 \sqrt{\frac{\xi}{\lambda}}M_{pl}^{-1}.
\end{equation}

\noindent Normalising the non-minimal coupling and the inflaton self-coupling using $\xi = 1.2 \times 10^{9}$ and $\lambda = 0.1$ respectively, we can write this as

\begin{equation}\label{eqn:q192}
\lambda_{min} = 6.5 \times 10^{-13}  \left(\frac{0.1}{\lambda}\right)^{1/2} \left(\frac{\xi}{10^{9}}\right)^{\frac{1}{2}} \GeV^{-1}.
\end{equation}

\noindent Taking $\lambda = 0.1$ and $\xi = 1.2 \times 10^{9}$ we find that the smallest physical wavelength in this model which we expect to produce non-linear perturbations in the inflaton condensate is $\lambda_{min} = 7.16 \times 10^{-13} \GeV^{-1}$. 

\noindent The Hubble parameter at the end of slow roll inflation is 

\begin{equation}\label{eqn:q193}
H^{2} = \frac{V_{E}}{3 M_{pl}^{2}},
\end{equation}

\noindent where we take $V_{E}$ to be the plateau approximation of the potential in the Einstein frame

\begin{equation}\label{eqn:q194}
V_{E} = \frac{\lambda M_{pl}^{4}}{4 \xi^{2}}. 
\end{equation}

\noindent Substituting (\ref{eqn:q194}) into (\ref{eqn:q193}) we find that

\begin{equation}\label{eqn:q195}
H^{2} = \frac{\lambda}{12 \xi^{2}}M_{pl}^{2}.
\end{equation}

\noindent This means that 

\begin{equation}\label{eqn:q196}
H^{-1} = \left(\frac{12}{\lambda}\right)^{\frac{1}{2}} \frac{\xi}{M_{pl}} = 4.56 \times 10^{-9} \left(\frac{0.1}{\lambda}\right)^{1/2} \left(\frac{\xi}{10^{9}} \right) \GeV^{-1},
\end{equation}

\noindent using the normalisation of (\ref{eqn:q192}). Taking $\lambda = 0.1$ and $\xi = 1.2\times 10^{9}$, we find $H^{-1} = 5.47 \times 10^{-9} \GeV^{-1} $, and the maximum physical perturbation wavelength is then

\begin{equation}\label{eqn:q197}
\lambda_{max} = 2\pi H^{-1} = 2.87 \times 10^{-8}\left(\frac{0.1}{\lambda}\right)^{1/2}\left( \frac{\xi}{10^{9}}\right) \GeV^{-1}.
\end{equation}

\noindent For $\lambda = 0.1$ and $\xi = 1.2 \times 10^{9}$, this gives the maximum physical wavelength of the perturbations which will become non-linear in this model to be $\lambda_{max}= 3.4 \times 10^{-8} \GeV^{-1}$.

We find therefore that strong perturbation growth occurs for modes with wavelengths in the range $7.16 \times 10^{-13} \GeV^{-1} < \lambda  < 3.4 \times 10^{-8} \GeV^{-1}$, using $\lambda = 0.1$ and  $\xi = 1.2 \times 10^{9}$ as our parameter estimates for the non-minimally coupled Palatini inflation model. The radii of all of the Q-ball solutions presented in Section \ref{section:46} fall within this range. Most notably, the $\phi_{0} = 3.2217991 \times 10^{17}\GeV$ Q-ball has a radius $r_{X} = 8.89\times 10^{-10} \GeV^{-1}$, which is comfortably within the range of mode wavelengths we expect to see undergoing tachyonic growth. It is therefore possible that  $\phi_{0} \sim 10^{17}-10^{18} \GeV$ Q-balls could form in this model as a result of the fragmentation of the inflaton condensate due to tachyonic preheating, assuming that all other conditions for fragmentation are met, which means that non-minimally coupled Palatini Q-balls are physically feasible for the model parameters considered here, although a numerical simulation of the dynamics and fragmentation of the inflaton condensate in an expanding background would be needed in order to establish tachyonic preheating as a formation mechanism for these non-minimally coupled Palatini Q-balls properly.

\section{The Effects of Curvature on Non-Minimally Coupled Palatini Q-balls}\label{section:49}

In this analysis we have studied Q-balls in flat space, as they have been derived historically. In this section we consider the possible effects of curvature on these Q-ball solutions, in order to ascertain the validity of the flat space solutions. As a first pass approximation, we conduct this analysis as though the Q-balls form in flat space and then curvature is switched on. In this approximation we can consider the Q-ball solution to have a spherically symmetric energy density. This means that the spacetime outside of the Q-ball is described by the Schwarzschild metric \\

\begin{equation}\label{eqn:q198}
ds^{2} = \left(1 - \frac{2 G M(r)}{r} \right) dt^{2} -  \left(1 - \frac{2 G M(r)}{r} \right)^{-1} dr^{2} - r^{2} d \Omega^{2},
\end{equation}

\noindent where $M(r)$ is the mass-energy contained within radius $r$. Let the radius of the Q-ball in this scenario be $R$. If the radius of the Q-ball is smaller than its Schwarzschild radius, $r_{S} = 2G M\left(r \right)$, then the Q-ball will collapse to form a black hole. The condition to ensure that the Q-balls are not signifcantly affected by the curvature of the surrounding spacetime is therefore

\begin{equation}\label{eqn:q199}
R > 2 G M(R) \equiv r_{S},
\end{equation}
\noindent assuming that the mass-energy $M\left(r \right)$ increases faster with $r$ than $r$ increases. Essentially this means that the flat space Q-ball solutions are a good approximation of non-minimally coupled Palatini Q-balls provided that the Q-balls are sufficiently large compared to their Schwarzschild radii $r_{S}$. 

\noindent Setting $R = r_{Z}$ and $M\left(r \right) = E$ from Table \ref{table:41}, we find that this condition is easily satisfied for all of the Q-balls which have been generated numerically in this model (see Table \ref{table:44}). This means that if curvature is switched on, all of the Q-ball solutions presented in Table \ref{table:41} remain as Q-balls and do not collapse to black holes. As a first approximation, this means that the flat space description is reasonable. It has been shown for conventional canonical scalar field theories that spatial curvature can affect the size of Q-balls \cite{multamaki022}, and the expression for the energy of the Q-balls will be different in curved space, since the energy density is derived from the energy-momentum tensor. In order to determine the full effects of gravity on these Q-balls, the Q-ball equation and the energy and charge of the Q-balls would need to be derived in a spherically symmetric spacetime. However, for the purposes of this thesis the flat space Q-ball solutions can be used to approximate the effects of curvature on Q-balls in a non-minimally coupled Palatini inflation framework.

\begin{table}[H]
\begin{center}
\begin{adjustbox}{max width=\textwidth}
\begin{tabular}{| c | c | c | c | c |}
\hline
$\omega/\omega_{c}$ & $\phi_{0}/\GeV$ & $r_{Z}/\GeV^{-1}$ & $r_{S}/\GeV^{-1}$ & $r_{S}/r_{Z}$ \\
\hline
$0.707155$ & $3.221799 \times 10^{17}$ & $8.94 \times 10^{-10}$ & $4.61 \times 10^{-11}$ & $0.052$ \\
\hline
$0.709$ & $1.3464098 \times 10^{16}$ & $3.79 \times 10^{-11}$ & $3.31 \times 10^{-15}$ & $8.74 \times 10^{-5}$ \\
\hline
$0.71$ & $8.855792 \times 10^{15}$ & $2.5 \times 10^{-11}$ & $9.33 \times 10^{-16}$ & $3.73 \times 10^{-5}$ \\
\hline
$0.72$ & $1.960795 \times 10^{15}$ & $6.09 \times 10^{-12}$ & $9.65 \times 10^{-18}$ & $1.58 \times 10^{-6}$ \\
\hline
$0.73$ & $1.090258 \times 10^{15}$ & $3.73 \times 10^{-12}$ & $1.57 \times 10^{-18}$ & $4.22 \times 10^{-7}$ \\
\hline
$0.74$ & $7.45339 \times 10^{14}$ & $2.72 \times 10^{-12}$ & $4.89 \times 10^{-19}$ & $1.80 \times 10^{-7}$ \\
\hline
$0.75$ & $5.61953 \times 10^{14}$ & $2.28 \times 10^{-12}$ & $2.04 \times 10^{-19}$ & $8.96 \times 10^{-8}$ \\
\hline
$0.80$ & $2.29632 \times 10^{14}$ & $1.47 \times 10^{-12}$ & $1.39 \times 10^{-20}$ & $9.48 \times 10^{-9}$ \\
\hline
$0.85$ & $1.16877 \times 10^{14}$ & $1.49 \times 10^{-12}$ & $2.66 \times 10^{-21}$ & $1.79 \times 10^{-9}$ \\
\hline
$0.89$ & $4.7918 \times 10^{13}$ & $2.70 \times 10^{-12}$ & $1.08 \times 10^{-21}$ & $4.00 \times 10^{-10}$ \\
\hline
\end{tabular}
\end{adjustbox}
\caption{Table showing the radii $r_{Z}$, Schwarzschild radii $r_{S}$, and the ratio of the two, $r_{S}/r_{Z}$, for all of the Q-ball solutions presented.}
\label{table:44}
\end{center}
\end{table}

\noindent This analysis raises the question of whether a Q-ball which collapses to a black hole could be produced in this model. In order to determine this we consider the ratio of Schwarzschild radius to Q-ball radius, $r_{S}/r_{Z}$. The largest Q-ball presented in Table \ref{table:41} is the $\omega = 0.707155\omega_{c}$ Q-ball which has $\phi_{0} = 3.22 \times 10^{17}\GeV$, a radius of $r_{Z} = 8.94 \times 10^{-10}\GeV^{-1}$and a mass-energy $M = E_{Z} = 3.34 \times 10^{27}\GeV$. Using $G = 6.9 \times 10^{-39}\GeV^{-2}$, its Schwarzschild radius is $r_{S} = 4.64 \times 10^{-11}\GeV^{-1}$ and the curvature ratio is $r_{S}/r_{Z} = 0.052$. In order to form a black hole, the Q-ball would thus need to be of larger $\phi_{0}$ than this. From Section \ref{section:47} we know that $E \propto \sqrt{\lambda}\phi_{0}^{3}$ and $r_{Q} \propto \sqrt{\lambda} \phi_{0}$, and we have confirmed that these relations approximately hold for the numerically calculated values of $E_{Z}$ and $r_{Z}$. Since $r_{S} \sim M = E_{Z}$, $r_{S} \propto \sqrt{\lambda}\phi_{0}^{3}$, and we therefore have that $r_{S}/r_{Z} \propto \phi_{0}^{2}$, independent of the inflaton self-coupling, $\lambda$. We can normalise this proportionality using the $\phi_{0} = 3.22 \times 10^{17}\GeV$ Q-ball to obtain

\begin{equation}\label{eqn:q200}
\frac{r_{S}}{r_{Z}} = 0.052 \left(\frac{\phi_{0}}{3.22 \times 10^{17} \GeV} \right)^{2}.
\end{equation}

\noindent This is in good agreement with the results of Table \ref{table:44}.

From this, a Q-ball would form a black hole once $\frac{r_{S}}{r_{Z}} = 1$ if curvature is included. Setting (\ref{eqn:q200}) equal to unity gives the value of $\phi_{0}$ which would result in a black hole, which we denote by  $\phi_{0, c}$, to be

\begin{equation}\label{eqn:q201}
\phi_{0, c} = 1.41 \times 10^{18}\GeV.
\end{equation}

\noindent This means that Q-balls of the $\phi_{0}$ we would expect from Q-balls formed from tachyonic preheating at the end of Palatini inflation could form black holes directly if the effects of curvature are taken into account.

We can use the proportionality of $E_{Z}$ and $r_{Z}$ to $\phi_{0}$ to estimate the mass-energy and radius of this Q-ball before it collapses to a black hole. We have that $E_{Z} \propto \phi_{0}^{3}$ and $r_{Z} \propto \phi_{0}$, and can write

\begin{equation}\label{eqn:q202}
\frac{E_{Z,c}}{E_{Z}} = \frac{\phi_{0,c}^{3}}{\phi_{0}^{3}},
\end{equation}

\begin{equation}\label{eqn:q203}
\frac{r_{Z,c}}{r_{Z}} = \frac{\phi_{0,c}}{\phi_{0}}.
\end{equation}

\noindent Normalising these using the $\phi_{0} = 3.22 \times 10^{17}\GeV$ Q-ball gives

\begin{equation}\label{eqn:q204}
E_{Z,c} = 3.34 \times 10^{27}\left(\frac{1.41 \times 10^{18}}{3.22 \times 10^{17}}\right)^{3} \GeV \Rightarrow E_{Z,c} = 2.80 \times 10^{29}\GeV,
\end{equation}

\begin{equation}\label{eqn:q205}
r_{Z,c} = 8.94 \times 10^{-10}\left(\frac{1.41 \times 10^{18}}{3.22 \times 10^{17}}\right) \GeV^{-1} \Rightarrow r_{Z,c} = 3.91 \times 10^{-9}\GeV,
\end{equation}

\noindent where if calculated using a curved space Q-ball solution, $\phi_{0,c}, E_{Z,c}$ and $r_{Z,c}$ would likely be slightly smaller. 

We can therefore say that Q-balls of $\phi_{0} \sim 10^{18}\GeV$ could form black holes in curved space in this model, and that these Q-balls could feasibly be produced from the fragmentation of the inflaton condensate from tachyonic preheating at the end of Palatini inflation. 

As aforementioned, this analysis considers the effects on flat space Q-balls if curvature is switched on once the Q-balls have formed, and the Q-ball equation would need to be derived and solved in curved space to obtain an accurate picture of how gravity affects the properties of non-minimally coupled Palatini Q-balls. We expect however that the overall effect of curvature on the properties of the Q-balls will be small since the deviation of the metric from the flat space metric is likely to be small, and the estimate of $\phi_{0,c}$ in the flat space case should therefore be reasonable.

To illustrate this consider the fact that the ratio of the Schwarzschild radius to the Q-ball radius is proportional to $\phi_{0}^{2}$ (\ref{eqn:q200}). If we reparameterise the metric of the spacetime surrounding the Q-balls using this ratio we find that we can rewrite (\ref{eqn:q198}) as

\begin{equation}\label{eqn:q206}
ds^{2} \sim \left(1 - \frac{\phi_{0}^{2}}{\phi_{0, c}^{2}} \right) dt^{2} -  \left(1 - \frac{\phi_{0}^{2}}{\phi_{0,c}^{2}} \right)^{-1} dr^{2} - r^{2} d \Omega^{2},
\end{equation}

\noindent which means that for a small patch of spacetime at the surface of the Q-ball, the local deviation of the metric from the flat space metric is proportional to $\phi_{0}^{2}$. For example, if $\phi_{0} = \phi_{0,c}/2$, then the radius of the Q-ball is $R = 4 r_{S}$, and the local deviation from the flat space metric is $25\%$. Since Q-balls are composed of a scalar field it is not strictly correct to model them as solid spheres in curved space, and we therefore cannot say that the metric inside the Q-balls automatically corresponds to the flat space metric. The mass-energy of the Q-balls in Schwarzschild space falls off as $M\left(r \right) \propto r^{3}$, and therefore the deviation of the metric from flat space inside the Q-ball will be more like $M\left(r \right)/r \propto r^{2}$, therefore

\begin{equation}\label{eqn:q207}
ds^{2} \sim \left(1 - \frac{2G r^{2}}{r_{S}^{2}} \right) dt^{2} -  \left(1 - \frac{2G r^{2}}{r_{S}^{2}} \right)^{-1} dr^{2} - r^{2} d \Omega^{2},
\end{equation}

\noindent which means that the deviation from the flat space metric inside the Q-balls will be small and will decrease further very rapidly as we get closer to the centre of the Q-ball. For $\phi_{0} < \phi_{0,c}/2$ we can therefore estimate that the presence of curvature will not significantly affect the properties of the Q-balls, and that the flat space approximation is a reasonable estimate for the properties of the Q-balls presented in Section \ref{section:46}. We also estimate that the prediction of the $\phi_{0}$ at which the Q-balls will start to collapse to black holes, $\phi_{0,c}$, will not be altered significantly when curvature is taken into account. It is likely that the presence of curvature may make the formation of black holes easier, since gravitational attraction may decrease the radius at which the gravitational force is ceases to be balanced by the gradient pressure of the scalar field inside the Q-ball. The formation of Q-ball derived black holes would also be affected by the effects of curvature, requiring treatment in curved spacetime, or Schwarzschild de-Sitter for considering the effects of expansion. We do not consider these effects in this thesis, in order to investigate the quality of the flat space Q-ball solutions as an approximation for the formation of Q-ball derived black holes, however a numerical simulation of the tachyonic preheating dynamics in this model of non-minimally coupled Palatini inflatonic Q-balls could be extended to explore the effects of curvature and expansion on the formation of Q-ball derived black holes. 

\section{Reheating via Q-ball Evaporation}\label{section:481}

In the following sections we discuss some cosmological consequences which could arise as a result of Q-ball formation at the end of inflation.

The formation of Q-balls at the end of non-minimally coupled Palatini inflation could theoretically lead to an early period of matter domination in the form of solitonic matter \cite{amin10,lozanov14}. This epoch of the energy density being largely composed of solitonic matter would lead to a different reheating mechanism and post-inflationary evolution in comparison to the usual decay of a homogeneous inflaton condensate. This is due to fact that the inflaton scalars are bound in compact objects. In general, the inflaton field decays throughout the volume of the inflationary condensate and can be modelled as a volume effect. When the scalars are bound in solitons, reheating occurs as a surface effect, with the scalars decaying perturbatively and the decay products evaporating away \cite{enqvist022} - \cite{mazumdar02} from the surface of the Q-ball. In this scenario, the evaporation rate is initially suppressed by the surface area to volume ratio of the object, therefore the larger the Q-ball the larger the suppression of the evaporation rate from its surface. In the case of decay of the inflaton to fermion pairs, the decay rate is further suppressed due to the effects of Pauli blocking \cite{cohen86,enqvist02}. This effect is due to the fact that as the scalars decay to fermion pairs, the energy levels of the fermions fill up. No more decays can then proceed until there are energy levels available for the decay products to occupy, or fermion pairs have evaporated away. This suppression of the decay rate leads to a lower reheating temperature. This slower reheating also means that there are a greater number of e-folds of expansion following inflation in these models, which in principle could lead to a shorter duration of inflation depending on the precise nature of the decay processes and the size of the Q-balls.

\section{Reheating from Q-ball Derived Black Holes and Associated Cosmological Implications}\label{section:410}

The black holes which could result from the gravitational collapse of non-minimally coupled Palatini Q-balls in this model are generally light black holes. The $\phi_{0}$ of the Q-balls which could evolve to black holes start at $\phi_{0} = 1.41 \times 10^{18}\GeV$, and this Q-ball has a mass-energy of $M = E_{Z} = 2.80 \times 10^{29}\GeV$. Using the conversion of $1\GeV = 1.79 \times 10^{-27}$kg  gives a mass for this Q-ball of $M_{BH} = 501.2$kg. Black holes of mass around $500$kg have a very high temperature, which means that the energy of the particles emitted by the black hole is also very high and thus the black hole can energetically access a large range of evaporation channels. $M_{BH} \approx 500$kg black holes can produce all the particles of the Standard Model as they decay. It is therefore possible that the Universe could be reheated via the decay of these Q-ball black holes if a sufficient number could form such that they came to dominate the energy density of the Universe. In this section we explore the possibility of reheating through Q-ball black hole decay, and whether this is compatible with the observed scalar spectral index.

\noindent The time taken for a black hole of mass $M \lesssim 10^{8} \, \kg$, with a black hole temperature is greater than $\mathcal{O}\left(100\right) \GeV$, to evaporate, $t_{ev}$, is given by \cite{evap}

\begin{equation} \label{eqn:q208}
t_{ev} = \frac{40960 \pi}{27 g_{*\,BH}} G^{2} M^{3},
\end{equation}

\noindent where $g_{*\,BH} = 108.5$ is the number of spin degrees of freedom for the Standard Model. We can equate this to the age for a matter-dominated Universe at reheating, $t = (2/3)H(T_{R})^{-1}$ (where the expansion rate at reheating is $H(T_{R}) = k_{T_{R}} T_{R}^{2}/M_{Pl}$, $k_{T_{R}} \approx 3.3$), to obtain the expression for the reheating temperature if the decay of these black holes were to reheat the Universe. This is

\begin{equation} \label{eqn:q209}
T_{R} \approx \left(\frac{2}{3}\right)^{1/2} \left(\frac{27 g_{*\,BH}}{40960 \pi k_{T}}\right)^{1/2} \frac{8 \pi M_{Pl}^{5/2}}{M^{3/2}} =  1.145 \times 10^{6} \left(\frac{1 \, {\rm kg}}{M}\right)^{3/2} \GeV,
\end{equation}

\noindent where $M = M_{BH}$ is the mass-energy of the black hole.

\noindent In Section \ref{section:49} we showed that the Q-ball with the smallest $\phi_{0}$ which could form a black hole, once curvature is included, is a Q-ball with $\phi_{0} = 1.41 \times 10^{18} \GeV$ and  $M = E = 2.80\times 10^{29} \GeV$. Taking $\lambda = 0.1$, we have that, using $1\kg = 5.62 \times 10^{26}\GeV$, the mass of this black hole is $M = 498\kg$, and substituting this into (\ref{eqn:q209}) we obtain the reheating temperature

\begin{equation}\label{eqn:q210}
T_{R} = 102.6 \GeV,
\end{equation}

\noindent which is a very low reheating temperature. We now want to establish whether reheating temperatures this low can produce an acceptable value of the scalar spectral index, in order to determine whether reheating via the decay of Q-ball black holes is a viable reheating mechanism.

In order to calculate the scalar spectral index we need to find the number of e-folds of inflation at the pivot scale, $N_{\ast}$. We use the pivot scale used by the Planck satellite, $k_{\ast} = 0.05\Mpc^{-1}$, and the wavelength of the pivot scale is then

\begin{equation}\label{eqn:q211}
\lambda_{\ast} = \frac{2\pi}{k_{\ast}} = 125.7\Mpc.
\end{equation} 

\noindent We assume that there is sufficient formation of Q-ball black holes that they come to dominate the energy density of the Universe, and that the Universe enters into a period of early matter domination after inflation until $T = T_{R}$. At this point the black holes evaporate and reheat the Universe, transitioning the energy density into an era of radiation domination. During inflation, we have that (\ref{eqn:31})

\begin{equation}\label{eqn:q212}
\lambda \left(t \right) = \frac{a}{a_{0}}\lambda_{\ast} = \frac{a}{a_{end}}\frac{a_{end}}{a_{R}}\frac{a_{R}}{a_{0}}\lambda_{\ast},
\end{equation}

\noindent where the $R$ subscripts correspond to the value of the quantity at reheating, "$0$" subscripts correspond to present-day values and "$end$" refers to the end of slow-roll inflation. We have that $a/a_{end} = e^{-N}$ and 

\begin{equation}\label{eqn:q213}
\frac{a_{end}}{a_{R}} = \left(\frac{\rho_{R}}{\rho_{end}} \right)^{\frac{1}{3}},
\end{equation}

\noindent during matter domination where

\begin{equation}\label{eqn:q214}
\rho_{end} = V_{E} \approx \frac{\lambda M_{pl}^{4}}{4\xi^{2}},
\end{equation}

\noindent is the energy density at the end of inflation, dominated by the inflaton potential, and 

\begin{equation}\label{eqn:q215}
\rho_{R} = \frac{\pi^{2}}{30}g\left(T_{R}\right) T_{R}^{4},
\end{equation}

\noindent is the energy density of radiation.

Using the fact that the entropy remains constant during slow roll inflation we have that (from Section \ref{section:10}),

\begin{equation}\label{eqn:q216}
a^{3}g_{s}\left(T \right) T^{3} = constant,
\end{equation}

\noindent we can write

\begin{equation}\label{eqn:q217}
\frac{a_{R}}{a_{0}} =  \left(\frac{g_{s}\left(T_{0}\right)}{g_{s}\left(T_{R}\right)}\right)^{\frac{1}{3}} \frac{T_{0}}{T_{R}}.
\end{equation}

\noindent Substituting (\ref{eqn:q217}) into (\ref{eqn:q212}), the wavelength as a function of time is 

\begin{equation}\label{eqn:q218}
\lambda \left(t \right) = \left(\frac{g_{s}\left(T_{0}\right)}{g_{s}\left(T_{R}\right)}\right)^{\frac{1}{3}} \frac{T_{0}}{T_{R}} \left(\frac{\rho_{R}}{\rho_{end}} \right)^{\frac{1}{3}}e^{-N }\lambda_{\ast}.
\end{equation}

\noindent At horizon exit during inflation, we have that 

\begin{equation}\label{eqn:q219}
\lambda = H^{-1} = \left(\frac{V_{E}}{3M_{pl}^{2}}\right)^{-\frac{1}{2}}.
\end{equation}

\noindent At the pivot scale we therefore have

\begin{equation}\label{eqn:q220}
\left(\frac{V_{E}}{3M_{pl}^{2}}\right)^{-\frac{1}{2}} = \left(\frac{g_{s}\left(T_{0}\right)}{g_{s}\left(T_{R}\right)}\right)^{\frac{1}{3}} \frac{T_{0}}{T_{R}} \left(\frac{\rho_{R}}{\rho_{end}} \right)^{\frac{1}{3}}e^{-N_{\ast} }\lambda_{\ast},
\end{equation}

\noindent equating (\ref{eqn:q218}) and (\ref{eqn:q219}). Rearranging (\ref{eqn:q220}) gives the expression for the number of e-folds at the pivot scale

\begin{equation}\label{eqn:q221}
N_{\ast} = ln\left[ \left(\frac{g_{s}\left(T_{0}\right)}{g_{s}\left(T_{R}\right)}\right)^{\frac{1}{3}} \frac{T_{0}}{T_{R}} \left(\frac{\pi^{2}g\left(T_{R}\right)T_{R}^{4}}{30V_{E}} \right)^{\frac{1}{3}}\left(\frac{V_{E}}{3M_{pl}^{2}}\right)^{\frac{1}{2}}\lambda_{\ast}\right].
\end{equation}

\noindent Here $T_{R} = 102.6\GeV$ is the reheating temperature, $T_{0} = 2.4 \times 10^{-13}\GeV$ is the temperature of the CMB, $g_{s}\left(T_{R}\right) = g\left(T_{R}\right) \approx 100$ and $g_{s}\left(T_{0}\right) = 3.91$. For $\lambda = 0.1$ and $\xi = 1.2163 \times 10^{9}$, $V_{E} = 5.61 \times 10^{53}\GeV^{4}$, and using the conversion $1\Mpc = 1.56 \times 10^{38}\GeV^{-1}$ we have $\lambda_{\ast} = 1.96 \times 10^{40}\GeV^{-1}$. 

Substituting these numbers into (\ref{eqn:q221}) gives the number of e-folds at the pivot scale to be

\begin{equation}\label{eqn:q222}
N_{\ast} = 43.1,
\end{equation}

\noindent which by (\ref{eqn:q44}) gives a value for the scalar spectral index of $n_{s} = 0.9536$, which is excluded by the $2-\sigma$ lower bound from the Planck satellite $n_{s} > 0.9565$ \cite{Planck18}. This means that reheating via the decay of Q-ball black holes in the case where the Q-ball derived black holes dominate the energy density at early times is not a viable reheating mechanism for non-minimally coupled Palatini inflation because it predicts a scalar spectral index which is too small.

An alternative scenario for a Q-ball black hole post-inflation cosmology can be realised if only a small fraction of the Q-balls collapse to black holes, leaving a small initial fraction of energy density as Q-ball derived black holes while the rest is converted to radiation as the remaining Q-balls decay. This results in a long period of radiation domination before the Q-ball black holes come to dominate at a later time. This means that there will be a greater number of e-folds of inflation at the time the pivot scale exits the horizon, which would give a larger value for the scalar spectral index, and could potentially bring the prediction for $n_{s}$ into the acceptable bounds on the scalar spectral index from the Planck results. 

In this scenario we use the assumption of instantaneous reheating, whereby the energy density of the inflaton is all immediately converted to radiation at the end of slow-roll inflation, such that

\begin{equation}\label{eqn:q223}
\rho_{end} = 3M_{pl}^{2}H^{2} = V_{E} = \rho_{R} = \frac{\pi^{2}}{30}g\left(T_{R}\right) T_{R}^{4}.
\end{equation}

\noindent Using $V_{E} = 5.61 \times 10^{53}\GeV^{4}$, for $g\left(T_{R}\right) \approx 100$, this equality gives a reheating temperature of

\begin{equation}\label{eqn:q224}
T_{R} = 1.14 \times 10^{13}\GeV,
\end{equation}

\noindent which is significantly higher than the reheating temperature in the case of reheating from Q-ball derived black hole evaporation after a period of matter domination from these black holes.

\noindent The number of e-folds at the pivot scale $k_{\ast} = 0.05Mpc^{-1}$, $\lambda_{\ast} = 1.96 \times 10^{40}\GeV^{-1}$ is derived similarly to the case of Q-ball black hole domination. The wavelength of the inflaton fluctuations as a function of time is (\ref{eqn:31})

\begin{equation}\label{eqn:q225}
\lambda \left( t \right) = \frac{a}{a_{0}}\lambda_{\ast} = \frac{a}{a_{end}} \frac{a_{end}}{a_{0}}\lambda_{\ast}, 
\end{equation}

\noindent with $a/a_{end} = e^{-N}$ and 

\begin{equation}\label{eqn:q226}
\frac{a_{end}}{a_{0}} = \left(\frac{\rho_{0}}{\rho_{end}} \right)^{\frac{1}{3}},
\end{equation}

\noindent we have that

\begin{equation}\label{eqn:q227}
\lambda\left(t \right) = \left(\frac{\rho_{0}}{\rho_{end}} \right)^{\frac{1}{3}}e^{-N} \lambda_{\ast}. 
\end{equation}

\noindent Using the fact that $\lambda = H^{-1}$ at horizon exit, and (\ref{eqn:q220}) we have that

\begin{equation}\label{eqn:q228}
e^{N_{\ast}} = \left( \frac{g_{s}\left(T_{0}\right)}{g_{s}\left(T_{end}\right)}\right)^{\frac{1}{3}}\frac{T_{0}}{T_{end}}\left(\frac{V_{E}}{3M_{pl}^{2}}\right)^{\frac{1}{2}}\lambda_{\ast}.
\end{equation}

\noindent Since we are assuming instantaneous reheating, this means that the Q-balls which do not collapse to black holes decay instantly to radiation. So we can say $T_{end} = T_{R} = 1.14 \times 10^{13}\GeV$, and using $T_{0} = 2.4 \times 10^{-13}\GeV$, $g_{s}\left(T_{R}\right) \approx 100$, $g_{s}\left(T_{0}\right) = 3.91$ and $V_{E} = 5.61 \times 10^{53}\GeV^{4}$ we find that 

\begin{equation}\label{eqn:q229}
N_{\ast} = 51.6,
\end{equation}

\noindent which using (\ref{eqn:q44}) gives a value for the scalar spectral index of $n_{s} = 0.9612$, which is within the $1-\sigma$ bound of the Planck value of $n_{s}$ \cite{Planck18}. Thus in general, $43.1 < N_{\ast} < 51.6$, when a density of black holes from Q-balls is included. This means that although reheating via the evaporation of Q-ball black holes is not viable in the case of complete black hole matter domination at early times, an acceptable value of the scalar spectral index can be generated in the case where a small fraction of Q-balls collapse to black holes and come to dominate the energy density of the Universe at a later time once reheating has proceeded through the instant decay of the remaining Q-balls to radiation, and the radiation energy density has decayed away with expansion. This provides a small insight into some of the rich implications that non-minimally coupled Palatini Q-balls could have for cosmology, since we have demonstrated that for a certain mass range of the inflaton, (\ref{eqn:q108}), the model can inflate successfully, may produce Q-balls, and could also seed Primordial Black Holes from the collapse of Q-balls for sufficiently large $\phi_{0}$.\\

\subsection{Q-ball Black Hole Mass Dependence on the Inflaton Self-Coupling, and Contraints from Baryogenesis}\label{section:4101}

Thus far we have explored a fairly minimal model of inflation and Q-balls which has the potential to be a self-sufficient basis for a cosmological model. We are, however, yet to discuss a possible mechanism for baryogenesis in the case of Q-balls from non-minimally coupled Palatini inflation. A complete study of the generation and transfer of the baryon asymmetry is beyond the scope of this thesis, however there are important aspects of the model in relation to baryogenesis which can be discussed.

\noindent Since these Q-ball black holes start at $M \sim 500$kg, as discussed in Section \ref{section:49} these Q-balls can produce all the particles of the Standard Model upon evaporation. It is therefore reasonable to think that there might be some decay channel accessible for these black holes which could seed a lepton number asymmetry, this could then be the basis for baryogenesis via thermal leptogenesis in this model if the asymmetry in lepton number could be transferred to a baryon number asymmetry through sphaleron processes. Alternatively, electroweak baryogenesis could occur if there is a first-order electroweak symmetry breaking transition. The issue with this is that the temperature required for sphalerons to instigate baryogenesis, or for a first-order phase transition to occur, must be at least equal to the temperature of electroweak symmetry breaking $T_{EW} = 159.5\GeV$ \cite{ew}. Thus the case which gives a reheating temperature of around $T_{R} \approx 100\GeV$ is too low for electroweak symmetry breaking, and therefore insufficient for the precipitation of the baryon asymmetry by sphalerons. If there were a way to raise the reheating temperature in the case of reheating via Q-ball derived black hole decay, then it would be more feasible to embed non-minimally coupled Palatini Q-balls into a cosmological model which includes the generation of the baryon asymmetry. \\

In order to explore this idea, we first note that, as found in Section \ref{section:47}, that the mass-energy of the Q-ball derived black holes depends on the square root of the inflaton self-coupling, $M = E_{Z} \propto \sqrt{\lambda}\phi_{0}^{3}$, and we find that for the $\phi_{0} = 1.41 \times 10^{18}\GeV$ Q-ball derived black hole the reheating temperature is insufficient for baryogenesis for $\lambda = 0.1$. From (\ref{eqn:q187}), we can write 

\begin{equation}\label{eqn:q230}
M \propto \sqrt{\lambda}\phi_{0}^{3},
\end{equation}

\noindent and for the same $\phi_{0} = 1.41 \times 10^{18}\GeV$, for smaller inflaton self-coupling, say $\lambda = 0.01$, we therefore have

\begin{equation}\label{eqn:q231}
M_{new} = \sqrt{0.01}\frac{M}{\sqrt{0.1}} = \sqrt{0.1}M.
\end{equation}

\noindent For $M = E = 2.8 \times 10^{29}\GeV$, this gives a mass energy for the $\phi_{0} = 1.41 \times 10^{18}\GeV$ Q-ball derived black hole of

\begin{equation}\label{eqn:q232}
M_{new} = 8.85 \times 10^{28}\GeV,
\end{equation}

\noindent for an inflaton self-coupling of $\lambda = 0.01$. Using the conversion $1\kg = 5.62 \times 10^{26}\GeV$, this gives a mass in kilograms of

\begin{equation}\label{eqn:q233}
M_{new} = 158\kg.
\end{equation}

\noindent Using (\ref{eqn:q209}), we find that the reheating temperature in this case is

\begin{equation}\label{eqn:q234}
T_{R} = 574\GeV.
\end{equation}

\noindent This shows that lowering the strength of the inflaton self-coupling results in a smaller mass for the black holes and subsequently a higher reheating temperature. For $\lambda = 0.01$, the reheating temperature is sufficiently high that a lepton number asymmetry generated by interactions of the decay products of these black holes could be transferred to the Standard Model particle content via sphalerons, or a first-order electroweak transition and electroweak baryogenesis could occur. This shows that the reheating temperature from Q-ball derived black hole evaporation can be adjusted using the inflaton self-coupling, and that if $\lambda$ is small enough the model can reheat to a $T_{R} > T_{EW}$ sufficient for baryogenesis. This allows more scope to embed this model of non-minimally coupled Palatini Q-balls with an early era of  Q-ball derived black hole domination and reheating into a more complete cosmological model which can account for the observed baryon asymmetry today.

There is also the possibility of baryogenesis through the evaporation of the primordial black holes formed from collapsing Q-balls themselves, with a non-zero baryon number being produced through Hawking radiation. The possibility of a baryon asymmetry being produced through the evaporation of black holes was originally proposed by Hawking in 1974 \cite{hawkingbh} and has been explored in a number of particle physics models in conjunction with other aspects of early Universe cosmology such as inflation, electroweak symmetry breaking and dark matter generation (see for example \cite{barrowbh} - \cite{hook}) through either an electroweak or gravitational \cite{hooman} baryogenesis mechanism. Baryogenesis from evaporation of Q-ball derived black holes also allows for the possibility of extending the non-minimally coupled Palatini inflation model with Q-balls into a more complete model of cosmology, with the potential for embedding into Beyond the Standard Model particle models. 

\subsection{Gravitational Waves from Black Holes via the Poltergeist Mechanism}\label{section:4102}

It may also be possible to produce gravitational waves from the decay of primordial black holes through a phenomenon known as the Poltergeist mechanism \cite{inomata20}. If there is an early era of matter domination in the form of primordial black holes, these primordial black holes can rapidly decay, converting sub-horizon density perturbations into sub-horizon perturbations in radiation density. These perturbations in the radiation density can generate pressure waves within the thermal bath, which in turn generate gravitational waves. These gravitational waves may be observable in the next generation of detection experiments (DECIGO and LISA \cite{decigo1} - \cite{lisa3}), for black holes in the mass range $2-400$kg, provided that the spread of mass is in the range $\sigma\left(\delta M/M\right) \lesssim 0.01$ \cite{inomata20} . 

The upper bound of this mass range is slightly smaller than the mass of the lowest $\phi_{0}$ Q-ball black hole predicted in this model, which starts around $M = 500$kg. It should be noted however that these mass predictions come from approximating the effects of curvature on flat space Q-balls, rather than from the direct formation of black holes - either directly from the collapse of overdensities within the inflaton condensate or from mergers of Q-balls post-formation \cite{kzpbh5} - in a curved spacetime, so it is likely that if the effects of gravity are taken into account, smaller Q-balls of a lower mass could be formed, and the Schwarzschild limit for these Palatini Q-balls may be lowered. \\

We also demonstrated in Section \ref{section:4101}, that Q-ball black holes of a lower mass can be generated for a smaller inflaton self-coupling, $\lambda$, in a regime where the Q-ball black holes dominate the energy density at early times and then rapidly decay to radiation later on, so it is possible in this scenario that the Poltergeist mechanism could produce observable gravitational waves, although a numerical simulation of Q-ball formation in this model would be necessary to determine this formally.

\section{Summary and Discussion}\label{section:411}

In this chapter we have presented a study of Q-balls within the framework of a non-minimally coupled Palatini inflation model. The underlying scalar field theory is comprised of a complex scalar field charged under a global $U(1)$ symmetry with a coupling to the Ricci scalar in a simple $\phi^{2} + \phi^{4}$ potential in the Jordan frame which fulfills the role of the inflaton. In the Einstein frame the non-minimal coupling of the inflaton field to gravity is recast as a non-canonical kinetic term in the inflaton action. The predictions for the inflationary observables match the standard predictions for Palatini inflation, and the non-minimal coupling is calculated to be $\xi = 1.2163 \times 10^{9}$. We find that the presence of the inflaton mass term during inflation does not affect the form of the slow-roll parameters, or the observables, but it does directly affect the value of the inflaton field at the end of slow-roll inflation, where the inflaton at the end of inflation in this model is predicted to be $\phi \sim \left(1-3 \right)M_{pl}$. We also use the fact that a positive gradient is a necessary condition on the inflaton potential in order for inflation to proceed successfully to establish an upper bound on the inflaton mass squared.

The Q-balls in this model correspond to a complex scalar with a non-canonical kinetic term, which constitute a new type of Q-ball not previously studied. We derived the Q-ball equation analytically in the Einstein frame, by extremising the Q-ball action in flat space with respect to the conserved $U\left(1 \right)$ charge, and found that the form of the Q-ball equation in non-minimally coupled Palatini gravity is different from the conventional case in that there is an additional gradient squared term and a dependence on the conformal factor on the right hand side, and we find that this Q-ball equation (\ref{eqn:q77}) is in exact agreement with the equation for non-canonical Q-balls presented in \cite{lennon}, which came out shortly after the article detailing the research this chapter is based on was released \cite{usqballs}. This demonstrates that non-minimally coupled Palatini Q-balls are a new class of Q-ball with different underlying dynamics to that of conventional Q-balls.

Despite this, we find that the existence condition for these Q-balls is the same as in the conventional case, $\omega < m$, and that Coleman's mechanical analogy for Q-ball existence can still be applied, although the underlying dynamics are not the same. We then used this to derive a range of inflaton masses squared for which the inflaton potential is both compatible with the existence of Q-balls and can inflate successfully, which we refer to as the Q-ball window. This is significant because it can allow for suitable candidates for the inflaton field in order to produce Q-balls to be distinguished.

We solved the Q-ball equation numerically for a fixed inflaton mass within the Q-ball window for a range of $\omega < m$. We then used the location of the zeroes of the Q-ball equation, corresponding to the minima of the effective scalar potential, in order to narrow the range of field values for which Q-ball solutions can form for each $\omega$, these values being denoted by $\phi_{0}$. A code was then run between the upper and lower bounds of $\phi_{0}$ for each $\omega$ until a Q-ball solution was found. We presented ten Q-ball solutions in the range $\omega = \left(0.707155 - 0.89 \right)\omega_{c}$, and have calculated the important properties of these Q-balls including the radii, the charge of the Q-balls and the global energy of the Q-balls. We established that the radius of this type of Q-ball is proportional to $\phi_{0}$, the energy of the Q-balls is proportional to $\phi_{0}^{3}$, and that the $\omega$ parameter is approximately equal to the ratio of the Q-ball energy to the Q-ball charge $E/Q$. We find that these relations hold very strongly for the larger $\phi_{0}$ Q-balls and the relations become less well-defined for the smaller $\phi_{0}$ Q-balls. We also confirm that the Q-balls are absolutely stable, $E < mQ$, with the exception of the $\omega = 0.89\omega_{c}$ Q-ball, which is very close to the existence limit of $\omega < m$. This is a particularly significant result for the $\phi_{0} \sim 10^{17}\GeV$ Q-ball, since this shows that Q-balls of the size we expect to form from the inflaton condensate at the end of slow-roll inflation in non-minimally coupled Palatini gravity exist, and could therefore survive long enough to affect the post-inflationary cosmology of this model.

We derived an analytical approximation to the Q-ball equation for $\phi >> M_{pl}/\sqrt{\xi} \approx \phi_{0}$ in the limit that $r < r_{Q}/2$. We find that the field profiles generated from the analytical approximations of the Q-ball solutions for large $\phi_{0}$ closely match the Q-ball profiles from the numerical solutions up to about $r = r_{Q}/2$, and become a decreasingly good approximation as the $\phi_{0}$ of the Q-balls decreases. We also use this approximation to derive analytical expressions for the energy and charge of the Q-balls in the small $r$ limit, and we use these as a test on the numerical calculation of the Q-ball properties, including a confirmation using the analytical solution that $E \propto \sqrt{\lambda}\phi_{0}^{3}$ and $r_{Q} \propto \sqrt{\lambda}\phi_{0}$, as predicted by the approximate proportionality of the numerical radii and energies. We closely followed a derivation presented in \cite{heeck21} applied to the case of non-minimally coupled Palatini Q-balls to verify that the energy and charge vary with $\omega$ together, as is the case for conventional Q-balls. We calculated that the energy-charge ratio is equal to $\omega$ plus an integral term which can be interpreted as a contribution from the surface energy of the Q-ball, as opposed to the exact prediction of $E/Q = \omega$ as the chemical potential of the theory in the conventional case, although we later confirm that for $\phi >> M_{pl}/\sqrt{\xi}$ the contribution from this term is very small and that $E/Q \approx \omega$, as confirmed by the numerical calculations of the energy and charge. We also find that this provides an alternative formulation of the absolute stability condition, $\omega < m$.

We speculate that these Q-balls could form from the fragmentation of the inflaton condensate at the end of slow roll inflation, possibly as a result of tachyonic preheating. We confirm that this is possible by following a calculation presented in \cite{rubio19} in order to determine the range of perturbation wavelengths which undergo tachyonic growth, $7.16 \times 10^{-13} \GeV^{-1} < \lambda < 3.4 \times 10^{-8} \GeV^{-1}$. We find that the largest perturbation length scale to undergo tachyonic amplification in this model is within the horizon during inflation, and that the range of perturbation wavelengths encapsulates the radii of all the Q-balls produced numerically in this model. This means that all of the Q-balls presented here could form from the growth of the inflaton field perturbations due to tachyonic preheating at the end of slow-roll inflation.

Due to the fact that at the end of inflation in this scenario the inflaton scalars are bound into compact objects, reheating will proceed differently than in conventional inflation when non-topological soliton solutions are not present. The production of stable Q-balls at the end of inflation leads to an early era of matter domination, where the energy density of the Universe is dominated by solitons for a period after inflation before the Universe reheats via the evaporation of the Q-balls to radiation. This can proceed very rapidly unless the decay of the Q-balls to fermions is suppressed by Pauli blocking. This can mean that more e-folds of expansion are required to reheat the model than in conventional inflation, which can alter the prediction of the reheating temperature, and subsequently the scalar spectral index.

We explored the effects of curvature on these Q-balls by considering the flat space Q-ball solutions in a Schwarzschild spacetime, and we find that these Q-balls will collapse to black holes of mass $\sim 500$kg for $\phi_{0} \geq \phi_{0,c} = 1.41 \times 10^{18}\GeV$ when $\lambda = 0.1$. This shows that Primordial Black Holes can form from the collapse of Q-balls of the field values expected from tachyonic preheating after Palatini inflation, making the proposed formation mechanism of these Q-balls and the resultant black holes to be physically feasible. It also raises the question of whether, in curved space, sufficiently large perturbations from tachyonic growth could directly seed primordial black holes even if the underlying scalar field theory does not admit non-topological soliton solutions. We also find that the flat space approximation of the curved space properties of these Q-balls/Q-ball derived black holes are reasonable for $\phi_{0} < \phi_{0,c}/2$, since below this limit the local deviation of the metric is less than $25\%$ from the flat space metric. We also confirm that the mass of these Q-ball black holes scales with $\sqrt{\lambda}$, and the ratio of the Schwarzschild and Q-ball radius is independent of the inflaton self-coupling. 

\noindent We examined the viability of reheating via the evaporation of Q-ball derived black holes in the event that they come to dominate the energy density after inflation and then decay to radiation at later times. We find that this reheats the Universe at a very low temperature $T_{R} \approx 100\GeV$, which produces an insufficient value for the scalar spectral index of $n_{s} = 0.9536$, which is excluded by the $2-\sigma$ Planck bound of $n_{s} = 0.9565$ \cite{Planck18}. An alternative scenario is where only a small fraction of the Q-balls produced collapse into black holes, and the remaining Q-balls rapidly decay to radiation, instantly reheating the Universe. This means that the Universe enters immediately into an era of radiation domination following inflation, and that the Q-ball black holes come to dominate the energy density at a later time. By this mechanism, reheating completes within a greater number of e-folds of expansion than the case of early Q-ball derived black hole matter domination. It can give a much large reheating temperature, up to $T_{R} = 1.14 \times 10^{13}\GeV$, and a scalar spectral index of $n_{s} = 0.9612$ which is within the $1-\sigma$ Planck bounds of $n_{s} = 0.9649 \pm 0.0042$ \cite{Planck18}. This shows that although reheating purely via the evaporation of Q-ball derived black holes is not a viable reheating mechanism, the model can still reheat successfully if a only small fraction of the Q-balls produced collapse to black holes.

\noindent Adjusting the value of the inflaton self-coupling can also adjust the reheating temperature. We found that, using the dependence of the black hole mass-energy on $\sqrt{\lambda}$, if the inflaton self-coupling is $\lambda = 0.01$, the mass of the black hole resulting from the $\phi_{0} = 1.41 \times 10^{18}\GeV$ Q-ball is $M = 158$kg, and this has a reheating temperature of $T_{R} = 574 \GeV$. This is significant because it shows that altering the value of the inflaton self-coupling for a given Q-ball of $\phi_{0} \gtrsim \phi_{0,c}$ can raise the reheating temperature, bringing it to a scale $T_{R} > T_{EW} = 159 \GeV$, greater than the electroweak phase transition temperature. This is the temperature at which sphalerons can exist, and can therefore be considered the temperature at which baryogenesis becomes viable in the model, which opens up the possibility of being able to generate the baryon asymmetry in this model in addition to inflation, Q-balls and Primordial Black Holes, forming a more complete cosmological model.

\noindent Altering the value of the inflaton self-coupling for a Q-ball of $\phi_{0} \gtrsim \phi_{0,c}$ also lowers the mass of the black hole produced by the collapse of the Q-ball. This can be significant because it has been shown that black holes in the mass range $2-400$kg may produce gravitational waves which are observable in the next generation of detection experiments via the Poltergeist mechanism. We have shown for $\lambda = 0.1$, the black holes produced from the Q-balls presented here start at $M \approx 500$kg which is slightly higher than the masses which can produce observable gravitational waves through the Poltergeist mechanism. It may be therefore that these Q-balls could produce lighter black holes in the $2-400$kg range for smaller $\lambda$, which is an exciting possibility for the observability of a model of non-minimally coupled Palatini inflation with Q-balls. 

\noindent It should be noted that the post-inflationary cosmology of this model is rich with other potential gravitational wave sources. It has previously been shown that non-linear eras in the early Universe can generate their own stochastic gravitational wave background (\cite{khlebnikov97} - \cite{amin14}); and that the rescattering of amplified modes during preheating (\cite{assadullahi09} - \cite{jedamzik102}); the inhomogeneous, anisotropic motions of the condensate associated with Q-ball formation (\cite{kusenko09,chiba10}); the fragmentation of an inflaton condensate \cite{gravfrag}; and fluctuations in the numbers of $\pm$Q-balls \cite{kzpbh1} - \cite{kzpbh4}, can also source gravitational waves. A sudden transition from an early matter dominated era to a radiation dominated era can also cause an enhancement of the gravitational waves induced at second order in the primordial curvature perturbations \cite{alabidi13}, \cite{kohri18}, \cite{inomata19}. This may be especially relevant in our case as a rapid decay of Q-balls can cause such a sudden change in the equation of state of the Universe \footnote{A recent work \cite{white21} has demonstrated that the decay of Q-balls from the Affleck-Dine mechanism may proceed in such a fashion and may result in an observable gravitational wave signature.}. Since the frequency and energy density of these gravitational wave signatures are largely model dependent and will be affected by subtleties in model evolution, it is difficult to speculate at present as to the potential observability of any gravitational wave signatures generated in this framework of Palatini Q-balls, however it is possible that the signals produced could be observable in future experiments able to detect higher frequencies of gravitational waves \cite{aggarwal18}.

As mentioned in the Section \ref{section:42}, Q-balls and possibly their decay products can be dark matter candidates. Although we do not address this possibility in depth in this thesis, if the Q-balls form only a very small component of the energy density at the end of inflation, then assuming that the Q-balls did not dissociate or dissipate due to thermal effects, they could decay at a later time or survive to the present day and contribute to the Dark Matter content of the Universe. Depending on the interactions of the inflaton with the particle content of the Standard Model, these objects could be observable with direct dark matter searches. 

Beyond the case of a complex inflaton and Q-balls, an additional motivation for this study of Q-ball solutions is the possible insights it may give into the case of a real inflaton and oscillons. A non-minimally coupled Palatini model of this type could also inflate successfully for the case of a real inflaton, and neutral oscillons could be formed from the fragmentation of such a condensate. There have been numerous works which explore the formation and existence of oscillons and related objects from condensate fragmentation \cite{amin11}, \cite{amin10} - \cite{cyncynates21}; the evolution of a subsequent period of oscillon domination \cite{lozanov14}, and the observational signatures these objects could leave \cite{antusch17},\cite{lozanov19}, and as such this is a many-faceted and active area of research in cosmology. However, unlike the case of Q-balls, there are no analytical solutions for oscillons. Given the similarity of the underlying physics of real and complex scalars, it is possible that the Q-ball properties will be similar to those of the corresponding oscillons and that an oscillon window for oscillon formation will exist, analogous to the Q-ball window. 

We have demonstrated in this chapter that a model of non-minimally coupled Palatini inflation with non-canonical Q-balls can produce a post-inflation cosmology potentially very different from the standard post-inflation evolution in conventional Palatini inflation, with possible consequences for observability and the arrival at today's cosmology which are broad and far-reaching. In order to determine the precise implications, a numerical simulation detailing the tachyonic growth of the inflaton perturbations, the fragmentation of the inflaton condensate, the formation and possible decay of the Q-balls, and the formation and evaporation of any of the Q-ball black holes in curved spacetime on a subsequent expanding background would need to be performed.

\chapter{Conclusion}

The main focus of the research presented in this thesis has been the application of non-minimally coupled, and $R^{2}$ term, inflation models in the Palatini formalism to other problems and phenomena in cosmology, as well as the compatibility of these models with Big Bang cosmology and particle physics in the broader sense. The non-minimal coupling of the inflaton to gravity, and any higher order curvature terms in the gravitational action, modify the dynamics of the scalar field, and as such the non-minimal coupling or higher order term is typically recast by performing a conformal transformation on the metric of the theory, such that the gravitational dynamics beyond the Einstein-Hilbert action manifest as non-canonical kinetic terms and a rescaled inflaton potential when examined in the Einstein frame - a frame defined such that the transformed gravitational action is Einstein-Hilbert, and the gravitational dynamics can be considered as equivalent to General Relativity. Non-minimally coupled and $R^{2}$ inflation models are therefore typically considered in the Einstein frame and we follow this approach in this thesis. The effect of these terms is that the inflaton potential becomes a plateau in the Einstein frame, and they therefore provide a very natural mechanism for producing a flat inflaton potential, as is favoured by observations of the inflationary observables. Non-minimally coupled and $R^{2}$ inflation models are therefore very favourable observationally, and we find that the results of the calculations of the predicted inflationary observables in the inflation models considered in this thesis are in good agreement with observations.

We primarily consider inflation in the Palatini formalism in this thesis. In the Palatini formalism, the conformal transformation of the Ricci scalar is much more straightforward - owing to the treatment of the spacetime metric $g_{\mu \nu}$ and the affine connection $\Gamma^{\rho}_{\mu \nu}$ as independent quantities - in comparison to the transformation in the metric formalism, and no additional non-canonical inflaton kinetic terms arise in the Einstein frame action as a result. This means that the canonical normalisation of the inflaton field is different depending on the formalism used, which results in the Einstein frame potential in terms of the canonically normalised inflaton being different, even if the Jordan frame inflaton potential is the same in both formalisms. Since the slow-roll parameters - and hence the inflationary observables built from them - depend explicitly on the inflaton potential, the expressions and the predictions for these quantities will therefore be different depending on the formalism used. Most notably, the Palatini formalism typically produces a heavily suppressed tensor-to-scalar ratio compared to the metric formalism, and also typically requires a larger value of the non-minimal coupling $\xi$ in order to produce the observed primordial curvature power spectrum $\mathcal{P}_{\mathcal{R}}$. 

In Chapter 4, we consider a $\phi^{2}$ Jordan frame inflaton potential in the Palatini formalism in a framework where there is an $R^{2}$ term in the gravitational action, setting the non-minimal coupling to zero in order to exclusively study the effects of the $R^{2}$ term on the $\phi^{2}$ model in the Palatini formalism. We were interested in whether the $\phi^{2}$ inflation model could be brought into observational favour - the original $\phi^{2}$ chaotic inflation model having been ruled out observationally on the basis of the tensor-to-scalar ratio being too large - by the addition of the $R^{2}$ term and the additional suppression of the tensor-to-scalar ratio by the Palatini formalism treatment. Another issue with the original $\phi^{2}$ chaotic inflation model is the fact that in order to complete inflation the inflaton must be super-Planckian, and we wanted to examine the constraints which would be placed on the model from the requirement that the inflaton be sub-Planckian for the duration of inflation. 

The transformation of the inflaton action to the Einstein frame if there is an $R^{2}$ term present is more involved that the case of a non-minimal coupling, requiring a transformation of the action to include an auxiliary field $\chi$, which must then be eliminated via its equations of motion to define the conformal transformation and recast the $R^{2}$ term into non-canonical inflaton kinetic terms. We find that the quartic kinetic term $\sim \left(\partial_{\mu}\phi\partial^{\mu}\phi\right)^{2}$ makes a negligible contribution to the dynamics of the inflaton during slow-roll, and so we neglect this term and proceed in the same way as for non-minimally coupled inflation. Canonically rescaling the inflaton field, we find that the inflaton dynamics are well-described by a threshold approximation utilising the dominant terms of the potential in the given regime of the field. Above the threshold, the field is on the plateau of the potential and below we find that the canonical scalar field is approximately the untransformed inflaton field, such that the dynamics of the potential are equivalent to the quadratic dependence in the Jordan frame, and we regard the frames to be dynamically equivalent below this point.

Imposing the constraint that the canonically normalised inflaton remains sub-Planckian during inflation places a constraint on the coefficient of the $R^{2}$ term to be $\alpha \gtrsim 10^{12}$. We therefore find that the issue of the too-large inflaton in $\phi^{2}$ inflation can be remedied in the Palatini formalism by adding an $R^{2}$ term to the gravitational part of the action with a coefficient of $\alpha \gtrsim 10^{12}$.

In addition to the constraint of a sub-Planckian inflaton, we also examined the possible effects of embedding this inflation model into a UV completion of gravity. An issue of inflation as an effective theory in quantum gravity is that higher-order corrections scaled by the scale of quantum gravity - the Planck mass in the Einstein frame - are introduced into the inflaton potential. As aforementioned, the inflationary observables depend explicitly on the inflaton potential, therefore any additional corrections to the inflaton potential will modify the predicted values of the inflationary observables and could possibly move a theory out of observational favour. If we want to embed inflation in a more complete cosmological model, and embed Big Bang cosmology into a more complete theory of physics, it is important to test the compatibility of inflation with a UV completed gravity framework. We therefore tested the robustness of the model with respect to observations when Planck-suppressed effective theory corrections are added to the inflaton potential. We find that the model remains consistent with observations provided that the shift on the $\eta$-parameter due to the introduction of the Planck-suppressed corrections from the UV completion of gravity does not exceed $\left| \Delta \eta \right| \leq 0.001$. 

Placing this constraint on the potential shift due to these Planck-suppressed corrections places the constraint on the coefficient of the $R^{2}$ term that $\alpha \gtrsim 10^{32}$ in order for the model to remain in agreement with observations when Planck-suppressed potential corrections from a UV completion of gravity are included. A similar constraint on the $\alpha$ parameter can be derived for the case of Planck-suppressed potential corrections with a broken shift symmetry, and in this case we find that the requirement for the model to remain consistent with observations is that $\alpha \gtrsim 10^{20}$. We note from these results that the $\phi^{2}$ model of Palatini inflation with an $R^{2}$ term is compatible with observations when embedded as an effective theory in quantum gravity framework provided that the contribution of the $R^{2}$ term to the inflaton action in the Jordan frame is very large. In the Einstein frame this can be interpreted as the conformal factor being very large, and the potential therefore being very flat and the inflaton kinetic term being heavily suppressed. 

Using the threshold approximation of the potential, we find that the number of e-folds of inflation at the pivot scale in this model can be expressed as a function of $\alpha$, and that for larger $\alpha$, the Planck pivot scale exits the horizon after fewer e-folds of inflation. For each of the derived lower bounds on $\alpha$ we calculate the scalar spectral index, the tensor-to-scalar ratio, the Hubble parameter, the number of e-folds at the pivot scale, the reheating temperature, the value of the inflaton field at the end of inflation, the unitarity cutoff in the Einstein frame, and the unitarity cutoff in the present vacuum. For all of the regimes of $\alpha$ we find that unitarity is conserved in the model, and that the scalar spectral index in each case is within the $1-\sigma$ bound on the scalar spectral index from the Planck observations (2018), with the exception of the case of Planck-suppressed potential corrections ($\alpha \gtrsim 10^{32}$), whereby the scalar spectral index is slightly outside of the $2-\sigma$ bounds on $n_{s}$ from the Planck results. It is possible that this result could be modified by the addition of other corrections, or by the combinatorial suppression factor in the $\sigma^{6}$ term from the UV completion of gravity.

In all cases the tensor-to-scalar ratio is heavily suppressed, with $r \sim 10^{-25} - 10^{-6}$, as is typical for Palatini inflation. This shows that the presence of the $R^{2}$ term in the Palatini formalism can bring the tensor-to-scalar ratio of the $\phi^{2}$ inflation model into observational favour in conjunction with a sub-Planckian inflaton, and possibly with the case of Planck-suppressed corrections for a sufficiently large $\alpha$. However, the predicted values of $r$ are below the observable limit of $r \sim 10^{-3}$ in the next generation of CMB experiments. 

We consider two possible reheating mechanisms in this model: reheating via the decay to right-handed neutrinos and reheating via Higgs portal annihilation. In both cases we use the Coleman-Weinberg corrections to the inflaton potential in order to estimate the shift of the potential due to the couplings between the inflaton and right-handed neutrinos, and the inflaton and the Higgs.  For each case we choose the mass scale of the decay channel as the renormalisation scale, and use the constraint on the $\eta$-shift due to the potential corrections, $\left| \Delta \eta \right| \leq 0.001$, in order to constrain the size of the couplings. We find that for Higgs portal annihilation, $\lambda_{\phi H} < 8.2 \times 10^{-7}$, and for decay to right-handed neutrinos $\lambda_{\phi N} < 1.1 \times 10^{-3}$.

In the case of reheating via Higgs portal annihilation, there are two different possibilities for how reheating can proceed; through rapid preheating or through fragmentation of the inflaton condensate. We find that Higgs portal annihilation is not a viable reheating mechanism for $\phi^{2}$ Palatini inflation with an $R^{2}$ term unless the annihilation rate exceeds the Hubble rate quickly enough that the model reheats immediately at the end of inflation, or the condensate fragments to oscillons. We find that the condition for the inflaton condensate to fragment is that $\alpha > 2.9 \times 10^{13}$ for the $\phi^{2}$ Palatini inflation model with an $R^{2}$ term, and this is therefore compatible with the inclusion of Planck-suppressed potential corrections but may not be satisfied for the minimum requirement of a sub-Planckian inflaton, $\alpha \gtrsim 10^{12}$, if $\alpha$ is close to its lower bound. Considering the fact that the $\alpha$ parameter is generally very large, we conclude that the condensate is likely to fragment, and that reheating via the Higgs portal then proceeds through the decay of inflatonic oscillons, in which case reheating is not instantaneous. Given that the condensate fragments, we find that the model can reheat successfully through the annihilation of oscillons for $\alpha \gtrsim 10^{20}, 10^{32}$, although given that the scalar spectral index $n_{s}$ is in tension with the $2-\sigma$ lower bound for the case of general Planck-suppressed corrections, $\alpha \gtrsim 10^{32}$, even for the case of instantaneous reheating, this suggests that reheating via Higgs portal annihilation is observationally disfavoured for this case.

For the case of reheating via the decay to right-handed neutrinos we find this can only reheat the model instantaneously for the case of general Planck-suppressed potential corrections, $\alpha \gtrsim 10^{32}$. This is because the minimum value of $\lambda_{\phi N}$ needed in order for the decay to proceed is only compatible with the upper bound on $\alpha$ for the case of general Planck-suppressed potential corrections. If the reheating is not instantaneous, then this reheating would proceed for a lower reheating temperature and consequently a lower $n_{s}$, which rules out this reheating mechanism for the case of general Planck-suppressed potential corrections, since the scalar spectral index is already in tension with the $2-\sigma$ Planck lower bound, but can allow the cases with smaller values of $\alpha$.

In Chapter 5 we consider an application of the Affleck-Dine mechanism to a scalar field in the early Universe with a $\left|\Phi\right|^{2} + \left|\Phi\right|^{4}$-type potential with $U(1)$ symmetry-breaking terms for a complex field. This form of the potential is naturally compatible with non-minimally coupled inflation models. In this work we consider quadratic $U(1)$-violating terms and set the cubic and quartic terms to zero in order to exclusively study the effects of the quadratic term. We use a threshold approximation for the potential and treat the model such that the Affleck-Dine dynamics of the field do not become significant until the potential is deep into the quadratic regime. We use this to derive an analytical approximation of the $U(1)$ asymmetry generated in the model. We then use this to derive expressions for the asymmetry generated within the coherently oscillating condensate in the $\Phi^{2}$ part of the potential - the condensate asymmetry - and the asymmetry transferred to the Standard Model by the decay of the condensate - the transferred asymmetry. We check the transferred asymmetry for both the case of no decay of the condensate and the condensate decaying. These two scenarios can be treated as the limit of the condensate decaying on timescales small and large compared to the scalar lifetime.

Dynamically the asymmetry is generated within the condensate as the inflaton scalars oscillate between their $\Phi$ and $\Phi^{\dagger}$ states and the condensate decays. This means that there is not an exact cancellation of the number density of $\Phi$ and $\Phi^{\dagger}$ scalars within the condensate, and allows for the generation of an asymmetry in $B$ within the inflaton condensate. As the condensate decays this also means that there isn't an exact cancellation of the decay products with the antiparticles when the scalars decay via a $B$-conserving interaction, resulting in an overall asymmetry in the $B$ number of the decay products of the inflaton field, which is then transferred to the Standard Model.

We find that our analytical approximation of the transferred asymmetry can generate the observed baryon-to-entropy ratio today, subject to certain constraints on $A^{\frac{1}{2}}/m_{\Phi}$, where $A$ is the $U(1)$ violating mass squared term. In the case where the lifetime of the scalars $\tau_{\Phi}$ is much larger than the oscillation time of the asymmetry $T_{asy}$, the generated asymmetry can easily be larger than the observed baryon-to-entropy ratio, and the suppression of the asymmetry over many oscillations is important in this case, whereas in the case where the condensate decays before the asymmetry can undergo many oscillations, the suppression of $A^{\frac{1}{2}}/m_{\Phi}$ must be much stronger to generate the observed baryon-to-entropy ratio.

We derive a constraint on $A^{\frac{1}{2}}/m_{\Phi}$ for which the threshold approximation of the asymmetry is valid from the fact that we require the quartic part of the potential to not make a significant contribution to the dynamics. This suppresses the dynamics of the angular component of the inflaton, and allows us to treat the Affleck-Dine dynamics of the field as being purely in the quadratic regime of the potential. This constraint on the validity of the threshold approximation is easily compatible with the constraints on $A^{\frac{1}{2}}/m_{\Phi}$ derived in order to produce the observed baryon-to-entropy ratio.

We analyse the behaviour of the condensate and transferred asymmetries from a numerical calculation, and we find that the analytical threshold approximation of the condensate and transferred asymmetries are in exact agreement with the numerically calculated asymmetries for $A^{\frac{1}{2}}/m_{\Phi} = 0.001$. 

We find that the baryon-to-entropy ratio calculated numerically is the same order of magnitude as the baryon-to-entropy ratio calculated using the analytical expressions, for both $\tau_{\Phi} = T_{asy}$ and $\tau_{\Phi} > T_{asy}$, and we demonstrate that the analytical approximation of the decay of the condensate asymmetry is a valid approximation of the behaviour of the asymmetry.

For a general inflaton decay process to a pair of fermions, we show that the asymmetry is unlikely to be washed out by inflaton exchange, and express this as a constraint on the reheating temperature. This is an estimate since a proper calculation would require a specific decay model of the inflaton, however we find that the constraint we have should be easily satisfied.

We verify that this model of Affleck-Dine baryogenesis with quadratic $B$-violating potential terms is compatible with the dynamics of non-minimally coupled inflation provided that the Affleck-Dine dynamics become dominant below the threshold, well after the non-minimally coupled dynamics have ceased to dominate the inflaton potential. We find that this requirement in both the metric and Palatini formalisms leads to the same constraint on the inflaton mass of $m_{\Phi} < 2.2 \times 10^{13}\GeV$, and we therefore find that this mechanism with quadratic $B$-violating terms is compatible with a non-minimally coupled inflation model where the inflaton also acts as the Affleck-Dine field.

We consider the baryon isocurvature perturbations generated in this model and use the limits on the observable baryon isocurvature fraction from the Planck satellite in order to constrain the size of the non-minimal coupling $\xi$ in both the metric and the Palatini formalisms. We find that baryon isocurvature perturbations are generally much smaller than the observational limit in Palatini inflation, whereas in metric inflation they require that $\lambda_{\Phi} \lesssim 10^{-4}$. This also means that baryon isocurvature perturbations are close to the present CMB bound on the isocurvature fraction in metric non-minimally coupled inflation if $\lambda_{\Phi} \sim 10^{-4}$.

We also consider the validity of the classical treatment of the inflaton for the calculation of the baryon asymmetry. We find that although the inflaton field is inherently quantum in nature, the baryon asymmetry is equivalent to its classical value due to spatial averaging, and therefore treating the inflaton field as classical gives the correct expression for the asymmetry.

In Chapter 6 we present a model of non-minimally coupled Palatini inflation with a $\left|\Phi\right|^{2} + \left|\Phi\right|^{4}$-type potential with Q-balls. We begin by deriving the slow-roll parameters and the inflationary observables in this model in the Einstein frame and find that they are the same as the typical expressions for $\phi^{4}$ Palatini inflation. We then derive an upper bound on the inflaton mass from the requirement that the gradient of the inflaton potential must be positive in this case in order for inflation to take place. This has not been previously considered.

We then derive the Q-ball equation in the Einstein frame for a scalar non-minimally coupled to gravity, and we find that the Q-ball equation in this case has an additional gradient squared term and an additional conformal factor dependence on the right-hand side as compared to the equation for conventional Q-balls. This shows that these non-minimally coupled Palatini Q-balls are dynamically different to the conventional case of Q-balls and correspond to a new class of Q-ball.

We find that we can still apply Coleman's mechanical analogy for the existence of Q-balls to these non-minimally coupled Palatini Q-balls, due to the fact that the inverted Q-ball potential is still compatible with the analogy despite the fact that the underlying dynamics are different. We find that the resulting existence condition for Q-balls in the non-minimally coupled Palatini case is the same as for conventional Q-balls, $\omega < m$. 

From the form of the Q-ball potential, we derive a mass range for the inflaton field within which the potential can produce Q-balls compatible with the upper bound on the inflaton mass needed for inflation to occur. We refer to this as the Q-ball window, and we then use this in order to narrow the parameter space within which we search for Q-ball solutions numerically.

We obtain ten Q-ball solutions for an inflaton mass of $m = 0.9\omega_{c}$ in the range of $\omega = (0.707155 - 0.89)\omega_{c}$, corresponding to $\phi_{0}$ in the range $10^{13}-10^{17}\GeV$ numerically. From here we numerically derive the key properties of these Q-balls, energy, charge and radius. We also derive an analytical solution of the Q-ball equation, and compare the properties of these Q-balls as calculated using the analytical approximation to the results calculated numerically. We find that these Q-balls are absolutely stable, with the exception of the $\omega = 0.89\omega_{c}$ Q-ball, and we also find that $E/Q \approx \omega$, in particular for the larger $\phi_{0}$ Q-balls. 

The analytical approximation of the behaviour of the Q-ball solutions is a good fit for the Q-ball solutions for larger $\phi_{0}$, corresponding to the limit $\phi_{0} >> M_{pl}/\sqrt{\xi}$. Since the analytical approximation is calculated in the plateau limit, the poor fit for the smaller $\phi_{0}$ solutions makes sense as these Q-balls form close to the point where the plateau approximation starts to break down. We also find that the approximation is a good estimate of the Q-ball properties for $r < r_{Q}/2$, since this is the limit the analytical approximations of energy and charge are taken to.

The analytical approximation predicts that $E, Q \propto \phi_{0}^{3}$ and $r \propto \phi_{0}$, which are both confirmed numerically. Using the analytical approximation we confirm that the $\omega$ parameter corresponds to the chemical potential of the system, as it does in the case of conventional Q-balls, and we show that $E/Q = \omega$ up to a small correction from the surface energy density of the Q-ball. Subsequently, we find that the existence condition for Q-balls, $\omega < m$, is also, to a good approximation, an expression of the absolute stability requirement for Q-balls.

We discuss the possibility that these Q-balls could form as a result of the fragmentation of the inflaton condensate from tachyonic preheating. We apply the analysis used in \cite{rubio19} to derive a range of perturbation wavelengths which would experience tachyonic growth in this model during tachyonic preheating, and we find that the radii of all the Q-ball solutions calculated in this model are contained within this range, $7.16 \times 10^{-13}\GeV^{-1} < \lambda < 3.4 \times 10^{-8}\GeV^{-1}$. This means that it is possible that these non-minimally coupled Palatini Q-balls could be formed from the fragmentation of the inflaton condensate as a result of tachyonic preheating.

It is possible that formation of Q-balls through this channel could result in an early epoch of matter domination in the form of Q-balls, in which case reheating in this model would occur via the evaporation of Q-balls. This perturbative decay of the Q-balls could potentially result in a slower reheating than the usual decay of a coherent inflaton condensate, due to the effects of Pauli blocking and also the suppression of the evaporation rate of the scalars from the size of the surface area of the Q-balls. This slower reheating could result in a lower reheating temperature, and therefore a shorter duration of inflation owing to the fact that a greater number of e-folds of expansion would be needed to complete reheating.

We also estimate the effects of curvature on these Q-balls. In order to do this, we consider the flat space Q-ball solutions used in this work in a Schwarzschild metric. We find that the radii of all of the Q-balls generated in this work are greater than their Schwarzschild radii, and that therefore all the Q-balls generated here are consistent flat space Q-balls and will not collapse to black holes. We find that the smallest $\phi_{0}$ which would be needed to generate a Q-ball which would collapse to a black hole is $\phi_{0, c} = 1.41 \times 10^{18}\GeV$, which is typically of the size we would expect the inflaton field to be at the end of non-minimally coupled Palatini inflation. This Q-ball has a predicted radius of $r = 3.91 \times 10^{-9}\GeV^{-1}$, which is within the range of perturbation wavelengths in this model which could experience tachyonic growth. We therefore have a model of non-minimally coupled Palatini inflation which could feasibly produce Q-balls which could seed primordial black holes, and that these Q-balls could form from the fragmentation of the inflaton condensate following tachyonic preheating. Calculating the mass-energy contained within the Schwarzschild radius of this Q-ball gives a black hole of mass $M \sim 500\kg$.

We find that for $\phi_{0} < \phi_{0,c}/2$, the effects of curvature in this approximation (on the flat space solutions) are small, with a maximum of $25\%$ local deviation from flat space within the metric at the surface of the Q-balls. Therefore the estimated value of $\phi_{0,c}$ should be close to the true value. A derivation of the Q-ball equation in curved space would be required in order to verify the effects of curvature on non-minimally coupled Palatini Q-balls and their dynamics.

We also speculate on the possibility of reheating the model via the decay of primordial black holes. Since the black holes which could form in this framework start at a mass of $M \sim 500\kg$, the temperature of these black holes is high enough that the decay of these black holes could produce all of the particle content of the Standard Model, and could provide a mechanism for reheating the Universe. We consider this possibility both in the event of an early period of matter domination of primordial black holes formed from collapsed Q-balls, and in the event that only a few of the Q-balls in the early period of Q-ball matter domination collapse to form black holes, with the primordial black holes making up a small fraction of the energy density. 

In the case of reheating from primordial black holes dominating the energy density after inflation, we find that the value of the scalar spectral index predicted in this scenario is excluded by the $2-\sigma$ lower bound on $n_{s}$ from the Planck results, and so is not a viable reheating mechanism for non-minimally coupled Palatini inflation with Q-balls. 

For the case where the dominant component of the energy density is still Q-ball matter after inflation, and only a small fraction of the Q-balls collapse to black holes, we expect that the reheating temperature can be much higher, with only a short period of black hole domination at late times, and that the predicted scalar spectral index therefore can be within the $1-\sigma$ bounds of the Planck results, $n_{s} = 0.9612$. This means that an early period of radiation domination due to Q-ball decay with a small fraction of Q-ball derived black holes may be a viable reheating model for non-minimally coupled Palatini inflation with Q-ball derived black holes. This is one example of the potential broader cosmological implications which could result from the presence of Q-balls in a non-minimally coupled Palatini inflation model.

We speculate on a possible channel of baryogenesis from this model, which could result from the generation of an asymmetry through the interactions of the particles created from the decay of these Q-ball derived black holes. Black hole decay in the case of an early era of primordial black hole domination gives a reheating temperature of around $T_{R} \approx 100\GeV$, which is too low for either electroweak baryogenesis or the transfer of an asymmetry to baryon number by sphalerons, both of which require $T > 159.5\GeV$. Using the proportionality of the Q-ball mass-energy to $\phi_{0}$, $M \propto \sqrt{\lambda}\phi_{0}^{3}$, we find that for a smaller inflaton self-coupling, $\lambda = 0.01$, the $\phi_{0,c} = 1.41 \times 10^{18}\GeV$ Q-ball produces a black hole of mass $M = 158\kg$, which has a reheating temperature of $T_{R} = 574\GeV$ and is sufficiently high for electroweak baryogenesis or the generation of sphalerons. This shows that the reheating temperature of this non-minimally coupled Palatini inflation with Q-balls can be adjusted by adjusting the inflaton self-coupling. Therefore, reheating through the decay of Q-ball derived black holes in the case where these black holes dominate the energy density of the Universe after inflation can be consistent with baryogenesis for a smaller inflaton self-coupling.

Finally, we discuss the possibility of observable gravitational wave signatures from the model. Specifically, we find that the generation of gravitational waves through the decay of primordial black holes via the Poltergeist mechanism \cite{inomata20} is a possibility in this model. These gravitational waves would be observable in the next generation of gravitational wave detection experiments, DECIGO and LISA, for black holes in the mass range $2 - 400\kg$. Although the upper bound here is slightly lower than the black holes formed from the collapse of Q-balls in this model when $\lambda = 0.1$, we have demonstrated that the mass of these black holes can be within the observable range for a smaller inflaton self-coupling, so there is a possibility of observable gravitational waves from black hole decay in this model for different parameter values. There is also the fact that these estimates have been made using the flat space Q-ball solutions, and it is therefore possible that the curved space Q-ball solution could produce black holes of smaller masses, and the analysis would need to be performed in curved space to assess this. 

There is also the possibility that future-observable gravitational waves could be generated from the decay of Q-balls themselves, mergers and interactions of Q-balls and possibly from the formation of the Q-balls, although a full numerical simulation of the formation and evolution of these Q-balls would need to be performed to ascertain the precise dynamics of these Q-balls and the subsequent possibilities for gravitational wave production.

In this thesis, we have presented three pieces of original research which propose solutions to problems in inflation or cosmology, or provide a basis for important phenomena in cosmology, within the framework of non-minimally coupled or $R^{2}$ inflation, primarily in the Palatini formulation of gravity. In every case we find that the conjunction of inflation with these other phenomena is at least feasible within a set of constraints on the model parameters. In addition to observational compatibility, in each case we have discussed the observability of the model, the compatibility of the model with the subsequent transition from the end of inflation to the standard cosmological evolution, and some aspect of the implications of the model or its place within cosmology in the broader sense. It is the hope of the author that the research presented has shown that inflation beyond a minimal coupling of the inflaton to gravity in the Palatini formalism is a strong candidate not just for inflation itself, and the observationally favourable status of the inflation model, but also as a basis for embedding inflation into the rest of cosmology, and indeed physics as a whole, in order to search for a more complete description of the origin of the Universe as we observe and understand it today.


\addcontentsline{toc}{chapter}{References}

\begin{thebibliography}{0}


\bibitem{2df} M.~ Colless et. al;
\textit{"The 2dF Galaxy Redshift Survey: Spectra and redshifts"};
Monthly Notices of the Royal Astronomical Society (2001), Volume 328, Issue 4, pp. 1039-1063;
doi: 10.1046/j.1365-8711.2001.04902.x;
[e-print: astro-ph/0106498].
M.~ Colless;\textit{"Cosmological results from the 2df galaxy redshift survey"};
Part of Measuring and modeling the universe. Proceedings, Symposium, Pasadena, USA, November 17-22, 2002, 196-206;
Contribution to: Carnegie Observatories Centennial Symposium. 2. Measuring and Modeling the Universe, 196-206;
[e-print: astro-ph/0305051].
Site: https://www.mso.anu.edu.au/2dFGRS/

\bibitem{sdss} D.~ J.~ Eisenstein et. al  (SDSS Collaboration);
\textit{"Detection of the Baryon Acoustic Peak in the Large-Scale Correlation Function of SDSS Luminous Red Galaxies"};
Astrophys. J. \textbf{633} (2005), 560-574;
doi: 10.1086/466512;
[e-print: astro-ph/0501171];
site: https://www.sdss.org/

\bibitem{cmb} A.~ A.~ Penzias and R.~ W.~ Wilson;
"\textit{A Measurement of Excess Antenna Temperature at 4080 Mc/s}"; 
The Astrophysical Journal. \textbf{142}(1965) (1): 419–421;
doi: 10.1086/148307.

\bibitem{gamow} G.~ Gamow;
"\textit{The Evolution of the Universe}";
Nature \textbf{162} (1948) 4122, 680-682;
doi: 10.1038/162680a0.


\bibitem{alpher} R.~ A.~ Alpher and R.~ Herman;
"\textit{Evolution of the Universe }'';
Nature \textbf{162} (1948) 4124, 774-775;
doi: 10.1038/162774b0.

\bibitem{dprw} R.~ H.~ Dicke, P.~ J.~ E.~ Peebles, P.~ G.~ Roll and D.~ T.~ Wilkinson;
"\textit{Cosmic Black-Body Radiation}";
Astrophysical Journal \textbf{142} (1965), 414-419;
doi: 10.1086/148306.



\bibitem{cmbtemp} J.~ C.~ Mather, D.~ J.~ Fixsen, R.~ A.~ Shafer, C.~ Mosier and D.~ T.~ Wilkinson
"\textit{Calibrator Design for the COBE Far-Infrared Absolute Spectrophotometer (FIRAS)}";
The Astrophysical Journal, \textbf{512} 2 (1999), 511-520;
doi: 10.1086/306805;
[e-print: astro-ph/9810373].

\bibitem{cobe} J.~ C.~ Mather et al;
\textit{"A Preliminary Measurement of the Cosmic Microwave Background Spectrum by the Cosmic Background Explorer (COBE) Satellite"};
Astrophysical Journal Letters \textbf{354} (1990), L37-L40;
doi: 10.1086/185717.

\bibitem{wmapcmb} \textbf{WMAP Collaboration}, C.~ L.~ Bennett et al;
\textit{"Nine-Year Wilkinson Microwave Anisotropy Probe (WMAP) Observations: Final Maps and Results"};
The Astrophysical Journal Supplement, \textbf{208} (2013), 2, article id. 20, 54;
doi: 10.1088/0067-0049/208/2/20;
[e-print: 1212.5225 [astro-ph]].

\bibitem{planckcmb}\textbf{Planck Collaboration}, P.~ A.~ R.~ Ade et. al;
\textit{"Planck Early Results. I. The Planck mission"};
Astronomy \& Astrophysics \textbf{536} (2011), A1;
doi: 10.1051/0004-6361/201116464;
[e-print: 1101.2022 [astro-ph.IM]].


\bibitem{hubble} E.~ Hubble;
\textit{"A relation between distance and radial velocity among extra-galactic nebulae"};
Proceedings of the National Academy of Sciences of the United States of America, \textbf{15} (1929) 3, 168-173;
doi: 10.1073/pnas.15.3.168.

\bibitem{planck184} N.~ Aghanim et al. [Planck Collaboration],
\textit{“Planck 2018 results. VI. Cosmological parameters”}, 
A\&A 641, \textbf{A6} (2020);
doi: 10.1051/0004-6361/201833910;
A\&A 652 \textbf{C4} (2021) (erratum);
doi: 10.1051/0004-6361/201833910e (erratum);
[e-print:1807.06209 [astro-ph.CO]].

\bibitem{riess} A.~ G.~ Riess et. al;
"\textit{Observational Evidence from Supernovae for an Accelerating Universe and a Cosmological Constant }";
The Astronomical Journal \textbf{116} (1998) 3, 1009-1038;
doi: 10.1086/300499;
[e-print: astro-ph/9805201].

\bibitem{perlmutter} S.~ Perlmutter et. al;
"\textit{Measurements of $\Omega$ and $\Lambda$ from 42 High-Redshift Supernovae }";
The Astrophysical Journal \textbf{517} (1999) 2, 565-586;
doi: 10.1086/307221;
[e-print: astro-ph/9812133].

\bibitem{riess20} A.~ G.~ Riess et. al.;
"\textit{Cosmic Distances Calibrated to $1\%$ Precision with Gaia EDR3 Parallaxes and Hubble Space Telescope Photometry of 75 Milky Way Cepheids Confirm Tension with $\Lambda$CDM}";
Astrophys. J. Lett. \textbf{908} (2021) 1, L6;
doi: 10.3847/2041-8213/abdbaf;
[e-print: 2012.08534 [astro-ph.CO]].


\bibitem{freedman21} W.~ L.~ Freedman;
"\textit{Measurements of the Hubble Constant: Tensions in Perspective}";
Astrophys. J. \textbf{919} (2021) 1, 16;
doi: 10.3847/1538-4357/ac0e95;
[e-print: 2106.15656 [astro-ph.CO]].

\bibitem{kolbturner} E.~ W.~ Kolb and M.~ S.~ Turner;
"\textit{The Early Universe}";
Frontiers in Physics \textit{69} (1990).

\bibitem{einsteinhilbert} D.~ Hilbert;
\textit{"Die Grundlagen der Physik" [Foundations of Physics]};
Nachrichten von der Gesellschaft der Wissenschaften zu Göttingen – Mathematisch-Physikalische Klasse (in German), \textit{3}(1915): 395–407.
R.~ P.~ Feynman; \textit{Feynman Lectures on Gravitation}; Addison-Wesley. p. 136, eq. (10.1.2). (1995). 

\bibitem{lythliddle}D.~ H.~ Lyth and A.~ R.~ Liddle;
"\textit{The primordial density perturbation}";
Cambridge University Press; 2009.

\bibitem{sdss06} \textbf{SDSS Collaboration}, M.~ Tegmark et. al;
"\textit{Cosmological Constraints from the SDSS Luminous Red Galaxies}";
Phys. Rev. D \textbf{74} (2006), 123507;
doi: 10.1103/PhysRevD.74.123507;
[e-print: astro-ph/0608632].



\bibitem{boomerang}\textbf{BOOMERANG Collaboration}, A.~ Melchiorri et al;
"\textit{A measurement of omega from the North American test flight of BOOMERANG}";
Astrophys. J. Lett. \textbf{536} (2000) L63-L66;
doi: 10.1086/312744;
[e-print: astro-ph/9911445].


\bibitem{wmap5}\textbf{WMAP Collaboration}, E.~ Komatsu;
\textit{"Five-Year Wilkinson Microwave Anisotropy Probe (WMAP) Observations: Cosmological Interpretation"};
Astrophys. J. Suppl. \textbf{180} (2009), 330-376;
doi: 10.1088/0067-0049/180/2/330;
[e-print: 0803.0547 [astro-ph]].


\bibitem{guth} A.~ H.~ Guth;
"\textit{The Inflationary Universe: A Possible Solution to the Horizon and Flatness Problems}";
Phys. Rev. D \textbf{23} (1981) 347-356, Adv. Ser. Astrophys. Cosmol. \textbf{3} (1987) 139-148 (reprint);
doi: 10.1103/PhysRevD.23.347.

\bibitem{lindeinf} A.~ D.~ Linde;
"\textit{A new inflationary universe scenario: A possible solution of the horizon, flatness, homogeneity, isotropy and primordial monopole problems }";
Physics Letters B \textit{108} (1982) 6, 389-393;
doi: 10.1016/0370-2693(82)91219-9.

\bibitem{starop1} A.~ A.~ Starobinsky;
"\textit{Spectrum of relic gravitational radiation and the early state of the universe}";
JETP Lett. \textbf{30} (1979) 682-685, Pisma Zh.Eksp.Teor.Fiz. \textbf{30} (1979) 719-723.

\bibitem{starop2}A.~ A.~ Starobinsky;
"\textit{A New Type of Isotropic Cosmological Models Without Singularity}";
Phys. Lett. B \textit{91} (1980) 99-102, Adv. Ser. Astrophys. Cosmol. \textit{3} (1987) 130-133;
doi: 10.1016/0370-2693(80)90670-X.


\bibitem{lazarides} G.~ Lazarides, Q.~ Shafi and T.~ F.~ Walsh;
"\textit{Superheavy Magnetic Monopole Hunt}";
Phys. Lett. B \textit{100} (1981), 21-24;
doi: 10.1016/0370-2693(81)90277-X.

\bibitem{baumann} D.~ Baumann;
"\textit{Cosmology}";
Lecture notes.


\bibitem{sakharov} A.~ D.~ Sakharov;
"\textit{Violation of CP Invariance, C asymmetry, and baryon asymmetry of the universe}";
Pisma Zh. Eksp. Teor. Fiz. \textbf{5} (1967) 32-35, JETP Lett. \textbf{5} (1967) 24-27, Sov. Phys. Usp. \textbf{34} (1991) 5, 392-393, Usp. Fiz. Nauk \textbf{161} (1991) 5, 61-64;
doi: 10.1070/PU1991v034n05ABEH002497.

\bibitem{clineb} J.~ M.~ Cline;
"\textit{Baryogenesis}";
Lecture notes from Les Houches Summer School, Session 86: Particle Physics and Cosmology: the Fabric of Spacetime, 7-11 Aug. 2006;
e-print: hep-ph/0609145v3.

\bibitem{buchbinder} I.~ L.~ Buchbinder, S.~ D.~ Odintsov and I.~ L.~ Shapiro;
"\textit{Effective Action in Quantum Gravity}";
CRC Press, 1992.

\bibitem{birrelldavies} N.~ D.~ Birrell and P.~ C.~ W.~ Davies;
"\textit{Quantum Fields in Curved Space}";
Cambridge University Press; 1982.



\bibitem{gravitation} C.~ W.~ Misner, K.~ S.~ Thorne and J.~ A.~ Wheeler;
"\textit{Gravitation}"; W. H Freeman and Company (1973).

\bibitem{kaiser} D.~ I.~ Kaiser;
"\textit{Conformal Transformations with Multiple Scalar Fields}";
Phys. Rev. D \textbf{81} (2010), 084044;
doi: 10.1103/PhysRevD.81.084044;
[e-print: 1003.1159 [gr-qc]].

\bibitem{bellido} J.~ Garcia-Bellido, D.~ G.~ Figueroa and J.~ Rubio;
"\textit{Preheating in the Standard Model with the Higgs-Inflaton coupled to gravity}";
Phys. Rev. D \textbf{79} (2009), 063531;
doi: 10.1103/PhysRevD.79.063531;
[e-print: 0812.4624 [hep-ph]].

\bibitem{Planck18} Planck Collaboration;
''\textit{Planck 2018 results. X. Constraints on inflation}'',
A\&A \textbf{641}, A10 (2020);
doi: 10.1051/0004-6361/201833887;
[e-print: 1807.06211 [astro-ph.CO]].


\bibitem{bicep} \textbf{BICEP/Keck Collaboration}, P.~A.~R.~ Ade et al;
"\textit{BICEP / Keck XIII: Improved Constraints on Primordial Gravitational Waves using Planck, WMAP, and BICEP/Keck Observations through the 2018 Observing Season}";
Phys. Rev. Lett. \textbf{127} (2021), 151301;
doi: 10.1103/PhysRevLett.127.151301;
[e-print: 2110.00483 [astro-ph.CO]].

\bibitem{cmbobserve} J.~ Martin, C.~ Ringeval and V.~ Vennin;
"\textit{How Well Can Future CMB Missions Constrain Cosmic Inflation?}";
JCAP \textbf{10} (2014), 038;
doi: 10.1088/1475-7516/2014/10/038;
[e-print: 1407.4034 [astro-ph.CO]].

\bibitem{litebird} M.~ Hazumi et al.; 
"\textit{LiteBIRD: A Satellite for the Studies of B-Mode Polarization and Inflation from Cosmic Back-ground Radiation Detection}”; 
J. Low. Temp. Phys. \textbf{194}, 5-6, 443 (2019); 
doi:10.1007/s10909-019-02150-5. \textbf{LiteBIRD Collaboration};
"\textit{Probing Cosmic Inflation with the LiteBIRD Cosmic Microwave Background Polarization Survey}";
e-print: 2202.02773 [astro-ph.IM].


\bibitem{maggiore} M.~ Maggiore;
"\textit{A Modern Introduction to Quantum Field Theory}";
Oxford Master Series in Statistical, Computational, and Theoretical Physics \textit{12} (2005).

\bibitem{lancasterblundell} T.~ Lancaster and S.~ J.~ Blundell;
\textit{Quantum Field Theory for the Gifted Amateur};
Oxford University Press (2014).

\bibitem{partialwave} L.~ Di Luzio, J.~ F.~ Kamenik and M.~ Nardecchia;
"\textit{Implications of perturbative unitarity for scalar di-boson resonance searches at LHC}";
Eur. Phys. J. C \textbf{77} (2017) 1, 30;
doi: 10.1140/epjc/s10052-017-4594-2;
[e-print: 1604.05746 [hep-ph]].


\bibitem{selfheal} U.~ Aydemir, M.~ M.~ Anber and J.~ F.~ Donoghue;
"\textit{Self-healing of unitarity in effective field theories and the onset of new physics}";
Phys. Rev. D \textbf{86} (2012), 014025;
doi: 10.1103/PhysRevD.86.014025;
[e-print: 1203.5153 [hep-ph]].

\bibitem{2loop} F.~ Bezrukov and M.~ Shaposhnikov;
"\textit{Standard Model Higgs boson mass from inflation: two loop analysis}";
JHEP \textbf{07} (2009), 089;
doi: 10.1088/1126-6708/2009/07/089;
[e-print: 0904.1537 [hep-ph]].

\bibitem{natjr} R.~ N.~ Lerner and J.~ McDonald;
"\textit{Higgs Inflation and Naturalness}";
JCAP \textit{04} (2010), 015;
doi: 10.1088/1475-7516/2010/04/015;
[e-print: 0912.5463 [hep-ph]].

\bibitem{scaleofqg} T.~ Han and S.~ Willenbrock;
"\textit{The Scale of Quantum Gravity}";
Phys. Lett. B \textbf{616} (2005), 215-220;
doi: 10.1016/j.physletb.2005.04.040;
[e-print: hep-ph/0404182].

\bibitem{cw} S.~ R.~ Coleman and E.~ J.~ Weinberg;
"\textit{Radiative Corrections as the Origin of Spontaneous Symmetry Breaking}";
Phys. Rev. D \textbf{7} (1973), 1888-1910;
doi: 10.1103/PhysRevD.7.1888.
E.~ J.~ Weinberg; "\textit{Radiative corrections as the origin of spontaneous symmetry breaking}"; PhD Thesis, e-Print: hep-th/0507214 [hep-th].

\bibitem{msher} M.~ Sher;
"\textit{Electroweak Higgs Potentials and Vacuum Stability}";
Phys. Rept. \textbf{179} (1989), 273-418;
doi: 10.1016/0370-1573(89)90061-6.
"\textit{Constraints on Higgs Boson Properties from the Higgs Potential}";
Perspectives on Higgs Physics, 44-78;
Advanced Series on Directions in High Energy Physics,
World Scientific (1993);
doi: 10.1142/9789814440783\_0002.

\bibitem{colemanqft} S.~ Coleman (Edited and typeset by Yuan-Sen Ting and Bryan Gin-ge Chen from scans of the handwritten
notes of Brian Hill. Numerous additional edits by Richard Sohn.)
"\textit{Notes from Sidney Coleman’s Physics 253a}";
[e-print: 1110.5013 [physics.ed-ph]].

\bibitem{coleman85} S.~Coleman;
"\textit{Q-balls}";
Nucl. Phys. B \textbf{262} (1985), 263-283
doi:10.1016/0550-3213(85)90286-X





\bibitem{linde83} A.~ D.~ Linde;
"\textit{Chaotic Inflation}'';
Phys.Lett.B \textbf{129} (1983), 177-181;
doi: 10.1016/0370-2693(83)90837-7.

\bibitem{linde07} A.~ D.~ Linde;
"\textit{Inflationary Cosmology}'';
Lect. Notes Phys. \textbf{738} (2008), 1-54;
doi: 10.1007/978-3-540-74353-8\_1;
Contribution to 22nd IAP Colloquium on Inflation + 25: The First 25 Years of Inflationary Cosmology;
[e-Print: 0705.0164 [hep-th]].


\bibitem{staro80} A.~ A.~ Starobinsky;
"\textit{A new type of isotropic cosmological models without singularity}";
Phys.Lett.B \textbf{91} (1980), 99-102;
Adv.Ser.Astrophys.Cosmol. \textbf{3} (1987), 130-133;
doi: 10.1016/0370-2693(80)90670-X;

\bibitem{kehagias14} A.~ Kehagias, A.~ M.~ Dizgah and A.~ Riotto;
"\textit{Remarks on the Starobinsky model of inflation and its descendants}'';
Phys. Rev. D \textbf{89} (2014) 4, 043527;
doi: 10.1103/PhysRevD.89.043527;
[e-Print: 1312.1155 [hep-th]].


\bibitem{enckell18} V.~ M.~ Enckell, K.~Enqvist, S.~Rasanen and L.~P.~Wahlman;
"\textit{Inflation with $R^{2}$ term in the Palatini formalism}'';
JCAP \textbf{02} (2019), 022;
doi: 10.1088/1475-7516/2019/02/022;
[e-print: 1810.05536 [gr-qc]].

\bibitem{antoniadis18} I.~ Antoniadis, A.~ Karam, A.~ Lykkas, T.~ Pappas, and K.~ Tamvakis;
"\textit{Palatini inflation in models with an $R^{2}$ term}'';
JCAP \textbf{11} (2018), 028;
doi: 10.1088/1475-7516/2018/11/028;
[e-print: 1810.10418 [gr-qc]].

\bibitem{enckell182} V.~ M.~ Enckell, K.~Enqvist, S.~Rasanen and L.~P.~Wahlman;
"\textit{Higgs-$R^{2}$ inflation - full slow-roll study at tree-level}'';
JCAP \textbf{01} (2020) 041;
doi: 10.1088/1475-7516/2020/01/041;
[e-print: 1812.08754 [astro-ph.CO]].

\bibitem{lykkas18} I.~ Antoniadis, A.~ Karam, A.~ Lykkas, T.~ Pappas, and K.~ Tamvakis; 
"\textit{Rescuing Quartic and Natural Inflation in the Palatini Formalism}'';
JCAP \textbf{03} (2019), 005;
doi: 10.1088/1475-7516/2019/03/005;
[e-print:1812.00847 [gr-qc]].

\bibitem{tenkanen19} T.~ Tenkanen; 
"\textit{Minimal Higgs inflation with an $R^{2}$ term in Palatini gravity}'';
Phys. Rev. D \textbf{99} (2019), 063528;
doi: 10.1103/physrevd.99.063528;
[e-print:1901.01794 [astro-ph.CO]].

\bibitem{transplanck} T.~ Tenkanen; 
"\textit{Trans-Planckian censorship, inflation, and dark matter}''; 
Phys. Rev. D \textbf{101} (2020), 063517;
doi: 10.1103/PhysRevD.101.063517;
[e-print:1910.00521 [astro-ph.CO]].


\bibitem{gialamas19} I.~ D.~ Gialamas and A.~ Lahanas; 
"\textit{Reheating in $R^{2}$ Palatini inflationary models}'';
Phys. Rev. D \textbf{101} (2020), 084007;
doi: 10.1103/physrevd.101.084007; 
[e-print: 1911.11513 [gr-qc]].

\bibitem{tenkanengrav} T.~ Tenkanen; 
"\textit{Tracing the high energy theory of gravity: an introduction to Palatini inflation}'';
Gen. Rel. Grav. \textbf{52} (2020), no. 4 33;
doi: 10.1007/s10714-020-02682-2; 
[e-print: 2001.10135 [astro-ph.CO]].

\bibitem{tomberg20} T.~ Tenkanen and E.~ Tomberg;
"\textit{Initial conditions for plateau inflation: a case study}'';
JCAP \textbf{04} (2020), 050;
doi: 10.1088/1475-7516/2020/04/050;
[e-print:2002.02420 [astro-ph.CO]].

\bibitem{us1} A.~ Lloyd-Stubbs and J.~ McDonald; 
"\textit{Sub-Planckian $\phi^{2}$ inflation in the Palatini formulation of gravity with an $R^{2}$ term}'';
Phys. Rev. D \textbf{101} (2020), 123515;
doi: 10.1103/PhysRevD.101.123515;
[e-print: 2002.08324 [hep-ph]].

\bibitem{antoniadis20} I.~ Antoniadis, A.~ Lykkas, and K.~ Tamvakis; 
"\textit{Constant-roll in the Palatini-$R^{2}$ models}'';
JCAP \textbf{04} (2020), no. 04 033;
doi: 10.1088/1475-7516/2020/04/033; 
[e-print: 2002.12681 [gr-qc]].

\bibitem{daspanda} N.~ Das and S.~ Panda; 
"\textit{Inflation in $f(R,h)$ theory formulated in the Palatini formalism}'';
JCAP \textbf{05} (2021), 019;
doi: 10.1088/1475-7516/2021/05/019;
[e-print: 2005.14054 [gr-qc]].

\bibitem{gialamas20} I.~ D.~ Gialamas, A.~ Karam, and A.~ Racioppi; 
"\textit{Dynamically induced Planck scale and inflation in the Palatini formulation}'';
JCAP \textbf{11} (2020), 014;
doi: 10.1088/1475-7516/2020/11/014;
[e-print: 2006.09124 [gr-qc]].

\bibitem{alinde} R.~ Kallosh and A.~ Linde;
"\textit{Planck, LHC, and $\alpha$-attractors}";
Phys. Rev. D \textbf{91} (2015), 083528;
doi: 10.1103/PhysRevD.91.083528;
[e-print: 1502.07733 [astro-ph.CO]].


\bibitem{infleff} D.~ Baumann and L.~ McAllister;
"\textit{Inflation and String Theory}'';
\textit{Inflation and String Theory};
Chapter 2,
Cambridge University press, Cambridge, England, 2015;
e-print: 1404.2601 [hep-th]].

\bibitem{convenhiggs} J.~ Rubio;
"\textit{Higgs Inflation}";
Front. Astron. Space Sci. \textbf{5} (2019), 50;
doi: 10.3389/fspas.2018.00050;
[e-print: 1807.02376 [hep-ph]].

\bibitem{traschen} J.~ H.~ Traschen and R.~ H.~ Brandenberger;
"\textit{Particle Production During Out-of-equilibrium Phase Transitions}'';
Phys. Rev. D \textbf{42} (1990), 2491-2504;
doi: 10.1103/PhysRevD.42.2491.

\bibitem{big3} L.~ Kofman, A.~ D.~ Linde, and A.~ A.~ Starobinsky;
"\textit{Reheating after inflation}''; 
Phys. Rev. Lett. \textbf{73} (1994), 3195-3198;
doi: 10.1103/PhysRevLett.73.3195;
[e-print: hep-th/9405187].

\bibitem{john01} J.~ McDonald;
"\textit{Inflaton Condensate Fragmentation in Hybrid Inflation Models}'';
Physical Review D \textbf{66} (2002), 043525;
doi: 10.1103/physrevd.66.043525;
[e-print: hep-ph/0105235].

\bibitem{amin11} M.~ A.~ Amin et al;
"\textit{Oscillons after Inflation}'';
Phys. Rev. Lett. \textbf{108} (2012), 241302;
doi: 10.1103/PhysRevLett.108.241302;
[arXiv: 1106.3335 [astro-ph.CO]].

\bibitem{lozanov17} K.~ D.~ Lozanov and M.~ A.~ Amin;
"\textit{Self-resonance after inflation: oscillons, transients and radiation domination}'';
Phys. Rev. D \textbf{97} (2018) 2, 023533;
doi: 10.1103/PhysRevD.97.023533;
[e-print: 1710.06851 [astro-ph.CO]].


\bibitem{osclife} J.~ Oll\'{e}, O.~ Pujolas and F.~ Rompineve;
"\textit{Recipes for Oscillon Longevity}";
JCAP \textbf{09} (2021), 015;
doi: 10.1088/1475-7516/2021/09/015;
[e-print: 2012.13409 [hep-ph]].

\bibitem{jinsu17} J.~ Kim and J.~ McDonald;
"\textit{Inflaton condensate fragmentation: Analytical conditions and application to $\alpha$-attractor models}'';
Phys. Rev. D \textbf{95} (2017), 123537;
doi: 10.1103/physrevd.95.123537;
[e-print: 1702.08777 [astro-ph]].


\bibitem{jinsu22} J.~ Kim and J.~ McDonald;
"\textit{General Analytical Conditions for Inflaton Fragmentation: Quick and Easy Tests for its Occurrence}'';
e-Print: 2111.12474 [astro-ph.CO].

\bibitem{tomberg211} A.~ Karam, E.~ Tomberg and H.~ Veermäe;
"\textit{Tachyonic preheating in Palatini $R^{2}$ inflation}'';
JCAP \textbf{06} (2021), 023;
doi: 10.1088/1475-7516/2021/06/023;
[e-Print: 2102.02712 [astro-ph.CO]].

\bibitem{higgsann} J.~ McDonald;
"\textit{Gauge Singlet Scalars as Cold Dark Matter}";
Phys. Rev. D \textbf{50} (1994), 3637-3649;
doi: 10.1103/PhysRevD.50.3637;
[e-print: hep-ph/0702143].

\bibitem{heisig} J.~ Heisig, M.~ Kr\"amer, E.~ Madge and A.~ M\"uck;
``\textit{Probing Higgs-portal dark matter with vector-boson fusion}'';
JHEP \textbf{03} (2020), 183;
doi:10.1007/JHEP03(2020)183;
[arXiv:1912.08472 [hep-ph]].


\bibitem{nrdecay} D.~ Suematsu;
"\textit{Leptogenesis in a neutrino mass model coupled with inflaton}";
Phys. Lett. B \textbf{760} (2016), 538-543;
doi: 10.1016/j.physletb.2016.07.048;
[e-print: 1606.07884 [hep-ph]].

\bibitem{bekov} S.~ Bekov, K.~ Myrzakulov, R.~ Myrzakulov and D.~ Sáez-Chillón Gómez;
"\textit{General Slow-Roll Inflation in f (R) Gravity under the Palatini Approach}";
Symmetry \textbf{12} (2020) 12, 1958;
doi: 10.3390/sym12121958;
[e-print: 2010.12360 [gr-qc]].


\bibitem{amin10} M.~ A.~ Amin, R.~ Easther and H.~ Finkel;
"\textit{Inflaton Fragmentation and Oscillon Formation in Three Dimensions}";
JCAP \textbf{12} (2010), 001;
doi: 10.1088/1475-7516/2010/12/001;
[e-print: 1009.2505 [astro-ph.CO]].

\bibitem{amin102} M.~ A.~ Amin
"\textit{Inflaton fragmentation: Emergence of pseudo-stable inflaton lumps (oscillons) after inflation}'';
e-print: 1006.3075v2 [astro-ph.CO], 13th September 2010.

\bibitem{kasuya03} S.~ Kasuya, M.~ Kawasaki and F.~ Takahashi;
"\textit{I-balls}'';
Phys. Lett. B. \textbf{559} (2003), 99;
doi: 10.1016/S0370-2693(03)00344-7;
[e-print: hep-ph/0209358].

\bibitem{hertzberg10} M.~ P.~ Hertzberg;
"\textit{Quantum Radiation of Oscillons}'';
Phys. Rev. D \textbf{82} (2010), 045022;
doi: 10.1103/PhysRevD.82.045022;
[e-print: 1003.3459 [hep-th]].


\bibitem{cyncynates21} D.~ Cyncynates and T.~ Giurgica-Tiron;
"\textit{The Structure of the Oscillon: The dynamics of attractive self-interaction}'';
e-print: 2104.02069v2 [hep-ph] (10th May 2021).


\bibitem{antusch17} S.~ Antusch, F.~ Cefala and S.~ Orani;
"\textit{Gravitational waves from oscillons after inflation}'';
Phys. Rev. Lett. \textbf{118} (2017) 1, 011303, Phys. Rev. Lett. \textbf{120} (2018) 21, 219901 (erratum);
doi: 10.1103/PhysRevLett.118.011303, 10.1103/PhysRevLett.120.219901 (erratum);
[e-print: 1607.01314 [astro-ph.CO]].

\bibitem{lozanov19} K.~ D.~ Lozanov and M.~ A.~ Amin;
"\textit{Gravitational perturbations from oscillons and transients after inflation}'';
Phys. Rev. D \textbf{99} (2019) 12, 123504;
doi: 10.1103/PhysRevD.99.123504;
[e-print: 1902.06736 [astro-ph.CO]].



\bibitem{khlebnikov97} S.~ Y.~ Khlebnikov and I.~ I.~ Tkachev;
"\textit{Relic gravitational waves produced after preheating}'';
Phys. Rev. D \textbf{56} (1997), 653-660;
doi: 10.1103/PhysRevD.56.653;
[e-print: hep-ph/9701423 [hep-ph]].

\bibitem{easther06} R.~ Easther, J.~ T.~ Giblin, Jr. and E.~ A.~ Lim
"\textit{Gravitational Wave Production At The End Of Inflation}'';
Phys. Rev. Lett. \textbf{99} (2007), 221301;
doi: 10.1103/PhysRevLett.99.221301;
[e-print: astro-ph/0612294 [astro-ph]].

\bibitem{amin14} M.~ A.~ Amin et. al;
"\textit{Nonperturbative Dynamics Of Reheating After Inflation: A Review}'';
Int. J. Mod. Phys. D \textbf{24} (2015), 1530003;
doi: 10.1142/S0218271815300037;
[e-print: 1410.3808v3 [hep-ph]]

\bibitem{assadullahi09} H.~ Assadullahi and D.~ Wands;
"\textit{Gravitational waves from an early matter era}'';
Phys. Rev. D \textbf{79} (2009), 083511;
doi: 10.1103/PhysRevD.79.083511;
[e-print: 0901.0989 [astro-ph.CO]].

\bibitem{assadullahi10} H.~ Assadullahi and D.~ Wands;
"\textit{Constraints on primordial density perturbations from induced gravitational waves}'';
Phys. Rev. D \textbf{81} (2010), 023527;
doi: 10.1103/PhysRevD.81.023527;
[e-print: 0907.4073 [astro-ph.CO]].

\bibitem{jedamzik10} K.~ Jedamzik, M.~ Lemoine and J.~ Martin;
"\textit{Generation of gravitational waves during early structure formation between cosmic inflation and reheating}'';
JCAP \textbf{04} (2010) 021;
doi: 10.1088/1475-7516/2010/04/021;
[e-print: 1002.3278 [astro-ph.CO]].

\bibitem{jedamzik102}  K.~ Jedamzik, M.~ Lemoine and J.~ Martin;
"\textit{Collapse of Small-Scale Density Perturbations during Preheating in Single Field Inflation}'';
JCAP \textbf{09} (2010), 034;
doi: 10.1088/1475-7516/2010/09/034;
[e-print: 1002.3039 [astro-ph.CO]].

\bibitem{alabidi13} L.~ Alabidi et al;
"\textit{Observable induced gravitational waves from an early matter phase}'',
JCAP \textbf{05} (2013) 033;
doi: 10.1088/1475-7516/2013/05/033;
[e-print: 1303.4519 [astro-ph.CO]].


\bibitem{aggarwal18} N.~ Aggarwal et. al;
"\textit{Challenges and Opportunities of Gravitational Wave Searches at MHz to GHz Frequencies}'';
Summary of the \textit{Challenges and opportunities of high-frequency gravitational wave detection} workshop at ICTP Trieste, October 2019; 
e-print:2011.12414v1 [gr-qc].




\bibitem{affleckdine} I.~ Affleck and M.~ Dine;
"\textit{A New Mechanism for Baryogenesis}'';
Nucl. Phys. B \textbf{249} (1985), 361-380;
doi: 10.1016/0550-3213(85)90021-5.

\bibitem{dine952} M.~ Dine, L.~ Randall and S.~ D.~ Thomas;
"\textit{Supersymmetry breaking in the early universe}'';
Phys. Rev. Lett. \textbf{75} (1995), 398-401;
doi: 10.1103/PhysRevLett.75.398;
[e-print: hep-ph/9503303].

\bibitem{dine95} M.~ Dine, L.~ Randall and S.~ D.~ Thomas;
"\textit{Baryogenesis from flat directions of the supersymmetric standard model}'';
Nucl. Phys. B \textbf{458} (1996), 291-326;
doi: 10.1016/0550-3213(95)00538-2;
[e-print: hep-ph/9507453].

\bibitem{enqvist98} K.~ Enqvist and J.~ McDonald;
"\textit{Q-balls and baryogenesis in the MSSM}'';
Phys. Lett. B \textbf{425} (1998), 309-321
doi: 10.1016/S0370-2693(98)00271-8
[e-print: hep-ph/9711514]

\bibitem{john99} K.~ Enqvist and J.~ McDonald;
"\textit{The dynamics of Affleck-Dine condensate collapse}'';
Nucl. Phys. B \textbf{570} (2000) 407-422, Nucl. Phys. B \textbf{582} (2000) 763-763 (erratum);
doi: 10.1016/S0550-3213(99)00776-2, 10.1016/S0550-3213(00)00304-7 (erratum);
[e-print: hep-ph/9908316 [hep-ph]].

\bibitem{kasuya002} S.~ Kasuya and M.~ Kawasaki;
"\textit{Q-ball formation in the gravity-mediated SUSY breaking scenario}'';
Phys. Rev. D \textbf{62} (2000), 023512
doi: 10.1103/PhysRevD.62.023512
[e-print: hep-ph/0002285]

\bibitem{enqvist00} K.~ Enqvist et. al;
"\textit{Numerical simulations of fragmentation of the Affleck-Dine condensate}'';
Phys. Rev. D \textbf{63} (2001), 083501;
doi: 10.1103/PhysRevD.63.083501;
[e-print: hep-ph/0011134 [hep-ph]].

\bibitem{kasuya012} S.~ Kasuya and M.~ Kawasaki;
"\textit{Q ball formation: Obstacle to Affleck-Dine baryogenesis in the gauge mediated SUSY breaking?}'';
Phys. Rev. D \textbf{64} (2001), 123515;
doi: 10.1103/PhysRevD.64.123515;
[e-print: hep-ph/0106119 [hep-ph]].

\bibitem{kasuya001} S.~ Kasuya and M.~ Kawasaki;
"\textit{Q-ball formation through the Affleck-Dine mechanism}'';
Phys. Rev. D \textbf{61} (2000), 041301(R)
doi: 10.1103/PhysRevD.61.041301
[e-print: hep-ph/9909509]

\bibitem{fujii01} M.~ Fujii and K.~ Hamaguchi;
"\textit{Higgsino and Wino Dark Matter from Q-ball Decay in Affleck-Dine Baryogenesis}'';
Phys. Lett. B \textbf{525} (2002), 143-149;
doi: 10.1016/S0370-2693(01)01412-5;
[e-print: hep-ph/0110072 [hep-ph]].

\bibitem{kasuyaq} S.~ Kasuya;
"\textit{Affleck-Dine baryogenesis and the Q ball dark matter in the gauge mediated SUSY breaking}'';
Contribution to: COSMO 2001;
January 2002;
e-print: hep-ph/0201185 [hep-ph].

\bibitem{tsumagari09} M.~ Tsumagari;
"\textit{Affleck-Dine dynamics, Q-ball formation and thermalisation}'';
Phys. Rev. D \textbf{80} (2009), 085010;
doi: 10.1103/PhysRevD.80.085010;
[e-print: 0907.4197 [hep-th]].

\bibitem{doddato11} F.~ Doddato and J.~McDonald;
"\textit{Affleck-Dine Baryogenesis, Condensate Fragmentation and Gravitino Dark Matter in Gauge-Mediation with a Large Messenger Mass}'';
JCAP \textbf{06} (2011), 008;
doi: 10.1088/1475-7516/2011/06/008;
[e-print: 1101.5328 [hep-ph]].

\bibitem{doddato113} F.~ Doddato and J.~McDonald;
"\textit{New Q-ball Solutions in Gauge-Mediation, Affleck-Dine Baryogenesis and Gravitino Dark Matter}'';
JCAP \textbf{06} (2012), 031;
doi: 10.1088/1475-7516/2012/06/031;
[e-print: 1111.2305 [hep-ph]].

\bibitem{charng} Y.~ Y.~ Charng, D.~-S.~ Lee, C.~ N.~ Leung and K.~-W.~ Ng;
"\textit{Affleck-Dine Baryogenesis, Split Supersymmetry, and Inflation}'';
Phys. Rev. D \textbf{80} (2009), 063519;
doi: 10.1103/PhysRevD.80.063519;
[e-print: 0802.1328 [hep-ph]].

\bibitem{karouby} M.~ P.~ Hertzberg and J.~ Karouby;
"\textit{Baryogenesis from the Inflaton Field}'';
Phys. Lett. B \textbf{737} (2014), 34-38;
doi: 10.1016/j.physletb.2014.08.021;
[e-print: 1309.0007 [hep-ph]].
"\textit{Generating the Observed Baryon Asymmetry from the Inflaton Field}'';
Phys. Rev. D \textbf{89} (2014) 6, 063523;
doi: 10.1103/PhysRevD.89.063523;
[e-print: 1309.0010 [hep-ph]].

\bibitem{takeda} N.~ Takeda;
"\textit{Inflatonic baryogenesis with large tensor mode}'';
Phys. Lett. B \textbf{746} (2015), 368-371;
doi: 10.1016/j.physletb.2015.05.039;
[e-print: 1405.1959 [astro-ph.CO]].


\bibitem{lin} C.~ M.~ Lin and K.~ Kohri;
"\textit{Inflaton as the Affleck-Dine Baryogenesis Field in Hilltop Supernatural Inflation}'';
Phys. Rev. D \textbf{102} (2020) 4, 043511;
doi: 10.1103/PhysRevD.102.043511;
[e-print: 2003.13963 [hep-ph]].

\bibitem{babichev1} E.~ Babichev, D.~ Gorbunov and S.~ Ramazanov;
"\textit{Dark Matter and Baryon Asymmetry from the very Dawn of Universe}";
Phys. Rev. D \textbf{97} (2018) 12, 123543;
doi: 10.1103/PhysRevD.97.123543;
[e-print: 1805.05904 [astro-ph.CO]].


\bibitem{babichev2} E.~ Babichev, D.~ Gorbunov and S.~ Ramazanov;
"\textit{Affleck–Dine baryogenesis via mass splitting}'';
Phys. Lett. B \textbf{792} (2019), 228-232;
doi: 10.1016/j.physletb.2019.03.046;
[e-print: 1809.08108 [astro-ph.CO]].


\bibitem{cline1} J.~ M.~ Cline, M.~ Puel and T.~ Toma;
"\textit{Affleck-Dine inflation}'';
Phys. Rev. D \textbf{101} (2020) 4, 043014;
doi: 10.1103/PhysRevD.101.043014;
[e-print: 1909.12300 [hep-ph]].

\bibitem{cline2} J.~ M.~ Cline, M.~ Puel and T.~ Toma;
\textit{A little theory of everything, with heavy neutral leptons}'';
JHEP \textbf{05} (2020), 039;
doi: 10.1007/JHEP05(2020)039;
[e-print: 2001.11505 [hep-ph]].



\bibitem{sphalerons} V.~ A.~ Kuzmin, V.~ A.~ Rubakov and M.~ E.~ Shaposhnikov;
"\textit{On the Anomalous Electroweak Baryon Number Nonconservation in the Early Universe}'';
Phys. Lett. B \textbf{155} (1985), 36;
doi: 10.1016/0370-2693(85)91028-7.

\bibitem{maybe} M.~ Fukugita and T.~ Yanagida;
"\textit{Baryogenesis Without Grand Unification}'';
Phys. Lett. B \textbf{174} (1986), 45-47;
doi: 10.1016/0370-2693(86)91126-3.

\bibitem{usad} A.~ Lloyd-Stubbs and J.~ McDonald;
"\textit{A Minimal Approach to Baryogenesis via Affleck-Dine and Inflaton Mass Terms}";
Phys. Rev. D \textbf{103} (2021), 123514;
doi: 10.1103/PhysRevD.103.123514;
[e-print: 2008.04339 [hep-ph]].


\bibitem{mo1} R.~ N.~ Mohapatra and N.~ Okada;
"\textit{Affleck-Dine baryogenesis with observable neutron-antineutron oscillation}";
Phys. Rev. D \textbf{104} (2021) 5, 055030;
doi: 10.1103/PhysRevD.104.055030;
[e-print: 2107.01514 [hep-ph]].

\bibitem{mo2}R.~ N.~ Mohapatra and N.~ Okada;
"\textit{Unified model for inflation, pseudo-Goldstone dark matter, neutrino mass, and baryogenesis}";
Phys. Rev. D \textbf{105} (2022) 3, 035024;
doi: 10.1103/PhysRevD.105.035024;
[e-print: 2112.02069 [hep-ph]].


\bibitem{mo3}R.~ N.~ Mohapatra and N.~ Okada;
"\textit{Neutrino mass from Affleck-Dine leptogenesis and WIMP dark matter}";
JHEP \textbf{03} (2022), 092;
doi: 10.1007/JHEP03(2022)092;
[e-print: 2201.06151 [hep-ph]].

\bibitem{mo4}R.~ N.~ Mohapatra and N.~ Okada;
"\textit{Affleck-Dine Leptogenesis with One Loop Neutrino Mass and strong CP}";
[e-print: 2207.10619 [hep-ph]].


\bibitem{turner} M.~ S.~ Turner;
"\textit{Coherent scalar-field oscillations in an expanding universe}'';
Phys. Rev. D \textbf{28} (1983), 1243;
doi: 10.1103/PhysRevD.28.1243.


\bibitem{kofmanre}  L.~ Kofman, A.~ D.~ Linde and A.~ A.~ Starobinsky;
"\textit{To-wards the theory of reheating after inflation}';'
Phys. Rev. D \textbf{56} (1997), 3258;
doi: 10.1103/PhysRevD.56.3258;
[e-print: hep-ph/9704452].

\bibitem{nakayama} K.~ Nakayama, S.~ Saito, Y.~ Suwa and J.~ Yokoyama;
"\textit{Probing reheating temperature of the universe with gravitational wave background}'';
JCAP \textbf{06} (2008), 020;
doi: 10.1088/1475-7516/2008/06/020;
[e-print: 0804.1827 [astro-ph]].

\bibitem{infnr} T.~ Fukuyama, T.~ Kikuchi and W.~ Naylor;
"\textit{Reheating temperature and the right-handed neutrino mass}";
Phys Lett B \textbf{632}, 2–3 (2006), 349-351;
doi: 10.1016/j.physletb.2005.10.066.


\bibitem{iso1} K.~ Enqvist and J.~ McDonald;
"\textit{Observable isocurvature fluctuations from the Affleck-Dine condensate}";
Phys. Rev. Lett. \textbf{83} (1999), 2510-2513;
doi: 10.1103/PhysRevLett.83.2510;
[e-print: hep-ph/9811412].

\bibitem{iso2} K.~ Koyama and J.~ Soda;
"\textit{Baryon Isocurvature Perturbation in the Affleck-Dine Baryogenesis mechanism}";
Phys. Rev. Lett. \textbf{82} (1999), 2632-2635;
doi: 10.1103/PhysRevLett.82.2632;
[e-print: astro-ph/9810006].

\bibitem{iso3} K.~ Enqvist and J.~ McDonald;
"\textit{Inflationary Affleck-Dine scalar dynamics and isocurvature perturbations}";
Phys. Rev. D \textbf{62} (2000), 043502;
doi: 10.1103/PhysRevD.62.043502;
[e-print: hep-ph/9912478].

\bibitem{iso4} M.~ Kawasaki and F.~ Takahashi;
"\textit{Adiabatic and isocurvature fluctuations of Affleck-Dine field in D-term inflation model}";
Phys. Lett. B \textbf{516} (2001), 388-394;
doi: 10.1016/S0370-2693(01)00957-1;
[e-print: hep-ph/0105134].

\bibitem{iso5} S.~ Kasuya, M.~ Kawasaki and F.~ Takahashi;
"\textit{Isocurvature fluctuations in Affleck-Dine mechanism and constraints on inflation models}"; 
JCAP \textbf{10} (2008), 017;
doi: 10.1088/1475-7516/2008/10/017;
[e-print: 0805.4245 [hep-ph]].

\bibitem{davidson} S.~ Davidson;
"\textit{Axions: Bose Einstein condensate or classical field?}'';
Astropart. Phys. \textbf{65} (2015), 101-107;
doi: 10.1016/j.astropartphys.2014.12.007;
[e-print: 1405.1139 [hep-ph]].

\bibitem{lozanovre} K.~ D.~ Lozanov;
 ' ' \textit{Lectures on Reheating After Inflation}'',
e-print: 1907.04402 [astro-ph.CO].



 \bibitem{rosen} G.~ Rosen;
"\textit{Particlelike Solutions to Nonlinear Complex Scalar Field Theories with Positive-Definite Energy Densities}";
Journal of Mathematical Physics. 9 (7): 996–998. 

\bibitem{friedberg76} R.~ Friedberg, T.~ D.~ Lee and A.~ Sirlin;
"\textit{Class of scalar-field soliton solutions in three space dimensions}'';
Phys. Rev. D \textbf{13} (1976), 2739
doi: 10.1103/PhysRevD.13.2739


\bibitem{lee92} T.~ D.~ Lee and Y.~ Pang;
"\textit{Nontopological solitons}'';
Phys. Rept. \textbf{221} (1992), 251-350
doi: 10.1016/0370-1573(92)90064-7


\bibitem{kusenko97} A.~ Kusenko
"\textit{Solitons in the supersymmetric extensions of the standard model}'';
Phys. Lett. B \textbf{405} (1997) 108
doi: 10.1016/S0370-2693(97)00584-4
[e-print: hep-ph/9704273 [hep-ph]]


\bibitem{kusenko98} A.~ Kusenko and M.~ E.~ Shaposhnikov
"\textit{Supersymmetric Q-balls as dark matter}'';
Phys. Lett. B. \textbf{418} (1998), 46-54
doi: 10.1016/S0370-2693(97)01375-0
[e-print: hep-ph/9709492]

\bibitem{troitsky15} S.~ Troitsky;
"\textit{Supermassive dark-matter Q-balls in galactic centers?}'';
JCAP \textbf{11} (2016), 027;
doi: 10.1088/1475-7516/2016/11/027;
[arXiv: 1510.07132 [hep-ph]].

\bibitem{kusenko01} A.~ Kusenko and P.~ J.~ Steinhart;
"\textit{Q-ball candidates for self-interacting dark matter}'';
Phys. Rev. Lett. \textbf{87} (2001), 141301;
doi: 10.1103/PhysRevLett.87.141301;
[e-print: astro-ph/0106008 [astro-ph]].

\bibitem{enqvist01} K.~ Enqvist et al.;
"\textit{Constraints on Self-Interacting Q-ball Dark Matter}'';
Phys. Lett. B \textbf{526} (2002), 9-18;
doi: 10.1016/S0370-2693(01)01500-3;
[e-print: hep-ph/0111348 [hep-ph]].

\bibitem{doddato12} F.~ Doddato and J.~ M.~ McDonald;
"\textit{Dark Matter Gravitinos and Baryons via Q-ball decay in the Gauge-Mediated MSSM}'';
JCAP \textbf{07} (2013), 004;
doi: 10.1088/1475-7516/2013/07/004;
[e-print: 1211.1892 [hep-ph]].

\bibitem{kasuya152} S.~ Kasuya, E.~ Kawakami and M.~ Kawasaki;
"\textit{Axino dark matter and baryon number asymmetry production by the Q-ball decay in gauge mediation}'';
JCAP \textbf{03} (2016), 011;
doi: 10.1088/1475-7516/2016/03/011;
[e-print: 1511.05655 [hep-ph]].

\bibitem{kawasaki16}  J.~ P.~ Hong, M.~ Kawasaki and M.~ Yamada;
"\textit{Charged Q-ball Dark Matter from B and L direction}'';
 JCAP \textbf{08} (2016), 053;
doi: 10.1088/1475-7516/2016/08/053;
[e-print: 1604.04352 [hep-ph]].

\bibitem{kawasaki17} J.~ P.~ Hong and M.~ Kawasaki;
"\textit{New type of charged Q -ball dark matter in gauge mediated SUSY breaking models}'';
Phys.Rev.D \textbf{95} (2017) 12, 123532;
doi: 10.1103/PhysRevD.95.123532;
[e-print: 1702.00889 [hep-ph]].

\bibitem{shoemaker092} I.~ M.~ Shoemaker;
"\textit{Properties of Q-ball dark matter: moving away from flat directions}'';
Report appearence ASP conference Series (SnowPAC 2009 Proceedings);
e-print: 0906.1834 [hep-ph].

\bibitem{kusenko21} M.~ M.~ Flores and A.~ Kusenko;
"\textit{Primordial black holes as a dark matter candidate in theories with supersymmetry and inflation}'';
e-print: 2108.08416 [hep-ph].

\bibitem{kasuya004} S.~ Kasuya and M.~ Kawasaki;
"\textit{A New type of stable Q balls in the gauge mediated SUSY breaking}'';
 Phys. Rev. Lett. \textbf{85} (2000), 2677-2680;
doi: 10.1103/PhysRevLett.85.2677;
[e-print: hep-ph/0006128 [hep-ph]].


\bibitem{multamaki01} T.~ Multamaki;
"\textit{Excited Q balls in the MSSM with gravity mediated supersymmetry breaking}'';
Phys. Lett. B \textbf{511} (2001), 92-100;
doi: 10.1016/S0370-2693(01)00613-X;
[e-print: hep-ph/0102339 [hep-ph]].

\bibitem{kasuya012} S.~ Kasuya and M.~ Kawasaki;
"\textit{Q ball formation: Obstacle to Affleck-Dine baryogenesis in the gauge mediated SUSY breaking?}'',
Phys. Rev. D \textbf{64} (2001), 123515;
doi: 10.1103/PhysRevD.64.123515;
[e-print: hep-ph/0106119 [hep-ph]].

\bibitem{multamaki02} T.~ Multamaki and I.~ Vilja;
"\textit{Simulations of Q-Ball Formation}'';
Phys. Lett. B \textbf{535} (2002), 170-176;
doi: 10.1016/S0370-2693(02)01730-6;
[e-print: hep-ph/0203195v3]].

\bibitem{palti04} E.~ Palti, P.~ M.~ Saffin and E.~ J.~ Copeland;
"\textit{Dynamics of Q-balls in an expanding universe}'';
Phys. Rev. D \textbf{70} (2004), 083520;
doi: 10.1103/PhysRevD.70.083520;
[e-print: hep-th/0405081 [hep-th]].

\bibitem{campanelli08} L.~ Campanelli and M.~ Ruggieri;
"\textit{Supersymmetric Q-balls: A numerical study}'';
Phys. Rev. D \textbf{77} (2008), 043504;
doi: 10.1103/PhysRevD.77.043504;
[e-print: 0712.3669 [hep-th]].

\bibitem{john05} M.~ Broadhead and J.~ McDonald
"\textit{Simulations of the end of supersymmetric hybrid inflation and nontopological soliton formation}'';
Phys. Rev. D \textbf{72} (2005), 043519;
doi: 10.1103/PhysRevD.72.043519;
[e-print: hep-ph/0503081].


\bibitem{hiramatsu10} T.~ Hiramatsu, M.~ Kawasaki and F.~ Takahashi
"\textit{Numerical study of Q-ball formation in gravity mediation}'';
JCAP \textbf{06} (2010) 008;
doi: 10.1088/1475-7516/2010/06/008;
[e-print: 1003.1779 [hep-ph]];


\bibitem{kofman01} L.~ Kofman
"\textit{Tachyonic Preheating}'';
e-print: hep-ph/0107280, 26th July 2001;
Review article based on invited talk at the Conference PASCOS2001


\bibitem{felder00} G.~ Felder, J.~ Garcia-Bellido, P.~ B.~ Greene, L.~ Kofman, A.~ Linde, and I.~ Tkachev;
"\textit{Dynamics of Symmetry Breaking and Tachyonic Preheating}'';
Phys. Rev. Lett. \textbf{87} (2001), 011601;
doi: 10.1103/PhysRevLett.87.011601;
[e-print: hep-ph/0012142 ].


\bibitem{felder01} G.~ Felder, L.~ Kofman and A.~ Linde;
"\textit{Tachyonic instability and dynamics of spontaneous symmetry breaking}'';
Phys. Rev. D \textbf{64} (2001), 123517;
doi: 10.1103/physrevd.64.123517;
[e-print: hep-th/0106179 ].


\bibitem{copeland02} E.~ J.~ Copeland, S.~ Pascoli and A.~ Rajantie;
 "\textit{Dynamics of tachyonic preheating after hybrid inflation}";
Phys. Rev. D \textbf{65} (2002), 103517;
doi:10.1103/physrevd.65.103517;
[e-print: 0202031 [hep-ph]].


\bibitem{tomberg21} E.~ Tomberg and H.~ Veerm{\"a}e;
"\textit{Tachyonic Preheating in Plateau Inflation}'';
JCAP \textbf{12} (2021), 035;
[e-print: 2108.10767 [astro-ph.CO]].


\bibitem{multamaki022} T.~ Multamaki and I.~ Vilja;
"\textit{Limits on Q-ball size due to gravity}'';
Phys. Lett. B \textbf{542} (2002), 137-146;
doi: 10.1016/S0370-2693(02)02274-8;
[e-print: hep-ph/0205302].

\bibitem{tamaki11} T.~ Tamaki and N.~ Sakai;
"\textit{How does gravity save or kill Q-balls?}'';
Phys. Rev. D \textbf{83} (2011), 044027;
doi: 10.1103/PhysRevD.83.044027;
[e-print: 1105.2932 [gr-qc]].

\bibitem{rubio19} J. ~Rubio and E. ~Tomberg
"\textit{Preheating in Palatini Higgs inflation}";
JCAP, \textbf{04} (2019), 021;
doi: 10.1088/1475-7516/2019/04/021;
[e-print: 1902.10148 [hep-ph]].

\bibitem{heeck21} J. ~ Heeck; A.~ Rajaraman, R.~ Riley, and C.~ B.~ Verhaaren;
"\textit{Understanding Q-balls beyond the thin-wall limit}";
Phys. Rev. D, Volume 103, Issue 4 (9th February 2021);
doi:10.1103/physrevd.103.045008;
[e-print: 2009.08462 [hep-th]].


\bibitem{lozanov14} K.~ D.~ Lozanov and M.~ A.~ Amin;
"\textit{End of inflation, oscillons and matter-antimatter asymmetry}'';
Phys. Rev. D \textbf{90} (2014) 8, 083528;
doi: 10.1103/PhysRevD.90.083528;
[e-print: 1408.1811 [hep-ph]].



\bibitem{enqvist022} K.~ Enqvist, S.~ Kasuya and A.~ Mazumdar;
"\textit{Reheating as a Surface Effect}";
Phys. Rev. Lett. \textbf{89} (2002,) 091301;
doi: 10.1103/PhysRevLett.89.091301;
[e-print: hep-ph/0204270 [hep-ph]].

\bibitem{cohen86} A.~ G.~ Cohen et al:
"\textit{The Evaporation of Q-balls}'';
Nucl. Phys. B 272 (1986), 301-321;
doi: 10.1016/0550-3213(86)90004-0.

\bibitem{enqvist02} K.~ Enqvist, S.~ Kasuya and A.~ Mazumdar;
"\textit{Inflatonic solitons in running mass inflation}'';
Phys. Rev. D \textbf{66} (2002), 043505;
doi: 10.1103/PhysRevD.66.043505;
[e-print: hep-ph/0206272].


\bibitem{laine98} M.~ Laine and M.~ E.~ Shaposhnikov;
"\textit{Thermodynamics of Non-Topological Solitons}'';
Nucl. Phys. B \textbf{532} (1998), 376-404;
doi: 10.1016/S0550-3213(98)00474-X;
[e-print: hep-ph/9804237 [hep-ph]].

\bibitem{mazumdar02} A.~ Mazumdar
"\textit{Inflatonic Q-ball evaporation: A new paradigm for reheating the Universe}'';
Talk given at SUSY-02, DESY;
[e-print: hep-ph/0211233].

\bibitem{evap} I.~ Baldes, Q.~ Decant, D.~ C.~ Hooper, and L.~ Lopez-Honorez;
"\textit{Non-Cold Dark Matter from Primordial Black Hole Evaporation}'';
JCAP \textbf{08} (2020) 045;
doi: 10.1088/1475-7516/2020/08/045;
[e-print: 2004.14773 [astro-ph.CO]].

\bibitem{ew} M.~ D'Onofrio and K.~ Rummukainen;
"\textit{The Standard Model cross-over on the lattice}";
Phys. Rev. D \textbf{93} (2016) 2, 025003;
doi: 10.1103/PhysRevD.93.025003;
[e-print: 1508.07161 [hep-ph]].

\bibitem{hawkingbh} S.~ W.~ Hawking;
"\textit{Black Hole Explosions?}";
Nature \textbf{248} (1974), 30-31;
doi: 10.1038/248030a0.

\bibitem{barrowbh} J.~ D.~ Barrow, E.~ J.~ Copeland, E.~ W.~ Kolb and A.~ R.~ Liddle;
"\textit{Baryogenesis in extended inflation. II. Baryogenesis via primordial black holes}";
Phys. Rev. D \textbf{43} (1991), 984-994;
doi: 10.1103/PhysRevD.43.984.

\bibitem{majumdar} A.~ S.~ Majumdar, P.~ Das Gupta and R.~ P.~ Saxena;
"\textit{Baryogenesis from black hole evaporation}";
Int. J. Mod. Phys. D \textbf{4} (1995), 517-529;
doi: 10.1142/S0218271895000363.

\bibitem{bhb} Y.~ Nagatani;
"\textit{Black hole baryogenesis}";
e-print: hep-ph/9805455.

\bibitem{ewbhb} Y.~ Nagatani;
"\textit{Electroweak baryogenesis by black holes}";
 Phys. Rev. D \textbf{59} (1999), 041301;
doi: 10.1103/PhysRevD.59.041301;
[e-print: hep-ph/9811485].

\bibitem{dolgov} A.~ D.~ Dolgov, P.~ D.~ Naselsky and I.~ D.~ Novikov;
"\textit{Gravitational waves, baryogenesis, and dark matter from primordial black holes}";
e-print: astro-ph/0009407.

\bibitem{pbhb} D.~ Baumann, P.~ J.~ Steinhardt and N.~ Turok;
"\textit{Primordial Black Hole Baryogenesis}";
e-print: hep-th/0703250.

\bibitem{fujita} T.~ Fujita, M.~ Kawasaki, K.~ Harigaya and R.~ Matsuda;
"\textit{Baryon Asymmetry, Dark Matter, and Density Perturbation from PBH}";
Phys. Rev. D \textbf{89} (2014) 10, 103501;
doi: 10.1103/PhysRevD.89.103501;
[e-print: 1401.1909 [astro-ph.CO]].

\bibitem{hook} A.~ Hook;
"\textit{Baryogenesis from Hawking Radiation}";
Phys. Rev. D \textbf{90} (2014) 8, 083535;
doi: 10.1103/PhysRevD.90.083535;
[e-print: 1404.0113 [hep-ph]].



\bibitem{hooman} H.~ Davoudiasl, , R.~ Kitano, G.~ D.~ Kribs, H.~ Murayama and P.~ J.~ Steinhardt;
"\textit{Gravitational Baryogenesis}";
Phys. Rev. Lett. \textbf{93} (2004), 201301;
doi: 10.1103/PhysRevLett.93.201301;
[e-print: hep-ph/0403019].




\bibitem{inomata20} K.~ Inomata, M.~ Kawasaki, K.~ Mukaida, T.~ Terada, and T.~ T.~ Yanagida,
"\textit{Gravitational Wave Production right after a Primordial Black Hole Evaporation}'';
Phys. Rev. D \textbf{101} (2020) 12, 123533;
doi: 10.1103/PhysRevD.101.123533;
[e-print: 2003.10455 [astro-ph.CO]].

\bibitem{decigo1} B.~ S.~ Sathyaprakash and B.~ F.~ Schutz;
"\textit{Physics, Astrophysics and Cosmology with Gravitational Waves}'';
Living Rev. Rel. \textbf{12}, 2 (2009), 
[e-print:0903.0338 [gr-qc]].

\bibitem{decigo2} C.~ J.~ Moore, R.~ H.~ Cole, and C.~ P.~ L.~ Berry, 
"\textit{Gravitational-wave sensitivity curves}'';
Class. Quant. Grav. \textbf{32}, 015014 (2015), 
[e-print:1408.0740 [gr-qc]].

\bibitem{lisa1} H.~ Audley et al.
"\textit{Laser Interferometer Space Antenna}'';
(LISA), (2017);
[e-print:1702.00786 [astro-ph.IM]].

\bibitem{lisa2} N.~ Seto, S.~ Kawamura, and T.~ Nakamura, 
"\textit{Possibility of Direct Measurement of the Acceleration of the Universe Using 0.1 Hz Band Laser Interferometer Gravitational Wave Antenna in Space}'';
Phys. Rev. Lett. \textbf{87}, 221103 (2001), 
[e-print:astro-ph/0108011 [astro-ph]].

\bibitem{lisa3} K.~ Yagi and N.~ Seto, 
"\textit{Detector configuration of DECIGO/BBO and identification of cosmological neutron-star binaries}'';
Phys. Rev. D \textbf{83}, 044011 (2011), 
[Erratum:Phys. Rev.D95,no.10,109901(2017)], 
[e-print:1101.3940 [astro-ph.CO]].

\bibitem{kzpbh5} M.~M.~Flores and A.~Kusenko,
"\textit{Spins of primordial black holes formed in different cosmological scenarios}'';
Phys. Rev. D \textbf{104} (2021) no.6, 063008
doi:10.1103/PhysRevD.104.063008
[e-print:2106.03237 [astro-ph.CO]].

\bibitem{lennon} O.~ Lennon;
"\textit{Non-Canonical Q-balls}";
e-print: 2112.12547 [hep-ph].

\bibitem{usqballs} A.~ K.~ Lloyd-Stubbs and J.~ McDonald;
"\textit{Q-balls in Non-Minimally coupled Palatini inflation and their implications for cosmology}";
Phys. Rev. D \textbf{105} (2022) 10, 103532;
doi: 10.1103/PhysRevD.105.103532;
[e-print: 2112.09121 [hep-th]].

\bibitem{kusenko09} A.~ Kusenko, A.~ Mazumdar and T.~ Multamaki;
"\textit{Gravitational waves from the fragmentation of a supersymmetric condensate}'';
Phys. Rev. D \textbf{79} (2009), 124034;
doi: 10.1103/PhysRevD.79.124034;
[e-print: 0902.2197 [astro-ph.CO]].

\bibitem{chiba10} T.~ Chiba, K.~ Komada and M.~ Yamaguchi;
"\textit{Gravitational waves from Q-ball formation}'';
Phys. Rev. D \textbf{81} (2010), 083503;
doi: 10.1103/PhysRevD.81.083503;
[e-print: 0912.3585 [astro-ph.CO]].



\bibitem{gravfrag}  A.~Kusenko and A.~Mazumdar;
"\textit{Gravitational waves from fragmentation of a primordial scalar condensate into Q-balls}'';
Phys. Rev. Lett. \textbf{101}, 211301 (2008);
doi:10.1103/PhysRevLett.101.211301;
[e-print:0807.4554 [astro-ph]].

\bibitem{kzpbh1} E.~Cotner, A.~Kusenko and V.~Takhistov;
"\textit{Primordial Black Holes from Inflaton Fragmentation into Oscillons}'';
Phys. Rev. D \textbf{98} (2018) no.8, 083513;
doi:10.1103/PhysRevD.98.083513;
[e-print:1801.03321 [astro-ph.CO]].

\bibitem{kzpbh2} E.~Cotner and A.~Kusenko;
"\textit{Primordial black holes from supersymmetry in the early universe}'';
Phys. Rev. Lett. \textbf{119} (2017) no.3, 031103;
doi:10.1103/PhysRevLett.119.031103;
[e-print:1612.02529 [astro-ph.CO]].

\bibitem{kzpbh3} E.~Cotner and A.~Kusenko;
"\textit{Primordial black holes from scalar field evolution in the early universe}'';
Phys. Rev. D \textbf{96} (2017) no.10, 103002;
doi:10.1103/PhysRevD.96.103002;
[e-print:1706.09003 [astro-ph.CO]].

\bibitem{kzpbh4} E.~Cotner, A.~Kusenko, M.~Sasaki and V.~Takhistov,
"\textit{Analytic Description of Primordial Black Hole Formation from Scalar Field Fragmentation}'';
JCAP \textbf{10} (2019), 077
doi:10.1088/1475-7516/2019/10/077
[e-print:1907.10613 [astro-ph.CO]].


\bibitem{kohri18} K.~ Kohri and T.~ Terada;
"\textit{Semianalytic Calculation of Gravitational Wave Spectrum Nonlinearly Induced from Primordial Curvature Perturbations}'';
Phys. Rev. D \textbf{97} (2018) 12, 123532;
doi: 10.1103/PhysRevD.97.123532;
[e-print: 1804.08577 [gr-qc]].

\bibitem{inomata19} K.~ Inomata et. al;
"\textit{Enhancement of Gravitational Waves Induced by Scalar Perturbations due to a Sudden Transition from an Early Matter Era to the Radiation Era}'';
Phys. Rev. D \textbf{100} (2019) 4, 043532;
doi: 10.1103/PhysRevD.100.043532;
[e-print: 1904.12879 [astro-ph.CO]].


\bibitem{white21}  G.~White, L.~Pearce, D.~Vagie and A.~Kusenko;
"\textit{Detectable Gravitational Wave Signals from Affleck-Dine Baryogenesis}'';
Phys. Rev. Lett. \textbf{127}, 18 (2021);
doi:10.1103/PhysRevLett.127.181601;
[e-print: 2105.11655 [hep-ph]].




\end{thebibliography}
\end{document}